\documentclass[a4paper, 12pt, english, twoside]{report}
\usepackage[english]{babel}
\usepackage{graphicx,rotating,epsfig}
\usepackage{fancyhdr}
\fancypagestyle{plain}{%
\fancyhf{}
\fancyfoot[LE,RO]{\thepage}

}
\begin{document}

\begin{center}
\begin{flushright}
DESY-THESIS-2005-042\\
VIN-T-2005-001\\
21st December 2005\\
\end{flushright}

\thispagestyle{empty}

\vspace*{1cm}

{\huge \bf Studies of Gauge Boson Production with a\\

$\gamma\gamma$-collider at TESLA\\}

\vspace{1cm}

{\Large D I S S E R T A T I O N\\}

zur Erlangung des akademischen Grades\\

doctor rerum naturalium\\

{\bf (Dr. rer. nat.)\\

\vspace*{1cm}

im Fach Physik}\\

\vspace{1cm}

{\bf eingereicht an der}\\

\noindent

{\bf    Mathematisch--Naturwissenschaftliche Fakult\"at I\\
der {Humboldt-Universit\"at zu Berlin}}

von\\

{\Large Dipl.\,Phys.\,Jadranka Sekaric\\}


\end{center}

\vspace{1cm}

\begin{flushleft}

Pr\"asident der Humboldt-Universit\"at zu Berlin\\
Prof. Dr. J\"urgen Mlynek\\


Dekan der Mathematisch--Naturwissenschaftliche Fakult\"at I\\
Prof. T. Buckhout, PhD\\

\vspace{1cm}

\begin{tabular}{l l}

 Gutachter: & 1.~Prof.\,Dr. Hermann Kolanoski\\

 & 2.~Prof.\,Dr. Nikolaj Pavel\\

 & 3.~Dr.\,Klaus M\"onig\\

\end{tabular}

\vspace{1cm}

Tag der m\"undlichen Pr\"ufung:\hspace{7cm} 2 August 2005\\

\end{flushleft}

\thispagestyle{empty}



\setcounter{page}{2}
\section*{Abstract}
In absence of the Standard Model Higgs boson the interaction among the gauge
bosons becomes strong at high energies ($\sim$ 1TeV) and influences the
couplings between them. Each trilinear and quartic gauge boson vertex is
characterized by a set of couplings which are expected to
deviate from their Standard Model values already at energies lower than the
energy scale of the New Physics. The precise measurement of gauge boson
couplings can provide clues to the mechanism of the electroweak symmetry
breaking and their anomalous values can be a sign of a New Physics effect
beyond the Standard Model. Estimation of the precision with which we can
measure the deviations of the charged trilinear gauge boson couplings (TGCs) with a photon collider at TESLA is one of the two topics covered by this theses. Deviations are denoted as $\Delta\kappa_\gamma$ and $\Delta\lambda_\gamma$ (anomalous TGCs) and represent the strength of the gauge boson couplings at the $WW\gamma$ vertices.
The single $W$ boson production in $\gamma e$
collisions is studied via $\gamma e^{-} \rightarrow W^{-} \nu_{e}$ in two
different operating ${\gamma}{e}$ modes - real and parasitic. While the first
assumes the electron and photon beam in the collision, the second one is
considered as a background to $\gamma\gamma$ collisions: the interaction
between photons (from each side) and unconverted electrons. The $W$ boson pair
production is studied in ${\gamma\gamma}$ collisions via $\gamma\gamma
\rightarrow W^{+}W^{-}$ in two different initial polarization states,
$J_{Z}=0$ and $|J_{Z}|=2$. While the first polarization state assumes the
colliding photons have the same helicities, the second one assumes opposite
photon helicities. The $W$ bosons from both channels are reconstructed from
hadronic final states, as two jets in ${\gamma}{e}$ collisions and as four
jets in $\gamma\gamma$ collisions. The study includes the influence of
low-energy events $\gamma\gamma\rightarrow hadrons$ (pileup) on the signal and
estimated background in both channels.
\par
The error estimation of the measurement of the TGCs is performed using a
binned $\chi^{2}$ and binned maximum Likelihood fit of reweighted event
distributions of variables which are sensitive to the anomalous couplings. It
was found that the error of the $\kappa_\gamma$ and $\lambda_\gamma$
measurement at a photon collider at TESLA is of ${\cal O}(10^{-3}-10^{-4})$,
depending on the $\gamma\gamma$/$\gamma e$ channel, mode and the coupling
under the consideration. Compared to the measurements at Tevatron and LEP
experiments this is about one to two orders of magnitude higher accuracy.
The influence of some systematic errors such those from the $W$ mass
measurement, uncertainties due to the beam energy and polarization and the
effects of the background is also estimated.
\par
Another part of this theses covers the optimization of the
$\gamma\gamma$-detector in the forward region. Detailed 'incoherent
particle-particle' and 'coherent particle-beam' interactions have been
simulated in order to estimate the low-energy background contribution to the
tracking devices. The $\gamma\gamma$-detector is optimized in the way to
minimize the direct and the backscattered background in its forward region.
For that purpose, the beam pipes in the region between the interaction
point and the electromagnetic calorimeter are surrounded by a tungsten mask.
The space around the beam pipes in the region where the electromagnetic and
hadronic calorimeters are positioned is filled by graphite and tungsten, while
the outgoing electron beam pipes are made of graphite. With this design of the
forward region the estimated background that enters into the main tracking system, the time projection chamber (TPC) and the vertex detector (VTX) is brought to the level that provides small occupancies in TPC and VTX.  
\newpage 
\setcounter{page}{3}
\thispagestyle{empty}


%
%
\section*{Zusammenfassung}

Existiert das Standardmodell Higgs Boson nicht, wird die Wechselwirkung der
Eichbosonen bei hohen Energien ($\sim$ 1TeV) stark und beeinflusst die
Kopplungen der Eichbosonen untereinander. Jeder Drei-Eichboson und
Vier-Eichboson Vertex wird durch einen Satz von Kopplungen
charakterisiert, von denen erwarted wird, schon bei Energien unterhalb der
Energieskala f\"ur neue Physik von den Standardmodell-Werten abzuweichen. Die
pr\"azise Messung von Eichbosonkopplungen gibt Anhaltspunkte zum Mechanismus
der elektroschwachen Symmetriebrechung und deren anomale Werte k\"onnen ein
Zeichen f\"ur neue Physik jenseits des Standardmodells sein. Die Absch\"atzung
der Genauigkeit, mit der wir die Abweichung der Drei-Eichboson Kopplungen
(trilinear gauge couplings - TGC's), bezeichnet als $\Delta\kappa_{\gamma}$ und
$\Delta\lambda_{\gamma}$ (anomale TGC's), an $WW\gamma$ Vertices am
Photon-Collider bei TESLA messen k\"onnen ist eines der Themen, das in dieser
Arbeit behandelt wird. Die Einzel-$W$-Boson Produktion in $\gamma e$
Kollisionen \"uber den Prozess $\gamma e^{-} \rightarrow W^{-}\nu_{e}$ wurde
in zwei $\gamma e$ Operationsmodi studiert - real und parasit\"ar. W\"ahrend im
realen Modus ein Elektronen- und ein Photonenstrahl kollidieren, ist der
parasitische Modus ein Untergrund bei der Kollision zweier Photonenstrahlen:
Wechselwirkung zwischen Photonen von beiden Seiten und nicht konvertierten
Elektronen. Die W-Boson Paarproduktion in $\gamma\gamma$ Kollisionen \"uber
$\gamma\gamma \rightarrow W^{+}W^{-}$ wurde f\"ur zwei
Polarisationsanfangszust\"ande betrachtet, $J_{z} = 0$ and $|J_{z}| = 2$. Im
ersten Polarisationszustand haben die kollidierenden Photonen dieselbe
Helizit\"at, im zweiten die entgegengesetzte. Die $W$-Bosonen aus beiden
Kan\"alen werden aus hadronischen Endzust\"anden rekonstruiert, zwei Jets in
$\gamma e$ Kollisionen und vier Jets in $\gamma\gamma$ Kollisionen. Weiterhin
wurde der Einfluss niederenergetischer $\gamma\gamma \rightarrow Hadronen$
Ereignisse (``pileup'') auf das Signal und den erwartet Untergrund in beiden
Kan\"alen studiert.
\par
Die Fehlerabsch\"atzung f\"ur die Messung der TGC's erfolgte mittels "binned
$\chi^{2}$"- und "binned maximum Likelihood"- Anpassung von umgewichteten
Ereignisverteilungen von Variablen, die sensitiv auf anomale Kopplungen sind.
Es stellte sich heraus, dass die Sensitivit\"at f\"ur $\kappa_{\gamma}$- und
$\lambda_{\gamma}$-Messungen am Photon-Collider bei TESLA in der
Gr\"o{\ss}enordnung $10^{-3}-10^{-4}$ ist, abh\"angig vom betrachteten
$\gamma\gamma/\gamma e$-Kanal, dem Operationsmodus und der betrachteten
Kopplung. Verglichen mit den Messungen am Tevatron und bei LEP-Experimenten
ist dies eine um ein bis zwei Gr\"o{\ss}enordnungen h\"ohere Genauigkeit.
Einfl\"usse einiger systematischer Fehler wie ein Effekt von der W-Boson
Massenbestimmung, Unbestimmtheiten von Strahlenergie und -polarisation und der
Effekte des Untergrundes wurden au{\ss}erdem abgesch\"atzt.
\par
Ein weiterer Teil dieser Arbeit behandelt die Optimierung des
$\gamma\gamma$-Detektors in dessen Vorw\"artsregion. Detaillierte
``inkoh\"arente Teilchen-Teilchen''- und ``koh\"arente
Teilchen-Strahl''-Wechselwirkungen wurden simuliert, um den Beitrag zum
niederenergetischen Untergrund in den Spurkammern abzusch\"atzen. Der
$\gamma\gamma$-Detektor wurde optimiert mit dem Ziel, den direkten und
r\"uckgestreuten Untergrund in der Vorw\"artsregion des Detektors zu
minimieren. Zu diesem Zweck sind die Laser-Strahlrohre in dem Bereich zwischen
Wechselwirkungspunkt und elektromagnetischen Kalorimeter von einer
Wolfram-Abschirmung umgeben. Der Zwischenraum au{\ss}erhalb der Strahlrohre im
Bereich des elektromagnetischen und des hadronischen Kalorimeters ist mit
Graphit und Wolfram gef\"ullt. Die herausf\"uhrenden Elektronenstrahlrohre
bestehen ganz aus Graphit. Mit diesem Design der Vorw\"artsregion l\"asst sich
der Untergrund in der ``Time Projection Chamber'' und im Vertex-Detektor auf ein Ma{\ss} reduzieren,
das eine effiziente Signalauslese erm\"oglicht.

\thispagestyle{empty}
\noindent
\begin{flushleft}
{\bf{\Huge Acknowledgments}} \\
\end{flushleft}
\vspace{2.in}
\par
\textit{I had a pleasure to work within the TESLA group in DESY-Zeuthen among the people who provided me a very pleasant atmosphere and helpful assistance during my stay here.}
\textit{First of all, I want to thank my advisor Klaus M\"onig for his continuous support and guidance trough the period of my theses work-his personality is truly an inspiration to me.}
\par
\textit{Also, I would like to thank Wolfgang Kilian and Sven-Olaf Moch for many useful advises, to Ralph and Ekatarina for their technical support, and to Wolfgang Lohmann and Achim Stahl for their real interest in teaching the students. My special thank to Predrag for his unselfish support during all my critical phases.}
\par
\textit{Besides, too many people have impacted my life to mention them all. They have contributed not only to my successful completion of this work but also to the enjoyable experience the past years have been. I am grateful to all of them.}
\par
\textit{I would not be here without support from my family and friends, in particular my mother and my sister. They not only stand in behind me in any of my decisions but have always encouraged me to follow my dreams, whether those take me far away from them.}
\vspace{.5in}
\begin{center}
Also I am very thankfull to Deutsches Elektronen-Synchrotron (DESY) in Zeuthen for the financial support they provided me during my PhD studies and to the Institute for Nuclear Sciences 'VINCA' in Belgrade for granting me with a privilege to work in such scientific environment as it is DESY.
\end{center}

\pagenumbering{arabic}
\setcounter{page}{6}
\tableofcontents
\chapter{Introduction}
The common efforts on the $e^{+}e^{-}$ linear collider project during the last decade from the North American physicist (Next Linear Collider, NLC Collaboration) and Japanese physicists (Global Linear Collider, GLC Collaboration) from the one side \cite{nlc_jlc} and European physicists from the other side (Teraelectronvolt Energy Superconducting Linear Accelerator, TESLA Collaboration) \cite{tesla} converged to the final solution called the International Linear Collider (ILC) in 2004 \cite{ilc}. The ILC is foreseen to operate at high center-of-mass energies up to $\sqrt{s_{e^{\pm}e^{-}}}=500$ GeV for a first stage and then be upgraded up to $1-1.5$ TeV. The polarized beams and the high luminosities achievable at ILC provide a high event statistics enabling a very precise measurement of diverse physics parameters of the Standard Model or discovery of New Physics effects. Particularly, the ILC is expected to be highly sensitive to the measurement of the trilinear gauge boson coupling parameters (TGCs). The order of magnitude with which the TGCs can be measured at ILC is $\sim 10^{-3}-10^{-4}$. The TGC studies and measurements that have been done up to now concern the sensitivities to the TGCs in $e^{+}e^{-}$, $p\bar{p}$ and $pp$ collisions. The estimation of sensitivities to the measurement of the TGCs in $\gamma e$ and $\gamma\gamma$ collisions at a photon collider at TESLA are the subject of this theses.
\section{Technical aspects}
%
\begin{figure}[htb]
\begin{center}
\epsfxsize=6.5in
\epsfysize=1.5in
\epsfbox{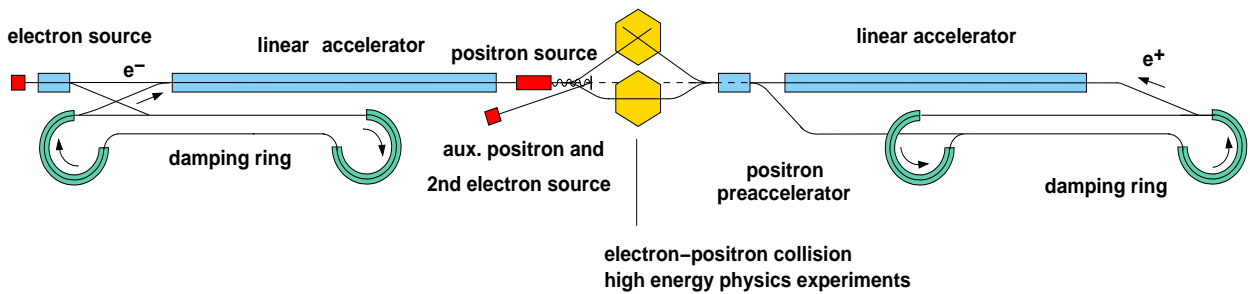}
\caption[bla]{Sketch of the overall layout of TESLA.}
\label{fig:tesla}
\end{center}
\end{figure}
\par
The sketch of the TESLA $e^{+}e^{-}$-collider is shown in Fig.~\ref{fig:tesla} with the overall length of 33 km. It consists of two linacs for acceleration of electron and/or positron beams, two damping rings to ensure a small beam emittance and the two interaction regions - one forseen for $e^{+}e^{-}$ collisions and another one with about 34 mrad crossing angle suitable for $\gamma e$ and $\gamma\gamma$ collisions. The TESLA linacs use 9-cell superconducting niobium cavities cooled by superfluid Helium to the temperature of $T=2$ K operating at a frequency of 1.3 GHz (so called '\textit{cold technology}') \cite{tdr2} instead of the so called '\textit{warm technology}' proposed by the NLC and SLC\footnote{Stanford Linear Collider.} Collaborations that uses copper cavities and operate at a frequency of 11.4 GHz. A center-of-mass energy of 500 GeV with a cold technology is reachable if the accelerating gradient is at least 23.4 MV/m and this value is already reached ($\approx$ 35 MV/m \cite{upgrade}) giving the possibility to upgrade TESLA to higher center-of-mass energies of 800 GeV. For the realization of the $\gamma e$ and $\gamma\gamma$ colliders two high energy and highly polarized electron beams will be used. The Compton backscattering of the circularly polarized laser photons off the high energy electrons results in the beam of high energy photons. In the collision of a high energy photon with a high energy electron or of the two high energy photons the $\gamma e$ and $\gamma\gamma$ center-of-mass energies of $\sqrt{s_{\gamma e}}\approx 0.9\sqrt{s_{e^{-}e^{-}}}$ and $\sqrt{s_{\gamma\gamma}}\approx 0.8\sqrt{s_{e^{-}e^{-}}}$ are reachable. The $\gamma e $ and $\gamma\gamma$ luminosities ($L_{\gamma e,\gamma\gamma}$) depend on $e^{-}\rightarrow \gamma$ conversion efficiency $k$ according to the relation $L_{\gamma e,\gamma\gamma}\approx k^{2}L^{geom}_{e^{-}e^{-}}\approx k^{2}\frac{1}{2}L_{e^{+}e^{-}}$ (see below) \cite{valery}. The maximal $k$ value is $\approx 0.63$ but taking the contribution from the high energy peak only ($z=E_{\gamma}/E_{e}\geq 0.8z_{max}\footnote{$z_{max}$ denotes the position of the maximum peak in the photon energy spectrum.}$), $k\approx 0.3$ gives the approximative relation between the luminosity of the $\gamma e$/$\gamma\gamma$ collisions and the $e^{+}e^{-}$ luminosity as $L_{\gamma e,\gamma\gamma}\approx 0.045L_{e^{+}e^{-}}$. The luminosity of the $e^{+}e^{-}$ collisions $L_{e^{+}e^{-}}$ is defined as:
\begin{equation}
L_{e^{+}e^{-}}\approx \frac{n_{b}N_{e}^{2}f_{r}}{4\pi\sigma_{x}\sigma_{y}}\times H_{D}
\label{eq:luminosity}
\end{equation}
and its high value at TESLA can be realized if the repetition rate $f_{r}$ is high enough while having a small horizontal $\sigma_{x}$ and vertical $\sigma_{y}$ beam sizes at the interaction point i.e. having small beam emittances at the interaction point, with a number of bunches per train $n_{b}$, number of electrons per bunch $N_{e}$ and a disruption enhancement factor $H_{D}$ which is typically $\approx$ 2 for $e^{+}e^{-}$ collisions. At $\gamma\gamma$-collider, $H_{D}=$ 1 due to the neutral colliding beams. Using a smaller horizontal beam emittance $\epsilon_{x}$ and a smaller $\beta_{x}$-function at the interaction point i.e. a smaller bunch size than at $e^{+}e^{-}$ collisions it is possible to obtain the relation $L_{\gamma e,\gamma\gamma}\approx\frac{1}{3}L_{e^{+}e^{-}}$ at a high energy peak. The needed beam parameters that ensure the previous luminosities, allowed by the superconducting technology are shown in Table \ref{tab:parameters} - $\sigma_{x}$ takes a value of 88 nm at a photon collider instead of 553 nm at an $e^{+}e^{-}$-collider and $\sigma_{y}$ takes a value of 4.3 nm instead of 5 nm at an $e^{+}e^{-}$-collider.
\begin{table}[htb]
\begin{center}
\begin{tabular}{|c||c|c|} \hline
Parameter & & TESLA-500 GeV, $\gamma\gamma$ \\ \hline\hline
Repetition rate & $f_{r}$[Hz] & 5 \\
N$^{o}$ of bunches per train & $n_{b}$ & 2820 \\
Bunch spacing & $\Delta t_{b}$[ns] & 337 \\
N$^{o}$ of $e^{-}$ per bunch & $N_{e}$[10$^{10}$] & 2 \\
Beta function & $\beta_{x}/\beta_{y}$[mm] & 1.5/0.3 \\ 
Emittance & $[\gamma\epsilon_{x}/\gamma\epsilon_{y}]/10^{-6}$[m$\cdot$rad] & 2.5/0.03 \\
Beam size  & $\sigma_{x}/\sigma_{y}$[nm] & 88/4.3 \\ 
Bunch length & $\sigma_{z}$[mm] & 0.3 \\ \hline
${\gamma e}$ luminosity ($z>0.8z_{max}$) & $L_{\gamma e}$[10$^{34}$cm$^{-2}$s$^{-1}$] & 0.94 \\ \hline
${\gamma\gamma}$ luminosity ($z>0.8z_{max}$) & $L_{\gamma\gamma}$[10$^{34}$cm$^{-2}$s$^{-1}$] & 1.1 \\ \hline
\end{tabular}
\end{center}
\caption{Parameters of the $\gamma\gamma$-collider based on TESLA. $z_{max}$ denotes the position of the maximum peak in the $\gamma e/\gamma\gamma$ energy spectrum.}
\label{tab:parameters}
\end{table}
\par
Such high luminosities and high $\gamma e^{-}\rightarrow W^{-}\nu_{e}$ and $\gamma\gamma\rightarrow W^{+}W^{-}$ production cross-sections make a photon collider an ideal $W$ boson factory and thus, an ideal place for testing the electroweak sector of the Standard Model \cite{sm}. Due to the non-Abelian nature of the gauge group which describes the electroweak interactions, the Standard Model predicts that the electroweak gauge bosons interact between themselves, giving rise to vertices with three or four gauge bosons. Each vertex is described by a set of dimensionless couplings, denoted as TGCs or QGCs (triple or quartic gauge boson couplings, respectively). Their strengths, predicted by the Standard Model applying the principle of gauge symmetry, can be directly measured in experiments. Their measurement can establish a stringent test of the electroweak theory but also probes possible extensions in the bosonic sector.
\par
Having two important conditions fullfiled, high center-of-mass energy and high luminosities, a precision of TGCs measurements that can be achieved at a future linear collider is higher, compared to the experiments at LEP and Tevatron as it is presented in this theses. The estimation of the precision of the measurement of $C-$ and $P-$ conserving TGCs, $\kappa_{\gamma}$ and $\lambda_{\gamma}$, describing the $\gamma WW$ vertices in the single $W$ boson production in $\gamma e^{-}\rightarrow W^{-}\nu_{e}$ and in the $W$ boson pair production in $\gamma\gamma\rightarrow W^{+}W^{-}$ at $\sqrt{s_{e^{-}e^{-}}}=500$ GeV using simulated events is the subject of this theses. 
\par
Although the Standard Model is very successful in describing the fundamental particles and their interactions, the importance of precise measurements of TGCs comes from the fact that the Standard Model as it is today, is still incomplete and needs new discoveries or some extensions to explain the origin of the mass. The Higgs boson \cite{higgsboson} of the Standard Model explains nicely how the heavy boson and fermion masses are created trough the mechanism of the electroweak symmetry breaking. The unitarity violated by the gauge boson scattering amplitudes is restored again introducing the Higgs boson in the Standard Model. But if the Higgs boson would not be discovered at the future high energy experiments (or it is too heavy), some mechanism responsible for the electroweak symmetry breaking and the mass generation should exist to replace the existing Standard Model prediction. One possible solution beyond the Standard Model is the mechanism of the strong electroweak symmetry breaking \cite{strong} where the symmetry is dynamically broken through the elastic scattering of the gauge bosons at high energies [$>1$ TeV]. In this scenario, the Standard Model represents the low energy limit of a larger theory. The unitarity is restored and the New Physics effects should appear already below the energy scale of the New Physics which is approximated to be about 3 TeV i.e. before the new resonances appear in the spectrum. It is expected that these effects are reflected in the values of the trilinear gauge boson couplings leading to their deviations $\Delta\kappa_{\gamma}$ and $\Delta\lambda_{\gamma}$ of ${\cal O}(\sim 10^{-3})$ from the Standard Model values. Since the deviations decrease as the energy scale of the New Physics increases, their observation needs a more precise measurements than those at Large Electron-Positron collider (LEP) and Tevatron.
\par
On the other hand, the TGCs may receive contributions from loops which cause deviations from their tree-level values. These corrections depend on the New Physics scenario behind \cite{mssm,gigaz} that might not be visible directly, but produces measurable effects contained in the radiative corrections. The magnitude of these corrections is of ${\cal O}(10^{-2}-10^{-3})$ and thus, a better precision of their measurement is desirable.
\par
Due to high $\gamma\gamma/\gamma e$ luminosities and high $W$ boson production cross-sections in $\gamma e^{-}\rightarrow W^{-}\nu_{e}$ and $\gamma\gamma\rightarrow W^{+}W^{-}$, a photon collider at TESLA gives a possibility to reach a higher accuracy of TGC measurements than of ${\cal O}(10^{-3})$. This is one to two orders of magnitude higher accuracy than at LEP and Tevatron. For the TGCs estimations the hadronic decay channels of the $W$ boson are described by several kinematical variables which are sensitive to the anomalous TGCs ($\kappa_{\gamma}\neq 1$ $\Rightarrow$ $\Delta\kappa_{\gamma}\neq 0$ and $\lambda_{\gamma}\neq 0$ $\Rightarrow$ $\Delta\lambda_{\gamma}\neq 0$ ). These are mainly the different angular distributions of the $W$ boson that contain the information about the TGCs. The multi-dimensional Standard Model angular distributions reweighted by the function dependent on the anomalous TGCs and fitted to the Standard Model prediction give the information on the achievable sensitivities to the corresponding trilinear gauge coupling.
\par
For the precise measurement of the TGCs or of any other physics parameter, the reconstructed variables relevant for their estimation should be measured with the highest possible precision. That strongly depends on the detector performances i.e. on the occupancy of the sub-detectors, and on the read-out efficiency. The low occupancy and fast read-out of the sub-detectors are desirable for an efficient event reconstruction avoiding the overlap of the physics events from more bunches. If in addition, the background events produced in the ``parasitic'' interactions at the interaction region are added, the occupancy of a sub-detector increases if the read-out is not fast enough. 
\par
The ``parasitic'' interactions are the ``beam-beam'' induced interactions, mainly $\gamma\gamma,\gamma e$ and $e^{-}e^{-}$ collisions resulting in several orders of magnitude higher low-energy background ($e^{+}e^{-}$ pairs, electrons and photons) than the interesting physics events, disabling the efficient data taking. High occupancy, increased by background, results with a larger error on the track reconstruction and thus, on the physics analysis. If the read-out speed is restricted by a chosen technology, the occupancy still can be decreased at the acceptable level. At a photon collider the ``beam-beam'' effects are somewhat different than at an $e^{+}e^{-}$-collider due to the different interactions at the two interaction regions. Thus, the low-energy background at a photon collider is larger than at a $e^{+}e^{-}$-collider. To prevent the entering of low-energy particles into the tracker devices, i.e. to reduce the occupancy of trackers, the redesign of the forward $\gamma\gamma$-detector region is done in the way that the background contribution to the time projection chamber and the vertex detector is brought to the level of the $e^{+}e^{-}$-collider. Since the proposed $\gamma\gamma$-detector matches in many points with those proposed for the TESLA $e^{+}e^{-}$-collider, this means that the amount of the received background is acceptable after the new design is imposed, resulting with an occupancy of $< 1\%$ at each of the two previously mentioned tracking sub-detectors.
\section{Outline of the theses}
The theses starts with a theoretical introduction of the gauge boson couplings and the electroweak interactions in Chapter 2 introducing the two possible scenarios for the electroweak symmetry breaking (EWSB) mechanism: within the Standard Model introducing the Higgs boson and beyond the Standard Model introducing the strong interaction at TeV energy scales. The two different parameterizations of the TGCs are presented depending on the EWSB scenario. The recent results from the direct measurements of the trilinear gauge couplings at LEP and Tevatron, used measurement techniques and analyzed final states are reviewed in Chapter 3. The Chapter 4 gives a description of the photon collider concept, explaining the Compton backscattering process as well as the main characteristics of the obtainable photon spectra: the final photon polarization and the photon energies. The spectral luminosities and their experimental measurements are also described in this Chapter. Description of the $\gamma\gamma$ detector together with the detailed detector study that concerns the optimization and design of the forward region in order to decrease the background occupancy of the tracking devices is contained in Chapter 5. The theory of the single $W$ boson and the $W$ boson pair production in $\gamma e$ and $\gamma\gamma$ collisions and the contribution from the radiative corrections is discussed in Chapter 6. In order to estimate the sensitivities to the measurement of the TGCs the separation of the signal events $\gamma e^{-}\rightarrow W^{-}\nu_{e}$ and $\gamma\gamma\rightarrow W^{+}W^{-}$ from the different background events is presented in Chapter 7. Depending on the polarization of the initial electron and photon beam the following cases are considered: the initial state $|J_{Z}|=3/2$ for the $\gamma e$ collisions in the two different $\gamma e$ modes, real and parasitic, and the $J_{Z}=0$ and the $|J_{Z}|=2$ state for the $\gamma\gamma$ collisions. For the analysis only the hadronic decay channels of the $W$ boson with two (in the single $W$ boson production) and four (in the $W$ boson pair production) jets in the final state are considered. The influence of the low energy $\gamma\gamma\rightarrow hadrons$ events (pileup) is taken into account too, as an unavoidable background contribution at a photon collider. The discussion of systematic errors arising from the backgrounds, beam energy and beam polarization uncertainties is also given in Chapter 7. Finally, the results presented in this thesis are summarized in Chapter 8.

\chapter{Theoretical Context}
\section{Gauge bosons in the Standard Model}
To the best of our present knowledge, all the elementary building blocks of matter in the Standard Model are fermions, i.e. particles with spin 1/2. These elementary fermions, divided into leptons and quarks, are grouped into three families. Each individual family is principally characterized by the masses of the particles it contains. The first family consists of the electron (\textit{$e^{-}$}) and its neutrino (\textit{$\nu_{e}$}) and the up (\textit{u}) and down (\textit{d}) quark. The second family consists of the muon (\textit{$\mu^{-}$}) and its neutrino (\textit{$\nu_{\mu}$}) and the charm (\textit{c}) and strange (\textit{s}) quark. The third family consists of the tau (\textit{$\tau^{-}$}) and its neutrino (\textit{$\nu_{\tau}$}) and the top (\textit{t}) and bottom (\textit{b}) quark. To each particle corresponds an antiparticle with a same mass, spin, isospin and eigenparity as a particle. It differs from the particle in the sign of its electric charge and in the signs of all its other additive quantum numbers. All particles are grouped into singlets or doublets of the weak isospin $\vec{T_{w}}$. Left-handed particles are grouped into doublets with isospin $\vec{T_{w}}=1/2$ while right-handed particles are grouped into singlets with isospin $\vec{T_{w}}=0$. The grouping into the three families, shown in Table \ref{tab:particles}, reflects the behavior of the fermions in the four known interactions, namely the \textit{strong}, the \textit{electromagnetic}, the \textit{weak} and the \textit{gravitational} interactions, while the electromagnetic and weak interactions are unified into the \textit{electroweak} interaction.
\begin{table}[h]
\begin{center}
\begin{tabular}{|c|c||c|c||c|c|} \hline
 \multicolumn{2}{|c||}{} & \multicolumn{2}{|c||}{\textbf{Quarks}} & \multicolumn{2}{|c|}{\textbf{Leptons}} \\ \hline\hline
 \textbf{Family} & $\begin{array}{c}
            1 \\
            2 \\
            3 \\
          \end{array}$ 
        & $\begin{array}{c}
            u \\
            c \\
            t \\
          \end{array}$ 
        & $\begin{array}{c}
            d \\
            s \\
            b \\
          \end{array}$ 
        & $\begin{array}{c}
            e \\
            \mu \\
            \tau \\
          \end{array}$ 
        & $\begin{array}{c}
            \nu_{e} \\
            \nu_{\mu} \\
            \nu_{\tau} \\
          \end{array}$ \\ \hline\hline
 \multicolumn{2}{|c||}{\textbf{Strong Interaction}} & {\textit{yes}} & {\textit{yes}} & {\textit{no}} & {\textit{no}} \\ \hline
 \multicolumn{2}{|c||}{\textbf{Strong Interaction}} & \multicolumn{2}{|c||}{\textit{color triplets}} & \multicolumn{2}{|c|}{\textit{color singlets}} \\ \hline\hline
 \multicolumn{2}{|c||}{\textbf{Electromagnetic Int.}} & {\textit{yes}} & {\textit{yes}} & {\textit{yes}} & {\textit{no}} \\ \hline
 \multicolumn{2}{|c||}{\textbf{Electromagnetic Int.}} & {\textit{Q=2/3}} & {\textit{Q=-1/3}} & {\textit{Q=-1}} & {\textit{Q=0}} \\ \hline\hline
 \multicolumn{2}{|c||}{\textbf{Weak Interaction}} & {\textit{yes}} & {\textit{yes}} & {\textit{yes}} & {\textit{yes}} \\ \hline
 \multicolumn{2}{|c||}{\textbf{Weak Interaction}} & \multicolumn{4}{|c|} {$\psi_{L}$ doublets; $\psi_{R}$ singlets} \\ \hline
 \multicolumn{2}{|c||}{\textbf{Weak Isospin $[\vec{T_{w}}]$}} & \multicolumn{4}{|c|} {$\vec{T}_{w}({\psi_{L}})=1/2$; $\vec{T}_{w}({\psi_{R}})=0$} \\ \hline\hline
 \multicolumn{2}{|c||}{\textbf{Gravitation}} & {\textit{yes}} & {\textit{yes}} & {\textit{yes}} & {\textit{yes}} \\ \hline
\end{tabular}
\end{center}
\caption{Elementary particles and their interactions.}
\label{tab:particles}
\end{table}
All interactions are based on exchanging field quanta called \textit{bosons}, and their structures are obtained in gauge theories from symmetry principles. To each interaction corresponds an underlying symmetry group: $SU(3)_{c}$ to the strong interaction and $SU(2)_{L}\times U(1)_{Y}$ to the electromagnetic and weak interaction. Particles, arranged in multiplets with respect to a symmetry group, interact between themselves in the way that they can be transformed by appropriate interaction into any other member of the same multiplet. Every fermion is a singlet with respect to the electromagnetic interaction, left-handed fermions are grouped into doublets while right-handed fermions are grouped into singlets, with respect to the weak interaction. Quarks are grouped into triplets while the leptons are grouped into singlets with respect to the strong interaction.
\par
\begin{figure}[htb]
\begin{center}
\epsfxsize=2.5in
\epsfysize=1.25in
\epsfbox{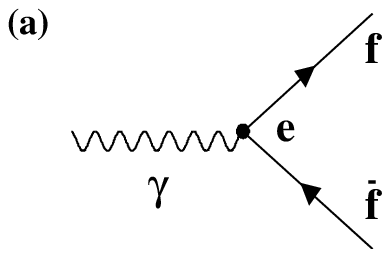}
\epsfxsize=2.5in
\epsfysize=1.25in
\epsfbox{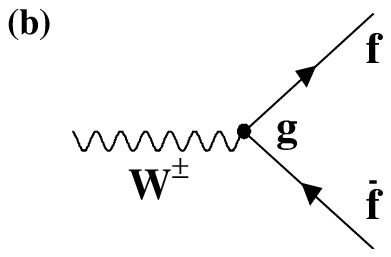}
\epsfxsize=2.5in
\epsfysize=1.25in
\epsfbox{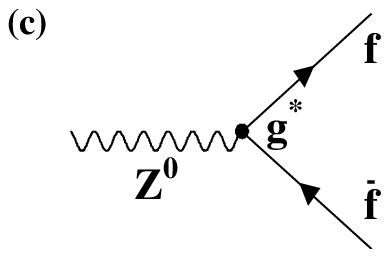}
\epsfxsize=2.5in
\epsfysize=1.25in
\epsfbox{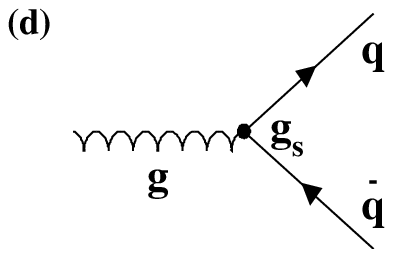}
\caption[bla]{The gauge boson - fermion vertices in the Standard Model with signed strengths for the electroweak interactions ($a$,$b$,$c$) and for the strong interaction ($d$). ($a$) The coupling $e$ is equal to the electric charge if $f$ are the charged fermions. ($b$) $g=e/\sin\theta_{w}$ for $f$ being the components of the same multiplet. ($c$) $g^{*}=(gg^{'}/e)[T^{f}_{3}-Q_{f}\sin^{2}\theta_{w}]$ where $T^{f}_{3}$ is a third component of the $\vec{T}_{w}$ and $Q$ is an electric charge of the fermion defined in Table \ref{tab:particles}. For $f = e_{L},\mu_{L},\tau_{L},d_{L},s_{L},b_{L} \rightarrow T^{f}_{3}=-1/2$. For $f = \nu_{eL},\nu_{\mu L},\nu_{\tau L},u_{L},c_{L},t_{L} \rightarrow T^{f}_{3}=+1/2$. For $f=f_{R} \rightarrow T^{f}_{3}=0$. ($f_{L}\equiv\bar{f}_{R}$).}
\label{fig:verteces1}
\end{center}
\end{figure}

The electroweak interaction is mediated either by massless gauge boson, photon ($\gamma$), massive neutral gauge boson $Z^{0}$ or massive charged gauge bosons $W^{\pm}$. The $\gamma$ couples to the electric charge, therefore to all charged fermions without altering the charge. The $Z^{0}$ boson couples to all fermions without altering the charge while the $W^{\pm}$ is the only gauge boson which can alter the charge of fermions and their flavors. The strong interaction is mediated by eight electrically neutral but color charged massless gauge bosons called gluons (\textit{g}), and therefore they couple only to the colored fermions - quarks. All bosons couple to the fermions via a space-time point interaction, called a \textit{vertex}, characterized by its \textit{coupling strength}, which is basically the elementary building block of any interaction. The coupling strengths of the electroweak interaction contain the constants $g^{'}$ and $g$, expressed in terms of the electric charge \textit{e} and weak mixing angle \textit{$\theta_{w}$}\footnote{The value of the weak mixing angle is measured experimentally to be $\sin^{2}\theta_{w}\approx 0.23$.} ($g^{'}=e/\cos\theta_{w}$ and $g=e/\sin\theta_{w}$). The couplings $g^{'}$ and $g$ correspond to $U(1)_{Y}$ and $SU(2)_{L}$ local gauge transformations, respectively. The coupling of colored fermions to colored gauge bosons in the strong interaction is expressed via the strong coupling constant $g_{s}$. The gauge boson - fermion vertices in corresponding interactions are shown in Fig.~\ref{fig:verteces1}.
At a level of experimental precision the charged leptons couple in the same way to the massive gauge bosons, $Z^{0}$ and $W^{\pm}$. This is referred to the \textit{$e-\mu-\tau$} universality. A similar property also holds for quarks.
\par
\begin{figure}[htb]
\begin{center}
\epsfxsize=2.5in
\epsfysize=1.25in
\epsfbox{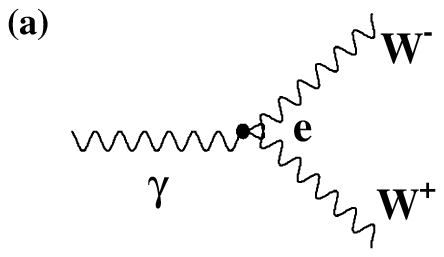}
\epsfxsize=2.5in
\epsfysize=1.25in
\epsfbox{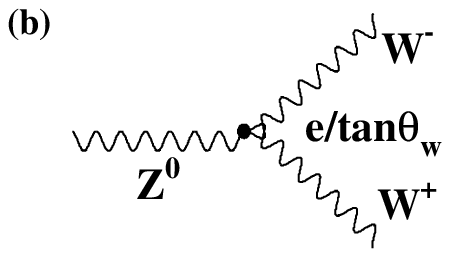}
\epsfxsize=2.5in
\epsfysize=1.25in
\epsfbox{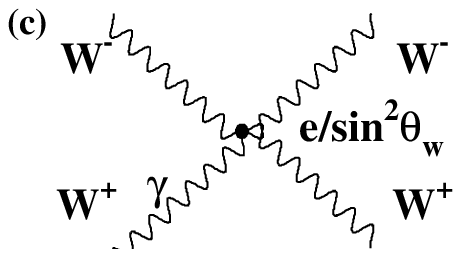}
\epsfxsize=2.5in
\epsfysize=1.25in
\epsfbox{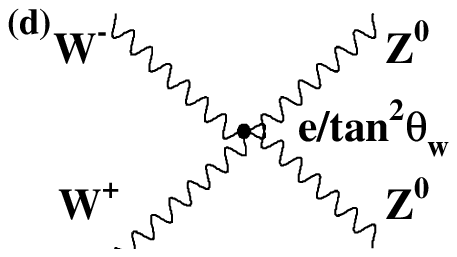}
\epsfxsize=2.5in
\epsfysize=1.25in
\epsfbox{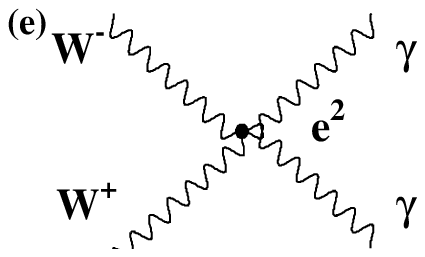}
\epsfxsize=2.5in
\epsfysize=1.25in
\epsfbox{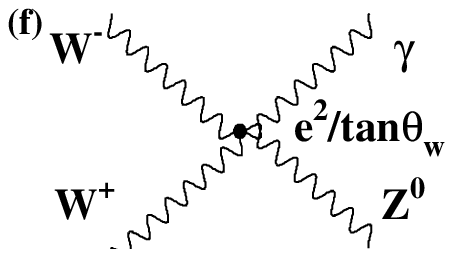}
\epsfxsize=2.5in
\epsfysize=1.25in
\epsfbox{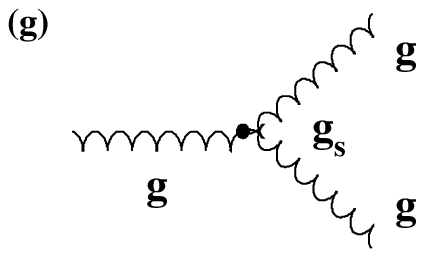}
\epsfxsize=2.5in
\epsfysize=1.25in
\epsfbox{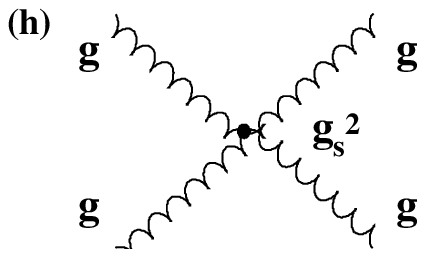}
\caption[bla]{(\textit{a,b}): Triple and (\textit{c,d,e,f}): quartic gauge boson vertices as a consequence of the electroweak gauge boson self-couplings. (\textit{g}): Triple and (\textit{h}): quartic gauge boson vertices as a consequence of the gauge boson self-couplings in a strong interaction.}
\label{fig:verteces2}
\end{center}
\end{figure}

The high precision experiments performed with $e^{+}e^{-}$ collisions at LEP and SLC have indicated the existence of bosonic self-interactions that are predicted by the Standard Model. These couplings that give rise to three and four gauge boson vertices are shown in Fig.~\ref{fig:verteces2}. In the Standard Model the photon self-couplings are not allowed. 
\section{Gauge Theories}
The high energy behavior of the $e-\nu$ scattering cross-section with a point-like interaction imposes the need for renormalisability of the Fermi theory of weak interaction. For high energies, a breakdown of perturbation theory occurs i.e., all terms in the perturbation series contribute equally. Furthermore, the divergences that appear in radiative corrections cannot be removed by renormalizing a finite number of parameters - at each order of the perturbation series, new types of divergences occur. The elimination of divergences requires the introduction into the theory of an infinite number of constants which would have to be determined experimentally. On the other hand, in a renormalizable theory a finite number of parameters must be experimentally inferred in order to remove all divergences in all orders of perturbation theory. A renormalization of the fundamental parameters such as coupling strengths and masses, is always necessary to obtain a one-to-one correspondence between theoretically calculated and experimentally measured values. The significance of a theory is called into question if it is not renormalizable. The fact that gauge theories are always renormalisable if they are free from anomalies makes these theories so important.
\subsection{Gauge Principle} 
The dynamics of an interaction that is described by gauge transformations, is determined by an underlying symmetry group. There are two types of symmetries, \textit{global} and \textit{local}. The term global refers to the space and time independence of a transformation. If an electron field is given as $\psi_{e}(x)$ (\textit{singlet}) and \textit{e} is its charge, then the electron field $\psi^{'}_{e}(x)$, transformed by the transformation:
\begin{equation}
{{\psi}^{'}}_{e}(x)=e^{ie\rho} {\psi}_{e}(x)
\label{eq:gp1}
\end{equation}
with space-time independent phase $e^{ie\rho}$ and a constant $\rho$, is equivalent to the field ${\psi_{e}}(x)$ i.e. both fields satisfy the same Dirac equation of motion and the field $\psi_{e}(x)$ transforms the same way everywhere. In other words, the equivalence of fields $\psi_{e}(x)$ and ${{\psi}^{'}}_{e}(x)$ is the consequence of the invariance of the equation of motion under the global transformation (\ref{eq:gp1}).
\par
The requirement for invariance under local transformations forces the existence of an interaction. If the phase is a function of (\textit{x}):
\begin{equation}
{{\psi}^{'}}_{e}(x)=e^{ie\rho(x)} {\psi}_{e}(x)
\label{eq:gp2}
\end{equation}
 the transformation may have different values at different space-time points and the fields $\psi_{e}(x)$ and ${{\psi}^{'}}_{e}(x)$ do not satisfy the same Dirac equation of motion i.e. it is not invariant under the transformation (\ref{eq:gp2}). Introducing the photon field $A_{\mu}(x)$ that transforms as:
\begin{equation}
{A_{\mu}}^{'}(x)={A_{\mu}}(x)+\partial_{\mu}\rho(x),
\label{eq:gp2a}
\end{equation}
the Dirac equation of motion becomes invariant, having the covariant derivative $D_{\mu}$ defined as $D_{\mu}=\partial_{\mu} - ieA_{\mu}(x)$ where $e$ is a coupling constant of the field $A_{\mu}(x)$. In this way the electromagnetic interaction is introduced and represents the element of the interaction. Under the simultaneous transformations (\ref{eq:gp2}) and (\ref{eq:gp2a}), replacing $\partial_{\mu}\rightarrow D_{\mu}$, that together form a \textit{gauge transformation}, the invariance is kept. The phase transformation (\ref{eq:gp2}) is a local one-dimensional unitary transformation, classified as a $U(1)$ transformation and the photon field $A_{\mu}(x)$ is the gauge field of the electromagnetic interaction that couples with a strength $e$.
\par
In order to obtain an interaction which transforms the particle identities as it is the case in the weak interaction, the pure phase transformation (\ref{eq:gp2}) is insufficient. Replacing the field $\psi_{e}(x)$ by a two-dimensional vector $\psi_{D}(x)$ (\textit{doublet}) (with fields $\psi_{\nu}(x)$ and $\psi_{e}(x)$):
\begin{equation}
\psi_{D}(x)
=\left(\begin{array}{c}
\psi_{\nu}(x) \\
\psi_{e}(x) \\
\end{array} \right)
=\left(\begin{array}{c}
\nu \\
 e \\
\end{array} \right)
\label{eq:gp3}
\end{equation}
the local phase transformation named $SU(2)$ has a following form:
\begin{equation}
{{\psi}^{'}}_{D}(x)=e^{-i\vec{\alpha}(x)\frac{\vec{\tau}}{2}} {\psi}_{D}(x) 
\simeq \left(1 - i\vec{\alpha}(x)\frac{\vec{\tau}}{2}\right) {\psi}_{D}(x)
\label{eq:gp4}
\end{equation}
where $\frac{{\tau}_{i}}{2}$ ($i=1,2,3$) are the generators of the $SU(2)$ transformations, $\tau_{i}$ are Pauli matrices and $\alpha_{i}(x)$ are arbitrary scalar functions of the space-time coordinate $x^{\mu}$. Since the equation of motion for ${{\psi}^{'}}_{D}(x)$ is not invariant under the transformation (\ref{eq:gp4}), the covariant derivative that insures the invariance of a Lagrangian, is $D_{\mu}=\partial_{\mu}+\frac{ig}{2}\vec{W}_{\mu}(x)\vec{\tau}$, introducing the new field $\vec{W}_{\mu}(x)$ that describes three new particles and transforms as: 
\begin{equation}
{\vec{{W}_{\mu}}^{'}(x)}=\vec{W}_{\mu}(x) + 
\frac{1}{g}\partial_{\mu}\vec{\alpha}(x) - 
\vec{W}_{\mu}(x)\times\vec{\alpha}(x)
\label{eq:gp5}
\end{equation}
i.e.,
\begin{equation}
{{W}_{\mu}}^{i'}(x)={{W}_{\mu}}^{i}(x) + \frac{1}{g}\partial_{\mu}{\alpha_{i}}(x) - \sum_{jk}\varepsilon_{ijk} {\vec{W}_{\mu}}^{j}(x) \alpha_{k}(x)
\label{eq:gp6}
\end{equation}
where \textit{g} is the coupling constant. The field transformation given by (\ref{eq:gp5}) is a consequence of the fact that the $SU(2)$ generators $\frac{{\tau}_{i}}{2}$ do not commute, i.e, $[\tau_{i},\tau_{j}]=2i\varepsilon_{ijk}\tau_{k}$ with $\varepsilon_{ijk}$ being the antisymmetric tensor \footnote{If two indexes are identical $\varepsilon_{ijk}=0$ and $\varepsilon_{ijk}=+1(-1)$ for even (odd) permutations of the indexes.}. A gauge theory in which not all transformations commute with each other is a \textit{non-Abelian gauge theory} or \textit{Yang-Mills theory}. The Lagrangian from which the equations of motion then may be derived is given by:
\begin{equation}
{\cal L} = \bar{\psi_{D}}(i\gamma^{\mu}\partial_{\mu}-m){\psi_{D}} -
\frac{g}{2}\bar{\psi_{D}}\gamma^{\mu}\vec{W_{\mu}}\vec{\tau}\psi_{D} -
\frac{1}{4}\vec{W^{\mu\nu}}\vec{W_{\mu\nu}}=
\cal L_{F} + \cal L_{F-B} + \cal L_{B}
\label{eq:gp7}
\end{equation}
where ${W_{\mu\nu}}^{i}$ is the field strength tensor defined as:
\begin{equation}
{W_{\mu\nu}}^{i} = \partial_{\mu}{W_{\nu}}^{i} - \partial_{\nu}{W_{\mu}}^{i} + g\sum_{jk}\varepsilon_{ijk}{W_{\mu}}^{j}{W_{\nu}}^{k}.
\label{eq:gp8}
\end{equation}
$\cal L_{F}$, $\cal L_{F-B}$ and $\cal L_{B}$ describe the interactions between fermions, fermions and bosons, and only bosons, respectively. By decomposition of $\cal L_{F-B}$ and introducing the fields ${W_{\mu}}^{+}=\frac{1}{\sqrt{2}}({W_{\mu}}^{1}-i{W_{\mu}}^{2})$ and ${W_{\mu}}^{-}=\frac{1}{\sqrt{2}}({W_{\mu}}^{1}+i{W_{\mu}}^{2})$, $\cal L_{F-B}$ becomes:
\begin{equation}
{\cal L_{F-B}} = \frac{-g}{2}(\bar{\nu}\gamma^{\mu}{W_{\mu}^{3}}\nu - \bar{e}\gamma^{\mu}{W_{\mu}^{3}}e) -
\frac{-g}{\sqrt{2}}(\bar{\nu}\gamma^{\mu}{W_{\mu}^{+}}e + \bar{e}\gamma^{\mu}{W_{\mu}^{-}}\nu),
\label{eq:gp9}
\end{equation}
where the weak field ${W_{\mu}}^{3}$ participates in the interaction without altering the fermion electric charge while fields ${W_{\mu}}^{\pm}$ alter the fermion charge i.e. cause conversion within the same family, between electrons and neutrinos.
\par
The last term in (\ref{eq:gp7}), absent in the case of the electromagnetic interaction\footnote{That implies the non-existence of the photon self-couplings in the Standard Model.}, describes the interaction among the weak fields. Introducing (\ref{eq:gp8}) into ($-\frac{1}{4}W^{i}_{\mu\nu}W^{i\mu\nu}$) of (\ref{eq:gp7}) and keeping only the interaction terms, the gauge boson self-interactions are introduced as:
\begin{equation}
{\cal L}_{GAUGE} = -{\frac{1}{2}}g(\partial_{\mu}{W_{\nu}}^{i}-\partial_{\nu}{W_{\mu}}^{i}) \varepsilon_{ijk} W^{j\mu}W^{k\nu} + 
\frac{1}{4}g^{2}\varepsilon_{ijk}\varepsilon_{i\lambda\rho}{W_{\mu}}^{j}{W_{\nu}}^{k}W^{\lambda\mu}W^{\rho\nu}
\label{eq:gp10}
\end{equation}
where the first term describes the trilinear gauge boson interactions and the second term describes the quartic gauge boson interactions.
\par
Generally, the covariant derivative $D_{\mu}$ that forms the gauge invariant Lagrangian is a fundamental part of a gauge principle and determines the interaction through the local gauge invariance requirements. The general feature of gauge theories is that they assume the symmetry between the fermions within the same family and predict the existence of massless bosons. However, these bosons may obtain a mass through the mechanism of \textit{spontaneous symmetry breaking} (SSB) leaving the photon massless and keeping the gauge invariance. SSB also explains the different characteristics of charged and neutral leptons, without touching upon the gauge principle.
\section{The Glashow-Weinberg-Salam theory of the electroweak interactions}
Due to the fact that the leptons in the same $SU(2)$ doublet carry different electric charges, the electromagnetic and the weak interaction cannot be treated separately. Instead of describing both interactions either as independent gauge theories or by a product ${SU(2)_{weak}} \times {U(1)_{EM}}$, that would lead to the equality of the electric charges of electron and neutrino, the ${SU(2)_{L}}\times{U(1)_{Y}}$ introduces the weak and the electromagnetic interaction as different components of a single gauge theory, the \textit{Electroweak Theory} of the Standard Model. This is also called the \textit{Glashow-Weinberg-Salam} (GWS) theory of the Standard Model. Both symmetry groups, $SU(2)_{L}$ and $U(1)_{Y}$ are chiral gauge groups since they transform the right-handed and left-handed fermionic components differently. The GWS theory does not necessarily involve right-handed neutrinos, and their inclusion in the spectrum is allowed, but not required. The ${SU(2)_{L}}\times{U(1)_{Y}}$ symmetry is spontaneously broken to the unbroken $U(1)_{EM}$ symmetry in order to provide masses to the massive gauge bosons, $W^{\pm}$ and $Z$ leaving the photon massless. The local gauge transformations of the left-handed doublets $\Psi_{D_{L}}$ and right-handed singlets $\Psi_{S_{R}}$ are given by:
\begin{flushleft} under $U(1)_{Y}$:  \end{flushleft}
\begin{equation}
{\Psi_{D_{L}}}^{'} = e^{-iY_{L}{\frac{\beta(x)}{2}}} {\Psi}_{D_{L}},
\label{eq:gws11}
\end{equation}
\begin{equation}
{\Psi_{S_{R}}}^{'} = e^{-iY_{R}{\frac{\beta(x)}{2}}} {\Psi}_{S_{R}},
\label{eq:gws12}
\end{equation}
\begin{flushleft} under $SU(2)_{L}$:  \end{flushleft}
\begin{equation}
{\Psi_{D_{L}}}^{'} = e^{-i\vec{\alpha}(x)\frac{\vec{\tau}}{2}} {\Psi}_{D_{L}},
\label{eq:gws13}
\end{equation}
\begin{equation}
{\Psi_{S_{R}}}^{'}={\Psi_{S_{R}}},
\label{eq:gws14}
\end{equation}
with
$$
\Psi_{D_{L}}=\left( \begin{array}{c}
              \frac{1}{2}(1-\gamma_{5})\psi_{\nu} \\
              \frac{1}{2}(1-\gamma_{5})\psi_{e}\\
             \end{array} \right),
\Psi_{S_{R}} = \frac{1}{2}(1+\gamma_{5})\psi_{e},
$$
where $\psi_{e}$ and $\psi_{\nu}$ are the electron and neutrino field operators, and $1/2(1\pm\gamma_{5})$ are the projection operators. The $Y_{L}$ and $Y_{R}$ are the weak hypercharges of the different fields with respect to $U(1)$ and satisfy the relation $Y=2(Q+T_{3})$ ($Q$ stands for electric charge and $T_{3}$ is the third component of the weak isospin). The covariant derivative that insures the invariance under the local ${SU(2)_{L}}\times{U(1)_{Y}}$ transformations is given by:
\begin{flushleft} for left-handed doublets ${\Psi}_{D_{L}}$ \end{flushleft}
\begin{equation}
D_{\mu}=\partial_{\mu} + {\frac{1}{2}}ig^{'}Y_{L}B_{\mu}(x)+
{\frac{1}{2}}ig\vec{\tau}\vec{W}_{\mu}(x),
\label{eq:gws15}
\end{equation}
\begin{flushleft} and for right-handed singlets ${\Psi_{e_{R}}}$ \end{flushleft}
\begin{equation}
D_{\mu}=\partial_{\mu} + {\frac{1}{2}}ig^{'}Y_{R}B_{\mu}(x),
\label{eq:gws16}
\end{equation}
where $B_{\mu}(x)$ and $\vec{W}_{\mu}(x)$ are massless gauge fields of the $U(1)$ and the $SU(2)$ transformations, respectively, coupled to the massless fermions with $U(1)_{Y}$ coupling constant $g^{'}$ and $SU(2)_{L}$ coupling constant $g$. In order to preserve the invariance, $B_{\mu}(x)$ and $\vec{W}_{\mu}(x)$ follow the transformations:
\begin{equation}
{B_{\mu}}^{'} = B_{\mu}(x) + {\frac{1}{{g}^{'}}}\partial_{\mu}\beta(x)
\label{eq:gws17}
\end{equation}
\begin{equation}
{\vec{W}_{\mu}^{'}} = \vec{W}_{\mu}(x) + {\frac{1}{{g}^{'}}}\partial_{\mu}\vec{\alpha}(x) - \vec{W}_{\mu}(x) \times \vec{\alpha}(x).
\label{eq:gws18}
\end{equation}
The kinetic term of the electroweak Lagrangian that includes the gauge bosons is defined as:
\begin{equation}
{\cal L}_{GAUGE} = {-\frac{1}{4}}\vec{W}_{\mu\nu}\vec{W}^{\mu\nu} - 
{\frac{1}{4}}B_{\mu\nu}B^{\mu\nu},
\label{eq:gws19}
\end{equation}
with $B_{\mu\nu}=\partial_{\mu}B_{\nu}-\partial_{\nu}B_{\mu}$ and $\vec{W}_{\mu\nu}=\partial_{\mu}\vec{W}_{\nu}-\partial_{\nu}\vec{W}_{\mu}$ and it does not permit any gauge boson mass term. Since the fields $B_{\mu}$ and $W_{\mu}^{3}$ do not have any direct physical interpretation, their linear combinations:
\begin{equation}
A_{\mu} = B_{\mu}\cos\theta_{w} + {W_{\mu}}^{3}\sin\theta_{w}
\label{eq:gws20}
\end{equation}
\begin{equation}
Z_{\mu} = -B_{\mu}\sin\theta_{w} + {W_{\mu}}^{3}\cos\theta_{w}
\label{eq:gws21}
\end{equation}
introduce the field $A_{\mu}$, required to have properties of the electromagnetic field and the neutral vector field $Z_{\mu}$, identified as a new neutral boson field assigned to the weak interaction. Linear combinations of $W_{\mu}^{1}$ and $W_{\mu}^{2}$ fields result in physical fields $W_{\mu}^{\pm}$ identified as charged boson fields. The weak mixing angle $\theta_{w}$ (\textit{Weinberg angle}) is defined in terms of the two coupling constants, $g^{'}$ and $g$, as $\tan\theta_{W}=g^{'}/g$. The electromagnetic coupling constant $e$ is related to $g^{'}$ and $g$ as $e = g\sin \theta_{w}$ and $e = g^{'}\cos\theta_{w}$.
\par
After inserting the definition of the field tensors $B_{\mu\nu}$ and $\vec{W}_{\mu\nu}$ and expressing the fields $W^{i}_{\mu}$ and $B_{\mu}$ in terms of the physical fields ${W_{\mu}}^{\pm}$, $Z_{\mu}$ and $A_{\mu}$ (with the transformations\footnote{The fields $B_{\mu}$ and $W_{\mu}^{3}$ are orthogonal in order to prevent any $A$-$Z$ mixing term to appear in the Lagrangian (\ref{eq:gws19}).} $B_{\mu}=A_{\mu}\cos\theta_{w}-Z_{\mu}\sin\theta_{w}$ and $W_{\mu}^{3}=A_{\mu}\sin\theta_{w}+Z_{\mu}\cos\theta_{w}$), the three gauge boson self-interaction from (\ref{eq:gp10}) can be expressed as:
\begin{equation}
\begin{array}{ccl}
{\cal L}_{TGC} & = & -{\frac{1}{2}}g(\partial_{\mu}{W_{\nu}^{i}}-\partial_{\nu}{W_{\mu}^{i}}) \varepsilon_{ijk} W^{j\mu}W^{k\nu} \\
 & = & \frac{-1}{2}g{\hat{{W}_{\mu\nu}^{i}}} \varepsilon_{ijk} W^{j\mu}W^{k\nu} \\
& = & -g{\hat{{W}_{\mu\nu}^{i}}}\varepsilon_{ij}W^{j\mu}W^{3\nu} - {\frac{1}{2}}g{\hat{{W}_{\mu\nu}^{3}}}\varepsilon_{ij}W^{i\mu}W^{j\nu} \\
& = & -g[{\hat{{W}_{\mu\nu}^{1}}}W^{2\mu} - {\hat{{W}_{\mu\nu}^{2}}}W^{1\mu}]W^{3\nu} - {\frac{1}{2}}g{\hat{{W}_{\mu\nu}^{3}}}[W^{1\mu}W^{2\nu}-W^{2\mu}W^{1\nu}] \\
& = & ig [{\hat{{W}_{\mu\nu}^{+}}}W^{-\mu} - {\hat{{W}_{\mu\nu}^{-}}}W^{+\mu}]W^{3\nu} + i{\frac{1}{2}}g{\hat{{W}_{\mu\nu}^{3}}}[W^{-\mu}W^{+\nu}-W^{+\mu}W^{-\nu}] \\
& = & ig ({\hat{{W}_{\mu\nu}^{+}}}W^{-\mu} - {\hat{{W}_{\mu\nu}^{-}}}W^{+\mu})W^{3\nu} + ig{\hat{{W}_{\mu\nu}^{3}}}W^{-\mu}W^{+\nu} \\
& = & ig \sin{\theta_{w}}({\hat{W}_{\mu\nu}}^{-}W^{+\mu} -
{\hat{W}_{\mu\nu}}^{+}W^{-\mu})A^{\mu} + ig \sin{\theta_{w}{\hat{A}_{\mu\nu}}}
W^{-\mu}W^{+\nu} \\
& + & ig \cos{\theta_{w}} ({\hat{W}_{\mu\nu}}^{-}W^{+\mu} -
{\hat{W}_{\mu\nu}}^{+}W^{-\mu})Z^{\nu} + ig \cos{\theta_{w}}{\hat{Z}_{\mu\nu}} W^{-\mu}W^{+\nu}
\label{eq:gws23}
\end{array}{}
\end{equation}
where $i,j =1,2$, ${\hat{{V}}_{\mu\nu}}=\partial_{\mu}V_{\nu}-\partial_{\nu}V_{\mu}$ and $V_{\mu}=W_{\mu},A_{\mu},Z_{\mu}$. The first and the second terms in the previous expression describe the $\gamma WW$ vertex with coupling strengths $e=g\sin{\theta_{w}}$ while the third and the fourth terms describe the $ZWW$ vertex with coupling strengths $g\cos{\theta_{w}}=e/\tan{\theta_{w}}$.  
\section{The mechanism of Spontaneous Electroweak Symmetry Breaking} 
The biggest mystery in particle physics is the origin of electroweak symmetry breaking (EWSB) and the mass generation mechanism for fermions. The Standard Model gives the solution by introducing an effective Higgs potential $V(\Phi)$:
\begin{equation}
V(\Phi)=\lambda(|\Phi|^{2}-v^{2}/2)^{2},
\label{eq:ewsb24}
\end{equation}
where the scalar field $\Phi$ is defined as a Higgs doublet:
\begin{equation}
\Phi=e^\frac{(i\omega^{i}\tau^{i})}{2v}\left(\begin{array}{c}
0 \\
(v+h)/\sqrt{2}, \\
\end{array} \right),
\label{eq:ewsb25}
\end{equation}
where $\omega^{i}$, $(i=1,2,3)$ are the Goldstone bosons and $h$ is the physical Standard Model Higgs boson of mass $m_{h}=v\sqrt{2\lambda}$. The vacuum expectation value $v\approx 246$ GeV, acquired by the scalar field $\Phi$, sets the mass scale for the electroweak gauge bosons and also for all fermions through Yukawa interactions:  
\begin{equation}
M_{W}={\frac{1}{2}}gv,
M_{Z}={\frac{1}{2}}{\sqrt{(g^{'2}+g^{2})}},
M_{\gamma}=0,
M_{fermions}={\frac{g_{f}v}{\sqrt{2}}}
\label{eq:ewsb26}
\end{equation}
where $g_{f}$ is a free parameter of the theory, called \textit{Yukawa coupling}. However, due to the unknown parameter $\lambda$ in the Higgs potential and the Higgs boson mass $m_{h}$ is a free parameter of the theory. The experimental searches at LEP2 have set the limit on its mass to be $m_{h}>113.5$ GeV at a $95\%$ confidence level (CL). The upper limit on the Standard Model Higgs boson mass is restricted by the theory and its value is approximatively 800 GeV. For heavy Higgs boson masses (above 800 GeV) the Standard Model is not a consistent effective theory anymore. If no light Higgs boson is found in experiment, new strong dynamics must take its place in order to restore the unitarity at high energies [$\sim 1$ TeV]. The mechanism responsible for mass generation in the effective description without the Standard Model Higgs boson is called dynamical or strong electroweak symmetry breaking (SEWSB) \cite{strong}. In this case, the Higgs-less Standard Model is considered as a low-energy approximation of another (enlarged) theory. Conversely, if a light Higgs boson exists, the Standard Model may nevertheless be incomplete, and New Physics could appear at higher energies. The effects of this larger theory are contained in the effective low energy Lagrangian expanded in powers of $(1/\Lambda_{NP})$, where $\Lambda_{NP}$ is the energy scale of the New Physics, as:
\begin{equation}
{\cal L}_{eff}=\sum_{n \ge 0}\sum_{i}\frac{\alpha_{i}^{n}}{{\Lambda_{NP}}^{n}}{O}_{i}^{(n+4)}=
{\cal L}_{eff}^{SM}+\sum_{n \ge 1}\sum_{i}\frac{\alpha_{i}^{n}}{{\Lambda_{NP}}^{n}}{O}_{i}^{(n+4)}.
\label{eq:ewsb27}
\end{equation}
The coefficients $\alpha_{i}$ are obtained from the parameters of the high energy theory and parameterize  all possible effects at low energies. Thus, the effective Lagrangian (\ref{eq:ewsb27}) parameterizes in a model-independent way, the low energy effects of the New Physics to be found at higher energies. Since the scale $\Lambda_{NP}$ is above the energies available in the experiments, the assumption is that the New Physics effects are not observed directly but they affect measured observables through virtual, e.g. in the gauge boson self-interactions. 
\par
In order to define the effective Lagrangian (\ref{eq:ewsb27}), it is necessary to specify the symmetry and the particle content of the low energy theory. In the case of SEWSB, the assumption is that only $SU(2)_{L}\times U(1)_{Y}$ gauge fields, fermions and Goldstone bosons are present. In the light Higgs boson case, the low-energy spectrum is augmented by the Higgs boson, but the effective Lagrangian formalism is essentially unchanged. For the study of the gauge boson self-interactions, the relevant terms in the effective Lagrangian (\ref{eq:ewsb27}) are those that produce vertices with three or four gauge bosons. The operators that contain fermions do not contribute to the trilinear and quartic gauge boson vertices while the operators that contain scalars may contribute since they may require a vacuum expectation value for the spontaneous EWSB. The effective Lagrangian that parameterizes the most general Lorentz invariant $WWV$ vertex ($V=Z,\gamma$), involving two $W$ bosons is defined as \cite{hagiwara}:
\begin{equation}
\begin{array}{ccl}
\frac{{\cal L}_{eff}^{WWV}}{g_{WWV}} & = & i g_{1}^{V} (W_{\mu\nu}^{*}W^{\mu}V^{\nu} - W_{\mu}^{*}V_{\nu}W^{\mu\nu})+i{\kappa}_{V}W_{\mu}^{*}W_{\nu}V^{\mu\nu} \\
  & + & i\frac{\lambda_{V}}{M_{W}^{2}} W_{\lambda,\mu}^{*}W_{\nu}^{\mu}V^{\nu\lambda} - g_{4}^{V}W_{\mu}^{*}W_{\nu}(\partial^{\mu}V^{\nu} + \partial^{\nu}V^{\mu}) \\
  & + & g_{5}^{V}\epsilon^{\mu\nu\lambda\rho}(W_{\mu}^{*}\partial_{\lambda}W_{\nu}-\partial_{\lambda}W^{*}_{\mu}W_{\nu})V_{\rho} \\
  & + & i\tilde{\kappa}_{V}W^{*}_{\mu}W_{\nu}\tilde{V}^{\mu\nu} + i\frac{\tilde{\lambda}_{V}}{M_{W}^{2}}W^{*}_{\lambda\mu}W^{\mu}_{\nu}\tilde{V}^{\nu\lambda},
\label{eq:ewsb28}
\end{array}{}
\end{equation}
where $\epsilon_{\mu\nu\lambda\rho}$ is the fully antisymmetric $\epsilon$ - tensor, $W$ denotes the $W$ boson field, $V$ denotes the photon or $Z$ boson field, $V_{\mu\nu}=\partial_{\mu}V_{\nu}-\partial_{\nu}V_{\mu}$, $W_{\mu\nu}=\partial_{\mu}W_{\nu}-\partial_{\nu}W_{\mu}$, $\tilde{V}_{\mu\nu}=1/2(\epsilon_{\mu\nu\lambda\rho}V^{\lambda\rho})$, $g_{WW\gamma}=-e$ and $g_{WWZ}=-e\cot\theta_{w}$. The fourteen coupling parameters of $WWV$ vertices are grouped according to their symmetries as $C$ and $P$ conserving couplings ($g_{1}^{V}, {\kappa}_{V}$ and $\lambda_{V}$), $C,P$ violating but $CP$ conserving couplings ($g_{5}^{V}$) and $CP$ violating couplings ($g_{4}^{V}, \tilde{\kappa}_{V}$ and $\tilde{\lambda}_{V}$). In the Standard Model all couplings vanish ($g_{5}^{V}=g_{4}^{V}=\tilde{\kappa}_{V}=\tilde{\lambda}_{V}=0$) except $g_{1}^{V}=\kappa_{V}=1$. The value of $g_{1}^{\gamma}$ is fixed by the electro-magnetic gauge invariance ($g_{1}^{\gamma}=1$ for on-shell photons) while the value of $g_{1}^{Z}$ may differ from its Standard Model value. Considering the $C$ and $P$ conserving couplings only, the deviations from the Standard Model values are denoted as the \textit{anomalous trilinear gauge couplings} (TGCs) $\Delta g_{1}^{Z}=(g_{1}^{Z}-1)$, $\Delta\kappa_{\gamma}=(\kappa_{\gamma}-1)$, $\Delta\kappa_{Z}=(\kappa_{Z}-1)$, $\Delta\lambda_{\gamma}=(\lambda_{\gamma}-0)=\lambda_{\gamma}$ and $\Delta\lambda_{Z}=(\lambda_{Z}-0)=\lambda_{Z}$. Besides, the $C$ and $P$ conserving terms in ${\cal L}_{eff}^{WWV}$ correspond to the lowest order terms in the expansion of the $W$ - $\gamma$ interactions and determine the charge $Q_{W}$, the magnetic dipole moment $\mu_{W}$ and the electric quadrupole moment $q_{W}$ of the $W$ boson:
\begin{eqnarray*}
& &
Q_{W}=eg_{1}^{\gamma}, \\
& &
\mu_{W}=\frac{e}{2M_{W}}(g_{1}^{\gamma}+\kappa_{\gamma}+\lambda_{\gamma}), \\
& &
q_{W}=-\frac{e}{2M_{W}^{2}}(\kappa_{\gamma}-\lambda_{\gamma}).
\end{eqnarray*}
\par
For the electroweak Fermi theory it is known that the effective Lagrangian of the theory breaks when $q^{2}$ in the propagator approaches $M^{2}_{W}$ (i.e. when the $W$ boson can be directly produced) and violates unitarity for $q^{2}>M^{2}_{W}/(g^{2}/4\pi)$. Unitarity is restored again by propagator (form-factor) effects and the scale at which the unitarity is violated gives an upper bound for the $W$ boson mass. Similar theoretical arguments suggest that the anomalous TGCs are at most of ${\cal O}(M^{2}_{W}/\Lambda^{2}_{NP})$. Thus, for $\Lambda_{NP} \sim 1$ TeV the anomalous TGCs are expected to be of ${\cal O}$(10$^{-2}$). If large values of anomalous TGCs would be observed, it would imply that the New Physics responsible for them is likely to be found directly below the TeV scale. As the New Physics scale increases, the effects on anomalous TGCs are less then ${\cal O}$(10$^{-2}$) and their observation needs more and more precise measurements.
\par
Approaching to $\Lambda_{NP}$ all terms in (\ref{eq:ewsb28}) become equally important and ${\cal L}_{eff}^{WWV}$ at energies of $\Lambda_{NP}$ or above requires an unitarization procedure which is model-dependent. This can be done by modifying the particle spectrum or by replacing any TGCs in (\ref{eq:ewsb28}) with appropriate form-factors \cite{haywood}:
\begin{equation}
\alpha\rightarrow \frac{\alpha_{0}}{(1+\hat{s}/\Lambda_{NP}^{2})^{n}},
\label{eq:ewsb29}
\end{equation} 
where $\sqrt{\hat{s}}$ is an effective center-of-mass energy and $n$ is chosen to ensure unitarity in the way that all couplings have the same high energy behavior. For the TGCs studies at the Tevatron (i.e. in $p\bar{p}$ collisions), where the center-of-mass energies are in a wide range, the value of $n=2$ is used for the $WW\gamma$ and $WWZ$ effective couplings. At LEP experiments this TGC modification was not needed due to the fixed center-of-mass energy.
\par
If a light Higgs boson exists, the New Physics is described using a \textbf{linear realization of the symmetry}. This is a direct extension of the Standard Model formalism. If the light Higgs boson is absent or sufficiently heavy then the effective Lagrangian should be expressed using a \textbf{nonlinear realization of the symmetry}.
\subsection{Linear Realization of the Symmetry Breaking} 
Within the linear realization of the symmetry, the underlying physics is expected to be weakly coupled and the assumption is that there are light scalars existing. Including the scalar Higgs doublet field $\Phi$ (\ref{eq:ewsb25}) in the low energy particle content the effective Lagrangian (\ref{eq:ewsb28}) can be written in terms of the operator coefficients $\alpha_{i}$, using the LEP parameterization where the New Physics scale $\Lambda_{NP}$ is replaced by the $W$ boson mass:
\begin{equation}
\begin{array}{ccl}
{\cal L}_{eff}^{TGC} & = & ig^{'}\frac{\alpha_{B\phi}}{M^{2}_{W}}(D_{\mu}\Phi)^{*}B^{\mu\nu}(D_{\nu}\Phi)+ig\frac{\alpha_{W\phi}}{M^{2}_{W}}(D_{\mu}\Phi)^{*}\vec{\tau}\hat{\vec{W}}^{\mu\nu}(D_{\nu}\Phi) \\
& + & g\frac{\alpha_{W}}{6M^{2}_{W}}\hat{\vec{W}}^{\mu}_{\nu}
(\hat{\vec{W}}^{\nu}_{\rho} \times \hat{\vec{W}}^{\rho}_{\mu})
\label{eq:ewsb30}
\end{array}{}
\end{equation} 
and contains three $C$ and $P$ conserving operators. The corresponding couplings are $\alpha_{B\phi}, \alpha_{W\phi}$ and $\alpha_{W}$. Replacing the Higgs field $\Phi$ with its vacuum expectation value $\Phi \rightarrow (0,v/\sqrt{2})$ and neglecting higher order terms in (\ref{eq:ewsb30}), the following relations between $\alpha_{i}$ coefficients and coupling parameters in (\ref{eq:ewsb28}) are obtained: 
\begin{equation}
\Delta g^{Z}_{1}=\frac{\alpha_{W\phi}}{\cos^{2}\theta_{W}},
\label{eq:ewsb31}
\end{equation}
\begin{equation}
\Delta\kappa_{\gamma}=-\frac{\cos^{2}\theta_{W}}{\sin^{2}\theta_{W}}(\Delta\kappa_{Z}-\Delta g^{Z}_{1})=\alpha_{W\phi}+\alpha_{B\phi},
\label{eq:ewsb32}
\end{equation} 
\begin{equation}
\lambda_{\gamma}=\lambda_{Z}=\alpha_{W}.
\label{eq:ewsb33}
\end{equation} 
Higher order terms would lead to $\lambda_{\gamma} \neq \lambda_{Z}$. In the linear realization of the symmetry the expected size of the anomalous TGCs would be $\sim 1/(16\pi^{2})(M_{W}^{2}/\Lambda_{NP}^{2})$ and increasing the scale $\Lambda_{NP}$ the coefficients $\alpha_{i}$ are expected to decrease as $(M_{W}^{2}/\Lambda_{NP}^{2})$. Thus, only small $\Lambda_{NP}$ scales would be accessible experimentally. In $W$ pair production processes, constant anomalous TGCs lead to a rapid growth of the vector boson pair production cross-section violating the unitarity at $\sqrt{s}=\Lambda_{U}$. Thus, the unitarity relation for each coupling is given as \cite{unitarity}:
\begin{equation}
|\alpha_{W}|\approx 19 \left(\frac{M_{W}}{\Lambda_{U}}\right)^{2};
|\alpha_{W\phi}|\approx 15.5 \left(\frac{M_{W}}{\Lambda_{U}}\right)^{2};
|\alpha_{B\phi}|\approx 49 \left(\frac{M_{W}}{\Lambda_{U}}\right)^{2}.
\label{eq:ewsb34}
\end{equation} 
Taking $\Lambda_{U}=\Lambda_{NP}=1$ TeV the corresponding TGC values are $|\alpha_{W}| \approx 0.12$, $|\alpha_{W\phi}| \approx 0.1$ and $|\alpha_{B\phi}| \approx 0.3$ which are larger than the LEP2 sensitivity. Thus, LEP2 is sensitive to the New Physics scale below 1 TeV but no deviation from Standard Model predictions was found. A sensitivity to smaller values of an anomalous TGC is equivalent to a sensitivity to potentially higher values of the corresponding $\Lambda_{NP}$.  
\subsection{Non-Linear Realization of Symmetry Breaking} 
In the absence of the Standard Model Higgs boson the effective Lagrangian becomes invariant if the $SU(2)_{L}\times U(1)_{Y}$ symmetry is realized non-linearly \cite{non_linear1,non_linear2}. Instead of the Higgs field, would-be Goldstone bosons (WBGBs) are included in the particle content. Without the Higgs boson the low energy Lagrangian violates the unitarity at a scale $\Lambda_{U}=\Lambda_{NP}\sim 1.2$ TeV \cite{scale}. This estimation of the $\Lambda_{NP}$ scale uses the analogy with low-energy QCD which describes $\pi\pi$ scattering below the $\rho$ resonance threshold and with chiral perturbation theory, representing the rough order of magnitude since it is unknown. Thus, the New Physics effects should appear at a scale below $\approx$3 TeV. 
\par
In order to probe the mechanism of the EWSB it is important to explore the dynamics of the weak gauge bosons since their longitudinal modes stem from the Goldstone bosons which are the remnants of the symmetry breaking sector. Before the opening up of new thresholds for resonances it is expected that the dynamics of the symmetry breaking sector affects the self-couplings of the gauge bosons. The description of these self-couplings relies on an effective Lagrangian adapted from pion physics where the symmetry breaking is non-linearly realized. The effective Lagrangian is organized as a set of operators whose leading order (in an energy expansion) operators reproduce the Higgs-less Standard Model while the symmetry breaking scenario of the the New Physics is described by the next-to-leading (NLO) operators.
\par
The starting point in the construction of the electroweak chiral Lagrangian (EChL) is the lowest order term of the non-linear $\sigma$-model (NLSM) \cite{non_linear1,nlsm} Lagrangian that describes the dynamics of the Nambu-Goldstone bosons (NGB) which are the low energy excitations of the vacuum in any theory with a spontaneously broken symmetry. The lowest order term of the NLSM Lagrangian is invariant under global $SU(2)_{L}\times SU(2)_{R}$ transformations and possesses a custodial $SU(2)_{C}$ symmetry responsible for the tree-level relation $\rho =M_{W}^{2}/M_{Z}^{2}\cos^{2}\theta_{w}=1$. Including the gauge fields through covariant derivatives, the obtained EChL, valid below the threshold for resonance production, is $SU(2)_{L}\times U(1)_{Y}$ invariant and can be written as:
\begin{equation}
{\cal L}_{eff}=\frac{v^{2}}{4}Tr[(D^{\mu}\Sigma)^{*}(D_{\mu}\Sigma)]+
\frac{1}{2}Tr[W_{\mu\nu}W^{\mu\nu}+B_{\mu\nu}B^{\mu\nu}],
\label{eq:ewsb35}
\end{equation}
where $\Sigma=\exp(i\vec{\omega}\cdot \vec{\sigma}/v)$ describes the Goldstone bosons $\omega_{i}$ with the built-in custodial $SU(2)_{C}$ symmetry. The $SU(2)_{L}\times U(1)_{Y}$ covariant derivative is defined as:
\begin{equation}
D_{\mu}\Sigma = \partial_{\mu}\Sigma + \frac{i}{2}gW^{i}_{\mu}\sigma^{i}\Sigma - 
\frac{i}{2}g^{'}B_{\mu}\Sigma\sigma_{3},
\label{eq:ewsb36}
\end{equation} 
where $\sigma^{i}$ are Pauli matrices, $W_{\mu\nu}=\partial_{\mu}W_{\nu}-\partial_{\nu}W_{\mu} - g(W_{\mu},W_{\nu})$, $B_{\mu\nu}=\partial_{\mu}B_{\nu}-\partial_{\nu}B_{\mu}$, $W_{\mu}=-(i/2)W^{i}_{\mu}\sigma^{i}$, $B_{\mu}=-(i/2)B_{\mu}\sigma^{3}$ and $B_{\mu\nu}=-(i/2)B_{\mu\nu}\sigma^{3}$. The coefficient $c_{n}(\Lambda_{NP})$ of any operator in the EChL that involves $b$ WBGB $b$, $d$ derivatives and $\omega$ gauge fields, can be estimated using the principle of "Naive Dimensional Analysis" (NDA) \cite{nda}. It is equivalent to the size of the coefficient $c_{n}(\Lambda_{NP})$ in front of the corresponding operator:
\begin{equation}
c_{n}(\Lambda_{NP}) \sim v^{2}\Lambda_{NP}^{2} 
\left(\frac{1}{v}\right)^{b} \left(\frac{1}{\Lambda_{NP}}\right)^{d} \left(\frac{g}{\Lambda_{NP}}\right)^{\omega},
\label{eq:ewsb37}
\end{equation}
and gives the size of ${\cal O}$($M_{W}^{2}/\Lambda_{NP}^{2}$) to $\Delta g^{V}_{1}$ and $\Delta\kappa_{V}$. The corresponding size of the couplings $\lambda_{V}$ is of ${\cal O}$($M_{W}^{4}/\Lambda_{NP}^{4}$). Introducing the $U(1)_{Y}$ gauge field the global $SU(2)_{L}\times SU(2)_{R}$ and $SU(2)_{C}$ symmetries are broken. The breaking of the $SU(2)_{C}$ symmetry introduces corrections to the $\rho$ parameter which are very small (as they would be in the case without a global symmetry breaking), $\rho\approx 1+{\cal O}(g^{'2})$ and we may assume that the $SU(2)_{C}$ symmetry is (approximatively) conserved.
\par
The physical amplitudes, derived from the EChL using the Feynman rules, describe the low energy dynamics of the WBGBs associated to the global spontaneous symmetry breaking $SU(2)_{L}\times SU(2)_{R} \rightarrow SU(2)_{C}$. That symmetry breaking provides masses to the electroweak gauge bosons through the WBGB - gauge bosons interaction without spoiling the $SU(2)_{L}\times U(1)_{Y}$ symmetry. The predictions of the EChL (\ref{eq:ewsb35}) are model-independent and they are known as Low-Energy Equivalence Theorems (LET) \cite{let}.
\par
After the renormalization of the EChL the effective Lagrangian contains parameters that absorb all the divergences at one-loop from EChL. A complete set of possible one-loop counter-terms for the EChL contains the set of fourteen $SU(2)_{L}\times U(1)_{Y}$ invariant and $CP$-invariant terms up to dimension-4 while only five of them are contained in the $SU(2)_{L}\times SU(2)_{R}$ invariant EChL:
\begin{equation}
\begin{array}{ccl}
& &
{\cal L}_{1}=\alpha_{1}^{'}\frac{igg^{'}}{2}B_{\mu\nu}Tr[TW_{\mu\nu}] \\
& &
{\cal L}_{2}=\alpha_{2}^{'}\frac{ig^{'}}{2}B_{\mu\nu}Tr[T[V^{\mu},V^{\nu}]] \\
& &
{\cal L}_{3}=\alpha_{3}^{'}gTr[W_{\mu\nu}[V^{\mu},V^{\nu}]] \\
& &
{\cal L}_{4}=\alpha_{4}^{'}Tr[V_{\mu}V_{\nu}]^{2} \\
& &
{\cal L}_{5}=\alpha_{5}^{'}Tr[V_{\mu}V^{\mu}]^{2} \\
\label{eq:ewsb38}
\end{array}
\end{equation}
where $T=\Sigma\sigma_{3}\Sigma^{\dagger}$ and $V_{\mu}=(D_{\mu}\Sigma)\Sigma^{\dagger}$. The parameters $\alpha_{i}^{'}$ represent the dimensionless couplings and take different values for different EWSB scenarios providing a parameterization of the unknown dynamics and thus, they are model-dependent. The operators ${\cal L}_{4}$ and ${\cal L}_{5}$ contribute to the quartic gauge boson vertices while ${\cal L}_{1}$, ${\cal L}_{2}$ and ${\cal L}_{3}$ contribute to the trilinear gauge boson vertices. The coefficients $\alpha_{i}$ are related to the New Physics scale by NDA as:
\begin{equation}
\alpha_{i}^{'}\approx \frac{\alpha_{i}}{16\pi^{2}} = \left(\frac{v}{\Lambda_{i}}\right)^{2},
\label{eq:ewsb39}
\end{equation} 
where the coefficients $\alpha_{i}$ are expected to be of ${\cal O}$(1) and $v=246$ GeV is the Fermi scale fixed by low energy weak interactions. The relation between the $\alpha_{i}$ coefficients and the TGCs given in (\ref{eq:ewsb28}) can be obtained setting the Goldstone fields $\omega_{i}$ to zero ($\Sigma\rightarrow 1$):
\begin{equation}
\begin{array}{ccl}
& &
\Delta g^{\gamma}_{1}=(g^{\gamma}_{1}-1) = 0, \\
& &
\Delta g^{Z}_{1}=(g^{Z}_{1}-1)=\frac{\alpha_{1}}{16\pi^{2}}\frac{e^{2}}{\cos^{2}\theta_{w}(\cos^{2}\theta_{w}-\sin^{2}\theta_{w})} + \frac{\alpha_{3}}{16\pi^{2}}\frac{e^{2}}{\sin^{2}\theta_{w} \cos^{2}\theta_{w}}, \\
& &
\Delta\kappa_{\gamma}=(\kappa_{\gamma}-1)=\frac{(\alpha_{2}+\alpha_{3}-\alpha_{1})}{16\pi^{2}}\frac{e^{2}}{\sin^{2}\theta_{w}}, \\
& &
\Delta\kappa_{Z}=(\kappa_{Z}-1)= \frac{\alpha_{1}}{16\pi^{2}}\frac{2e^{2}}{(\cos^{2}\theta_{w}-\sin^{2}\theta_{w})}-\frac{\alpha_{2}}{16\pi^{2}}\frac{e^{2}}{\cos^{2}\theta_{w}}+\frac{\alpha_{3}}{16\pi^{2}}\frac{e^{2}}{\sin^{2}\theta_{w}},\\
& &
\lambda_{\gamma}=\lambda_{Z}=0
\label{eq:ewsb40}
\end{array}{}
\end{equation}
and satisfy the relations (\ref{eq:ewsb32},\ref{eq:ewsb33}). From previous expressions it is clear that the anomalous couplings $\lambda_{V}$ are not induced by the NLO operators. They represent the transverse weak bosons and therefore do not probe efficiently the EWSB sector as evidenced by the fact that they do not involve the Goldstone bosons. Basically, within the EChL approach there is only one effective coupling $\Delta\kappa_{\gamma}$ that can be reached in $\gamma\gamma$ collisions, as long as $(v/\Lambda_{i})^{4}$ terms are discarded.
\par
\begin{figure}[htb]
\begin{center}
\epsfxsize=2.0in
\epsfysize=1.0in
\epsfbox{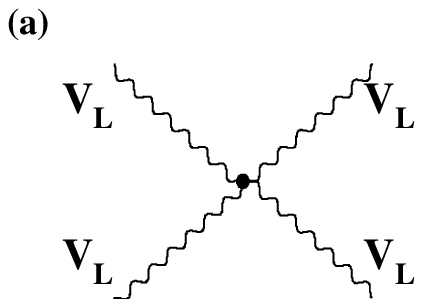}
\epsfxsize=2.0in
\epsfysize=1.0in
\epsfbox{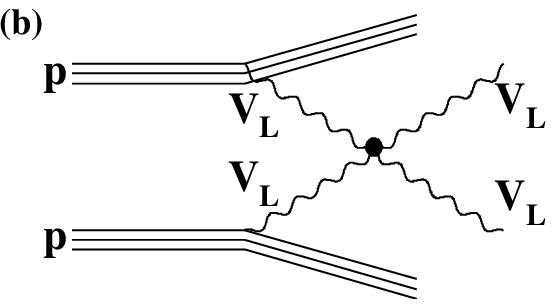}
\epsfxsize=2.0in
\epsfysize=1.0in
\epsfbox{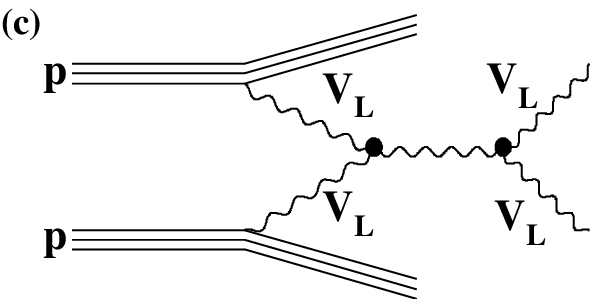}
\caption[bla]{The QGV contribution to the elastic scattering diagrams for $V_{L}V_{L}\rightarrow V_{L}V_{L}$ at (\textit{a}): $e^{+}e^{-}$ and $\gamma\gamma$ collisions ($V_{L}=W,Z$) and (\textit{b}): at $pp$ collisions ($V_{L}=W,Z,\gamma$) with a quartic gauge boson vertices. (\textit{c}): The TGV contribution to the elastic scattering diagrams ($a$) and ($b$). The same diagram stands if $pp$ collisions are replaced with $e^{+}e^{-}$. The index ``L'' denotes the longitudinal component of the gauge boson.}
\label{fig:ewsb}
\end{center}
\end{figure}

As it was mentioned before, in the SEWSB scenario the elastic scattering of the electroweak gauge bosons $V_{L}V_{L}$ ($V=W,Z$) corresponds to $\pi\pi$ scattering in QCD. In particular, the longitudinal components of gauge bosons are closely related with the WBGBs associated to the spontaneous global symmetry breaking of $SU(2)_{L}\times SU(2)_{R}$ to $SU(2)_{C}$. This relation, stated by LET \cite{let}, is more evident at high energies. Thus, studying the interactions of longitudinal gauge bosons, the unknown WBGBs dynamics can be probed. The diagrams shown in Fig.~\ref{fig:ewsb}\,$a,b$ represent the quartic gauge boson vertices in the elastic scattering of the gauge bosons and the diagram in Fig.~\ref{fig:ewsb}\,$c$ represents the trilinear gauge boson vertices.
\subsection{Technicolor model} 
If the Higgs boson is not discovered, the scenario of the strong EWSB at a scale of order of approximatively 1 TeV is an alternative possibility to restore unitarity and to explain the mass generation by the spontaneous symmetry breaking mechanism. The most studied class of theories that involve the elastic scattering of the gauge bosons is Technicolor \cite{techno1,techno2,techno3}. The generic prediction of Technicolor theories is the existence of a vector resonance ($|J_{Z}|=1$) with a mass below 2 TeV which unitarizes the $WW$ scattering cross-section. Scalar ($J_{Z}=0$) and tensor ($|J_{Z}|=2$) resonances are also possible \cite{poulose} along with light pseudo-Goldstone bosons which can be produced in pairs or in association with other particles \cite{others}. At the energies below the resonance threshold, the effects of the SEWSB could influence the values of trilinear and quartic gauge couplings, as an indirect evidence of resonance existence at higher energies.
\par
The main problem of this approach is the explanation of the fermionic masses since it is difficult to avoid flavor-changing neutral currents. The $SU(2)_{L}\times SU(2)_{R}$ Technicolor theory for dynamical EWSB is a well defined model but needs Extended Technicolor to give fermions their masses \cite{exttechno1,exttechno2}. 

\chapter{Measurements of Trilinear Gauge Couplings}
The measurement of the gauge boson self-couplings either provides a fundamental test of the non-Abelian structure of the electroweak theory or can give an information about the New Physics effects at high energies. Knowledge of the electroweak sector of the Standard Model can be improved only through new direct measurements of the coupling parameters ($g_{1}^{Z},\kappa_{Z,\gamma},\lambda_{Z,\gamma},$ etc.) with a higher accuracy then it is achieved in the experiments that have been performed till today, at LEP2 and at Tevatron.
\section{Measurements of TGCs at Lepton Colliders}
\begin{figure}[htb]
\begin{center}
\epsfxsize=2.5in
\epsfysize=1.5in
\epsfbox{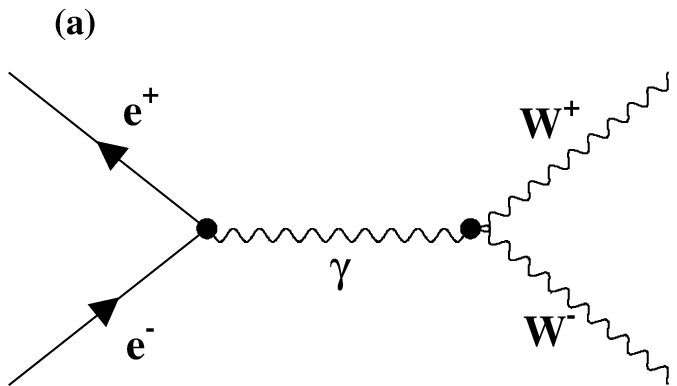}
\epsfxsize=2.5in
\epsfysize=1.5in
\epsfbox{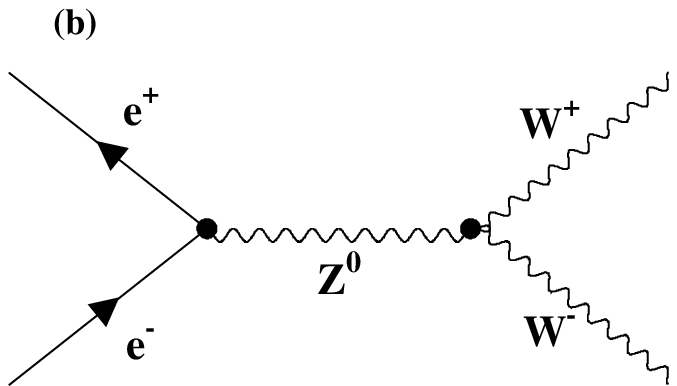}
\epsfxsize=3.5in
\epsfysize=2.in
\epsfbox{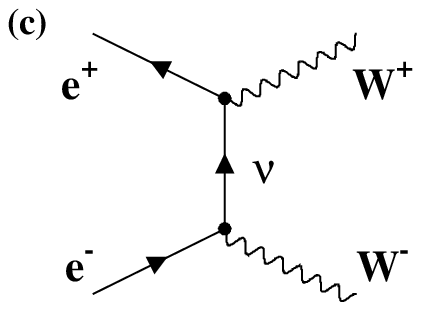}
\caption[bla]
{Feynman diagrams for the process $e^{+}e^{-} \rightarrow W^{+}W^{-}$.}
\label{fig:tgc1}
\end{center}
\end{figure}

\begin{figure}[htb]
\begin{center}
\epsfxsize=2.5in
\epsfysize=1.5in
\epsfbox{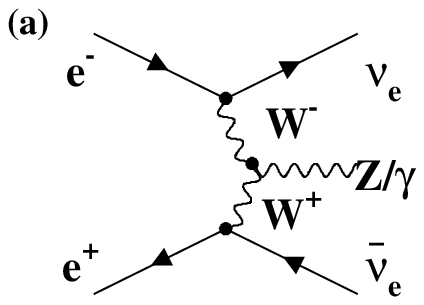}
\epsfxsize=2.5in
\epsfysize=1.5in
\epsfbox{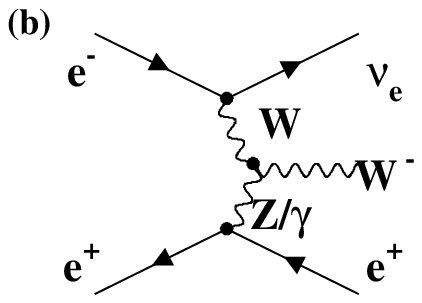}
\caption[bla]
{The $W^{+}W^{-} \rightarrow \gamma$ fusion diagrams.}
\label{fig:tgc2}
\end{center}
\end{figure}

The main processes sensitive to TGCs at LEP2 are either of the $1 \rightarrow 2$ type, annihilation of a fermion pair into a boson followed by emission of two bosons (Fig.~\ref{fig:tgc1}), or of the $2 \rightarrow 1$ type, i.e. fusion of two bosons into a single one (Fig.~\ref{fig:tgc2}). 
The best suited process for the measurements of TGCs at LEP2 is $e^{+}e^{-} \rightarrow W^{+}W^{-} \rightarrow f_{1}\bar{f_{2}}f_{3}\bar{f_{4}}$ which proceeds via three diagrams shown in Fig.~\ref{fig:tgc1}. Depending on the $W$ boson decay modes the possible final states are purely leptonic where both $W$ bosons decay into lepton pair ($l\bar{\nu_{l}}\bar{l}\nu_{l}$), purely hadronic where both $W$ bosons decay into quark-antiquark pairs ($q\bar{q}^{'}q^{'}\bar{q}=4$ jets) or semi-leptonic where one $W$ boson decays into quark-antiquark pairs while the other $W$ boson decays into a lepton pair ($q\bar{q}^{'}l\nu_{l}=2$ jets). Since the jet charge-flavor identification is inefficient, the four-jet final state events are characterized by existing ambiguities in quark-antiquark angular distributions; on the other hand, the cross-section for the leptonic decay mode is much less then for the two others, the semi-leptonic $WW$ decay channel is the most favored channel for studying TGCs at $e^{+}e^{-}$ collisions containing the maximum kinematical information. Nevertheless, at LEP2 all mentioned four-fermion final states are used for estimation of TGCs. 
\par
\begin{figure}[htb]
\begin{center}
\epsfxsize=5.5in
\epsfysize=3.0in
\epsfbox{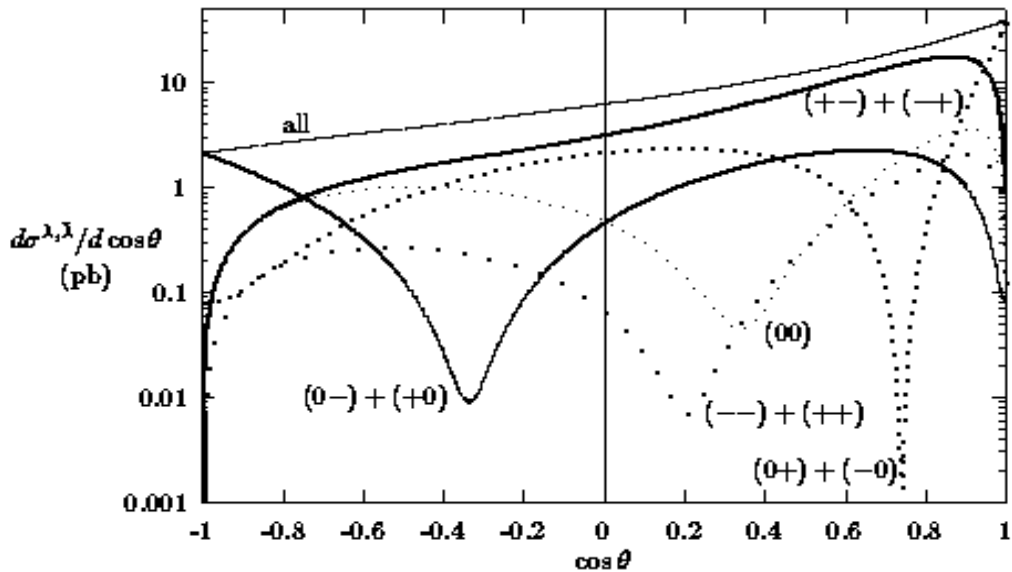}
\caption[bla]{Angular distributions over the $W$ boson production angle for different $WW$ helicity combinations ($\lambda,\bar{\lambda}$) for $e^{+}e^{-} \rightarrow W^{+}W^{-}$ at $\sqrt{s}=200$ GeV.}
\label{fig:tgc3}
\end{center}
\end{figure}

The differential $W^{+}W^{-}$ production cross-section is defined as \cite{bilenky}:
\begin{equation}
\frac{d\sigma (e^{+}e^{-}\rightarrow W^{+}W^{-})}{d\cos\theta}=\frac{\beta}{32\pi s} \sum_{\eta\lambda\bar{\lambda}}
|F_{\lambda\bar{\lambda}}^{\eta}(s,\cos\theta)|^{2} 
\label{eq:diff_eeww1}
\end{equation}
with $\cos\theta$ being the production angle of the $W$ boson measured with respect to the electron beam direction, $\sqrt{s}$ is the center-of-mass energy, $\eta$ is the helicity of the electron (positron), $\lambda$ and $\bar{\lambda}$ are the corresponding helicities of the two $W$ bosons while $\beta$ is the $W$ boson velocity defined as $\beta=\sqrt{1-4M_{W}^{2}/s}$. The $W$ bosons with the helicity $\pm$1 are transversally polarized while the $W$ bosons with the helicity $0$ are longitudinally polarized. Hence, there are nine possible $WW$ helicity combinations: $(++),(--),(00),(+-),(-+),(+0),(0+)(-0)$ and $(0-)$. Due to the angular momentum conservation the $WW$ helicity combinations $(\pm\mp)$ do not contribute to the \textit{s}-channel $Z/\gamma$-exchange diagrams with $|J|=0,1$. The $WW$ helicity combinations $(\pm\mp)$ contribute only to the \textit{t}-channel $\nu$-exchange diagram which allows the state with $|J|=2$. The contribution of each $WW$ helicity combination to the differential cross-section is contained in the helicity amplitudes \cite{rujula} described by the function $F_{\lambda\bar{\lambda}}^{\sigma}$ of equation (\ref{eq:diff_eeww1}), shown in Fig.~\ref{fig:tgc3}.
\par
\begin{figure}[htb]
\begin{center}
\epsfxsize=3.5in
\epsfysize=2.0in
\epsfbox{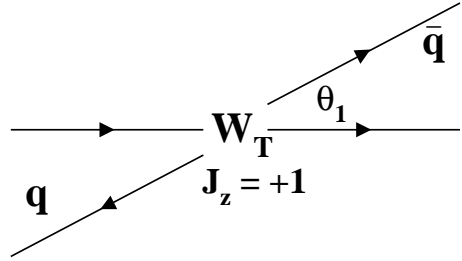}
\caption[bla]{Hadronic decay channel of the transversally polarized $W$ boson produced in the state $J_{z}=+1$.}
\label{fig:qqhelicities}
\end{center}
\end{figure}

Due to the $V-A$ structure of the $W$-fermion vertices the decay angular distributions of the fermion ($f$) and anti-fermion ($\bar{f}$) in the rest frame of the parent $W$ boson are sensitive to the different $W$ boson helicities. If the $W$ boson flies along the $z$-axis after the $e^{+}e^{-}$ collision, the $W^{\pm}$ rest frame is obtained by boosting along the $z$-axis. The decay angular dependence of the fermion and anti-fermion is identical for the $W^{-}$ and $W^{+}$ boson decays. The fermion and the anti-fermion are produced back-to-back in the rest frame of the parent $W$ boson and thus, the transformation between their decay angles is:
\begin{equation}
\begin{array}{ccl}
\theta_{f}\leftrightarrow \pi-\theta_{\bar{f}}, & & 
\phi_{f}\leftrightarrow \pi+\phi_{\bar{f}}
\label{eq:ang_dep}
\end{array}{}
\end{equation}
where $\theta_{f}$ is the production angle of the fermion in the rest frame of the parent $W$ boson (polar decay angle) and $\phi$ is the azimuthal decay angle of the fermion with respect to a plane defined by the $W$ boson and the beam axis. In the hadronic $W$ boson decay channels the relations (\ref{eq:ang_dep}) lead to the ambiguities ($\cos{\theta}_{f}$) $\leftrightarrow$ ($-\cos{\theta}_{\bar{f}}$) and ($\phi_{f}$) $\leftrightarrow$ ($\phi_{\bar{f}}+\pi$) caused by difficulties to distinguish between the quark and the anti-quark. These angular distributions are folded like:
\begin{equation}
\begin{array}{ccl}
\frac{dN}{d\theta_{folded}}=\frac{dN(\theta)}{d\theta}+\frac{dN(\pi-\theta)}{d\theta}, & & 0 < \theta < \pi \\
\frac{dN}{d\pi_{folded}}=\frac{dN(\phi)}{d\phi}+\frac{dN(\phi +\pi)}{d\phi}, & & 0 < \phi < 2\pi
\label{eq:folded}
\end{array}{}
\end{equation}
for each $W$ boson i.e. in the hadronic channel two sets of folded decay angles exist, $(\cos{\theta}_{1},\phi_{1})_{folded}$ and $(\cos{\theta}_{2},\phi_{2})_{folded}$. For example, this means that the quark and anti-quark angular distributions are $(dN/d\cos\theta_{1})\sim (1-\cos\theta_{1})^{2}$ and $(dN/d\cos\theta_{1})\sim (1+\cos\theta_{1})^{2}$ respectively, if the parent $W$ boson is produced in the $J_{z}=+1$ state (transversally polarized $W$ boson, $W_{T}$) and the angle $\theta_{1}$ is the angle between the $W$ boson direction and the anti-quark as it is shown in Fig.~\ref{fig:qqhelicities}. For the decay of the $W$ boson produced in the $J_{z}=-1$ state, the quark and anti-quark distributions are just exchanged. The folded distributions are then obtained summing the quark and anti-quark event distributions since the charge sign cannot be determined efficiently. This results in $(1+\cos^{2}\theta_{1,2})$-like distributions for the hadronic decay of the transversal $W$ bosons, i.e. for $J_{z}=\pm1$ $W$ bosons. Thus, the two transversal $W$ boson helicity states cannot be distinguished.
\par
Since there are five angles that describe the process $e^{+}e^{-} \rightarrow W^{+}W^{-}$ $\rightarrow$ $f_{1}\bar{f_{2}}f_{3}\bar{f_{4}}$, the general form of the differential cross-section expression \cite{bilenky} is given as:
\begin{equation} 
\begin{array}{ccl}
\frac{d^5\sigma(e^{+}e^{-}\rightarrow W^{+}W^{-}\rightarrow f_{1}\bar{f_{2}}f_{3}\bar{f_{4}})}{d\cos\theta d\cos{\theta}_{1}d\phi_{1} d\cos{\theta}_{2}d\phi_{2}} & = &
\frac{\beta}{32\pi s}\left(\frac{3}{8\pi}\right)^{2} 
B\left(W \rightarrow f_{1}\bar{f_{2}}\right) B\left(W \rightarrow f_{3}\bar{f_{4}}\right) \\
& \times & \sum_{\sigma\lambda{\bar\lambda}\lambda^{'}{\bar\lambda^{'}}}
F_{\lambda,\bar{\lambda}}^{\eta}(s,\cos\theta) 
F_{\lambda^{'},{\bar\lambda^{'}}}^{*\eta}(s,\cos\theta) \\
& \times & D_{\lambda\lambda^{'}}(\theta_{1},\phi_{1}) 
D_{\bar{\lambda},{\bar\lambda^{'}}}(\pi - \theta_{2},\phi_{2}+\pi),
\label{eq:diff_eeww3}
\end{array}{}
\end{equation}
which represents the five-fold differential angular distribution for the four-fermion production. The decay functions $D_{\lambda\lambda^{'}}$ are given in \cite{bilenky} and (\ref{eq:app1}) and describe the angular dependences in the rest frame of the parent $W$ boson. Integrating (\ref{eq:diff_eeww3}) over the azimuthal angles $\phi_{1}$ and $\phi_{2}$ the three-fold differential angular distribution can be derived:
\begin{equation}
\begin{array}{ccl}
\frac{{d^{3}\sigma(e^{+}e^{-}\rightarrow W^{+}W^{-}\rightarrow f_{1}\bar{f_{2}}f_{3}\bar{f_{4}})}}{d\cos\theta d\cos{\theta}_{1} d\cos{\theta}_{2}} & = &
\left(\frac{3}{4}\right)^{2}[\frac{d\sigma_{TT}}{d\cos\theta}\frac{1}{4}(1+\cos^{2}\theta_{1})(1+\cos^{2}\theta_{2}) \\
 & + & \frac{d\sigma_{LL}}{d\cos\theta}\sin^{2}\theta_{1}\sin^{2}\theta_{2} \\
 & + & \frac{d\sigma_{LT}}{d\cos\theta}\frac{1}{2}(1+\cos^{2}\theta_{2})\sin^{2}\theta_{1} + \frac{d\sigma_{TL}}{d\cos\theta}\frac{1}{2}(1+\cos^{2}\theta_{1})\sin^{2}\theta_{2}]
\label{eq:diff_eeww2}
\end{array}{}
\end{equation}
where $\theta_{1,2}$ is the fermion production angle in the rest frame of the parent $W$ boson.

($d\sigma_{TT}/d{\cos\theta}$) in (\ref{eq:diff_eeww2}) is the differential cross-section for the production of transversally polarized $W$ bosons ($\pm\pm,\pm\mp$) distributed as $(1+\cos^{2}\theta_{1,2})$ and ($d\sigma_{LL}/d\cos\theta$) is the differential cross-section for the production of longitudinally polarized $W$ bosons ($00$) distributed as $(\sin^{2}\theta_{1,2})$. ($d\sigma_{LT,TL}/d\cos\theta$) is the differential cross-section for the production of $W$ bosons with mixed helicities ($\pm 0,0\pm$) distributed as $(1+\cos^{2}\theta_{1,2})\sin^{2}\theta_{2,1}$.
\par
Anomalous TGCs give a contribution to the different helicity amplitudes which are proportional to $s$ or to $\sqrt{s}$ and hence, the sensitivity to TGCs depends on the center-of-mass energy. The sensitivity will increase if the center-of-mass energy increases. As it was mentioned, $WW$ helicity combinations with angular momentum $|J|=1$ receive contributions only from $Z/\gamma$ \textit{s}-channel exchange diagrams while the neutrino \textit{t}-channel exchange is canceled \cite{gounaris}. In the presence of anomalous couplings this cancellation does not occur and the interference with large amplitudes for $(+-)$ and/or $(-+)$ (from \textit{t}-channel exchange) can amplify the effects of anomalous couplings. 
\subsection{Analysis of final states at LEP2 experiments}
$WW$ pair production in $e^{+}e^{-}$ collisions results in three different four-fermion final states used in the analysis for the extraction of TGCs, $g_{1}^{Z},\kappa_{\gamma}$ and $\lambda_{\gamma}$. Since the $WW\gamma$ and $WWZ$ vertices cannot be distinguished if unpolarized $e^{+}e^{-}$ beams are used as it was at LEP, the $SU(2)_{L}\times U(1)_{Y}$ relations (\ref{eq:ewsb31}-\ref{eq:ewsb33}) have to be applied. On the other hand, using polarized beams enables us to treat these vertices separately\footnote{Changing the beam polarizations the contribution from one or another vertex is favorized.} and thus, the corresponding couplings independently. The LEP2 data, taken by the ALEPH, DELPHI, L3 and OPAL Collaborations at different center-of-mass energies up to $\sqrt{s}=209$ GeV are analyzed in terms of the five angles describing $WW$ pair production and decay or using the calculated cross-section from the collected events. Table \ref{tab:final1} shows the availability of angular information in different $WW$ final states. 
\begin{table}[h]
\begin{center}
\begin{tabular}{|c||c||c|} \hline
$WW$ decay channels & Decay fraction & Angular information  \\ \hline\hline
$jje\nu_{e}$ & $\approx 44/3\%$ & $\cos\theta$,($\cos\theta_{l},\phi_{l}$), $(\cos\theta_{j},\phi_{j})_{f}$\\ \hline
$jj\mu\nu_{\mu}$ & $\approx 44/3\%$ & $\cos\theta$,($\cos\theta_{l},\phi_{l}$), $(\cos\theta_{j},\phi_{j})_{f}$ \\ \hline
$jjl\tau\nu_{\tau}$ & $\approx 44/3\%$ & $\cos\theta$,($\cos\theta_{l},\phi_{l}$), $(\cos\theta_{j},\phi_{j})_{f}$ \\ \hline
$jjjj$ & $\approx 46\%$ & $|\cos\theta|$, $(\cos\theta_{j1},\phi_{j1})_{f}$, $(\cos\theta_{j2},\phi_{j2})_{f}$\\ \hline
$l\nu l\nu$ & $\approx 10\%$ & $\cos\theta$, $(\cos\theta_{1},\phi_{1})_{f}$, $(\cos\theta_{2},\phi_{2})_{f}$ \\ \hline
\end{tabular}
\end{center}
\caption{Availability of angular information in different $WW$ final states. $\theta$ is the $W$ boson production angle, $(\cos\theta_{l,j},\phi_{l,j})$ denotes the $W$ boson decay angles for leptons and jets, respectively. $(\cos\theta_{j},\phi_{j})_{f}$ implies the ambiguities ($\cos\theta_{j} \rightarrow -\cos\theta_{j}$) and ($\phi_{j} \rightarrow \phi_{j}+\pi$) incurred by the inability to distinguish quark from anti-quark jets. Notation '$f$' stands for '$folded$' distributions.}
\label{tab:final1}
\end{table}
\begin{enumerate}
\item \textbf{$WW \rightarrow q\bar{q}^{'}l\nu_{l}$} \\
with $l=e,\mu$ or $\tau$ (where $e$ and $\mu$ come either from the $W$ boson decay or from the cascade decay of the $W$ boson through a $\tau$ lepton) is the most prominent channel for TGC extraction in $e^{+}e^{-}$ collisions due to the large branching ratio and available kinematical informations. This is the main channel for measurement of TGCs despite the two-fold ambiguity on the decay angles of one of the $W$ boson. These events are characterized by the presence of a high energy isolated lepton, two jets and high missing transverse momentum resulting from the neutrino. For $q\bar{q}^{'}\tau\nu_{\tau}$ events, the lepton candidates are constructed by looking for an isolated $e$ or $\mu$ or a low multiplicity jet belonging to the $\tau$ lepton. If such a lepton is found the rest of the event is forced into two jets and a constrained fit is performed \cite{delphi1,aleph1,opal1}. The $q\bar{q}(\gamma)$ events and four-fermion final state events with two quarks and two leptons of the same flavor represent the main background.
\item \textbf{$WW \rightarrow q\bar{q}^{'}q^{'}\bar{q}$} \\
This decay channel is the most problematic in despite of the high production cross-section. The ambiguities in decay angles and the high background are the main reasons why this channel is inferior compared to the semi-leptonic $WW$ decay channel. These events are characterized by four well separated hadronic jets and hence with large visible energy and invariant mass. The missing energy is rather small coming from the initial state radiation. The large multiplicity ($N_{charged+neutral}>20$) of the signal event helps to reject the background events coming from QCD background ($e^{+}e^{-}\rightarrow q\bar{q}\gamma$) and the semi-leptonic $WW$ decay channel. The large sphericity $S$ of the $WW$ events is used to reject more QCD events which are rather characterized by two back-to-back jets with $S$ close to zero. The charges of two pairs of jets are evaluated, based on a jet-charge technique \cite{aleph1,opal2,l32,delphi2} to assign a charge to the reconstructed $W$ boson. The efficiency of the jet-charge reconstruction is not high\footnote{This is due to the jet-algorithm and the tracking efficiency being less than 100$\%$.} and thus not used in the study presented in this theses. Background coming from $Z(\gamma)\rightarrow q\bar{q}(\gamma)$ events is suppressed by imposing the cut on an estimate of the effective collision energy $\sqrt{s^{'}}$ in the $q\bar{q}\gamma$ final state after initial state radiation \cite{delphi2}. The rest of the event is forced into four jets and a constrained fit is performed.
\item \textbf{$WW \rightarrow l\bar{\nu}_{l}\bar{l}\nu_{l}$} \\
These events are characterized by the presence of two high energetic leptons, low multiplicity ($N_{charged+neutral}<5$) of the jets produced by the decay of the $\tau$ hadronic decay channel and a large missing momentum due to the neutrinos. The analysis is restricted to the case where both leptons are electrons or muons while a two-folded ambiguity on the neutrino momenta remains. These events are easy for the identification but the background contribution is large and mainly comes from $e^{+}e^{-},\gamma\gamma\rightarrow Z(\gamma)\rightarrow q\bar{q}(\gamma)$ events. The leptonic $\tau$ decay from $l\nu\tau\nu_{\tau}$ contributes as a non-negligible background too, since electrons or muons could originate from tau decay.
\item \textbf{$WW \rightarrow \gamma/Z$} ($e^{+}e^{-} \rightarrow \bar{\nu}\nu\gamma, \bar{\nu}\nu Z,$) \\
These are the $W$ boson fusion processes schematically shown in Fig.~\ref{fig:tgc2}\,$a$. The $WW \rightarrow \gamma$ channel \cite{aleph3} allows to explore the $WW\gamma$ vertex independently, unlike in the $1 \rightarrow 2$ processes where the mixing of $Z/\gamma$ couplings occurs. Since the $W$ bosons are exchanged at low momentum transfer this channel is mainly sensitive to $\Delta\kappa_{\gamma}$ because contributions from $\lambda_{\gamma}$ contain higher powers of the $W$ boson momenta. The signature of these events is a high energy isolated photon having the two relevant observables for the analysis, the energy and the production angle with no reconstructed charged particle tracks. The main background comes from the $Z$ production in $e^{+}e^{-} \rightarrow \bar{\nu}\nu\gamma$ process via \textit{t}-channel exchange. The background photons are coming from the initial state radiation (ISR) and hence, they are preferentially emitted at low angles, close to the beam while the photons from the fusion are rather emitted at large angles. The energy peak of the ISR photons is at a value of $E_{\gamma}=(s-M_{Z}^{2})/(2\sqrt{s})$. In order to exclude the region of the $Z$ peak return only the events with a high or low energy photon, emitted at large angles ($|\cos\theta_{\gamma}|<0.9$) are chosen for the analysis. This signal events can be identified with a small background contribution.
\item \textbf{$WZ/\gamma \rightarrow W$} ($e^{+}e^{-} \rightarrow W^{\mp}e^{\pm}\nu_{e}$) \\
shown in Fig.~\ref{fig:tgc2}\,$b$ represents the single $W$ boson production used for the TGC measurements. Similarly to the previous case, this channel is mainly sensitive to $\Delta\kappa_{\gamma}$. Since the $e^{\pm}$ is emitted at very low angles, only the $W^{\mp}$ boson decay products are detected. In the leptonic $W$ boson decay channel a single high energy charged isolated lepton ($e$ or $\mu$) or $\tau$ jet with a low multiplicity, with a missing momentum due to the neutrino, are required. The main background comes from $Zee\rightarrow \nu_{\mu,\tau}\bar{\nu}_{\mu,\tau}ee$ events. Concerning the hadronic $W$ boson decay channel, the event is characterized by two acoplanar jets with a mass around the $W$ boson mass. The main background comes from the semi-leptonic $WW$ decay channels. The common backgrounds, two photon and two fermion events with an initial state radiation, are rejected by the cut on the transverse missing momentum. More about selection cuts and estimated efficiencies can be found in \cite{aleph1,aleph4}. For the TGCs estimation only the the total cross-section is used.
\end{enumerate}
\subsection{Measurement Techniques at LEP2 experiments}
Three different methods were proposed for measurement of TGCs at LEP2: the spin density matrix method, the maximum likelihood method and the method of optimal observables. The unbinned maximum likelihood method was not used in the LEP analysis due to the difficulties to include the acceptance corrections. The choice of the method is based on having in mind to use as much of the available angular data for each event as possible and that the expected LEP2 data will not be sufficient to perform the $\chi^{2}$ fit of binned five angular distributions.
\begin{enumerate}
\item \textbf{The spin density matrix (SDM) method} \\
gives the most direct insight on the TGCs although it can be applied only to the semi-leptonic $WW$ decay channel. The SDM elements are defined as the normalized products of the helicity amplitudes and describe the polarization of the $W$ boson pair. The SDM elements for the two-particle joint density matrix are defined as:
\begin{equation}
\rho_{\lambda_{-}\lambda_{-}^{'}\lambda_{+}\lambda_{+}^{'}}= 
\frac{{\sum_{\eta}}F_{\lambda_{-}\lambda_{+}}^{\eta} (F_{\lambda_{-}^{'}\lambda_{+}^{'}}^{\eta})^{*}}{\sum_{\eta\lambda_{+}\lambda_{-}}
|F_{\lambda_{-}\lambda_{+}}^{\eta}|^{2}}
\label{eq:sdm1}
\end{equation}
where $F_{\lambda_{-}\lambda_{+}}^{\eta}$ is the helicity amplitude for a given electron (positron) helicity $\eta$ and $WW$ helicities $\lambda_{-}$ and $\lambda_{+}$. The total number of SDM elements for the $WW$ system is $80$ and can be reduced to $9$ for each $W$ boson, 
considering the one-particle SDM which is defined as the sum of the two-particle SDM over all helicity states of the other $W$ boson, i.e.:
\begin{equation}
\rho_{\lambda_{-}\lambda_{-}^{'}}^{W^{-}}(\cos\theta)=
\sum_{\lambda_{+}} \rho_{\lambda_{-}\lambda_{-}^{'}\lambda_{+}\lambda_{+}^{'}}
\label{eq:sdm2}
\end{equation}
and
\begin{equation}
\rho_{\lambda_{+}\lambda_{+}^{'}}^{W^{+}}(\cos\theta)=
\sum_{\lambda_{-}} \rho_{\lambda_{-}\lambda_{-}^{'}\lambda_{+}\lambda_{+}^{'}}.
\label{eq:sdm3}
\end{equation}
Since the diagonal matrix elements are real they can be interpreted as the probability to produce the $W^{\pm}$ boson with the helicity $\lambda_{\pm}$. The non-diagonal elements can be complex and describe the interference effects between the different helicity states. Using the one-particle SDM elements instead of using the two-particle SDM elements neglects the spin correlations between two different $W$ bosons, but the expected loss in the sensitivity is small.
\par
The analysis with the SDM method proceeds in two steps: first, the experimental SDM elements are determined from the angular distribution in bins of the $W$ boson production angle and second, the different predictions of theoretical models are fitted to the experimental distributions using a $\chi^{2}$ minimization fit. The SDM elements are calculated using the orthogonality of the $W$ boson decay functions $D_{\lambda_{-}\lambda_{-}^{'}}$ and $D_{\lambda_{+}\lambda_{+}^{'}}$ and integrating over the $W$ boson decay angles using the projection operators $\Lambda_{\lambda_{\pm}\lambda_{\pm}^{'}}^{W^{\pm}}$ as:
\begin{equation}
\begin{array}{ccl}
\rho_{\lambda_{\pm}\lambda_{\pm}^{'}}^{W^{\pm}}\frac{d\sigma(e^{+}e^{-} \rightarrow W^{+}W^{-})}{d\cos\theta} 
& = & \frac{1}{B_{q\bar{q}^{'}l\bar{\nu}}} \int{\frac{d^{3}\sigma(W^{+}W^{-} \rightarrow q\bar{q}^{'}l\bar{\nu})}{d\cos\theta d\cos\theta_{l}d\phi_{l}}} \\
& \times & \Lambda_{\lambda_{\pm}\lambda_{\pm}^{'}}^{W^{\pm}}(\theta_{l},\phi_{l}) d\cos\theta_{l} d\phi_{l}
\label{eq:sdm4}
\end{array}{}
\end{equation}
where $B_{q\bar{q}^{'}l\bar{\nu}}$ represents the branching ratio for the semi-leptonic decay channel. Expressions for $\Lambda_{\lambda_{\pm}\lambda_{\pm}^{'}}^{W^{\pm}}$ are given in \cite{operators}.
\item \textbf{The optimal Observables method} \\
does not use the direct fit of the angular distributions but uses a function which is a series expansion of the differential cross-section with respect to the anomalous couplings. The optimal observables are quantities with maximal sensitivity to the unknown coupling parameter. Assuming that the helicity amplitudes are quadratic in the TGC parameters, the differential cross-section may be written as:
\begin{equation}
\frac{d\sigma(\Omega,\vec{P})}{d\Omega}=S_{0}(\Omega)+\sum_{i}S_{1,i}(\Omega)P_{i}+
\sum_{i,j}S_{2,ij}(\Omega)P_{i}P_{j}
\label{eq:oo1}
\end{equation}
where $\Omega$ represents the phase space variables, the coefficients $S_{0}, S_{1,i}$ and $S_{2,ij}$ are functions of the phase-space variables $\Omega=(\cos\theta, \cos\theta_{l,j}, \phi_{l,j})$ and $P_{i,j}$ is the set of chosen couplings which have a zero value in the Standard Model. Taking into account only the linear expansion the optimal observables are defined as:
\begin{equation}
{\cal O}_{i}(\Omega)=\frac{S_{1,i}(\Omega)}{S_{0}(\Omega)}
\label{eq:oo2}
\end{equation}
and measured, while their mean values $\langle{\cal O}_{i}\rangle$ are evaluated to first order in the $P_{i}$ as \cite{optimal}:
\begin{equation}
\langle{\cal O}_{i}\rangle = {\langle{\cal O}_{i}\rangle}_{0}+\sum_{j}c_{ij}P_{j}
\label{eq:oo3}
\end{equation}
which give the optimal sensitivity under the assumption that the parameters under the investigation are small deviations from the Standard Model predictions. Since ${\langle{\cal O}_{i}\rangle}_{0}$ and $c_{ij}$ can be calculated \cite{papadopulus}, the couplings $P_{j}$ can be extracted from (\ref{eq:oo3}). This is an economical binning method in which the real data points are used to divide the phase space into an equal number of multi-dimensional bins since $\Omega=(\cos\theta, \cos\theta_{l,j}, \phi_{l,j})$. This is the most commonly used method for analysis of the semi-leptonic and hadronic $WW$ decay channels, although the physical information is completely hidden in the optimal variables.
\end{enumerate}
\subsection{Results of the TGCs measurements at LEP experiments}
\begin{figure}[p]
\begin{center}
\epsfxsize=6.0in
\epsfysize=5.5in
\epsfbox{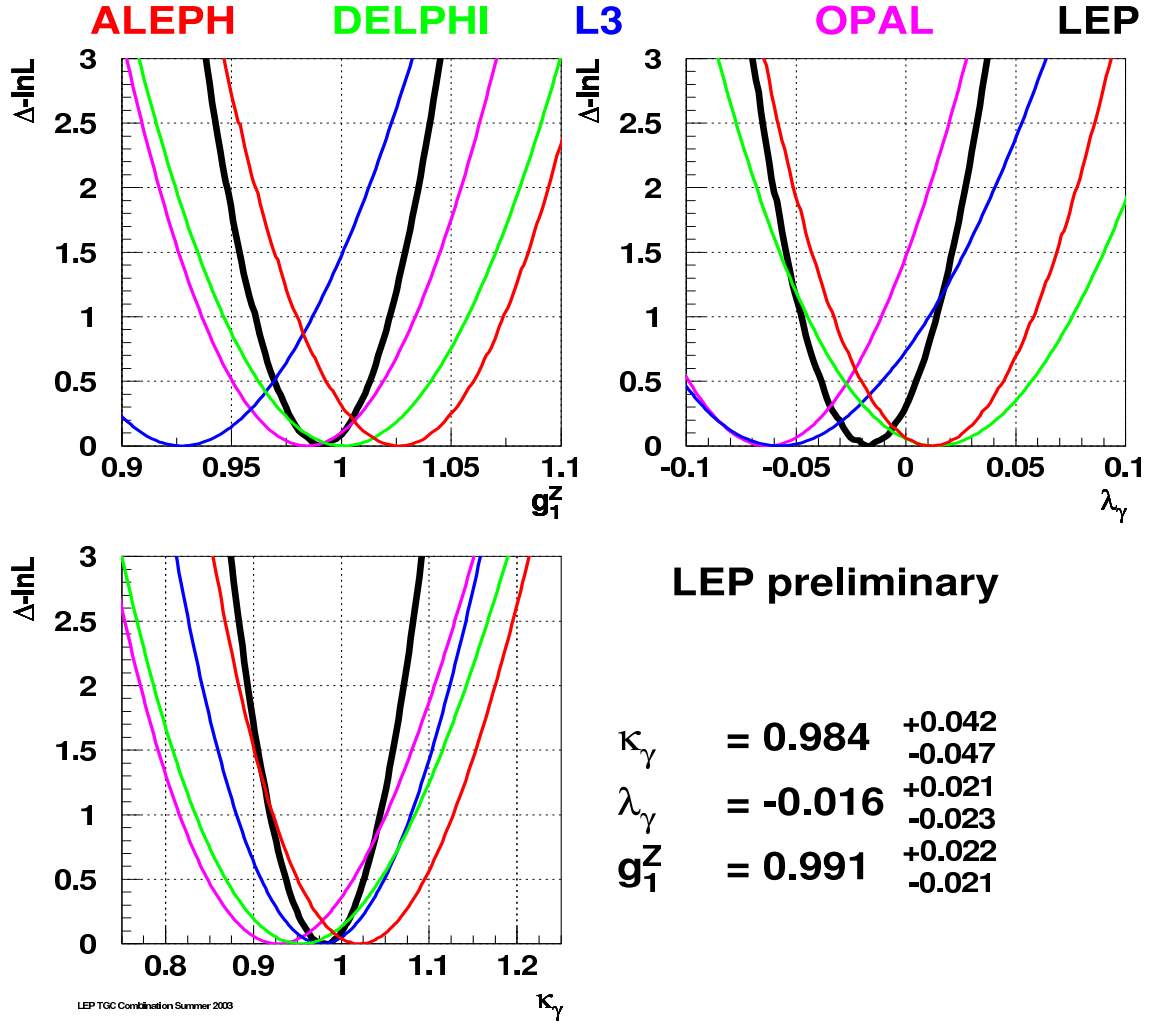}
\caption[bla]{The log${\cal L}$ functions for $g_{1}^{Z},\kappa_{\gamma}$ and $\lambda_{\gamma}$ from the single-parameter fit. The LEP2 combined results are presented by black curve. The minimal TGC values are pointed in the right lower corner.} 
\label{fig:loglh}
\end{center}
\end{figure}

Since LEP started to operate, the data are collected at different center-of-mass energies and analyzed periodically, updating the previous results. The most recent preliminary updated LEP2 results on TGCs measurement are published in \cite{lep2005} combining almost all data collected by DELPHI, OPAL, L3 and ALEPH, up to $\sqrt{s}=209$ GeV. Significant deviations from the Standard Model predictions for any of the studied coupling are not observed, so far.
\begin{table}[htb]
\begin{center}
\begin{tabular}{|c||c|c|c|} \hline
Single-parameter fit & \multicolumn{3}{c|}{Measurement}\\ \hline
Experiment & ${g_{1}^{Z}}$ & ${\kappa_{\gamma}}$ & $\lambda_{\gamma}$ \\ \hline\hline
ALEPH & $1.026^{+0.034}_{-0.033}$ & $1.022^{+0.073}_{-0.072}$ & $0.012^{+0.033}_{-0.032}$ \\ \hline
DELPHI & $1.002^{+0.038}_{-0.040}$ & $0.955^{+0.090}_{-0.086}$ & $0.014^{+0.044}_{-0.042}$ \\ \hline
L3 & $0.928^{+0.042}_{-0.041}$ & $0.922^{+0.071}_{-0.069}$ & $-0.058^{+0.047}_{-0.044}$ \\ \hline
OPAL & $0.985^{+0.035}_{-0.034}$ & $0.929^{+0.085}_{-0.081}$ & $-0.063^{+0.036}_{-0.036}$ \\ \hline\hline
Single-parameter fit & \multicolumn{3}{c|}{68$\%$ CL limits}\\ \hline\hline
LEP Combined & $0.991^{+0.022}_{-0.021}$ & $0.984^{+0.042}_{-0.047}$  & $-0.016^{+0.021}_{-0.023}$ \\ \hline\hline
Single-parameter fit & \multicolumn{3}{c|}{95$\%$ CL limits}\\ \hline\hline
LEP Combined & [0.949, 1.034] & [0.895, 1.069]  & [-0.059, 0.026] \\ \hline
\end{tabular}
\end{center}
\caption{Results on the TGC parameters $g_{1}^{Z},\kappa_{\gamma}$ and $\lambda_{\gamma}$, obtained from the single-parameter fit at four LEP2 experiments. The statistical and systematic errors are included. The results from the combined fit are given at 68$\%$ CL and 95$\%$ CL.}
\label{tab:lep1}
\end{table}
Table \ref{tab:lep1} shows the results obtained from the single-parameter fits from each of LEP experiments and the combination performed by the LEP Electroweak Working Group analyzing the different $WW$ decay channels, typically the semi-leptonic and fully hadronic. In the single-parameter fit, one of three charged TGCs measured at LEP (${g_{1}^{Z}}, {\kappa_{\gamma}}$ or $\lambda_{\gamma}$) is varied while the others are fixed at their Standard Model values. In Fig.~\ref{fig:loglh} the individual log${\cal L}$ functions for each TGC and from each of the LEP2 experiments are plotted as well as their combined results obtained from the single-parameter fit.
\par
\begin{figure}[p]
\begin{center}
\epsfxsize=6.0in
\epsfysize=5.5in
\epsfbox{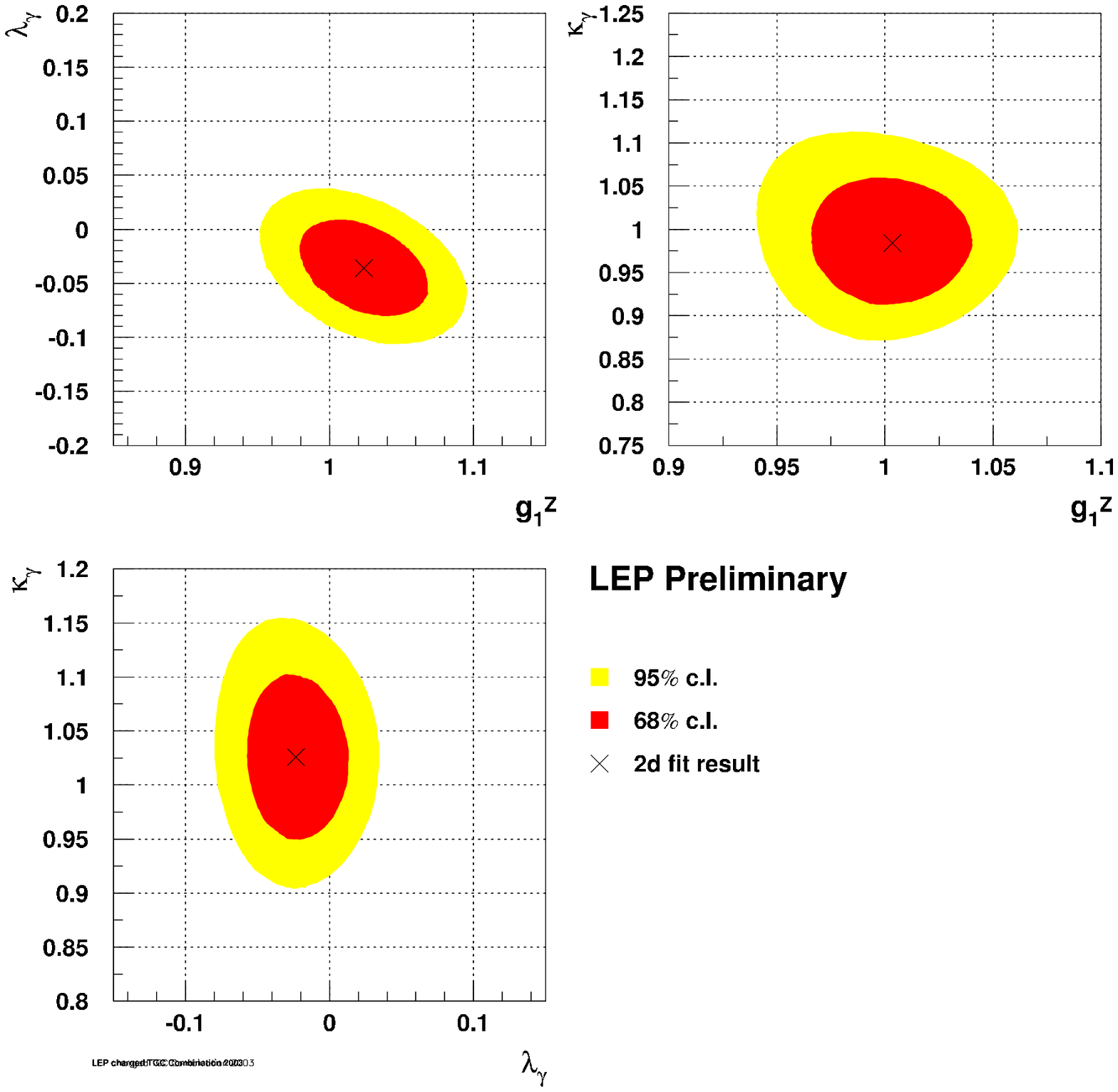}
\caption[bla]{The $68\%$ CL and $95\%$ CL contour plots for the three two-parameter fits to $g^{Z}_{1}-\lambda_{\gamma}$, $g^{Z}_{1}-\kappa_{\gamma}$ and $\lambda_{\gamma}-\kappa_{\gamma}$ including the systematic errors. The cross indicates the fitted TGC value.} 
\label{fig:lep_cont}
\end{center}
\end{figure}

In the two-parameter fit (Table \ref{tab:lep2}), two of three coupling parameters are varied simultaneously while the third one kept its Standard Model value. The results from a two-parameter fit of TGCs at $68\%$ CL and $95\%$ CL are shown in Fig.~\ref{fig:lep_cont}, presented in the $g^{Z}_{1}-\lambda_{\gamma}$, $g^{Z}_{1}-\kappa_{\gamma}$ and $\lambda_{\gamma}-\kappa_{\gamma}$ planes. The $SU(2)_{L}\times U(1)_{Y}$ relations between the coupling parameters are assumed in all fits.
\begin{table}[htb]
\begin{center}
\begin{tabular}{|c||c|c|} \hline
 Parameter & 68$\%$ CL & 95$\%$ \\ \hline\hline
  $\begin{array}{c} 
  g_{1}^{Z} \\ 
  \kappa_{\gamma} \\
\end{array}$ 
& $\begin{array}{c} 
  1.004^{+0.024}_{-0.025} \\ 
  0.984^{+0.049}_{-0.049} \\
\end{array}$
& $\begin{array}{c} 
  +0.954, +1.050 \\ 
  +0.894, +1.084 \\
\end{array}$ \\ \hline
$\begin{array}{c} 
  g_{1}^{Z} \\ 
  \lambda_{\gamma} \\
\end{array}$  
& $\begin{array}{c} 
  1.024^{+0.029}_{-0.029} \\ 
  -0.036^{+0.029}_{-0.029} \\
\end{array}$
& $\begin{array}{c} 
  +0.966, +1.081 \\ 
  -0.093, +0.022 \\
\end{array}$ \\ \hline
$\begin{array}{c} 
  \kappa_{\gamma} \\ 
  \lambda_{\gamma} \\
\end{array}$
& $\begin{array}{c} 
  1.026^{+0.048}_{-0.051} \\ 
  -0.024^{+0.025}_{-0.021} \\
\end{array}$ 
& $\begin{array}{c} 
  +0.928, +1.127 \\ 
  -0.068, +0.023 \\
\end{array}$ \\ \hline
\end{tabular}
\end{center}
\caption{The 68$\%$ CL and 95$\%$ CL results on the TGC parameters $g_{1}^{Z},\kappa_{\gamma}$ and $\lambda_{\gamma}$, obtained from the combined two-parameter fit at LEP2. The statistical and systematic errors are included.}
\label{tab:lep2}
\end{table}
\section{Measurements of TGCs at Hadron Colliders}
\begin{figure}[htb]
\begin{center}
\epsfxsize=2.25in
\epsfysize=1.25in
\epsfbox{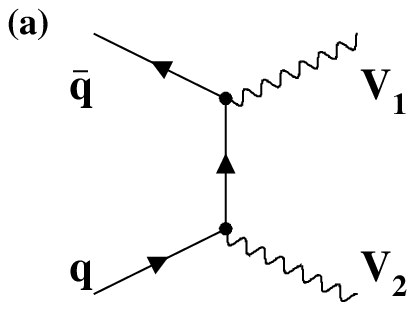}
\epsfxsize=2.25in
\epsfysize=1.25in
\epsfbox{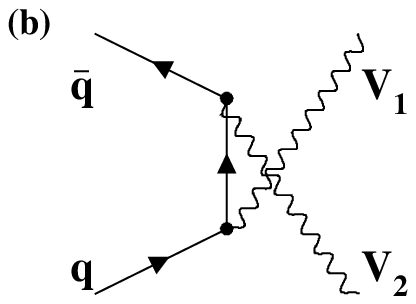}
\epsfxsize=1.5in
\epsfysize=1.in
\epsfbox{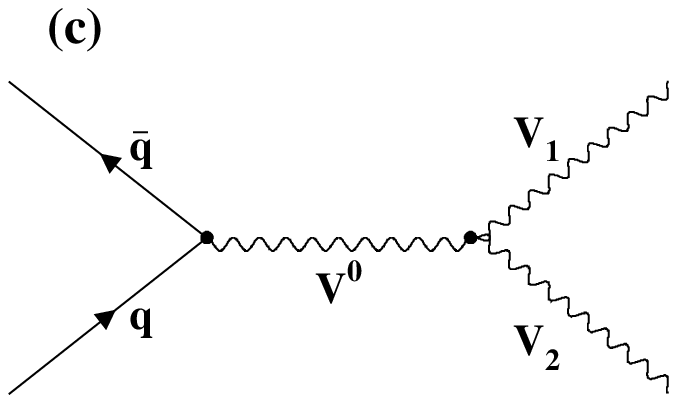}
\caption[bla]{Feynman diagrams for vector boson pair production in $p\bar{p}$ collisions. $V_{0}$=$V_{1}$=$W$ and $V_{2}$=$\gamma$ for $\gamma W$ production. $V_{0}$=$\gamma$ or $Z$, $V_{1}$=$W^{+}$ and $V_{2}$=$W^{-}$ for $WW$ production. $V_{0}$=$V_{1}$=$W$ and $V_{2}$=$Z$ for $WZ$ production. $V_{0}$=$\gamma$ or $Z$, $V_{1}$=$Z$ and $V_{2}$=$\gamma$ for $Z\gamma$ production.}
\label{fig:vvv1}
\end{center}
\end{figure}
\begin{figure}[htb]
\begin{center}
\epsfxsize=6.0in
\epsfysize=4.0in
\epsfbox{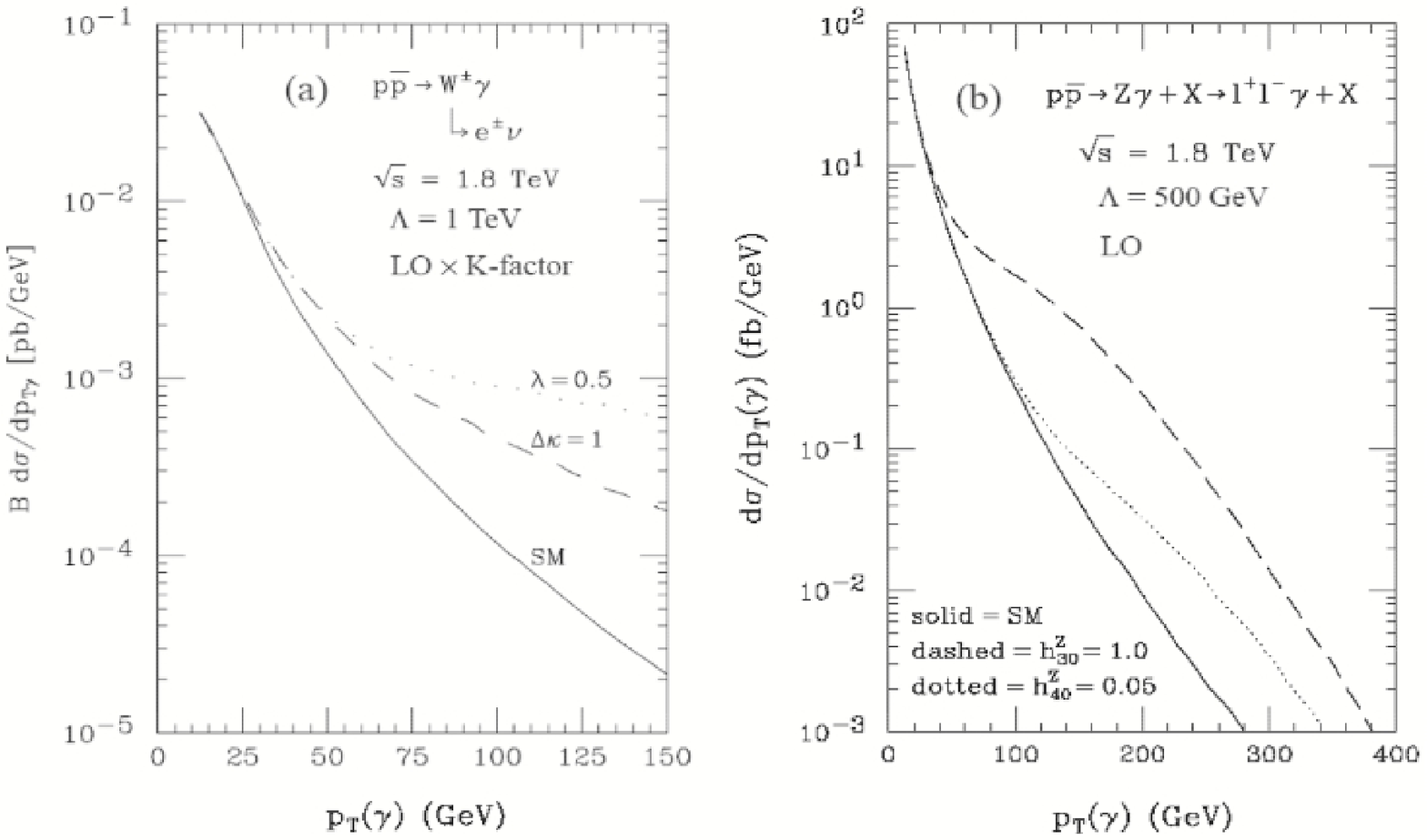}
\caption[bla]{Distributions of the photon $p_{T}$ for $W\gamma$ and $Z\gamma$ production at the Tevatron in the Standard Model and in the presence of the anomalous couplings.}
\label{fig:vvv2}
\end{center}
\end{figure}
The largest hadron collider so far, where the TGCs have been measured is the Tevatron at Fermilab, using $p\bar{p}$ collisions at $\sqrt{s}\approx 2$ TeV. Di-boson production channels in $p\bar{p}$ collisions at two Tevatron experiments, CDF and D0 are shown in Fig.~\ref{fig:vvv1} resulting in $W\gamma, WZ$ and $W^{+}W^{-}$ final states. The TGCs measurements are accessible in the \textit{s}-channel process shown in Fig.~\ref{fig:vvv1}\,$c$ providing the study of $WW\gamma$ and $WWZ$ trilinear gauge vertices allowed by the Standard Model at tree level. These vertices have been directly probed at the Tevatron using the collected data in Run I with integrated luminosity of 100 pb$^{-1}$. Among all di-boson final states the process with a final state $W\gamma$ has the largest production cross-section ($\approx 110$ pb) while the production cross-section for the production of two massive gauge bosons varies between 5 and 10 pb. 
\par
The feature of the $p\bar{p} \rightarrow W\gamma$ process in the Standard Model is the existence of the radiation zeros \cite{zeros} at corresponding $\theta$ values (angle between a photon and incoming quark in the center-of -mass system). For those angles the helicity amplitudes \cite{zero} vanish. In the presence of anomalous TGCs these amplitudes are finite increasing the average photon transverse momentum $p_{T}$. Thus, the photon $p_{T}$ distribution is sensitive to the anomalous couplings as it is shown in Fig.~\ref{fig:vvv2}. This quantity is used at Tevatron for searches of anomalous TGCs.
\subsection{Analysis of final states at CDF and D0 experiments}
In the Tevatron Run I analysis of di-boson events the following final states are considered: $W\gamma \rightarrow l\nu\gamma$, $WW \rightarrow l\nu l\nu$, $WW/WZ \rightarrow q\bar{q}l\nu$, $WZ \rightarrow q\bar{q}l^{+}l^{-}$, $Z\gamma \rightarrow l^{+}l^{-}\gamma$ and $Z\gamma \rightarrow \nu\bar{\nu}\gamma$. Except for the $Z\gamma \rightarrow \nu\bar{\nu}\gamma$, only the $W/Z$ decays into the electron and/or muon have been studied since these events are characterized with the isolated leptons with a high $p_{T}$.
\begin{enumerate}
\item \textbf{$W\gamma \rightarrow l\nu_{l}\gamma$} \\
events are characterized with a high $p_{T}$ electron or muon, a large missing transverse energy indicating the presence of the $W$ boson and a high $p_{T}$ isolated photon. The requirements on the pseudo-rapidity\footnote{$\eta = -\ln(\tan\theta/2)$, where $\theta$ is the polar angle with respect to the beam axis.} range for the lepton and photon detection and on the photon $p_{T}$ are given in \cite{wudka}. The separation between the photon and lepton is large enough to reduce the background contributing from radiative $W$ decays shown in Fig.~\ref{fig:vvv3}. An improvement of $\approx 10\%$ in sensitivity to the anomalous couplings is achieved by D0 experiment imposing a cut on the $W$ boson transverse mass $M_{T}$, defined as $M_{T}=\sqrt{2p_{T}^{e}p_{T}^{\nu}(1-\cos\phi^{e\nu})}$ where $\phi^{e\nu}$ is the azimuthal angle between the lepton and neutrino. The dominant background comes from $W+jet$ production where the jet fluctuates to a neutral meson such as a $\pi^{0}$ and decays into two photons. In both experiments, limits on the TGC parameters are obtained from binned maximum likelihood fit to the photon $p_{T}$ distribution. More details on selection criteria could be found in \cite{d0_gamaW} for D0 and in \cite{cdf_gamaW} for CDF. 
\begin{figure}[htb]
\begin{center}
\epsfxsize=1.75in
\epsfysize=1.in
\epsfbox{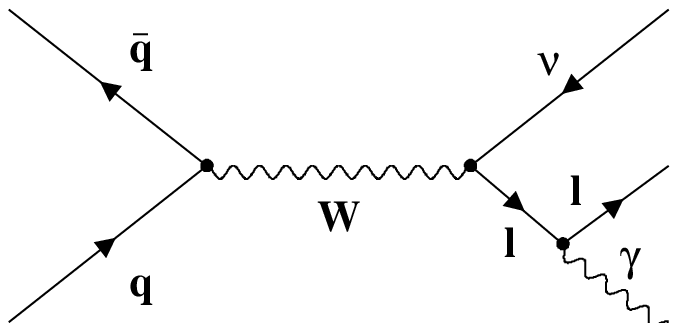}
\caption[bla]{Feynman diagram for the $W$ boson production with radiative $W$ boson decay: the charged lepton radiates a photon by bremsstrahlung leading to the same final state $l\nu\gamma$ as the diagram in Fig.~\ref{fig:vvv1}$c$.}
\label{fig:vvv3}
\end{center}
\end{figure}
\item \textbf{$WW \rightarrow l\nu l^{'}\nu^{'}$} ($e\nu e\nu, e\nu\mu\nu$ and $\mu\nu\mu\nu$) \\
events are characterized by two isolated leptons and missing transverse energy. The selection criteria concerning the pseudo-rapidity for the lepton are similar as used for $l\nu\gamma$ final state. In addition, the events that contain hadronic energy in the calorimeters are rejected in order to suppress the background coming from $p\bar{p} \rightarrow t\bar{t}+X \rightarrow W^{+}W^{-}b\bar{b}+X$ and the cut on the visible and missing transverse energies are imposed. CDF used the total cross-section to set the limits in TGCs while D0 used a binned likelihood fit to the measured $p_{T}$ or $E_{T}$ of two leptons in each event. The kinematic information used in the analysis at D0 provides tighter constraints on anomalous couplings. The different selection criteria at D0 \cite{d0_llnunu} and CDF \cite{cdf_llnunu} are shortly described in \cite{wudka}.
\item \textbf{$WW/WZ \rightarrow q\bar{q}l\nu$} and \textbf{$WZ \rightarrow q\bar{q}l^{+}l^{-}$} \\
events are characterized with two jets with invariant mass consistent with the $W$ boson or the $Z$ boson mass, a high $p_{T}$ lepton and missing $E_{T}$ due to the neutrino for the $q\bar{q}l\nu$ final state or, two high $p_{T}$ leptons and two jets for the $q\bar{q}l^{+}l^{-}$ final state. The main difference in the analysis of these channels at CDF and D0 is that at CDF a cut is imposed on the reconstructed boson transverse momentum and the limits on TGCs are extracted from the number of events which survived the cut while D0 uses a binned likelihood fit to the $W$ boson $p_{T}$ distribution. Other applied cuts are similar for both experiments and can be found in \cite{cdf_jjlnu} for CDF and in \cite{d0_jjlnu} for D0. The main background originate from $W+ \leq 2$ jets with $W \rightarrow e\nu$ in D0, and multi-jet production where one jet is misidentified as an electron and it is small in the region sensitive to the anomalous TGCs.
\item \textbf{$Z\gamma \rightarrow l^{+}l^{-}\gamma$} ($l=e,\mu$) \\
events are characterized with two leptons and the high energy photon. The photon selection criteria are similar to those for the $W\gamma$. More detailed description of the analysis for D0 are given in \cite{d0_Zgama} and in \cite{cdf_Zgama} for CDF. The main source of the background comes from $Z+jet$ production where the jet fakes a photon or an electron and from QCD multi-jet and direct photon production where one or more jets are misidentified as electrons or photons. Limits on the anomalous TGCs are obtained using a binned likelihood fit to the photon $E_{T}$ distribution.
\item \textbf{$Z\gamma \rightarrow \nu\bar{\nu}\gamma$} \\
was measured for the first time by D0 and this final state shows a higher sensitivity to anomalous TGCs (to $ZZ\gamma$ and $Z\gamma\gamma$) than $Z\gamma \rightarrow l^{+}l^{-}\gamma$. This final state is characterized with large missing energy due to the two neutrinos and high $E_{T}$ isolated photon. The main advantages of this channel are the following: the radiative decay backgrounds from the charged lepton is not present as it is in $Z \rightarrow l^{+}l^{-}$ and the branching ratio $B(Z\rightarrow \nu\bar{\nu})/B(Z \rightarrow l^{+}l^{-}) \approx 3$. The additional backgrounds (cosmic-ray muons, beam-halo muons, QCD processes, etc.) and the impossibility to reconstruct the $Z$ boson mass are the main disadvantages. Using appropriate selection criteria \cite{d0_nunug,d0_nunug1} the background can be significantly reduced and the limits on the anomalous TGCs are obtained from a maximum likelihood fit to the photon $E_{T}$ spectrum representing the most stringent limits obtained from any decay channel.
\end{enumerate}
\subsection{Results of the TGC measurements at the CDF and D0 experiments}
In both of the two Tevatron experiments limits on the coupling parameters are obtained from a binned maximum likelihood fit to some distribution depending on the final state. In the case of $W\gamma, l\nu l^{'}\nu^{'}$ and $q\bar{q}l\nu$-$q\bar{q}l^{+}l^{-}$ final states the fitted distribution is the $p_{T}$ distribution of the photon, two leptons or of the $W$ boson, respectively. In the case of $l^{+}l^{-}\gamma$ and $\nu\bar{\nu}\gamma$ final states the fitted distribution is the photon $E_{T}$ distribution. Any excess of the number of events in bins at high $p_{T}$ or $E_{T}$ would be a consequence of an anomalous TGC presence and it was not observed in data, so far.
\par
The main difference between the TGCs measurements at LEP and Tevatron is the TGC form factor dependence introduced in the case of $p\bar{p}$ collisions to avoid the unitarity violation at high enough energies. As it was given by equation (\ref{eq:ewsb29}), the unitarity violation is avoided by replacing the TGC parameters ($\Delta\kappa_{\gamma},\lambda_{\gamma}$, etc.) in the following way:
\begin{equation}
TGC(\hat{s})\rightarrow \frac{TGC}{(1+\hat{s}/\Lambda_{FF}^{2})^{2}}
\label{eq:scale}
\end{equation}
where $\Lambda_{FF}$ is the energy scale while $\hat{s}/\Lambda_{FF}^{2}$ introduce the corrections to the TGCs. In the Tevatron TGC analysis the quantity $(1+\hat{s}/\Lambda_{FF}^{2})$ is greater than one and hence any limit obtained for a TGC parameter provides an upper bound on the sensitivity to the corresponding parameter TGC($\hat{s}$). The maximum scale which can be probed with the Tevatron Run I data is of order of $2$ TeV. 
\par
\begin{figure}[htb]
\begin{center}
\epsfxsize=4.in
\epsfysize=2.0in
\epsfbox{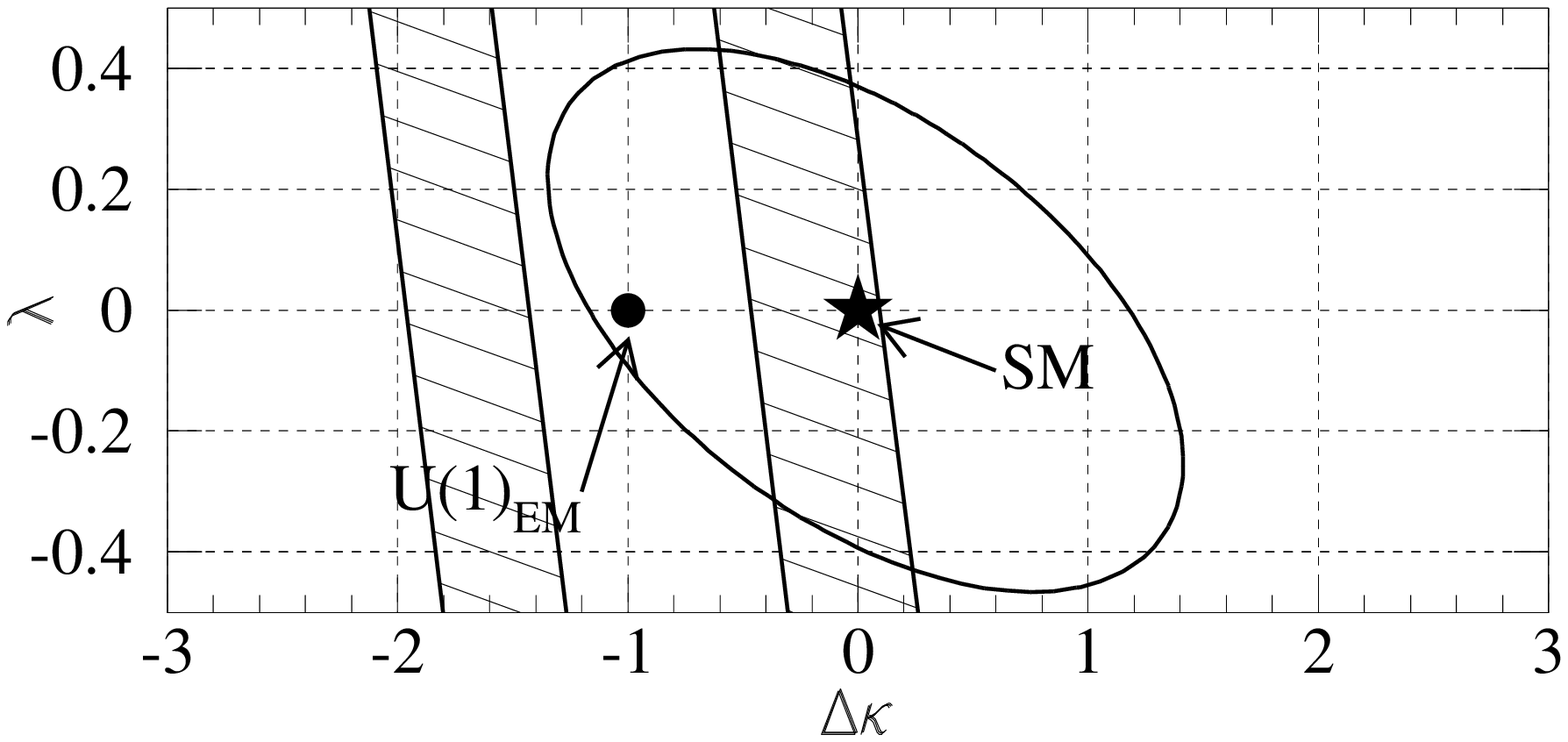}
\epsfxsize=2.0in
\epsfysize=2.0in
\epsfbox{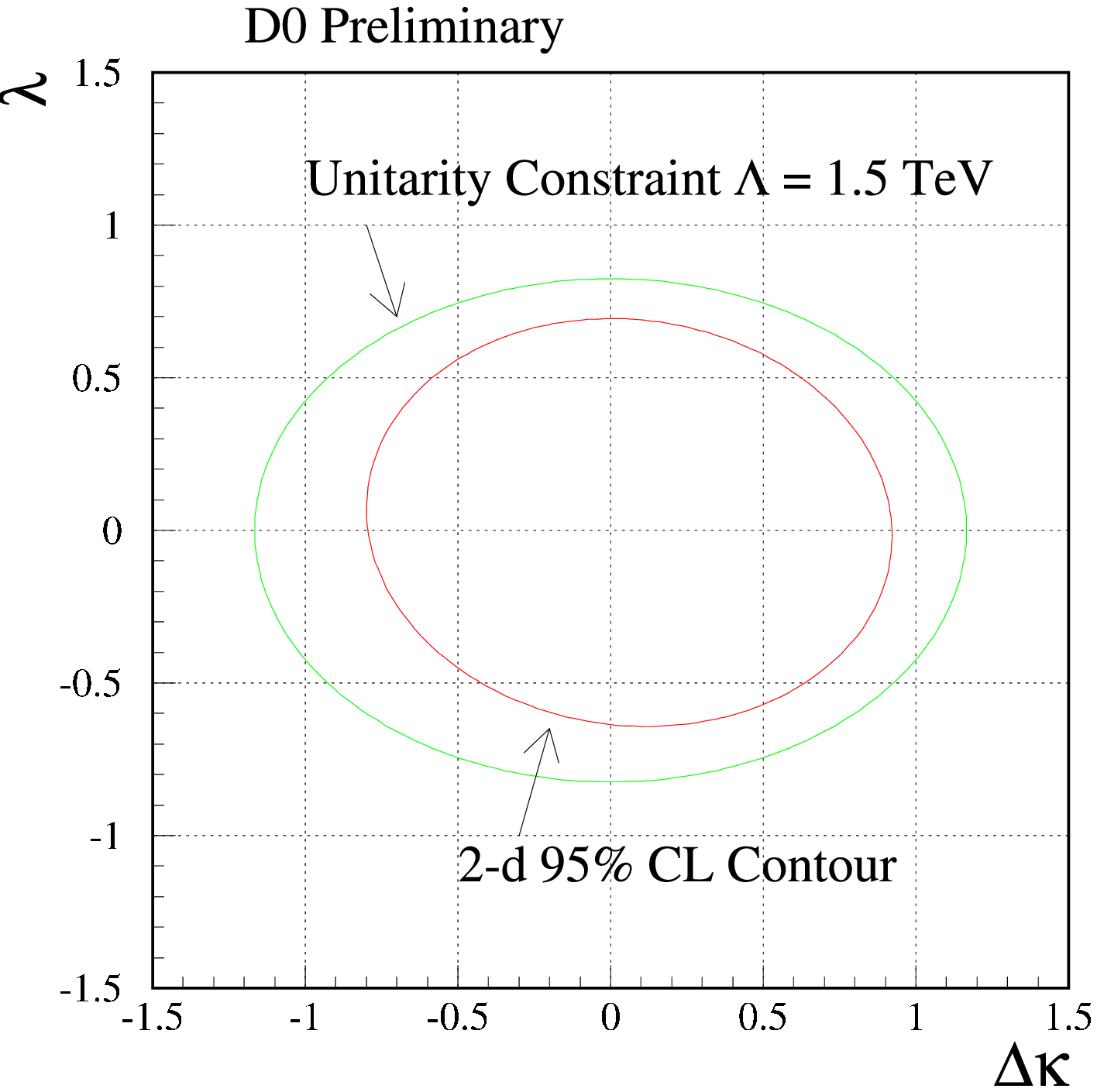}
\caption[bla]{($a$) Limits on anomalous $WW\gamma$ coupling parameters from a maximum likelihood fit to the photon $p_{T}$ spectrum (Run 1a and Run 1b D0 data combined) analyzing the $l\nu\gamma$ final state. The contour is the 95$\%$ CL limit obtained from the single-parameter fit. The shaded regions are the allowed regions from the CLEO measurement of $b\rightarrow s\gamma$. ($b$) Limits on anomalous $WW\gamma/WWZ$ coupling parameters from a maximum likelihood fit to two-lepton $E_{T}$ spectra analyzing $l\nu l^{'}\nu^{'}$ final state. The contour is the 95$\%$ CL limit obtained from single-parameter fit for $\Lambda_{FF}=1.5$ TeV.}
\label{fig:d0}
\end{center}
\end{figure}

The limits on anomalous $WW\gamma$ coupling parameters obtained analyzing $l\nu\gamma$ events collected by D0 are shown in Fig.~\ref{fig:d0}\,$a$, while Fig.~\ref{fig:d0}\,$b$ represents the limits on $\Delta\kappa_{\gamma}$ and $\lambda_{\gamma}$ obtained analyzing $l\nu l^{'}\nu^{'}$ events (where $ll^{'}=e$ or $\mu$) using the constraints $\Delta\kappa_{\gamma}=\Delta\kappa_{Z}$ and $\lambda_{\gamma}=\lambda_{Z}$. The measurement of TGCs using $q\bar{q}l\nu$ and $q\bar{q}l^{+}l^{-}$ final states also use the previous constraints.
\par
The TGC measurements for different final states are summarized in Table \ref{tab:cdf_d0} assuming the energy scale $\Lambda_{FF}=1.5$ TeV for $l\nu\gamma$ and $l\nu l^{'}\nu^{'}$ final states while for the final states $q\bar{q}l\nu$ and $q\bar{q}l^{+}l^{-}$, the value of $\Lambda_{FF}=2$ TeV is used.  
\par
\begin{table}[h]
\begin{center}
\begin{tabular}{|c||c|c|} \hline
 Final State & CDF & D0 \\ \hline\hline
 $W\gamma \rightarrow l\nu\gamma$ & 
$\begin{array}{c} 
-1.8<\Delta\kappa_{\gamma}<2.0 \\ 
-0.70<\Delta\lambda_{\gamma}<0.60 \\
\end{array}$
& $\begin{array}{c} 
-0.93<\Delta\kappa_{\gamma}<0.94 \\
-0.31<\Delta\lambda_{\gamma}<0.29 \\
\end{array}$ \\ \hline
 $WW \rightarrow l\nu l^{'}\nu^{'}$ & & 
$\begin{array}{c} 
-0.62<\Delta\kappa<0.77 \\
-0.52<\Delta\lambda<0.56 \\
\end{array}$ \\ \hline
 $WW/WZ \rightarrow q\bar{q}l\nu,q\bar{q}l^{+}l^{-}$ & 
$\begin{array}{c} 
-0.49<\Delta\kappa<0.54 \\ 
-0.35<\Delta\lambda<0.32 \\
\end{array}$ 
& $\begin{array}{c} 
-0.43<\Delta\kappa<0.59 \\
-0.33<\Delta\lambda<0.36 \\
\end{array}$ \\ \hline
\end{tabular}
\end{center}
\caption{The $95\%$ CL limits on TGCs measured at CDF and D0. $\Lambda_{FF}=1.5$ TeV for $l\nu\gamma$ and $l\nu l^{'}\nu^{'}$ final states. $\Lambda_{FF}=2$ TeV for $q\bar{q}l\nu$ and $q\bar{q}l^{+}l^{-}$ final states assuming $\Delta\kappa_{\gamma}=\Delta\kappa_{Z}$ and $\lambda_{\gamma}=\lambda_{Z}$ for $l\nu l^{'}\nu^{'}$, $q\bar{q}l\nu$, and $q\bar{q}l^{+}l^{-}$ final states.}
\label{tab:cdf_d0}
\end{table}
Table \ref{tab:d0} shows D0 limits on anomalous couplings at $95\%$ CL for two different $\Lambda_{FF}$ values \cite{d0_jjlnu} from a simultaneous fit to the $W\gamma, WW \rightarrow l\nu l^{'}\nu^{'}$, $WZ\rightarrow e\nu ee, \mu\nu ee$ and $WW/WZ \rightarrow q\bar{q}e\nu$ channels using Run I data.
\begin{table}[h]
\begin{center}
\begin{tabular}{|c||c|c|} \hline
 Coupling & $\Lambda_{FF}=1.5$ TeV  & $\Lambda_{FF}=2$ TeV \\ \hline\hline
 $\Delta\kappa_{\gamma}=\Delta\kappa_{Z}$ & -0.27, 0.42 & -0.25, 0.39 \\ \hline
 $\lambda_{\gamma}=\lambda_{Z}$ & -0.20, 0.20 & -0.18, 0.19 \\ \hline
\end{tabular}
\end{center}
\caption{D0 results: lower and upper TGC limits at $95\%$ CL from combined single-parameter fit using $W\gamma$, $WW\rightarrow$ dilepton ($l\nu l^{'}\nu^{'}$), $WZ\rightarrow$ trilepton ($e\nu ee, \mu\nu ee$), $WW/WZ\rightarrow q\bar{q}e\nu$ and $WW/WZ\rightarrow q\bar{q}\mu\nu$ data. Two different $\Lambda_{FF}$ values are used, assuming that $\Delta\kappa_{\gamma}=\Delta\kappa_{Z}$ and $\lambda_{\gamma}=\lambda_{Z}$.}
\label{tab:d0}
\end{table}
\par
Comparing the results from Table \ref{tab:lep1} or \ref{tab:lep2} with the results from Table \ref{tab:d0}, the TGCs at LEP experiments are measured with larger precision than the ones at the Tevatron experiments. The limits on TGCs from Tevatron in Run I, analyzing the data with integrated luminosity of 0.1 fb$^{-1}$, should be improved with the data collected in Run II. The expected luminosity in 2005 is approximately 1 fb$^{-1}$ which is a factor of 10 larger than at Run I. It should increase the sensitivity of the TGC measurements by a factor of 3.
\section{Measurements of TGCs at Future Colliders}
So far, the best limits for the TGCs parameters are done by LEP2 experiments. In the next several years the new data with increased luminosity and at higher energies will be collected at the Large Hadron Collider (LHC at CERN) and the International Linear Collider (ILC). These data are expected to be more sensitive to the effects that generate deviations from the Standard Model TGC values. 
\subsection{Measurements of TGCs at Large Hadron Collider}
\begin{figure}[htb]
\begin{center}
\epsfxsize=3.0in
\epsfysize=2.75in
\epsfbox{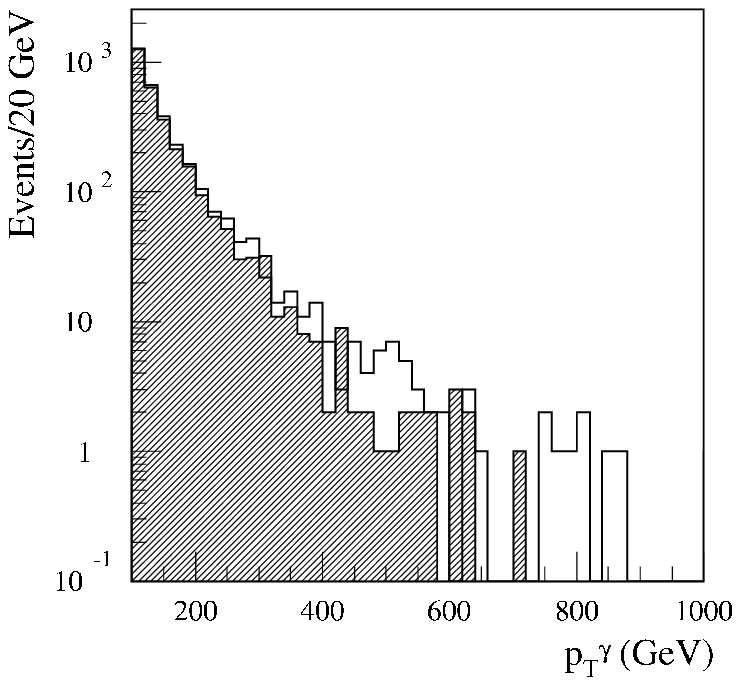}
\epsfxsize=3.0in
\epsfysize=2.75in
\epsfbox{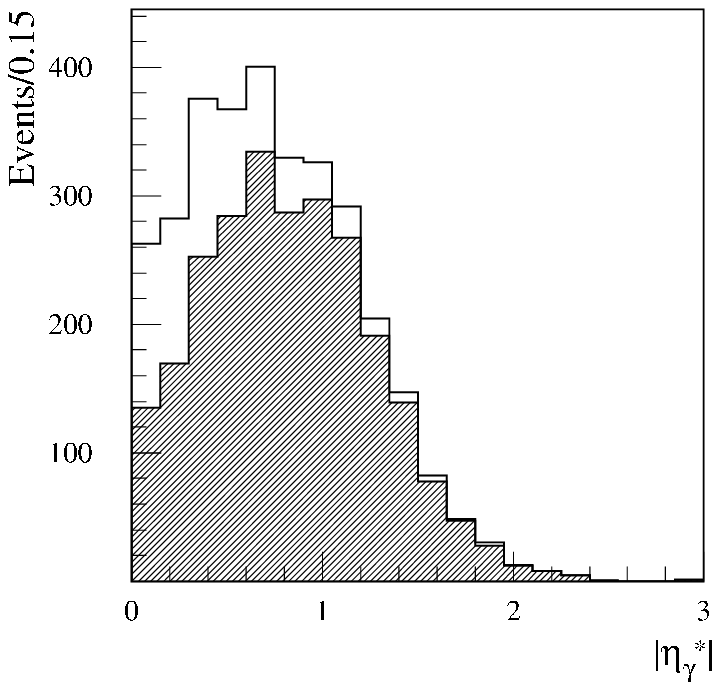}
\caption[bla]{(\textit{Left}): $p_{T}^{\gamma}$ and (\textit{Right}): $|\eta_{\gamma}|$ distributions from $W\gamma$ events for an integrated luminosity of 30 fb$^{-1}$. Shaded histograms represent the Standard Model contributions while the white histograms are the event distributions in the presence of anomalous $\lambda_{\gamma}=0.01$ (\textit{left}) and anomalous $\Delta\kappa_{\gamma}=0.2$ (\textit{right}).}
\label{fig:lhc1}
\end{center}
\end{figure}
\begin{figure}[p]
\begin{center}
\epsfxsize=3.0in
\epsfysize=3.0in
\epsfbox{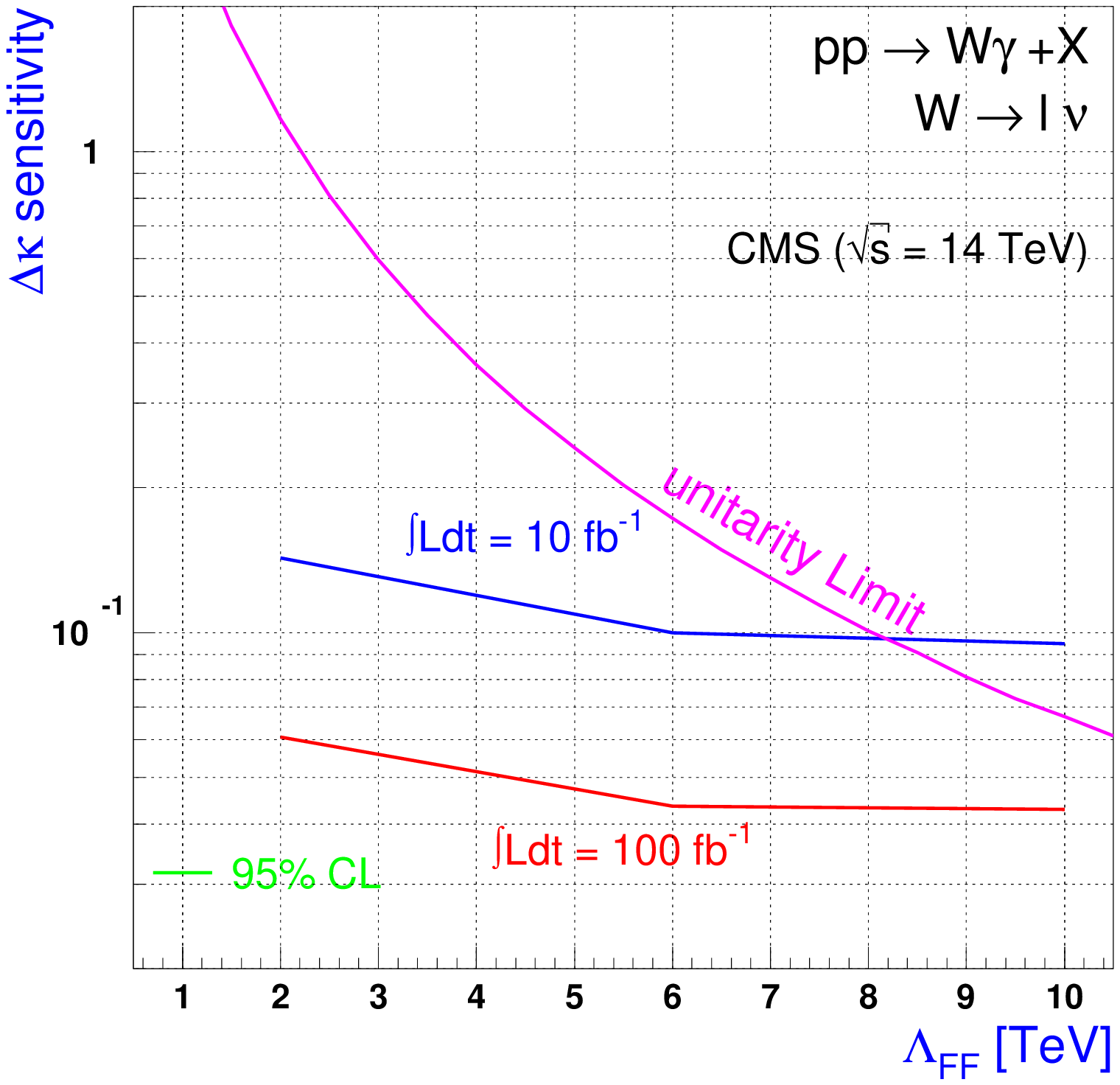}
\epsfxsize=3.0in
\epsfysize=3.0in
\epsfbox{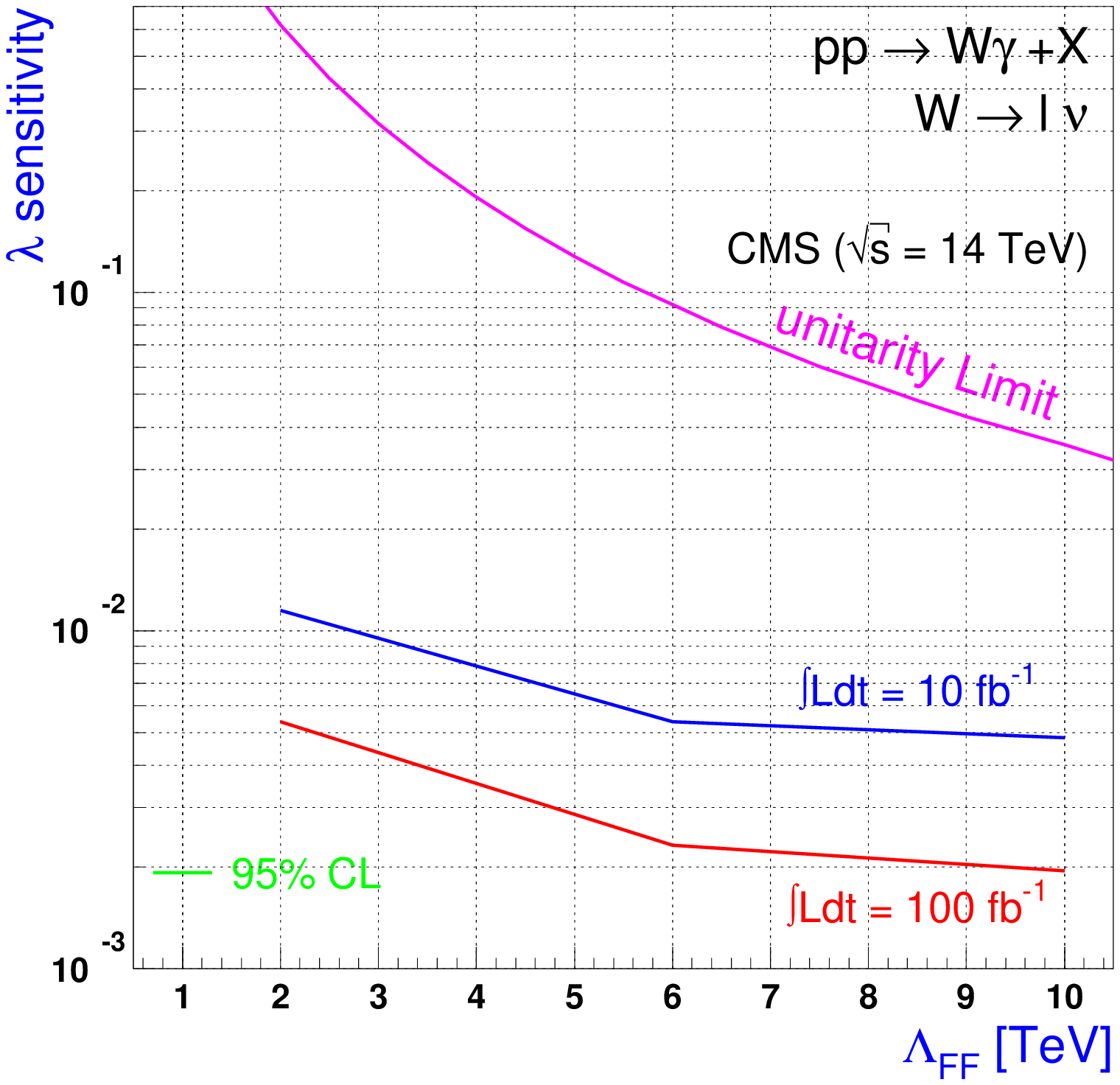}
\epsfxsize=3.0in
\epsfysize=3.0in
\epsfbox{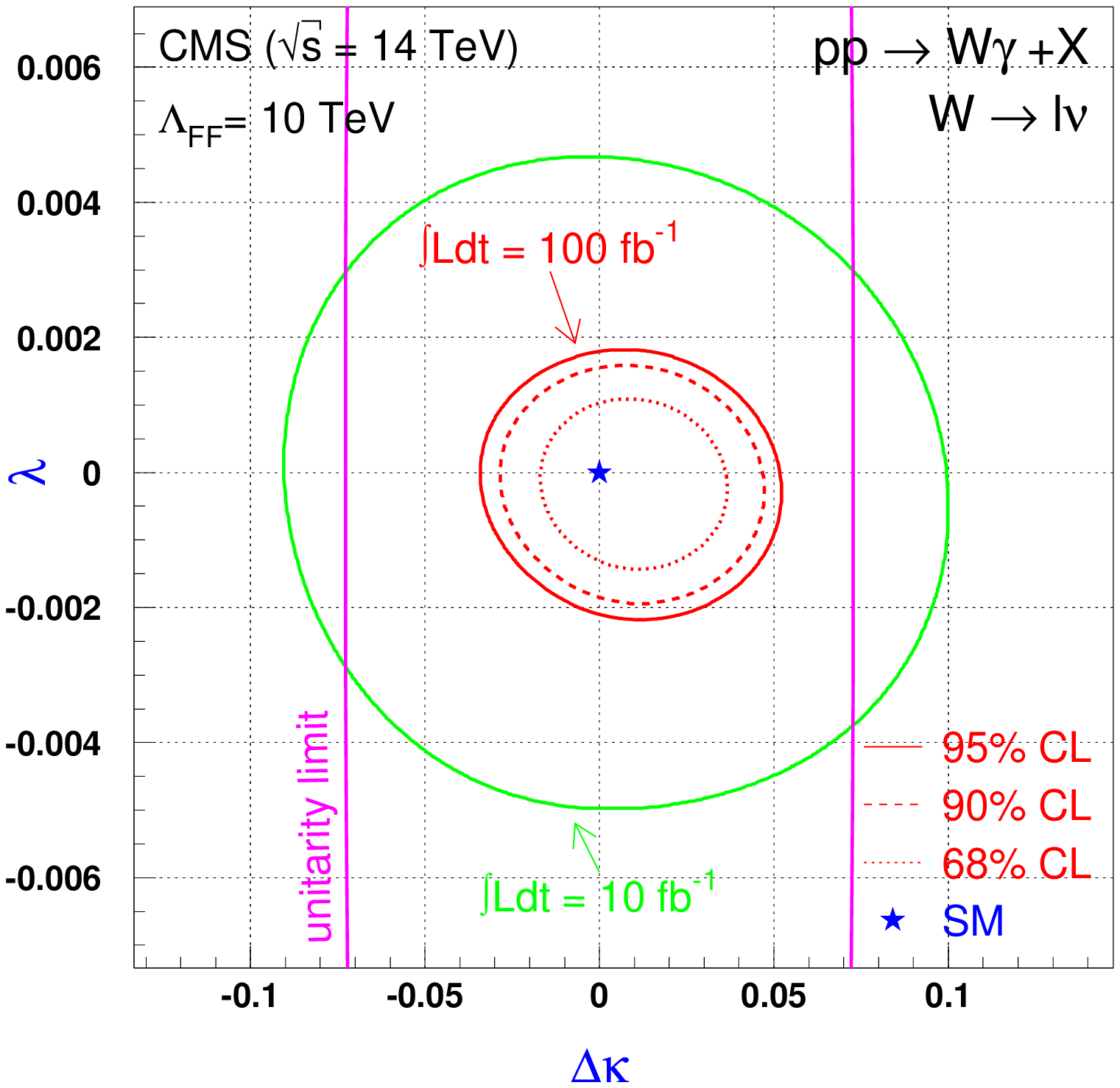}
\caption[bla]{CMS estimations - (\textit{Upper plots}): The expected 95$\%$ CL limits on the anomalous $WW\gamma$ couplings together with the corresponding unitarity limits obtained from two-parameter fits as a function of the form-factor scale $\Lambda_{FF}$, for two different integrated luminosities: 10 fb$^{-1}$ and 100 fb$^{-1}$. (\textit{Lower plot}): $\Delta\kappa_{\gamma}$-$\lambda_{\gamma}$ contour plot for luminosities of 10 fb$^{-1}$ and 100 fb$^{-1}$ assuming the form-factor scale $\Lambda_{FF}=10$ TeV.}
\label{fig:lhc2}
\end{center}
\end{figure}

At hadron collisions the extraction of TGC deviations from the Standard Model predictions is complicated due to the large contributions generated by the QCD corrections \cite{lhc}. The expected limits on TGCs from the reactions $pp\rightarrow W\gamma,WZ$ at the Large Hadron Collider (LHC) are estimated to be of order of $10^{-2}-10^{-3}$ \cite{lhc1}, using the fully leptonic final states ($l=e,\nu$) of $W\gamma$ and $WZ$ production. To probe the $WW\gamma$ couplings in $W\gamma$ analysis at ATLAS and CMS, two sets of the variables can be used: ($M_{W\gamma}, |\eta_{\gamma}|$) and ($p_{T}^{\gamma},\theta_{l}$), where $\theta_{l}$ is the production angle of the charged lepton in the $W$ boson rest frame, $p_{T}^{\gamma}$ is the photon transverse momentum, $M_{W\gamma}$ is the invariant mass of the $W\gamma$ system and $\eta_{\gamma}$ is the rapidity of the photon with respect to the beam direction in the $W\gamma$ system. In the presence of anomalous TGCs, the $p_{T}^{\gamma}$ and $|\eta_{\gamma}|$ event distributions will be enhanced by the events at high $p_{T}^{\gamma}$ values and in the regions with smaller rapidities as it is shown in Fig.~\ref{fig:lhc1}. The distributions of these variables are fitted by a binned maximum-likelihood function in combination with the cross-section information. Similar sets can be used for the $WZ$ final state ($p_{T}^{Z},\theta_{l}$). Table \ref{tab:lhc} contains the ATLAS expected limits on the TGCs at 95$\%$ CL obtained from the single-parameter fits analyzing the $W\gamma$ final state at generator-level and assuming an integrated luminosity of 30 fb$^{-1}$. The CMS experiment estimated the sensitivity to the $WW\gamma$ TGCs for different $\Lambda_{FF}$ assuming an integrated luminosities of 10 fb$^{-1}$ and 100 fb$^{-1}$. A lower plot in Fig.~\ref{fig:lhc2} represents the contour plot in the $\Delta\kappa_{\gamma}$-$\lambda_{\gamma}$ plane estimated by two-parameter fit for the form-factor scale of $\Lambda_{FF}=10$ TeV.
\begin{table}[htb]
\begin{center}
\begin{tabular}{|c||c|c|} \hline
 Coupling & 95$\%$ CL ($M_{W\gamma}, |\eta_{\gamma}|$) & 95$\%$ CL ($p_{T}^{\gamma},\theta_{l}$) \\ \hline\hline
 $\Delta\kappa_{\gamma}\cdot 10^{-2}$ & 3.5  & 4.6 \\ \hline
 $\lambda_{\gamma}\cdot 10^{-2}$      & 0.25 & 0.27 \\ \hline
 $g_{1}^{Z}\cdot 10^{-2}$             & 0.78 & 0.89 \\ \hline
 $\Delta\kappa_{Z}\cdot 10^{-2}$      & 6.9  & 10.0 \\ \hline
 $\lambda_{Z}\cdot 10^{-2}$           & 0.58 & 0.71 \\ \hline
\end{tabular}
\end{center}
\caption{The estimated TGCs statistical precisions at 95$\%$ CL from $W\gamma$ final state with integrated luminosity of 30 fb$^{-1}$ at $\sqrt{s}=14$ TeV. The limits are presented for two different sets of variables used in the analysis. The form-factor scale $\Lambda_{FF}=10$ TeV.}
\label{tab:lhc}
\end{table}

\subsection{Measurements of TGCs at the International Linear Collider}
The polarized $e^{+}e^{-}$ beams foreseen for the International Linear Collider (ILC) are one of the main advantages compared to LEP. The polarized beams allow to disentangle the contributions from $WW\gamma$ and $WWZ$ couplings and to measure them independently of the $SU(2)_{L}\times U(1)_{Y}$ relations. The ILC beams with $\approx \pm$80$\%$ polarized electrons and $\approx \pm$60$\%$ polarized positrons suppress (for $e^{-}_{R}e^{+}_{L}$) or enhance (for $e^{-}_{L}e^{+}_{R}$) the contribution from the \textit{t}-channel $\nu$-exchange diagram shown in Fig.~\ref{fig:tgc1}\,$c$\footnote{A 100$\%$ $e^{-}_{R}e^{+}_{L}$ polarization would lead to the total \textit{t}-channel cancellation.}. In addition, the different coupling of left-/right-handed fermions to the $Z^{0}$ boson influences the contribution from the \textit{s}-channel $Z^{0}$-exchange diagram shown in Fig.~\ref{fig:tgc1}\,$b$ while the contribution from the $\gamma$-exchange diagram (Fig.~\ref{fig:tgc1}\,$a$) does not depend on the beam polarization comparing $e^{-}_{R}e^{+}_{L}$ and $e^{-}_{L}e^{+}_{R}$. 
\par
The analysis for ILC, using the TESLA parameters at $\sqrt{s_{e^{+}e^{-}}}=500$ and 800 GeV, is performed for different electron and positron polarizations splitting up the total luminosity of 500 fb$^{-1}$ equally on both polarizations. (R+L) means that half of the luminosity (250 fb$^{-1}$) corresponds to the 80$\%$ right-handed electrons (R) while the another half corresponds to the 80$\%$ left-handed electrons (L) leaving the positron beam unpolarized. The (RL+LR) combination denotes that in addition the positrons are 60$\%$ polarized with the opposite polarization to the electron polarization. To be able to disentangle between the $WW\gamma$ and $WWZ$ couplings and to get the maximal sensitivity both data sets are fitted at the same time. The limits for TGC measurements at ILC $e^{+}e^{-}$ collisions are estimated analyzing the semi-leptonic $WW$ decay channels using the SDM method for their extraction \cite{menges}. The results of the multi-parameter fits to the coupling parameters, with and without the $SU(2)_{L}\times U(1)_{Y}$ relation between the coupling parameters, are given in Table \ref{tab:tesla500} for $\sqrt{s_{e^{+}e^{-}}}=500$ GeV and in Table \ref{tab:tesla800} for $\sqrt{s_{e^{+}e^{-}}}=800$ GeV. The expected sensitivities to the TGCs are of the order of $10^{-3}$-$10^{-4}$ depending on the coupling. Generally, an increase of the $\sqrt{s_{e^{+}e^{-}}}$ and of the luminosity increases the sensitivity of the TGCs measurements, decreasing the correlations between the couplings. The use of both polarized beams also increases the sensitivity. For the given $\sqrt{s_{e^{+}e^{-}}}$ and the same beam polarization combinations, not using the $SU(2)_{L}\times U(1)_{Y}$ relation, leads to a decrease in the sensitivity of $\Delta g^{Z}_{1}$ and $\lambda_{\gamma}$.
\par
The estimated sensitivity limits are high enough to ensure that the International Linear Collider will be able to probe the New Physics effects at scales beyond its center-of-mass energies. The estimations of the TGCs in $\gamma e$ and $\gamma\gamma$ collisions at ILC using the TESLA parameters, are the subject of this theses.
\begin{table}[htb]
\begin{center}
\begin{tabular}{|c||c|c|c|} \hline
$\begin{array}{c}
\textbf{ILC(TESLA)} \\
\sqrt{s}=500 GeV \\
\end{array}$ &
$\begin{array}{c}
SU(2)_{L}\otimes U{1}_{Y} \\
(R+L) \\
\end{array}$ &
$\begin{array}{c}
SU(2)_{L}\otimes U{1}_{Y}\\
(RL+LR)\\
\end{array}$ &
$\begin{array}{c}
{no} SU(2)_{L}\otimes U{1}_{Y}\\
(RL+LR)\\
\end{array}$ \\ \hline\hline
$\Delta g^{Z}_{1}\cdot 10^{-4}$   & 4.3 & 3.1  & 30.0   \\ \hline
$\Delta\kappa_{\gamma}\cdot 10^{-4}$ & 4.6 & 3.5  & 3.6    \\ \hline
$\lambda_{\gamma}\cdot 10^{-4}$   & 5.5 & 4.6  & 11.0   \\ \hline
$\Delta\kappa_{Z}\cdot 10^{-4}$   &   &      & 5.5    \\ \hline
$\lambda_{Z}\cdot 10^{-4}$        &   &      & 12.0   \\ \hline
\end{tabular}
\end{center}
\caption{ILC expected limits on the TGCs obtained analyzing the semi-leptonic final states in $e^{+}e^{-} \rightarrow W^{+}W^{-}$ at generator-level at $\sqrt{s_{e^{+}e^{-}}}=500$ GeV with integrated luminosity of 500 fb$^{-1}$ with and without the $SU(2)_{L}\times U(1)_{Y}$ relation between the coupling parameters \cite{menges}. The TESLA parameters are used.}
\label{tab:tesla500}
\end{table}
\begin{table}[htb]
\begin{center}
\begin{tabular}{|c||c|c|c|} \hline
$\begin{array}{c}
\textbf{ILC(TESLA)} \\
\sqrt{s}=800 GeV \\
\end{array}$ &
$\begin{array}{c}
SU(2)_{L}\otimes U{1}_{Y} \\
(R+L) \\
\end{array}$ &
$\begin{array}{c}
SU(2)_{L}\otimes U{1}_{Y}\\
(RL+LR)\\
\end{array}$ &
$\begin{array}{c}
{no} SU(2)_{L}\otimes U{1}_{Y}\\
(RL+LR)\\
\end{array}$ \\ \hline\hline
$\Delta g^{Z}_{1}\cdot 10^{-4}$   & 2.4 & 1.7  & 15.9   \\ \hline
$\Delta\kappa_{\gamma}\cdot 10^{-4}$ & 2.4 & 2.0  & 2.1    \\ \hline
$\lambda_{\gamma}\cdot 10^{-4}$   & 2.9 & 2.4  & 3.3   \\ \hline
$\Delta\kappa_{Z}\cdot 10^{-4}$   &   &      & 2.1    \\ \hline
$\lambda_{Z}\cdot 10^{-4}$        &   &      & 3.3   \\ \hline
\end{tabular}
\end{center}
\caption{ILC expected limits on the TGCs obtained analyzing the semi-leptonic final states in $e^{+}e^{-} \rightarrow W^{+}W^{-}$ at generator-level at $\sqrt{s_{e^{+}e^{-}}}=800$ GeV with integrated luminosity of 500 fb$^{-1}$ with and without the $SU(2)_{L}\times U(1)_{Y}$ relation between the coupling parameters \cite{menges}. The TESLA parameters are used.}
\label{tab:tesla800}
\end{table}

\chapter{Photon Colliders}
\section{Compton (back)scattering}
\begin{figure}[htb]
\begin{center}
\vspace{-0.5in}
\epsfxsize=3.0in
\epsfysize=2.0in
\epsfbox{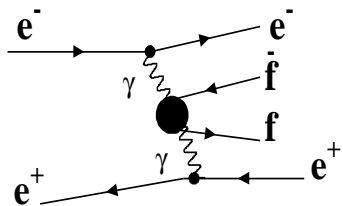}
\vspace{-0.25in}
\caption[bla]{Virtual bremsstrahlung diagram for photon emission from the $e^{+}e^{-}$ beams.}
\label{fig:bremss}
\end{center}
\end{figure}

So far, the only possibility to study two photon physics is in the collisions of the particles from $e^{+}e^{-}$ beams at LEP and other $e^{+}e^{-}$-colliders \cite{lowqcd}. In this scenario, the high energy electron beam radiates a virtual photon via \textit{bremsstrahlung}, that collides with another virtual photon emitted from the opposite beam as it is shown by diagram in Fig.~\ref{fig:bremss}.
The main disadvantages of such a method are a soft energy photon spectrum leading to decreasing photon flux with photon energy. The obtainable virtual photon spectrum is described by the Equivalent Photon Approximation (EPA) \cite{epa} as:
\begin{equation}
f_{\gamma/e}(y,Q^{2})=\frac{\alpha}{2\pi Q^{2}}\left[\frac{1+(1-y)^{2}}{y}-\frac{2m_{e}^{2}y}{Q^{2}}\right]
\label{eq:epa}
\end{equation}
that after the integration over the virtuality $Q^{2}$, leads to the well known Weizs\"{a}cker-Williams Approximation (WWA) \cite{wwa} for the photon spectrum in some interval of virtualities $Q^{2}_{min}$ and $Q^{2}_{max}$:
\begin{equation}
f_{\gamma/e}(y,Q^{2}_{min},Q^{2}_{max})=
{\frac{\alpha}{2\pi}} \left [{\frac{1+{(1-y)}^{2}}{y}} 
\ln{\frac{Q^{2}_{max}}{Q^{2}_{min}}} -2{m^{2}_{e}}y
({\frac{1}{Q^{2}_{min}}}-{\frac{1}{Q^{2}_{max}}})\right]
\label{eq:wwa}
\end{equation}
with $y$ being the energy fraction taken by the photon ($y=E_{\gamma}/E_{e}$) and $Q^{2}_{max,min}$ are the upper and lower boundaries on the virtuality $Q^{2}\approx 2E_{e}E_{e}^{'}(1-\cos\theta)$ where $E_{e}$ and $E_{e}^{'}$ are the electron energies before and after the scattering and $\theta$ is the angle of the scattered electron.
\par
On the other hand, the International Linear Collider (ILC) offers the opportunity to produce more energetic photons than those produced by virtual bremsstrahlung and thus, gives the opportunity to study $\gamma\gamma$ and $\gamma e$ collisions at high center-of-mass energies, close to those in $e^{+}e^{-}$ collisions. Unlike at the $e^{+}e^{-}$ storage rings, at ILC each bunch of electrons is used only once that makes it possible to convert them into high energy photons and to bring two of them into collision. The realization of the photon collider at TESLA is based on such a principle. A photon collider demands a high power laser \cite{tdr6} to produce high energy photon beams from one electron beam ($\gamma e$-collider) or both electron beams ($\gamma\gamma$-collider) via the process of Compton backscattering. In this way the photons with an energy of about 80$\%$ of the electron energy are available for a collisions. This provides a $\sqrt{s_{\gamma e}}\approx 0.9\sqrt{s_{e^{+}e^{-}}}$ at the $\gamma e^{-}$ collider and $\sqrt{s_{\gamma\gamma}}\approx 0.8\sqrt{s_{e^{+}e^{-}}}$ at the $\gamma\gamma$-collider. In contrast to the 'bremsstrahlung' two-photon physics, high center-of-mass energies at a photon collider will give the possibility to explore some physics not available at $e^{+}e^{-}$ collisions as the resonant production of neutral Higgs boson \cite{higgs}. Some SUSY particles can be studied better at a photon collider due to higher statistical accuracy or due to higher accessible masses \cite{susy}. However, $e^{+}e^{-}$ and a photon collider are complementary in searches and measurements of new physics phenomena \cite{zerwas,hagiw}.
\par
\begin{figure}[htb]
\begin{center}
\epsfxsize=6.0in
\epsfysize=3.0in
\epsfbox{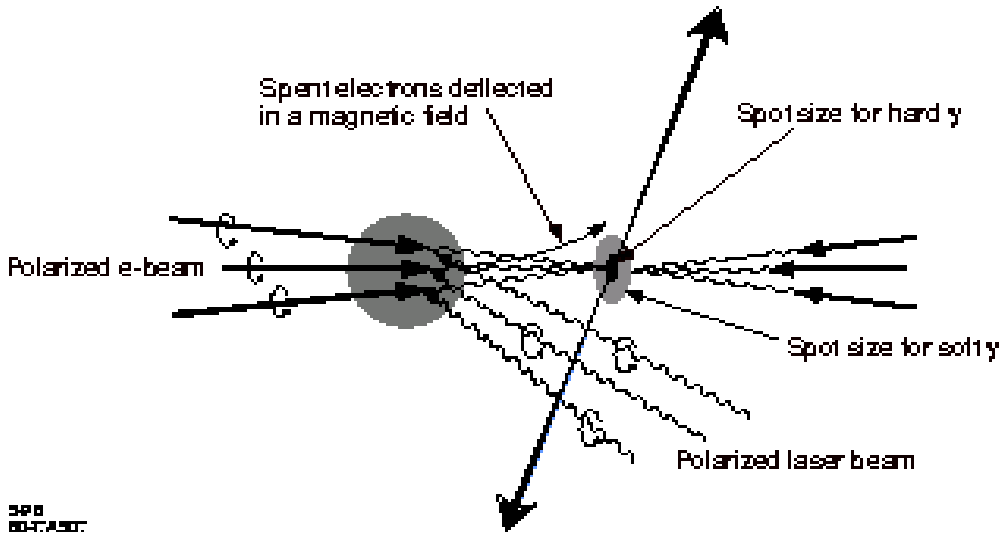}
\caption[bla]{The creation of the high energy photon beam by Compton backscattering of laser photons off beam electrons.}
\label{fig:compton1}
\end{center}
\end{figure}

The basic technique used for realization of a photon collider consists of using low energy, circularly polarized laser photons that collide with the longitudinally polarized high energy electrons as it is shown in Fig.~\ref{fig:compton1}.
The polarization of backscattered laser photons depends on the polarization of the initial laser photons and electrons and they receive a major fraction of the incoming electron energy. Hence, high energy $\gamma e$ and $\gamma\gamma$ interactions can be provided depending on whether only one or both electron beams are converted into high energy photons. While the high energy $\gamma\gamma$ collisions can be studied at the $\gamma\gamma$-collider if both electron beams are converted, the high energy $\gamma e$ collisions one can explore in two different modes. If only one electron beam is converted into high energy photons which collide with the opposite electron beam the $\gamma e$-collider works in the \textit{real mode}. In the case of the $\gamma\gamma$-collider, the $\gamma e$ collisions occur as a background to the $\gamma\gamma$ interactions, and the $\gamma e$-collider runs in the \textit{parasitic mode}.
\section{The characteristics of the photon spectrum}
Desirable characteristics of a photon spectrum are high energy and highly polarized photons. The electrons of energy $E_{e}\approx 250$ GeV are converted into high energy photons with an energy $E_{\gamma}\approx 0.8 E_{e}$ in a conversion region which is approximately 2 mm in front of the interaction point. If the initial energy of the laser photon is $\omega_{0}$ and the laser photon scatters off a high energy electron within the small collision angle $\theta$ called \textit{crossing angle} shown in Fig.~\ref{fig:compton2}, then the energy of the scattered photon $\omega$ and the photon scattering angle $\theta_{\gamma}$ are correlated as \cite{ginzburg}:
\begin{figure}[htb]
\begin{center}
\epsfxsize=3in
\epsfysize=1.25in
\epsfbox{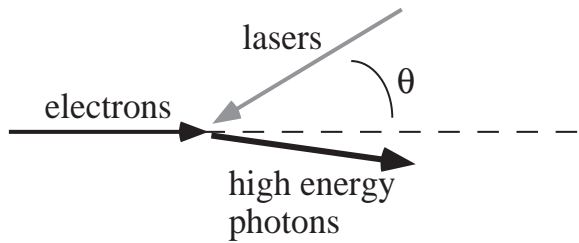}
\caption[bla]{Simplified scheme of Compton backscattering of laser photon off an electron within a small crossing angle $\theta$.}
\label{fig:compton2}
\end{center}
\end{figure}
\begin{equation}
\omega={\frac{\omega_{m}}{1+(\theta_{\gamma}/\theta_{0})^{2}}}.
\label{eq:omega}
\end{equation}
This means that converted photons with lower energies are backscattered at a larger angles relative to the direction of the incoming electrons and (\ref{eq:omega}) describes the photon's energy-angular correlation. The maximum energy of the scattered photon $\omega_{m}$ is given by:
\begin{equation}
\omega_{m}(>\omega)={\frac{x}{x+1}}E_{e}
\label{eq:omegamax}
\end{equation}
and $\theta_{0}$ is 
\begin{equation}
\theta_{0}={\frac{m_{e}c^{2}}{E_{e}}}\sqrt{1+x}\approx \frac{0.511\sqrt{1+x}}{E_{e}/\textrm{TeV}}[\mu\textrm{rad}].
\label{eq:theta0}
\end{equation}
The energy of scattered photons is influenced by the value of $x$ which is given by:
\begin{equation}
x={\frac{4E_{e}\omega_{0}}{m_{e}^{2}c^{4}}}{\cos^{2}}{\frac{\theta}{2}} =
19 \left[\frac{E_{e}}{\textrm{TeV}} \right] \left[\frac{\mu\textrm{m}}{\lambda} \right]
\label{eq:ixs}
\end{equation}
i.e. depends on the product $E_{e}\omega_{0}$ and on the crossing angle. For very small angles $\theta$ ($\cos^{2}\theta/2\approx 1$) the energy of scattered photon depends on the electron energy and on the laser wavelength $\lambda$. From (\ref{eq:omegamax}) it is clear that an increase of $x$ will result in a higher maximal energy of the scattered photons. On the other hand, the maximal energy of the scattered photons is limited by the $x$ value if the energy of a single created photon is high enough to create $e^{+}e^{-}$ pairs in a collision with laser photons. The $e^{+}e^{-}$ pair production decreases the spectral luminosity $dL/d(\sqrt{s_{\gamma\gamma,\gamma e}})$, leading to a loss of high energy photons. Thus, the upper limit on $x$ is given by the threshold for the $e^{+}e^{-}$ pair creation ($\omega_{m}\omega_{0}>m^{2}c^{4}$) which should not exceed the value of $4.8$. Having the beam electrons with an energy of 250 GeV and very small angles $\theta$ the corresponding wavelength of the laser photons, deduced from (\ref{eq:ixs}), is:
\begin{equation}
\lambda = 4.2 E_{e}[\textrm{TeV}] \approx 1.06 \mu\textrm{m}
\label{eq:lambda1}
\end{equation}
which corresponds to the wavelength of the solid state lasers based on Neodymium (Nd:Yag, Nd:Glass). Having the laser photons of $\lambda = 1.06\mu$m, their energy corresponds to:
\begin{equation}
\omega_{0} = \frac{0.3}{E_{e}}[\textrm{TeV}]\approx 1.17 \textrm{eV}.
\label{eq:lambda2}
\end{equation}
\par
\begin{figure}[htb]
\begin{flushleft}
\epsfxsize=3.0in
\epsfysize=3.0in
\epsfbox{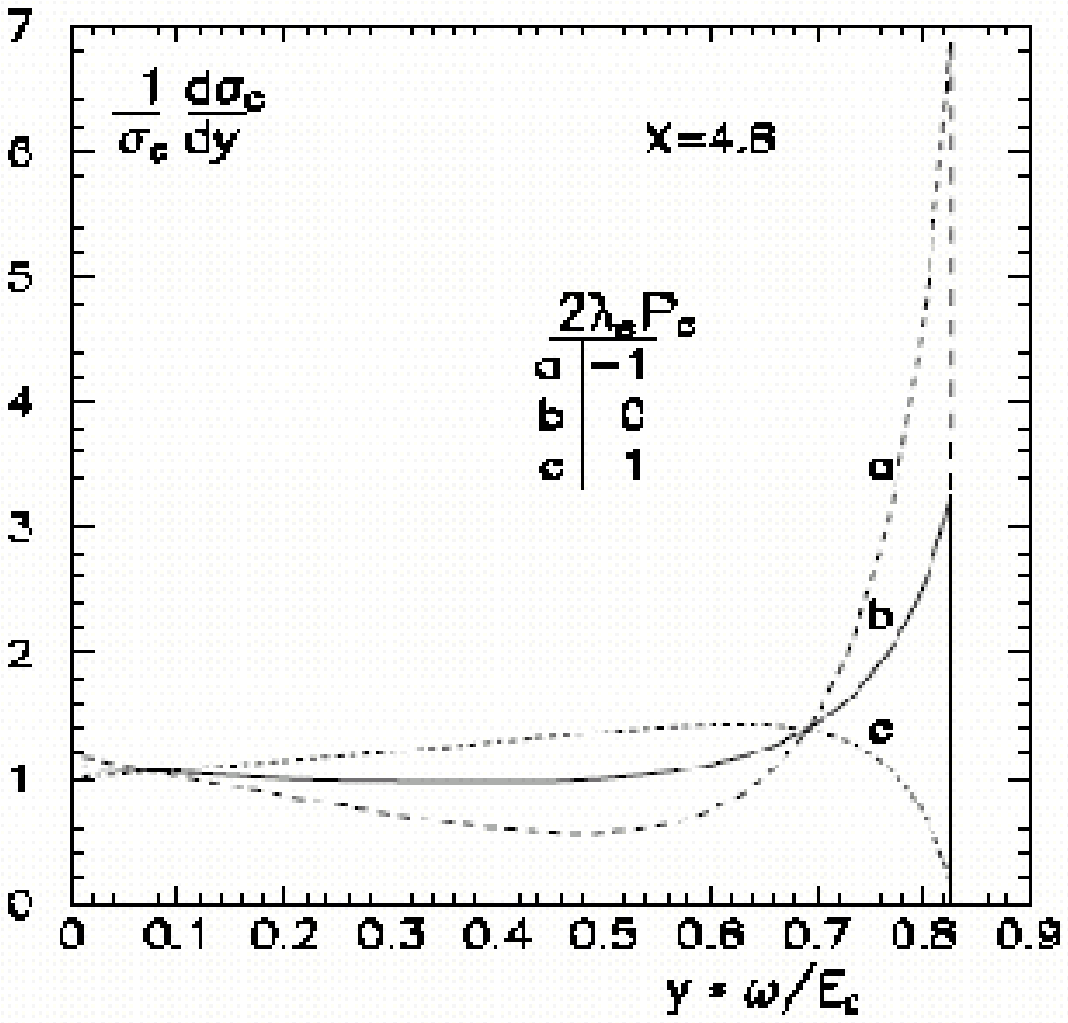}
\epsfxsize=3.0in
\epsfysize=3.0in
\epsfbox{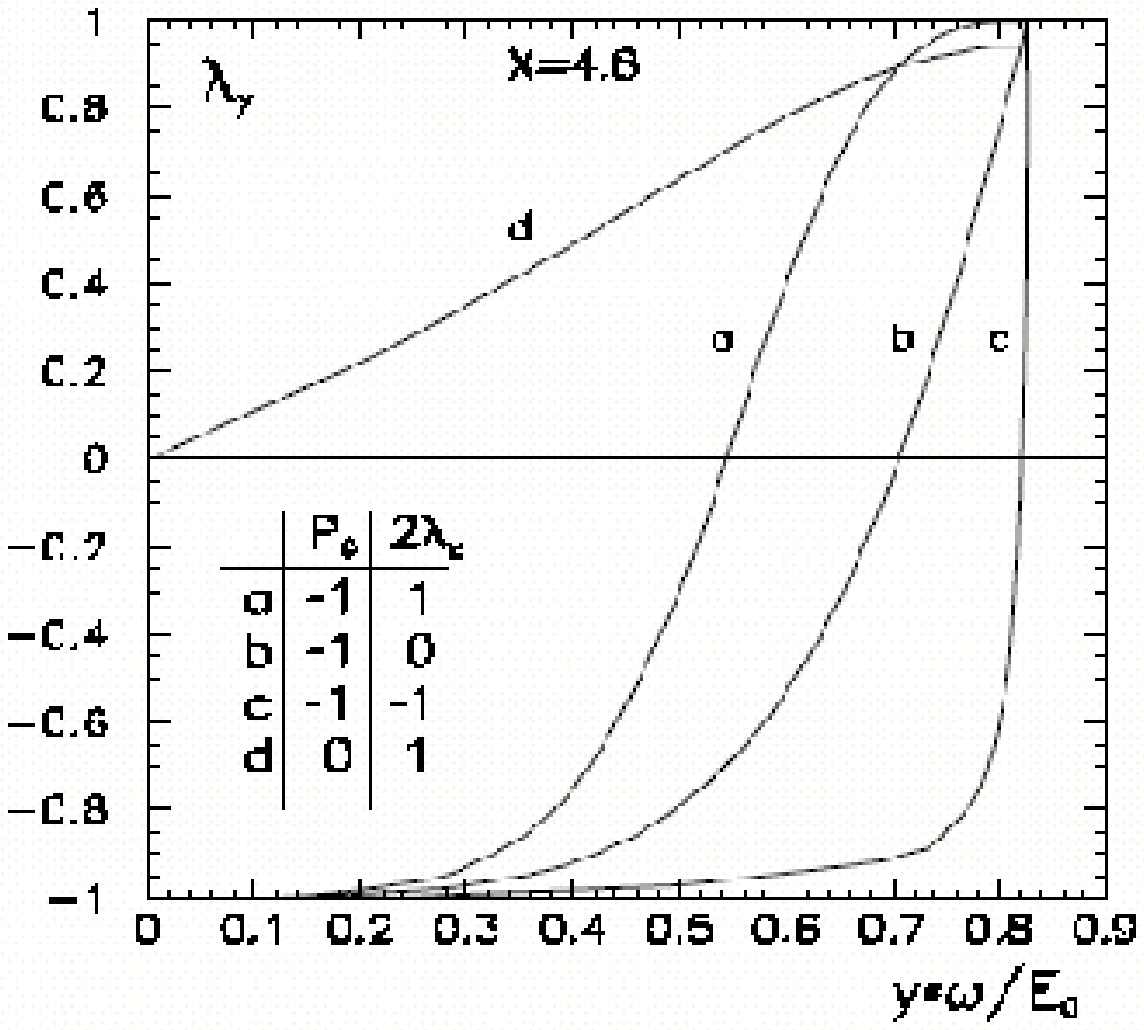}
\caption[bla]{(\textit{Left plot}): Spectra of the Compton scattered photons for different helicities of the laser photons and electrons ($E_{e}\equiv E_{0}$) for $x=$ 4.8. The solid line corresponds to $(2\lambda_{e}P_{c})=-1$, the dashed line corresponds to $(2\lambda_{e}P_{c})=0$ and dotted line corresponds to $(2\lambda_{e}P_{c})=+1$. (\textit{Right plot}): Mean circular polarization of the backscattered photons.}
\label{fig:spectra}
\end{flushleft}
\end{figure}

The energy spectrum of the scattered photons is defined by the Compton cross-section $\sigma_{c}$ and strongly depends on the mean laser photon helicity ($|P_{c}|\leq 1$) and mean electron helicity ($|\lambda_{e}|\leq 1/2$) via their product (2$\lambda_{e}P_{c}$) as:
\begin{equation}
{\frac{1}{\sigma_{c}}}{\frac{d\sigma_{c}}{dy}} = {\frac{2\sigma_{0}}{x\sigma_{c}}}
\left[{\frac{1}{1-y}} + 1-y -4r(1-r) + 2\lambda_{e}P_{c}x(1-2r)(2-y) \right]
\label{eq:compcross}
\end{equation}
where $y$ is the fraction of the electron energy taken by the photon ($y=\omega/E_{e}$) while $r$ and $\sigma_{0}$ are defined as:
\begin{equation}
r=\frac{y}{x(1-y)}, \sigma_{0}=\pi r_{e}^{2}=2.5\cdot10^{-25}\textrm{cm}^{2}.
\label{eq:rdef}
\end{equation}
Depending on the value of the product $(2\lambda_{e}P_{c})$, the energy spectrum given by (\ref{eq:compcross}) behaves differently, especially in the high energy region which is of the biggest interest for the present study. Taking the three different values of $(2\lambda_{e}P_{c})$ to be $\pm$1 and 0, the obtained energy spectra are shown in Fig.~\ref{fig:spectra}\,($left$). The most peaked spectrum in the high energy region corresponds to the combination with the opposite helicities of initial photons and electrons, i.e., to $(2\lambda_{e}P_{c})=-1$.
The backscattered photons with a maximal energy have the energy spectrum nearly two times higher if we use polarized photons and electrons with opposite helicities than in case of unpolarized initial photons and electrons. This improves the energy spread of the photon beam as it is clearly seen comparing the distributions $a$ and $b$ in Fig.~\ref{fig:spectra}\,($left$).
\par
The degree of the final photon circular polarization $\langle\lambda_{\gamma}\rangle$ is given by:
\begin{equation}
\langle \lambda_{\gamma}\rangle = 
\frac{-P_{c}(2r-1)[(1-y)^{-1}+(1-y)]+2\lambda_{e}xr[1+(1-y)(2r-1)^{2}]}
{(1-y)^{-1}+1-y-4r(1-r)-2\lambda_{e}P_{c}xr(2-y)(2r-1)}
\label{eq:10}
\end{equation}
and shown in Fig.~\ref{fig:spectra}\,($right$) for the optimal $x$ value. It depends on different $\lambda_{e}$ and $P_{c}$ values being zero only if both the initial photon and electron beams are unpolarized. In the high energy region $\langle\lambda_{\gamma}\rangle$ is very close to 100$\%$ (for $|P_{c}|\simeq$ 1), in a wide energy range only for the helicity combination $(2\lambda_{e}P_{c})=-1$. Even if the degree of the electron polarization is not maximal but $|P_{c}|=1$, $\langle\lambda_{\gamma}\rangle$ is still $\approx 100\%$ for photons with the maximum energy while the energy region with highly polarized photons is shifted to lower energies. If $(2\lambda_{e}P_{c})=+1$ or if the electron beam is unpolarized, the final photons are highly polarized but the degree of polarization decreases fast with an energy in the high energy region. Using unpolarized laser photons the backscattered photons are never maximally polarized.
\section{Multiple Scattering and Non-Linear Effects}
The expression for the maximum energy of the scattered photon given by (\ref{eq:omegamax}) is applicable in the so called ``linear case'', i.e. if one laser photon is scattered off one electron. In the reality, due to the high laser density, several laser photons can interact simultaneously with one electron or high energy photon (``non-linear case'') leading to the non-linear effects. These effects will be described later in the text.
The influence of the non-linear effects on the characteristics of $\gamma e$ and $\gamma\gamma$ collisions is mainly reflected in the photon energy spectrum distributions and in the spectral luminosity, while the polarization of backscattered photons propagating in the laser wave does not change for $x<$ 4.8 and $(2\lambda_{e}P_{c})=-1$. The behavior of the spectral $\gamma e$ and $\gamma\gamma$ luminosities is analogous to that of the photon energy spectrum inducing a widening of the spectra of high energy photons and a generation of additional peaks corresponding to radiation of higher harmonics due to multi-photon scattering.
\par
Moving through the laser field the multiple scattering occurs if the single high energy electron interacts consecutively with several laser photons  producing the low energy electrons and photons. The energy of scattered photons is smaller than desirable for high energy collisions and contributes to the low energy part of the photon spectrum leading to its broadening. The broadness of the photon spectrum induces the broadness of the spectral luminosities in the $\gamma\gamma$ and $\gamma e$ collisions. Due to the energy-angular correlation in Compton scattering mainly the high energy photons collide at the interaction point if $b$ is sufficiently large leading to narrow invariant $\gamma e$ and $\gamma\gamma$ mass spectra $W_{\gamma e,\gamma\gamma}$. Thus, the spectral luminosity depends on the distance $b$ between the conversion region and the interaction point. The decrease of the distance $b$ leads to a larger contribution of low energy photons to the luminosity spectra increasing the probability for their collisions at the interaction point.
\par
\begin{figure}[htb]
\begin{center}
\epsfxsize=6.25in
\epsfysize=3.0in
\epsfbox{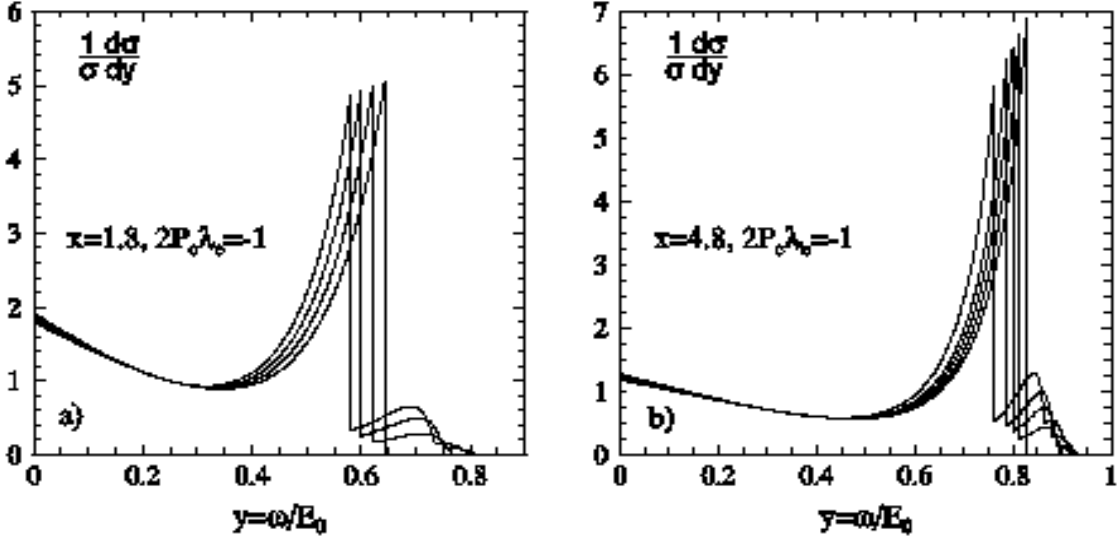}
\caption[bla]{The evolution of the Compton spectra for different values of the parameter $\xi^{2}$: 0, 0.1, 0.2, 0.3, 0.5, for curves from right to left, and $2\lambda_{e}P_{c}=-1$ taking (\textit{Left plot}): $x=$ 1.8 and (\textit{Right plot}): $x=$ 4.8. ($E_{e}\equiv E_{0}$).}
\label{fig:nonlinearity2}
\end{center}
\end{figure}

In order to reach a high probability for $e^{-}\rightarrow \gamma$ conversion the density of laser photons at the conversion region is so high that the multi-photon processes occur. As it was mentioned, a single high energy electron or high energy photon can interact simultaneously with several laser photons $\gamma_{0}$ as:
$$
e^{-} + n\gamma_{0} \rightarrow e^{-} + \gamma^{'}
$$
$$
\gamma + n\gamma_{0} \rightarrow e^{+}e^{-}.
$$
These interactions which represent non-linear effects, characterized by the \textit{nonlinearity} parameter $\xi^{2}$:
\begin{equation}
\xi^{2}=\left(\frac{eF\hbar}{m\omega_{0}c}\right)^{2} = 
n_{\gamma}\left(\frac{4\pi\alpha}{m^{2}\omega_{0}}\right)
\label{eq:nonlinearity1}
\end{equation}
that depends on the laser field strength $F$ i.e. on the density of the laser photons $n_{\gamma}$. Since the transverse motion of the scattered electron in a strong electromagnetic laser field increases its effective mass as $m^{2} \rightarrow m^{2}(1+\xi^{2})$, according to (\ref{eq:theta0}) the photon is scattered within larger angles $\theta_{\gamma}$. Since $x$ is modified to $x/(1+\xi^{2})$, the maximum energy fraction of the scattered photons (\ref{eq:omegamax}) is modified too, in the following way:
\begin{equation}
y_{max}=\frac{\omega_{m}}{E_{e}}= \frac{x}{(1+x+\xi^{2})},
\label{eq:shift}
\end{equation}
shifting the spectrum to lower energies. As the laser field increases, multi-photon scattering results in additional peaks (higher harmonics) at high $\omega$ as it is shown in Fig.~\ref{fig:nonlinearity2}, widening the spectra. As a consequence of widening, the height of the first harmonics decreases in comparison with that in Fig.~\ref{fig:spectra}\,$(left)$ ($\xi^{2}=0$) for the same value of $x$. Thus, a low $\xi^{2}$ value is desirable to keep a sharp edge of the luminosity spectrum and a small shift of $y_{max}$ to lower energies.
%
\section{$\gamma\gamma$ and $\gamma e$ Luminosity Measurements at a Photon Collider}
In the reality, the luminosity spectra at photon collider are broad and electrons and photons may have various polarizations. Due to energy-angular correlations and beam-beam induced interactions, $\gamma\gamma$ and $\gamma e$ luminosities in Compton backscattering cannot be described by convolution of some photon spectra. The processes at the conversion region and interaction point are very complex and decrease the accuracy of the luminosity prediction by simulation. Thus, all luminosity properties have to be measured experimentally. In general, the total number of events in $\gamma\gamma$ or $\gamma e$ collisions with transversally polarized photons is given by \cite{lumi}:
\begin{equation}
\begin{array}{ccl}
\dot{dN_{\gamma i}} & = & dL_{\gamma i}[d\sigma^{np} + (\lambda_{\gamma}\lambda_{i})d\sigma^{c}]
\label{eq:dngama}
\end{array}{}
\end{equation}
where $i$ in index denotes the particle of the opposite beam, either electron in the $\gamma e$ collisions or photon in the $\gamma\gamma$ collisions, $\sigma^{np}=\frac{1}{2}(\sigma_{a}+\sigma_{b})$ is the unpolarized cross-section for a given process, $\lambda_{\gamma,i}$ is the mean degree of circular polarization and $\sigma^{c}=\frac{1}{2}(\sigma_{a}-\sigma_{b})$. $\sigma_{a,b}$ are the polarized cross-sections for the two different initial helicity states, '$a$' if the helicities of interacting beam particles are the same and '$b$' if the helicities of interacting beam particles are the opposite. In case of $i=e$ ($\gamma e$ collisions) the parameters for the second particle in (\ref{eq:dngama}) are replaced by the electron spin vector, e.i. by double mean electron helicity $2\lambda_{e}$. In case of $\gamma\gamma$ collisions the equation (\ref{eq:dngama}) can be rearranged in such a way that $\dot{dN_{\gamma\gamma}}$ is expressed via $dL_{0}$ and $dL_{2}$ as:
$$
\dot{dN_{\gamma\gamma}}=(dL_{0}-dL_{2})d\sigma^{c},
dL_{0}\sim\frac{1}{2}(1+\lambda_{1}\lambda_{2}),
dL_{2}\sim\frac{1}{2}(1-\lambda_{1}\lambda_{2})
$$
where the indexes $0$ and $2$ in $dL_{0,2}$ denote the two different photon helicity combinations and $1$ and $2$ in $\lambda_{1,2}$ the helicities of the two colliding beams. Thus, for the estimation of $\dot{dN_{\gamma\gamma}}$ one should measure either $dL_{\gamma\gamma}$ and $(\lambda_{\gamma}\lambda_{\gamma})$ or $dL_{0}$ and $dL_{2}$ as a function of $\sqrt{s}$.
\subsection{Measurement of $\gamma\gamma$ Luminosity}
The main characteristics of the system produced in $\gamma\gamma$ collisions are its invariant mass $W_{\gamma\gamma}=\sqrt{4(\omega_{1}\omega_{2})}$ and rapidity $\eta=0.5\ln(\omega_{1}\omega_{2})$ where $\omega_{1,2}$ are given by (\ref{eq:omega}). The spectral luminosities, $dL/dW_{\gamma\gamma}d\eta$, should be measured for two different photon helicity combinations: $J_{Z}=0$ and $|J_{Z}|=2$. The best process for the $\gamma\gamma$ luminosity measurement is two fermion (lepton) production, i.e. $\gamma\gamma \rightarrow l^{+}l^{-}$, where $l^{\pm}$ could be either $e^{\pm}$ or $\mu^{\pm}$. Assuming that colliding photons are arbitrarily circularly polarized, with helicities $\lambda_{1}$ and $\lambda_{2}$ ($|\lambda_{1,2}|\neq 1$)\footnote{This corresponds to the luminosity spectrum with mixed helicities as it will be used further in theses.}, the two fermion production cross-section is given by:
\begin{equation}
\sigma_{\gamma\gamma \rightarrow l^{+}l^{-}} = 
{\frac{1+\lambda_{1}\lambda_{2}}{2}}\sigma_{0} + 
{\frac{1-\lambda_{1}\lambda_{2}}{2}}\sigma_{2}
\label{eq:total_lumi_gg}
\end{equation}
where $\sigma_{0}$ and $\sigma_{2}$ are the cross-sections for two different helicity combinations defined as:
\begin{equation}
\sigma_{0}(|\cos\theta|<a) 
\approx {\frac{4\pi \alpha^{2}}{W_{\gamma\gamma}^{2}}}
\frac{8m_{l}^{2}}{W_{\gamma\gamma}^{2}}
\left[\frac{1}{2}\ln\left(\frac{1+a}{1-a}\right)+\frac{a}{1-a^{2}}\right]
\label{eq:lumi1}
\end{equation}
and
\begin{equation}
\sigma_{2}(|\cos\theta|<a)
\approx {\frac{4\pi \alpha^{2}}{W_{\gamma\gamma}^{2}}}
\left[2\ln\left(\frac{1+a}{1-a}\right)-2a\right].
\label{eq:lumi2}
\end{equation}
Since $\sigma_{0}$ is largely suppressed by a factor $8m_{l}^{2}/W_{\gamma\gamma}^{2}$ compared to $\sigma_{2}$ ($\sigma_{0} \ll \sigma_{2}$), the number of collected events for $|J_{Z}|=2$ according to (\ref{eq:total_lumi_gg}), allows us to measure the luminosity $dL_{2}/dWd\eta$ as:
\begin{equation}
dN_{\gamma\gamma \rightarrow l^{+}l^{-}}
\approx dL\frac{1-\lambda_{1}\lambda_{2}}{2} \sigma_{2} \equiv dL_{2}\sigma_{2}
\label{eq:24}
\end{equation}
while the first term in (\ref{eq:total_lumi_gg}) for the $J_{Z}=0$ state is very close to zero due to the suppression factor.
\par
Inverting the helicity of one photon beam by simultaneously changing the helicity signs of the laser photons and the electrons, the spectrum of scattered photons remains the same but the product of $\lambda_{1}\lambda_{2}$ changes its sign. That allows to measure the spectral luminosity for the $J_{Z}=0$ state, before the helicity flipping.
\par
An additional process that can be used for the $\gamma\gamma$ luminosity measurement and as an independent check is $\gamma\gamma \rightarrow \l^{+}\l^{-}\l^{+}\l^{-}$. It can be measured with high accuracy only at large angles where the cross-section is smaller by $\approx 3$ orders of magnitude compared to the pair production cross-section.
\subsection{Measurement of $\gamma e$ Luminosity}
The main characteristics of the system produced in $\gamma e$ collisions is its invariant mass $W_{\gamma e}=\sqrt{4(\omega E_{e})}$. The spectral luminosity $dL/dW_{\gamma e}$, should be measured for two different photon-electron helicity combinations: $|J_{Z}|=1/2$ (same photon-electron helicities) and $|J_{Z}|=3/2$ (photon-electron opposite helicities) that strongly depend on the circular photon polarization $\lambda_{\gamma}$ and longitudinal electron polarization $\lambda_{e}$. The best channels for $\gamma e$ luminosity measurements are $\gamma e \rightarrow \gamma e$ and $\gamma e \rightarrow  eZ$. The cross-section for the QED process is given as:
\begin{equation}
d\sigma_{\gamma e \rightarrow \gamma e} =
(1-2\lambda_{e}\lambda_{\gamma})d\sigma_{3/2} +
(1+2\lambda_{e}\lambda_{\gamma})d\sigma_{1/2}
\label{eq:ge_lumi_1}
\end{equation}
where $d\sigma_{3/2}$ and $d\sigma_{1/2}$ are given by:
\begin{equation}
d\sigma_{3/2} = \frac{\pi\alpha^{2}}{2W_{\gamma e}}(1-\cos\theta_{\gamma})
d(\cos\theta_{\gamma})
\label{eq:ge_lumi_2}
\end{equation}
\begin{equation}
d\sigma_{1/2} = \frac{\pi\alpha^{2}}{2W_{\gamma e}}
(\frac{4}{1-\cos\theta_{\gamma}})d(\cos\theta_{\gamma})
\label{eq:ge_lumi_3}
\end{equation}
and the second term of (\ref{eq:ge_lumi_1}) is the dominating one. The measurement of $L_{1/2}$ and $L_{3/2}$ with a high accuracy ($\approx 1/\sqrt{N_{i}},i=1/2,3/2$) is possible in a similar way as in case of the $\gamma\gamma$ luminosities; for example, to measure the number of events for $|J_{Z}|=3/2$ ($N_{3/2}$) to deduce $L_{3/2}$ and to measure the number of events for $|J_{Z}|=1/2$ ($N_{1/2}$) by inverting the helicity of one beam to the opposite. In the parasitic $\gamma e$ mode, luminosities for each direction should be measured separately.

\chapter{Detector at a Photon Collider}
In general, the design of a detector depends on the physics program for which the detector is planed to be built but it is always desirable that the main particle characteristics like charge and momenta are measured with a high accuracy having the energy resolution as high as possible. If the physics program is oriented to searches for new particles a detector should cover as much as possible of the solid angle while the forward detectors are needed to cover the region close to the beam pipe.
\begin{figure}[htb]
\begin{center}
\epsfxsize=6.0in
\epsfysize=3.0in
\epsfbox{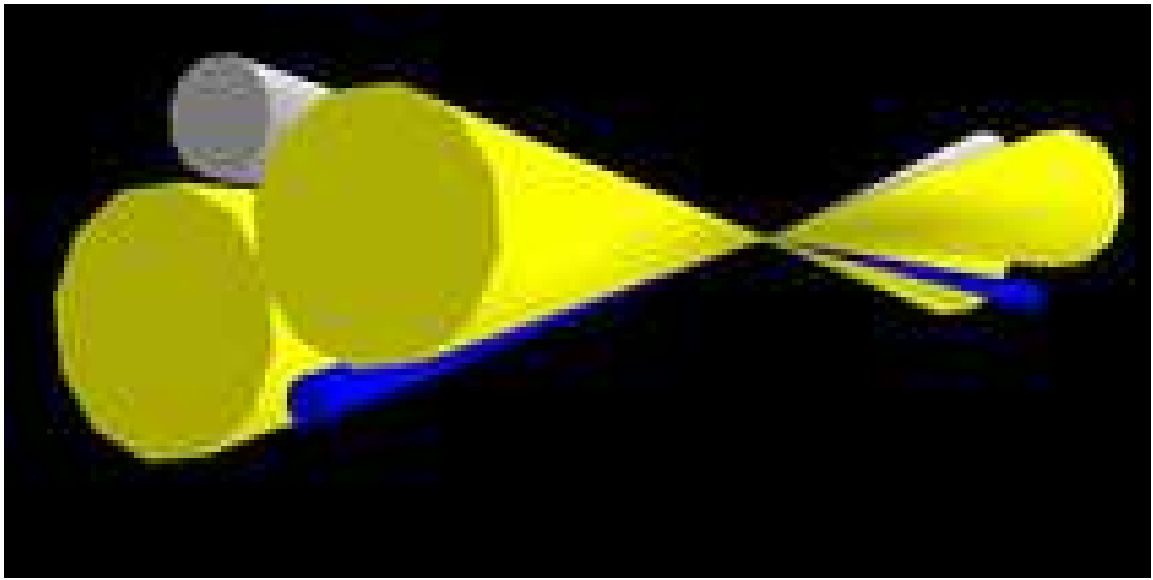}
\caption[bla]{Layout of the beam pipes. Yellow ones represent the incoming and outgoing laser beam pipes. The blue ones represent the incoming electron beam pipes while the grey represent the outgoing electron beam pipes.}
\label{fig:pipes}
\end{center}
\end{figure}
Since the study on a photon collider presented in this theses assumes that a photon collider is an extension of the TESLA $e^{+}e^{-}$-collider, the detector for a  photon collider matches in many points with the proposed TESLA detector \cite{tdr4}. 
\par
\begin{figure}[htb]
\begin{center}
\epsfxsize=5.0in
\epsfysize=3.5in
\epsfbox{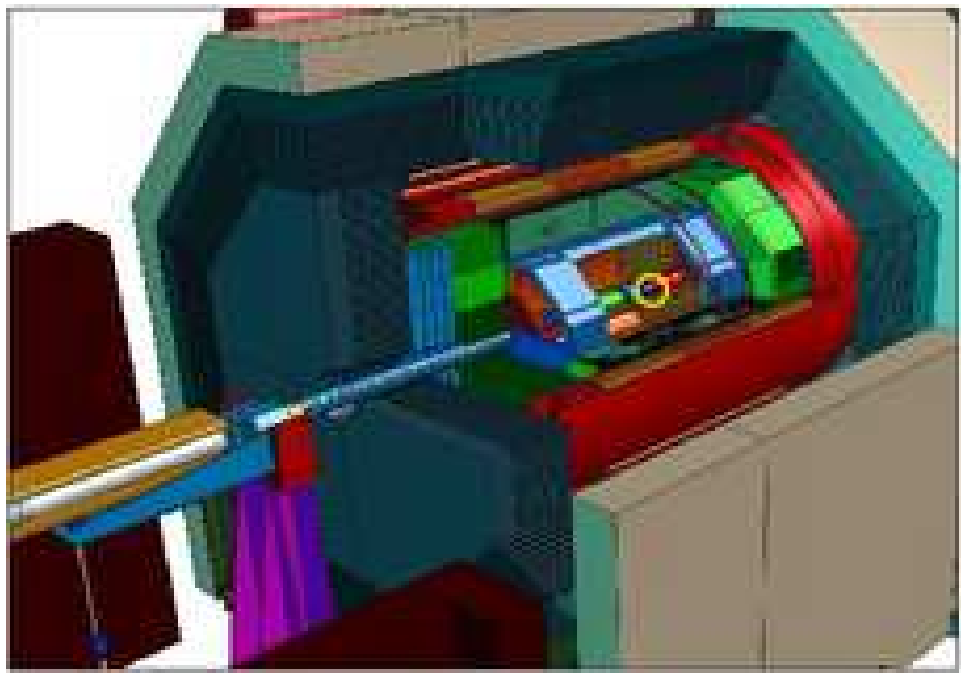}
\caption[bla]{TESLA detector.}
\label{fig:guido}
\end{center}
\end{figure}

The detector at TESLA, shown in Fig.~\ref{fig:guido}, is an universal detector planed to provide high precision measurements within the different physics topics in the Standard Model and beyond. The track momentum and spatial resolution, jet flavor tagging, energy flow and hermeticity are essential for the foreseen precision studies at TESLA so that the design of the detector is led by these requirements. The energy flow concept based on the optimization of the jet energy resolution measuring each particle individually with the corresponding sub-detector - charged particles with a tracker, photons with a electromagnetic calorimeter and neutral hadrons with a hadronic calorimeter is valid for a photon collider too. That means that the sub-detector components are mainly the same as for the $e^{+}e^{-}$-detector but with some exceptions. 
\par
Some parts inside the $\gamma\gamma$-detector had to be redesigned due to the implementation of the laser system within. The main difference consists of the absence of low angle calorimeters (low-angle tagger and low-angle luminosity calorimeter \cite{tdr4}) in the very forward region due to the lack of the space caused by the implementation of the laser beam pipes shown schematically in Fig.~\ref{fig:pipes} which are necessary for the realization of a photon collider. Thus, the minimal polar angle at a photon collider is 7$^{\circ}$. Two interaction regions, the first where the $e^{-}\rightarrow \gamma$ conversion occurs and the second interaction region $\approx$ 2 mm away from the conversion region are the main characteristics of a photon collider.
The distance $b$ between these two regions is limited by the repulsion of the two electron beams. Since both initial beams have the same charge sign\footnote{A photon collider is based on $e^{-}e^{-}$ beams.}, close to the interaction point they are influenced by the field of the ongoing (opposite) beam at a distance of approximatively 2-3 electron bunch lengths (less than 0.9 mm) in beam direction. Thus, the $e^{-}\rightarrow \gamma$ conversion should occur far enough before the repulsion forces start to act and before the contribution from the low energy photons (from the multiple scattering) becomes too large but close enough to avoid a decrease of the luminosity. Optimizing the value of $b\approx \sigma_{y}/\theta_{\gamma}$ it is chosen to be $b\approx$ 2 mm where $\sigma_{y}$ is the vertical size of an electron bunch and $\theta_{\gamma}$ is the angular spread of scattered photons\footnote{In Fig.~\ref{fig:compton2}, $\theta_{\gamma}$ is the angle between the $z$-axis (dashed line) and the high energy photons.}. In this way, the photon beam size at the interaction point receives equal contributions from the electron beam size and the angular spread of photons from Compton scattering.
\par
Such realization of the interaction region at a photon collider is a source of many interactions considered as a background that disturb an efficient track reconstruction. Thus a study on the optimization of the forward region has been done in order to maximize the shielding of the tracking devices and to bring the amount of produced background to a manageable level. Inside of the detector the tracking devices (vertex detector, forward, central and intermediate trackers as well as forward chambers), electromagnetic and hadronic calorimeters, coil and muon identifier (tail catcher) are positioned.
\section{Tracking System}
\begin{figure}[h]
\begin{center}
\epsfxsize=4.0in
\epsfysize=2.5in
\epsfbox{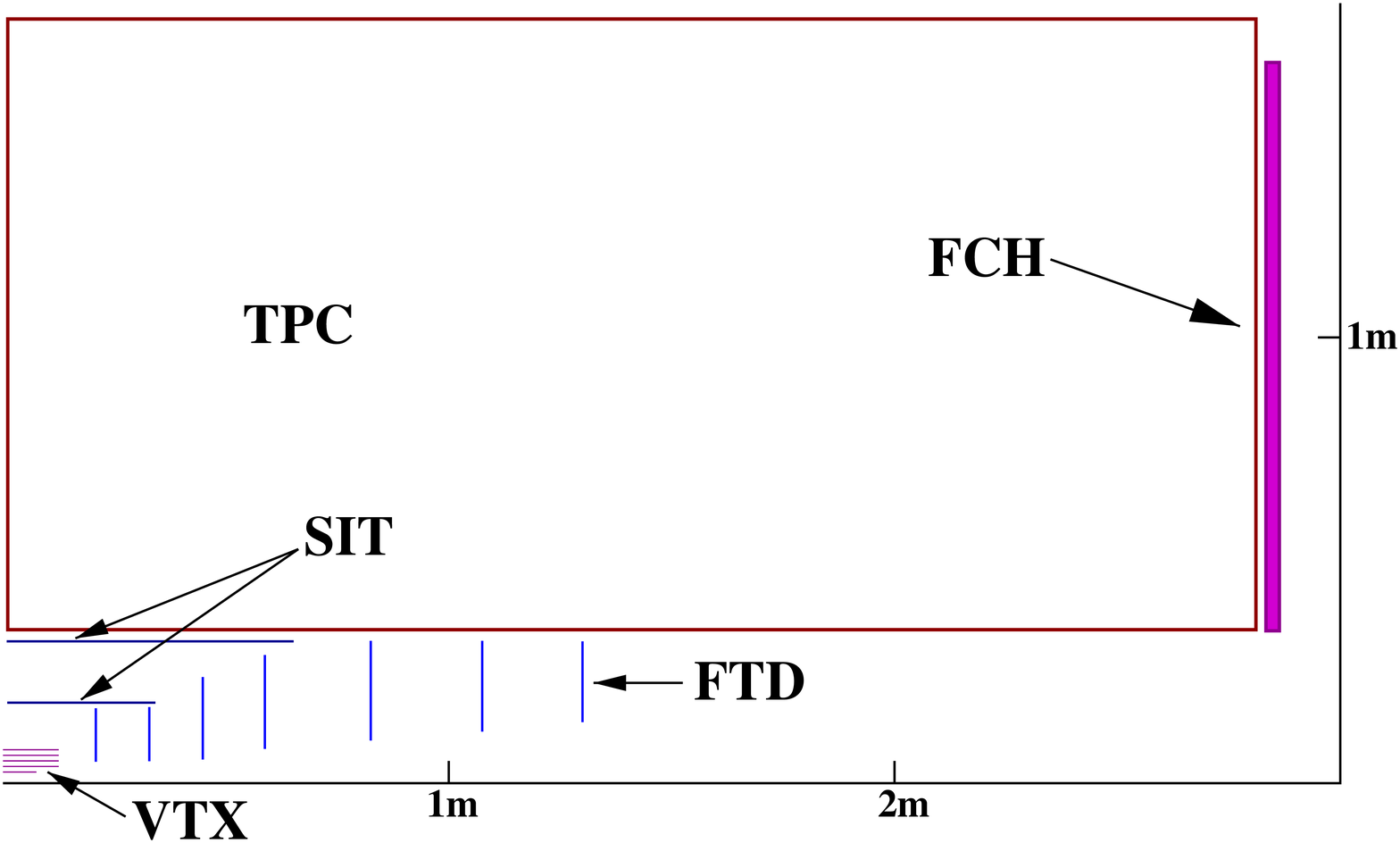}
\caption[bla]{The components of the tracking system; vertex detector (VTX), time projection chamber (TPC), forward tracking discs (FTD), intermediate silicon tracker (SIT) and forward chamber (FCH).}
\label{fig:tracking}
\end{center}
\end{figure}

The choice of the trackers is based on the requirement to achieve a high momentum resolution, to cover polar angles as low possible and to have high $b-$ and $c-$quark tagging capabilities with a minimum of material used for the construction. The whole tracking system consists of several single tracking devices as shown in Fig.~\ref{fig:tracking}.
\begin{figure}[p]
\begin{center}
\epsfxsize=4.0in
\epsfysize=3.0in
\epsfbox{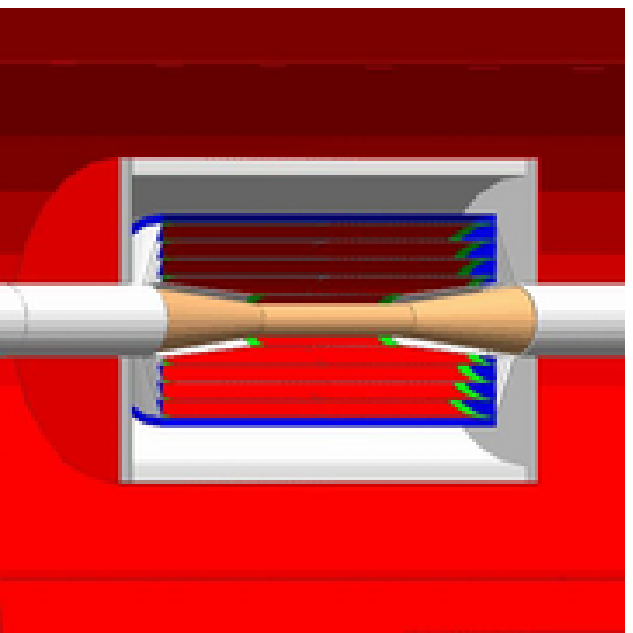}
\epsfxsize=4.0in
\epsfysize=3.0in
\epsfbox{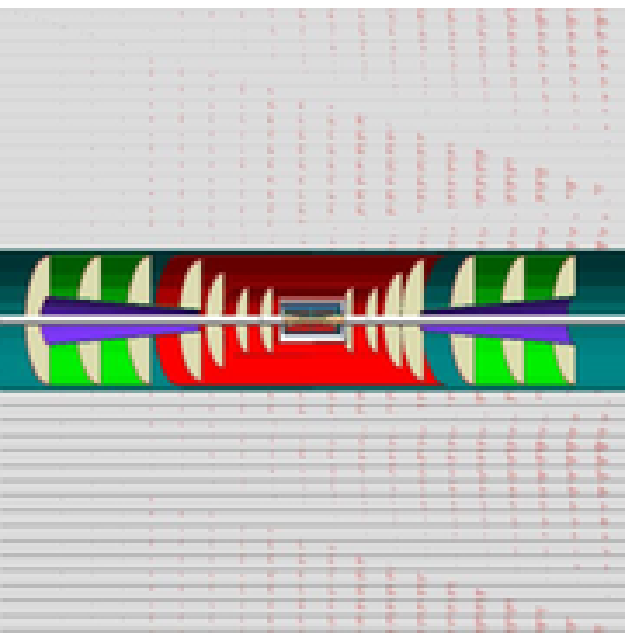}
\caption[bla]{(\textit{Upper plot}): Layout of the proposed vertex detector with five layers. (\textit{Lower plot}): Layout of the proposed two silicon intermediate trackers (cylindrically positioned around the VTX) and seven forward tracking disks, perpendicular to the beam axis.}
\label{fig:vertex}
\end{center}
\end{figure}
\par
Close to the beam pipe a very precise silicon micro-Vertex Detector (VTX) is positioned, based on the charge-coupled device (CCD) technology \cite{ccd}. The vertex detector is optimized to measure  secondary vertices and to provide a good $b-$ and $c-$quark tagging with a good $b/c$ separation and thus has a good impact parameter resolution of $\sim 2 \mu$m. This is important in the studies where the contribution from the low-energy $\gamma\gamma\rightarrow$hadrons events (pileup) has to be minimized separating the pileup tracks from physics tracks using the impact parameter information from the vertex detector. The resolutions in $r\phi$ and $rz$ used in the fast TESLA detector simulation SIMDET \cite{simdet3}  provide a precise measurement of the track impact parameters in the $r\phi$ and $rz$ projections. They are given by following expressions:
$$
\sigma_{r\phi}=\sqrt{A^{2}+\left(\frac{B}{p\sin^{2/3}\theta}\right)^{2}}
$$
and
$$
\sigma_{rz}=\sqrt{C^{2}+\left(\frac{D}{p\sin^{3/2}\theta}\right)^{2}}
$$
for a track with momentum $p$ in GeV and polar angle $\theta$. The parameters $A,B,C$ and $D$ for the CCD depend on the different angular regions covered by the detector layers and they are given in \cite{simdet3}. For the central region ($0.9<|\cos\theta|<0.928$) the parameters are $A=7.7\mu$m, $B=14.7\mu$m, $C=7.7\mu$m and $D=14.7\mu$m. The basic outline of the vertex detector are 5 layers, starting at a radius of 1.5 cm up to 6 cm, as it is shown in Fig.~\ref{fig:vertex}, that cover the angles down to $|\cos\theta|=0.96$. Since the first layer is close to the beam pipe the background coming from the beam-beam interactions is strongly peaked at this layer. Thus, the readout of the first layer is needed to be the fastest.
\par
The central tracking system is the Time Projection Chamber (TPC) \cite{tdr4,tpc} that provides the information on spatial coordinates of a particle and on its energy loss along the track. The inner radius of the TPC depends on the opening angle of the shielded area in the forward region of a photon collider (larger than 7$^{\circ}$) while the outer radius depends on the desired momentum resolution. The proposed one consists of a large tracking volume inside the high magnetic field of 4T to provide a momentum resolution of $\Delta(1/p_{t})<2\times 10^{-4}$ (GeV/c)$^{-1}$ and a dE/dx resolution less than 5$\%$. The momentum resolution at lower angles down to $|\cos\theta|=$ 0.99 ($\approx 8.1^{\circ}$) can be improved extending the central tracking system by implementing the forward chamber (FCH) \cite{tdr4} between the TPC endplate and the end-cap calorimeter. The forward chamber consists of straw-chambers in 6 planes with different wire orientations serving to resolve the track ambiguities.
\par
Between the vertex detector and the time projection chamber two additional trackers are positioned: the Silicon Intermediate Tracker (SIT) \cite{tdr4} that consists of two silicon cylinders starting at 16 cm and 30 cm with coverage angles down to 25$^{\circ}$ and seven Forward Tracking Disks (FTD) \cite{tdr4} that consist of three planes made of active pixel sensors and four silicon strip planes with coverage angles down to 5.7$^{\circ}$ ($|\cos\theta|=$ 0.995). Their role is to help the pattern recognition in linking the tracks found in the TPC and tracks found in the VTX with a required spatial resolution of 25$\mu$m for the forward tracking disks and $\sigma_{r\phi}\approx 10 \mu$m and $\sigma_{rz}\approx 50 \mu$m for the silicon intermediate tracker. In addition, the forward tracking disks are assumed to improve the angular resolution in the forward region since the combined VTX and TPC resolution degrades at low angles due to the shorter projected track length. Both, silicon intermediate tracker and forward tracking disks are shown in Fig.~\ref{fig:vertex}.
\section{Calorimeters}
\begin{figure}[htb]
\begin{center}
\epsfxsize=5.0in
\epsfysize=4.25in
\epsfbox{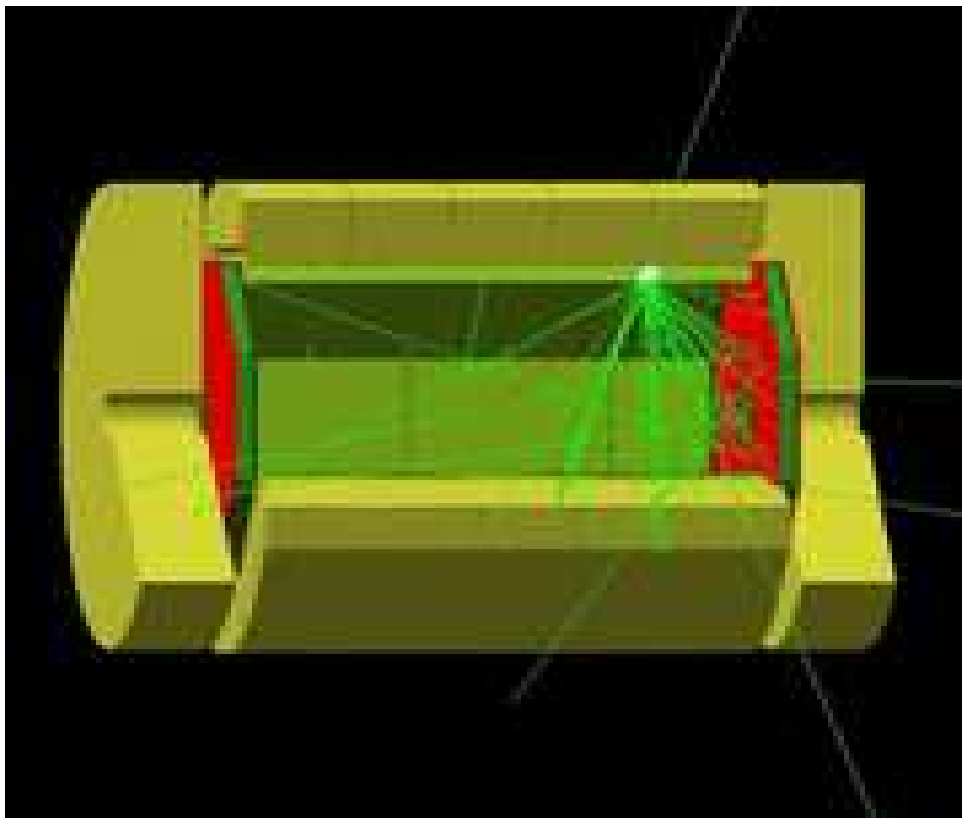}
\caption[bla]{A view of the electromagnetic and hadronic calorimeters.}
\label{fig:ecalhcal}
\end{center}
\end{figure}

The tracking system is surrounded by the calorimeter system as it is shown in Fig.~\ref{fig:ecalhcal} that consists of electromagnetic (ECAL) and hadronic calorimeter (HCAL) while the whole system together is located inside the strong magnetic field of 4 T. Both calorimeters measure the energy and the angles of photons and jets and allow the tracking of minimum ionizing particles in their volume. Thus, the main requirements on the calorimeter design are a good hermeticity down to small polar angles, excellent jet energy resolution, excellent angular resolution and a good time resolution to avoid event pileup. In the reality where the shower from photons, charged and neutral particles overlap in one cell the resolution of $\sigma(\textrm{E}_{jet})/\textrm{E}_{jet}=30\%/\sqrt{\textrm{E}_{jet}}$ would be possible and can be achieved having a dense calorimeter and sampling with high granularity and high transversal and longitudinal segmentation.
\par
Concerning the ECAL, the requirement to use a material with a small Moliere radius leads to tungsten as a solution for the absorber ($r_{Moliere}\approx9$ mm) while as the active material silicon is proposed. Tungsten provides a small ratio of the radiation length and hadronic interaction length which ensures the longitudinal separation of electromagnetic and hadronic energy depositions. Thus, the Si-W calorimeter (sandwich calorimeter) \cite{si_w} with a high segmentation is one proposal for the electromagnetic calorimeter but due to the large costs there is an alternative solution like Shashlik calorimeter \cite{shashlik}. It is a highly segmented calorimeter with longitudinal sampling in which the scintillation light is read-out via wavelength shifting fibers positioned perpendicularly to the absorber plates made of lead.
\par
A test of a segment of a sandwich calorimeter, done using the electron test beams at CERN in 2003, has shown the energy resolution as expected to be $\sigma(\textrm{E})/\textrm{E}=11.1\%/\sqrt{{\textrm{E}}}$ \cite{electr}. 
\par
The HCAL \cite{h_cal} is foreseen to be built as a tile sampling calorimeter of a material with a low magnetic permeability as an absorber (stainless steel or brass) and scintillator plates or gas as the active medium that will mainly depend on the chosen read-out concept. The expected energy resolution for single hadrons is $\sigma(\textrm{E})/\textrm{E}=35\%/\sqrt{\textrm{E[GeV]}}\otimes 3\%$.
\section{Interaction Region}
The interaction region at a photon collider is rather complicated due to the existence of two interaction regions: the conversion region where the laser photons backscatter off the high energetic electrons (the region of the Compton backscattering) and the interaction region where the two high energetic photons collide.
\begin{figure}[htb]
\begin{center}
\epsfxsize=6.0in
\epsfysize=3.0in
\epsfbox{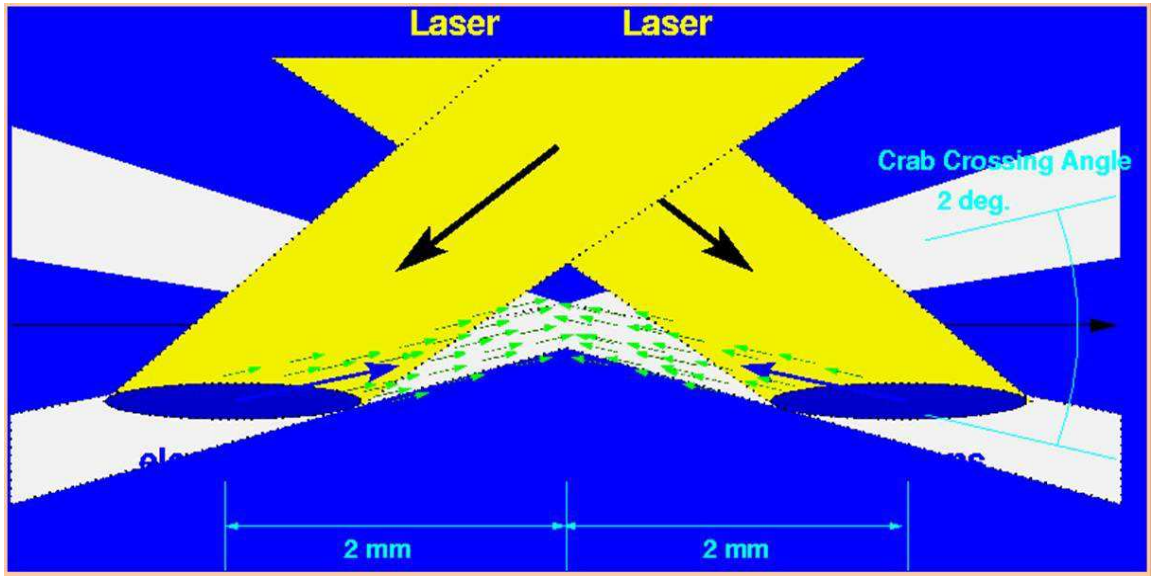}
\caption[bla]{Crab crossing scheme at the interaction point of a photon collider.}
\label{fig:crab}
\end{center}
\end{figure}
\par
After the Compton scattering the electrons travel in the direction of the interaction point with a wide energy spread having $E\approx(0.02-1)E_{e}$. Due to the energy-angular correlation the electron beams are disrupted already by the multiple Compton scattering. In addition, a lot of 'parasitic' interactions occur closely to the interaction point producing a large background mainly by $\gamma\gamma, \gamma e$ and $e^{-}e^{-}$ interactions. Photons and electrons can interact between themselves either incoherently, producing a significant amount of low energy $e^{+}e^{-}$ pairs or, interacting coherently with the collective field of the opposite electron bunch. In addition, the hard beamstrahlung photons can convert into $e^{+}e^{-}$ pairs in the presence of the strong external field. Produced low energy pairs together with low energy electrons from the multiple Compton scattering and beamstrahlung lead to the disruption of the electron beams. The removal of disrupted beams can be solved using a crab crossing scheme shown in Fig.~\ref{fig:crab}. The luminosity of the collision is not restricted in this scheme due to the tilted electron bunches that result in 'head-on' - like collisions. The minimal crab crossing angle ($\alpha_{c}$) i.e. the angle between incoming and outgoing electron beam pipes depends on the maximal disruption angle for soft electrons that is given by:
$$
\theta_{d}\sim 0.7 \sqrt{\left(\frac{4\pi r_{e}N}{\sigma_{z}\gamma_{min}}\right)}\approx 15 \textrm{mrad}
$$
and it is estimated to be $\approx$ 34 mrad taking into account the position of the quadrupoles. In this way the spent beams leave the interaction region outside the final quadrupoles. 
\par
For two different types of collider $\gamma\gamma$ and $\gamma e$, the collision effects induced by the $e^{-}e^{-}$ beam repulsion are somewhat different. The beam repulsion has rather positive consequences at $\gamma\gamma$-collider, suppressing the coherent pair creation from the beamstrahlung while the produced beamstrahlung photons have a smaller probability to collide with photons from the opposite side. In this way the contribution to the lower part of the luminosity spectrum is decreased. At the $\gamma e$-collider, the coherent pair creation is not suppressed as it is at the $\gamma\gamma$-collider and thus, the beamstrahlung is a significant source of low energy electrons and photons that influence the achievable luminosity and contribute to a larger background.
\section{Backgrounds}
All previously mentioned background sources that occur close to the interaction region increase the occupancy of sub-detectors. The low energy electrons with large disruption angles produced by multiple Compton scattering and beamstrahlung hitting the final quadrupoles, $e^{+}e^{-}$ pairs and low energy $\gamma\gamma$ interactions are the main background sources at the photon collider.
\subsection{Pair Creation}
%
\begin{figure}[htb]
\begin{center}
\epsfxsize=2.0in
\epsfysize=1.0in
\epsfbox{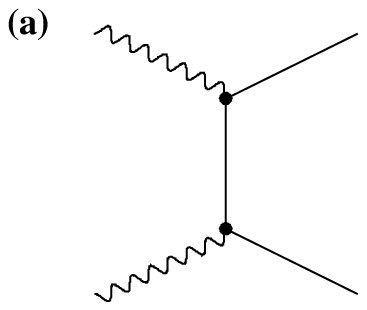}
\epsfxsize=2.0in
\epsfysize=1.0in
\epsfbox{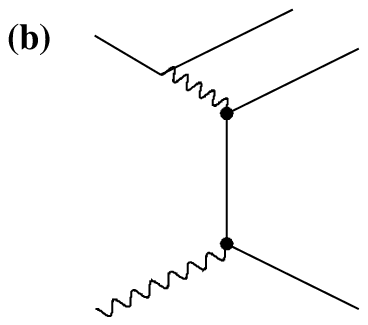}
\epsfxsize=2.0in
\epsfysize=1.0in
\epsfbox{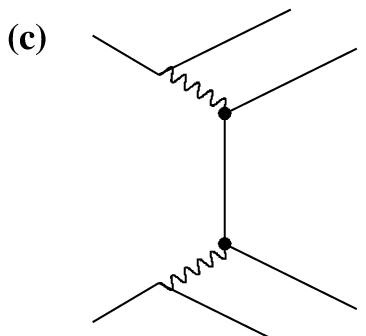}
\caption[bla]{Processes responsible for the incoherent $e^{+}e^{-}$ pair production - (\textit{a}): Breit-Wheeler process, (\textit{b}): Bethe-Heitler process and (\textit{c}): Landau-Lifshitz process.}
\label{fig:pairs}
\end{center}
\end{figure}
The main processes that contribute to the creation of low energy $e^{+}e^{-}$ pairs are the incoherent interactions between the individual particles, photons and electrons \cite{chen1,chen2}. The interaction of two real photons shown in Fig.~\ref{fig:pairs}\,$a$, is the Breit-Wheeler process ($\gamma\gamma\rightarrow e^{+}e^{-}$) in which the photons can originate either from the beamstrahlung or from the Compton scattering. In the Bethe-Heitler process ($e\gamma\rightarrow ee^{+}e^{-}$) in Fig.~\ref{fig:pairs}\,$b$, one photon is real (beamstrahlung or Compton) while the another one is virtual. The interaction of the two virtual photons in the Landau-Lifshitz process ($ee\rightarrow eee^{+}e^{-}$) shown in Fig.~\ref{fig:pairs}\,$c$ is suppressed at the $\gamma\gamma$-collider compared to the $e^{+}e^{-}$-collider due to the $e^{-}\rightarrow\gamma$ conversion. Most of the incoherent pairs are emitted in the forward direction and can be deflected by the opposite bunch field getting larger angles. These pairs mainly induce so called 'backscattered' background in the central tracking system hitting the inside  parts of the detector. A smaller amount of incoherent pairs is emitted with initially larger angles inducing so called 'direct' background that mostly hits the vertex detector and hardly can be avoided.
\par
Another possibility to create pairs are the coherent ``particle-beam'' interactions i.e. if the beamstrahlung photons (produced in the interaction of electrons and the collective beam field) or high energy Compton photons convert to $e^+e^-$ pairs in the strong external (electro-magnetic) field. The positrons from created pairs are focused by the opposite electron bunch and mainly end up in the beam pipe. The electrons get larger angles by the deflection off the opposite electron bunch field, hit the inner parts of the detector, inducing the backscattered background. In addition, there are the low energy electrons from multiple Compton scattering, electrons left after the emission of the beamstrahlung photons and electrons from incoherently created pairs that can be deflected at large angles by the field of the opposite bunch. A large amount of these interactions mainly induce a background in the central tracking system from backscattering inside the detector.
\begin{figure}[htb]
\begin{center}
\epsfxsize=3.0in
\epsfysize=2.5in
\epsfbox{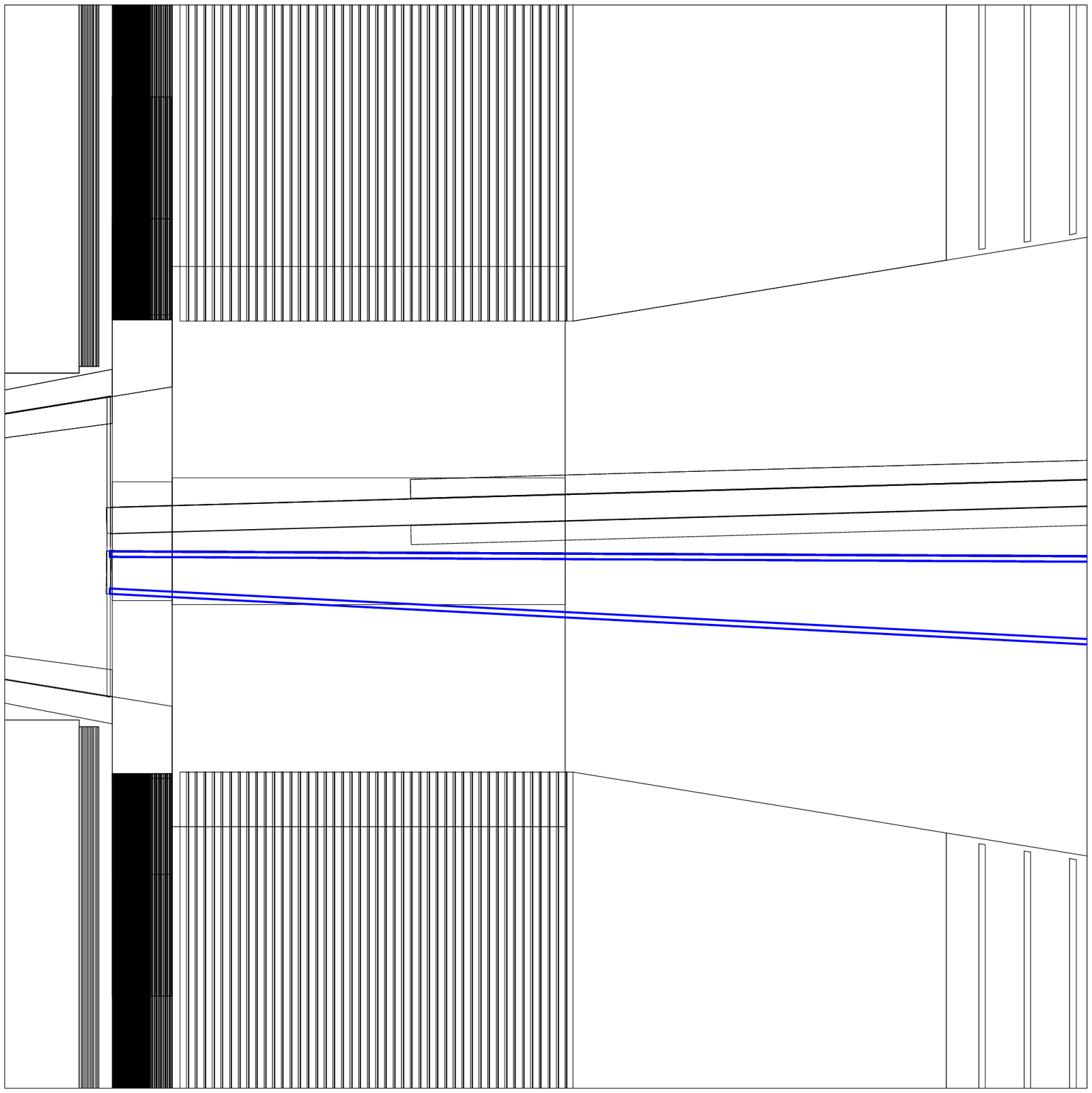}
\epsfxsize=3.0in
\epsfysize=2.5in
\epsfbox{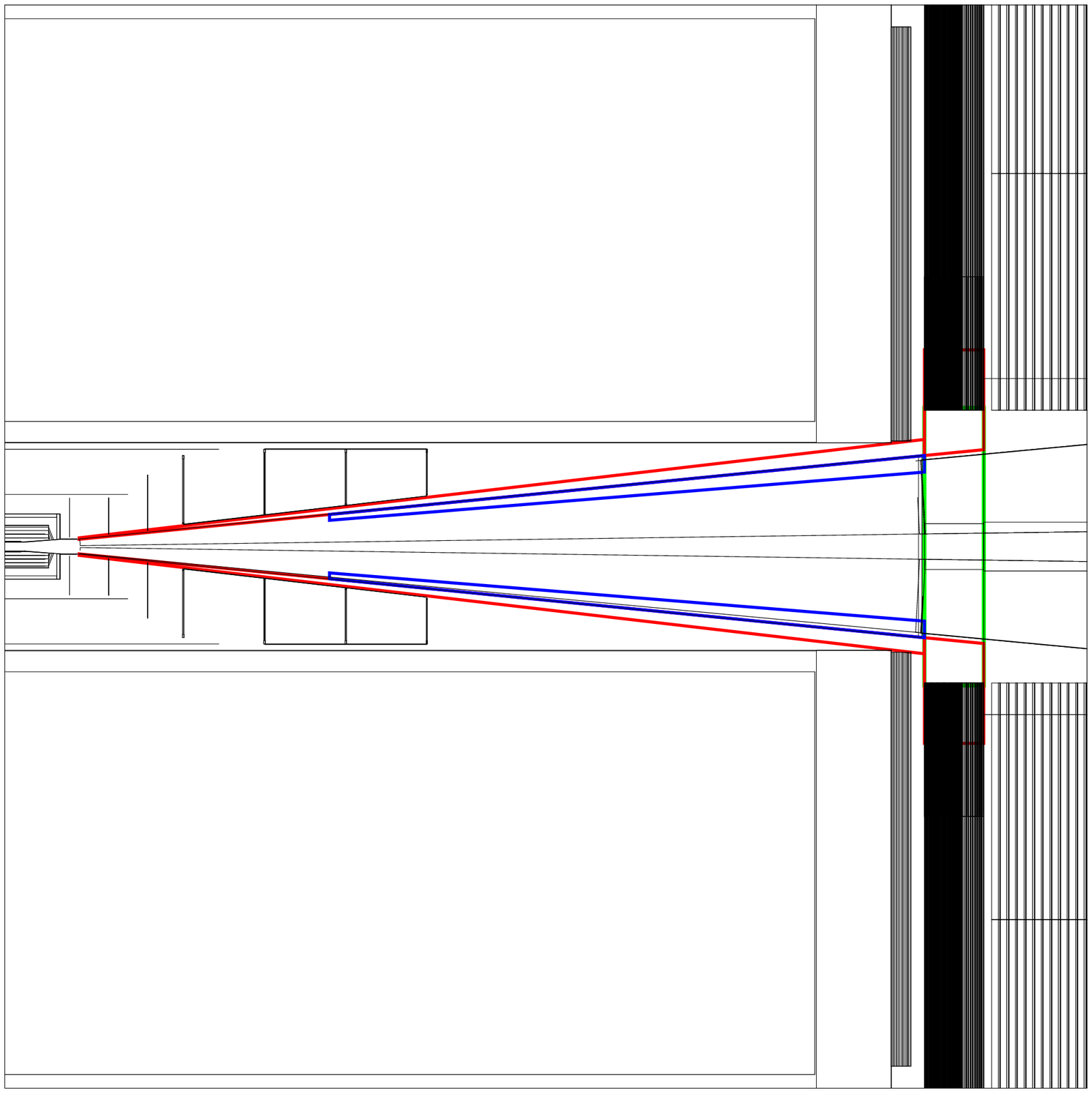}
\caption[bla]{({\textit{Left plot}}): The outgoing graphite electron beam pipe in the $x-z$ projection. ({\textit{Right plot}}): The $y-z$ projection of the inner forward region of the $\gamma\gamma$-detector. The two masks, outer (red, longer) and inner (blue, shorter) made of tungsten, shield the TPC from the backscattered background.}
\label{fig:aura}
\end{center}
\end{figure}
\begin{figure}[htb]
\begin{center}
\epsfxsize=3.0in
\epsfysize=2.5in
\epsfbox{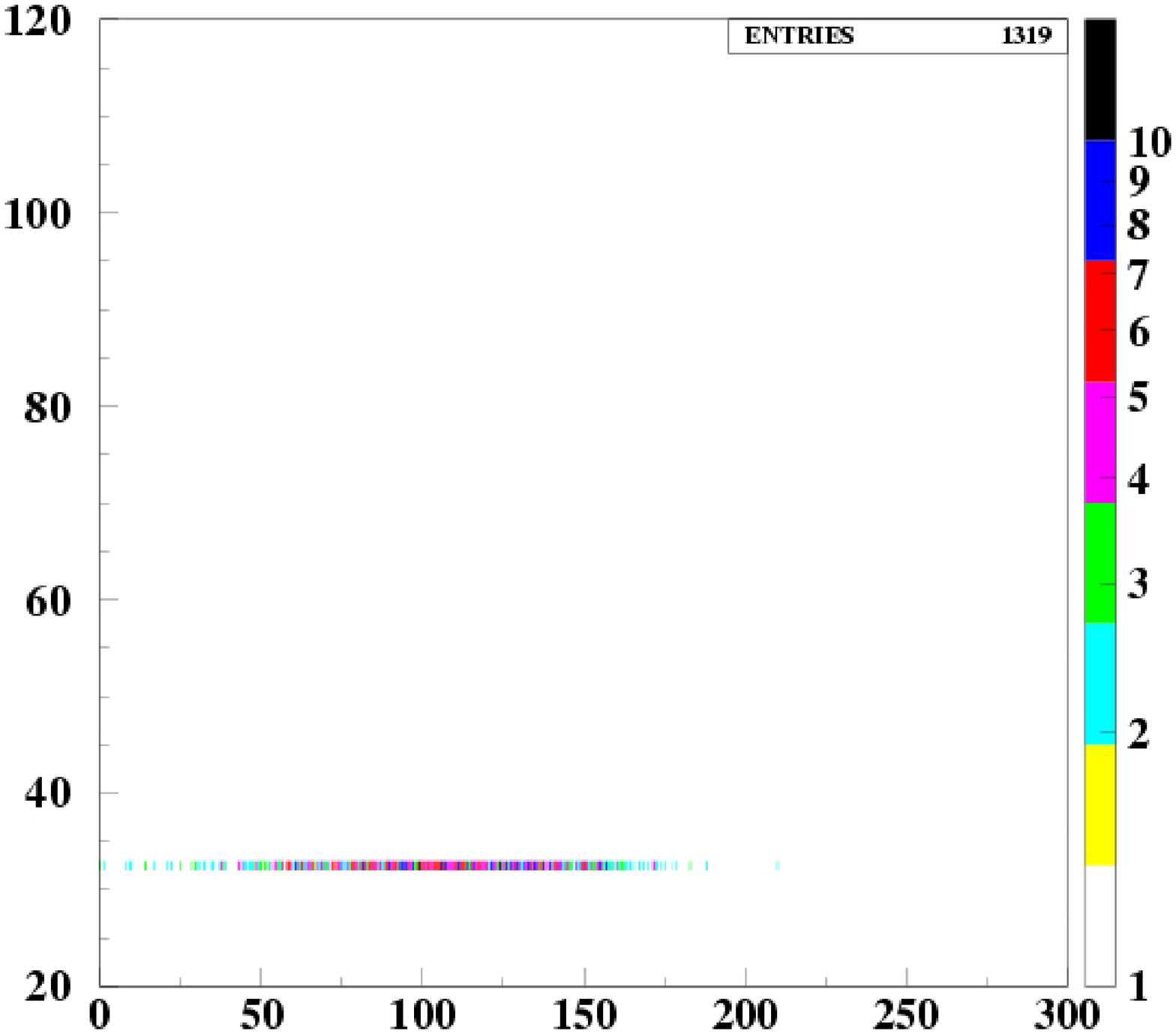}
\epsfxsize=3.0in
\epsfysize=2.5in
\epsfbox{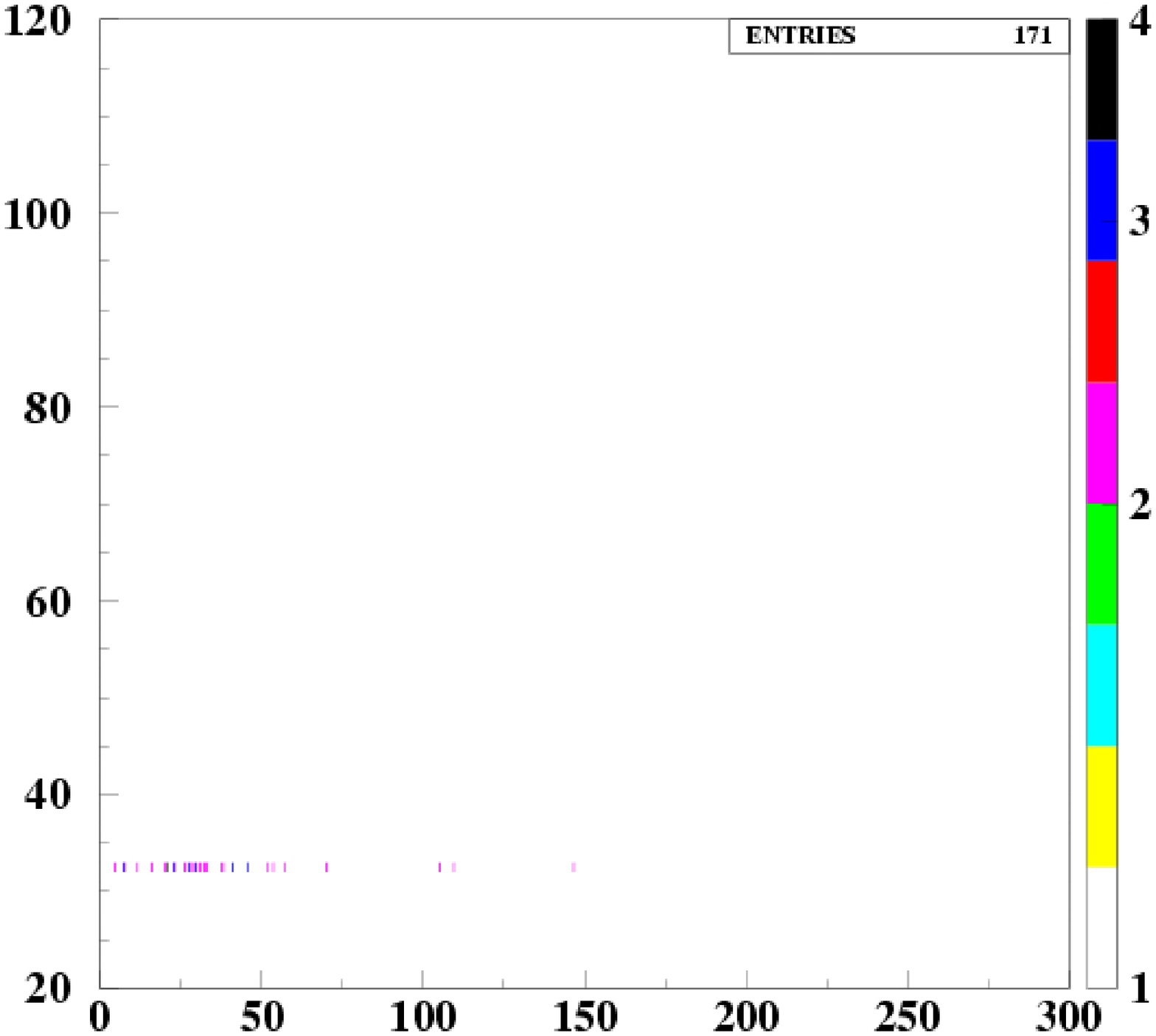}
\caption[bla]{The radial position $r$ ($y$-axis) of photons versus $|z|$ ($x$-axis) entering into the TPC, produced from particles created in the coherent interactions and scattered on the inside detector parts. ({\textit{Left plot}}): The photons in the TPC of the $\gamma\gamma$-detector with the design without the inner mask - only the outer mask is implemented. ({\textit{Right plot}}): The setup of the $\gamma\gamma$-detector where the inner mask is implemented in addition. The TPC receives about 8 times less photons. The number of entries should be multiplied by 5 to estimate this background per one bunch crossing.}
\label{fig:tpcs}
\end{center}
\end{figure}
\begin{figure}[htb]
\begin{center}
\epsfxsize=2.5in
\epsfysize=2.5in
\epsfbox{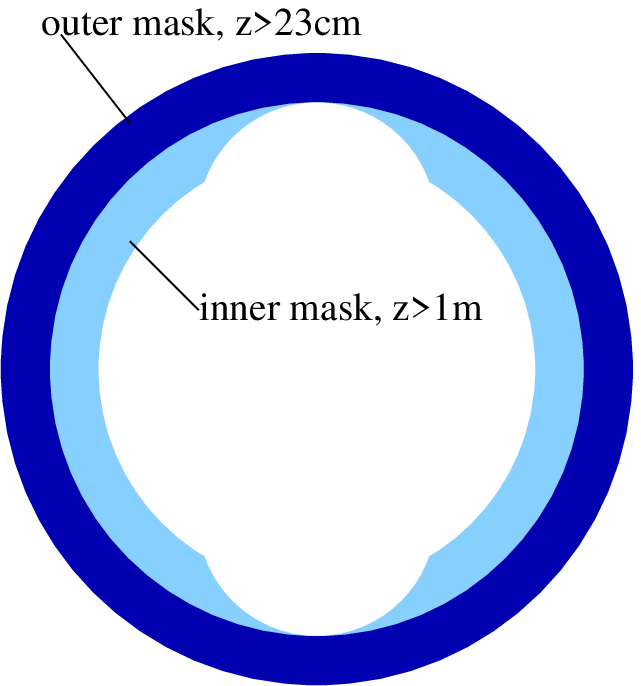}
\epsfxsize=6.5in
\epsfysize=2.5in
\epsfbox{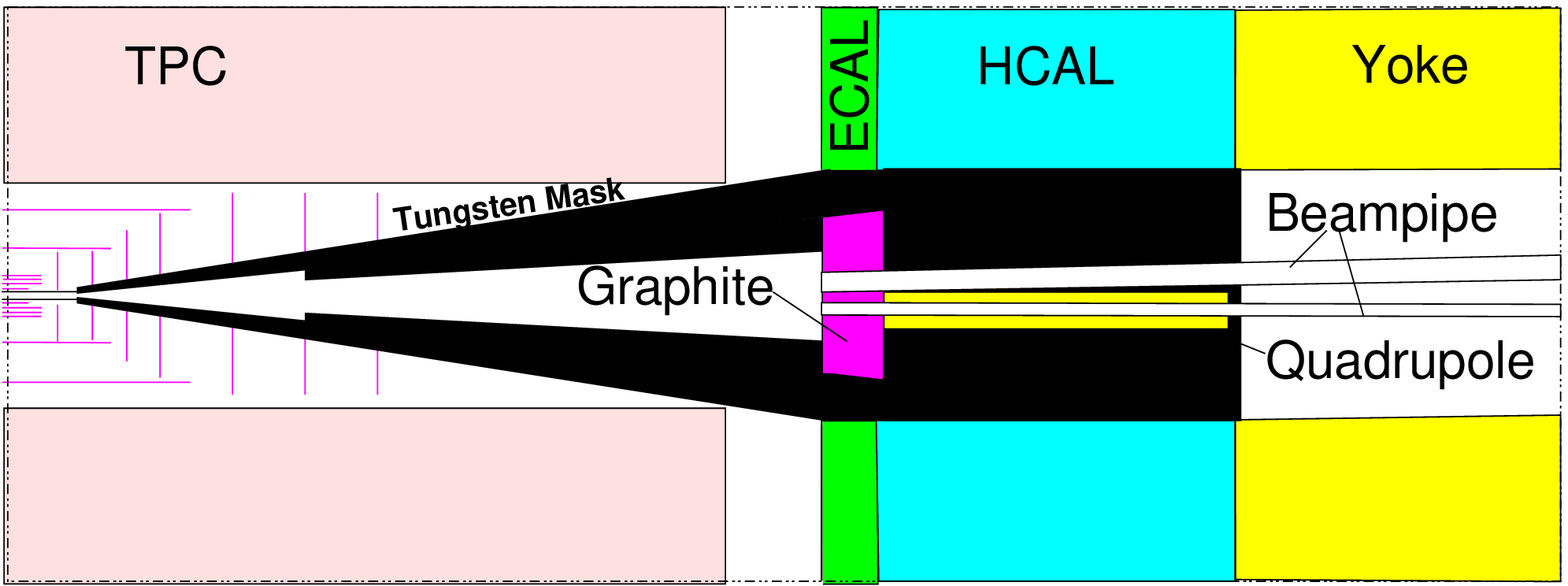}
\caption[bla]{(\textit{Upper plot}): The $y-x$ projection at $z=2.8$ m shows the presence of the two shielding masks - the outer and the inner one with a left space for a laser beam pipes. (\textit{Lower plot}): The $y-z$ projection of the inner forward region of the $\gamma\gamma$-detector. Black area are filled with tungsten.}
\label{fig:optimized}
\end{center}
\end{figure}
\begin{figure}[htb]
\begin{center}
\epsfxsize=3.in
\epsfysize=3.in
\epsfbox{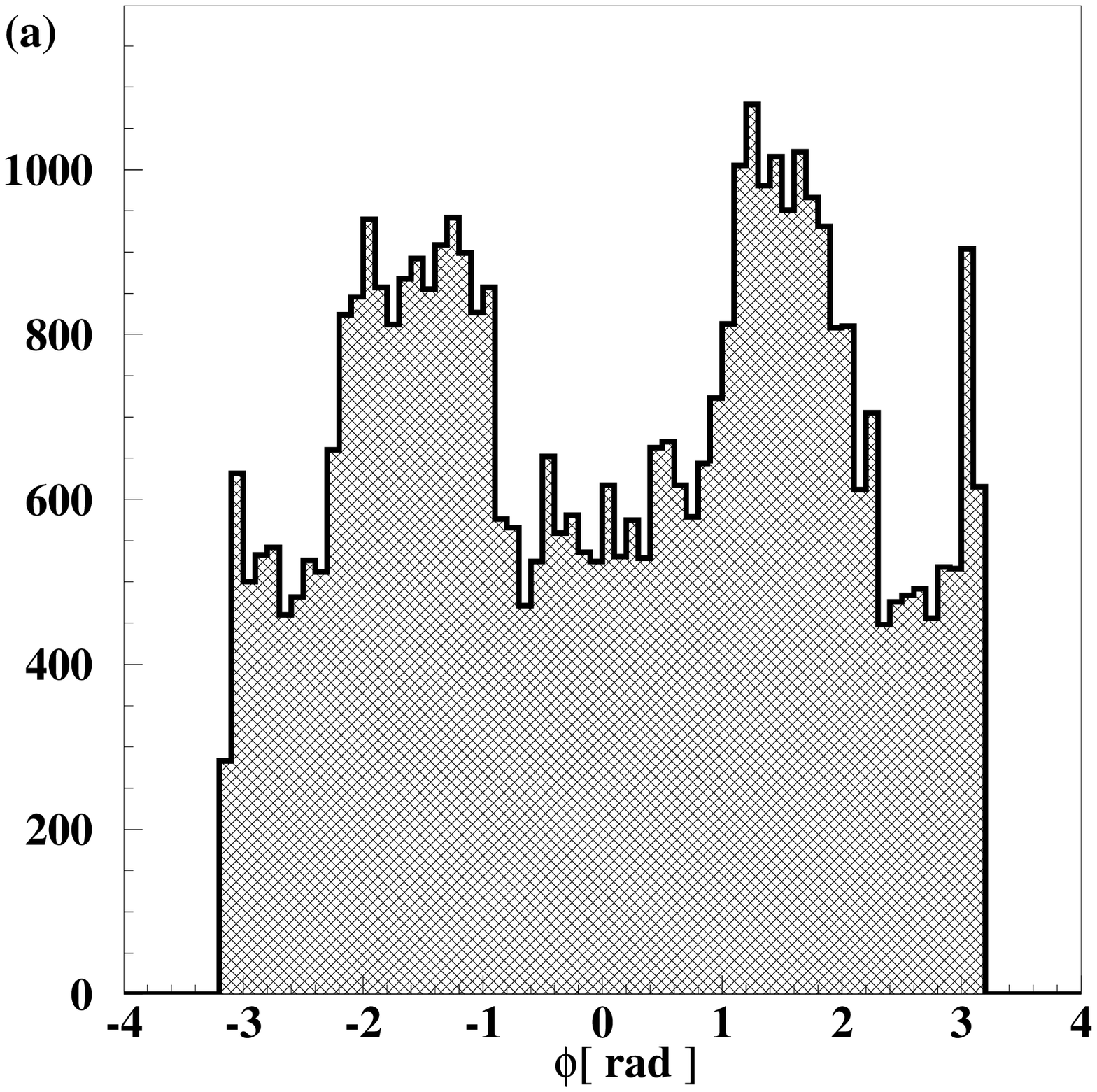}
\epsfxsize=3.in
\epsfysize=3.in
\epsfbox{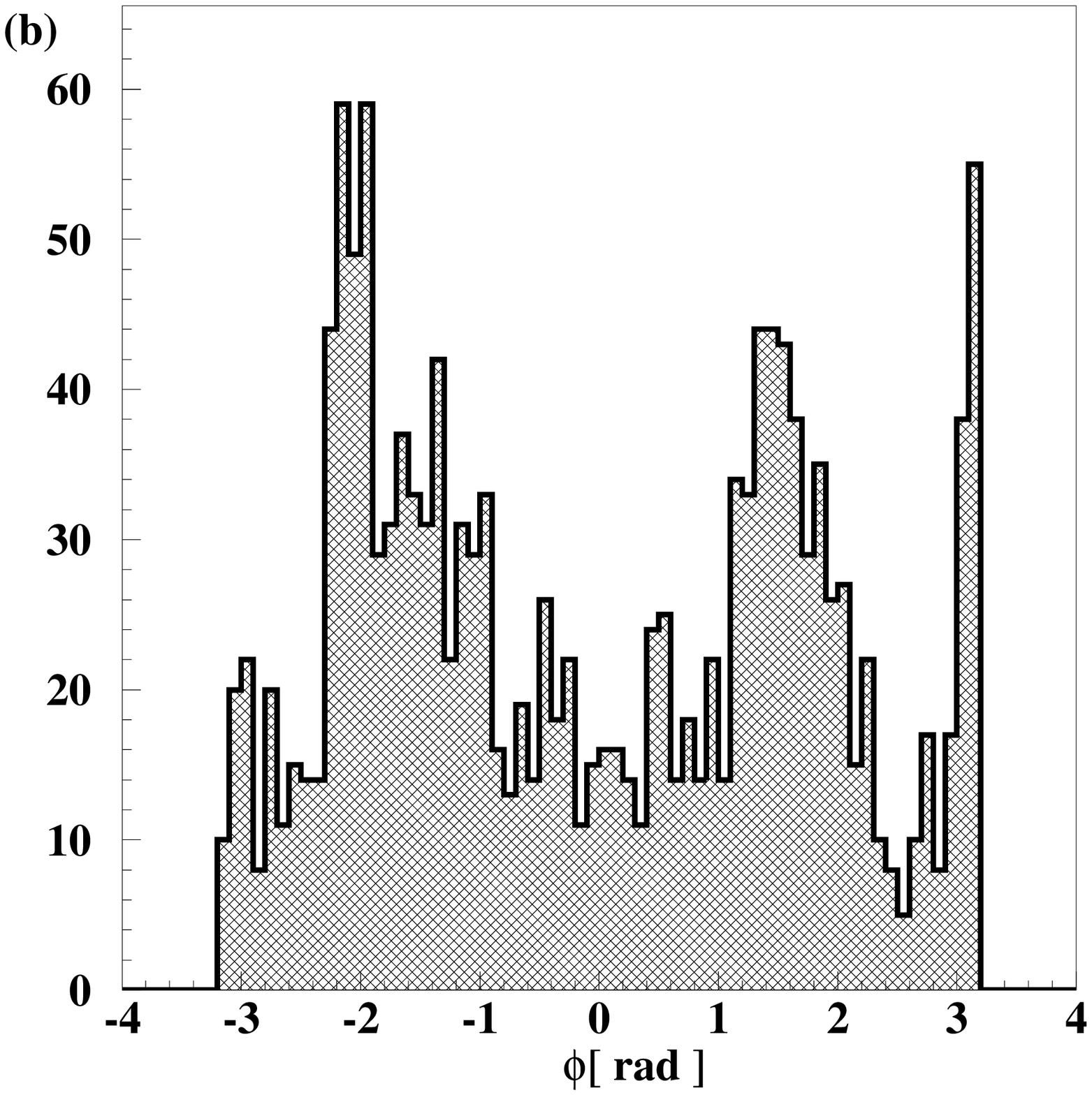}
\caption[bla]{Phi distributions of the particles (photons, electrons and positrons) from the incoherent interactions that (\textit{a}) enter the inner mask and that (\textit{b}) leave the inner mask.}
\label{fig:phi_photons}
\end{center}
\end{figure}
\begin{figure}[htb]
\begin{center}
\epsfxsize=3.0in
\epsfysize=3.0in
\epsfbox{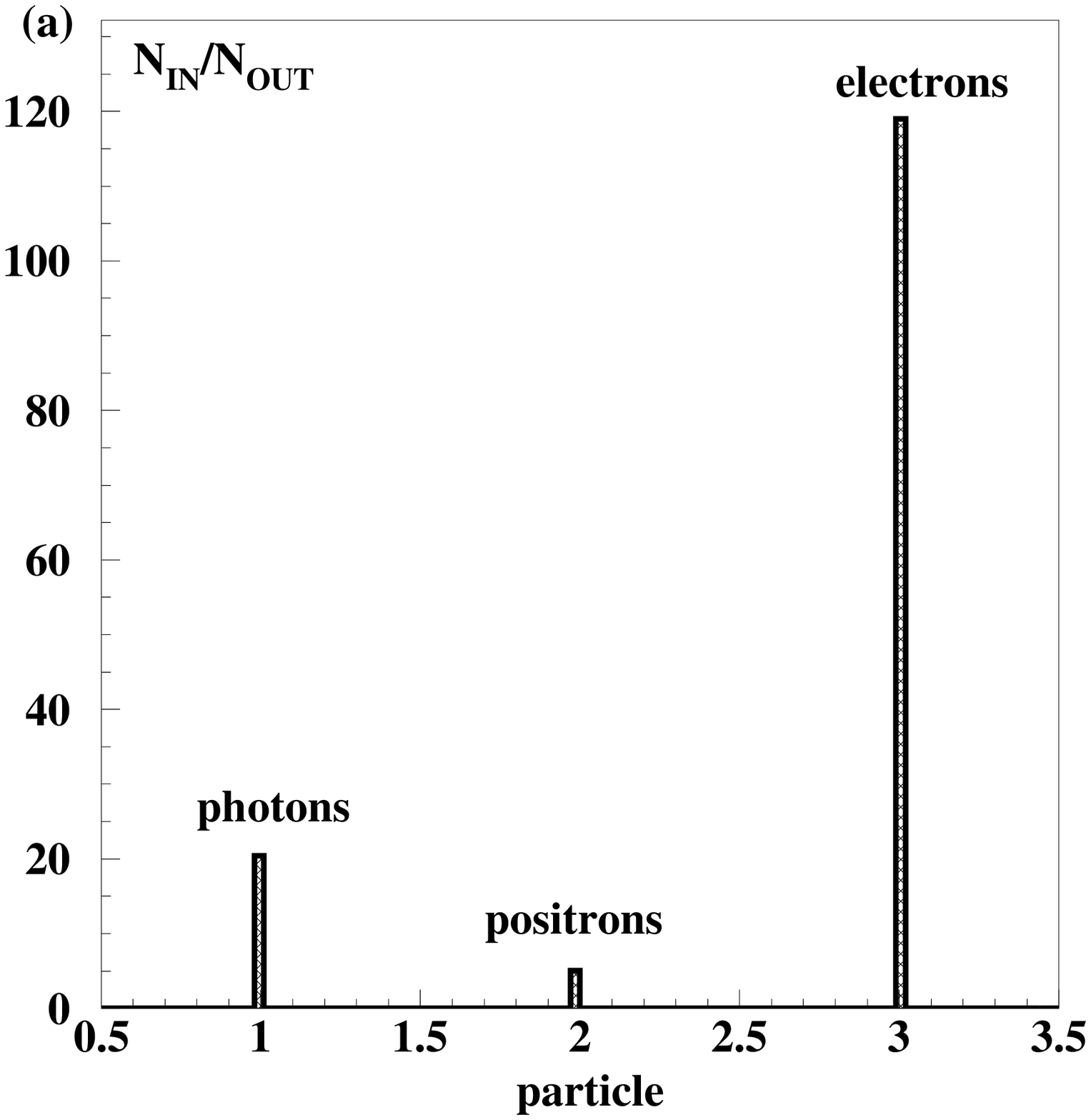}
\epsfxsize=3.0in
\epsfysize=3.0in
\epsfbox{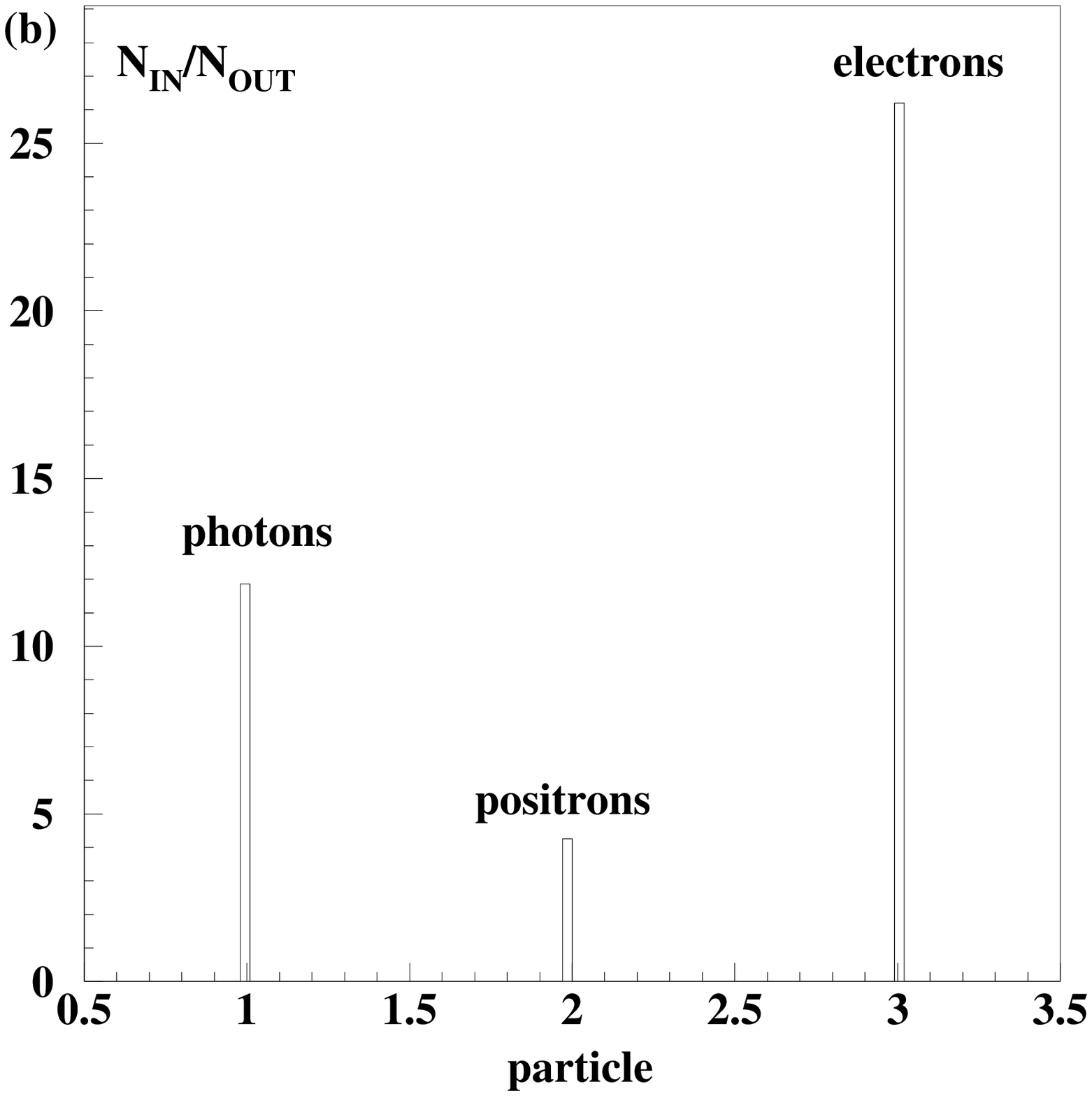}
\caption[bla]{(\textit{a}): Ratio of the photons, positrons and electrons from the incoherent interactions that enter and leave the inner mask. (\textit{b}): Ratio of the photons, positrons and electrons created in the coherent interactions that enter and leave the inner mask.}
\label{fig:ratio_inner}
\end{center}
\end{figure}
\begin{figure}[p]
\begin{center}
\epsfxsize=3.0in
\epsfysize=3.0in
\epsfbox{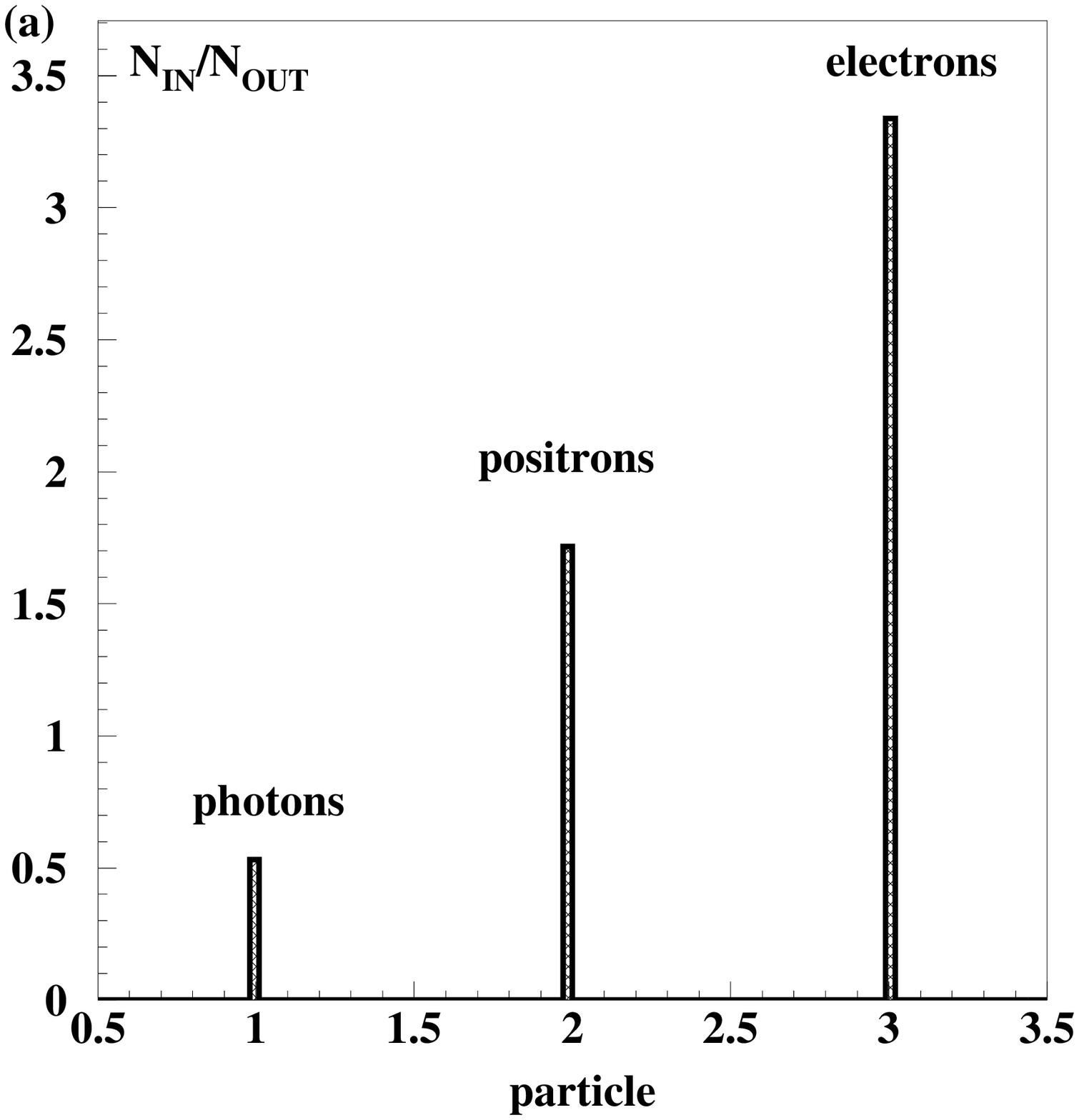}
\epsfxsize=3.0in
\epsfysize=3.0in
\epsfbox{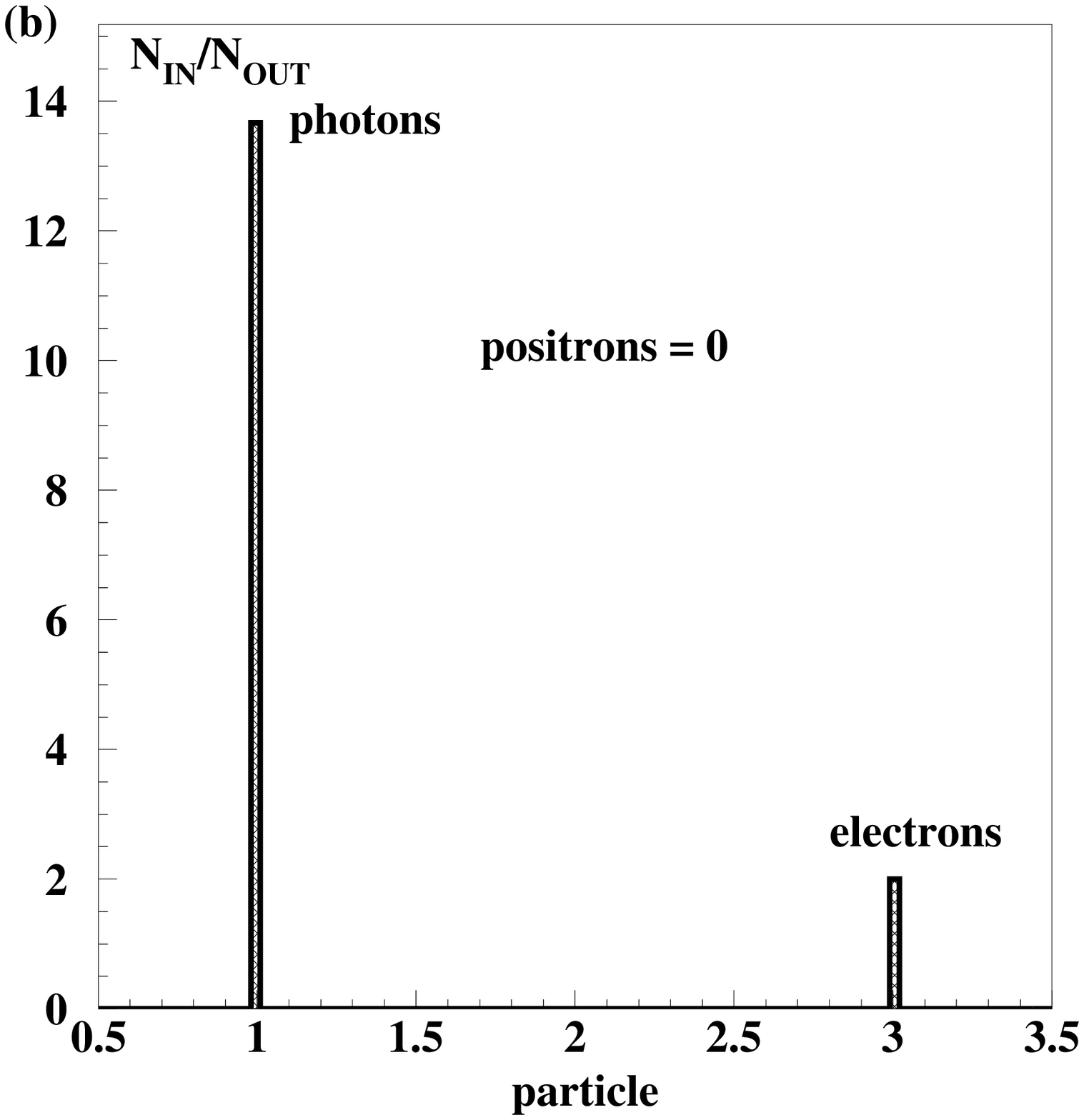}
\epsfxsize=3.0in
\epsfysize=3.0in
\epsfbox{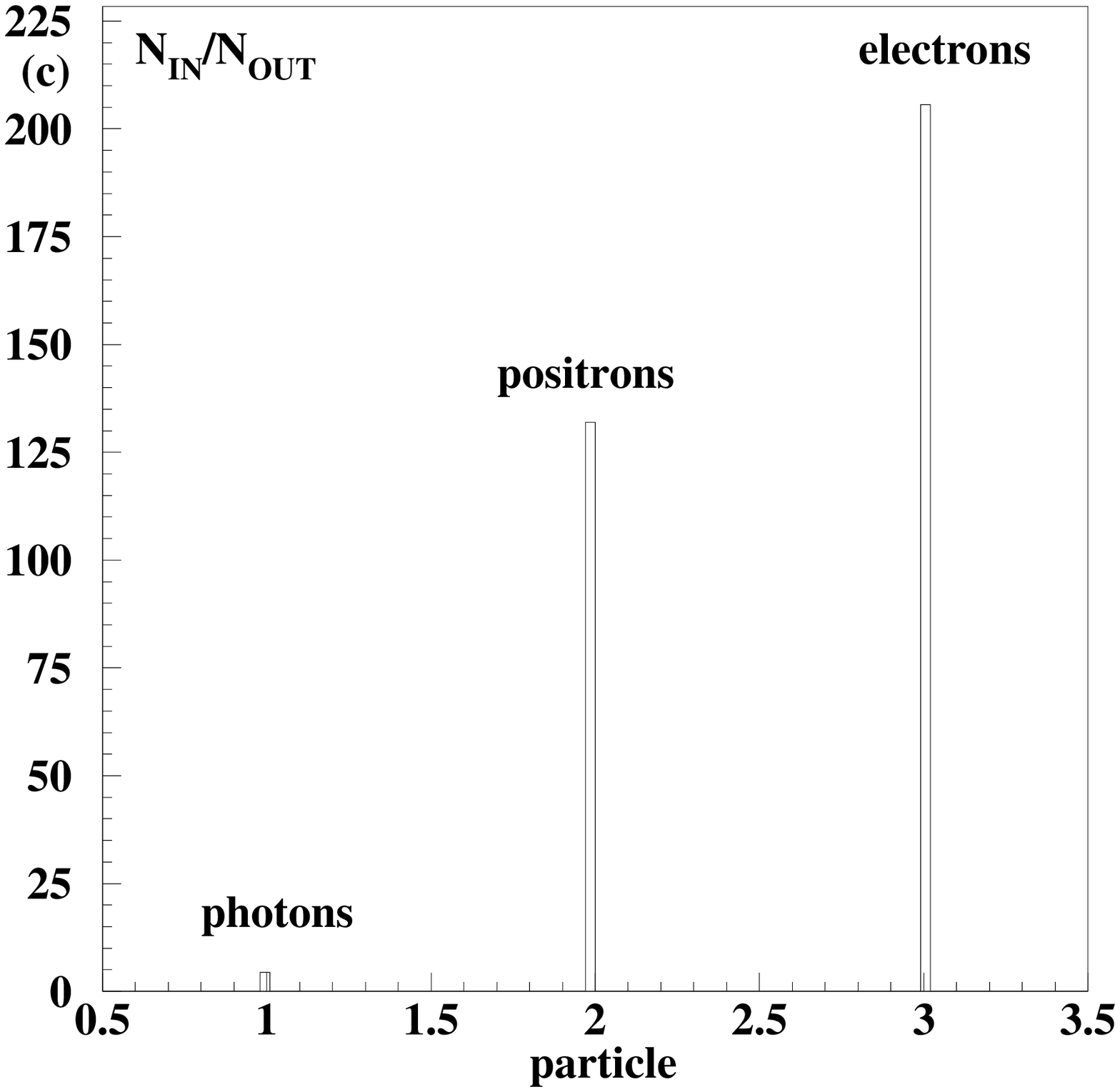}
\epsfxsize=3.0in
\epsfysize=3.0in
\epsfbox{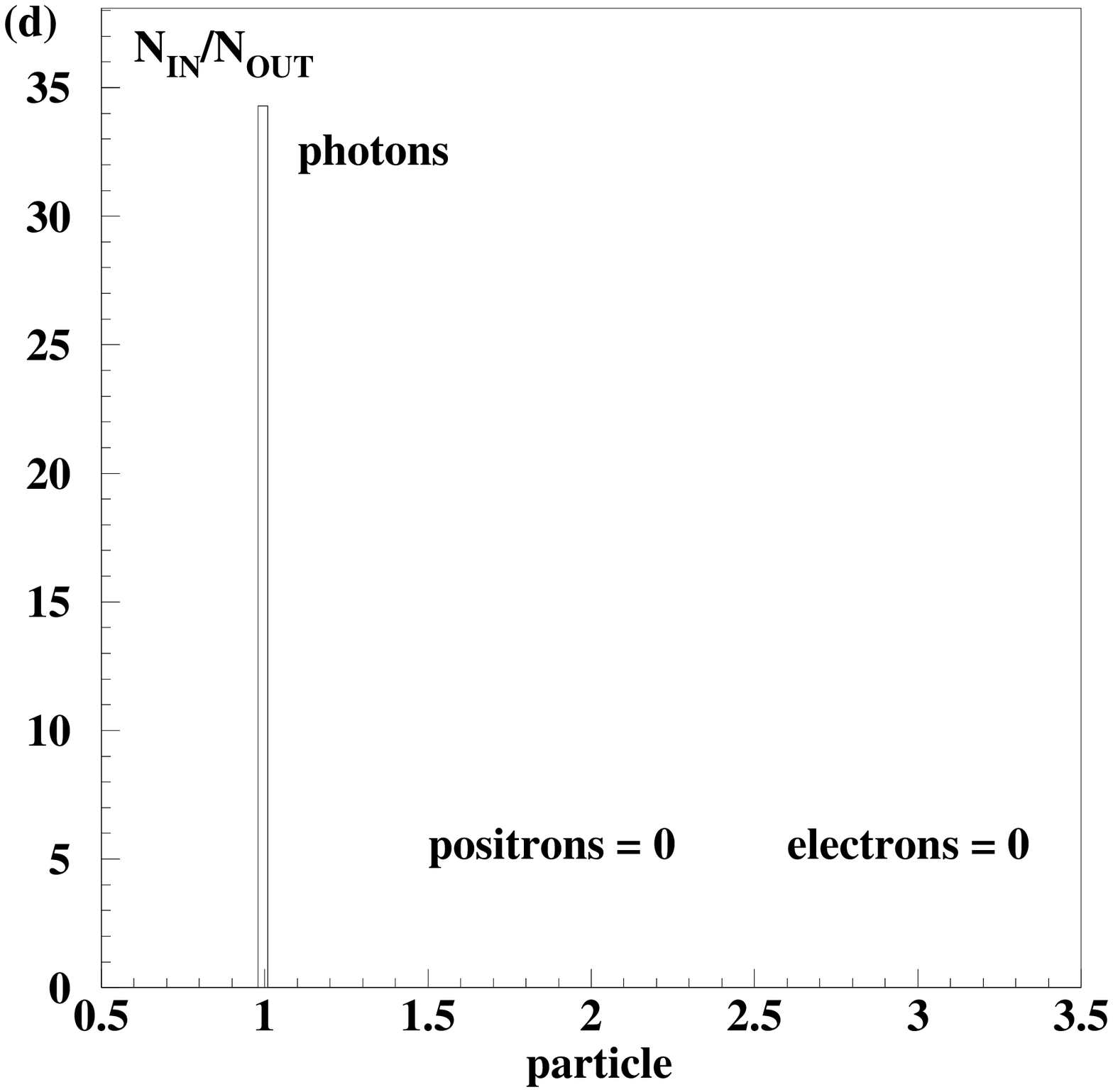}
\caption[bla]{Ratio of the photons, positrons and electrons from the incoherent interactions that (\textit{a}): enter and leave the outer mask in the range $z=23-100$ cm (\textit{b}): enter and leave the outer mask in the range $z=100-280$ cm. Ratio of the photons, positrons and electrons from the coherent interactions that (\textit{c}): enter and leave the outer mask in the range $z=23-100$ cm (\textit{d}): enter and leave the outer mask in the range $z=100-280$ cm.}
\label{fig:ratio_outer}
\end{center}
\end{figure}
\begin{figure}[htb]
\begin{center}
\epsfxsize=3.0in
\epsfysize=3.0in
\epsfbox{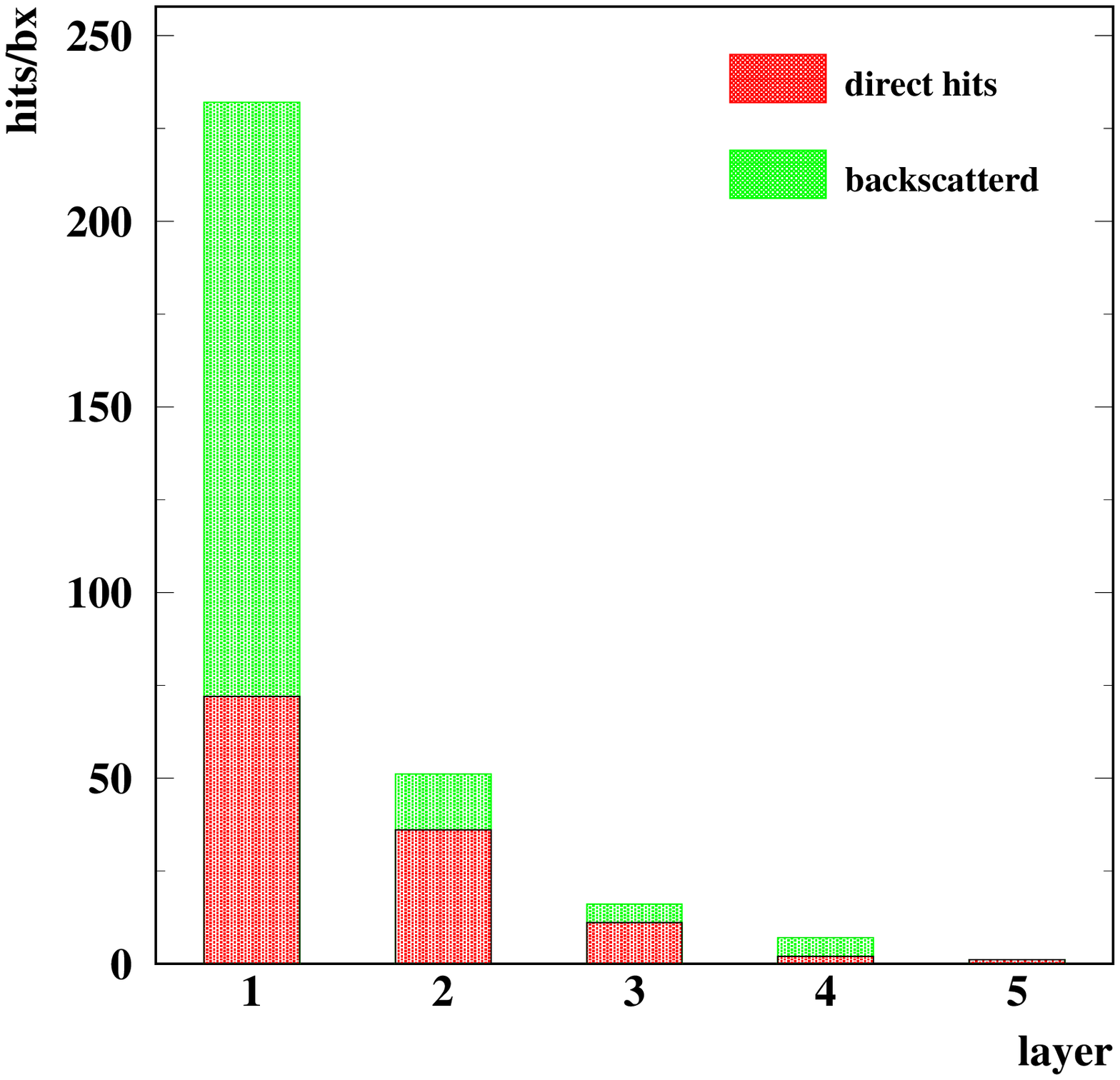}
\epsfxsize=3.0in
\epsfysize=3.0in
\epsfbox{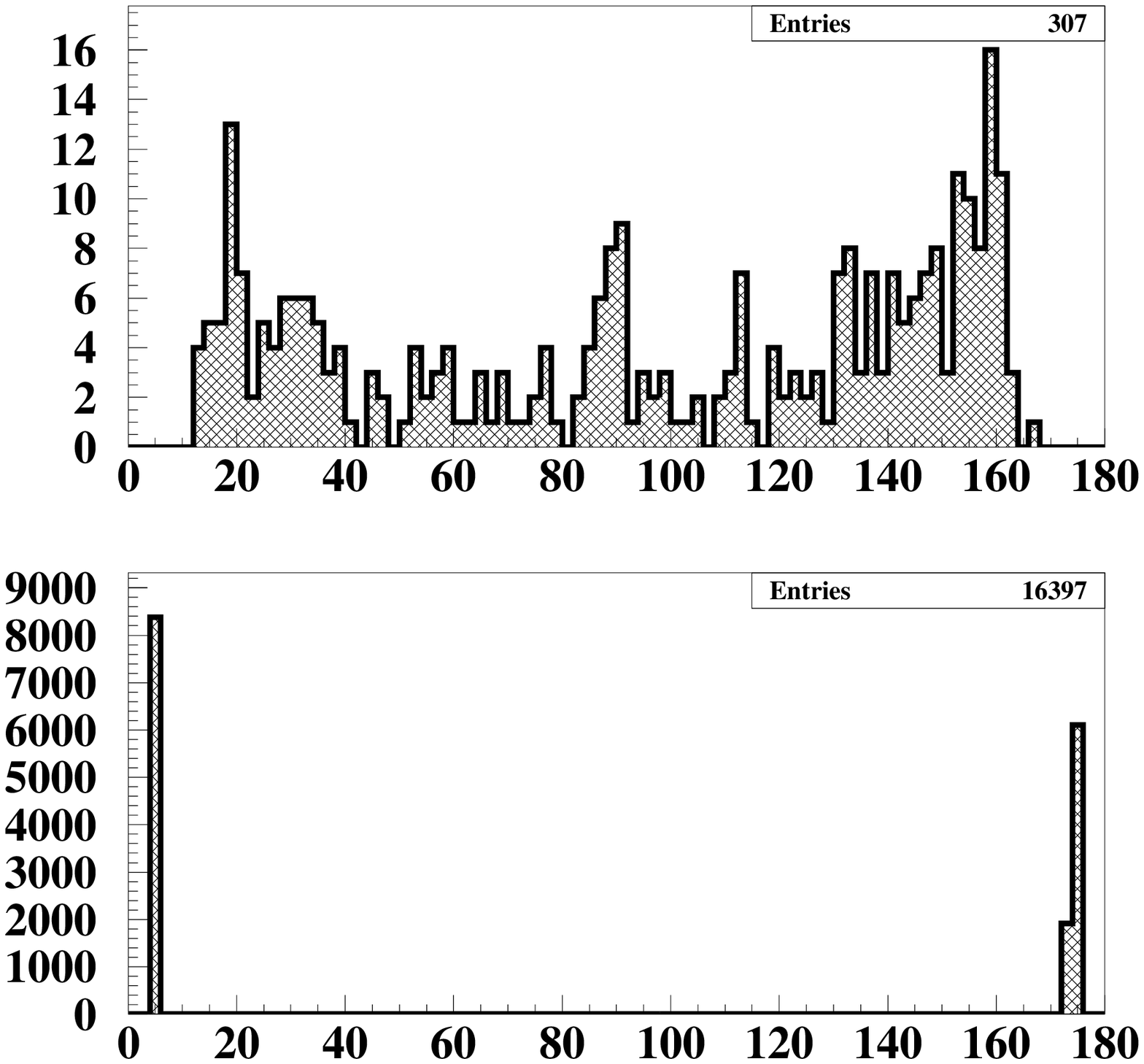}
\caption[bla]{(\textit{Left plot}): Hits in the vertex detector from the incoherent $e^+e^-$ pairs created at the interaction point. (\textit{Right upper plot}): Angular distribution of the $e^-$ and $e^+$ from incoherent interactions that hit the vertex detector. (\textit{Right lower plot}): Angular distribution of the $e^-$ and $e^+$ from incoherent interactions that hit the inner mask.}
\label{fig:hits}
\end{center}
\end{figure}
\par
The produced amount of background, both direct and backscattered, required an optimization of the $\gamma\gamma$-detector in the forward region. Detailed backgrounds have been simulated using CAIN \cite{cain}. For these simulations the 'incoherent particle - particle' and 'the coherent particle - beam' interactions have been considered. The simulation of the detector response to beam induced backgrounds is done using the TESLA simulation program BRAHMS \cite{brahms}, based on GEANT3 \cite{geant3} and it is optimized in the way to minimize the direct background as well as the background that comes from the backscattered particles. The absorption of $e^+e^-$ pairs by the detector parts creates a large amount of secondary particles which are a major background source for the detector ('backscattered' background). When an $e^+e^-$ pairs impinge on the beam pipe, ECAL or a mask, the produced photons enter the TPC volume and also may convert into $e^+e^-$ pairs inducing background tracks. In order to decrease the amount of secondary particles mainly in the central tracking system the space around the beam pipe in the region where ECAL and HCAL are positioned is filled by graphite and tungsten in the best estimated ratio. Several different configurations of the two absorbers have been tested changing the area sizes filled by two of them. The best background estimate resulted with a thickness of $\approx$ 9 cm of graphite in front of the zone of $\approx$ 60 cm filled with tungsten as shown in Fig.~\ref{fig:aura}\,$(left)$. The graphite as a low $Z$ absorber serves to reduce the backscattering of showers which develop when the pairs hit the tungsten mask or quadrupoles. Because of the same reasons the beam pipes in the region between the interaction point and ECAL are surrounded by a tungsten mask with pointing geometry and with a thickness of 2$\times$5 cm at $|z|$=2.8 m shielding the TPC. The mask consists of two parts: an outer and an inner mask, as it is shown in Fig.~\ref{fig:aura}\,$(right)$. The outer mask starts at a distance of 23 cm from the interaction point and the inner tungsten mask starts at a distance of 1 m from the interaction point - otherwise if it is closer, it receives too many direct hits in its forward part from background particles created at the interaction point, rising the amount of backscattered background in the TPC. The shielding effect of the two masks on the number of photons that enter in the TPC is shown in Fig.~\ref{fig:tpcs}. The outgoing electron beam pipes in Fig.~\ref{fig:aura}\,$(left)$ are made of graphite to absorb low energy particles from showers protecting the inner layers of the vertex detector from backscattered ones. The design of the forward region for a $\gamma\gamma$-detector is shown in Fig.~\ref{fig:optimized}. The upper plot is the $y-x$ projection of the two masks, outer and inner, at $|z|=2.8$ m showing the two half-concave places belonging to the inner mask, foreseen for the laser beam pipes. The inner mask absorbs a large amount of photons so that the photon distributions around $\pm\pi/2$ (close to the $y-$axis), where the mask is thin, are decreased to a safe level as it is shown in Fig.~\ref{fig:phi_photons}. The amount of the photons from the incoherent interactions is decreased approximatively 20 times and from the coherent interactions about 12 times by the inner mask as it is shown in Fig.~\ref{fig:ratio_inner}. The number of positrons that enter the inner mask is small compared to the number of electrons and photons since they mainly end up in the beam pipe focused by the opposite charge of the incoming bunch. 
\par
The efficiency of the outer mask for two different $z$ regions, unprotected by the inner mask ($|z|=23-100$ cm) and protected by the inner mask ($|z|=100-280$ cm), is shown in Fig.~\ref{fig:ratio_outer}. Electrons and positrons from both, incoherent and coherent interactions, are highely suppressed passing the two masks i.e. at $|z|>100$ cm. The amount of photons entering the TPC in that region is efficiently decreased as it is clear from Fig.~\ref{fig:tpcs}\,$b$ and photons enter the TPC only at $|z|<100$ cm. This photon contribution comes mainly from the region which is not protected by both masks. Photons originating from the coherent interactions are more efficiently decreased by the outer mask at $|z|<100$ cm than that one coming from incoherent interactions (Fig.~\ref{fig:ratio_outer}\,$a,c$). The number of photons from incoherent interactions is even doubled by interactions within the outer mask at $|z|<100$ cm.
\par
With such a forward region design of the $\gamma\gamma$-detector the estimated background in the time projection chamber and in the vertex detector is brought to the level of background of the $e^+e^-$-collider and should be manageable to provide an efficient read-out of the first vertex detector layer and a small occupancy of the time projection chamber. Quantitatively, that means that the TPC receives about 1825 photons per bunch crossing which will interact inside and the VTX receives about 300 hits per bunch crossing taking into account those from incoherent $e^+e^-$ pairs only. The last results in hit densities of less than 0.04 hits/mm$^{2}$ which is below the critical background level. The background contribution per layer of the vertex detector is shown in Fig.~\ref{fig:hits}. The first layer receives most of the background hits and there is no way to improve its protection without enhancing the background in the TPC, so far.
\par
The sample taken to estimate the background that comes from the coherent interactions corresponds to only 20$\%$ of the total background per bunch crossing due to the long computing time necessary for the simulation. For the TPC it is estimated to be about 171 photons and in the VTX about 18 hits in the fourth and fifth layer originating from only one track. The obtained numbers of photons and hits are than multiplied by five in order to estimate the full background per one bunch crossing in the TPC (855 photons/BX) and in the VTX (90 hits/BX). This introduces the statistical uncertainty of factor of two. 
The difference of about 970 photons in the TPC corresponds to the secondary photons produced from the incoherent $e^+e^-$ pairs after they hit the inside of the detector. 
\subsection{$\gamma\gamma\rightarrow$ hadrons}
\begin{figure}[htb]
\begin{center}
\epsfxsize=2.25in
\epsfysize=1.25in
\epsfbox{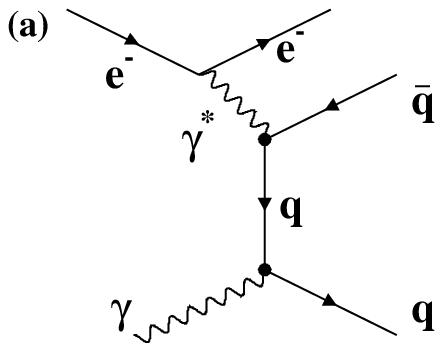}
\hspace{-0.5in}
\epsfxsize=2.25in
\epsfysize=1.25in
\epsfbox{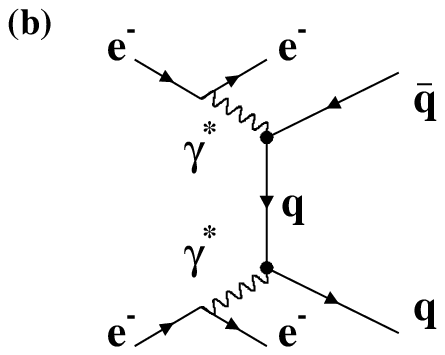}
\hspace{-0.5in}
\epsfxsize=2.25in
\epsfysize=1.25in
\epsfbox{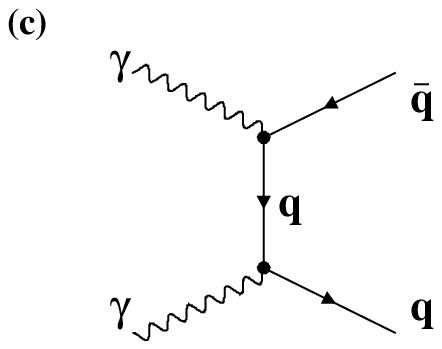}
\caption[bla]{Feynman diagrams for low energy $\gamma\gamma\rightarrow hadrons$ event production. $\gamma^{*}$ denotes a real or a virtual photon from the beamstrahlung while $\gamma$ represents a real photon from the Compton backscattering.}
\label{fig:lowgg}
\end{center}
\end{figure}

\begin{figure}[htb]
\begin{center}
\epsfxsize=3.0in
\epsfysize=3.0in
\epsfbox{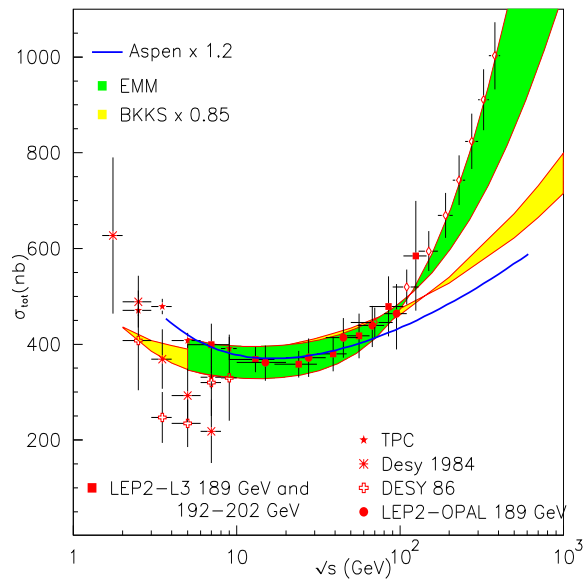}
\caption[bla]{The total $\gamma\gamma$ cross-section as a function of the center-of-mass energy compared with some model calculations. The BKKS band corresponds to different partonic densities for the photon \cite{bkks}, the EMM band corresponds to the different choices of parameters in the EMM model (Eikonal Mini-jet Models) \cite{emm} and the solid line corresponds to a ``proton-like'' model \cite{parton}. DESY data correspond to PLUTO experiment. Empty diamonds represent the pseudo-data of a future photon collider following the EMM predictions.}
\label{fig:gg_hadrons}
\end{center}
\end{figure}

\begin{figure}[htb]
\begin{center}
\epsfxsize=3.0in
\epsfysize=2.75in
\epsfbox{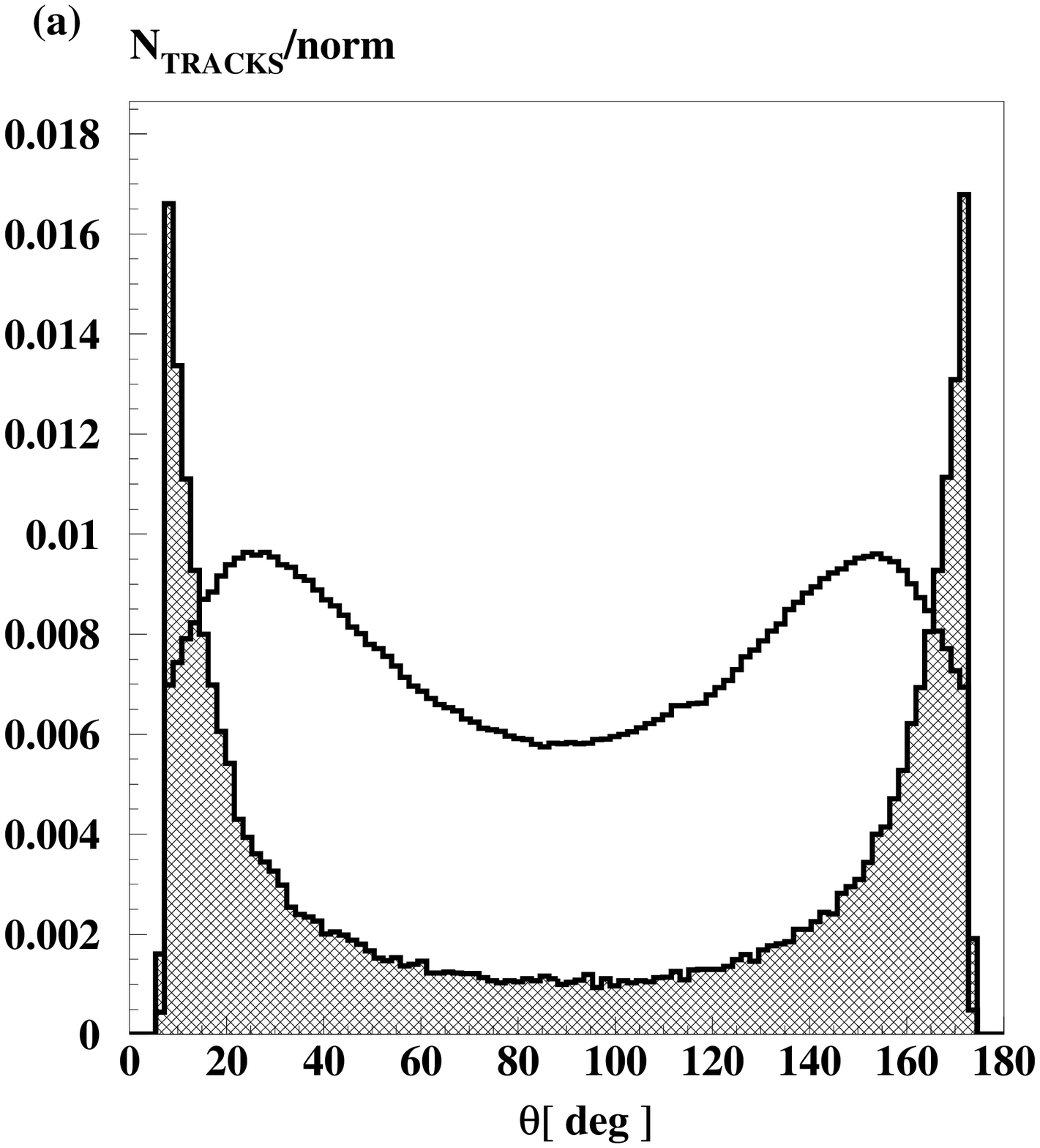}
\epsfxsize=3.0in
\epsfysize=2.75in
\epsfbox{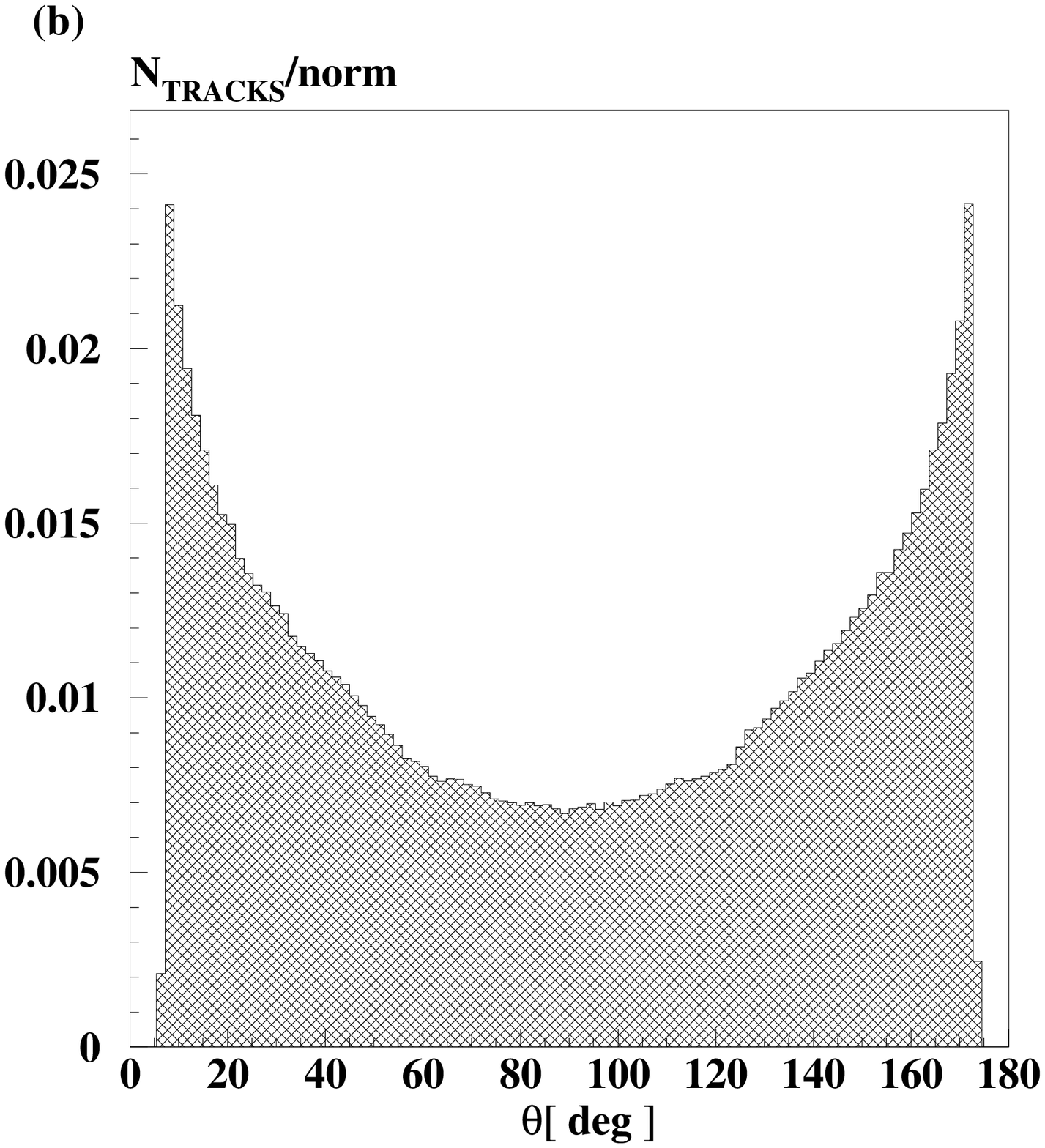}
\caption[bla]{(\textit{a}): The angular distributions of pileup tracks (hatched area) and signal tracks (solid line) and (\textit{b}): the angular distribution of signal tracks overlayed with pileup tracks after the detector simulation with included acceptance of a photon collider detector of 7$^{\circ}$. (An impact parameter cut is imposed accepting only the tracks with a transversal impact parameter less than 2$\sigma_{r\phi}$.)}
\label{fig:tracks}
\end{center}
\end{figure}

Low energy $\gamma\gamma\rightarrow hadrons$ interactions, so called pileup events, are the most specific background at a photon collider and their contribution to the high energy events has to be included. In that sense, each high energy event at a photon collider is overlayed with a corresponding number of the pileup events. The low energy photons originate mainly from the multiple electron scattering with laser photons in the conversion region and travel in the direction of the interaction point shifting the spectral luminosity of the high energy $\gamma\gamma$ and $\gamma e$ collisions contributing to lower energies. As it was already stated, their contribution can be decreased enlarging the distance $b$ between the conversion and the interaction point. The low energy photons are emitted at large angles but any compromising solution for $b$, in order to keep the desirable height of the luminosity peak, leads to their contribution of 0.1 - 2.5 events per bunch crossing at a luminosity of $L_{\gamma\gamma}\approx 0.25-4.5\cdot 10^{30}\textrm{cm}^{-2}$ per bunch crossing (bx). These numbers take into account the contribution from both, real and virtual photons from the diagrams in Fig.~\ref{fig:lowgg} and can be approximatively calculated from:
$$
N_{\gamma\gamma\rightarrow hadrons}[\textrm{bx}]^{-1}=L_{\gamma\gamma}\int\int {n_{1}(y_{1},Q^{2})n_{2}(y_{2},Q^{2}) \sigma(\gamma\gamma\rightarrow X)dy_{1}dy_{2}}
$$
where $n_{i}(y_{i},Q^{2})$ is either the virtual photon spectrum given by (\ref{eq:epa}) or the average beamstrahlung/Compton spectrum for the real photons given by (\ref{eq:epa})/(\ref{eq:compcross}) and $y$ is the energy fraction taken by the photon. Replacing $n_{i}$ by the corresponding spectrum and describing the cross-section $\sigma(\gamma\gamma\rightarrow X)$ by one of the parameterized models given in \cite{schulte1}, the numbers of low energy $\gamma\gamma$ events per bunch crossing are estimated to the previously mentioned ones. Using Telnov's spectra \cite{telnovv} this gives 1.2 events per bunch crossing at a $\gamma e$-collider ($L_{\gamma\gamma}\approx 2-3\cdot 10^{30}\textrm{cm}^{-2}\textrm{bx}^{-1}$) and 1.8 events per bunch crossing at a $\gamma\gamma$-collider ($L_{\gamma\gamma}\approx 3-4.5\cdot 10^{30}\textrm{cm}^{-2}\textrm{bx}^{-1}$) \cite{schulte}. The cross-section for these processes is about $400-600$ nb in the energy range of $\sqrt{s}=10-500$ GeV (Fig.~\ref{fig:gg_hadrons}). The events are induced by $t$-channel quark-exchange and most tracks per event are distributed over very low angles as it is illustrated in Fig.~\ref{fig:tracks}. Eventually, these tracks can be distinguished and rejected from the signal tracks if the kinematics of the signal tracks allows that. Otherwise, a large contribution from the pileup tracks will distort the angular distributions and influences the result of the analysis.

\chapter{W boson production at Photon Colliders}
So far, the TGCs have been measured mainly in diboson production with $e^{+}e^{-}$ collisions at LEP and $p\bar{p}$ collisions at Tevatron with a precision of order $\sim 10^{-2}-10^{-3}$ at LEP and $\sim 10^{-1}-10^{-2}$ at Tevatron, depending on the center-of-mass energy, achieved luminosity and the coupling under the consideration. In the future, operating at much higher center-of-mass energies and higher luminosities, using polarized beams, the International Linear Collider will give an opportunity to study the TGCs with a higher precision than LEP and Tevatron achieved. Due to the possibility to produce high energetic polarized photons in Compton backscattering, new channels for the TGCs studies will be opened via $\gamma\gamma$ and $\gamma e$ collisions. Since the photon couples to all charged particles, the pair production mostly goes via \textit{t}-channel exchange. The energy suppression of a \textit{t}-channel cross-section is smaller than for the \textit{s}-channel production. As a consequence, the cross-sections in $\gamma e$ and $\gamma\gamma$ collisions are larger than those for $e^{+}e^{-}$. The cross-section for the $W$ pair production at $\gamma\gamma$ collisions very quickly reaches a plateau of about $80$ pb and becomes constant at asymptotic energies:
\begin{equation}
\sigma(\gamma\gamma\rightarrow W^{+}W^{-})=\frac{8\pi\alpha^{2}}{M_{W}^{2}}
\label{eq:startgg}
\end{equation}
while the total cross-section for $W$ boson pair production at $e^{+}e^{-}$ collisions behaves as $1/s$ \cite{baillargeon}, i.e:
\begin{equation}
\sigma(e^{+}e^{-}\rightarrow W^{+}W^{-})  = 
\frac{\pi\alpha^{2}}{2s\sin^{4}\theta_{w}}\log(\frac{s}{M_{W}^{2}}).
\label{eq:startee}
\end{equation}
\begin{figure}[p]
\begin{center}
\epsfxsize=6.0in
\epsfysize=6.50in
\epsfbox{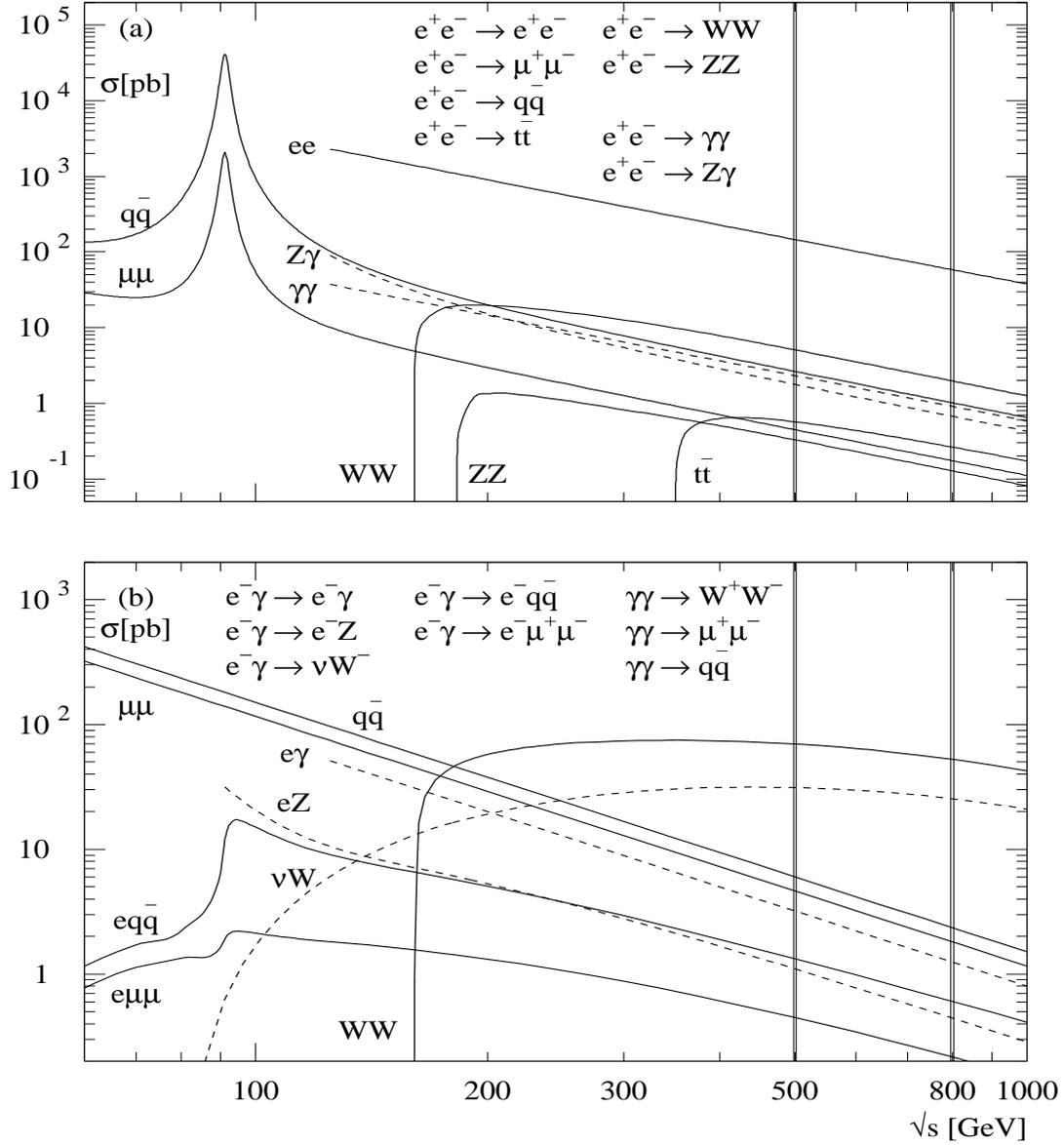}
\caption[bla]{Comparison of unpolarized cross-sections $e^{+}e^{-}$ (\textit{upper plot}) and $\gamma {e^{-}}$ and $\gamma\gamma$ (\textit{lower plot}) collisions. The applied angular cut on the $W$ boson production in $\gamma e$ and $\gamma\gamma$ interactions leads to the smooth decreas of the cross-section at high $\sqrt{s}$.}
\label{fig:cut_gg}
\end{center}
\end{figure}
Such a cross-section behavior for $\gamma\gamma$ reactions ensures a high event rate and makes photon collider attractive for anomalous TGC measurements at high center-of-mass energies. The unpolarized cross-section for single $W$ boson production ($\sigma_{W}$) in $\gamma e^{-}\rightarrow W^{-}\nu_{e}$ and $W$ boson pair production ($\sigma_{WW}$) in $\gamma\gamma\rightarrow W^{+}W^{-}$ at different $\sqrt{s}$ values is shown in Fig.~\ref{fig:cut_gg} in comparison with the corresponding pair production cross-sections in $e^{+}e^{-}$ collisions. In spite of the lower luminosities at a photon collider\footnote{$L_{\gamma\gamma,\gamma e}(z>0.8z_{max})\approx \frac{1}{3}L_{e^{+}e^{-}}$ (see caption of Table \ref{tab:parameters}).} the event rates would be somewhat higher than those at $e^{+}e^{-}$. At $\sqrt{s_{\gamma\gamma}}=500$ GeV, $\sigma_{WW}$ is $\approx 10$ times larger than $\sigma_{WW}$ at $e^{+}e^{-}$ collisions and the number of $\gamma\gamma\rightarrow W^{+}W^{-}$ events is $\approx 3$ times larger.
\par
\begin{figure}[htb]
\begin{center}
\epsfxsize=3.in
\epsfysize=1.75in
\epsfbox{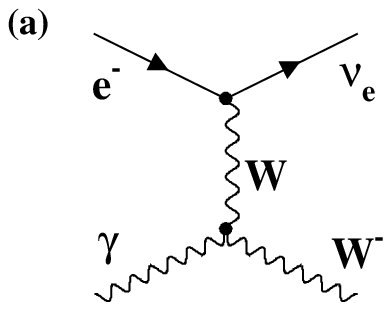}
\epsfxsize=2.5in
\epsfysize=1.5in
\epsfbox{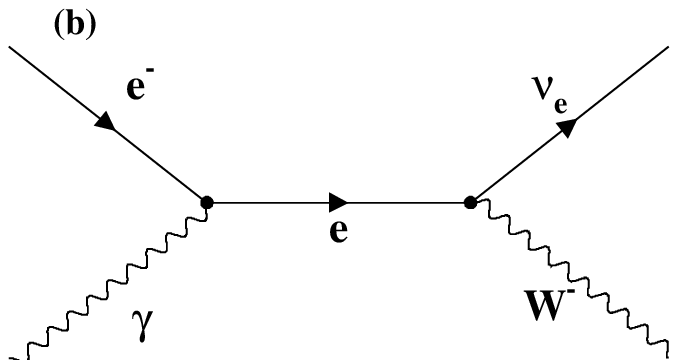}
\caption[bla]{(\textit{a}): Feynman \textit{t}-channel $W$-exchange and (\textit{b}): \textit{s}-channel $e$-exchange diagrams contributing to the single $W$ boson production in $\gamma e^{-}\rightarrow W^{-}\nu_{e}$.}
\label{fig:gewnu}
\end{center}
\end{figure}

The single $W$ boson production in $\gamma e^{-}\rightarrow W^{-}\nu_{e}$ and the $W$ boson pair production in $\gamma\gamma\rightarrow W^{+}W^{-}$ depends on the initial photon and electron beam polarizations. For each channel there are two possible initial states. In $\gamma e$ collisions there are states with $|J_{Z}|=1/2$, if the initial helicities of the photon and electron have the same sign, and $|J_{Z}|=3/2$ if the initial helicities of the photon and electron have the opposite sign. The lowest order dominating Feynman diagrams contributing to the single $W$ boson production in $\gamma e^{-}\rightarrow W^{-}\nu_{e}$ are shown in Fig.~\ref{fig:gewnu}.
The charged\footnote{Denotes the coupling of the photon to the charged gauge bosons.} $C$ and $P$ conserving TGCs, $\kappa_{\gamma}$ and $\lambda_{\gamma}$, contribute only through \textit{t}-channel \textit{W}-exchange, present at the $WW\gamma$ vertex. The beam electrons are required to be left-handed since the $W$ boson does not couple to right-handed electrons. On the other hand, the photons can be right-handed or left-handed leading to the initial states $|J_{Z}|=3/2$ and $|J_{Z}|=1/2$, respectively.
\par
\begin{figure}[htb]
\begin{center}
\epsfxsize=2.75in
\epsfysize=1.5in
\epsfbox{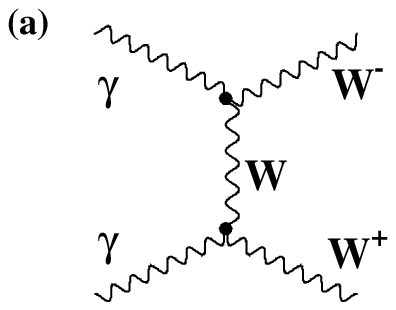}
\epsfxsize=2.75in
\epsfysize=1.5in
\epsfbox{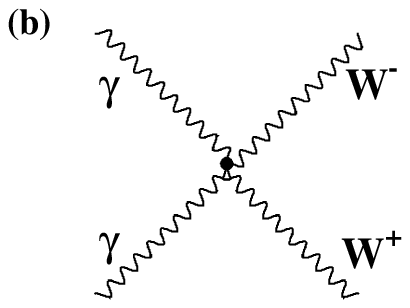}
\caption[bla]{Lowest order Feynman diagrams contributing to the $W$ boson pair production in $\gamma\gamma\rightarrow W^{+}W^{-}$. The diagram $(a)$ is dominating one.}
\label{fig:ggww}
\end{center}
\end{figure}

In $\gamma\gamma$ collisions it is possible to produce the $W$ boson pair having the two different $\gamma\gamma$ initial states: $J_{Z}=0$ if the initial photon helicities have the same sign ($\pm\pm$), and $|J_{Z}|=2$ if the initial photon helicities have the opposite signs ($\pm\mp$). A difference between the $W$ boson pair production in $e^{+}e^{-}$ and $\gamma\gamma$ collisions is that $\gamma\gamma$ collisions allow to access directly the state $J_{Z}=0$, that is not allowed at $e^{+}e^{-}$ collisions where the chirality highly suppresses this \textit{s}-channel production and only the $|J_{Z}|=1$ state is available. Besides, there is an additional state with $|J_{Z}|=2$ that can be used in the study of the New Physics effects if resonances of the strong EWSB exist. The lowest order Feynman diagrams contributing to the $W$ boson pair production in $\gamma\gamma\rightarrow W^{+}W^{-}$ are shown in Fig.~\ref{fig:ggww} where the diagram ($a$) is the dominating one. 
\par
In analogy with the $W$ boson pair production in $e^{+}e^{-}$ collisions the five-fold differential cross-section for $\gamma\gamma\rightarrow W^{+}W^{-}\rightarrow f_{1}\bar{f}_{2}f_{3}\bar{f}_{4}$ for definite photon helicities $\lambda_{1,2}$ and the $W$ boson helicities $\lambda_{+,-}$ can be defined in the same way as (\ref{eq:diff_eeww3}).
\par
In $W$ boson pair production the charged TGCs contribute through the dominating \textit{t}-channel \textit{W}-exchange , unlike in $e^{+}e^{-}$ where the mixing of $WW\gamma$ and $WWZ$ vertexes is present through the \textit{s}-channel exchange. Thus, at $\gamma e$ and $\gamma\gamma$ collisions, the independent\footnote{$WWZ$ couplings are absent and thus the $SU(2)_{L}\times U(1)_{Y}$ relation is not applicable.} measurement of the $C$ and $P$ conserving couplings $\kappa_{\gamma}$ and $\lambda_{\gamma}$ is allowed. Using polarized $e^{+}e^{-}$ beams it is also possible to suppress the $Z$-exchange diagram and to measure these couplings independently, as it will be explained later. The value of $g_{1}^{\gamma}$ is fixed by the electro-magnetic gauge invariance, i.e. $g_{1}^{\gamma}=1$. 
\section{Single $W$ boson production at a Photon Collider}
%
\begin{figure}[htb]
\begin{center}
\epsfxsize=3.0in
\epsfysize=3.0in
\epsfbox{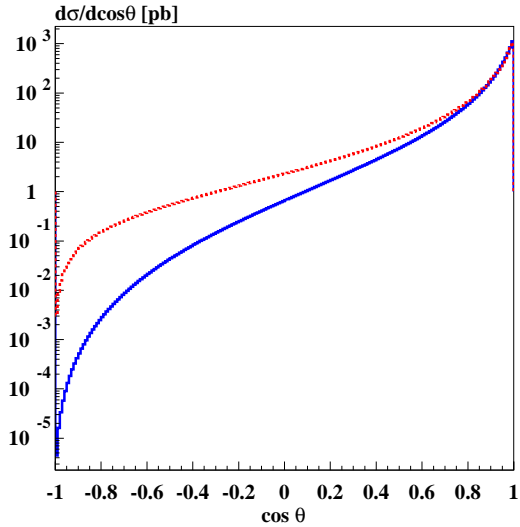}
\caption[bla]{The Standard Model differential cross-section distributions for $\gamma e\rightarrow W^{-}\nu_{e}$ with two different initial photon helicities - left-handed, i.e. $|J_{Z}|=1/2$ (dotted line) and right-handed, i.e. $|J_{Z}|=3/2$ (solid line) at $\sqrt{s_{\gamma e}}=450$ GeV, assuming 100$\%$ polarized beams. The contribution from the \textit{s}-channel is visible for left-handed photons leading to a larger cross-section.}
\label{fig:diffeg}
\end{center}
\end{figure}
\begin{figure}[htb]
\begin{center}
\epsfxsize=3.5in
\epsfysize=3.5in
\epsfbox{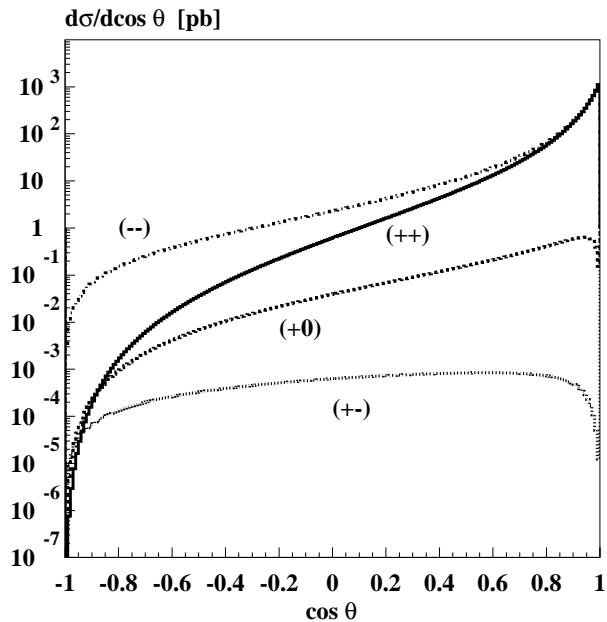}
\caption[bla]{Contribution of each $W$ boson helicity state for $|J_{Z}|=1/2$ and $|J_{Z}|=3/2$ to the Standard Model differential cross-section at $\sqrt{s_{\gamma e}}=450$ GeV, assuming 100$\%$ polarized beams. The angle $\theta$ is defined as the angle between the $\gamma$ beam and the outgoing $W$ boson. Notation: ($h_{\gamma},h_{W}$) = ($\gamma$ helicity, $W$ helicity).}
\label{fig:diffalleg}
\end{center}
\end{figure}

The differential $\gamma e\rightarrow W^{-}\nu_{e}$ cross-sections for the two different initial photon helicities, $|J_{Z}|=1/2$ and $|J_{Z}|=3/2$, at $\sqrt{s_{\gamma e}}=450$ GeV ($\approx 90\%\sqrt{s_{e^{-}e^{-}}}$) are shown in Fig.~\ref{fig:diffeg}, assuming left-handed electrons. Due to the \textit{t}-channel (Fig.~\ref{fig:gewnu}\,$a$) the differential cross-sections are peaked at small angles $\theta$, related to the photon beam direction. For $|J_{Z}|=1/2$, the \textit{s}-channel contribution leads to a higher differential cross-section while for the $|J_{Z}|=3/2$ state the \textit{s}-channel contribution is suppressed. The contribution of each $W$ boson helicity state to the differential cross-section is shown in Fig.~\ref{fig:diffalleg}. For both initial $J_{Z}$ states, the contributions from the transversal $W$ bosons with the same helicity as the initial photon are the dominating ones.
\begin{figure}[htb]
\begin{center}
\epsfxsize=3.0in
\epsfysize=3.0in
\epsfbox{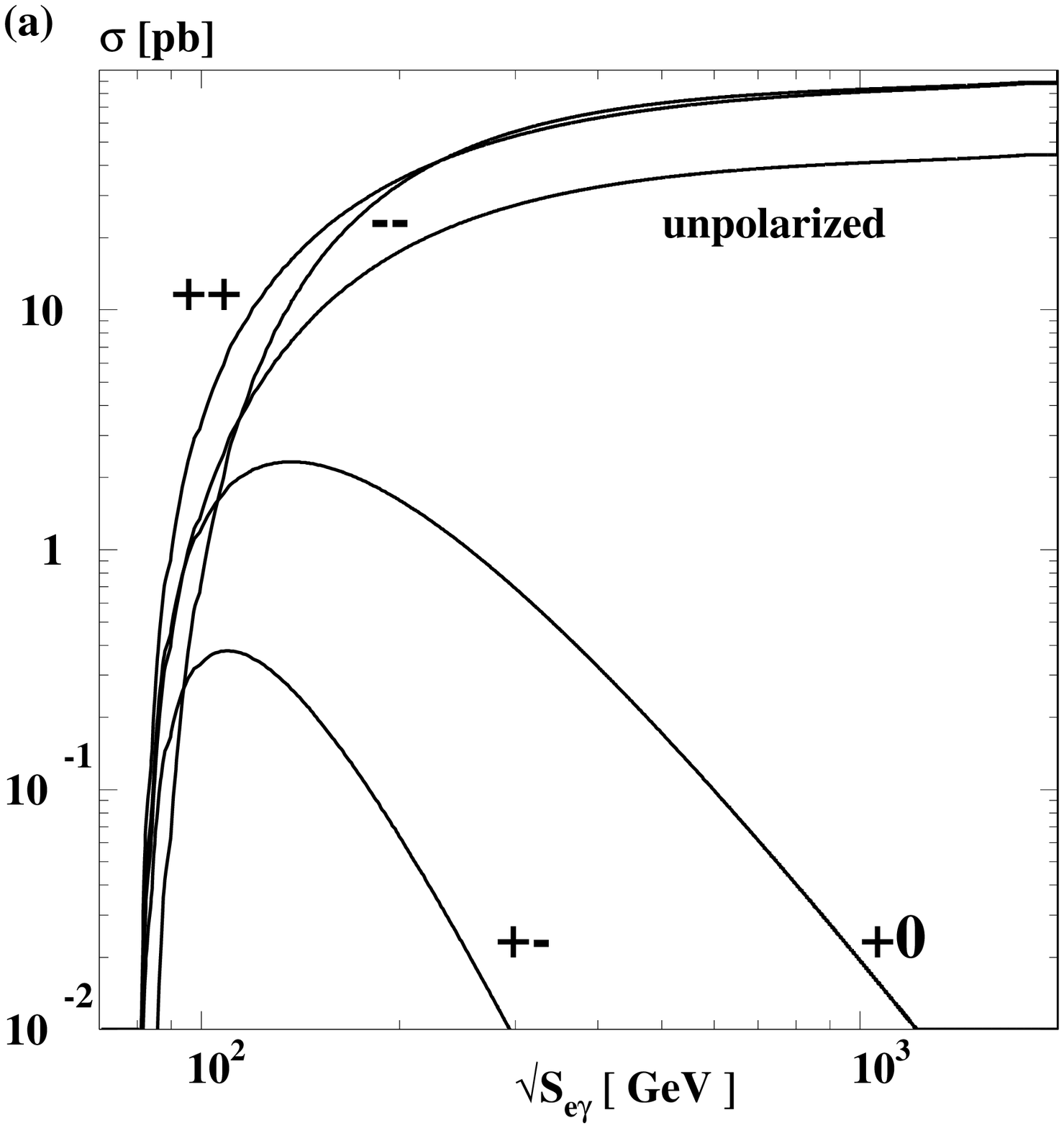}
\epsfxsize=3.0in
\epsfysize=3.0in
\epsfbox{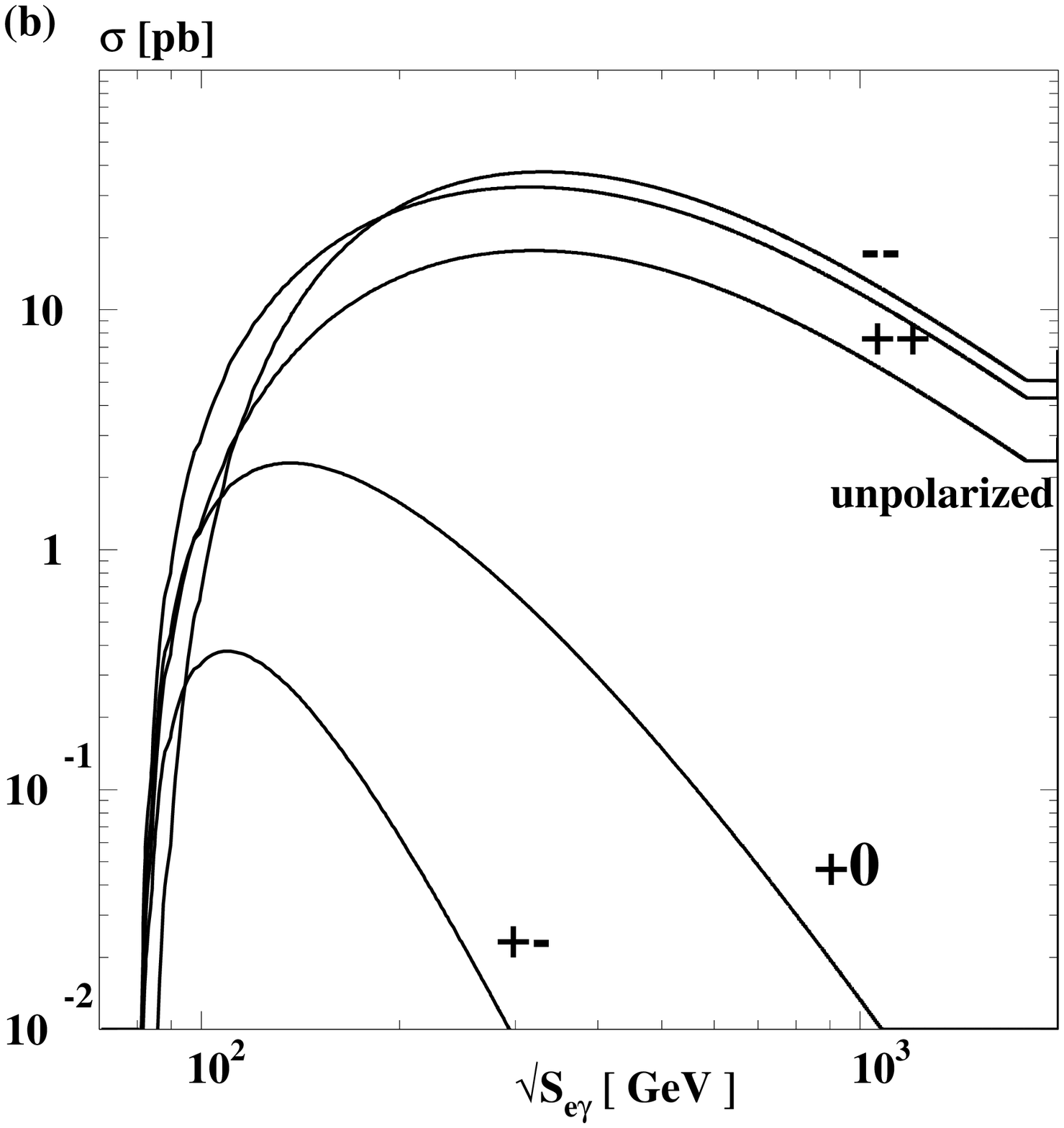}
\caption[bla]{Total lowest-order cross-sections as a function of ${\sqrt{s_{\gamma e}}}$ for different $W$ boson helicities assuming that the electron is left-handed and the beams are 100$\%$ polarized. (\textit{a}): without an angular cut. (\textit{b}): with an angular cut $20^{\circ} \leq \theta \leq 160^{\circ}$. Notation: ($h_{\gamma},h_{W}$) = ($\gamma$ helicity, $W$ helicity).}
\label{fig:fscmeg}
\end{center}
\end{figure}
\begin{table}[htb]
\begin{center}
\begin{tabular}{|c||c|c|c|c|c|c|} \hline
\multicolumn{2}{|c|}{} & \multicolumn{3}{|c|} {$J_{Z}=\frac{3}{2}$} & $J_{Z}=\frac{1}{2}$ \\ \hline\hline
 $\sqrt{s}$[GeV] & $\theta$ & $W_{T_{(+)}}$ & $W_{T_{(-)}}$ & $W_{L}$ & $W_{T_{(-)}}$ \\ \hline\hline
 450 & $\begin{array}{c}
        0^{\circ}-180^{\circ} \\
        20^{\circ}-160^{\circ} \\
        \end{array}$ 
        & $\begin{array}{c}
         66.6 \\
         29.1 \\
        \end{array}$
        & $\begin{array}{c}
        1.0\cdot 10^{-3} \\
        1.0\cdot 10^{-3} \\
         \end{array}$
        & $\begin{array}{c}
        2.3\cdot 10^{-1} \\
        2.1\cdot 10^{-1} \\
        \end{array}$
        & $\begin{array}{c}
        70.2 \\
        34.2 \\
        \end{array}$ \\ \hline\hline
900 & $\begin{array}{c}
        0^{\circ}-180^{\circ} \\
        20^{\circ}-160^{\circ}\\
        \end{array}$ 
        & $\begin{array}{c}
        79.7 \\
        13.7 \\
        \end{array}$
        & $\begin{array}{c}
        2.0\cdot 10^{-5} \\
        1.9\cdot 10^{-5} \\
         \end{array}$
        & $\begin{array}{c}
        2.7\cdot 10^{-2} \\
        2.0\cdot 10^{-2} \\
        \end{array}$
        & $\begin{array}{c}
        82.1 \\
        16.2 \\
        \end{array}$ \\ \hline
\end{tabular}
\end{center}
\caption{The Standard Model cross-sections in [pb] for production of different $W$ boson helicities in $|J_{Z}|=1/2$ and $|J_{Z}|=3/2$, in $\gamma e^{-}\rightarrow W^{-}\nu_{e}$ for two different $\sqrt{s}$, assuming 100$\%$ polarized beams. It is assumed that the electrons are left-handed. The sign in brackets denotes the $W$ boson helicity.}
\label{tab:cross_ge}
\end{table}
The state $|J_{Z}|=1/2$ in the Standard Model receives only a contribution from $W$ bosons with helicity $h_{W}=-1$. For the gauge boson helicity combinations ($h_{\gamma},h_{W}$)=(-1,+1) and (-1,0) the Standard Model amplitudes are equal to zero. For the initial $|J_{Z}|=3/2$ state the helicity conservation allows the contribution of all $W$ boson helicity states to the differential cross-sections. Different $W$ boson helicity states are contained in the differential cross-section distribution over the $W$ boson decay angle as:
\begin{equation}
{\frac{{d^2\sigma}(W\rightarrow f\bar{f}^{'})}{d\cos\theta d\cos{\theta}_{1}}} =
{\frac{3}{4}}\left[{\frac{1}{2}}{\frac{{d{\sigma}_{T}}}
{d\cos\theta}}(1+{\cos^{2}}{\theta_{1}})+{\frac{{d{\sigma}_{L}}}
{d\cos\theta}}{\sin^{2}}{\theta_{1}}\right]
\label{eq:decayeg}
\end{equation}
where $\theta$ denotes the production angle of the $W$ boson relative to the beam axis. The angle $\theta_{1}$ denotes the decay angle of the $W$ boson, i.e. the angle of the produced fermions relative to the $W$ boson direction in the rest frame of the $W$ boson. More detailed explanation of fermion-antifermion distributions in the rest frame of the $W$ boson is given in Chapter 2.
\par
${\frac{{d{\sigma}_{T}}}{d{\cos \theta}}}$ is the differential cross-section for the production of transversely polarized $W$ bosons distributed as ${(1+{\cos^{2}}{\theta_{1}})}$ and ${\frac{{d{\sigma}_{L}}}{d{\cos \theta}}}$ is the differential cross-section for longitudinal $W$ production, distributed as ${\sin^{2}}{\theta_{1}}$.
\par
The contribution of each $W$ boson helicity state to the total cross-section for different center-of-mass energies is shown in Fig.~\ref{fig:fscmeg}. At high energies the total cross-sections for different $(h_{\gamma},h_{W})$ combinations behave as \cite{denner}:
\begin{equation}
\begin{array}{lcl}
\sigma_{tot} (\pm 1,\pm 1) & \sim & \sigma_{0} \\
\sigma_{tot} (+1,-1) & \sim & (1/2)\sigma_{0}(M_{W}^{6}/s^{3}) \\
\sigma_{tot} (+1,0) & \sim & \sigma_{0}(M_{W}^{4}/s^{2})[2\log (s/M_{W}^{2})-5],
\label{eq:section}
\end{array}{}
\end{equation}
where $\sigma_{0}=\alpha^{2}\pi/M_{W}^{2}\sin^{2}\theta$. The helicity changing channels ($\Delta=h_{\gamma}-h_{W}\neq 0$) are suppressed by several orders of $M_{W}^{2}/s$ tending to zero rapidly. Thus, the dominating contribution to the total cross-section comes from $W_{T}$ bosons with the same helicity as the initial photon and it is about two orders of magnitude larger than the contribution from the $W_{L}$ bosons at $\sqrt{s}=450$ GeV. At higher energies, the cut on the $W$ boson production angle of $20^{\circ} \leq \theta \leq 160^{\circ}$ decreases the contribution from $W_{T}$ bosons with the same helicity as the initial photon without a significant influence on other $W$ boson helicity states (Table \ref{tab:cross_ge}). This can be used for the extraction of the $W_{L}$ bosons whose contribution does not decreases if an angular cut is imposed. 
\par
\begin{figure}[p]
\begin{center}
\epsfxsize=3.0in
\epsfysize=3.0in
\epsfbox{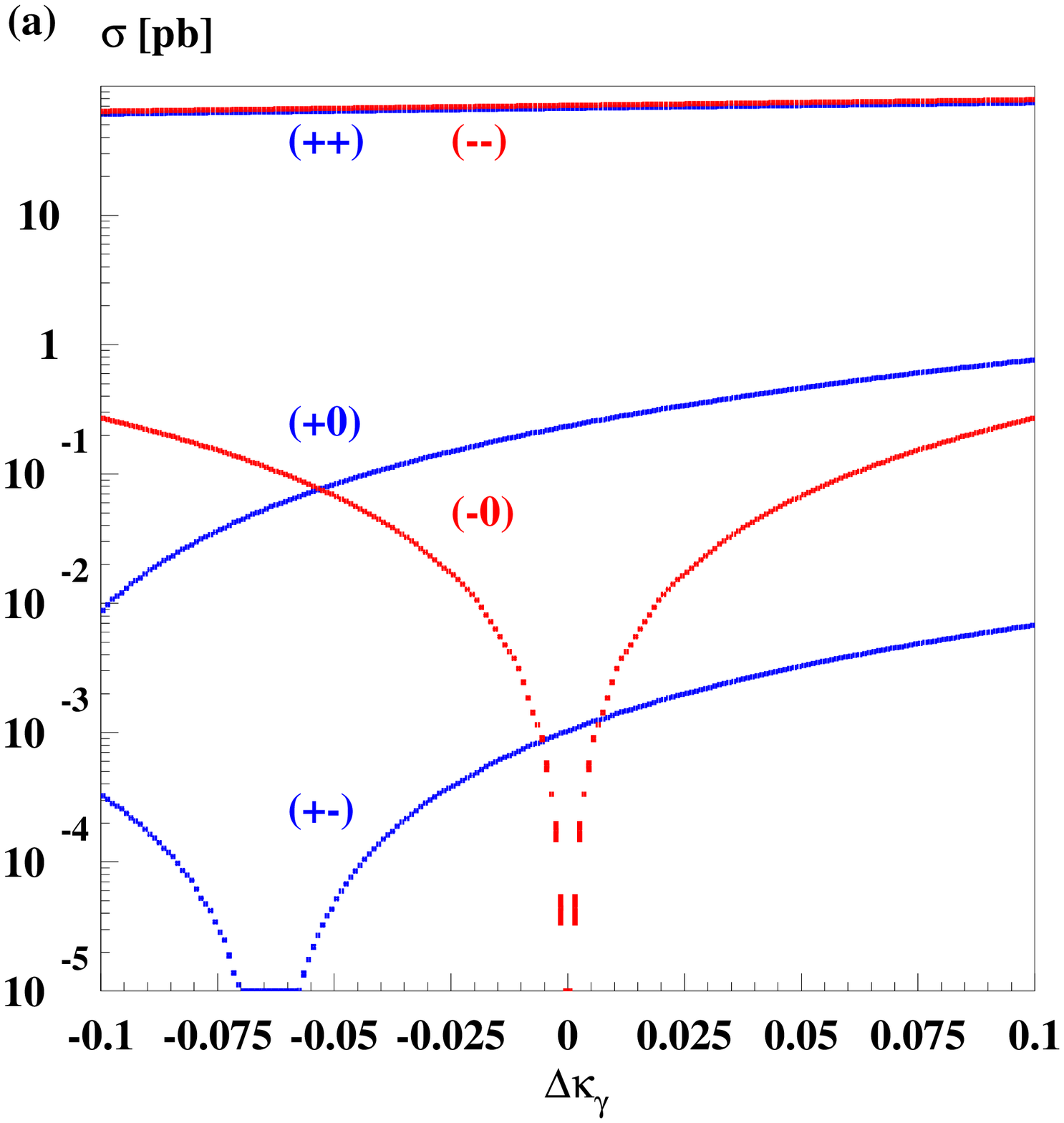}
\epsfxsize=3.0in
\epsfysize=3.0in
\epsfbox{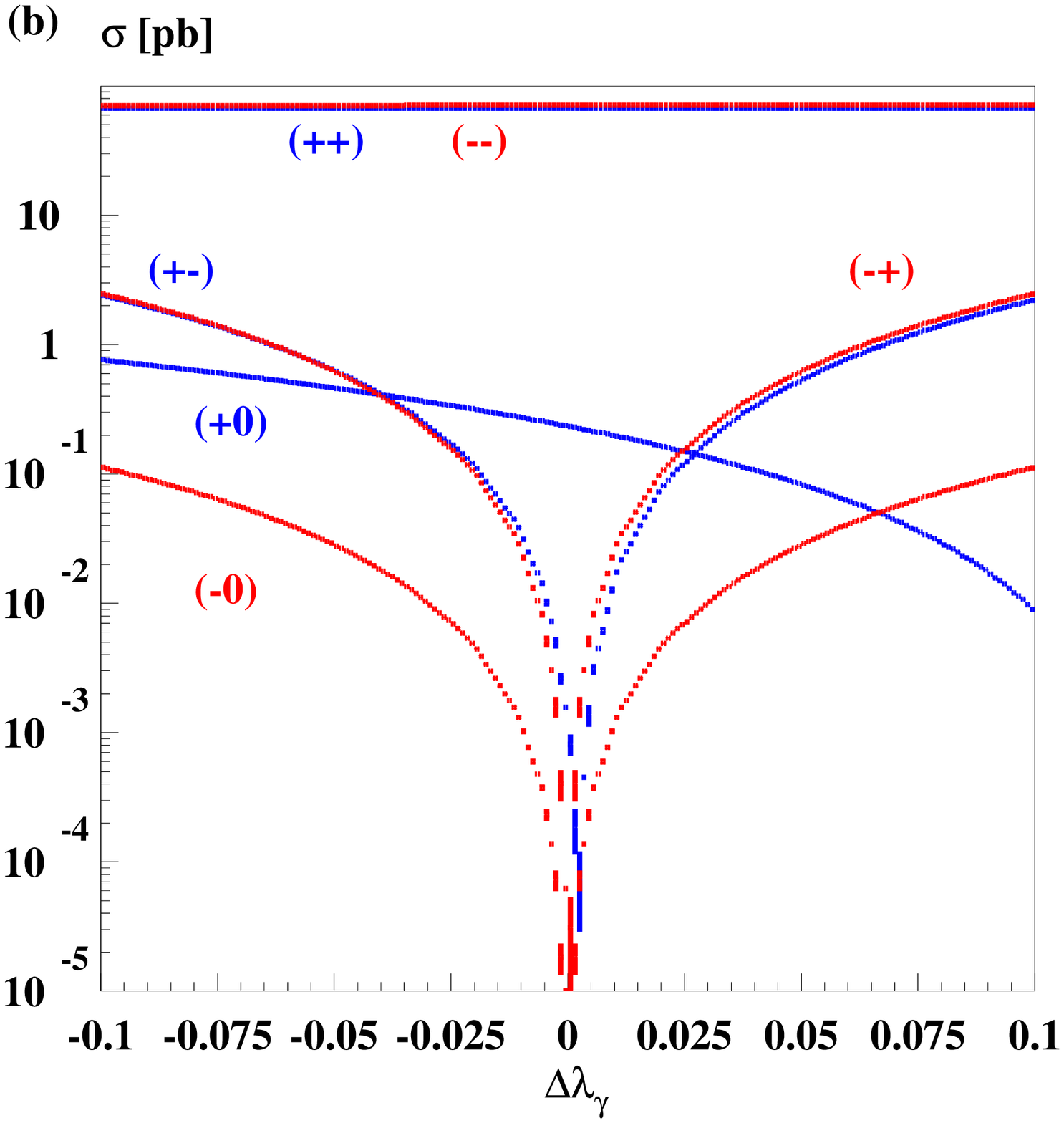}
\epsfxsize=3.0in
\epsfysize=3.0in
\epsfbox{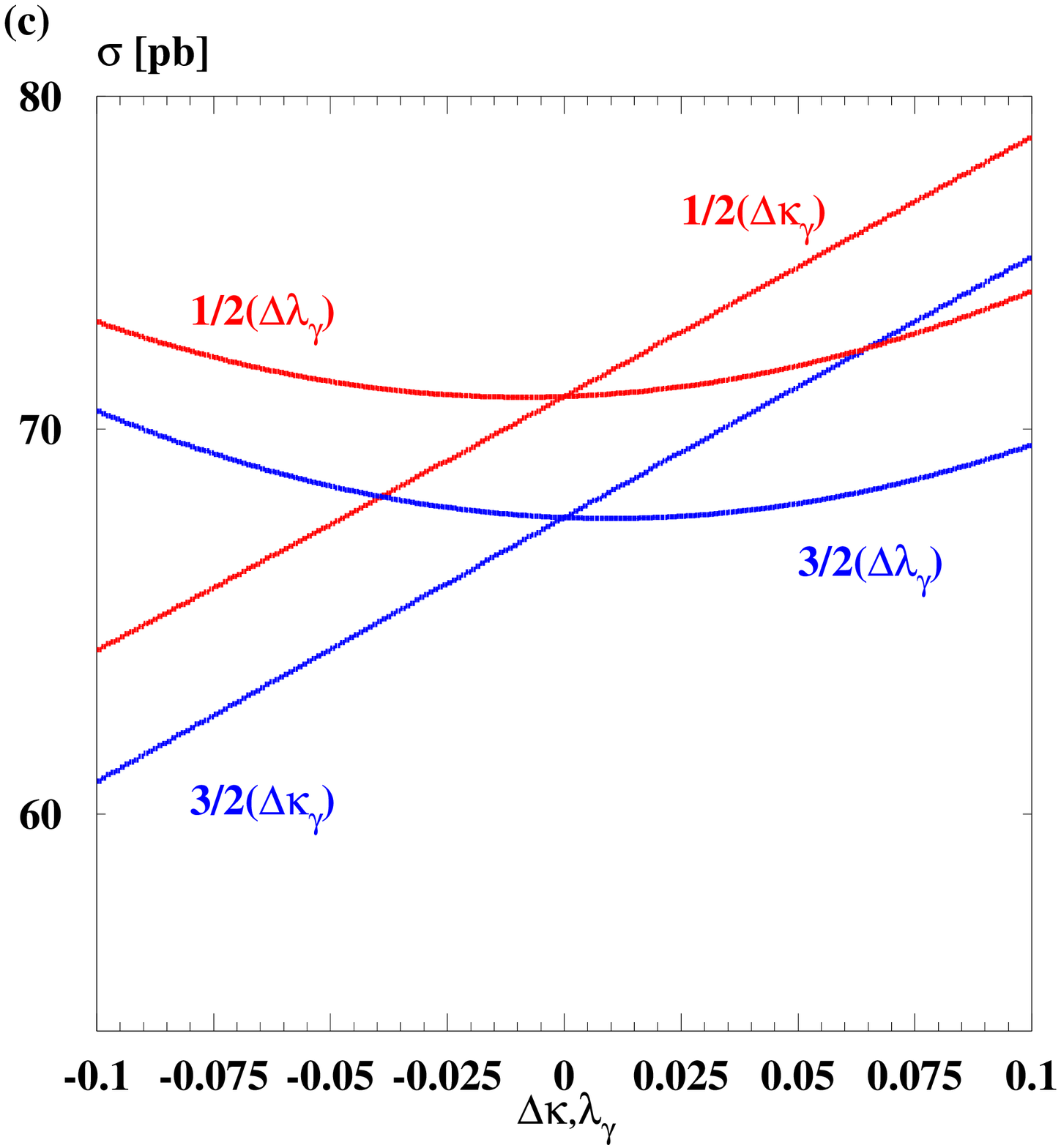}
\caption{Contribution of different $W$ boson helicity states for $|J_{Z}|=1/2$ and $|J_{Z}|=3/2$ in the presence of anomalous couplings (\textit{a}): ${\Delta}{\kappa}_{\gamma}$ and (\textit{b}): ${\Delta}{\lambda}_{\gamma}$ at $\sqrt{s_{\gamma e}}=450$ GeV, assuming 100$\%$ polarized beams. (\textit{c}): Total cross-section dependence on anomalous $\kappa_{\gamma}$ and $\lambda_{\gamma}$. Notation: ($h_{\gamma},h_{W}$) = ($\gamma$ helicity, $W$ helicity).}
\label{fig:kappa_lambda}
\end{center}
\end{figure}
\begin{figure}[p]
\begin{center}
\epsfxsize=3.0in \epsfysize=3.0in \epsfbox{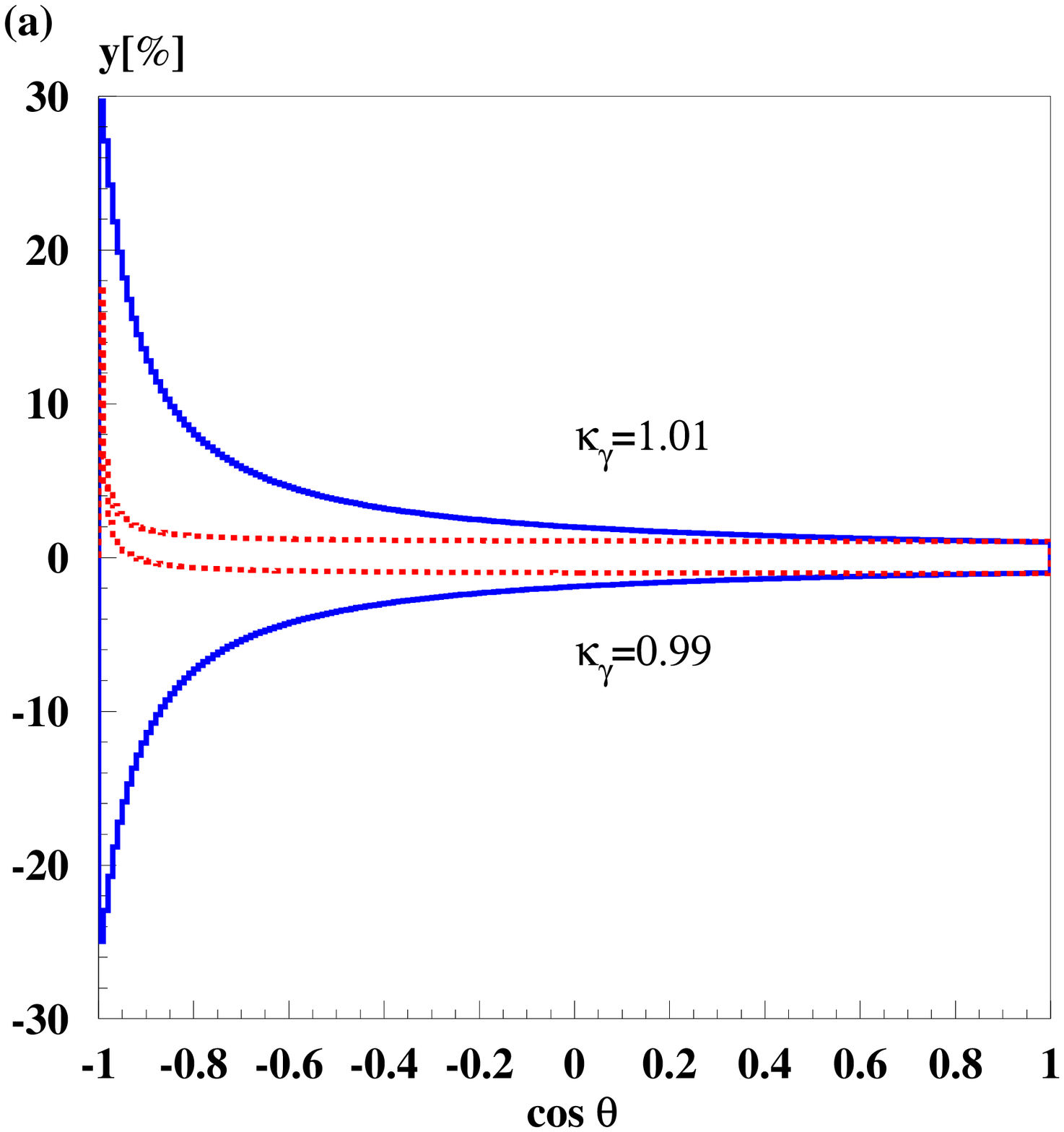}
\epsfxsize=3.0in \epsfysize=3.0in \epsfbox{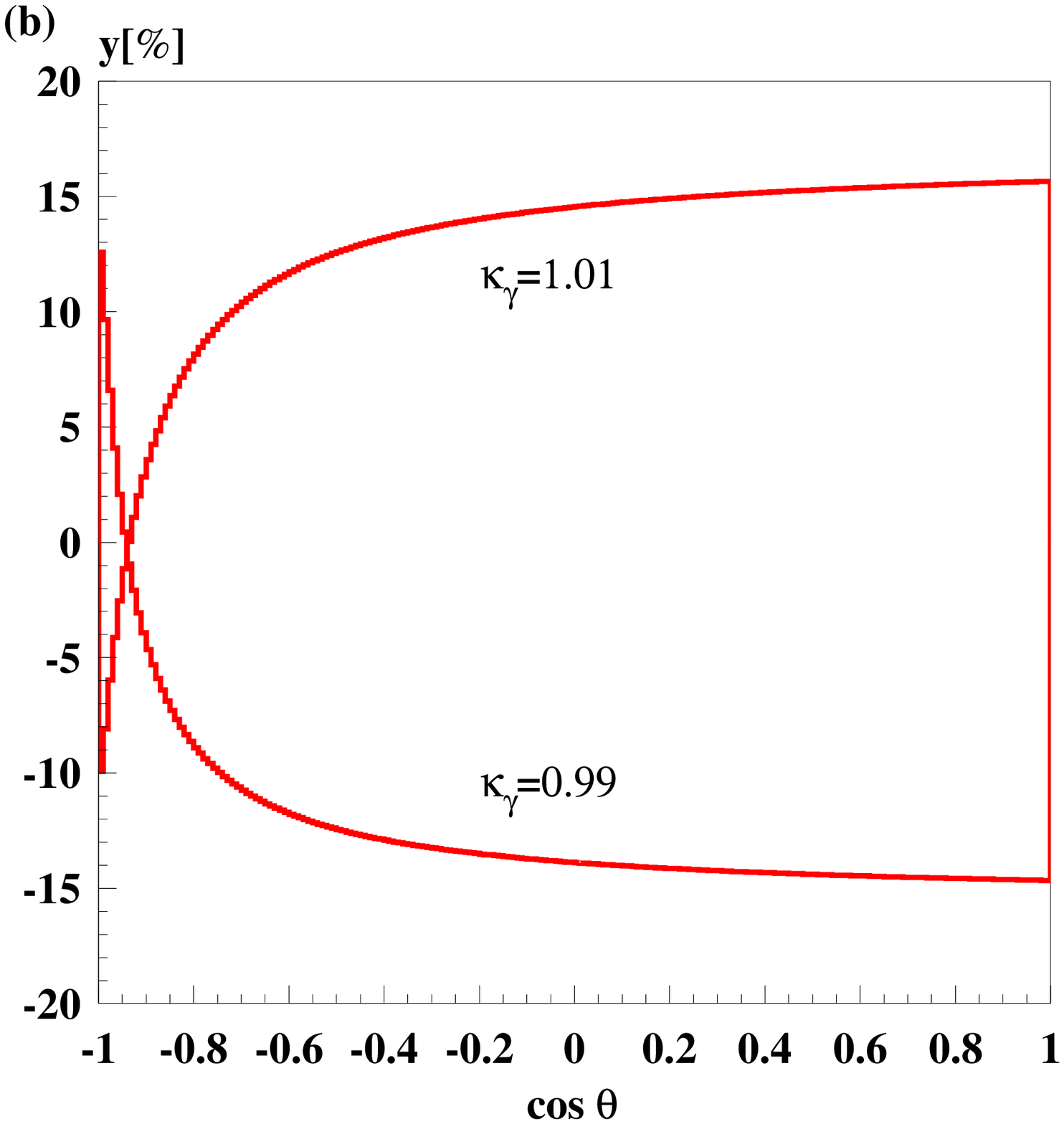}
\caption[bla]{(\textit{a}): Differential cross-section in the presence of anomalous TGCs for both initial photon helicity states - $|J_{Z}|=1/2$ (dotted lines) and $|J_{Z}|=3/2$ (solid lines), normalized to their Standard Model values at $\sqrt{s_{\gamma e}}=450$ GeV, assuming 100$\%$ polarized beams. $y=\frac{d\sigma_{TOT}^{AC}-d\sigma_{TOT}^{SM}}{d\sigma_{TOT}^{SM}}$. (\textit{b}): Deviation of the $W_{L}$ boson fraction in presence of anomalous TGCs from the fraction predicted by the Standard Model for ${\Delta}{\kappa}_{\gamma}={\pm}0.01$. $y=\frac{d\sigma_{LL}^{AC}-d\sigma_{LL}^{SM}}{d\sigma_{LL}^{SM}}$.}
\label{fig:kappa}
\end{center}
\end{figure}

Anomalous TGCs affect both the total production cross-section and the shape of the differential cross-section as a function of the $W$ boson production angle. The relative contributions of each helicity state of the $W$ boson to the total $W$ boson production cross-section in the presence of anomalous couplings ($\kappa_{\gamma}\neq$1 and $\lambda_{\gamma}\neq$0) at $\sqrt{s}=450$ GeV are shown in Fig.~\ref{fig:kappa_lambda}. In the Standard Model $\kappa_{\gamma}=1$ and $\lambda_{\gamma}=0$ and any deviation from these values is denoted as $\Delta\kappa_{\gamma}$ and $\Delta\lambda_{\gamma}$. The helicity non-changing channels are almost not sensitive to the anomalous TGCs. The most interesting channel for a study of New Physics effects, the production of $W_{L}$ bosons with initial right-handed photons, is sensitive to the anomalous TGCs. Fig.~\ref{fig:kappa}\,$a$ shows that the differential cross-section distribution in the backward\footnote{$W$ production angle is defined as the angle between the photon and the $W$ boson.} region is more sensitive to the presence of the anomalous coupling $\kappa_{\gamma}$ in the case of right-handed photons than for left-handed ones. The value $y$ represents the relative deviation in angular distribution in the presence of the anomalous $\kappa_{\gamma}$ related to the Standard Model prediction. Fig.~\ref{fig:kappa}\,$b$ shows the deviation of the $W_{L}$ boson angular distribution in the presence of the anomalous $\kappa_{\gamma}$ relative to the Standard Model prediction. The production of $W_{L}$ bosons in the presence of anomalous couplings will differ from the Standard Model prediction. This behavior comes from the fact that the information about the strong EWSB can be obtained through the study of Goldstone boson interactions which are the longitudinal component of the gauge bosons. Total and differential cross-sections are calculated on the basis of the formula given in \cite{denner} using helicity amplitudes in the presence of anomalous couplings from \cite{yehu}. In the $W$ boson production via $\gamma e$ collisions the favorable initial ``photon - electron'' helicity states are ``right-left'', respectively. Because of the missing \textit{s}-channel $e$-exchange in this state, the $W$ boson angular distributions show larger sensitivity to TGCs in the backward region than in the case with initial left-handed photons where the production of the $W_{L}$ bosons is suppressed.
\begin{figure}[p]
\begin{center}
\epsfxsize=5.in
\epsfysize=5.in
\epsfbox{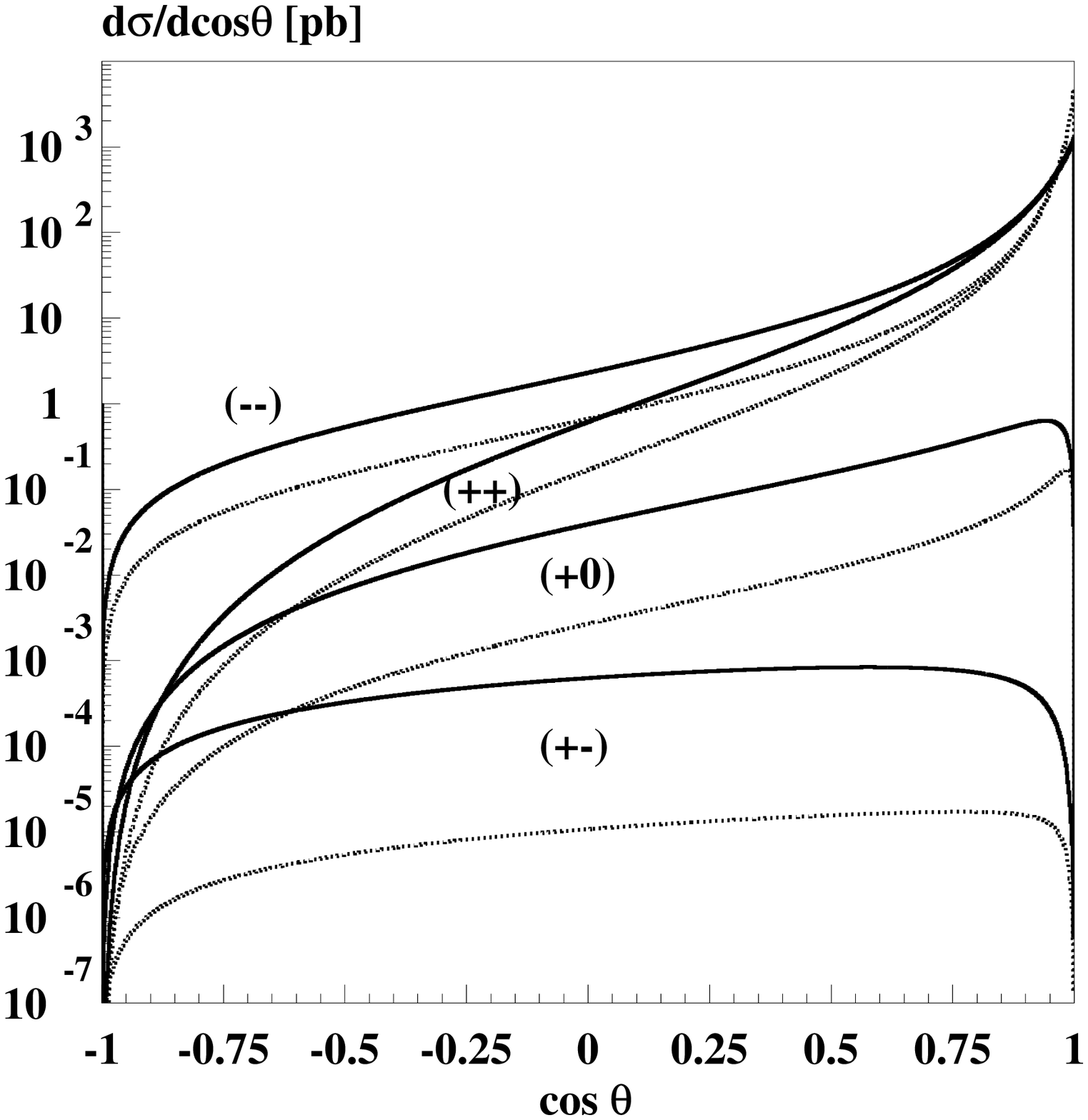}
\caption[bla]{Contribution of each $W$ boson helicity state for $|J_{Z}|=1/2$ and $|J_{Z}|=3/2$ to the Standard Model differential cross-section at two different $\sqrt{s}$. The solid lines correspond to the $\sqrt{s_{\gamma e}}=450$ GeV and dotted lines correspond to the $\sqrt{s_{\gamma e}}=900$ GeV, assuming 100$\%$ polarized beams. The angle $\theta$ is defined as the angle between the $\gamma$ beam and the outgoing $W$. Notation: ($h_{\gamma},h_{W}$) = ($\gamma$ helicity, $W$ helicity).}
\label{fig:born_450_900}
\end{center}
\end{figure}
\section{Initial State Radiation}
Having a collider that operates at high energies above 500 GeV requires to include the complete ${\cal O}(\alpha)$ corrections to the $W$ boson production cross-sections. The QED radiative corrections to the Born level cross-section consist of virtual corrections resulting from loop diagrams, and of a real photon emission leading to $\gamma e^{-}\rightarrow W^{-}\nu_{e}\gamma$. Most of these (bremsstrahlung) photons have a very small energy (``soft photons''), therefore it is not so easy to detect them while their contribution to the ${\cal O}(\alpha)$ corrections is significant. Since detectors are not able to detect soft photons, the events can be treated as those with visible photons i.e. with photons with an energy $E_{\gamma}$ above a certain threshold $\Delta E$ seen in the detector ($E_{\gamma}>\Delta E$) and with invisible photons i.e. with photons with an energy below a certain threshold $\Delta E$, not seen in the detector ($E_{\gamma}<\Delta E$).
\par
In general, the total production cross-section which includes the full ${\cal O}(\alpha)$ radiative corrections can be written as \cite{racoon}:
\begin{equation}
\begin{array}{lcl}
\int d\sigma_{corr} & = & \int d\sigma_{Born}+\int d\sigma_{virtual}+\int d\sigma_{real},
\label{eq:isr1}
\end{array}{}
\end{equation}
where the last two terms introduce the radiative corrections. Each of these two terms can be splitted into a part which is finite and a part which contains soft and collinear singularities\footnote{The singularities are related to the initial and final state radiation.} \cite{racoon}, so that (\ref{eq:isr1}) can be decomposed into: 
\begin{equation}
\begin{array}{lcl}
\int d\sigma_{corr} & = & \int d\sigma^{\gamma e^{-}\rightarrow W^{-}\nu_{e}}_{Born}+\int d\sigma^{\gamma e^{-}\rightarrow W^{-}\nu_{e}}_{virt,finite}+\int d\sigma^{\gamma e^{-}\rightarrow W^{-}\nu_{e}}_{virt+real,sing}+\int d\sigma^{\gamma e^{-}\rightarrow W^{-}\nu_{e}\gamma}_{finite},
\label{eq:isr2}
\end{array}{}
\end{equation}
where the last term is added since the experimental and theoretical cut-off $\Delta E$ might not be the same. The singularities arising from the virtual and real corrections are regularized introducing an infinitesimal photon mass and small fermion masses. They are canceled in the sum of the virtual and real corrections, integrated up to some cut-off energy $\Delta E_{theo}$ that should be small enough, so that the soft-photon approximation applies \cite{spa}. The remnants of this integration are the leading-logarithms $\sim (\alpha/\pi)\ln(s/m_{e}^{2})\ln(E/\Delta E)$. Thus, the cross-section for events in which the detector does not measure the radiated photon is given by a sum of the Born cross-section corrected by the virtual loops and the cross-section for the emission of a real photon with energy smaller than $\Delta E$, is given by:  
\begin{equation}
d\sigma_{measured}\sim d\sigma_{Born}\left[1-\frac{\alpha}{\pi}\log\left(\frac{s}{m_{e}^{2}}\right)
\log\left(\frac{E}{\Delta E}\right)^{2}+{\cal O}(\alpha^{2})\right]+
\int_{\Delta E_{theo}}^{\Delta E_{exp}}d\sigma_{finite}^{\gamma e^{-}\rightarrow W^{-}\nu_{e}\gamma}.
\label{eq:isr3}
\end{equation}
The last term in (\ref{eq:isr2}), included in the full ${\cal O}(\alpha)$ corrections, represents the emission of a visible (hard) photon with an energy above $\Delta E_{exp}$ and can be detected by a detector.
\par
The leading-logarithmic term contained in (\ref{eq:isr3}) arises if photons are radiated off the light charged particles and give rise to sizeable ${\cal O}(\alpha)$ corrections due to the difference in scale between the mass of the radiating particle and its energy ($s\gg m_{e}^{2}$). 
\par
The large leading-logarithmic corrections also can be calculated using the so-called structure-function method \cite{strfunc1,strfunc2} taken from QCD, where the ``soft-photon'' emission (the leading-logarithmic term (\ref{eq:isr3})) is included by means of exponentiation. Using this approach the corrected production cross-section is written as a convolution of the electron structure function and the lowest order differential cross-section at the reduced center-of-mass energy \cite{strfunc2,strfunc3} as:
$$
\sigma_{LL}^{\gamma e\rightarrow W\nu}=\int_{0}^{1} F_{e^{-}\rightarrow e^{-}}(x,s) \sigma_{Born}^{\gamma e\rightarrow W\nu}(sx)dx
$$
with $F_{e^{-}\rightarrow e^{-}}(x,s)$ being the electron distribution function that gives the probability of finding an electron with a longitudinal momentum fraction $x=p_{e^{-}}/E_{beam}$. 
\par
If the QED correction includes only terms $\sim((\alpha/\pi)\ln(s/m_{e}^{2}))^{n}$ this approach is called leading-logarithmic approximation (LLA). In order to achieve an accuracy at the 0.1$\%$ level a higher order corrections has to be taken into account. The Monte Carlo generators use the electron structure-function approach implemented up to ${\cal O}(\alpha^{3})$ in LLA. The expression of a leading-logarithmic structure function that includes ${\cal O}(\alpha^{3})$ terms can be found in \cite{order3}.
\subsection{Theoretical Prediction for the Standard Model Radiative ${\cal O}(\alpha)$ corrections}
The predicted full one-loop ${\cal O}(\alpha)$ radiative corrections to the Born level cross-section for the production of the single $W$ boson, from real soft-photon bremsstrahlung and virtual radiative corrections are calculated \cite{denner} in the soft-photon approximation. The cut-off dependent terms\footnote{These predictions assume a value of $\Delta E=0.05E_{beam}$ ($E_{beam}=250$ GeV).} and the leading logarithms caused by collinear photon emission ($\sim \log m_{e}$) denoted together as the QED corrections $\delta_{QED}$, are extracted from the full corrections $\delta$ in order to define the weak corrections as $\delta_{weak}=\delta-\delta_{QED}$, arising from the box diagrams and vertex corrections. 
\par
Assuming unpolarized electrons, the contribution of $\delta_{weak}$ to the total cross-section is below 5$\%$ up to energies of 1 TeV. Above 1 Tev, $\delta_{weak}$ for $h_{\gamma},h_{W}=(+1,-1)$ and $(+1,0)$ becomes large because the cross-section vanishes faster for $s\rightarrow \infty$ than the ${\cal O}(\alpha)$ corrections\footnote{In this case, the ${\cal O}(\alpha^{2})$ contributions become relevant.}. The cut on the $W$ boson production angle of $20^{\circ} <\theta <160^{\circ}$, enlarges $\delta_{weak}$ at the same $\sqrt{s}$. The reason for that is the fact that the cutting off the dominant forward peak (i.e. $W_{T}$ bosons which mostly contribute to the cross-section), the region where the influence of the radiative corrections is more important is left.  The large corrections to the total cross-section arise at high energies mainly from the non-photonic box diagrams. The corrections $\delta_{weak}$ to the differential cross-sections are of order $\approx 10\%$ and diverge when the Born cross-section is suppressed or tends to zero.
\par
These corrections are not included in the Monte Carlo simulation of $\gamma e^{-}\rightarrow W^{-}\nu_{e}$. For energies of interest in this study, it can be assumed that the $\delta_{weak}$ contribution is small, on the level of percent.
\section{$W$ boson pair production at a Photon Collider}
\begin{figure}[htb]
\begin{center}
\epsfxsize=3.5in
\epsfysize=3.5in
\epsfbox{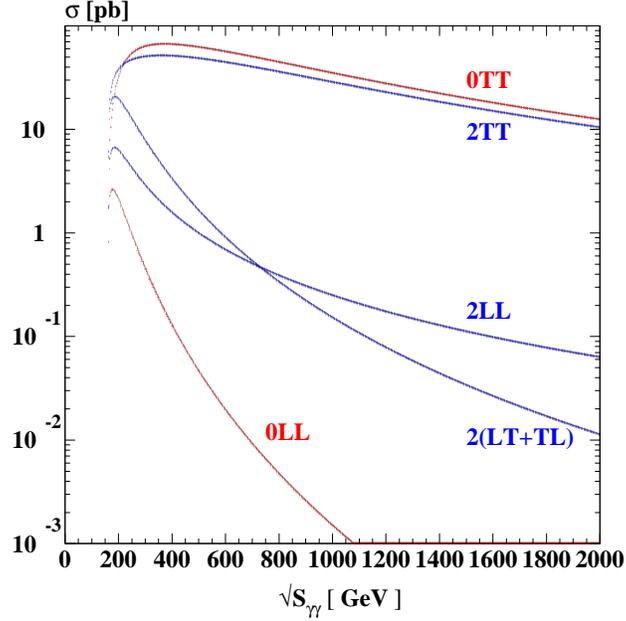}
\caption[bla]{Polarized Standard Model cross-section distributions for the initial state $J_{Z}=0$ at $\sqrt{s_{\gamma\gamma}}=400$ GeV integrated over the $W$ boson production angle $\theta$, taking $10^{\circ}<\theta<170^{\circ}$. The initial photon state $J_{Z}=0$ is denoted with ($0$) in front of the helicity state labeling ($LL,TT,(LT+TL)$) while the state $|J_{Z}|=2$ is denoted with ($2$). TT=($\pm\pm$) for $J_{Z}=0$ and TT=($\pm\pm$)+($\pm\mp$) for $|J_{Z}|=2$. LT+TL=($\pm 0$)+($0\pm$) and LL=($00$).}
\label{fig:scmgg}
\end{center}
\end{figure}

The Fig.~\ref{fig:scmgg} shows the contribution from each $W$ boson helicity state to the lowest order production cross-section in $\gamma\gamma$ collisions at $\sqrt{s_{\gamma\gamma}}=400$ GeV ($\approx 80\%\sqrt{s_{e^{-}e^{-}}}$), assuming that the photon beams are 100$\%$ polarized. Only $W$ bosons produced within the angle $\theta$ between 10$^{\circ}$ and 170$^{\circ}$ are taken into account. 
The contribution to the initial state $J_{Z}=0$ comes from longitudinally polarized $W$ bosons $(h_{W},h_{W})=(00)$ and transversally polarized $W$ bosons with the same helicities ($\pm\pm$). The $W$ boson helicity combinations ($\pm\mp$),($\pm 0$) and ($0\pm$) are not produced in the Standard Model while the presence of the New Physics associated with the symmetry breaking gives them a rise. At high energies, without the cut on the production angle, the cross-section for the production of transversally polarized $W$ bosons ($W_{T}W_{T}$) with the same helicities takes a constant value while the cross-section for the production of longitudinally polarized $W$ bosons ($W_{L}W_{L}$) decreases like $1/s^{2}$. With a finite cut on the $W$ boson production angle, the cross-section for the production of $W_{T}W_{T}$ bosons decreases as $1/s$ for large $s$ while the cross-section for the production of $W_{L}W_{L}$ bosons decreases as $1/s^{3}$. The initial state $|J_{Z}|=2$ receives a contribution from all possible $W$ boson helicity combinations: $W_{T}W_{T}$ bosons, including helicity combinations ($\pm\pm$) and ($\pm\mp$), $W_{L}W_{L}$ bosons and from $W$ bosons with mixed helicities $W_{L,T}W_{T,L}$, ($0\pm$) and ($\pm 0$). At asymptotic energies, without the cut on the production angle, the cross-section for the production of $W_{T}W_{T}$ bosons takes a constant value (as in the $J_{Z}=0$ state), while with a finite cut on the $W$ boson production angle, the cross-section decreases as $1/s$. The cross-section for the production of $W_{L}W_{L}$ bosons decreases as $1/s$ while the production of $W_{L,T}W_{T,L}$ bosons behaves as $1/s^{2}$, both independently on the angular cut. Generally, the cross-section behavior related to the angular cuts is similar to the $\gamma e$ case. Numerical values for the lowest order cross-sections at $\sqrt{s_{\gamma\gamma}}=400$ and 800 GeV  are calculated on the basis of the formulae given in \cite{denner_gg} and can be found in Table \ref{tab:cross_gg}. The calculated cross-sections at $\sqrt{s_{\gamma\gamma}}=0.5,1$ and 2 TeV can be found in \cite{denner_gg}.
\begin{table}[h]
\begin{center}
\begin{tabular}{|c||c|c|c|c|c|c|} \hline
 $\sqrt{s}$ & $\theta$ & 0TT & 0LL & 2TT & 2LL & 2(LT+TL) \\ \hline\hline
 400 & $\begin{array}{c}
        0^{\circ}-180^{\circ} \\
        10^{\circ}-170^{\circ}\\
        20^{\circ}-160^{\circ}\\
        \end{array}$ 
        & $\begin{array}{c}
        79.5 \\
        69.4 \\
        50.1 \\
        \end{array}$
        & $\begin{array}{c}
        1.5\cdot 10^{-1} \\
        1.3\cdot 10^{-1} \\
        0.98\cdot 10^{-1} \\
         \end{array}$
        & $\begin{array}{c}
        66.3 \\
        54.4 \\
        33.8 \\
        \end{array}$
        & $\begin{array}{c}
        1.6 \\
        1.6 \\
        1.6 \\
        \end{array}$
        & $\begin{array}{c}
        3.4 \\
        3.4 \\
        3.1 \\
        \end{array}$ \\ \hline\hline
800 & $\begin{array}{c}
        0^{\circ}-180^{\circ} \\
        10^{\circ}-170^{\circ}\\
        20^{\circ}-160^{\circ}\\
        \end{array}$ 
        & $\begin{array}{c}
        83.2 \\
        49.9 \\
        23.9
        \end{array}$
        & $\begin{array}{c}
        8.8\cdot 10^{-3} \\
        5.3\cdot 10^{-3} \\
        2.5\cdot 10^{-3} \\
         \end{array}$
        & $\begin{array}{c}
        75.3 \\
        41.4 \\
        15.8 \\
        \end{array}$
        & $\begin{array}{c}
        3.9\cdot 10^{-1} \\
        3.9\cdot 10^{-1} \\
        3.8\cdot 10^{-1} \\
        \end{array}$
        & $\begin{array}{c}
        3.6\cdot 10^{-1} \\
        3.5\cdot 10^{-1} \\
        2.8\cdot 10^{-1}        
        \end{array}$ \\ \hline
\end{tabular}
\end{center}
\caption{Lowest order integrated cross-sections in [pb] for the initial states $J_{Z}=0$ and $|J_{Z}|=2$ and for different $W$ boson helicity combinations. The cross-sections are calculated without and with a cut on the $W$ boson production angle. 0 and 2 denote the initial $J_{Z}$ state while TT=($\pm\pm$) for $J_{Z}=0$ and TT=($\pm\pm$)+($\pm\mp$) for $|J_{Z}|=2$. LT+TL=($\pm 0$)+($0\pm$) and LL=($00$).}
\label{tab:cross_gg}
\end{table}
\begin{figure}[p]
\begin{center}
\epsfxsize=2.5in
\epsfysize=2.5in
\epsfbox{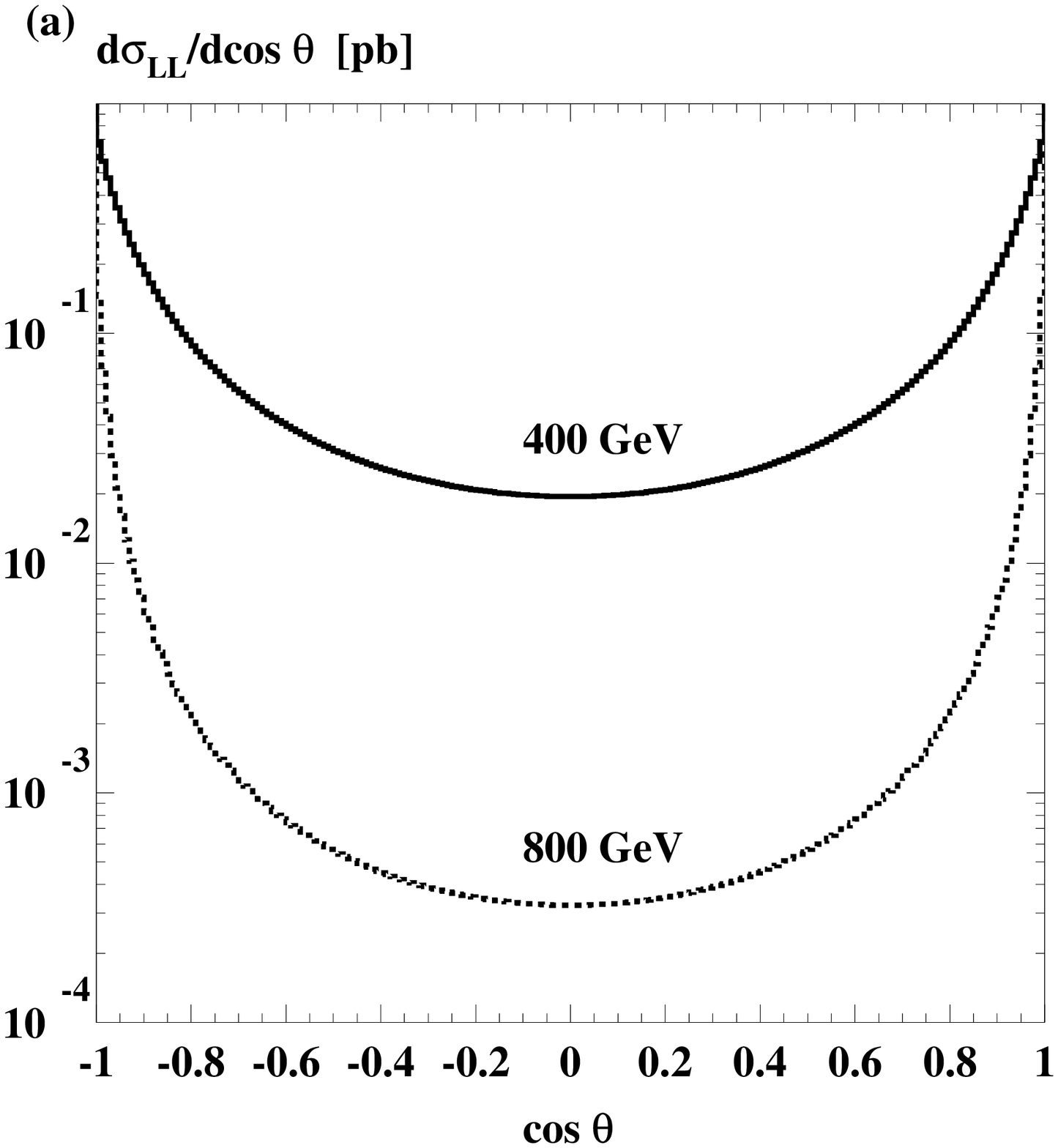}
\epsfxsize=2.5in
\epsfysize=2.5in
\epsfbox{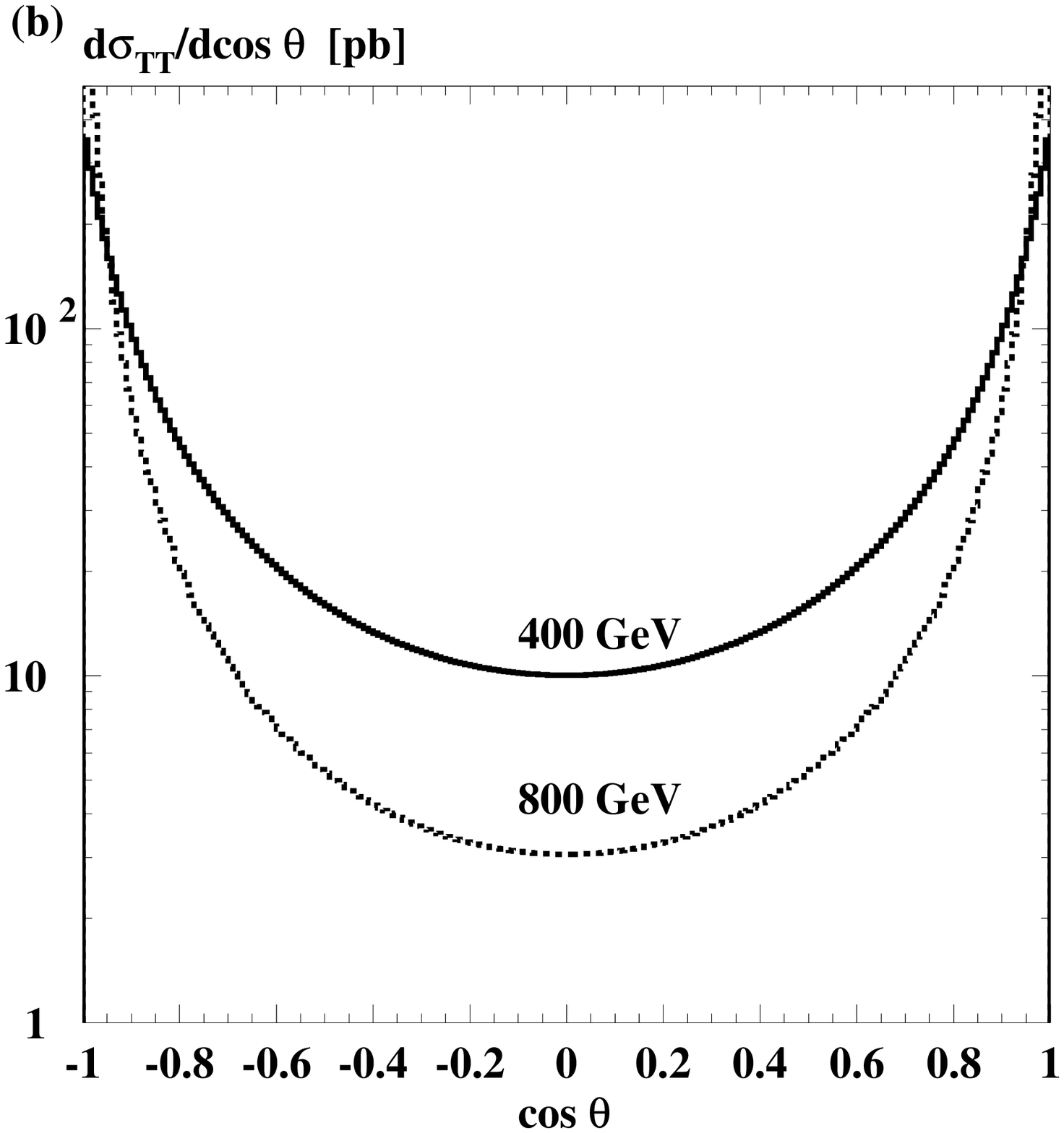}
\epsfxsize=2.5in
\epsfysize=2.5in
\epsfbox{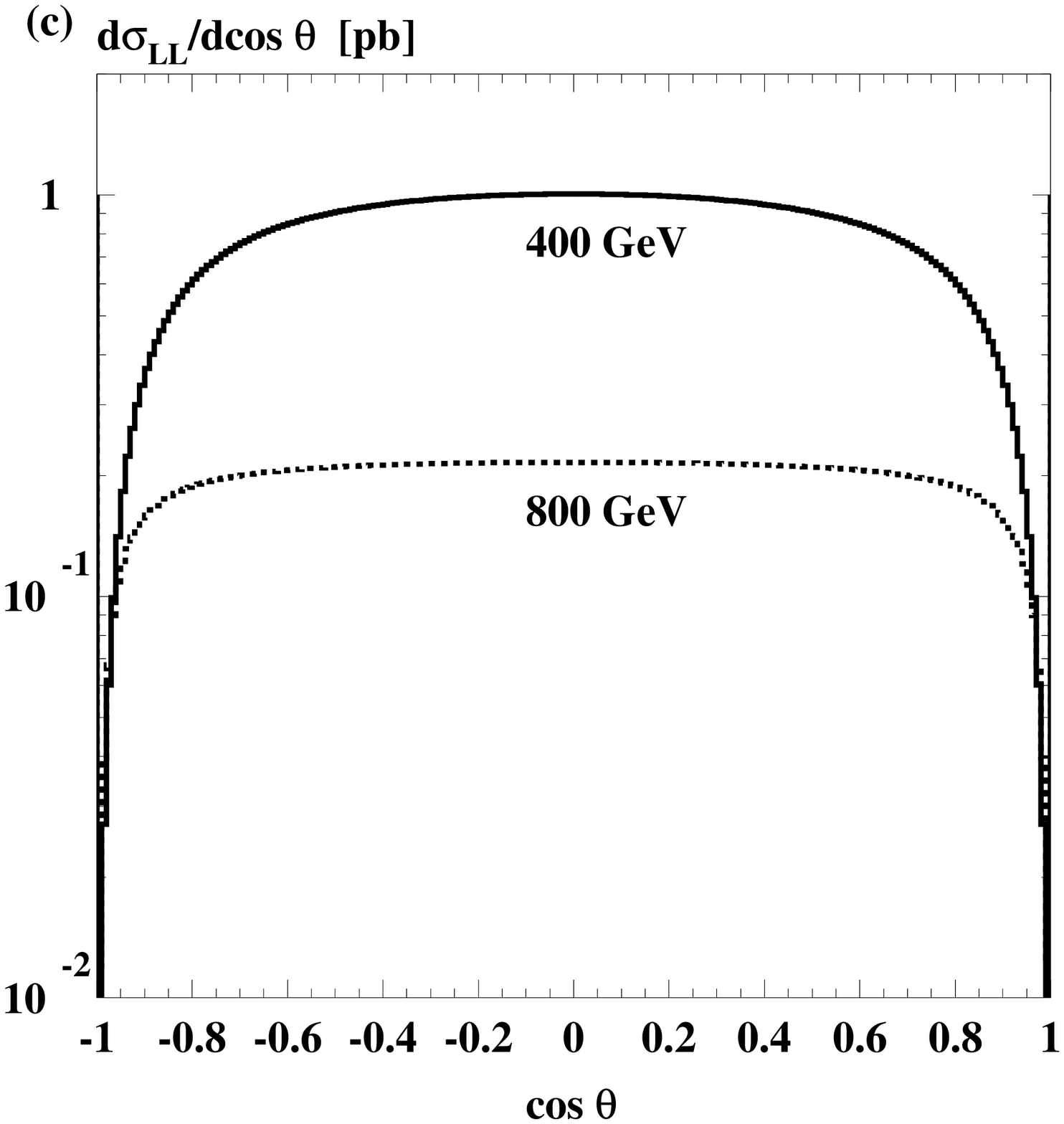}
\epsfxsize=2.5in
\epsfysize=2.5in
\epsfbox{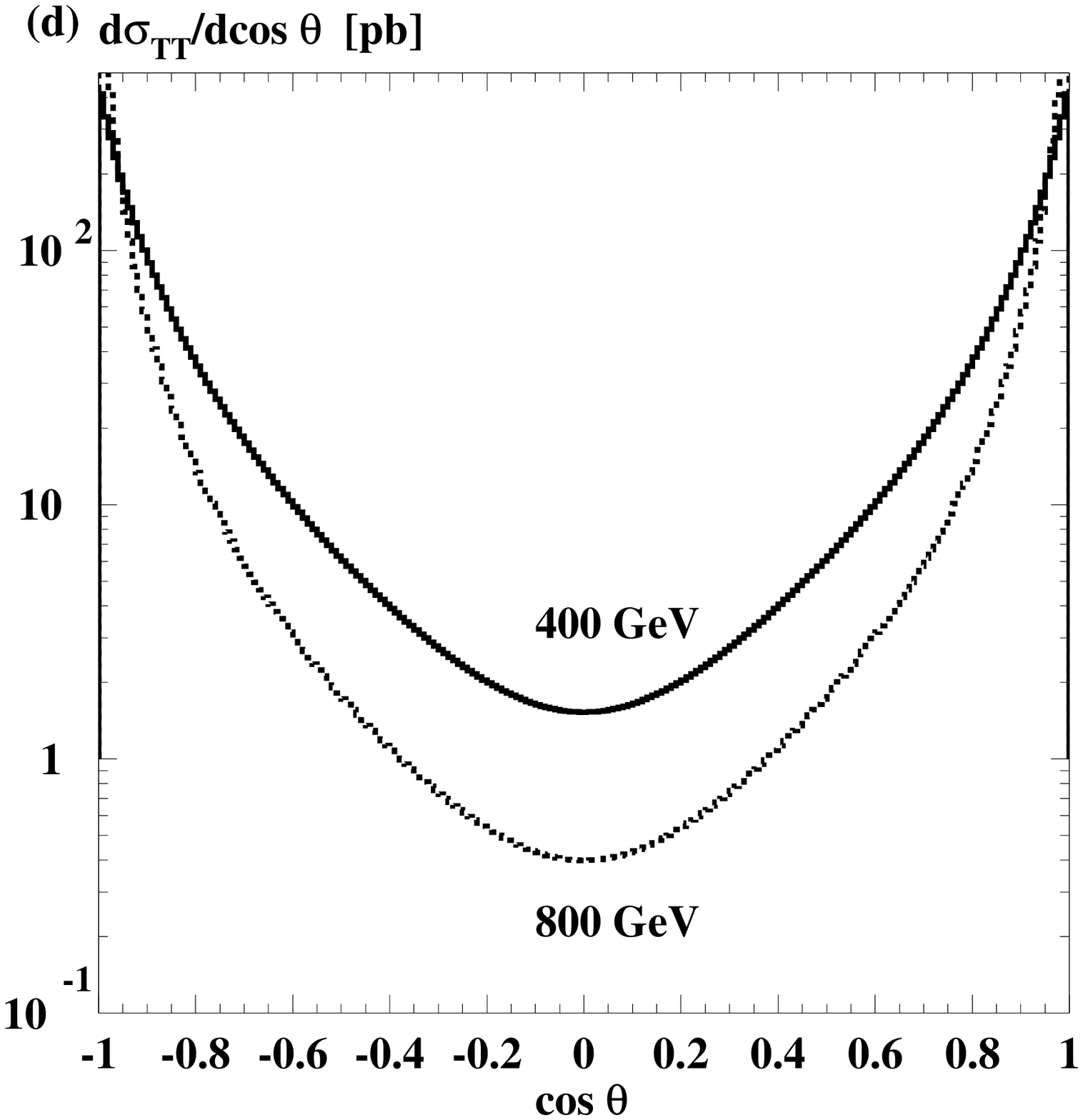}
\epsfxsize=2.5in
\epsfysize=2.5in
\epsfbox{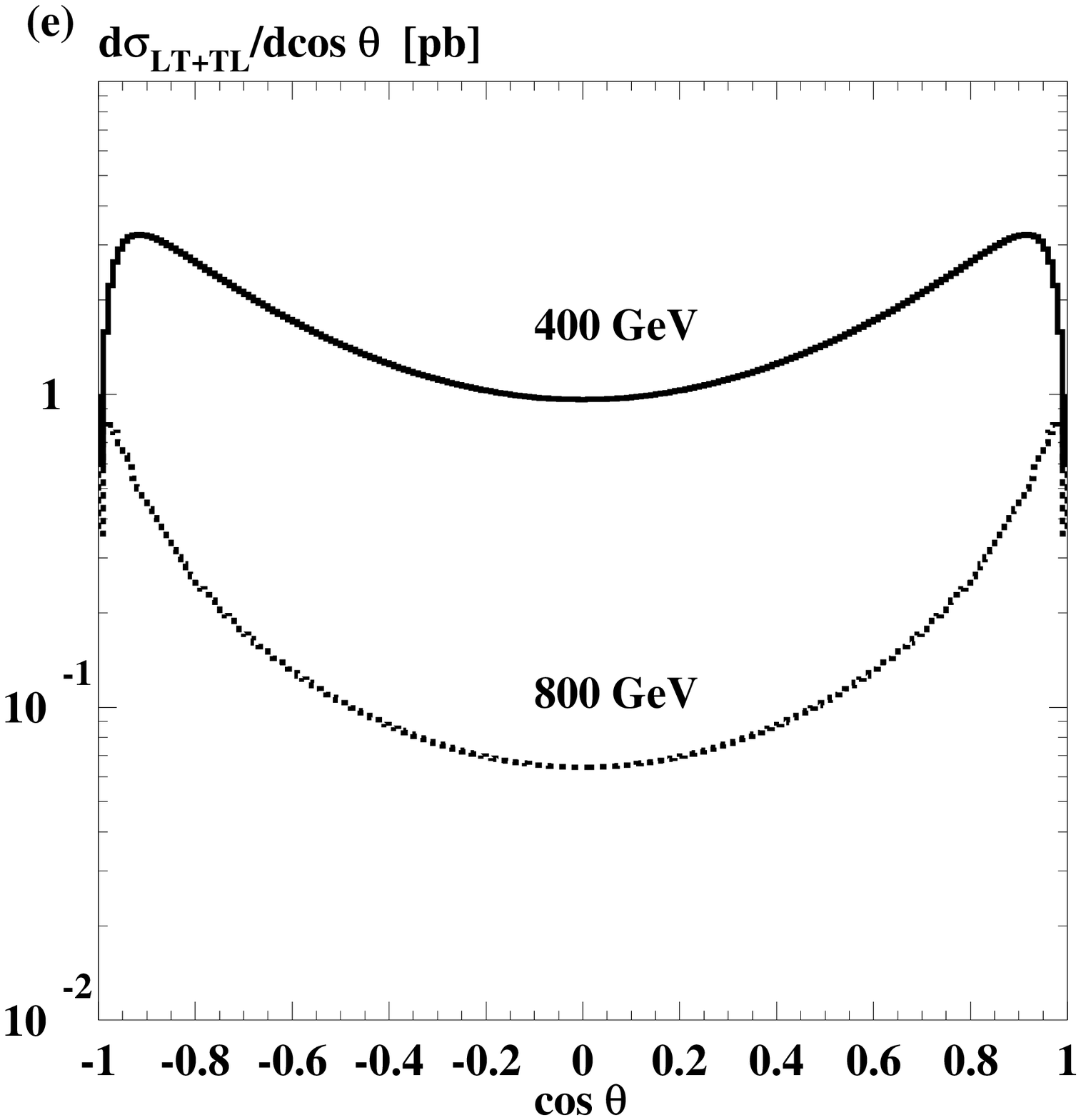}
\caption[bla]{The Standard Model differential cross-section distributions for the initial states (\textit{a-b}): $J_{Z}=0$ and (\textit{c-e}): $|J_{Z}|=2$ at $\sqrt{s_{\gamma\gamma}}=400$ and 800 GeV.}
\label{fig:diffgg}
\end{center}
\end{figure}
\par
The differential cross-section distributions for $J_{Z}=0$ and $|J_{Z}|=2$ at $\sqrt{s_{\gamma\gamma}}=400$ and 800 GeV are shown in Fig.~\ref{fig:diffgg}. Due to the dominating \textit{t}-channel exchange the production of $W_{T}W_{T}$ bosons is favorised in the forward-backward region, close to the beam axis. Increasing the center-of-mass energy for both $J_{Z}$ states, the production of $W_{T}W_{T}$ bosons becomes more pronounced in the forward-backward region than in the central region. Since the $W_{T}W_{T}$ fraction contributes in the largest amount to the total the cross-section, the cut on the production angle reduces their contribution and total cross-section decreases as $1/s$. Increasing $s$, the differential cross-section for the production of $W_{L}W_{L}$ bosons in the $J_{Z}=0$ state decreases faster in the central region than in the forward-backward region. For the $|J_{Z}|=2$ state the production of $W_{L}W_{L}$ bosons vanishes in the forward-backward region having a maximum at $\theta=90^{\circ}$. The production of $W_{L,T}W_{T,L}$ bosons also vanishes in the forward-backward region having a maximum at $|\cos\theta|=\beta$ ($\beta=\sqrt{1-4M_{W}^{2}/s}$) and a minimum at $\theta=90^{\circ}$. Increasing $s$, the central distribution decreases faster than in the forward-backward region. Differential and total cross-sections are calculated on the basis of the formula given in \cite{denner_gg}.
\par
\begin{figure}[p]
\begin{center}
\epsfxsize=3.0in
\epsfysize=3.0in
\epsfbox{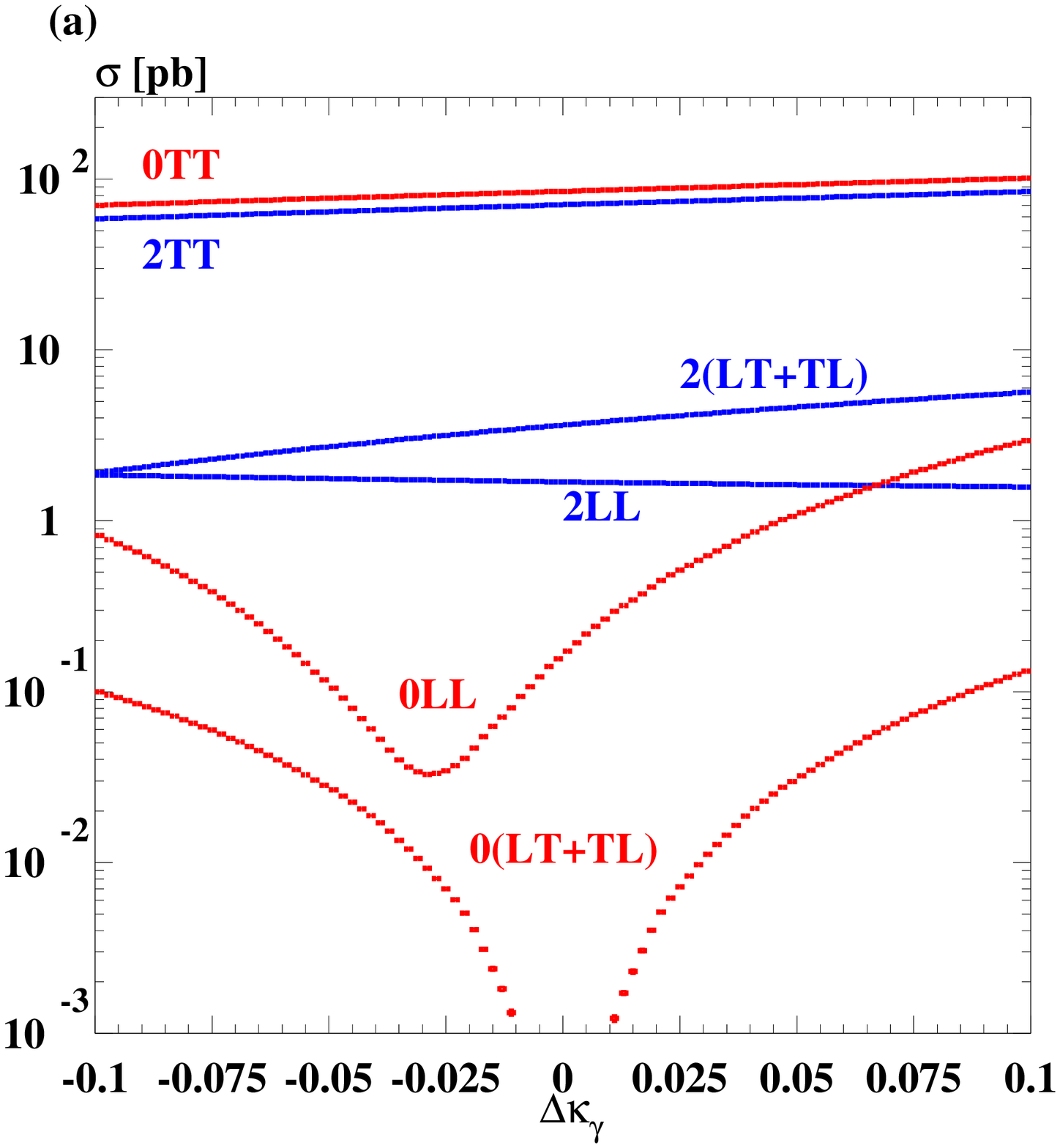}
\epsfxsize=3.0in
\epsfysize=3.0in
\epsfbox{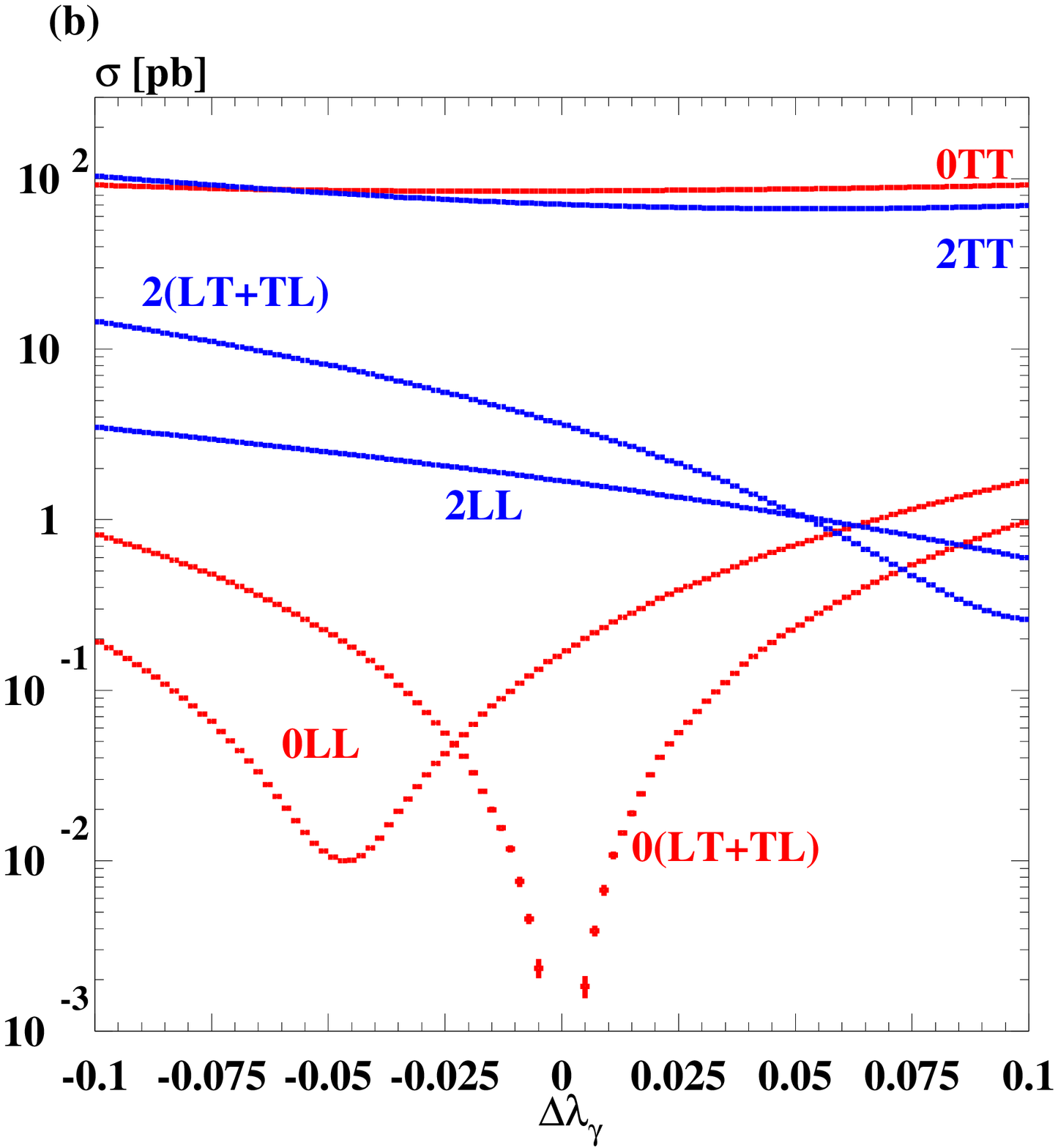}
\epsfxsize=3.0in
\epsfysize=3.0in
\epsfbox{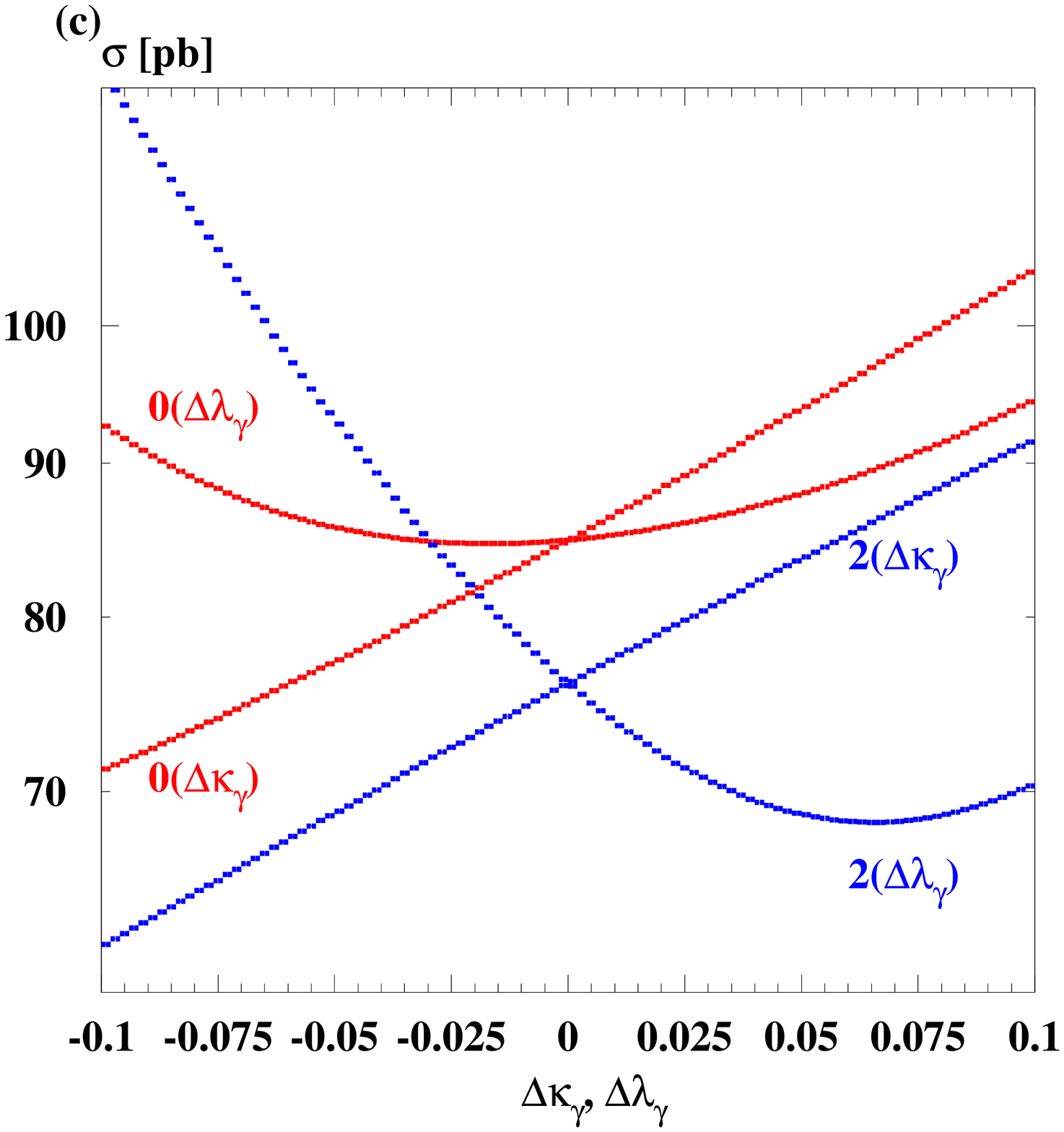}
\caption[bla]{Contribution of different $WW$ helicity states for $J_{Z}=0$ and $|J_{Z}|=2$ states in the presence of anomalous couplings (\textit{a}): ${\kappa}_{\gamma}$ and (\textit{b}): ${\lambda}_{\gamma}$ at $\sqrt{s_{\gamma\gamma}}=400$ GeV, assuming 100$\%$ photon polarizations. (\textit{c}): Total cross-section dependence on the anomalous $\kappa_{\gamma}$ and $\lambda_{\gamma}$. The deviations are denoted as $\Delta\kappa_{\gamma}$ and $\Delta\lambda_{\gamma}$. The initial photon state $J_{Z}=0$ is denoted with 0 in front of helicity labeling ($LL,TT,(LT+TL)$) while the state $|J_{Z}|=2$ is denoted with 2. TT=($\pm\pm$) for $J_{Z}=0$ and TT=($\pm\pm$)+($\pm\mp$) for $|J_{Z}|=2$. LT+TL=($\pm 0$)+($0\pm$) and LL=($00$).}
\label{fig:ac_kl}
\end{center}
\end{figure}
Introducing the anomalous TGCs $\kappa_{\gamma}$ and $\lambda_{\gamma}$, the total production cross-sections for different $W$ boson helicity combinations deviate from their Standard Model values. Depending on the initial $J_{Z}$ state and the $W$ boson helicities, the sensitivity to the anomalous values of $\kappa_{\gamma}$ and $\lambda_{\gamma}$ is different. Fig.~\ref{fig:ac_kl}\,$a,b$ shows that the anomalous values of $\kappa_{\gamma}$ and $\lambda_{\gamma}$ mostly affect the production of $W_{L}W_{L}$ bosons in the $J_{Z}=0$ state while the production of $W_{L}W_{L}$ bosons in the $|J_{Z}|=2$ is less sensitive to the anomalous TGCs with respect to the SM cross-section. The production of $W_{L,T}W_{T,L}$ bosons in the $J_{Z}=0$ state at the Standard Model level is zero while the anomalous $\lambda_{\gamma}$ contributes to the production cross-section more than the anomalous $\kappa_{\gamma}$. The production of the dominating $W_{T}W_{T}$ bosons is less sensitive to the anomalous TGCs. An anomalous $\kappa_{\gamma}$ influences their production in a similar way for both $J_{Z}$ states while the anomalous $\lambda_{\gamma}$ influences more their production in the $|J_{Z}|=2$ state. The total cross-section dependence on the anomalous $\kappa_{\gamma}$ and $\lambda_{\gamma}$ is shown in Fig.~\ref{fig:ac_kl}\,$c$.
\par
\begin{figure}[p]
\begin{center}
\epsfxsize=2.5in
\epsfysize=2.5in
\epsfbox{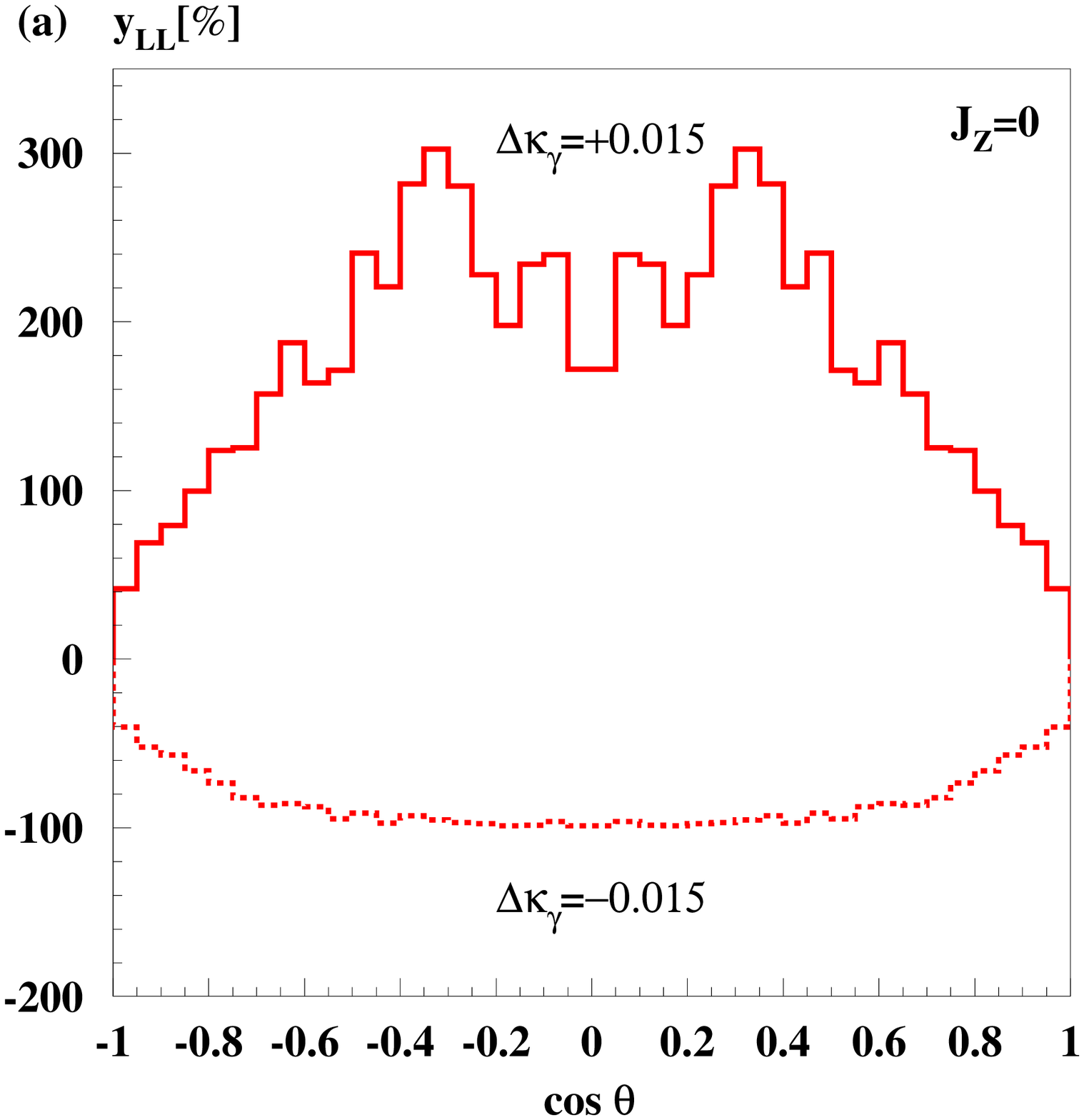}
\epsfxsize=2.5in
\epsfysize=2.5in
\epsfbox{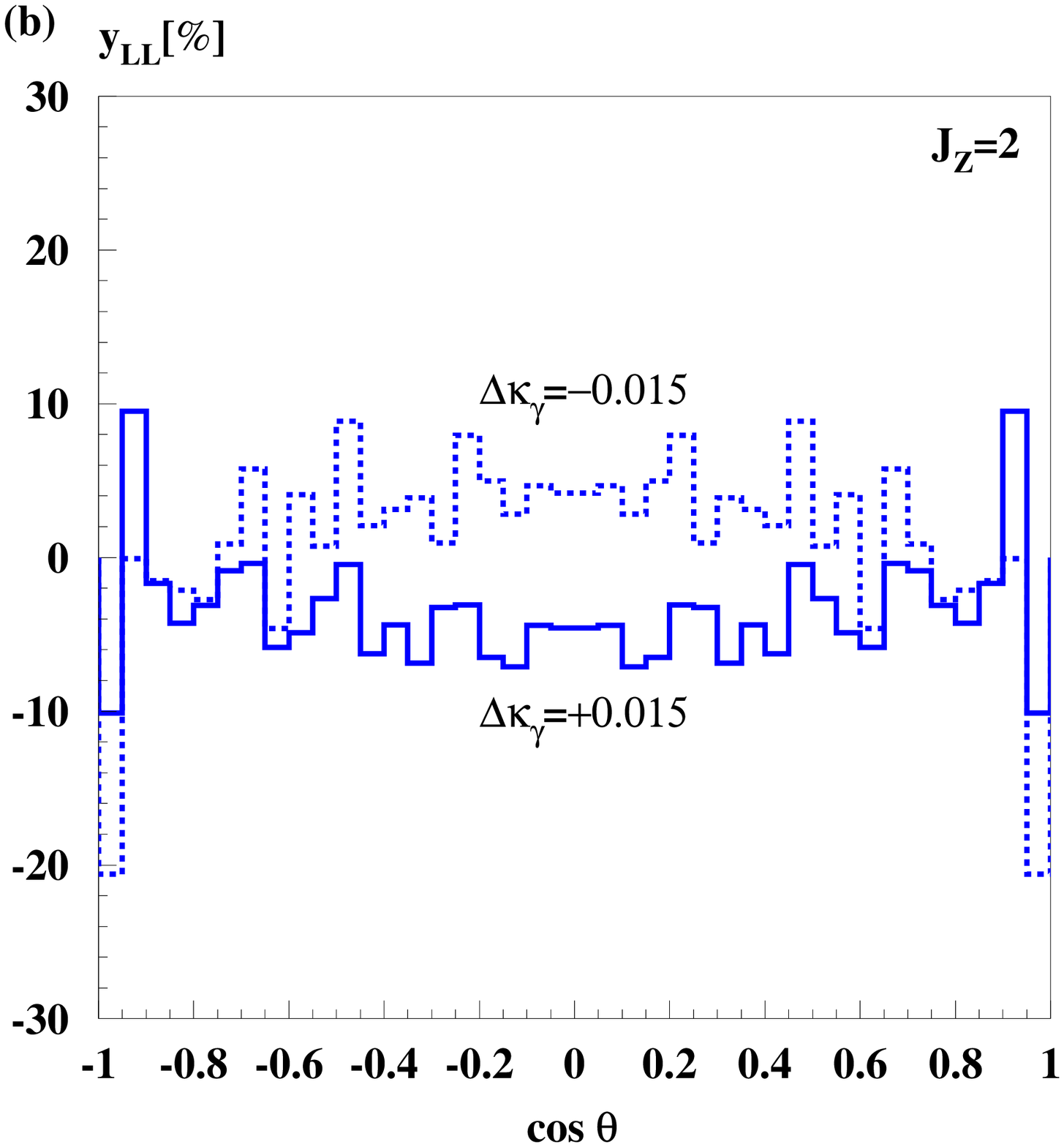}
\epsfxsize=2.5in
\epsfysize=2.5in
\epsfbox{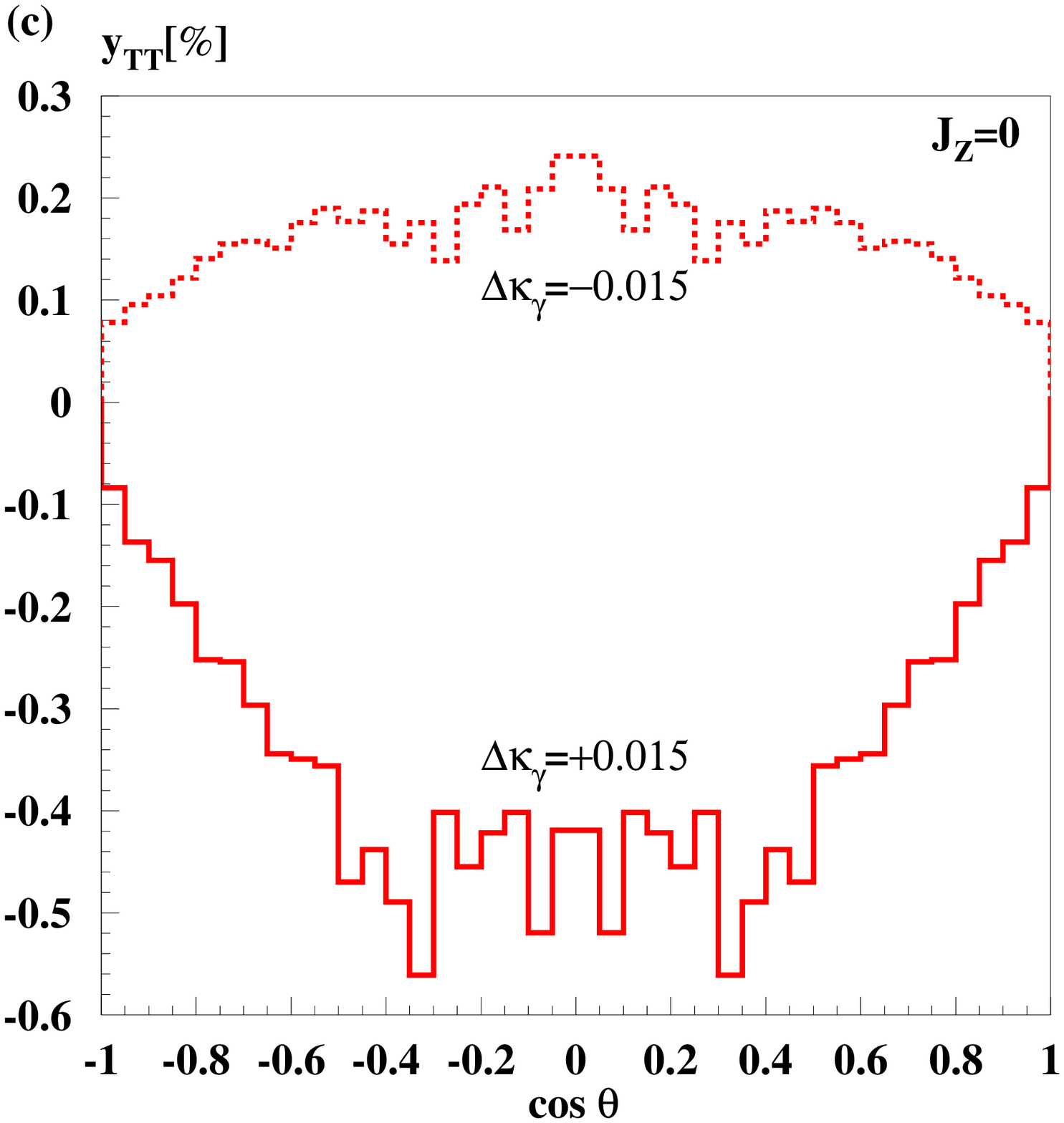}
\epsfxsize=2.5in
\epsfysize=2.5in
\epsfbox{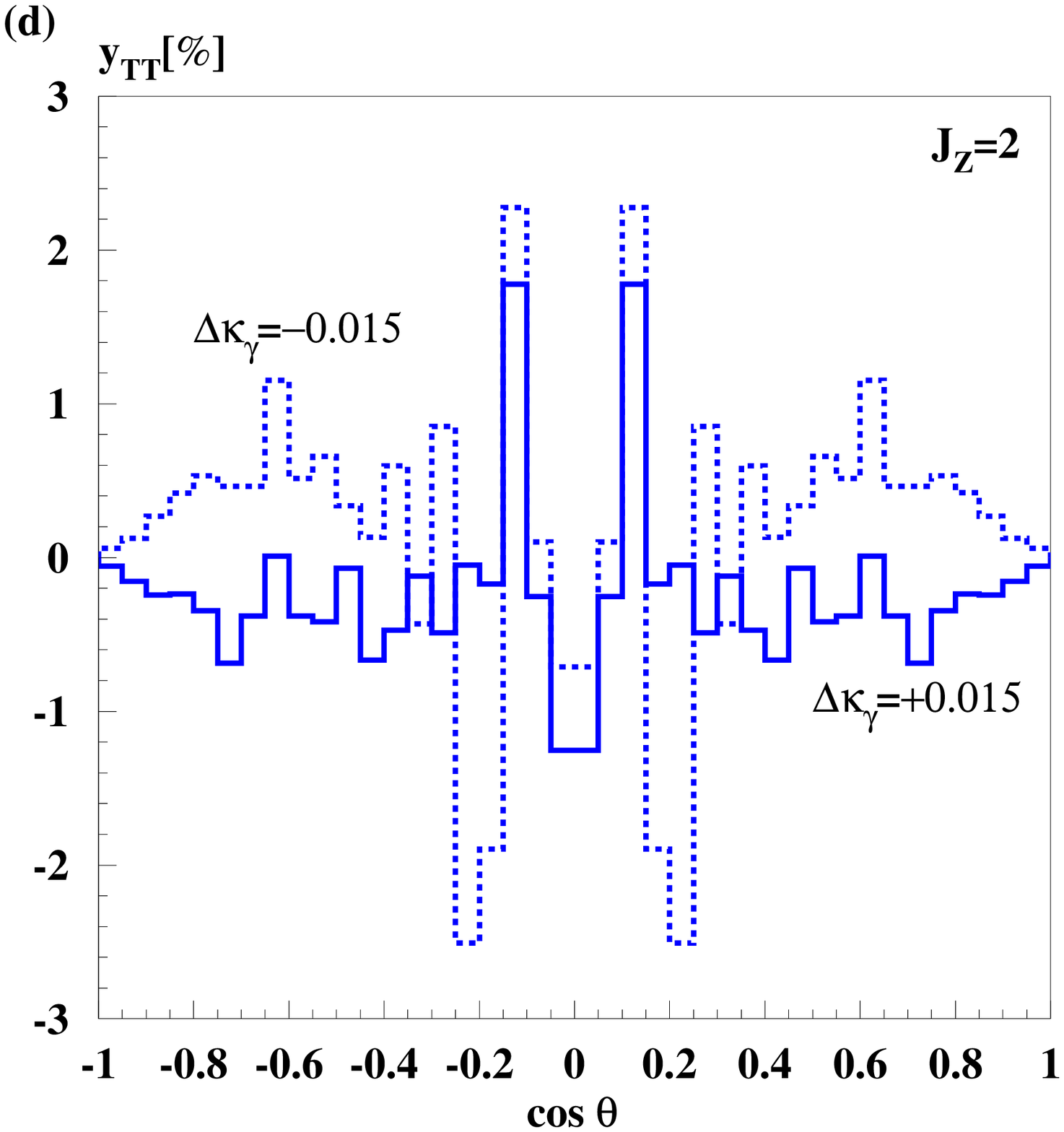}
\epsfxsize=2.5in
\epsfysize=2.5in
\epsfbox{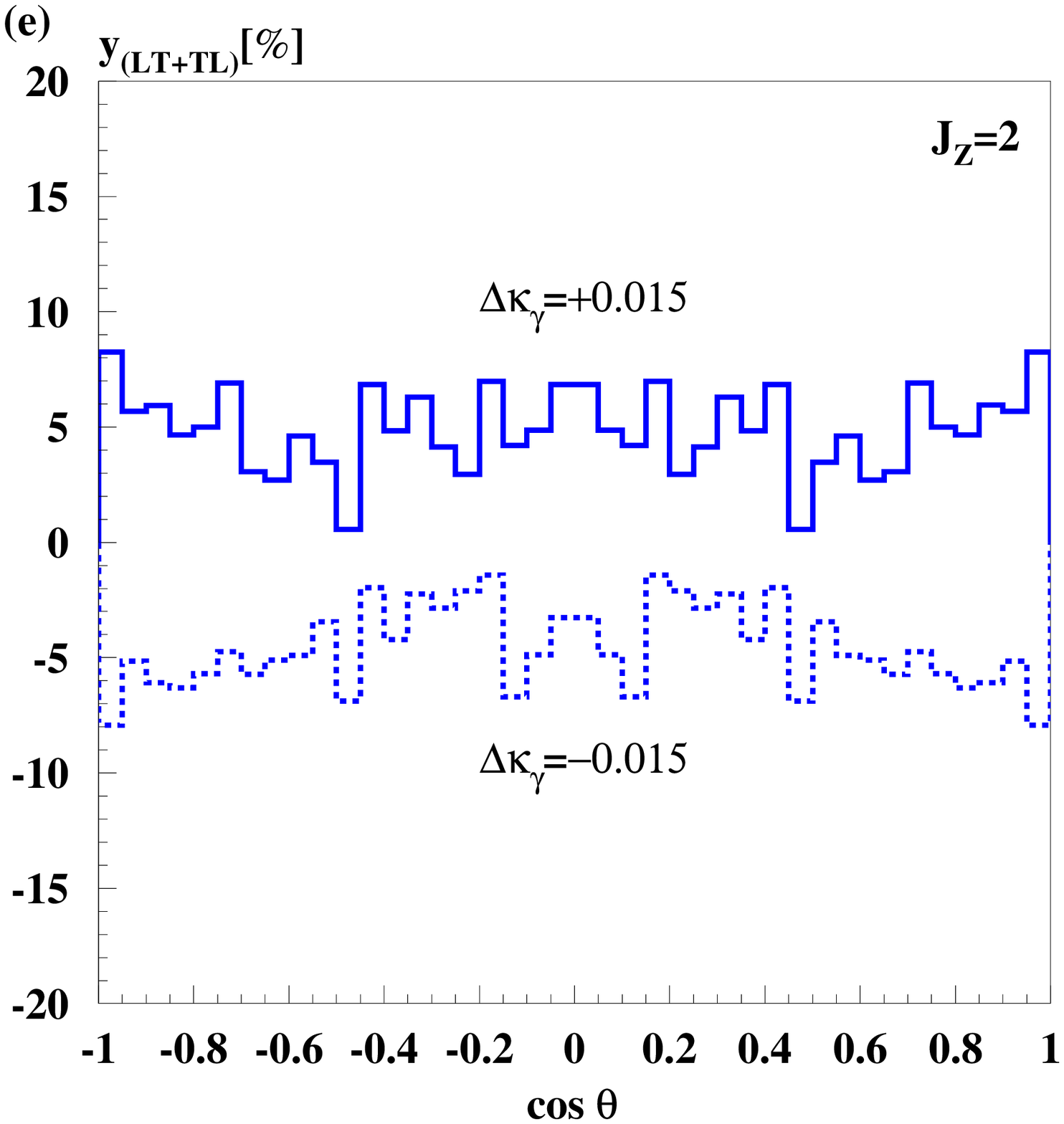}
\caption[bla]{Relative deviations of different $WW$ fractions from the Standard Model predictions (\textit{a,c}): in the $J_{Z}=0$ state and (\textit{b,d,e}): in the $|J_{Z}|=2$ state in the presence of anomalous coupling $\kappa_{\gamma}$ (${\Delta}{\kappa}_{\gamma}={\pm}0.015$) at $\sqrt{s_{\gamma\gamma}}=400$ GeV assuming 100$\%$ photon beam polarizations. Solid lines correspond to ${\Delta}{\kappa}_{\gamma}=+0.015$ and dotted lines correspond to ${\Delta}{\kappa}_{\gamma}=-0.015$ with ${\Delta}{\lambda}_{\gamma}=0$. $y_{(LL,TT,LT+TL)}=\frac{F^{AC}_{LL,TT,(LT+TL)}-F^{SM}_{LL,TT,(LT+TL)}}{F^{SM}_{LL,TT,(LT+TL)}}$, $F^{AC,SM}_{LL,TT,(LT+TL)}=\left[\frac{d\sigma_{LL,TT,(LT+TL)}}{d\sigma_{TOT}}\right]^{AC,SM}$.}
\label{fig:diff_kap}
\end{center}
\end{figure}
\begin{figure}[p]
\begin{center}
\epsfxsize=2.5in
\epsfysize=2.5in
\epsfbox{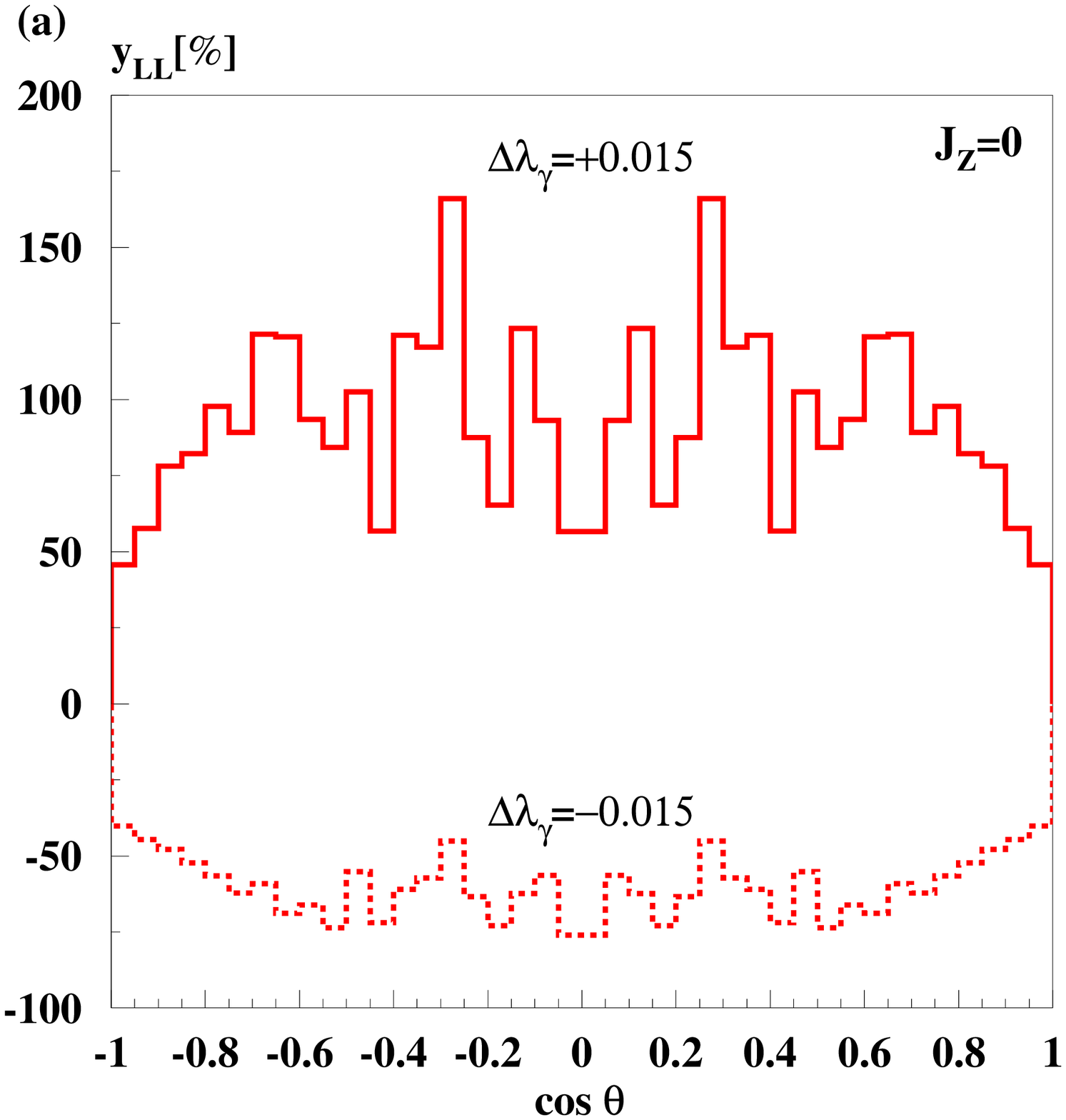}
\epsfxsize=2.5in
\epsfysize=2.5in
\epsfbox{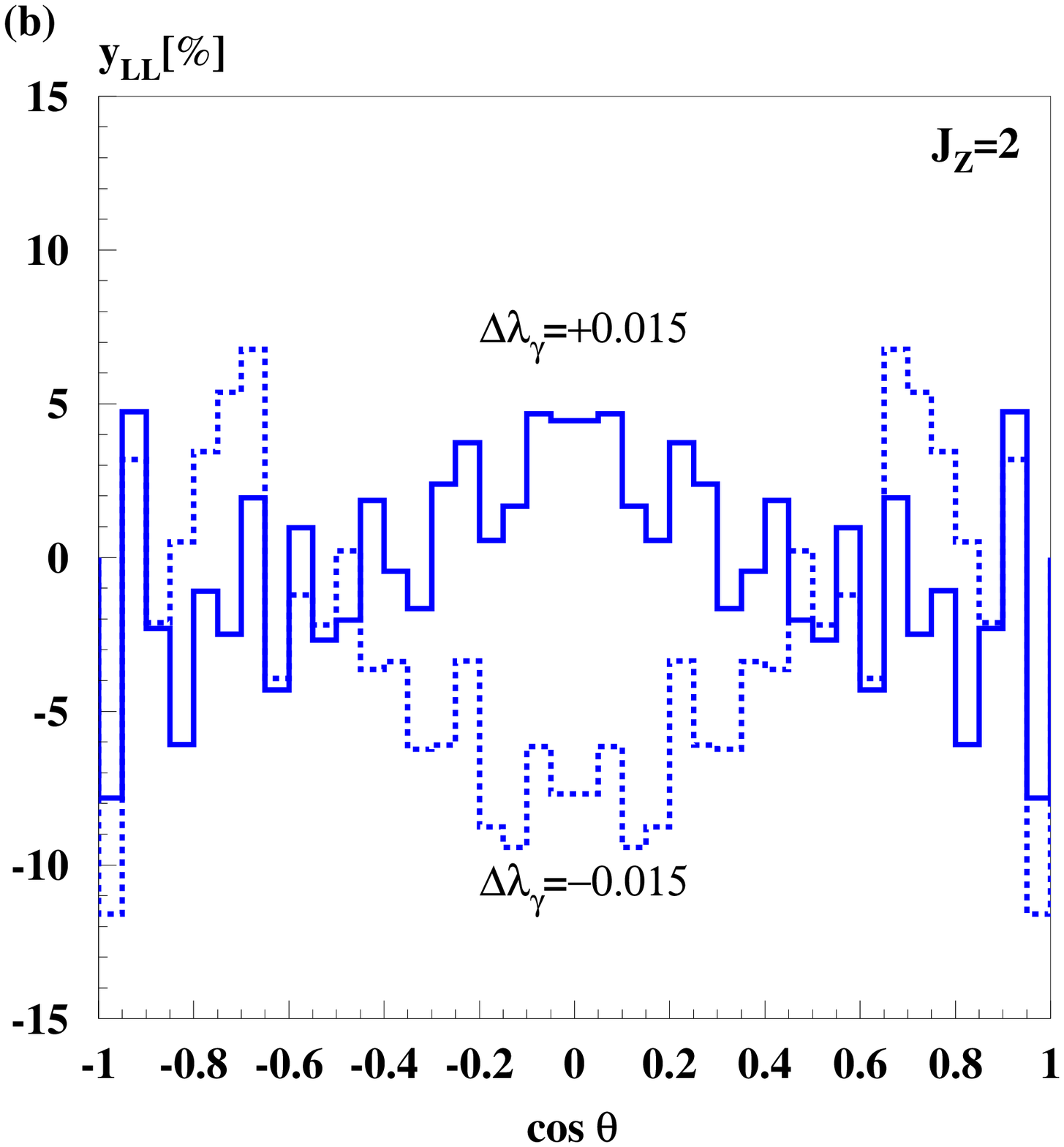}
\epsfxsize=2.5in
\epsfysize=2.5in
\epsfbox{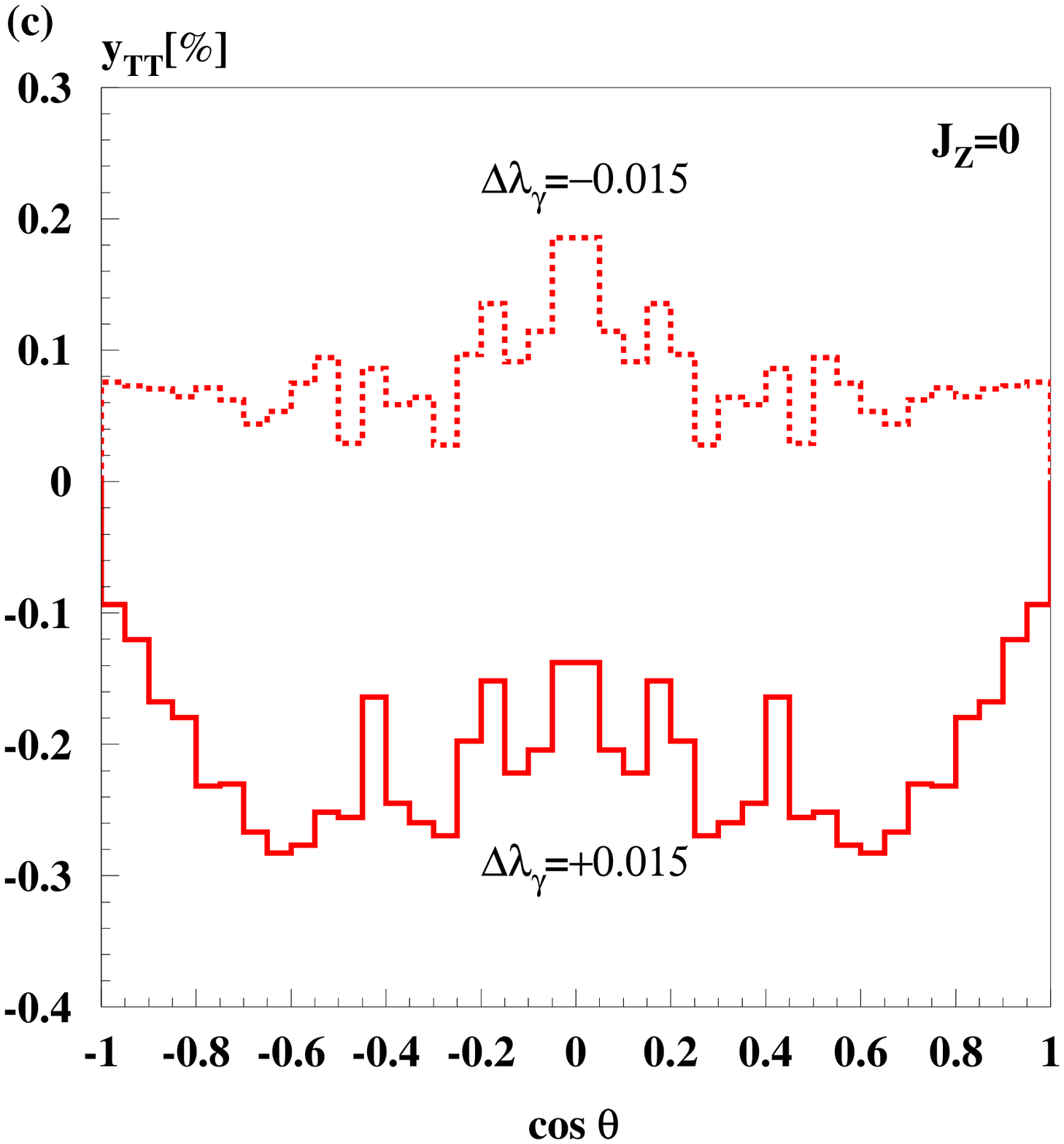}
\epsfxsize=2.5in
\epsfysize=2.5in
\epsfbox{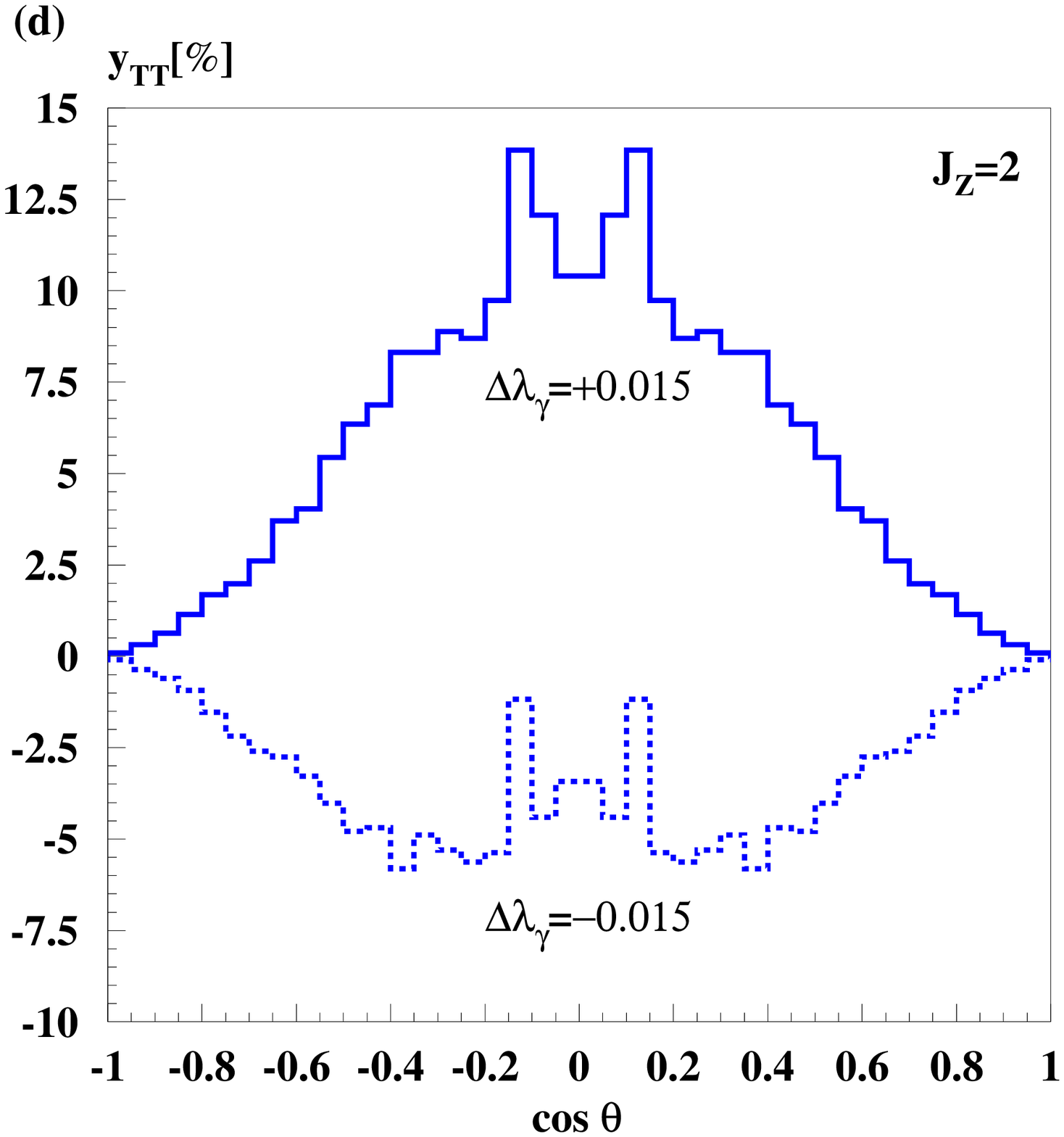}
\epsfxsize=2.5in
\epsfysize=2.5in
\epsfbox{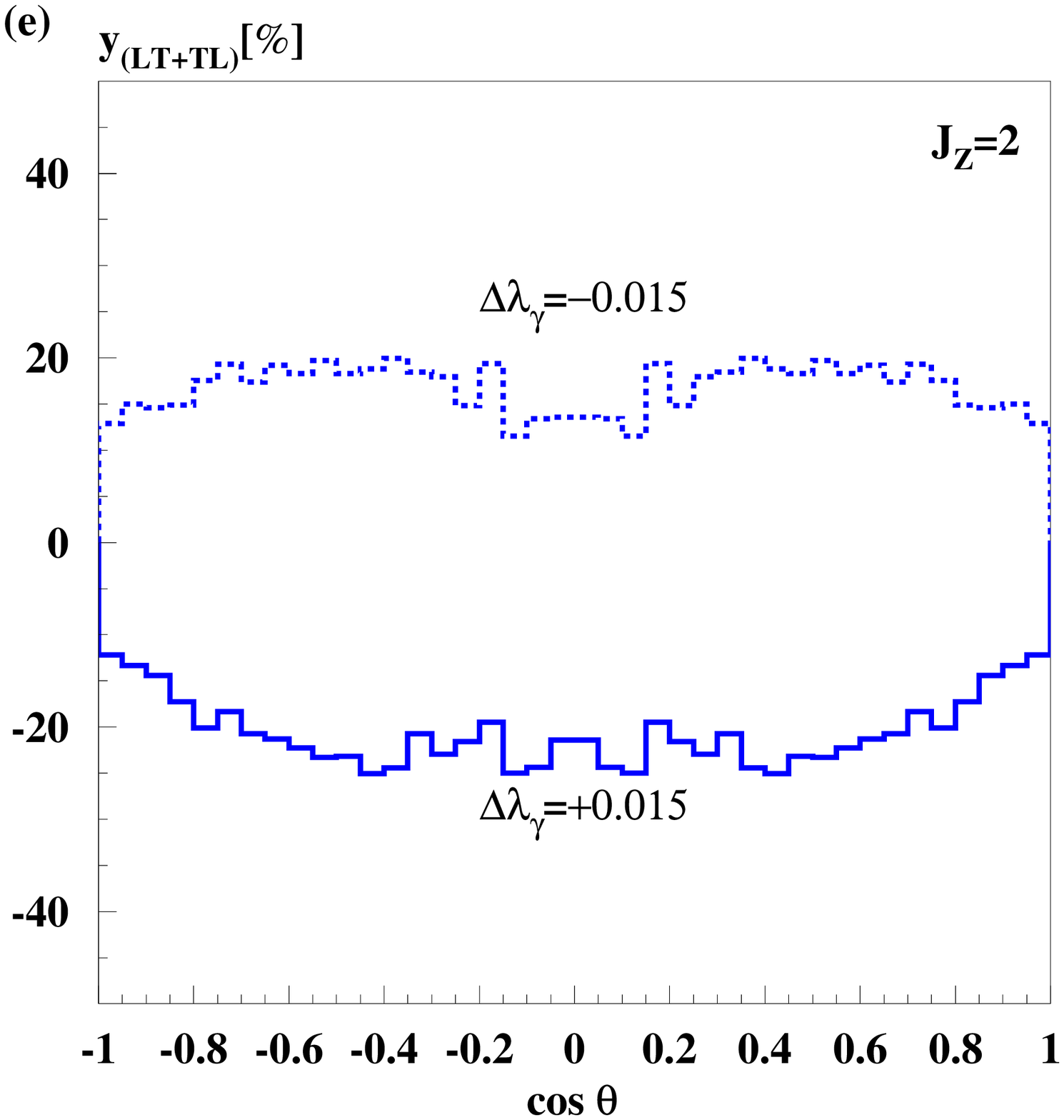}
\caption[bla]{Relative deviations of different $WW$ fractions from the Standard Model predictions (\textit{a,c}): in the $J_{Z}=0$ state and (\textit{b,d,e}): in the $|J_{Z}|=2$ state in presence of anomalous coupling $\lambda_{\gamma}$ (${\Delta}{\lambda}_{\gamma}={\pm}0.015$) at $\sqrt{s_{\gamma\gamma}}=400$ GeV assuming 100$\%$ photon beam polarizations. Solid lines correspond to ${\Delta}{\lambda}_{\gamma}=+0.015$ and dotted lines correspond to ${\Delta}{\lambda}_{\gamma}=-0.015$ with ${\Delta}{\kappa}_{\gamma}=0$. $y_{(LL,TT,LT+TL)}=\frac{F^{AC}_{LL,TT,(LT+TL)}-F^{SM}_{LL,TT,(LT+TL)}}{F^{SM}_{LL,TT,(LT+TL)}}$, $F^{AC,SM}_{LL,TT,(LT+TL)}=\left[\frac{d\sigma_{LL,TT,(LT+TL)}}{d\sigma_{TOT}}\right]^{AC,SM}$.}
\label{fig:diff_lam}
\end{center}
\end{figure}

The angular distributions for different $WW$ helicity combinations in the presence of anomalous TGCs are affected also and differ from those predicted by the Standard Model. The influence of $\Delta\kappa_{\gamma}$ and $\Delta\lambda_{\gamma}$ on the angular distributions of different $WW$ helicity combinations is shown in Fig.~\ref{fig:diff_kap} and Fig.~\ref{fig:diff_lam} respectively, as the relative deviation $y$ of angular distribution for certain $WW$ helicity fraction in the presence of anomalous TGCs from its Standard Model prediction. It is assumed that one of the two couplings, $\kappa_{\gamma}$ or $\lambda_{\gamma}$, deviates from its Standard Model value for $\Delta\kappa_{\gamma},\Delta\lambda_{\gamma}=\pm 0.015$ while the second coupling takes its Standard Model value. 
\par
In despite of the larger $W_{L}W_{L}$ contribution to the total cross-section in the $|J_{Z}|=2$ state than in the $J_{Z}=0$ state, the production of $W_{L}W_{L}$ bosons in the $J_{Z}=0$ is more influenced by $\Delta\kappa_{\gamma}$ than in the $|J_{Z}|=2$ state, increasing the $W_{L}W_{L}$ fraction up to 3 times related to the Standard Model, particularly in the central region. The production of $W_{T}W_{T}$ bosons for both $J_{Z}$ states in the presence of $\Delta\kappa_{\gamma}=\pm 0.015$ induces small deviations (below 1$\%$), i.e. the $W_{T}W_{T}$ fractions are not sensitive to the anomalous $\kappa_{\gamma}$. The production of $W_{L,T}W_{T,L}$ bosons in the $|J_{Z}|=2$ state is changed over the production angle up to 10$\%$ related to the Standard Model.
\par
\begin{figure}[htb]
\begin{center}
\epsfxsize=3.0in
\epsfysize=3.0in
\epsfbox{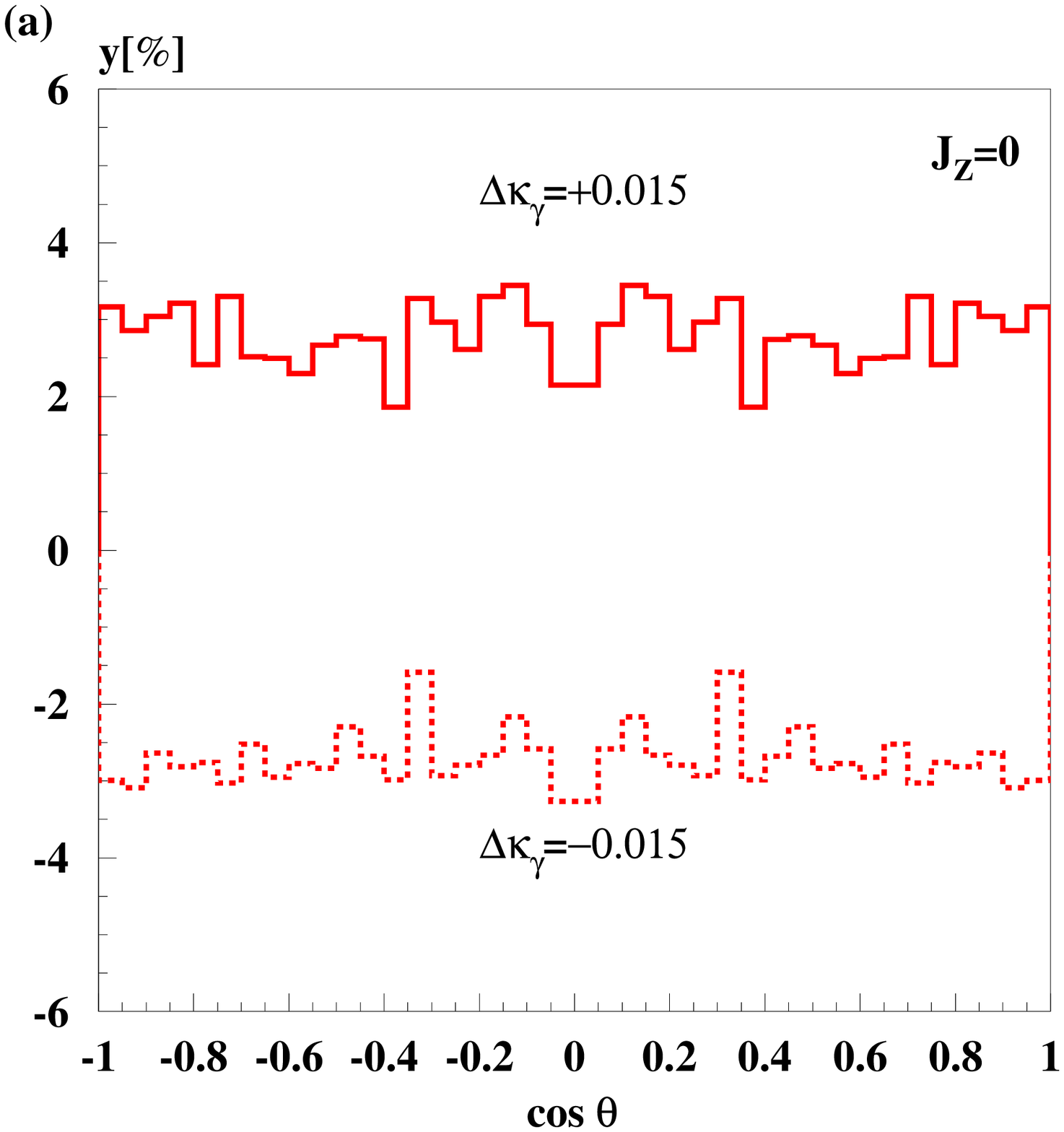}
\epsfxsize=3.0in
\epsfysize=3.0in
\epsfbox{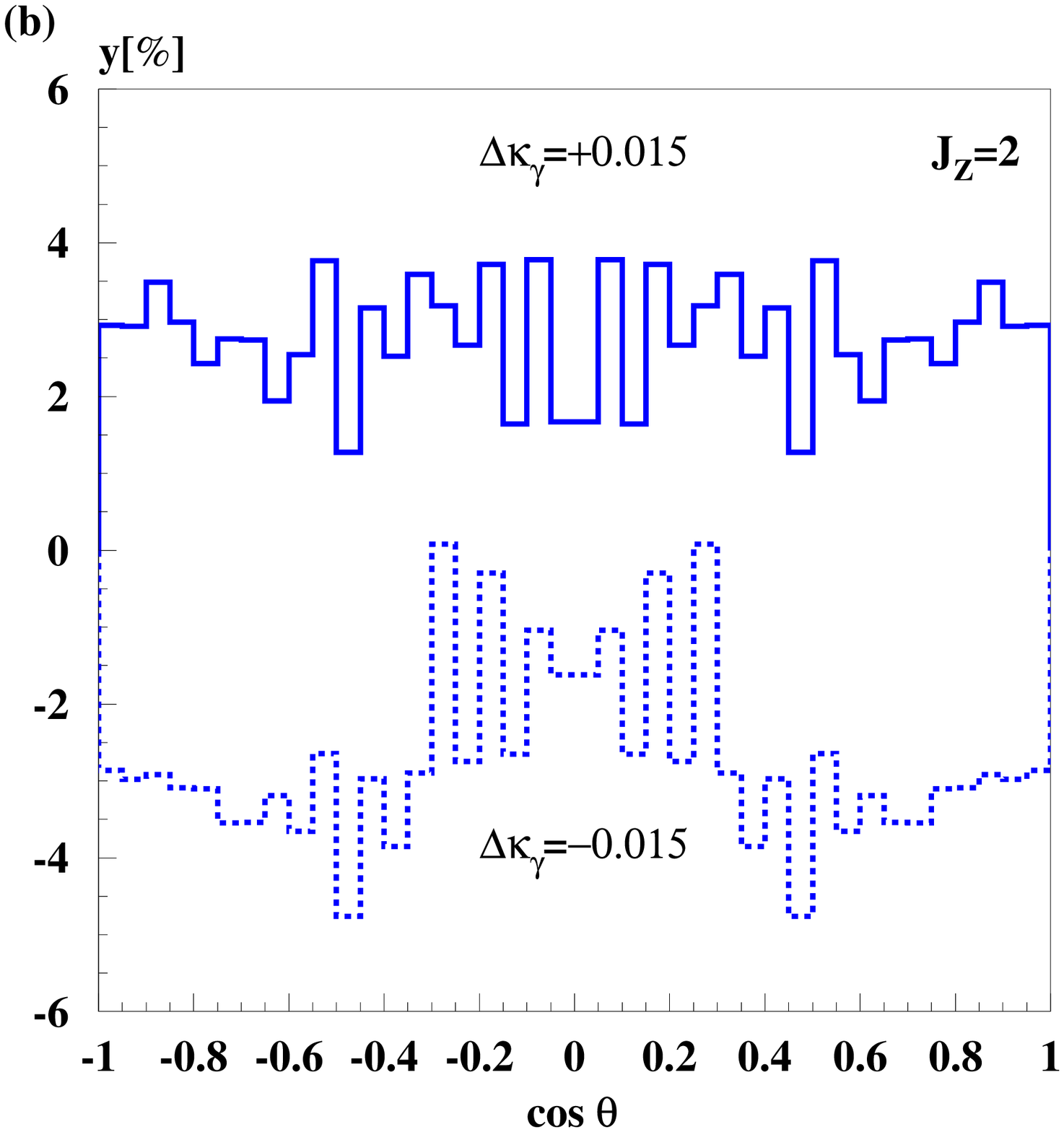}
\caption[bla]{Relative deviations of differential cross-sections from the Standard Model predictions in presence of anomalous coupling $\kappa_{\gamma}=\pm 1.015$ (\textit{a}): in the $J_{Z}=0$ state and (\textit{b}): in the $|J_{Z}|=2$ state at $\sqrt{s_{\gamma\gamma}}=400$ GeV, assuming 100$\%$ photon beam polarizations. Solid lines correspond to ${\Delta}{\kappa}_{\gamma}=+0.015$ and dotted lines correspond to ${\Delta}{\kappa}_{\gamma}=-0.015$ with ${\Delta}{\lambda}_{\gamma}=0$. All $WW$ helicity combinations are included. $y=\frac{\left[d\sigma_{TOT}^{AC}-d\sigma_{TOT}^{SM}\right]}{d\sigma_{TOT}^{SM}}$.}
\label{fig:kappa_gg}
\end{center}
\end{figure}

Relative deviations of the differential cross-sections for all $WW$ helicity combinations in the presence of $\Delta\kappa_{\gamma}=\pm 0.015$, in Fig.~\ref{fig:kappa_gg}, show that anomalous coupling $\kappa_{\gamma}=\pm 1.015$ causes the similar effects on the angular $WW$ distributions for both $J_{Z}$ states. The contribution from the $W_{L}W_{L}$ and $W_{L,T}W_{T,L}$ fraction are masked by the dominating $W_{T}W_{T}$ fraction. Thus, one should expect similar sensitivities for $\kappa_{\gamma}$ measurements in both $J_{Z}$ states.
\par
The influence of the anomalous coupling $\lambda_{\gamma}=\pm 0.015$ is somewhat different for the two different $J_{Z}$ states. Fig.~\ref{fig:diff_lam} corresponds to the Fig.~\ref{fig:diff_kap} except that $\Delta\kappa_{\gamma}$ is replaced by $\Delta\lambda_{\gamma}$ while $\kappa_{\gamma}$ keeps its Standard Model value. Similarly to the $\Delta\kappa_{\gamma}$ dependence, the $W_{L}W_{L}$ fraction in the $J_{Z}=0$ state is more affected by an anomalous $\lambda_{\gamma}$ than in the $|J_{Z}|=2$ state, relative to the Standard Model. The production of $W_{T}W_{T}$ in the $J_{Z}=0$ state is not sensitive to $\Delta\lambda_{\gamma}$ as in the $|J_{Z}|=2$ state, the $W_{T}W_{T}$ fraction deviates in the central region up to 10$\%$. The $W_{L,T}W_{T,L}$ fraction in the $|J_{Z}|=2$ state is influenced more by $\Delta\lambda_{\gamma}$ than by $\Delta\kappa_{\gamma}$, compared to the Standard Model.
\par
\begin{figure}[htb]
\begin{center}
\epsfxsize=3.0in
\epsfysize=3.0in
\epsfbox{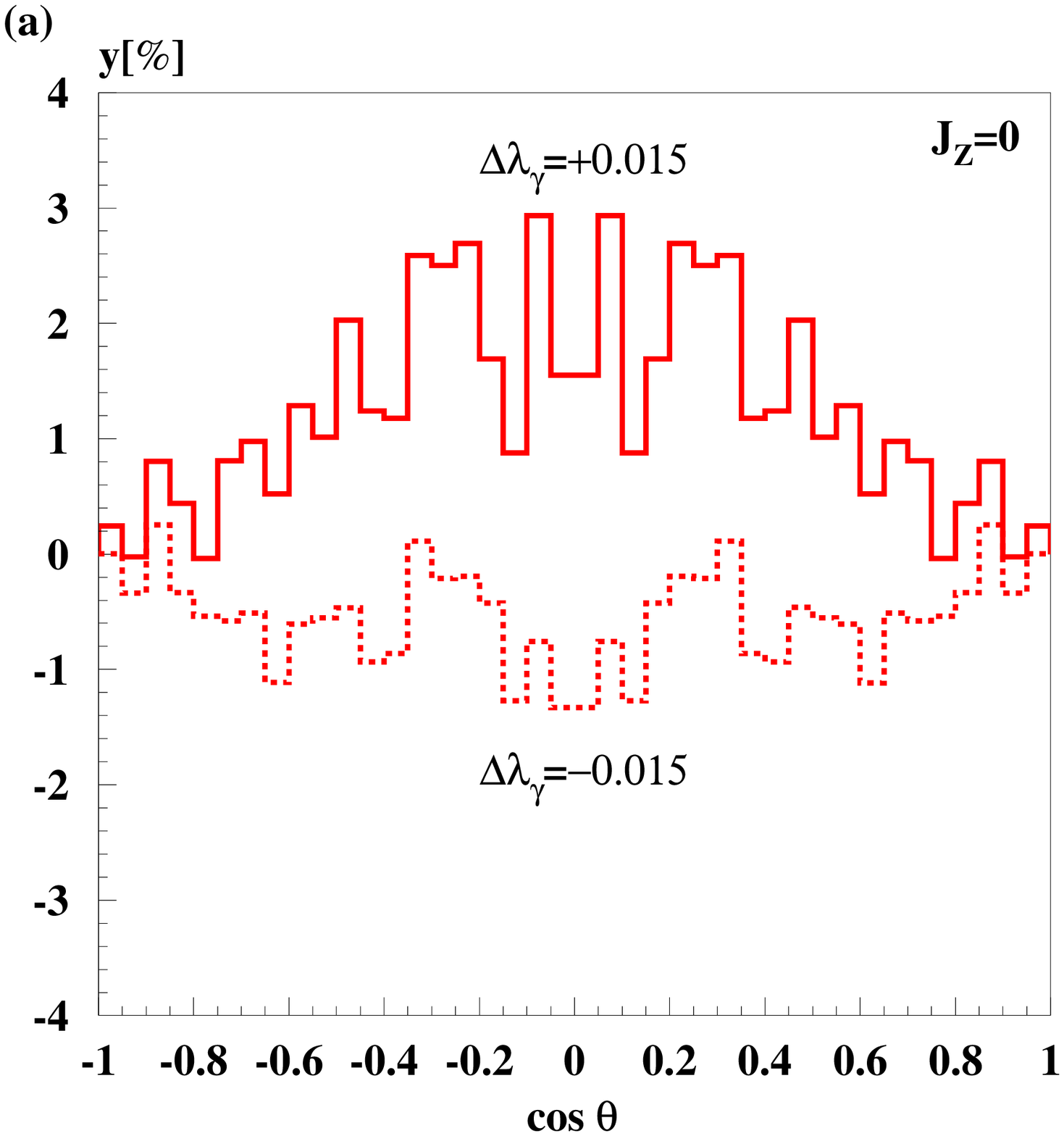}
\epsfxsize=3.0in
\epsfysize=3.0in
\epsfbox{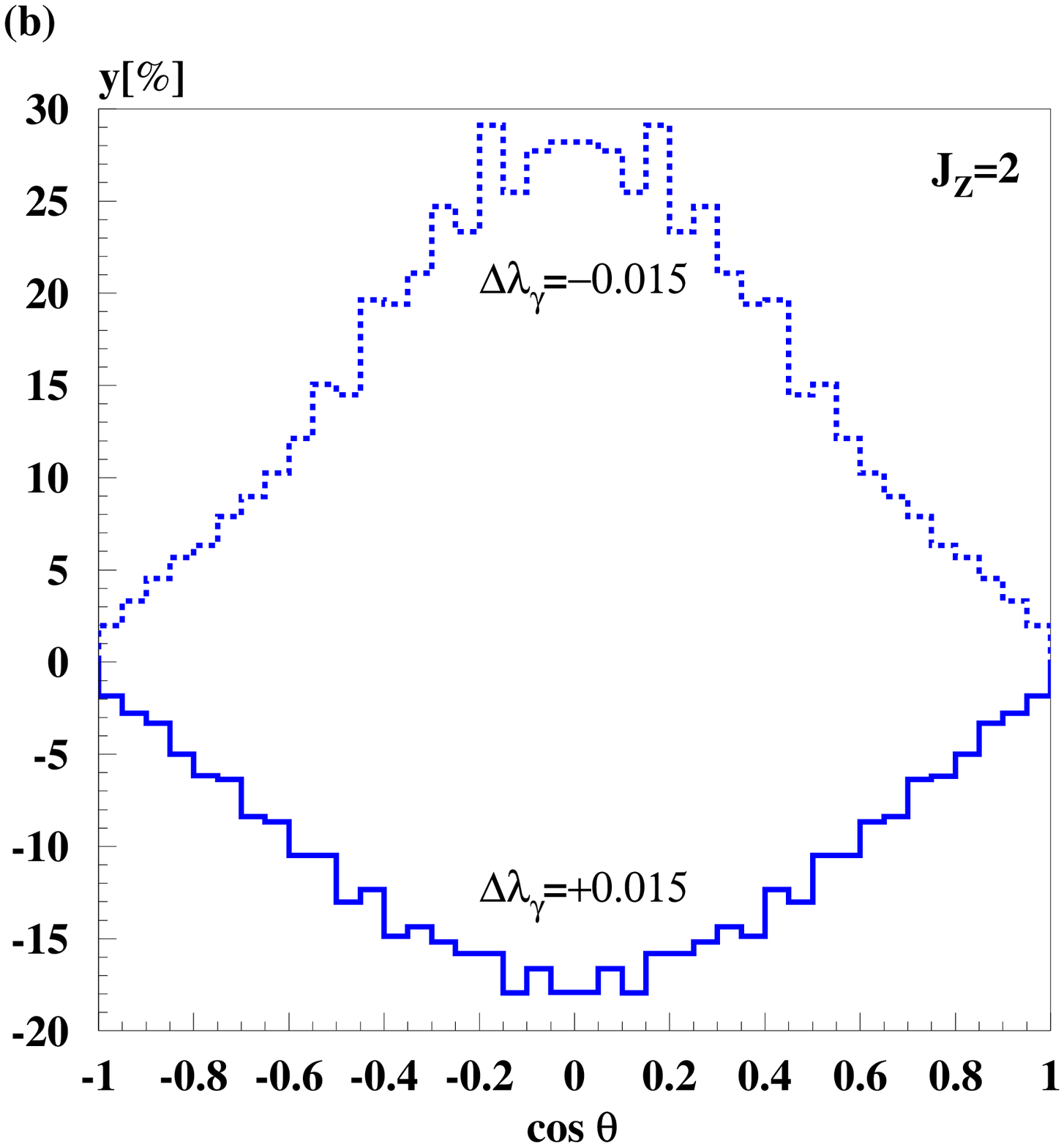}
\caption[bla]{Relative deviations of differential cross-sections from the Standard Model predictions in presence of anomalous coupling $\lambda_{\gamma}=\pm 0.015$ (\textit{a}): in the $J_{Z}=0$ state and (\textit{b}): in the $|J_{Z}|=2$ state at $\sqrt{s_{\gamma\gamma}}=400$ GeV, assuming the 100$\%$ photon beam polarizations. Solid lines correspond to ${\Delta}{\lambda}_{\gamma}=+0.015$ and dotted lines correspond to ${\Delta}{\lambda}_{\gamma}=-0.015$ with ${\Delta}{\kappa}_{\gamma}=0$. All $WW$ helicity combinations are included. $y=\frac{\left[d\sigma_{TOT}^{AC}-d\sigma_{TOT}^{SM}\right]}{d\sigma_{TOT}^{SM}}$.}
\label{fig:lambda_gg}
\end{center}
\end{figure}

Deviations of the differential cross-section for all $WW$ helicity combinations in the presence of $\Delta\lambda_{\gamma}=\pm 0.015$ from the Standard Model are shown in Fig.~\ref{fig:lambda_gg}. The anomalous coupling $\lambda_{\gamma}$ causes different effects on the angular $WW$ distributions in $J_{Z}=0$ and $|J_{Z}|=2$ states. For a given $\lambda_{\gamma}$, the deviations from the Standard Model are one order of magnitude larger in the $|J_{Z}|=2$ compared to the $J_{Z}=0$ state. Thus, one should expect a larger sensitivity for the $\lambda_{\gamma}$ measurements in the $|J_{Z}|=2$ state than in $J_{Z}=0$.
\par
Differential and total cross-sections as a function of the anomalous TGCs are calculated using the tree-level Monte Carlo generator WHIZARD \cite{whizard}.
\subsection{Theoretical Prediction for the Standard Model Radiative ${\cal O}(\alpha)$ corrections}
The prediction of the one-loop ${\cal O}(\alpha)$ radiative corrections to the $W$ boson pair production cross-section in $\gamma\gamma$ collisions from soft-photon bremsstrahlung, virtual radiative corrections and Higgs resonant ${\cal O}(\alpha)$ contribution\footnote{Only $J_{Z}=0$ state is affected by the Higgs resonance through the ${\cal O}(\alpha)$ corrections to the Higgs boson width.} are calculated \cite{denner_gg} in the soft-photon approximation. Since $\gamma\gamma\rightarrow W^{+}W^{-}$ does not involve external light charged particles, the large leading-logarithmic corrections are absent and thus, the photonic corrections are smaller than the weak corrections. The weak corrections, coming from fermionic and bosonic box diagrams and vertex corrections, are about $10\%$ at $\sqrt{s}\sim 1$ TeV. At high energies ($s\gg M_{W}^{2}$) they are dominated by logarithm terms like $\sim(\alpha/\pi)\log^{2}(s/M_{W}^{2})$.
\par
The complete radiative corrections to the total cross-section, including the correction due to the hard photon emission and non-QED corrections, integrated over $10^{\circ}<\theta <170^{\circ}$ for production of $W_{T}W_{T}$ bosons for both $J_{Z}$ states, are approximatively $-10\%$ at $\sqrt{s}\approx 500$ GeV. At higher energies ($\sqrt{s}\approx 2$ TeV), they increase to $-20\%$. The corrections to the total cross-section for production of $W_{L}W_{L}$ in the $|J_{Z}|=2$ state behave in a similar way while the production of $W_{L,T}W_{T,L}$ bosons has slightly larger corrections. The corrections to the total cross-section for production of $W_{L}W_{L}$ bosons are different for the $J_{Z}=0$ state. At energies below $1\,$TeV the corrections are dominated by the Higgs resonance inducing corrections to the total cross-section larger than $-50\%$, while at higher energies the corrections are proportional to $\sim (M_{H}^{2}/M_{W}^{2})$ and increase above $50\%$.
\par
Concerning the differential cross-sections the radiative corrections are of order of $10\%$ whenever the differential cross-section is sizeable. The corrections are smaller in the forward and backward regions where the production of $W_{T}W_{T}$ boson dominates while their contribution to the differential cross-sections becomes large for the suppressed or zero Standard Model cross-sections, especially for production of $W_{L}W_{L}$ bosons in the $J_{Z}=0$ state. The maximal corrections are usually obtained for central values of the $W$ boson production angle. These corrections are not included in the Monte Carlo simulation of $\gamma\gamma\rightarrow W^{+}W^{-}$.
\par
In the heavy-Higgs limit ($M_{H}\gg\sqrt{s}$) the non-QED corrections to the $W_{L}W_{L}$ production cross-section $\sim\textrm{log}(M_{H}^{2}/M_{W}^{2})$ or $(M_{H}^{2}/M_{W}^{2})$ are absent \cite{denner2}. In the Higgsless scenarios which is of interest in this study they are absorbed by the NLO operators in an effective approach.
\par
Thus, it can be assumed that the radiative corrections to the total $WW$ boson production cross-section are less than 10$\%$.

\chapter{Analysis}
The single $W$ boson production in $\gamma e^{-}\rightarrow W^{-}\nu_{e}$ interactions and $W$ boson pair production in $\gamma\gamma\rightarrow W^{-}W^{+}$ interactions, as well as corresponding background events, are studied on event samples generated with a tree-level Monte Carlo generator WHIZARD \cite{whizard}. Only the hadronic $W$ boson decay channels are considered (simulated), representing $\sim\,68\%$ of all single $W$ boson decay channels in $\gamma e$ collisions and $\sim\,45\%$ of all $WW$ decay channels in $\gamma\gamma$ collisions. The beam spectra in $\gamma e$- and $\gamma\gamma$-colliders at $\sqrt{s_{e^{-}e^{-}}}=500$ GeV are simulated with CIRCE2 \cite{circe2}. The response of the detector has been simulated with SIMDET \cite{simdet4}, a parametric Monte Carlo for the TESLA $e^{+}e^{-}$-detector.
\par
The hadronic decay channel of a single $W$ boson from $\gamma e^{-}$ collisions is used due to the impossibility to provide all necessary kinematical informations for its reconstruction if the semi-leptonic channel is used. The neutrino produced in a semi-leptonic channel and a variable energy spectrum leave a hadronic $W$ boson decay channel as the only possibility for its reconstruction. In the semi-leptonic decay channel of the $W$ boson pair, the reconstruction of $W$ bosons is still possible with some constraints but leading to worse resolutions. Thus, the hadronic $W$ boson decay channels are used in both cases.
\section{Simulation Tools}
\textbf{WHIZARD - W, HIggs, Z, And Respective Decays} \cite{whizard},
is a Monte Carlo generator for the calculation of multi-particle scattering cross-sections and simulated event samples in the Standard Model, created for studies that concern the physics at linear colliders. For the calculation of tree-level matrix elements WHIZARD uses the external generators O'Mega \cite{omega}, MadGraph \cite{madgraph} which support the beam polarizations and CompHEP \cite{comphep} where the beam polarizations are not taken into account. In this analysis mainly the O'Mega generator is used since it includes the Standard Model anomalous triple gauge boson couplings $g_{1}^{\gamma,Z},\kappa_{\gamma,Z}$ and $\lambda_{\gamma,Z}$. On the other hand, MadGraph can be used for an approximative simulation of QCD effects with the possibility to specify the order in the QCD coupling constant but the interference between different orders of QCD is not included. For the estimation of QCD backgrounds that originate from gluon splitting, the quark flavor summation is still not possible and the contribution from individual quark flavors should be calculated separately. The non-existence of full QCD calculations makes the estimation of the QCD effects incomplete.
\par
The beamstrahlung, the initial state radiation (ISR) and the beam spectra for the photon collider are included in WHIZARD. The ISR spectrum is calculated in the leading-logarithmic approximation taking the electron structure function from \cite{lipat} with the logarithmic terms up to ${\cal O}(\alpha^{3})$. 
\par
The event samples generated on the parton level are fragmented and hadronized with PYTHIA \cite{pythia} which uses the Lund string fragmentation scheme \cite{lund}, giving additional quarks, gluons and photons, radiated off the partons to form hadrons. The transition of the primary quarks to the observable hadrons starts with a parton shower i.e. a sequence of consecutive branchings of a mother parton into two daughter partons (some possible branchings of interest are $q\rightarrow qg$, $g\rightarrow gg$ and $g\rightarrow q\bar{q}$ included in PYTHIA). Each daughter is free to branch again in the same way, building a tree-like structure. Branching continues until a cut-off is reached, which is usually chosen to be of order of 1 GeV and gives soft daughter partons, almost collinear to a mother parton. This is not correct for the simulation of additional partons (gluons) with large transverse momentum leading to three or more distinct jets. A better description of branchings is achieved matching the parton shower and the pertubative matrix element calculations for three partons \cite{matching}. The additional weights derived from the three-parton matrix elements are included to the first branching in the shower. Besides the mass ordering in the parton shower, an angle ordering is applied leading to smaller opening angles in each branching relative to the previous one \cite{muller}. An approximative description of branching properties is to generate events with gluons and 'secondary' quarks in the final state using MadGraph (specifying the order in $\alpha_{s}$) and fragmenting them using the Lund string fragmentation scheme. The radiated hard gluons and 'secondary' quarks describe in a proper way distinct jets but an additional problem appears during the shower: the gluon emission is doubly counted since the gluons and 'secondary' quarks are regarded as partons.
\par
The hadronization in the Lund string scheme \cite{lund} assumes the existence of two quarks connected by the color flux tube like a rubber string and thus, it is called the string hadronization \cite{string}. When quark and anti-quark that form a string start to move apart, a color field between them increases and finally breaks a string. In fact, the color flux tube is interrupted by a virtual $q\bar{q}$ pair coming from the vacuum along the string. When a virtual $q\bar{q}$ pair fluctuates the color field is locally compensated\footnote{If a virtual $q\bar{q}$ pair has a same color as the quarks of the string.} and the string breaks into two pieces. This procedure repeats until the remaining energy of the strings is insufficient to transform a virtual $q\bar{q}$ pair into a real one. Then the string is transformed into hadrons. Hard gluons are included in the strings connecting two nearby quarks and appear as kinks in the string. The splitting of the string produces on-shell hadrons until the invariant mass of remaining string drops below a cut-off and the remainder is split into two on-shell hadrons.
\par
\textbf{CIRCE2} \cite{circe2}
is a fast parameterization of $\gamma e$ and $\gamma\gamma$ luminosity spectra at a photon collider, supported by the WHIZARD generator that uses adapted histograms of the luminosity distributions given by \cite{telnovv}. Simulated spectra include the effects of multiple interactions, non-linear effects taking into account an effective electron mass and multi-photon scattering, $e^{+}e^{-}$ pair creation in the conversion region for $x>$ 4.8, deflection by magnetic fields and synchrotron radiation in the region between the conversion region and the interaction point, coherent pair creation and beamstrahlung at the interaction point. In this study, it is assumed that the electron beams have 85$\%$ longitudinal polarization and that the laser photons have 100$\%$ circular polarization. A larger distance $b$ between the conversion region and the interaction point has been used for the simulation of $\gamma e$ spectra than for $\gamma\gamma$, in order to decrease the contribution of low energy photons to the $\gamma e$ luminosity spectrum.
\par
\begin{figure}[htb]
\begin{center}
\epsfxsize=3.0in
\epsfysize=3.0in
\epsfbox{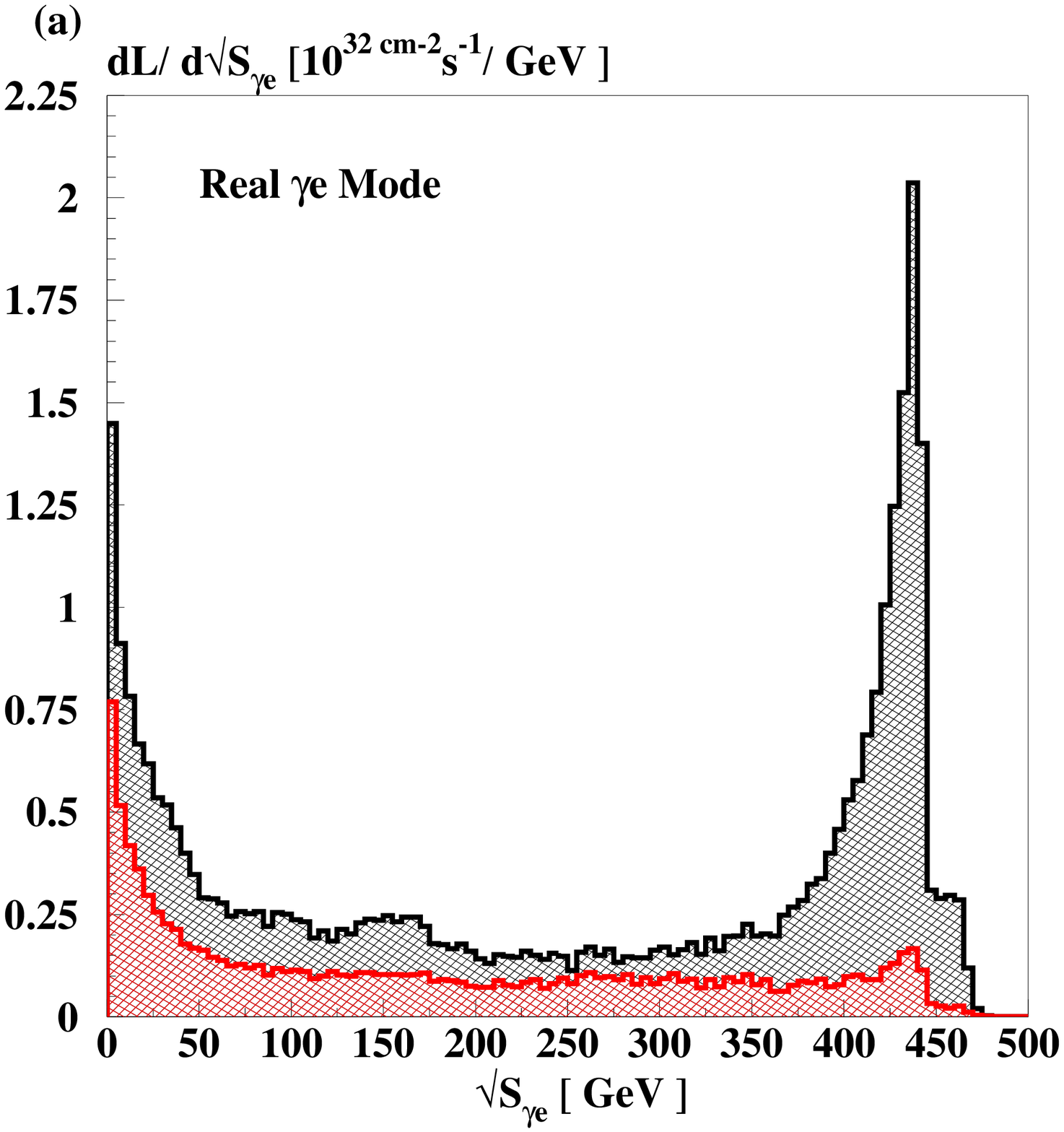}
\epsfxsize=3.0in
\epsfysize=3.0in
\epsfbox{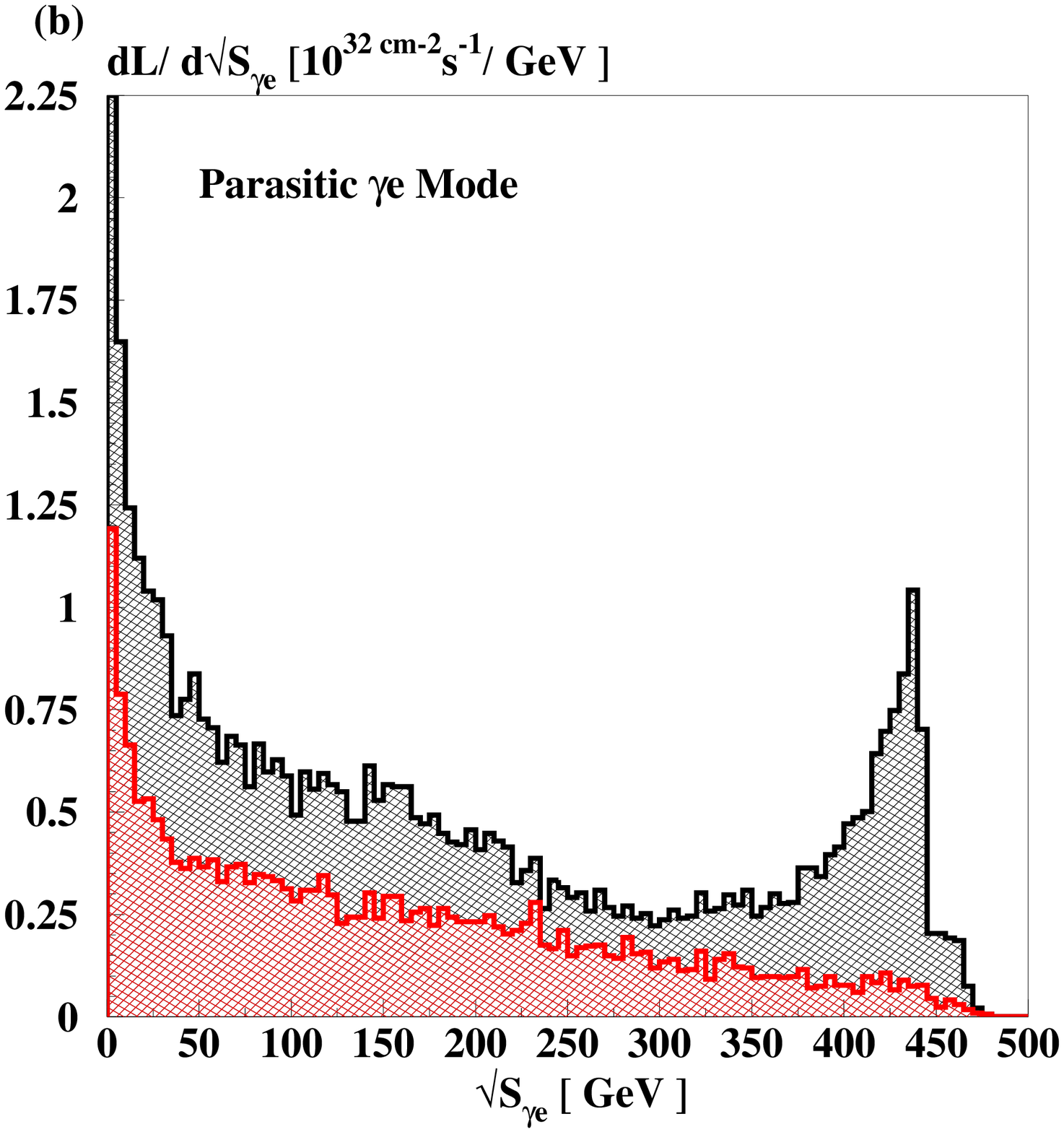}
\caption[bla]{$\gamma e$ luminosity spectra (\textit{a}): for the real and (\textit{b}): for the parasitic $\gamma e$ mode, simulated with CIRCE2 for ${\sqrt{s_{ee}}=500}\,{\rm GeV}$. Red areas represent the $|J_{Z}|=1/2$ contribution while grey areas represent the total luminosity for both $J_{z}$ states.}
\label{fig:lumi_ge}
\end{center}
\end{figure}

Since the beam spectra affect the sensitivity of experiments for the search of deviations from the Standard Model predictions, they have to be included into the simulation. The luminosity spectra used for the generation of $\gamma e^{-}\rightarrow W^{-}\nu_{e}$ events are shown in Fig.~\ref{fig:lumi_ge}. The number of events is normalized to the signal cross-section given by the generator.
Assuming an integrated luminosity of 71 fb$^{-1}$ in the region $\sqrt{s_{\gamma e}}\geq 0.8\sqrt{s_{\gamma e}(max)}$ about 3$\cdot 10^{6}$ $W$ bosons in hadronic decay channel will be produced in the real mode in the $|J_{Z}|=3/2$ state. Assuming an integrated luminosity of 52 fb$^{-1}$ in the region $\sqrt{s_{\gamma e}}\geq 0.8\sqrt{s_{\gamma e}(max)}$ about 2.5$\cdot 10^{6}$ $W$ bosons in the hadronic decay channel will be produced in the parasitic mode in the $|J_{Z}|=3/2$ state with highly polarized beams assuming 100$\%$ detector acceptance.
\par
\begin{figure}[htb]
\begin{center}
\epsfxsize=3.0in
\epsfysize=3.0in
\epsfbox{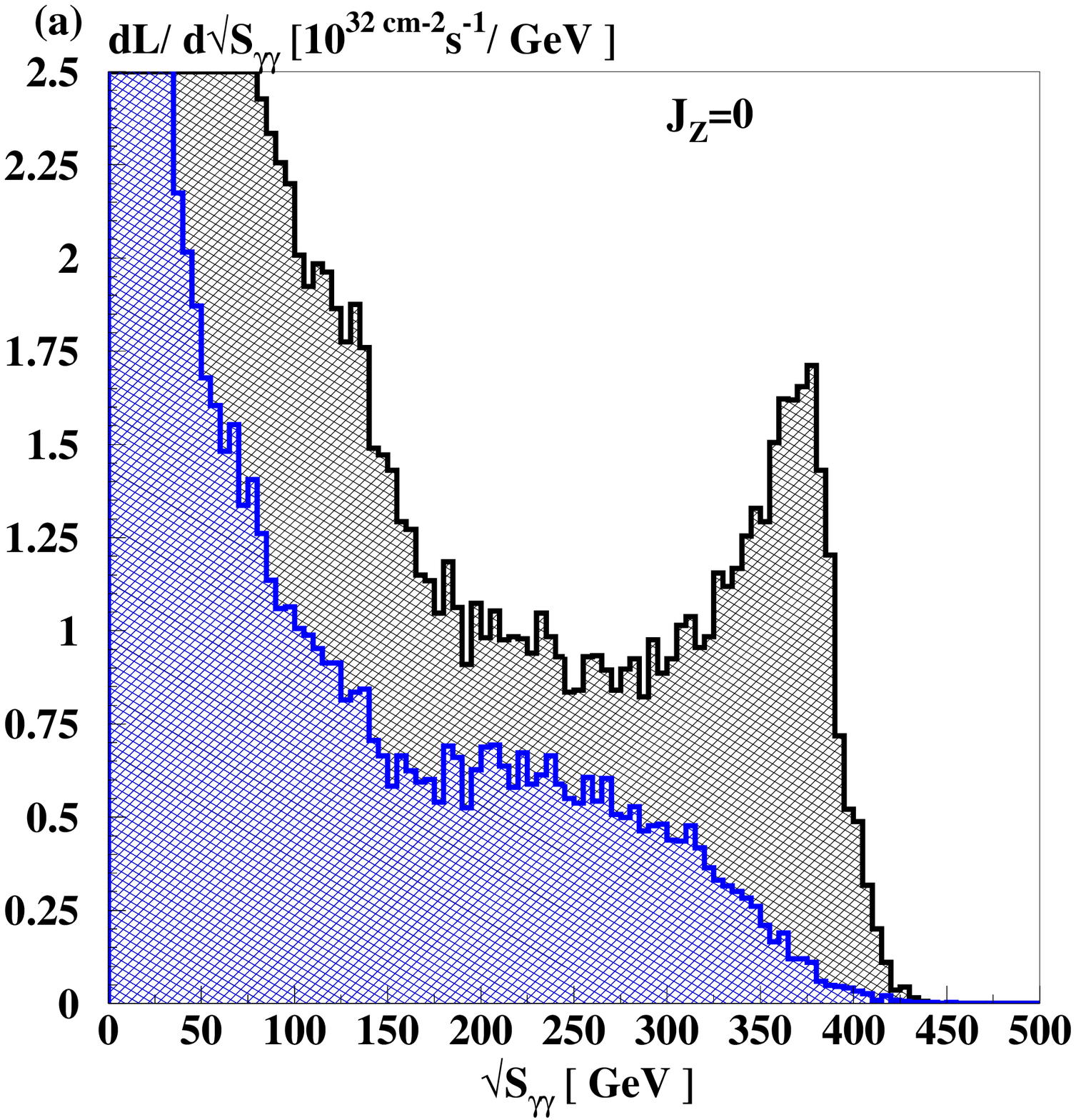}
\epsfxsize=3.0in
\epsfysize=3.0in
\epsfbox{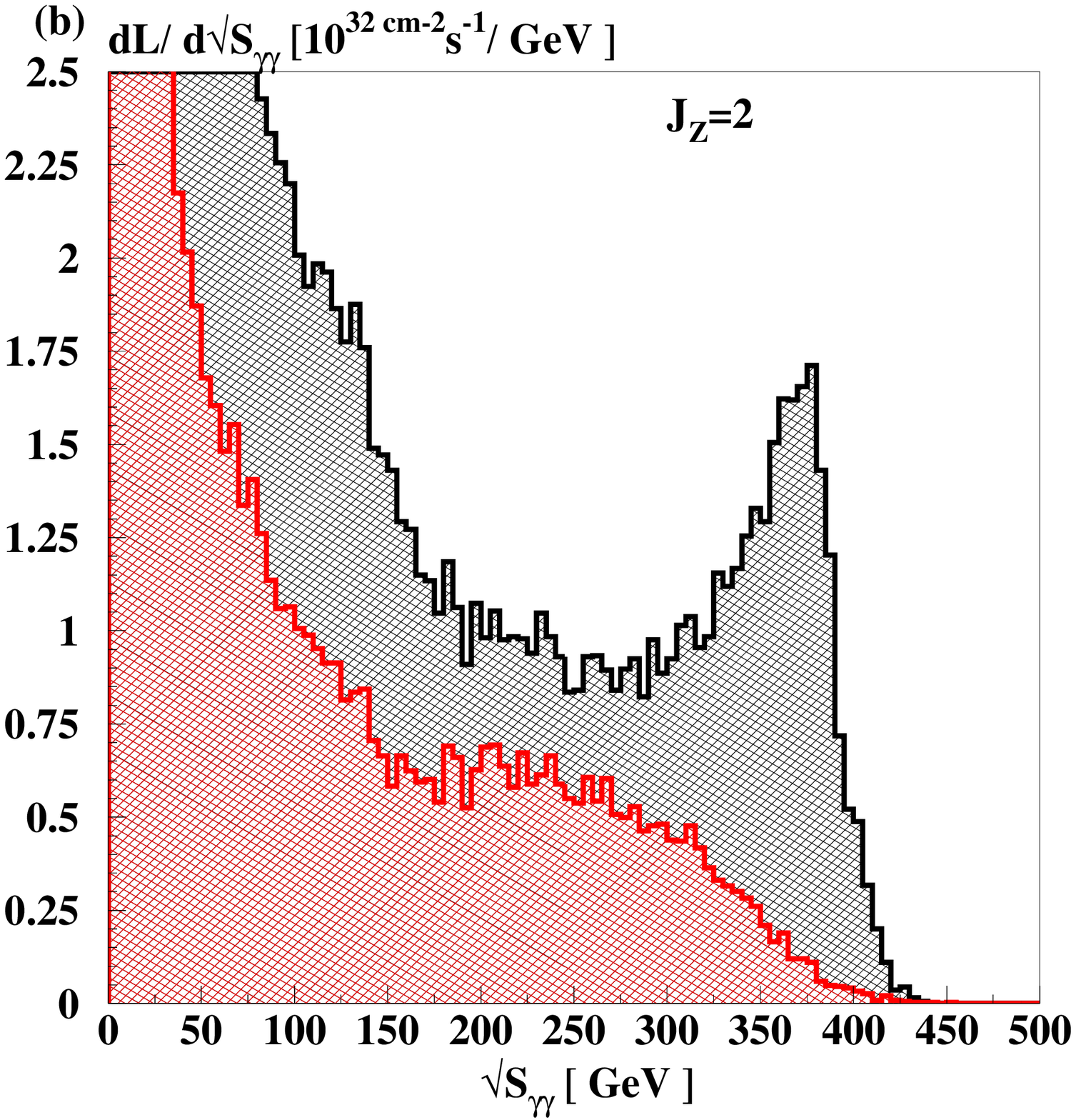}
\caption[bla]{$\gamma\gamma$ luminosity spectra (\textit{a}): for the dominating $J_{Z}=0$ state (grey is the total luminosity for both $J_{Z}$) with a $|J_{Z}|=2$ contribution in blue and (\textit{b}): for the dominating $|J_{Z}|=2$ state (grey is the total luminosity for both $J_{Z}$ states) with a $J_{Z}=0$ contribution in red, simulated with CIRCE2 for ${\sqrt{s_{ee}}=500}\,{\rm GeV}$.}
\label{fig:lumi_gg}
\end{center}
\end{figure}

The luminosity spectra used for the generation of $\gamma\gamma\rightarrow W^{-}W^{+}$ events are shown in Fig.~\ref{fig:lumi_gg}. The number of events is normalized to the signal cross-section given by the generator. Assuming an integrated luminosity of 127 fb$^{-1}$ in the region $\sqrt{s_{\gamma\gamma}}\geq 0.8\sqrt{s_{\gamma\gamma}(max)}$ about 4.3$\cdot 10^{6}$ $W$ boson pairs will be produced in the $J_{Z}=0$ state and about 4$\cdot 10^{6}$ $W$ boson pairs will be produced in the $|J_{Z}|=2$ state in the hadronic decay channel with highly polarized beams assuming 100$\%$ detector acceptance.
\par
All previously mentioned integrated luminosities correspond to one year (10$^{7}$s) of running of an $\gamma e/\gamma\gamma$-collider.
%
\par
\textbf{SIMDET Version 4} \cite{simdet4}
is a fast simulation tool for an $e^{+}e^{-}$ linear collider detector treating the detector respond in a realistic manner using a parameterization from the $ab$ $initio$ simulation from Monte Carlo program BRAHMS \cite{brahms} that uses GEANT3 \cite{geant3} as a detector simulation. The basic components of the simulated detector are a vertex detector, tracker system (central, intermediate and forward), electromagnetic and hadronic calorimeter, low-angle tagger \cite{tdr4} and low-angle luminosity calorimeter \cite{tdr4}. The output of the simulation is defined as an energy flow object, consisting of electrons, photons, muons, charged and neutral hadrons, and unresolved clusters that deposit energy in the calorimeters. SIMDET also supports external event generators that can be included as well as background events to be overlayed to each processed event. In this analysis only the energy flow objects with a polar angle above $7^{\circ}$ are taken for the $W$ boson reconstruction simulating the acceptance of the photon collider detector as the only difference to the $e^{+}e^{-}$-detector \cite{klaus1}. The low-angle tagger and low-angle luminosity calorimeter are not used since their covering angles are about 30 mrad and they are not foreseen to be implemented inside the $\gamma\gamma$-detector. The best estimates for the energy flow objects are stored such that the physics analysis package VECSUB \cite{vecsub} can be used directly.
\section{$\gamma e^{-}\rightarrow W^{-}\nu_{e}$}
\begin{figure}[h]
\begin{center}
\epsfxsize=2.5in
\epsfysize=1.25in
\epsfbox{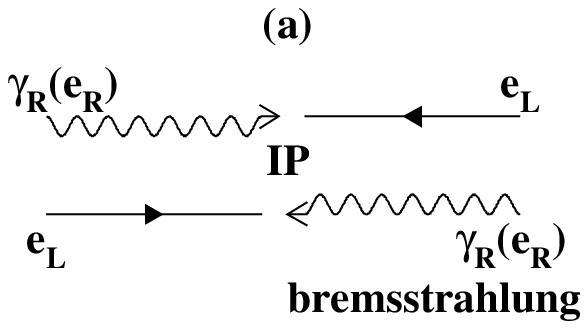}
\epsfxsize=2.5in
\epsfysize=1.25in
\epsfbox{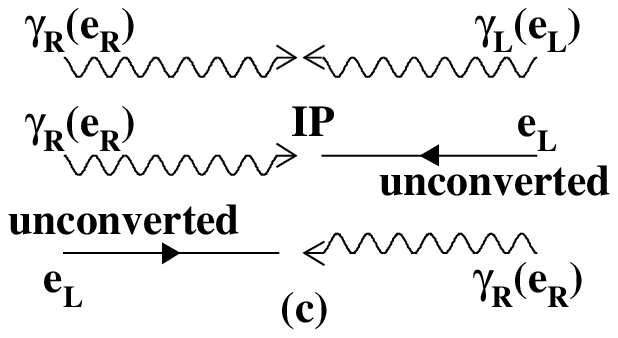}
\epsfxsize=2.5in
\epsfysize=1.25in
\epsfbox{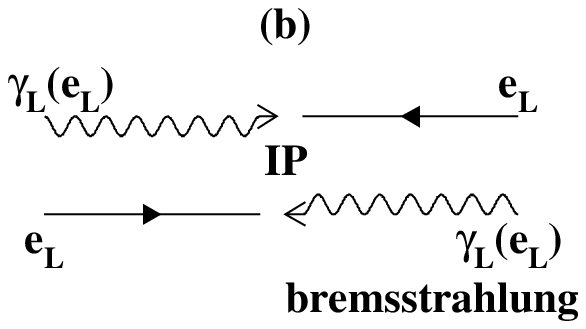}
\epsfxsize=2.5in
\epsfysize=1.25in
\epsfbox{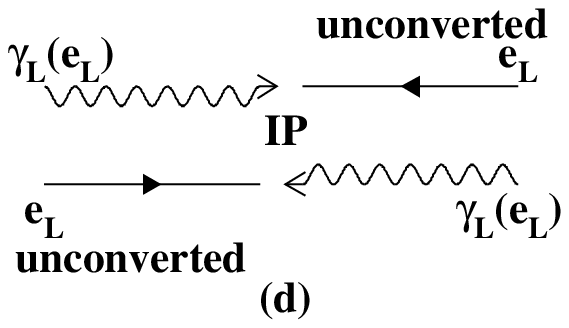}
\caption[bla]{The starting assumption is that the electrons are 85$\%$ polarized. (\textit{a-b}): Photon - electron interactions that contribute to the luminosity spectrum in the real mode. The upper diagrams in $(a)$ and $(b)$ contribute to the $\gamma e$ spectrum with photons from the Compton backscattering while the lower ones in $(a)$ and $(b)$ contribute to the $e\gamma$ spectrum where bremsstrahlung photons are emitted from the electron beams. (\textit{c-d}): Photon - electron interactions that contribute to the luminosity spectrum in the parasitic mode. The upper diagrams in $(c)$ are $\gamma\gamma$ collisions with the appropriate $J_{Z}$ state ($|J_{Z}|=2$) to ensure the high energy peak for $|J_{Z}|=3/2$ in $\gamma e$ collisions. The second one in $(c)$ and the first one in $(d)$ represent photon - electron interactions in the parasitic mode with unconverted electrons and contribute to the $\gamma e$ spectrum. The third one in $(c)$ and the second one in $(d)$ represent photon - electron interactions in the parasitic mode with unconverted electrons contributing to the $e\gamma$ spectrum.}
\label{fig:scheme_eg}
\end{center}
\end{figure}

In the case of $\gamma e^{-}\rightarrow W^{-}\nu_{e}$ interactions, hadronic $W$ boson events are characterized by two-jets in the final state ($q^{'}\bar{q}$) and missing momentum due to the neutrino. Only the initial state $|J_{Z}|=3/2$ is considered since it has been found to be more sensitive to anomalous couplings than the state $|J_{Z}|=1/2$. A first event sample is generated on parton level at a fixed center-of-mass energy of $\sqrt{s}=450$ GeV using 100$\%$ polarized beams and the ISR with a cut-off energy for the soft-photon radiation $\Delta E=1$ GeV. The second sample is generated on parton level too, using CIRCE2 beam spectra.
\begin{figure}[h]
\begin{center}
\epsfxsize=3.0in
\epsfysize=3.0in
\epsfbox{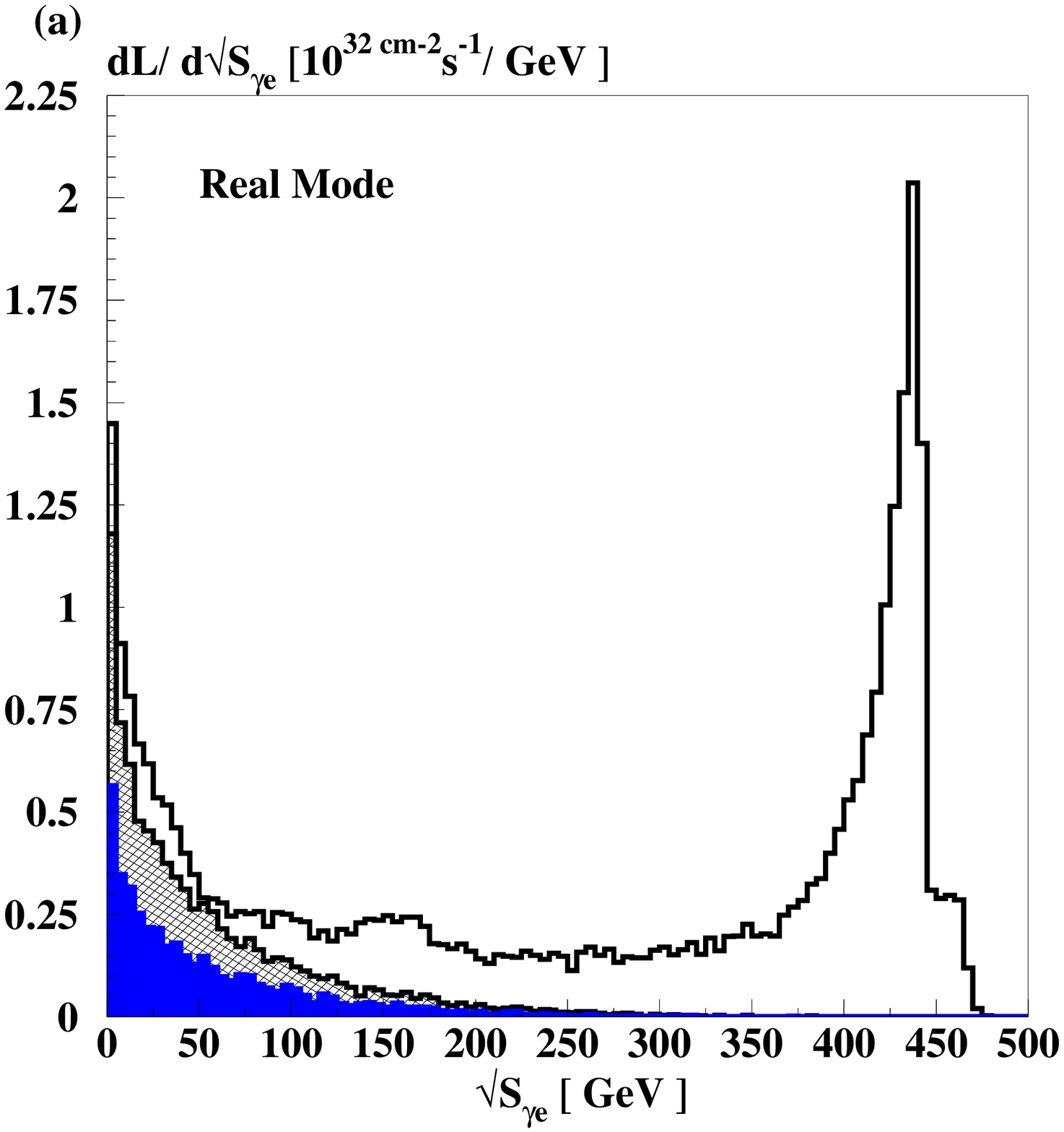}
\epsfxsize=3.0in
\epsfysize=3.0in
\epsfbox{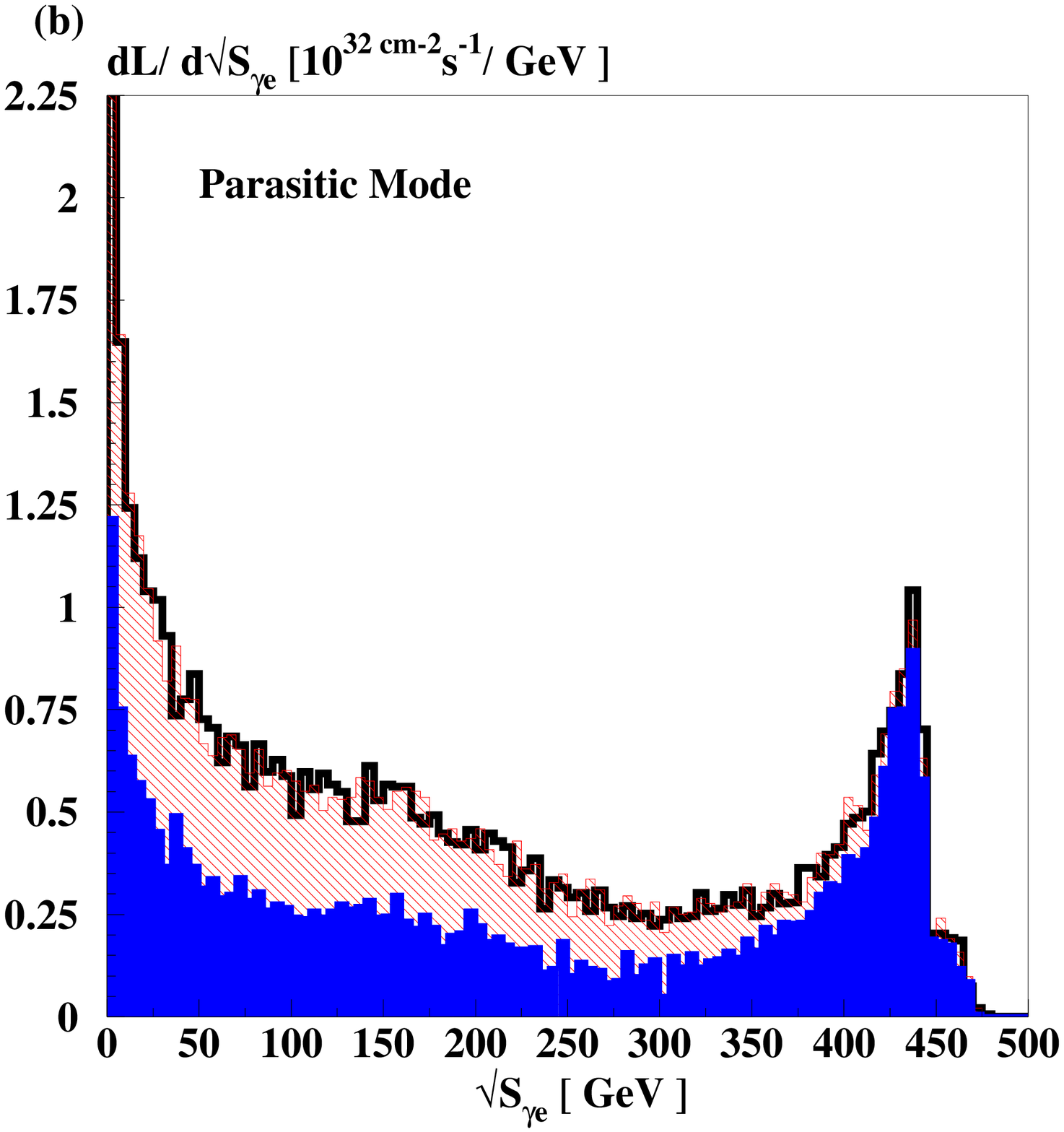}
\caption[bla]{(\textit{a}): $\gamma e$ luminosity spectrum (white) in the real mode - the grey colored area is the contribution from the total $e\gamma$ luminosity spectrum while the blue area is the contribution from the $|J_{Z}|=3/2$ state of the $e\gamma$ spectrum. (\textit{b}): $\gamma e$ luminosity spectrum (black solid line) in the parasitic mode - the red hatched area is the contribution from the total $e\gamma$ luminosity spectrum while the blue area is the contribution from the $|J_{Z}|=3/2$ state of the $e\gamma$ spectrum.}
\label{fig:egamma}
\end{center}
\end{figure}
In the parasitic mode where $\gamma\gamma$ spectra optimized for $|J_{Z}|=2$ are used (left-handed electrons from one side and right-handed electrons from the opposite side), in addition to $\gamma e$ collisions the contribution from collisions of unconverted electrons and Compton backscattered photons is encountered using the $e\gamma$ spectrum for event generation. The contributing $|J_{Z}|=3/2$ part of the $e\gamma$ spectrum is shown in Fig.~\ref{fig:egamma}\,$b$ in blue while the total $e\gamma$ luminosity spectrum is colored in red (hatched). Since the $\gamma e$ and $e\gamma$ luminosity spectra are the same the total number of events in the parasitic mode is the sum of both contributions according to their cross-sections. The ratio of the cross-sections is $\sigma_{\gamma e}/\sigma_{e\gamma}\approx$ 5.9 due to the small fraction of unconverted left-handed electrons in $e\gamma$ collisions. In the real mode, where only one electron beam is converted into high energy photons, the $\gamma e$ luminosity dominates over the $e\gamma$ luminosity almost in the whole energy region and $|J_{Z}|=3/2$ state dominates over the $|J_{Z}|=1/2$ state (Fig.~\ref{fig:lumi_ge}\,$a$). The Fig.~\ref{fig:egamma}\,$a$ shows that in the real mode the $e\gamma$ luminosity spectrum is positioned at low center-of-mass energies since the photons are coming from the bremsstrahlung and not from the Compton backscattering. This contribution is thus neglected. A schematic representation of $\gamma e$ and $e\gamma$ contributions in both modes is shown in Fig.~\ref{fig:scheme_eg}. 
\par
These events are fragmented and hadronized and passed the detector simulation with and without the corresponding number of pileup events. All events are generated using the O'Mega matrix element generator. The informations about the neutral particles from the calorimeter and tracks from the tracking detector are used to reconstruct the signal and background events. The considered backgrounds depend on the two different modes of the $\gamma e$-collider.
\par
For the real $\gamma e$ mode the considered backgrounds are the following:
\begin{enumerate}
\item ${e}{\gamma}{\rightarrow}{e}{Z} {\rightarrow}e {q}{\bar{q}}$ \\
events are characterized by two jets originating from the $Z$ boson and can be easily detected due to the high energy isolated electron in the final state. At high energies the $Z$ decay products are distributed along the incident electron, i.e. over the angular region opposite to the signal $W$ bosons \cite{renard}, while the electron can be lost in the beam-pipe or emitted in the dead region of the detector. In order to estimate the contribution of this background in case when the electron cannot be detected, these events are simulated with a kinematic cut which allows only production of electrons at low angles, below $15^\circ$. The preselection cut used to reduce the background contributing to this channel was to reject events with a high energetic electron ($\geq 100\, {\rm GeV}$) in the detector. By this cut $33\%$ of the background events are rejected not affecting the signal efficiency.
\item ${\gamma}(e^-){\gamma}{\rightarrow}{q}{\bar{q}}$ \\
($|J_{Z}|=2$) has a large cross-section compared to the signal and represents the interaction between a real, high energy photon and a virtual bremsstrahlung photon. These events are characterized with two-jets in the final state with low invariant masses and thus, the cut on the $W$ boson mass is found to be efficient to reject the largest part of this background.
\end{enumerate}
\par
Additional backgrounds considered for the parasitic $\gamma e$ mode are the following:
\begin{enumerate}
\item ${\gamma}{\gamma}{\rightarrow}{W}{W}$ \\
where one $W$ boson decays leptonically and the other $W$ boson decays hadronically. These events are characterized by two jets in the final state, missing momentum due to the neutrino and by an isolated lepton, $e^{\pm}$ or $\mu^{\pm}$ while the $\tau^{\pm}$ lepton can also decay hadronically. $e^{\pm}$ and $\mu^{\pm}$ originate either from the $W$ boson decay or from the cascade decay of the $W$ boson through a $\tau^{\pm}$ lepton. To reduce the background contributing from this channel in each event it is searched for a lepton in the detector with an energy higher than $5\,{\rm GeV}$. For these leptons a cone of $60^{\circ}$ is defined around their flight directions and the energies of all particles (excluding the lepton) are summed inside the cone. Events with energies smaller than $20\,{\rm GeV}$ were rejected. This cut rejects $\sim70\%$ (without pileup) i.e., $\sim60\%$ (with pileup) of the semileptonic $WW$ background events, not affecting the signal efficiency.
\item ${\gamma}{\gamma}{\rightarrow}{q}{\bar{q}}$ \\
($|J_{Z}|=2$) has a large cross-section compared to the signal and represents the interaction between the two real photons. These events are characterized by two jets in the final state.
\end{enumerate}
\par
Due to the different $\gamma\gamma$ luminosities in the two $\gamma e$ modes, the pileup contribution to each mode is different - 1.2 events per bunch crossing for the real mode and 1.8 events per bunch crossing for the parasitic mode and has been included in the simulation of signal and background \cite{schulte}.
\section{$\gamma\gamma\rightarrow W^{+}W^{-}$}
In the case of $\gamma\gamma\rightarrow W^{+}W^{-}$ interactions, the events of the hadronic $WW$ decay channel are characterized by four jets in the final state. For both initial states $J_{Z}=0$ and $|J_{Z}|=2$, the signal events are generated on parton level at a center-of-mass energy of $\sqrt{s}=400$ and 800 GeV, using 100$\%$ polarized beams. The signal events are also generated on parton level using CIRCE2 beam spectra with mixed $J_{Z}$ contributions as shown in Fig.~\ref{fig:lumi_gg}. This means that the total cross-section receives a contribution from both $J_{Z}$ states. If the $J_{Z}=0$ state is simulated it contains the $|J_{Z}|=2$ contribution (Fig.~\ref{fig:lumi_gg}\,$a$) and vice-versa (Fig.~\ref{fig:lumi_gg}\,$b$), if not stated differently. These events are fragmented and hadronized, and passed through the detector simulation with and without the pileup events. All signal events are generated using the O'Mega matrix element generator.
\par
For both initial $J_{Z}$ states the main background comes from $\gamma\gamma\rightarrow q\bar{q}$ events that can mimic the signal, i.e. the four-jet events, when gluons are radiated in the final state. The $\gamma\gamma\rightarrow q\bar{q}$ events are generated with the O'Mega matrix element generator for both $J_{Z}$ states. In the $|J_{Z}|=2$ state the QCD correction to the Born level cross-section is proportional to $\alpha_{s}$ as $(1+k\alpha_{s}/\pi)$ ($\sigma_{2}^{QCD}\sim\sigma_{2}^{0}(1+k\alpha_{s}/\pi)$, with $k$ being of ${\cal O}$(1)). The gluon radiation off the quarks in the final state is well described by Lund parton shower method implemented in PYTHIA\footnote{Matching between the parton shower and the matrix element calculations for three partons.} giving a correction of approximatively 4-5$\%$ to the Born level cross-section. In the used CIRCE2 beam spectrum in Fig.~\ref{fig:lumi_gg}\,$b$, the $|J_{Z}|=2$ state dominantly contributes to the total cross-section while the contribution from the $J_{Z}=0$ state is highly suppressed by a factor $(m_{q}^{2}/s)$ \cite{melles,tkabla} and does not contribute to the total cross-section at the tree-level.
\par
The next-to leading (NLO) QCD correction to the Born level cross-section, calculated using the three parton matrix elements for $\gamma\gamma\rightarrow q\bar{q}g$ in the $J_{Z}=0$ state, can be larger due to the suppression factor $(m_{q}^{2}/s)$ \cite{fadin}. The higher order radiative corrections $\sim(\alpha_{s}\log^{2}(s/m_{q}^{2}))^{n}$, so called the non-Sudakov double logarithms, are present at each order of perturbation theory. If resummed to all orders they remove that suppression \cite{removal}. At one-loop level, leading $\log^{2}(s/m_{q}^{2})$-terms can lead to a negative cross-section in some restricted phase-space regions (all three contributions, lowest order\footnote{The Born level contribution.}, virtual\footnote{The interference term between one-loop and the tree-level contribution.} and gluon emission\footnote{The tree-level contribution from quark pair production accompanied by gluon emission $\gamma\gamma\rightarrow q\bar{q}g$ well described by the parton shower model in PYTHIA.} are of the same order of magnitude if a small $y_{cut}$ is used) and thus, higher order contributions have to be included to give a well defined and positive cross-section, which is already restored at the two-loop level. This also could be cured if the full NLO calculations would be taken into account. From the theoretical predictions it is found that these corrections might be comparable or even larger than the Born level cross-section for heavy quark production ($c\bar{c}/b\bar{b}$ pair production) \cite{tkabla}. The heavy quark mass acts as an effective infrared cut-off for the double logarithms while for the light quark masses the $y_{cut}$ parameter regulates divergencies. In this analysis massless quarks are assumed, except the $b$ quarks which are assumed to be massive.
\par
Thus, the NLO QCD correction to the tree-level cross-section in the $J_{Z}=0$ state can be expressed as $\sigma_{0}^{QCD}\sim\sigma_{0}^{0}(1+j\alpha_{s}/\pi)$ with $j>k$, where $j$ is a correction that contains double logarithmic terms. The relative QCD correction in the $J_{Z}=0$ state is expected to be larger than in the $|J_{Z}|=2$ state and cannot be described just by the Lund parton shower model that gives a correction of 4-5$\%$.
\begin{figure}[h]
\begin{center}
\epsfxsize=2.5in
\epsfysize=1.25in
\epsfbox{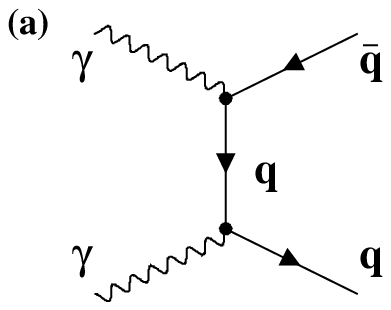}
\epsfxsize=2.5in
\epsfysize=1.25in
\epsfbox{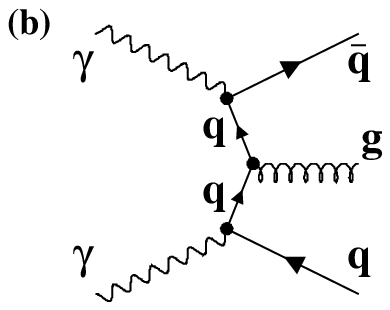}
\epsfxsize=2.5in
\epsfysize=1.25in
\epsfbox{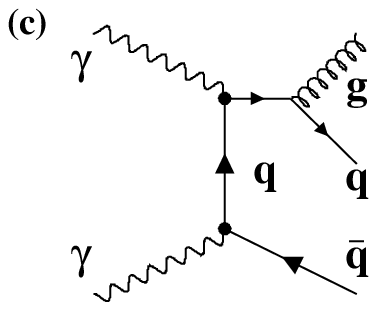}
\epsfxsize=2.5in
\epsfysize=1.25in
\epsfbox{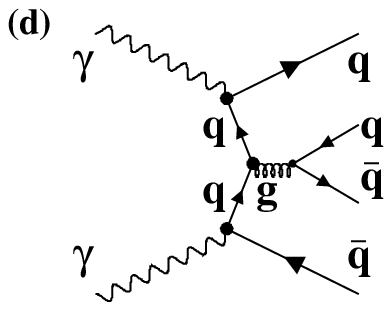}
\epsfxsize=2.5in
\epsfysize=1.25in
\epsfbox{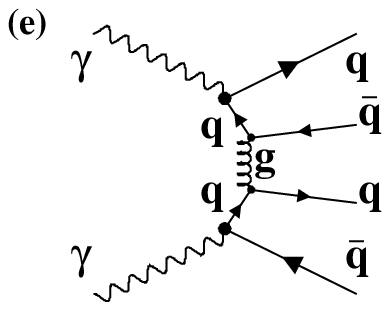}
\epsfxsize=2.5in
\epsfysize=1.25in
\epsfbox{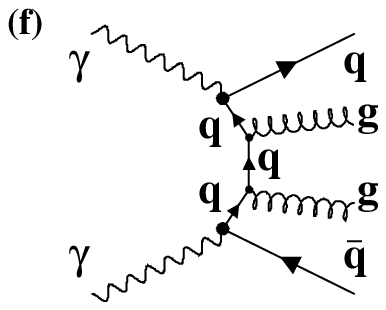}
\caption[bla]{(\textit{a}): Lowest order QCD diagram for $\gamma\gamma\rightarrow q\bar{q}$ simulated with O'Mega matrix element generator. (\textit{b-c}): Diagrams of ${\cal O}(\alpha_{s})$ leading to the three-jet final state and thus, not included into simulation. (\textit{d-f}): Diagrams of ${\cal O}(\alpha_{s}^{2})$ leading to the four-jet final state, simulated with MadGraph matrix element generator.}
\label{fig:qcd}
\end{center}
\end{figure}
\par
Due to the lack of generators that calculate all QCD diagrams together, in order to estimate $\sigma_{0}^{QCD}$, the QCD correction for $\gamma\gamma\rightarrow q\bar{q}$ in the $J_{Z}=0$ state is approximated by the diagrams of ${\cal O}$($\alpha_{s}^{2}$) in the following way: the QCD contribution from each quark flavor is simulated separately using the MadGraph matrix element generator which includes double logarithms and the pure $J_{Z}=0$ luminosity spectrum (without the $|J_{Z}|=2$ contribution). The correction includes the diagrams with an emission of one or two (virtual or real) gluons in the final state shown in Fig.~\ref{fig:qcd}\,$d,e,f$. These are events that result in four distinct jets in the final state and can mimic a signal events - two gluon emission $\gamma\gamma \rightarrow q\bar{q}gg$ and the production of 'secondary' quarks $\gamma\gamma \rightarrow q\bar{q}(g\rightarrow)q\bar{q}$, originating from gluon splitting. The $y_{cut}$ cut parameter ($(p_{a}+p_{b})^{2}>y_{cut}s$; $a,b=q,\bar{q},g/q,g/\bar{q}$) for a variable center-of-mass energy $s$ is defined by generating only events with invariant masses of each parton pair (two-quark, two-gluon or quark-gluon pairs) above 30 GeV. The events are fragmented in the same way as the background for the $|J_{Z}|=2$ state, i.e. a double counting of the gluon emission from the Lund parton shower is included. Then, the $\gamma\gamma\rightarrow q\bar{q}$ events generated with O'Mega generator in the $J_{Z}=0$ state with a $|J_{Z}|=2$ contribution and fragmented with PYTHIA, are added to the QCD ones.
\par
The tree-level diagram which is taken into account with O'Mega for both $J_{Z}$ states is shown in Fig.~\ref{fig:qcd}\,$a$. Diagrams of ${\cal O}(\alpha_{s}^{2})$ taken into account with MadGraph are shown in Fig.~\ref{fig:qcd}\,$d,e,f$ while the diagrams in Fig.~\ref{fig:qcd}\,$b,c$ are included by parton shower model. This QCD estimation is just an approximative one performed in order not to underestimate the background contribution in the $J_{Z}=0$ state. For the full simulation signal and background events are overlayed with 1.8 pileup events per bunch crossing.
\section{Pileup Rejection}
In order to simulate realistic physical conditions at photon colliders, the low energy $\gamma\gamma$ events that can produce hadrons (pileup) are mixed with high energy events taking them from a database \cite{schulte}. The pileup contribution depends on the running mode of the collider. For the real $\gamma e$ mode, the pileup contributes with 1.2 events per bunch crossing while in the parasitic $\gamma e$ mode, i.e. in the $\gamma\gamma$ mode, pileup contributes with 1.8 events per bunch crossing. As a consequence of their presence, the angular distributions of signal events are distorted and thus, the angular resolutions are worse. This is reflected in the estimated errors making them larger. Thus, the goal is to minimize the pileup contribution, to restore the corresponding $W$ boson distributions and to increase the efficiency of the signal.
\par  
In order to minimize the pileup contribution to the high energy signal tracks the first step in the separation procedure was to reject pileup tracks as much as possible. The measurement of the impact parameter of a particle along the beam axis ($z$ axis) with respect to the primary vertex is used for this purpose\footnote{This impact parameter is defined as a $z-$coordinate of the impact point in the $x-y$ plane.}. The logic comes from the fact that the pileup tracks do not originate from the same interaction point as the signal tracks. Thus, their impact parameters should differ from the impact parameters of signal tracks and that information can serve for their identification.
\begin{figure}[p]
\begin{center}
\epsfxsize=3.0in
\epsfysize=3.0in
\epsfbox{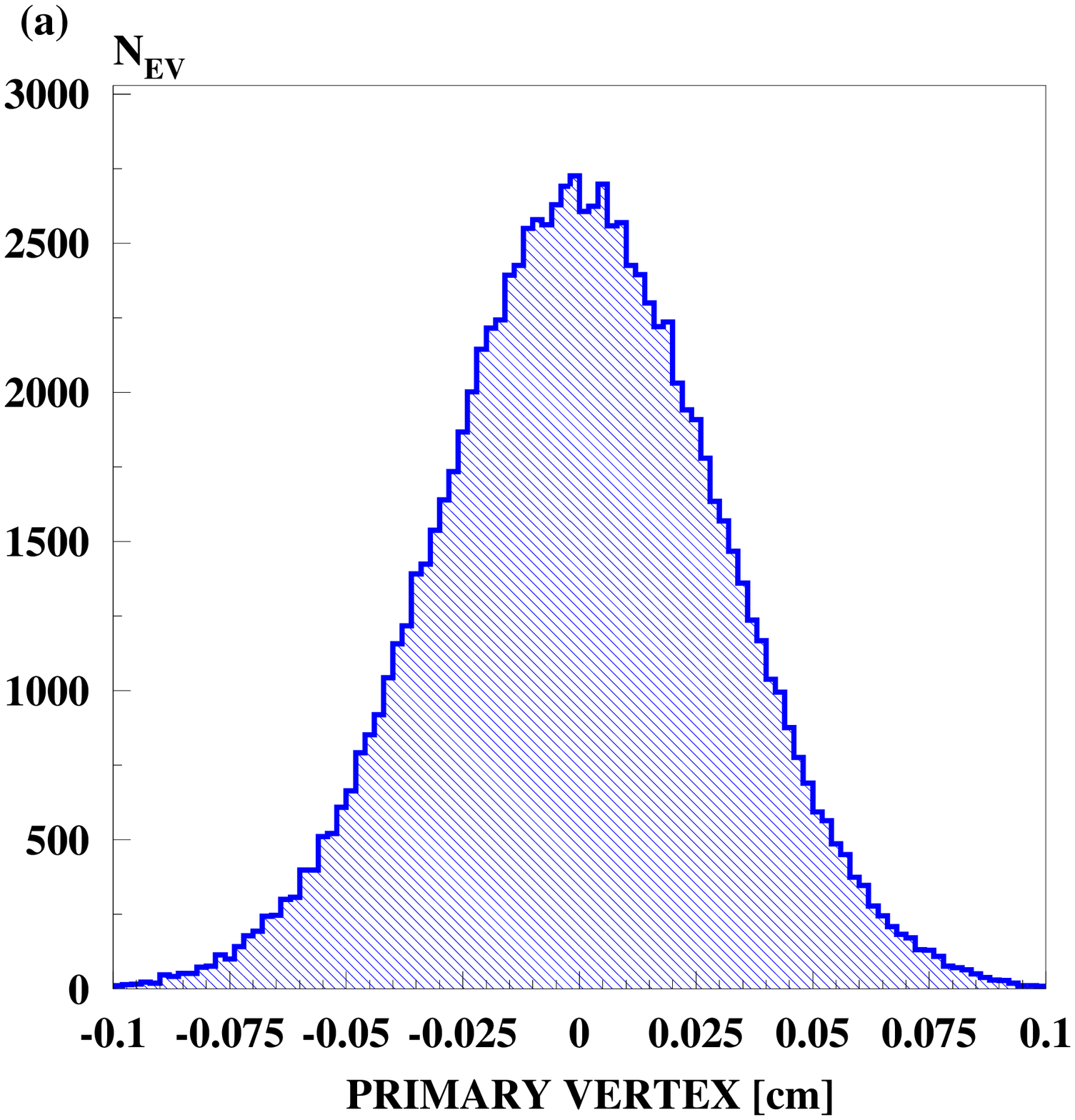}
\epsfxsize=3.0in
\epsfysize=3.0in
\epsfbox{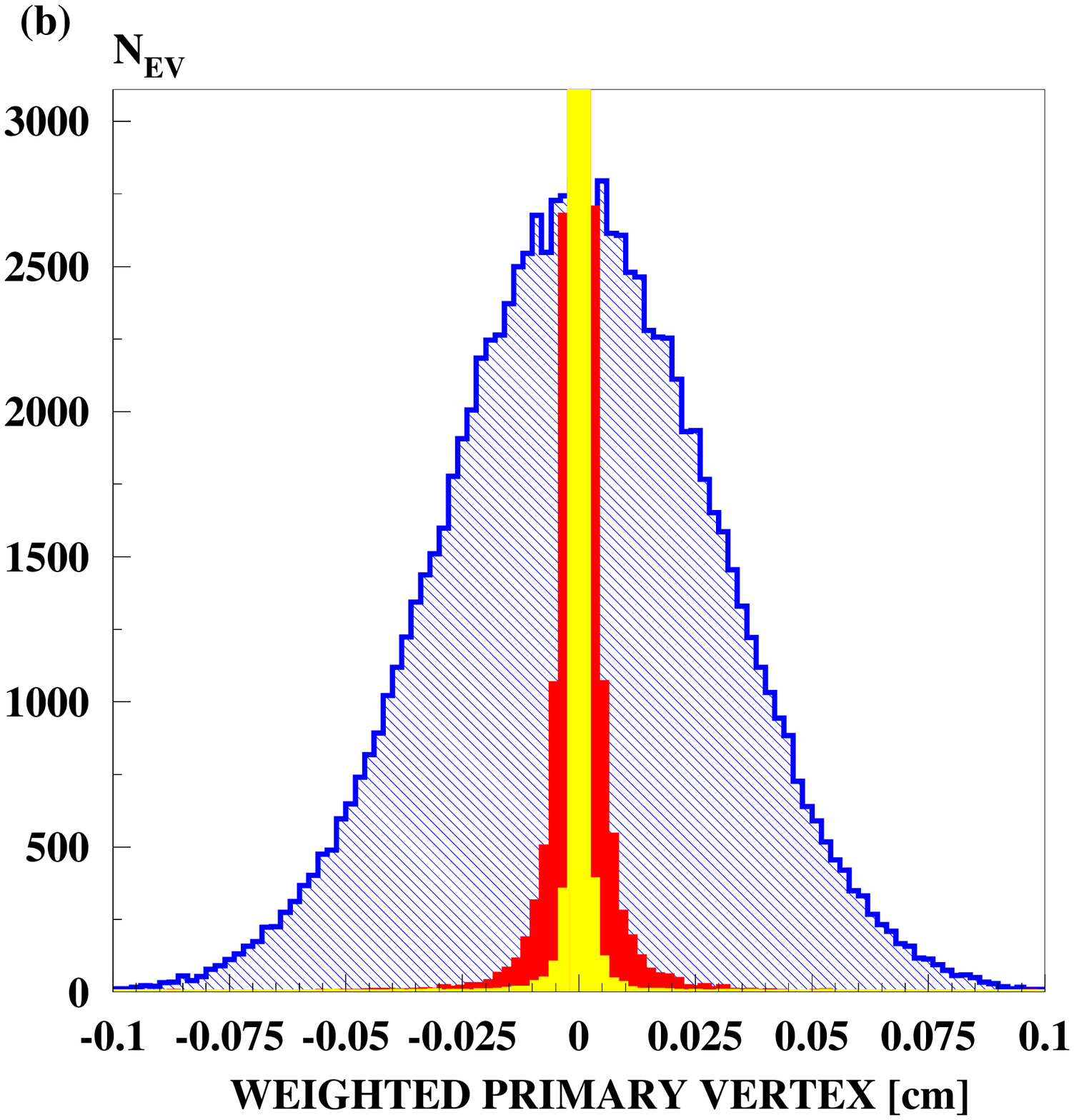}
\epsfxsize=3.0in
\epsfysize=3.0in
\epsfbox{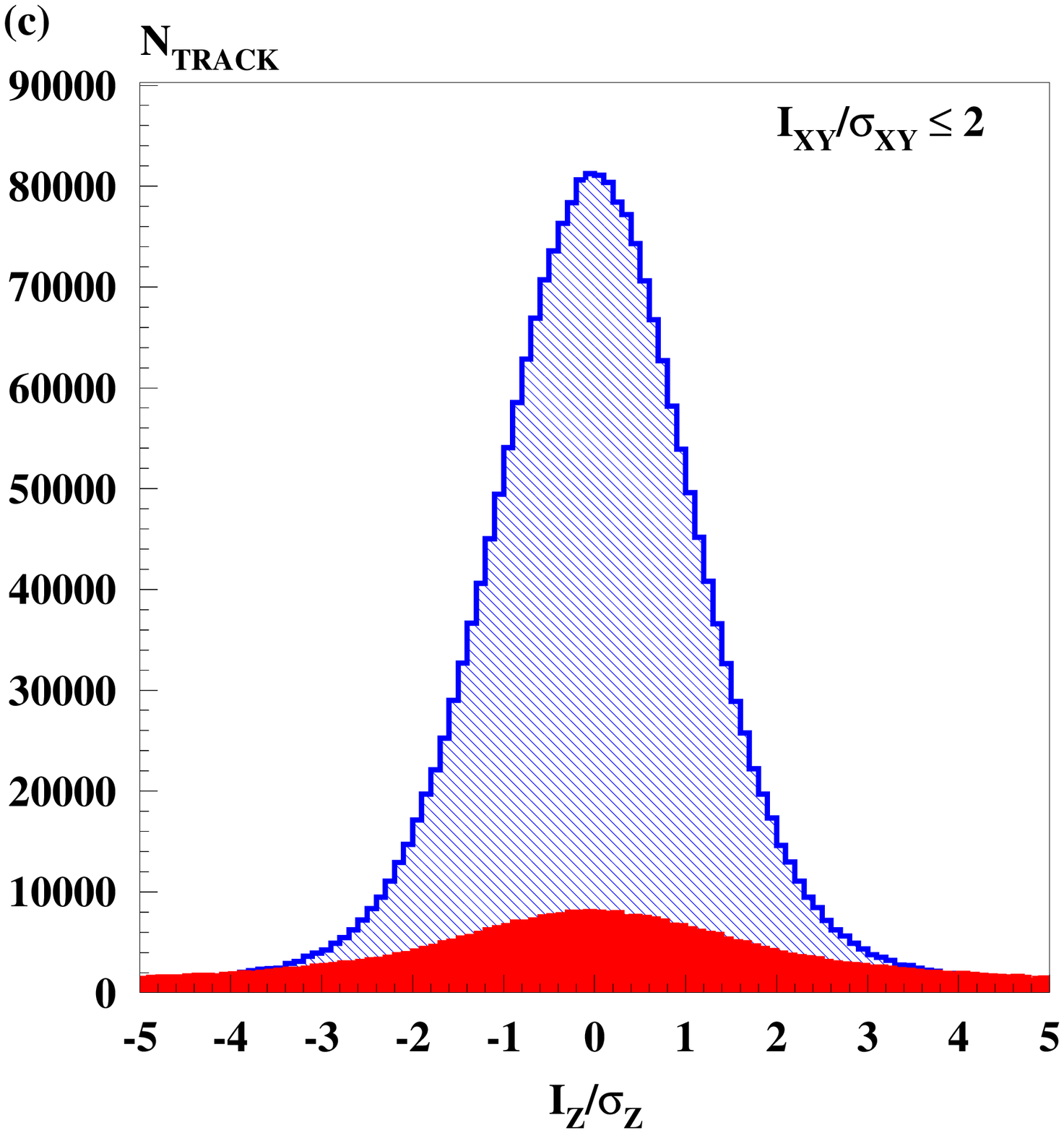}
\epsfxsize=3.0in
\epsfysize=3.0in
\epsfbox{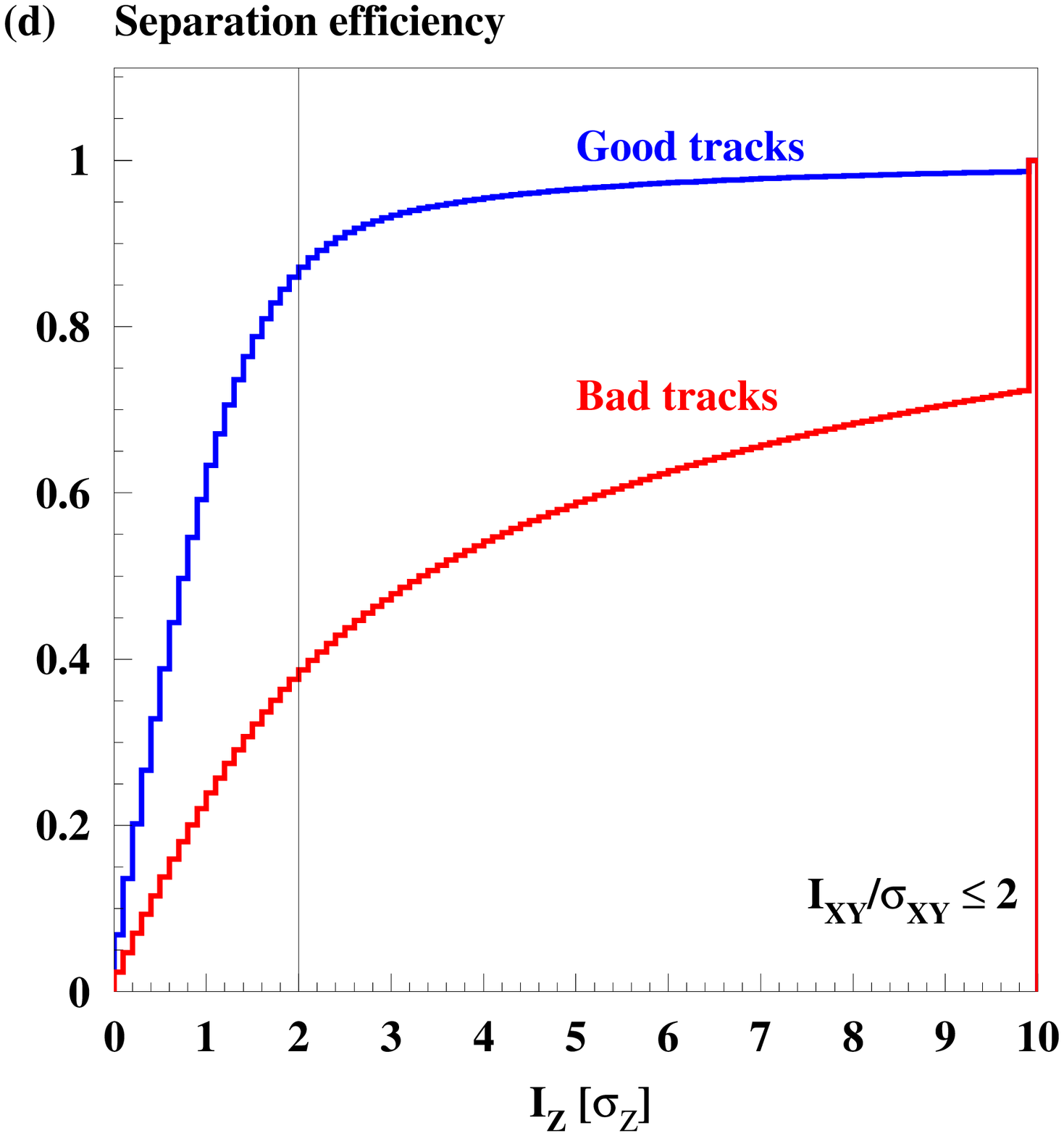}
\caption[bla]
{(\textit{a}): Simulated primary vertex distribution $PV^{sim}$ per event, projected onto the beam axis. (\textit{b}): Reconstructed primary vertex distribution $PV^{reco}$ per event projected onto the beam axis, using the measurements from the vertex detector. Plot shows the difference between $PV^{reco}$ per event if the pileup is taken into account (red) and without (yellow). (\textit{c}): Normalized $I_{Z}^{reco}$ of good (blue) and bad (red) tracks in event as a deviation from $PV^{reco}$. (\textit{d}): Separation efficiency for signal (good tracks, blue) and pileup (bad tracks, red) tracks for $I_{XY}\leq 2\sigma_{XY}$.}
\label{fig:bspot}
\end{center}
\end{figure}
\par
The beamspot length of 300 $\mu$m, foreseen for TESLA, is simulated and shown in Fig.~\ref{fig:bspot}\,$a$, representing the Gaussian distribution of primary vertices per event along the \textit{z}-axis, $PV^{sim}$. This distribution is used as a reference to check the 'quality' of reconstructed vertices using the information about track positions and momenta from the detector. Using the precise measurements from the vertex detector, first the primary vertex of an event is reconstructed as the momentum weighted average $z$-impact parameter, $PV^{reco}$ of all tracks in the event. $PV^{reco}$ per event shown in Fig.~\ref{fig:bspot}\,$b$, is in good agreement with $PV^{sim}$. The deviation from $PV^{reco}$ per event differs if there are pileup tracks leading to a broader distribution than if the events do not contain the pileup tracks, as it is shown in Fig.~\ref{fig:bspot}\,$b$. Using $PV^{reco}$, the impact parameters $I_{Z}^{reco}$ of all tracks are recalculated. Having in mind that the tracks can originate from secondary vertices too, only the tracks with transversal impact parameter $I_{XY}$ normalized to its error $\sigma_{XY}$ fullfilling $|I_{XY}/\sigma_{XY}|\leq$ 2 are assumed to originate from the primary vertex. That value came out as a result of the optimization that assumed the rejection of pileup tracks as much as possible while retaining many signal tracks. Their distribution in $I_{Z}^{reco}$ normalized to its error $\sigma_{Z}$, is shown in Fig.~\ref{fig:bspot}\,$c$. The high Gaussian distribution belongs to the signal tracks while the lower spreaded distribution comes from the pileup tracks. Selecting the tracks with $|I_{Z}^{reco}|\leq 2\sigma_{Z}$ a large fraction of pileup tracks can be rejected. The separation efficiency is shown in Fig.~\ref{fig:bspot}\,$d$; with this cut about $\sim 60-65\,{\%}$ of the pileup tracks and only $\sim 10-15\,{\%}$ of the signal tracks are rejected, for all considered $\gamma e$ and $\gamma\gamma$ modes. All tracks with $|I_{XY}/\sigma_{XY}|>$ 2 are accepted since they could originate from a secondary vertex of a signal. All neutral particles are accepted since information about them can not be obtained from the vertex detector.
\begin{figure}[p]
\begin{center}
\epsfxsize=3.0in
\epsfysize=3.0in
\epsfbox{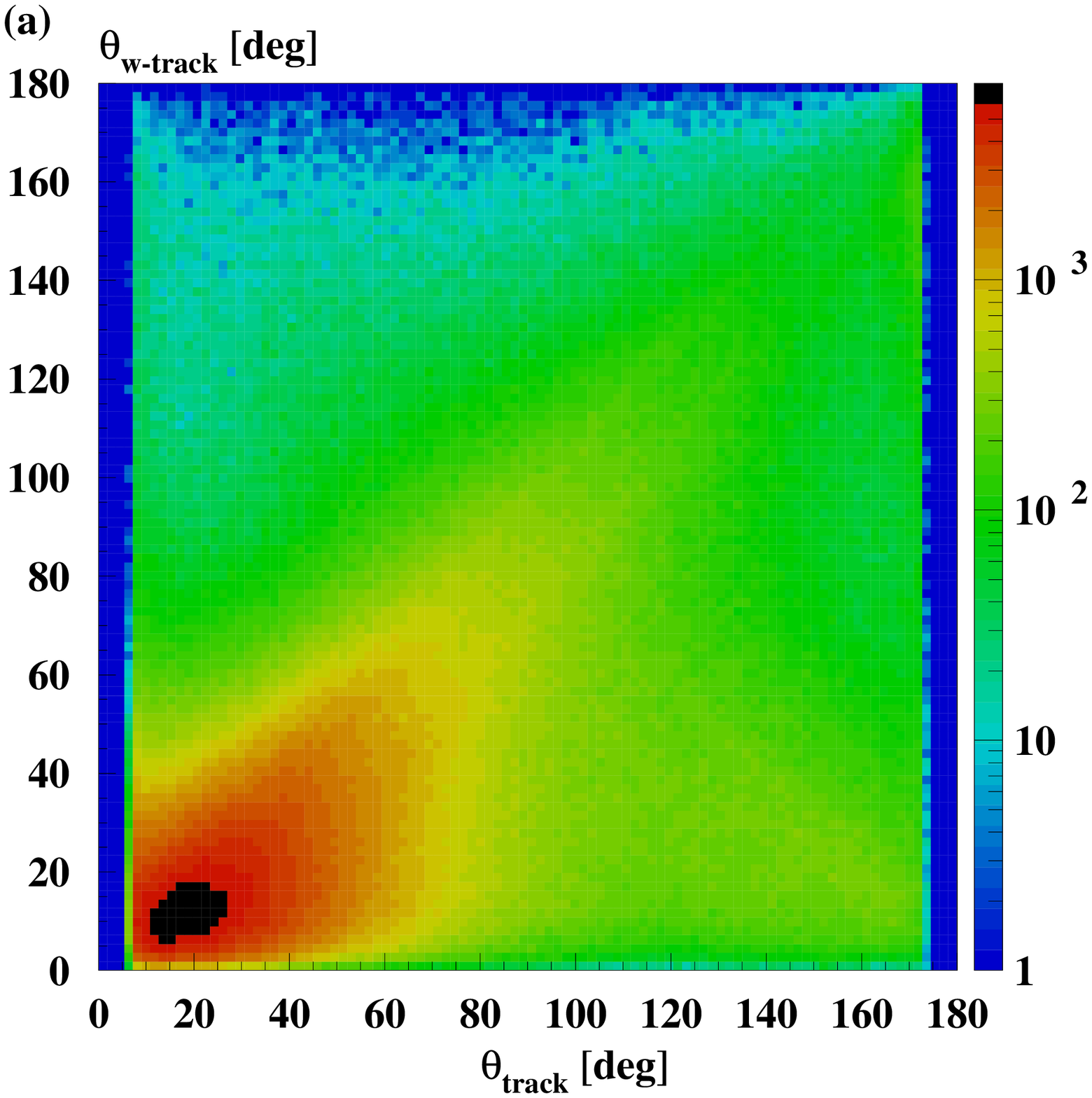}
\epsfxsize=3.0in
\epsfysize=3.0in
\epsfbox{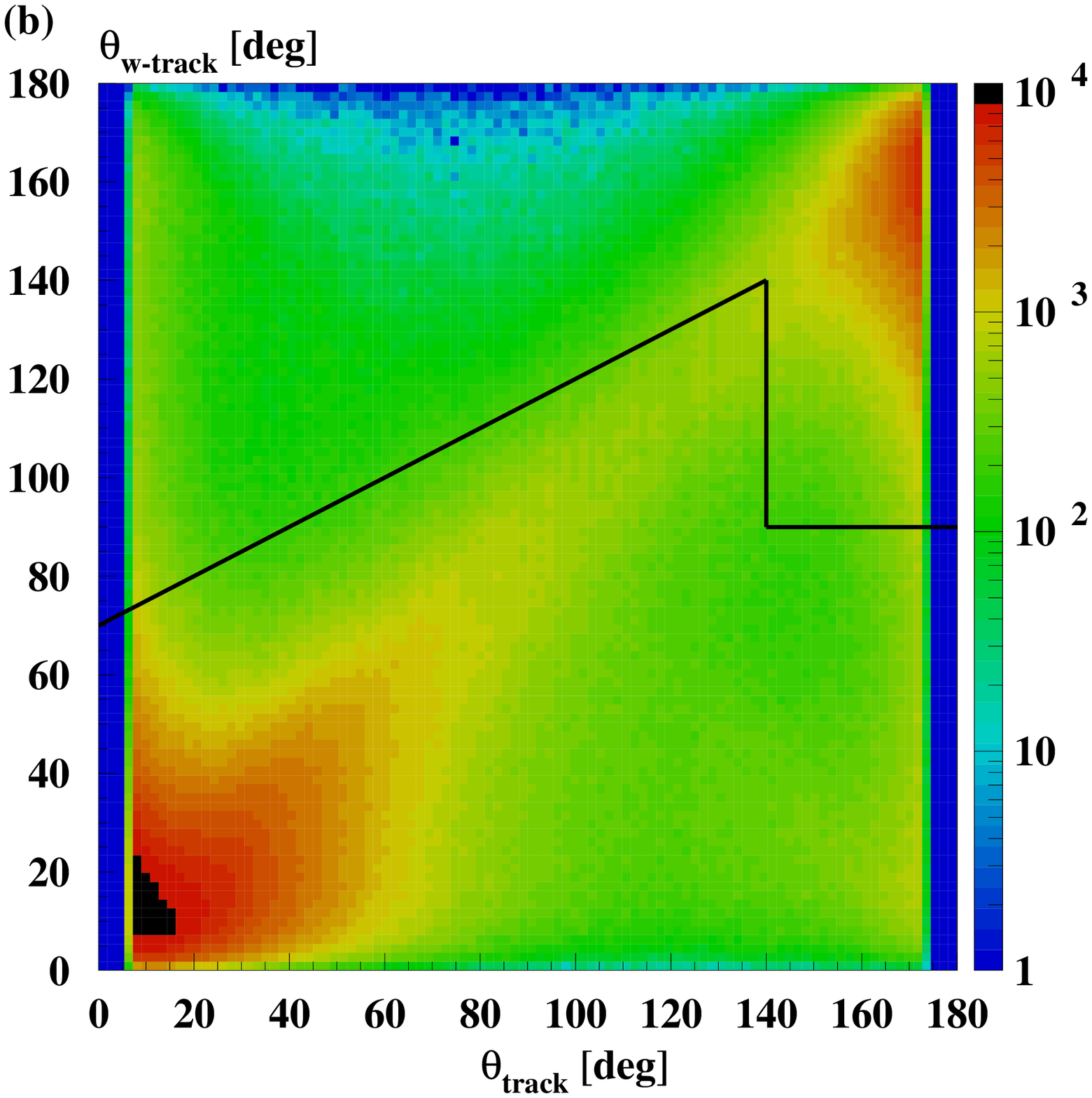}
\epsfxsize=3.0in
\epsfysize=3.0in
\epsfbox{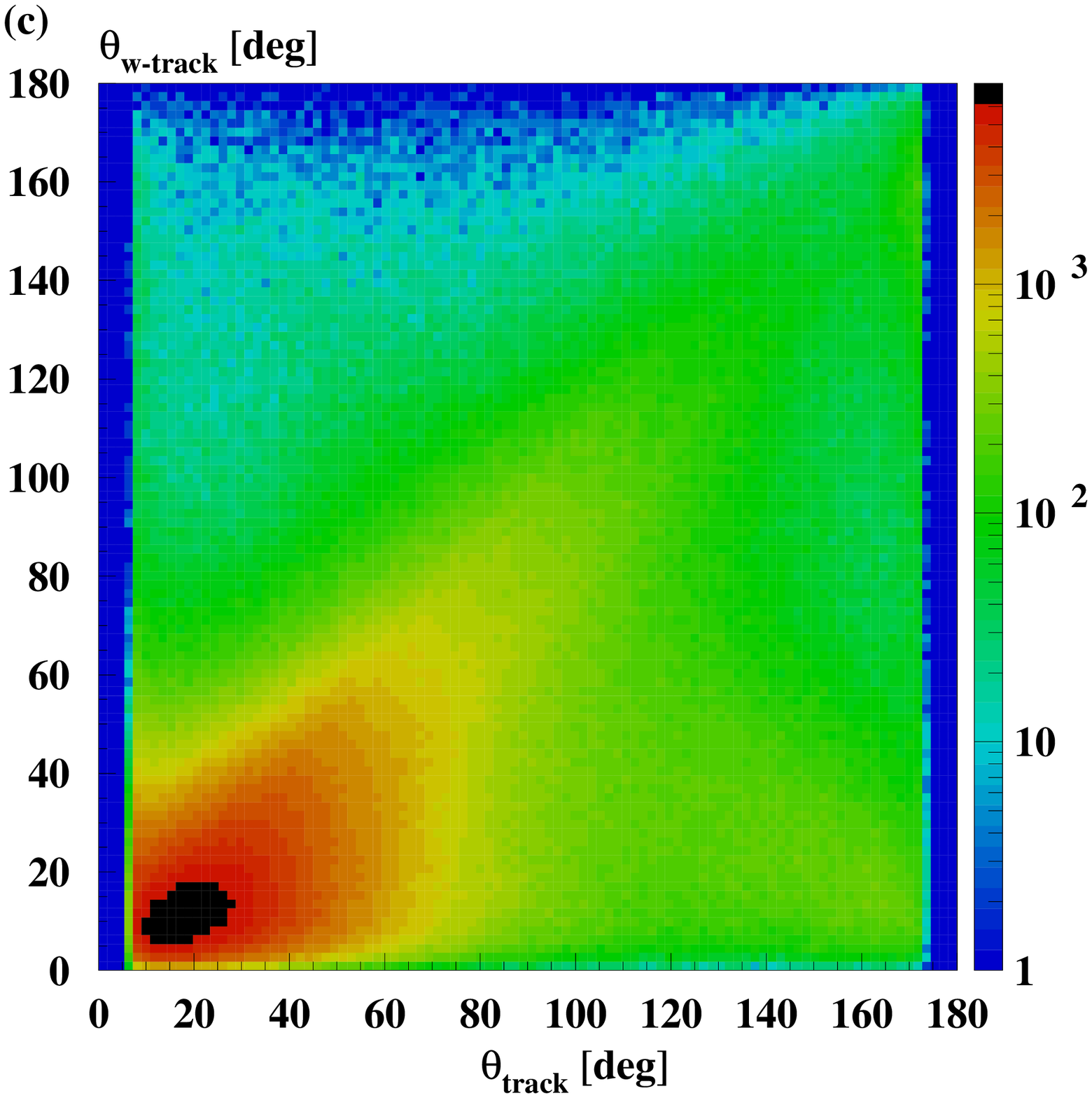}
\epsfxsize=3.0in
\epsfysize=3.0in
\epsfbox{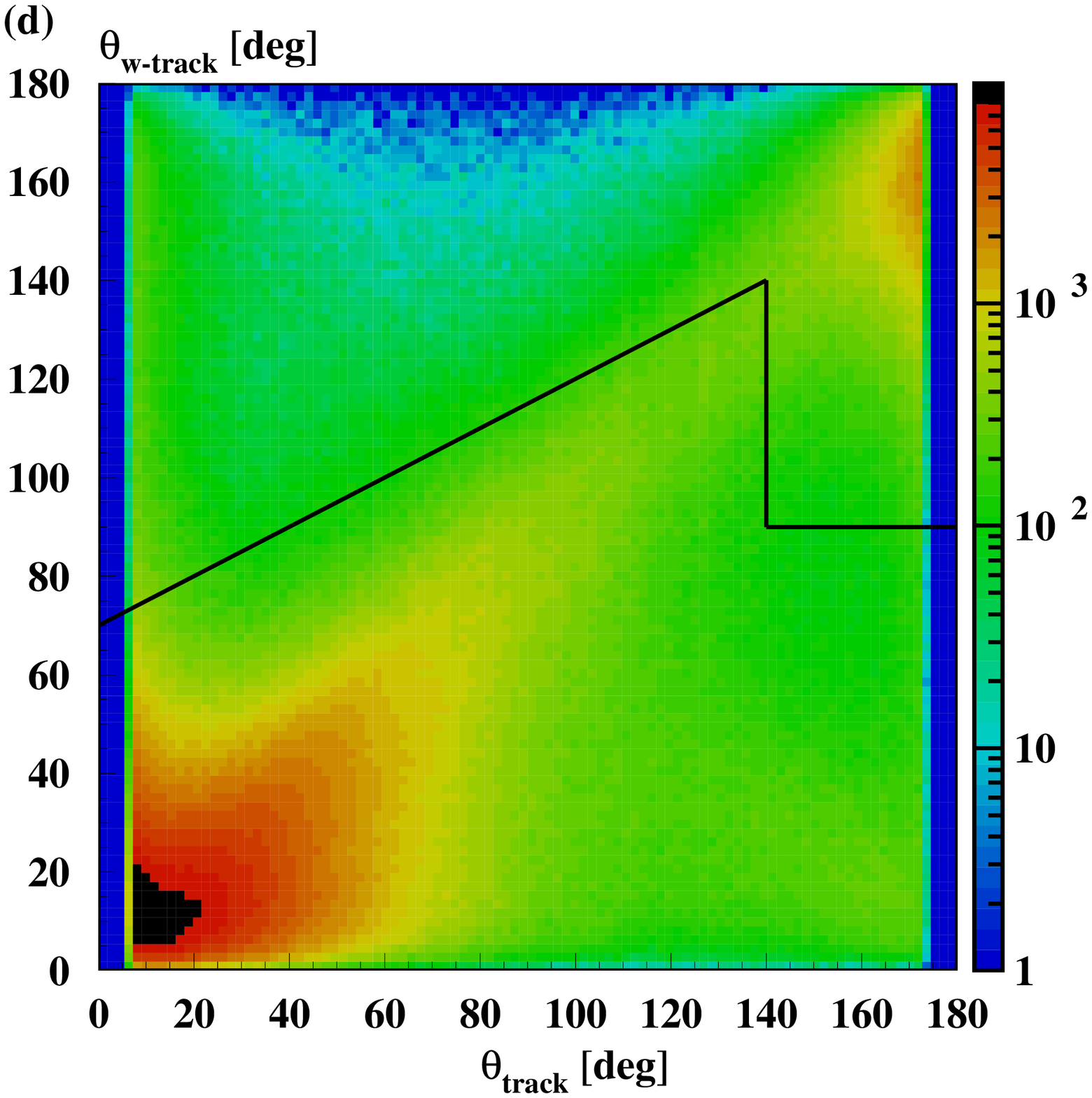}
\caption[bla]{Angle of the track/neutral particle with the beam axis versus their angle with the reconstructed $W$ direction (\textit{a}): for signal only in the parasitic $\gamma e$ mode, (\textit{b}): for signal plus pileup in the parasitic $\gamma e$ mode (\textit{c}): for signal only in the real $\gamma e$ mode, (\textit{d}): for signal plus pileup in the real $\gamma e$ mode. The tracks/neutral particles in the region above the line shown in (\textit{b}) and (\textit{d}) are rejected in the analysis.}
\label{fig:tracksge}
\end{center}
\end{figure}
\par
Concerning the $\gamma e$ modes, a further pileup rejection is based on reconstruction of the angle of each track/neutral (energy flow objects, EFOs) particle with respect to the \textit{z}-axis and the angle between the EFO and the flight direction of the reconstructed $W$ boson shown in Fig.~\ref{fig:tracksge}. Rejecting the EFOs positioned in the area shown in ($b$) for the parasitic $\gamma e$ mode and ($d$) for the real $\gamma e$ mode it is possible to decrease a part of the pileup contribution to the signal events. These are basically, the EFOs with a small angle related to the beam axis and large angle related to the $W$ boson direction. Those which are close to the $W$ boson flight direction can not be separated from signal tracks. Previously described pileup rejection in single $W$ production is not applicable for $W$ boson pair production since two $W$ bosons are in the opposite flight directions related to the beam axis. Consequently, the region which is excluded in the $\gamma e$ mode would be the region of other $W$ boson direction in the $\gamma\gamma$ mode.
\par
\begin{figure}[p]
\begin{center}
\epsfxsize=3.0in
\epsfysize=3.0in
\epsfbox{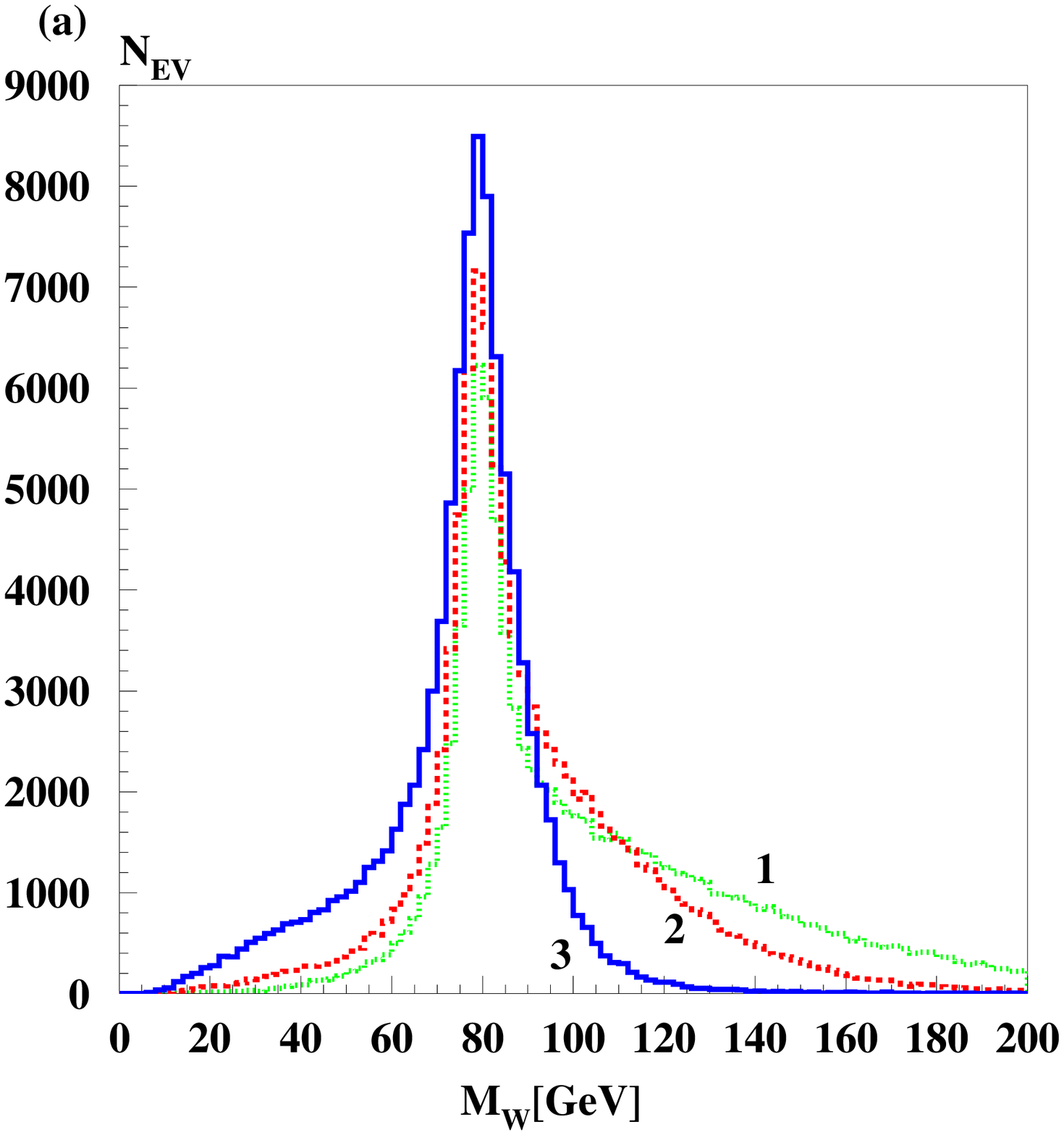}
\epsfxsize=3.0in
\epsfysize=3.0in
\epsfbox{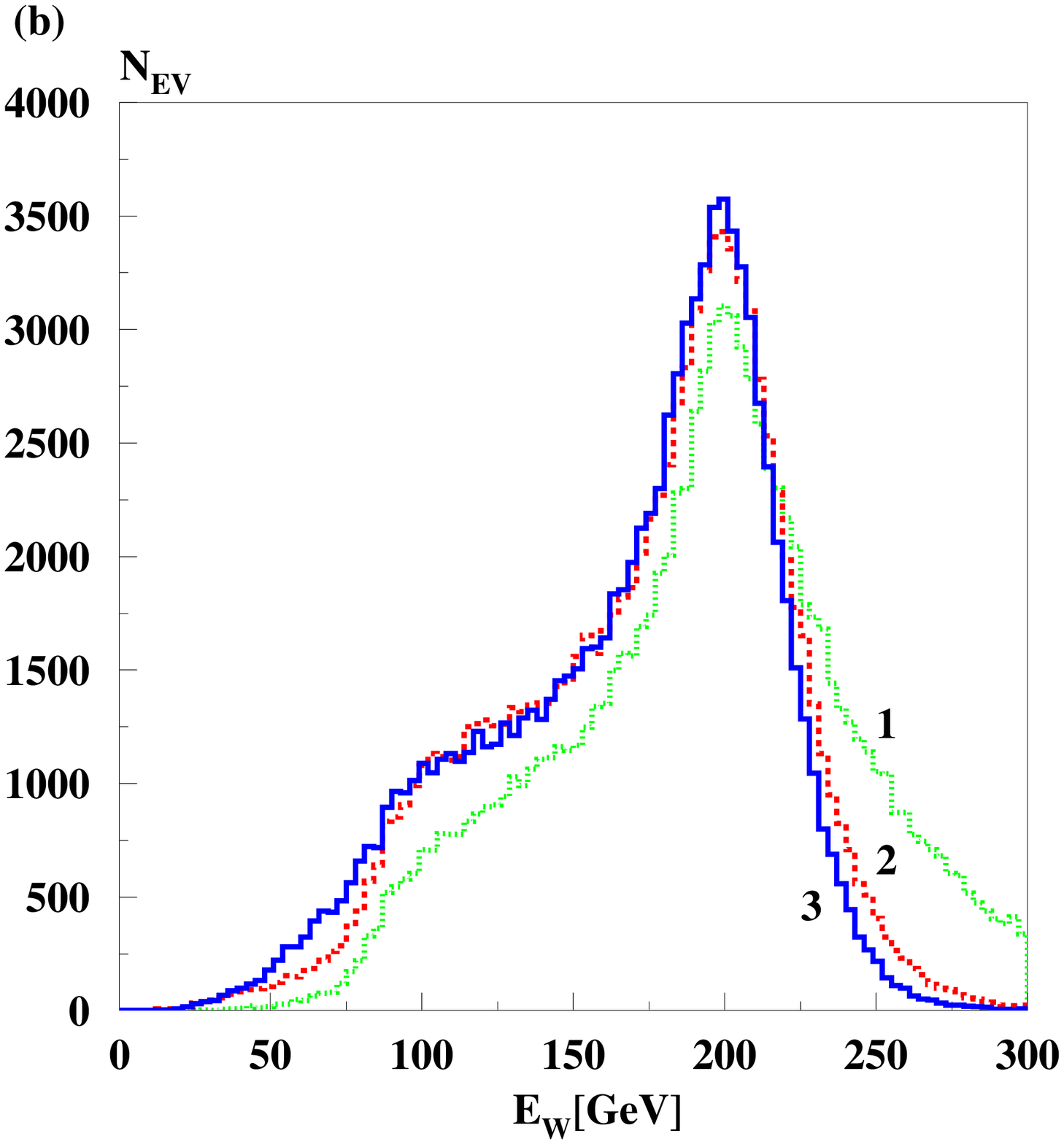}
\epsfxsize=3.0in
\epsfysize=3.0in
\epsfbox{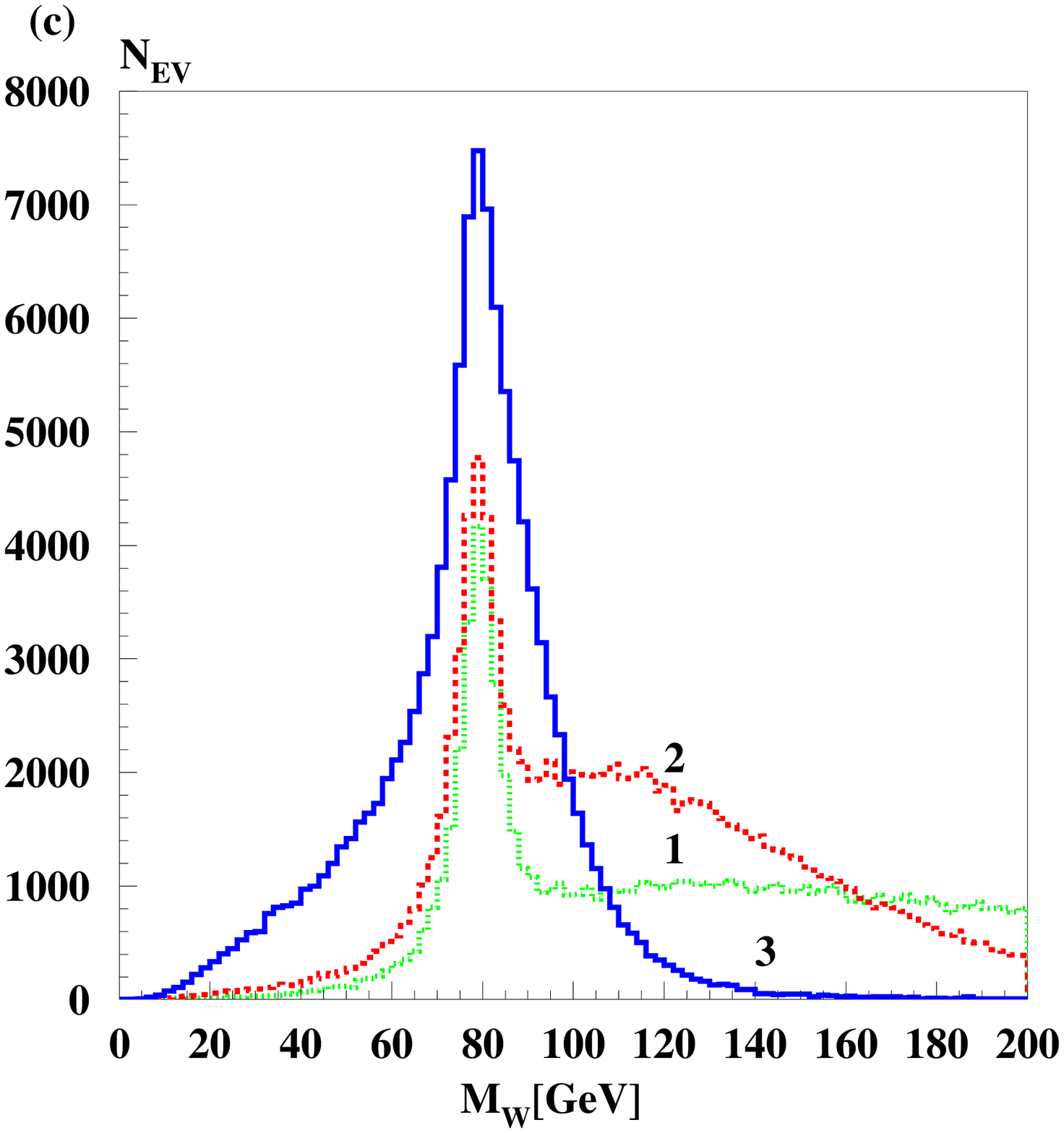}
\epsfxsize=3.0in
\epsfysize=3.0in
\epsfbox{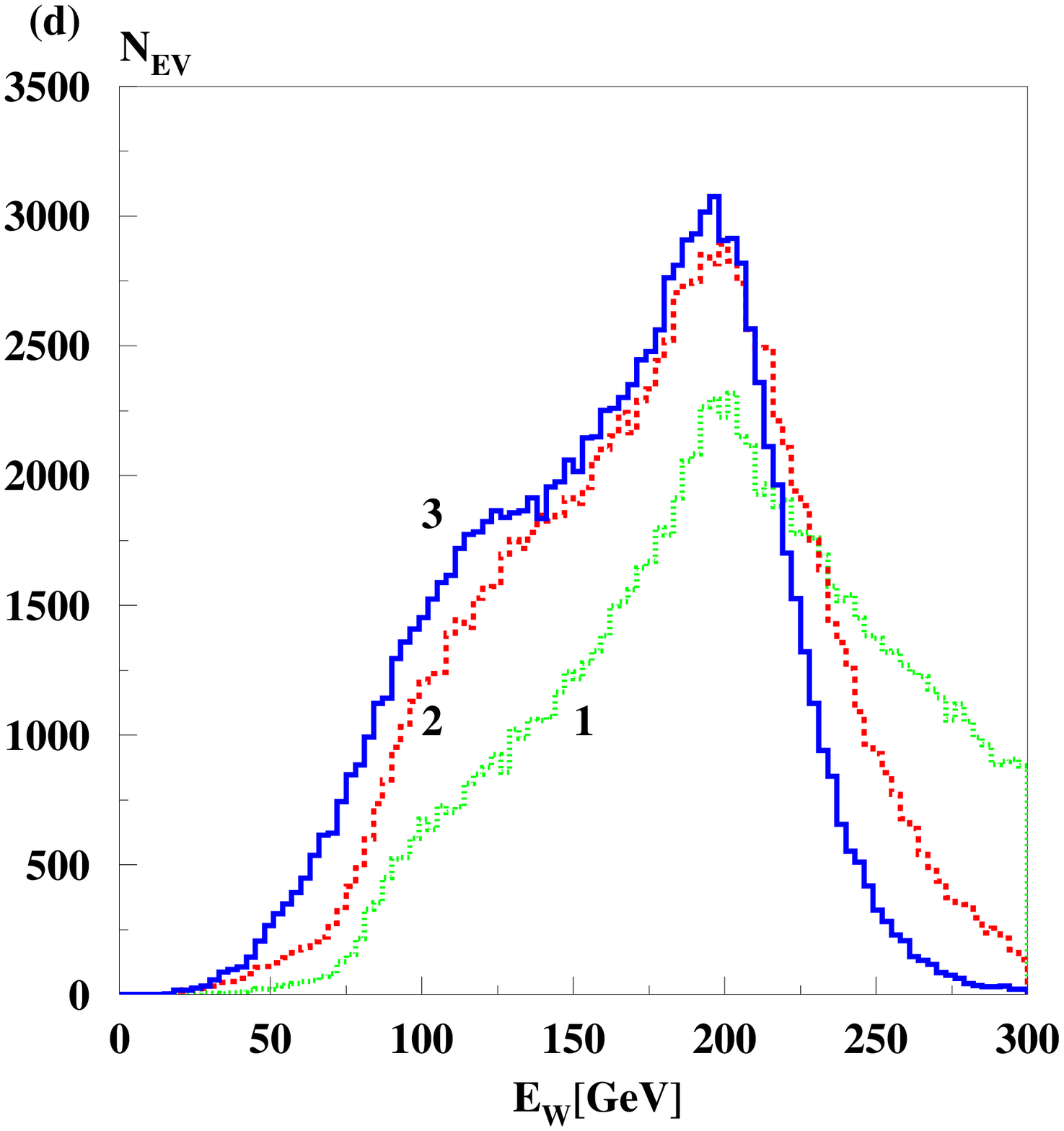}
\caption[bla]{(\textit{a}): Mass and (\textit{b}): energy distributions of the reconstructed $W$ boson for the real $\gamma e$ mode during the different steps in the track/neutral particle rejection. (\textit{c}): Mass and (\textit{d}): energy distributions of the reconstructed $W$ boson for the parasitic $\gamma e$ mode during the different steps in the track/neutral particle rejection. Initial shape (green-1) without any rejection, after the track rejection using \textit{$I_{Z}$} (red-2) and final shape (blue-3) after the track/neutral particle rejection shown in Fig.~\ref{fig:tracksge}\,$b,d$.}
\label{fig:comparison}
\end{center}
\end{figure}

The different steps during the separation procedure for the real and parasitic
$\gamma e$ mode are shown in Fig.~\ref{fig:comparison}. Adding the low energy pileup events, they contribute through the tail in the high mass and high energy region. 
\section{Event Selection}
After the pileup rejection, the remaining tracks of the event are collected and the analysis is done on the 'event level' applying several successive cuts. A cluster analysis i.e. determination of the number of jets present in the event and reconstruction of the corresponding jet axes is done using the Lund clustering algorithm \cite{algorithm} incorporated in the VECSUB package.
\par
\begin{figure}[p]
\begin{center}
\epsfxsize=6.5in
\epsfysize=6.5in
\epsfbox{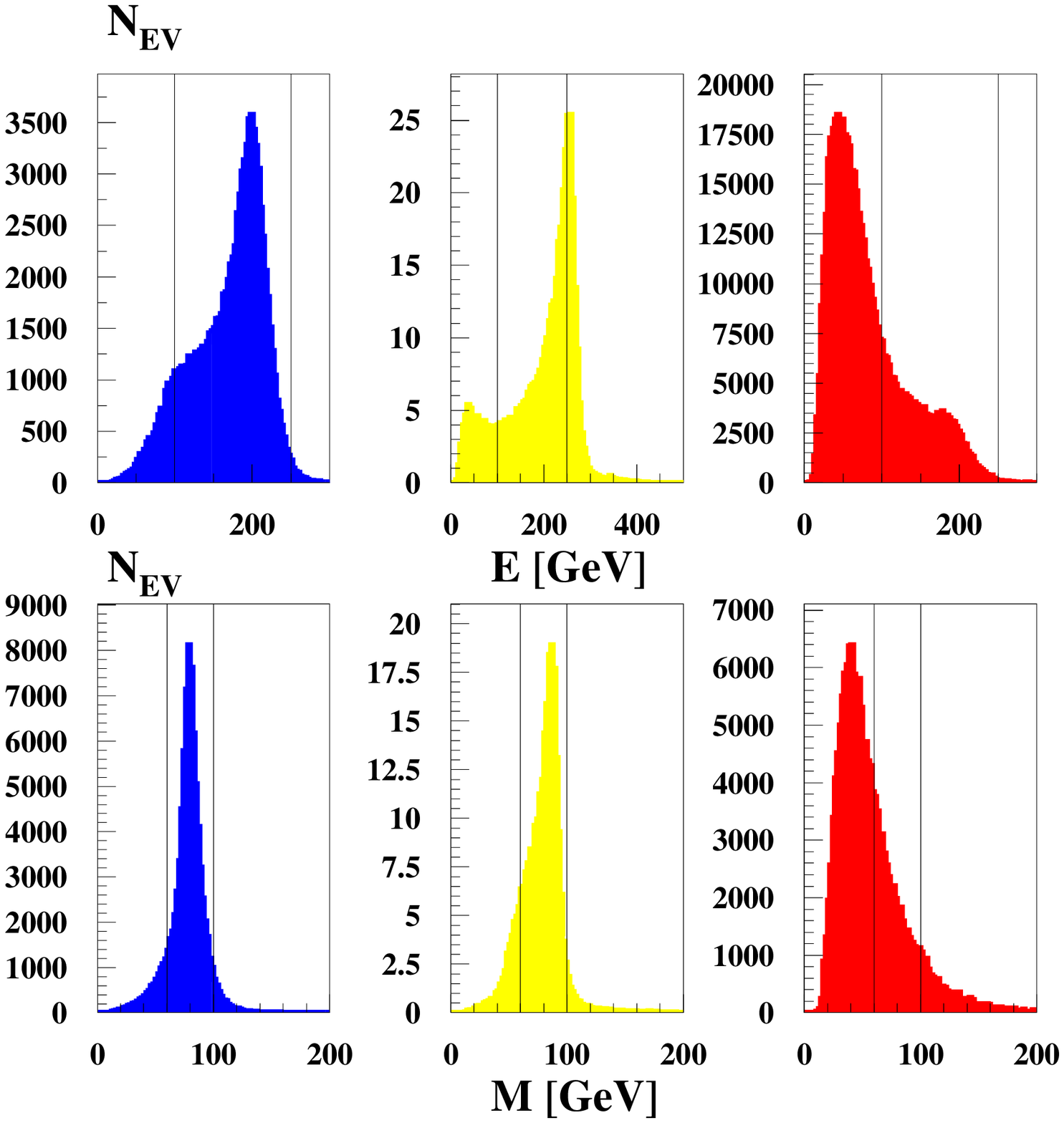}
\caption[bla]{Separation of the signal $\gamma e\rightarrow \nu_{e} W$ in the real $\gamma e$ mode (first column) from the background (second column ($\gamma e\rightarrow eZ$) and third column ($\gamma(e^{-})\gamma\rightarrow q\bar{q}$). First row: Energy spectrum of the hadronic final states - the energy cut selects the range between 100 GeV and 250 GeV. Second row: Mass spectrum of the hadronic final states - the mass cut selects the range between 60 GeV and 100 GeV. The ratio of background to signal events is $N_{B}/N_{S}\approx$ 5.4 before the cuts are applied, after the energy cut the ratio is $N_{B}/N_{S}\approx$ 1.8 and $N_{B}/N_{S}\approx$ 0.56 after the mass cut.}
\label{fig:cuts_real_ge}
\end{center}
\end{figure}
\begin{figure}[p]
\begin{center}
\epsfxsize=6.5in
\epsfysize=6.5in
\epsfbox{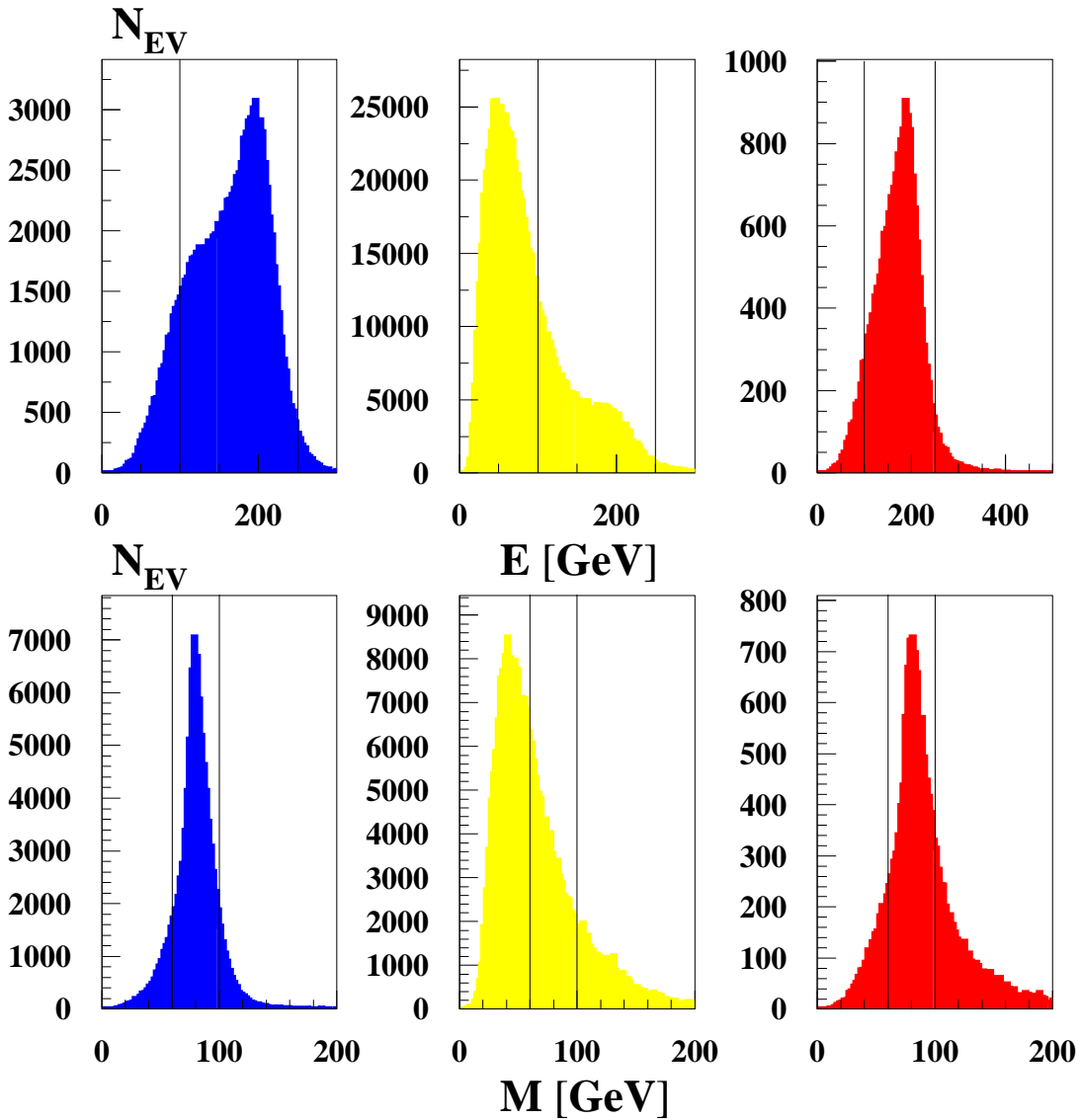}
\caption[bla]{Separation of the signal $\gamma e\rightarrow \nu_{e} W$ in the parasitic $\gamma e$ mode (first column) from the background (second column ($\gamma\gamma\rightarrow q\bar{q}$) and third column ($\gamma\gamma\rightarrow WW(q\bar{q}l\nu_{l})$) separation. First row: Energy spectrum of the hadronic final states - the energy cut selects the range between 100 GeV and 250 GeV. Second row: Mass spectrum of the hadronic final states - the mass cut selects the range between 60 GeV and 100 GeV. The ratio of background to signal events is $N_{B}/N_{S}\approx$ 6.9 before the cuts are applied, after the energy cut the ratio is $N_{B}/N_{S}\approx$ 2.6 and $N_{B}/N_{S}\approx$ 1.07 after the mass cut.}
\label{fig:cuts_parasitic_ge}
\end{center}
\end{figure}
\begin{figure}[htb]
\begin{center}
\epsfxsize=3.0in
\epsfysize=3.0in
\epsfbox{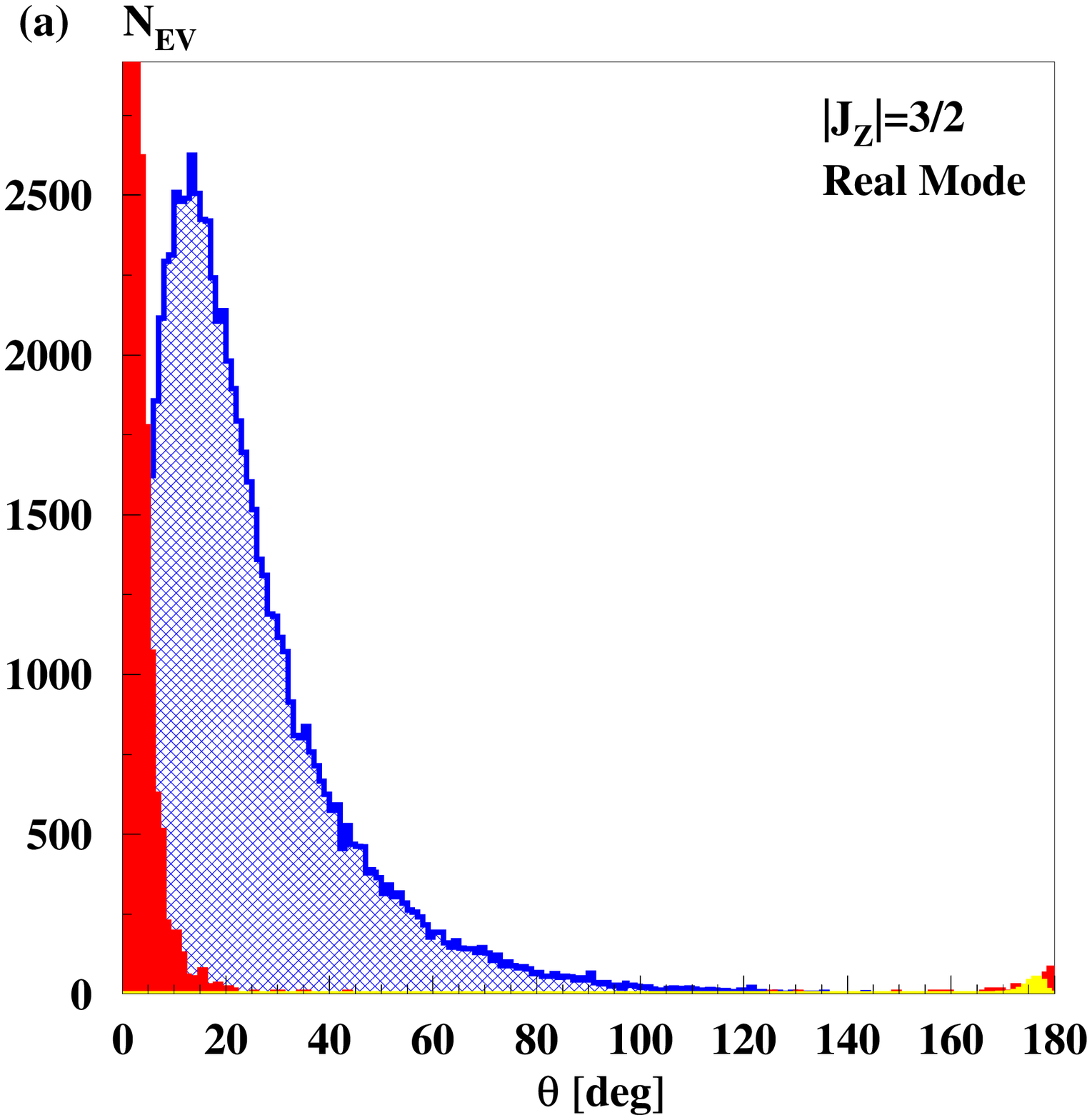}
\epsfxsize=3.0in
\epsfysize=3.0in
\epsfbox{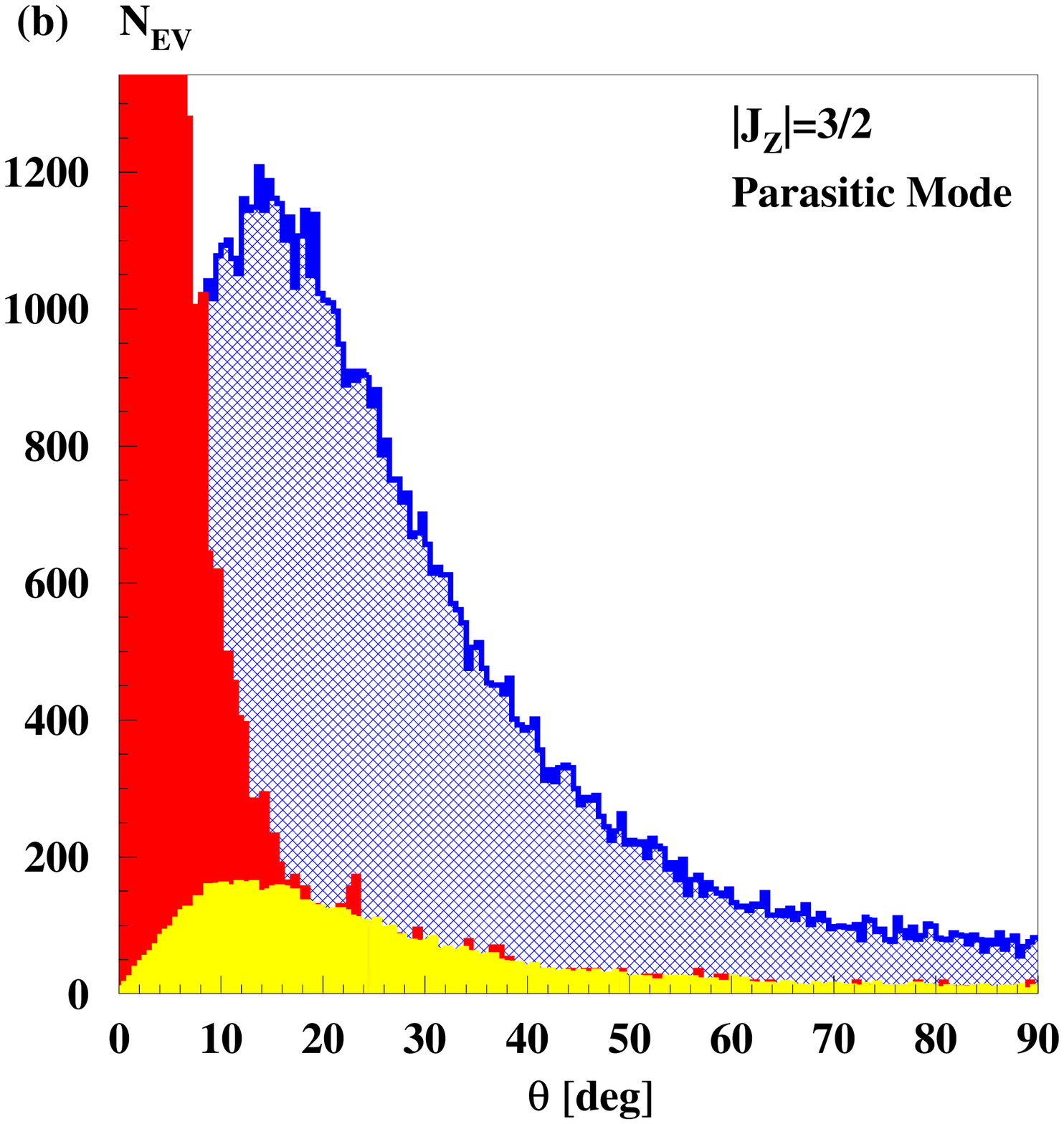}
\caption[bla]{Signal and background distributions for $\gamma e\rightarrow W\nu_{e}$ as a function of the $W$ production angle. The different processes are normalized to the same luminosity. The blue area represents the signal. (\textit{a}): The real $\gamma e$ mode. The red contribution corresponds to ${\gamma}(e^-){\gamma}{\rightarrow}{q}{\bar{q}}$ processes and the yellow one corresponds to $\gamma e\rightarrow eZ$. (\textit{b}): The parasitic $\gamma e$ mode. The yellow contribution corresponds to ${\gamma}{\gamma}{\rightarrow}{W}{W}$ while the red one corresponds to ${\gamma}{\gamma}{\rightarrow}{q}{\bar{q}}$ processes.}
\label{fig:compare_ge}
\end{center}
\end{figure}

In the single $W$ boson production channel (two-jet events), the background is rejected applying the same cuts for the real and parasitic $\gamma e$ mode. In order to separate the signal events from the background the events with a number of EFOs larger than 10 and number of charged tracks larger than 5 are accepted only. In addition to the vetos on high energy and isolated leptons cuts on two reconstructed variables, the energy ($100\,{\rm GeV} - 250\,{\rm GeV}$) and the mass ($60\,{\rm GeV} - 100\,{\rm GeV}$) of the reconstructed $W$ boson are applied and shown in Fig.~\ref{fig:cuts_real_ge} and Fig.~\ref{fig:cuts_parasitic_ge}. In the real $\gamma e$ mode the ratio of background to signal events is $N_{B}/N_{S}\approx$ 5.4 before the cuts are applied. After the energy cut the ratio is $N_{B}/N_{S}\approx$ 1.8 and $N_{B}/N_{S}\approx$ 0.56 after the mass cut. In the parasitic $\gamma e$ mode the ratio of background to signal events is $N_{B}/N_{S}\approx$ 6.9 before the cuts are applied. After the energy cut the ratio is $N_{B}/N_{S}\approx$ 2.6 and $N_{B}/N_{S}\approx$ 1.07 after the mass cut. The final angular distributions of signal and background events for both $\gamma e$ modes are shown in Fig.~\ref{fig:compare_ge}.
The efficiency obtained for the real mode is $73\,{\%}$ with a purity of $64\,{\%}$. In the parasitic mode, due to the fact that the pileup is larger than in the case of the real mode, the efficiency is $66\,{\%}$ with a purity of $49\,{\%}$. Background events are mostly distributed close to the beam pipe and an additional cut on the $W$ production angle is applied in order to increase the purity of the signal in both modes. Events in the region below $5^{\circ}$ are rejected leading to a purity of $95\,{\%}$ for the real mode and $72\,{\%}$ for the parasitic mode. This cut has only a small influence on the signal resulting in efficiencies of $70\,{\%}$ and $63\,{\%}$ for the real and parasitic mode, respectively.
\par
\begin{figure}[htb]
\begin{center}
\epsfxsize=3.0in
\epsfysize=3.0in
\epsfbox{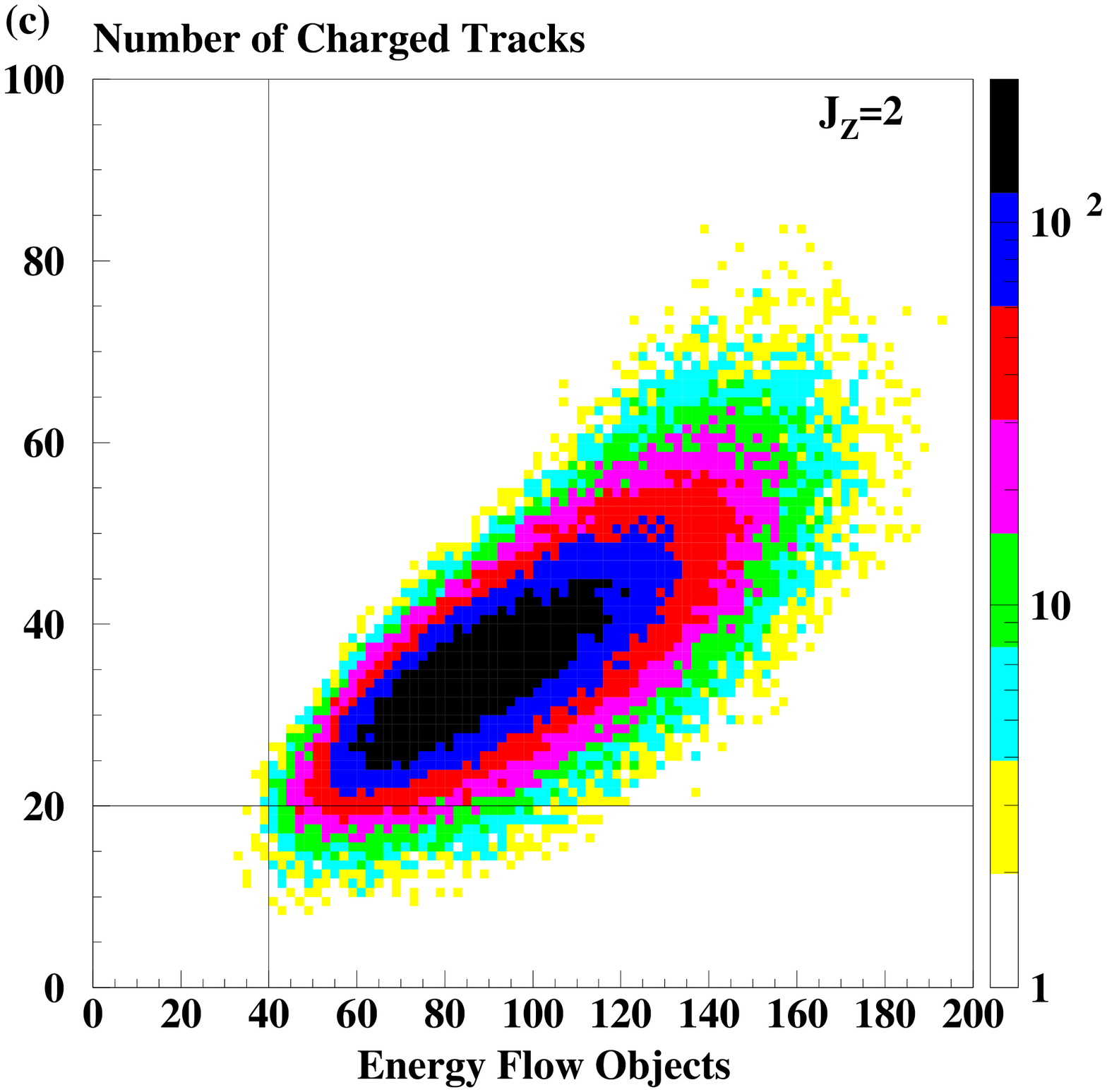}
\epsfxsize=3.0in
\epsfysize=3.0in
\epsfbox{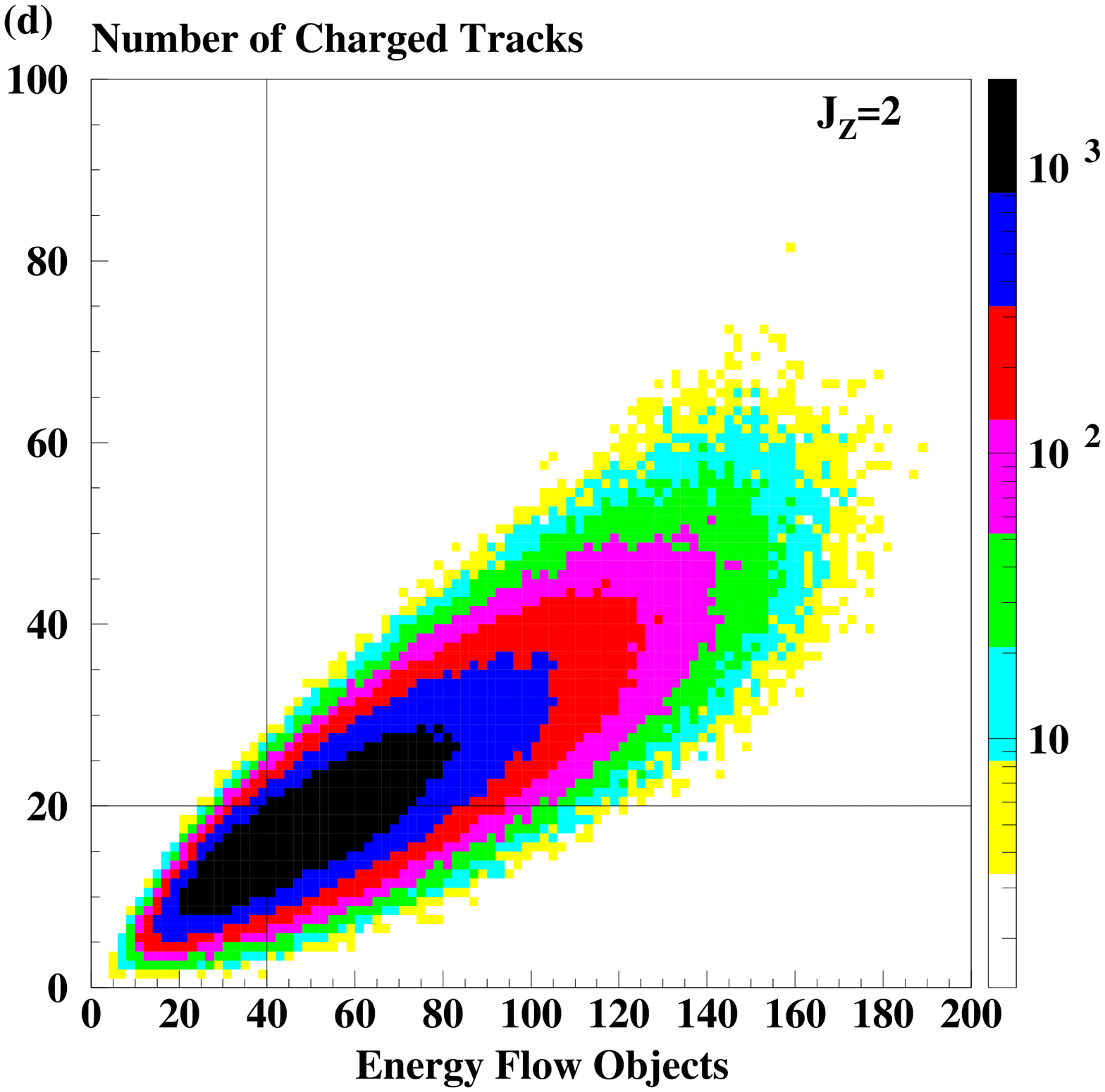}
\caption[bla]{Number of EFOs ($x$-axis) and tracks ($y$-axis) per event in $|J_{Z}|=2$ state for (\textit{a}): signal and (\textit{b}): background. Events with number of EFOs $<$ 40 and tracks $<$ 20 are rejected.}
\label{fig:nct_efo_gg}
\end{center}
\end{figure}
\begin{figure}[htb]
\begin{center}
\epsfxsize=3.0in
\epsfysize=3.0in
\epsfbox{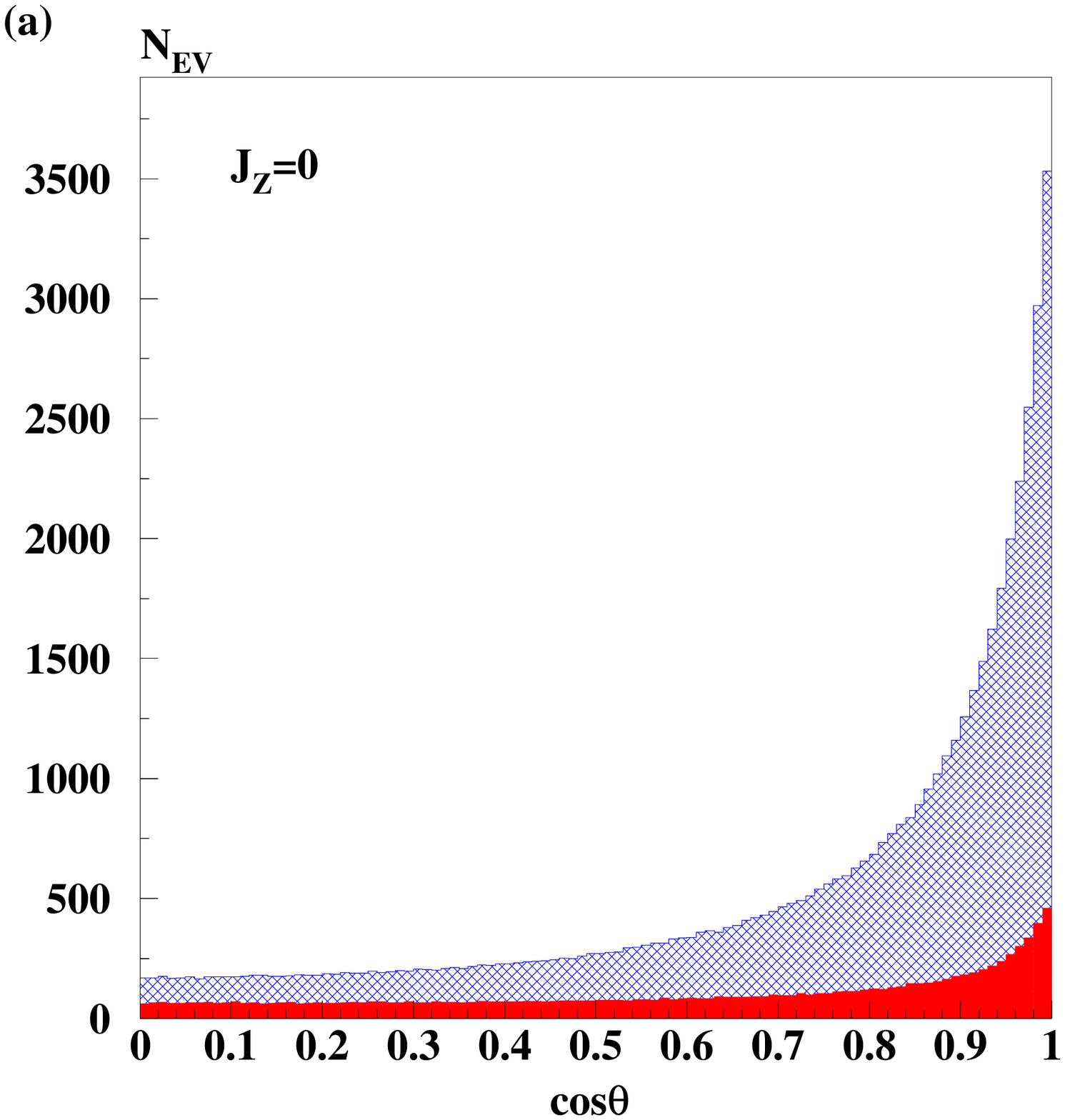}
\epsfxsize=3.0in
\epsfysize=3.0in
\epsfbox{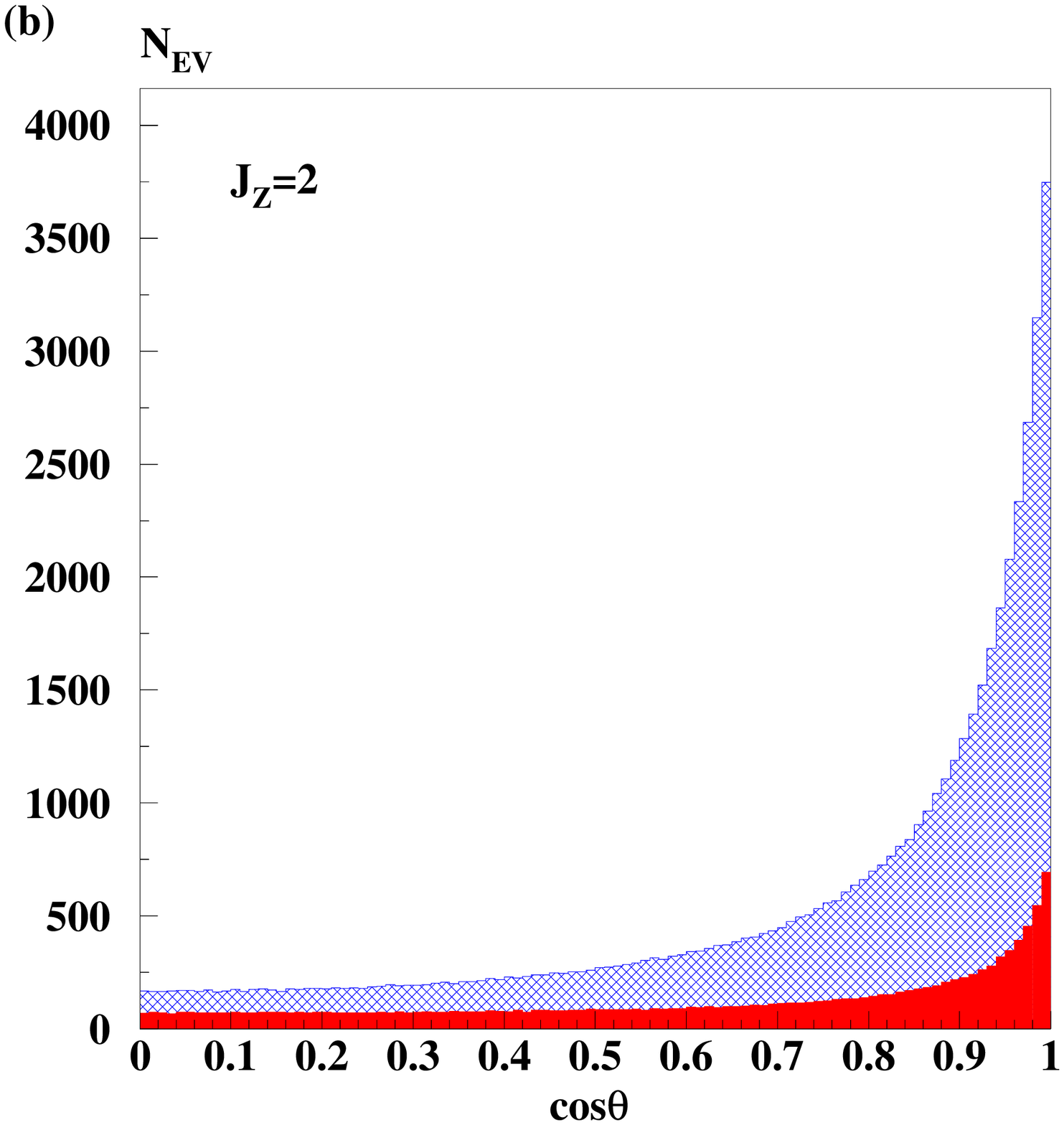}
\caption[bla]{Signal (blue) and background (red) event distributions as a function of the $W$ production angle in (\textit{a}): $J_{Z}=0$ state (\textit{a}) and (\textit{b}): $|J_{Z}|=2$ state. The different processes are normalized to the same luminosity.}
\label{fig:angles_gg}
\end{center}
\end{figure}

Concerning the $W$ boson pair production (four-jet events), after the rejection of the pileup tracks, the same cuts for background event rejection are applied for both $J_{Z}$ states. Four-jet events are characterized by a high multiplicity so that this feature is used as a first selection criterium to reject the background events. The events are accepted if their number of EFOs is larger than 40 and the number of charged tracks larger than 20. The two-dimensional distribution of the number of EFOs and tracks per event for the $|J_{Z}|=2$ state is shown in Fig.~\ref{fig:nct_efo_gg}\,$a$ for signal and in Fig.~\ref{fig:nct_efo_gg}\,$b$ for background.
\begin{figure}[htb]
\begin{center}
\epsfxsize=2.in
\epsfysize=2.in
\epsfbox{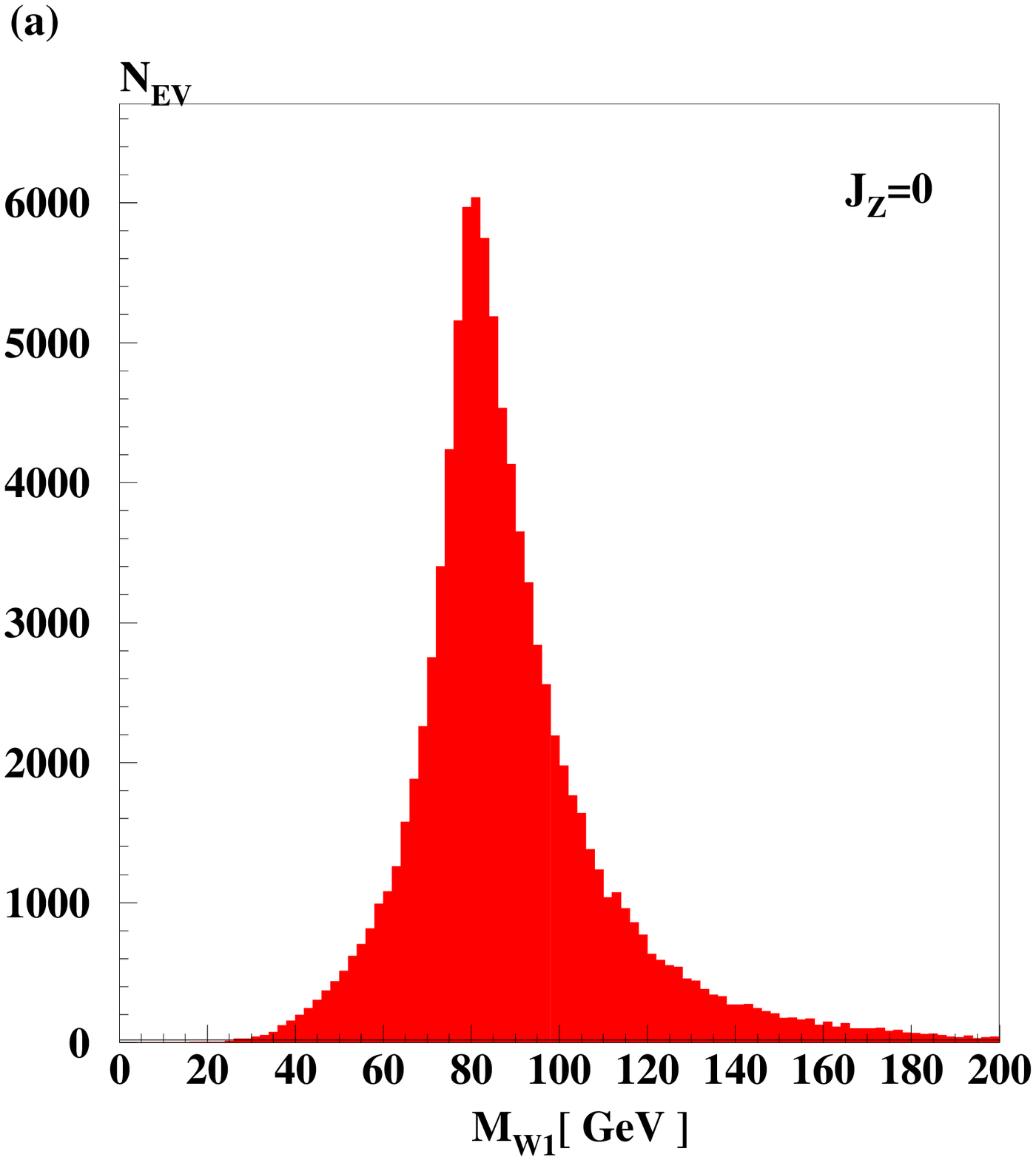}
\epsfxsize=2.in
\epsfysize=2.in
\epsfbox{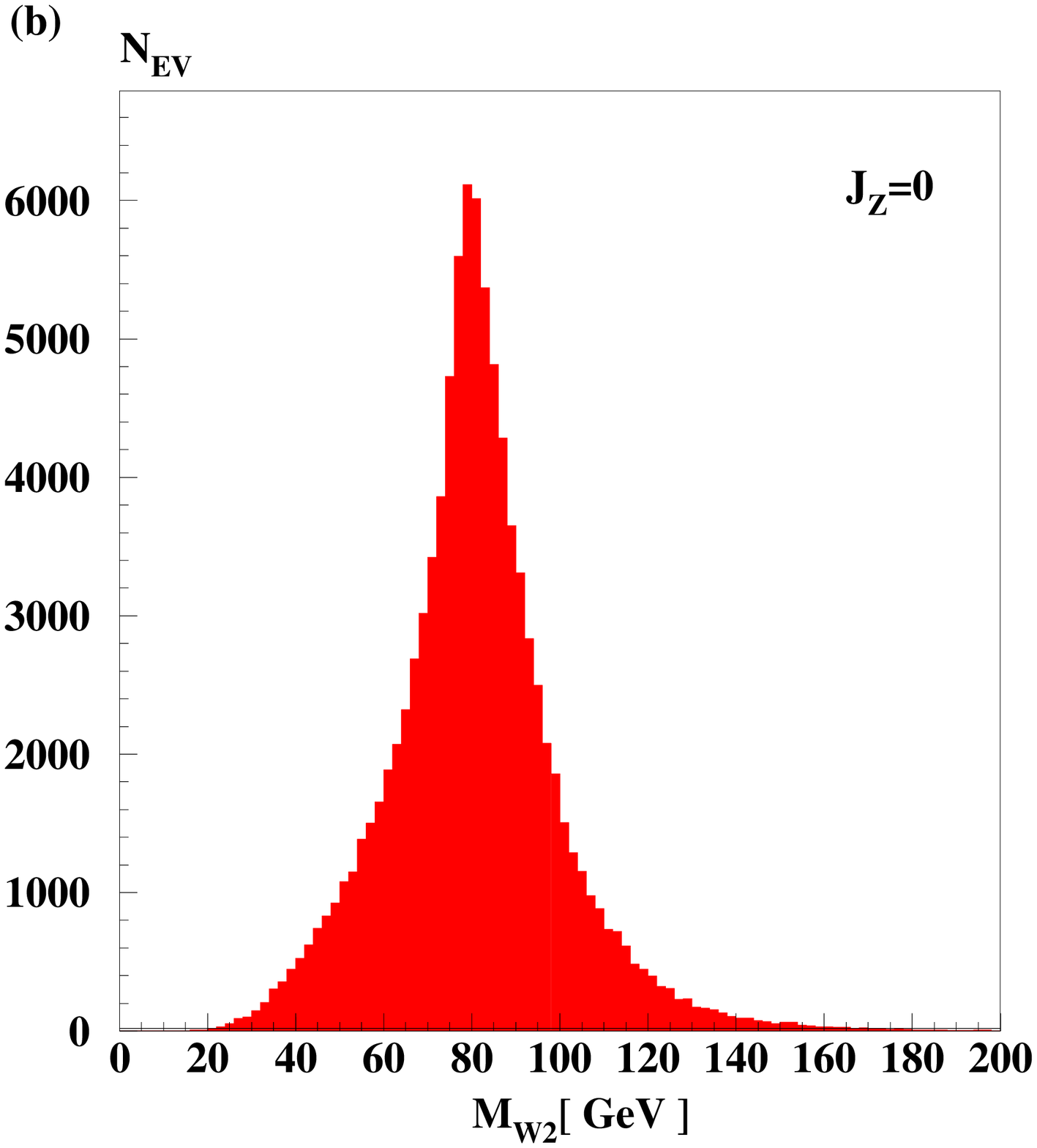}
\epsfxsize=2.in
\epsfysize=2.in
\epsfbox{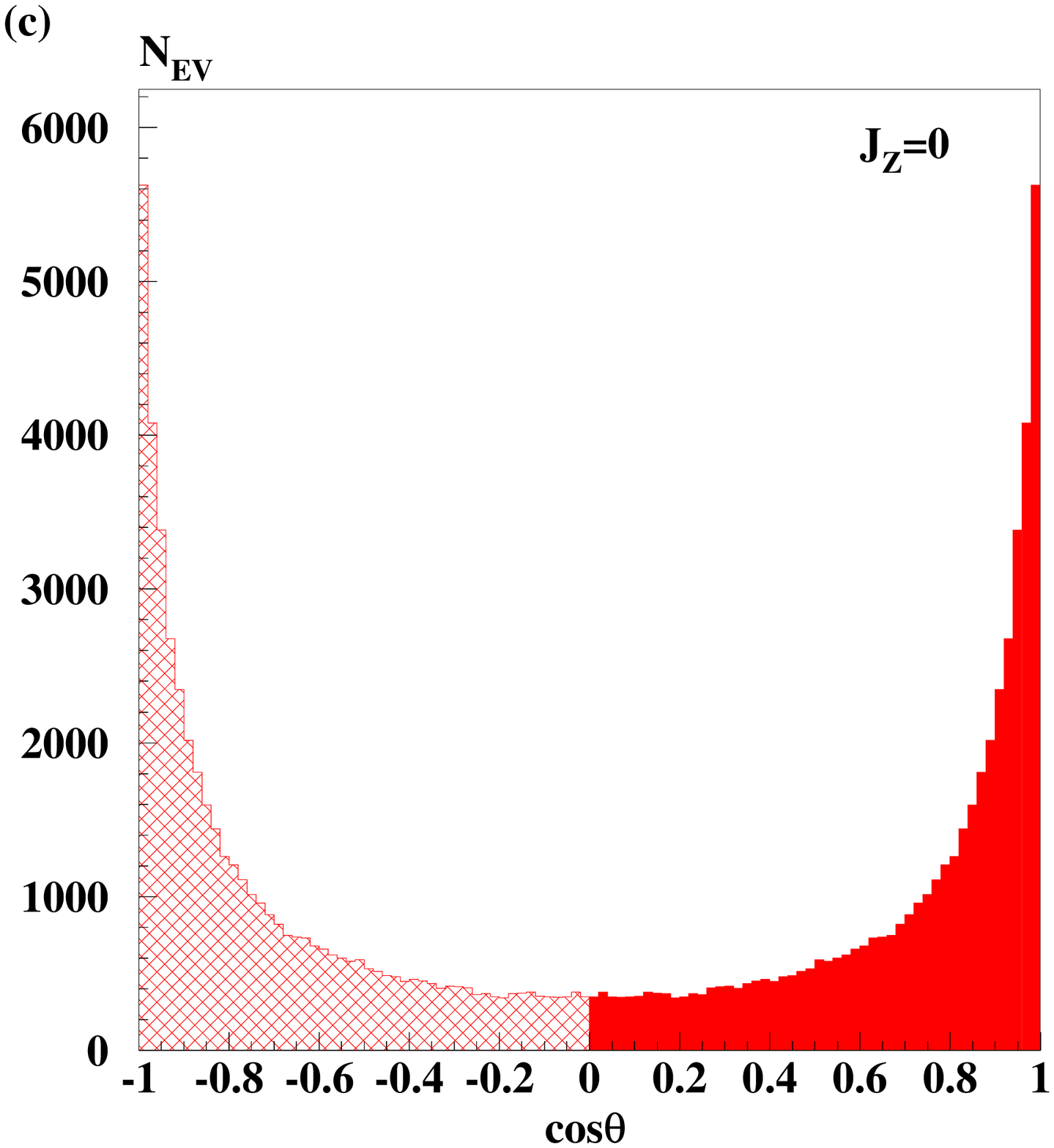}
\caption[bla]{(\textit{a-b}): Reconstructed $W$ boson masses in $J_{Z}=0$ state after the jet pairing. (\textit{c}): Reconstructed angular distributions for forward (full colored) and backward (hatched colored) $W$ boson in $J_{Z}=0$ state.}
\label{fig:pairing}
\end{center}
\end{figure}
\begin{figure}[htb]
\begin{center}
\epsfxsize=3.0in
\epsfysize=3.0in
\epsfbox{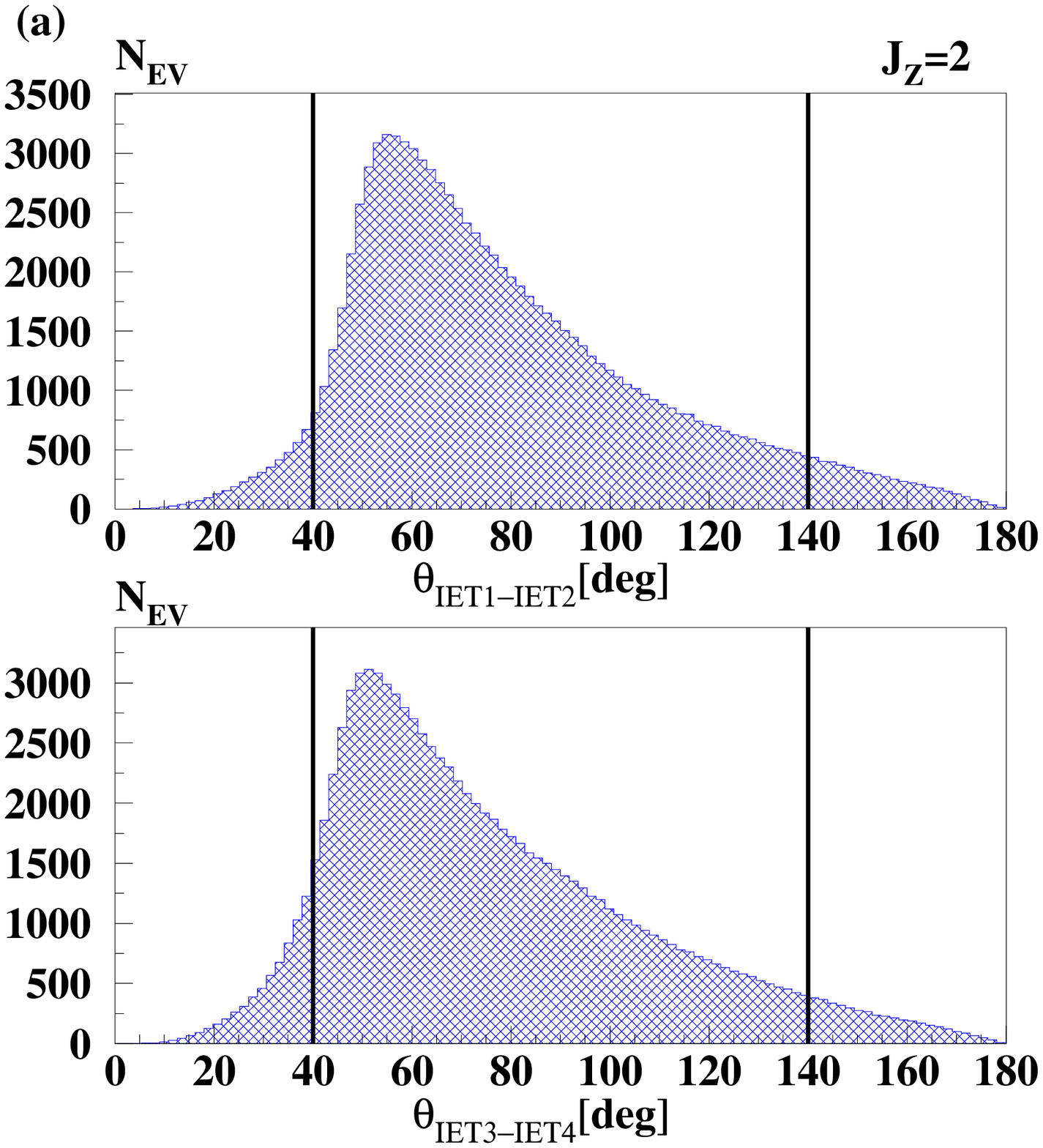}
\epsfxsize=3.0in
\epsfysize=3.0in
\epsfbox{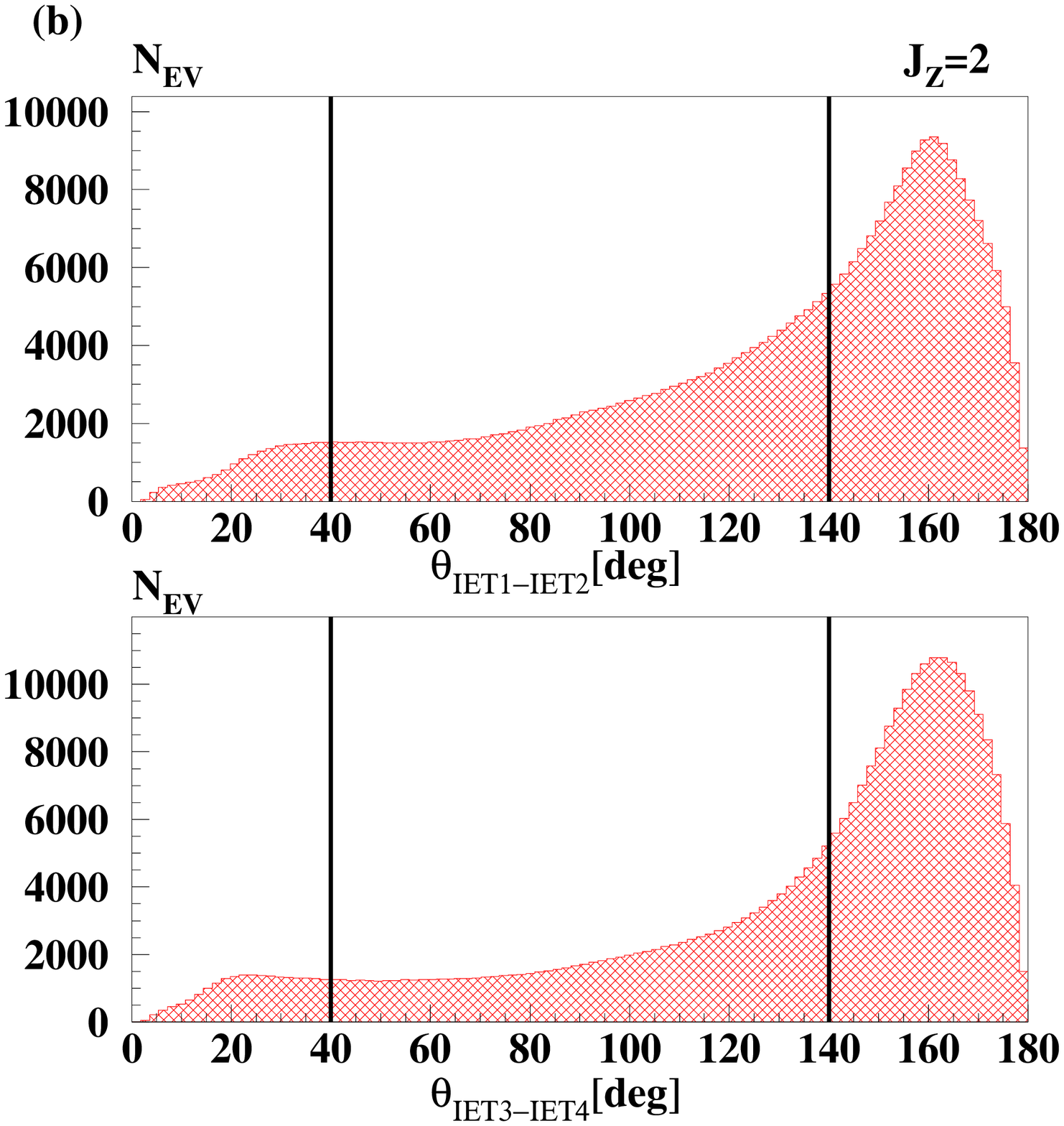}
\caption[bla]{Angles between the two jets within a $W_{F,B}$ boson in the $|J_{Z}|=2$ state for (\textit{a}): signal and (\textit{b}): background. Events out of the range $40^{\circ}<\theta_{jet1,3-jet2,4}<140^{\circ}$ are rejected leading to an improvement of the ratio $N_{B}/N_{S}\approx$ 3.3 to 0.85.}
\label{fig:coswg}
\end{center}
\end{figure}
\begin{figure}[htb]
\begin{center}
\epsfxsize=3.0in
\epsfysize=3.0in
\epsfbox{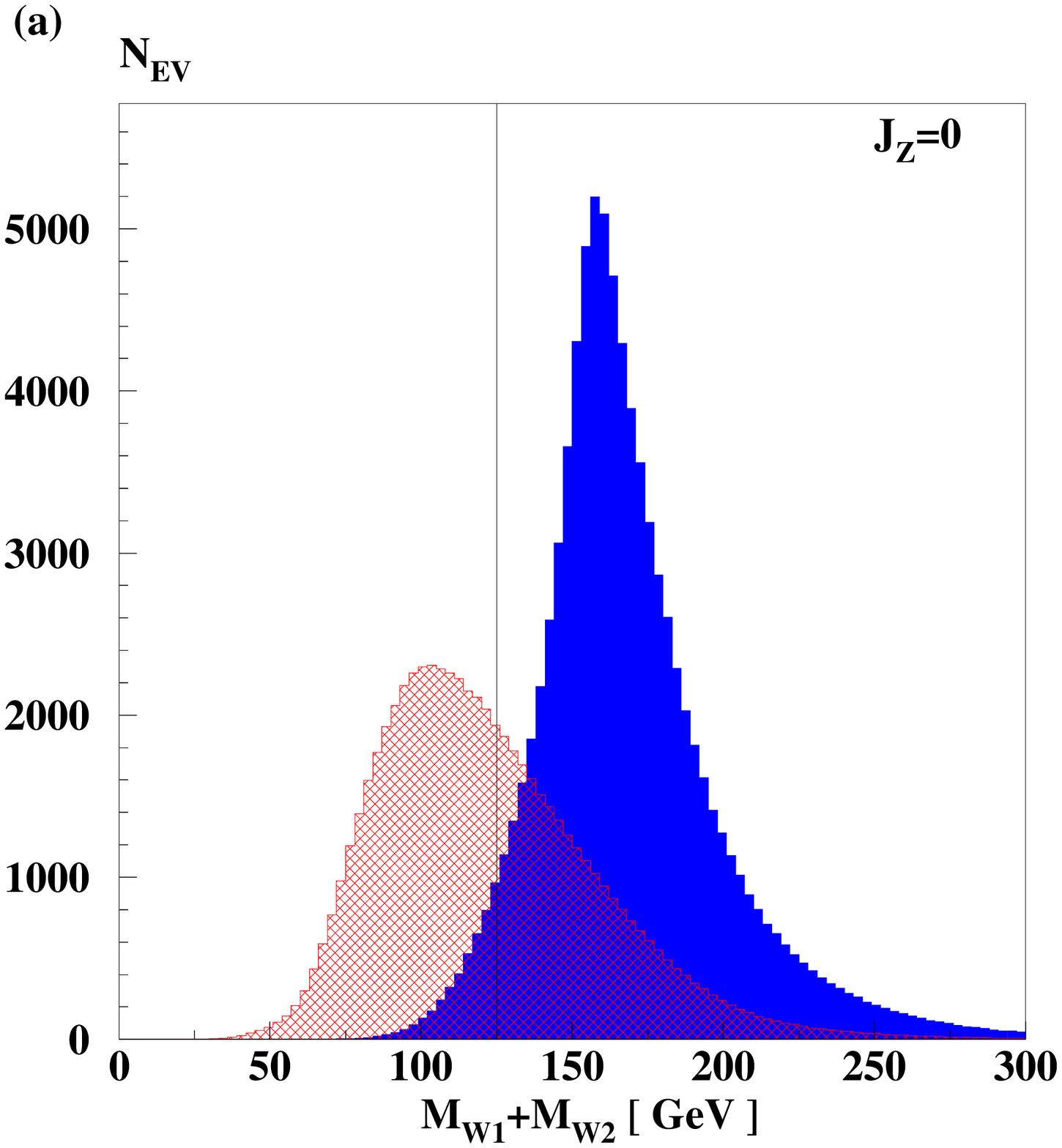}
\epsfxsize=3.0in
\epsfysize=3.0in
\epsfbox{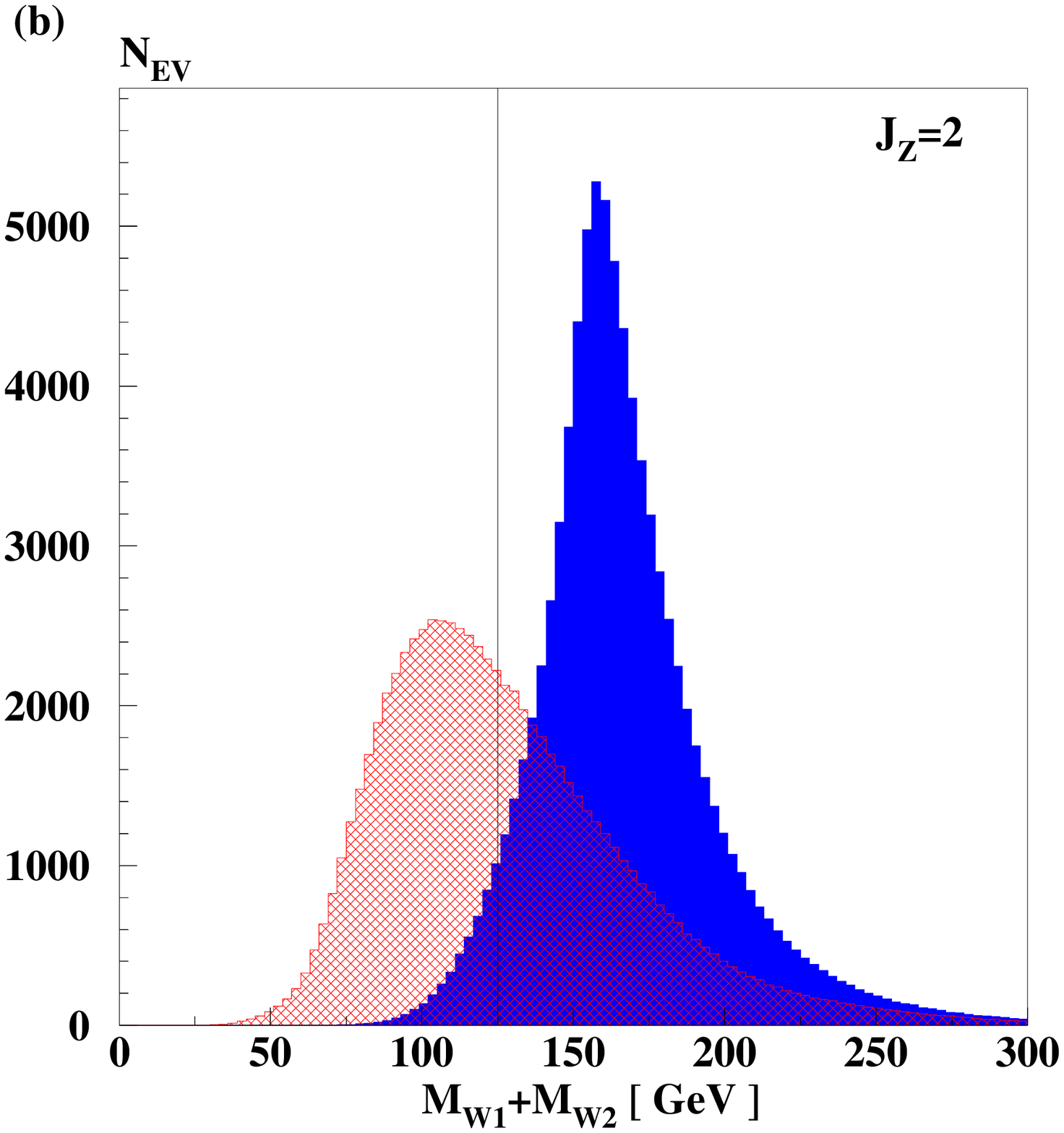}
\caption[bla]{Signal (blue) and background (red) $(M_{W1}+M_{W2})$ distributions in (\textit{a}): $J_{Z}=0$ state and in (\textit{b}): $|J_{Z}|=2$ state. Events below 125 GeV are selected for the rejection. The different processes are normalized to the same luminosity.}
\label{fig:masstot}
\end{center}
\end{figure}
About 97$\%$ of the signal events and about $\sim 46\%$ of background events survive this cut in both $J_{Z}$ states. The rest of the event is forced into four jets giving three possible combinations into $W$ bosons. The criteria to accept two jet pairs as two $W$ bosons with masses $M_{W1}$ and $M_{W2}$ was that the value $\chi_{min}=[M^{i}_{2jets}-M_{W}(80.423)]^{2}+[M^{j}_{2jets}-M_{W}(80.423)]^{2}$ ($i \neq j$) is minimal per event. An example of the accepted $W$ boson pairs in the $J_{Z}=0$ state are shown in Fig.~\ref{fig:pairing}\,$a,b$. Boosted to the center-of-mass system the two $W$ bosons are distributed back-to-back and defined as a forward $W$ boson, $W_{F}$ ($\cos\theta > 0$) and backward $W$ boson, $W_{B}$ ($\cos\theta < 0$) $W$ boson, as it is shown in Fig.~\ref{fig:pairing}\,$c$. For each $W_{F,B}$ boson the angle between the two jets boosted to the center-of-mass system is used as next selection criteria. The Fig.~\ref{fig:coswg} shows that accepting the events with $\theta_{J1-J2}$ and $\theta_{J3-J4}$ between 40$^{\circ}$ and 140$^{\circ}$ a large fraction of background events can be rejected since the signal four-jet events are more 'spherical' compared to background four-jet events. After this cut about 88$\%$ of signal events and about 10$\%$ of background events survive in both $J_{Z}$ states. The mass distribution of the two $W$ bosons $(M_{W1}+M_{W2})$ is plotted in Fig.~\ref{fig:masstot}. A cut on the mass of the two $W$ bosons is applied accepting the events in the region with a total mass $(M_{W1}+M_{W2})>125$ GeV leaving 84$\%$ of signal events and 4-5$\%$ from background events in both $J_{Z}$ states. Accepting the $W$ bosons with mass range of $M_{W1,W2}=60-100$ GeV the final angular distributions are shown in Fig.~\ref{fig:angles_gg}. The signal efficiencies are decreased drastically due to the insufficient pileup rejection which makes the $W$ boson mass distributions broader. About 53$\%$ of signal events and 1.8$\%$ from background events in both $J_{Z}$ states are left. The final ratio of signal to background events after previous cuts is $N_{S}/N_{B}\approx$ 4.3 in the $|J_{Z}|=2$ state while for the $N_{S}/N_{B}$ estimation in the $J_{Z}=0$ state the QCD events should be added. At this stage, in the $J_{Z}=0$ state $N_{S}/N_{B}\approx$ 5 taking into account only the background contribution at the tree-level; the contribution from the $\gamma\gamma\rightarrow q\bar{q}$ cross-section in the $J_{Z}=0$ state is suppressed as $\sim(m_{q}^{2}/s)$ and the estimated background comes from the $|J_{Z}|=2$ state only.
\begin{figure}[p]
\begin{center}
\epsfxsize=3.0in
\epsfysize=3.0in
\epsfbox{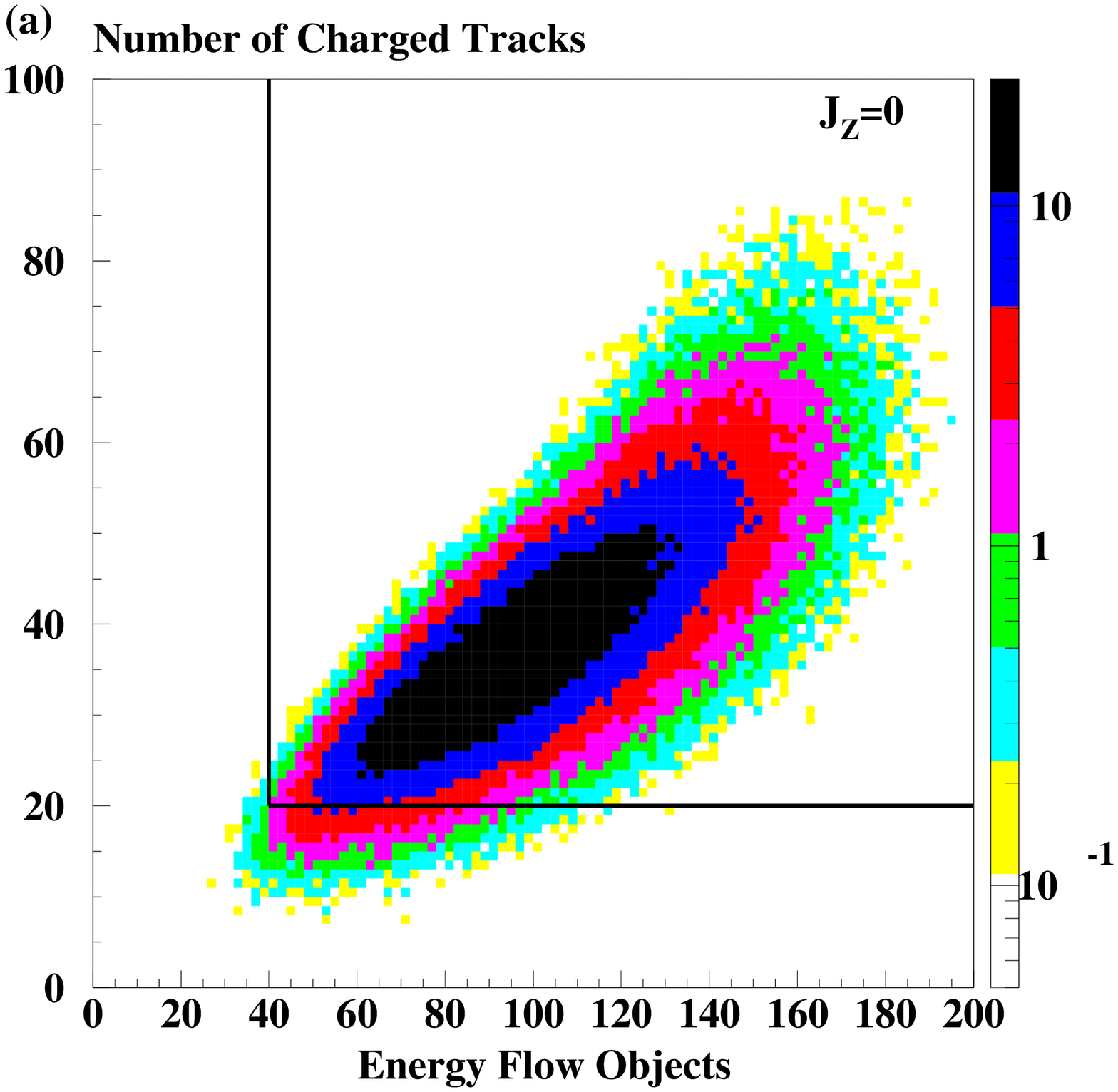}
\epsfxsize=3.0in
\epsfysize=3.0in
\epsfbox{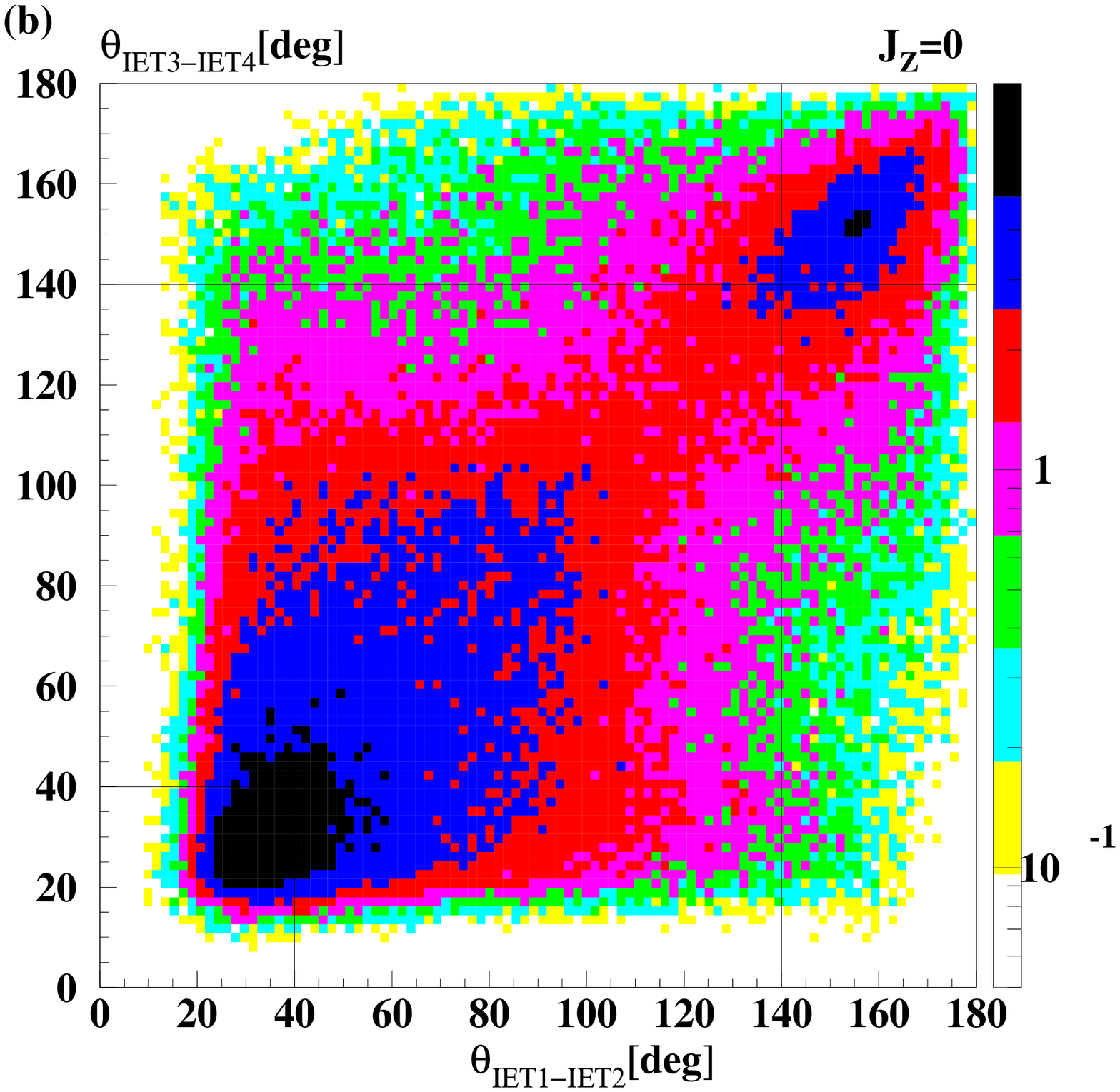}
\epsfxsize=3.0in
\epsfysize=3.0in
\epsfbox{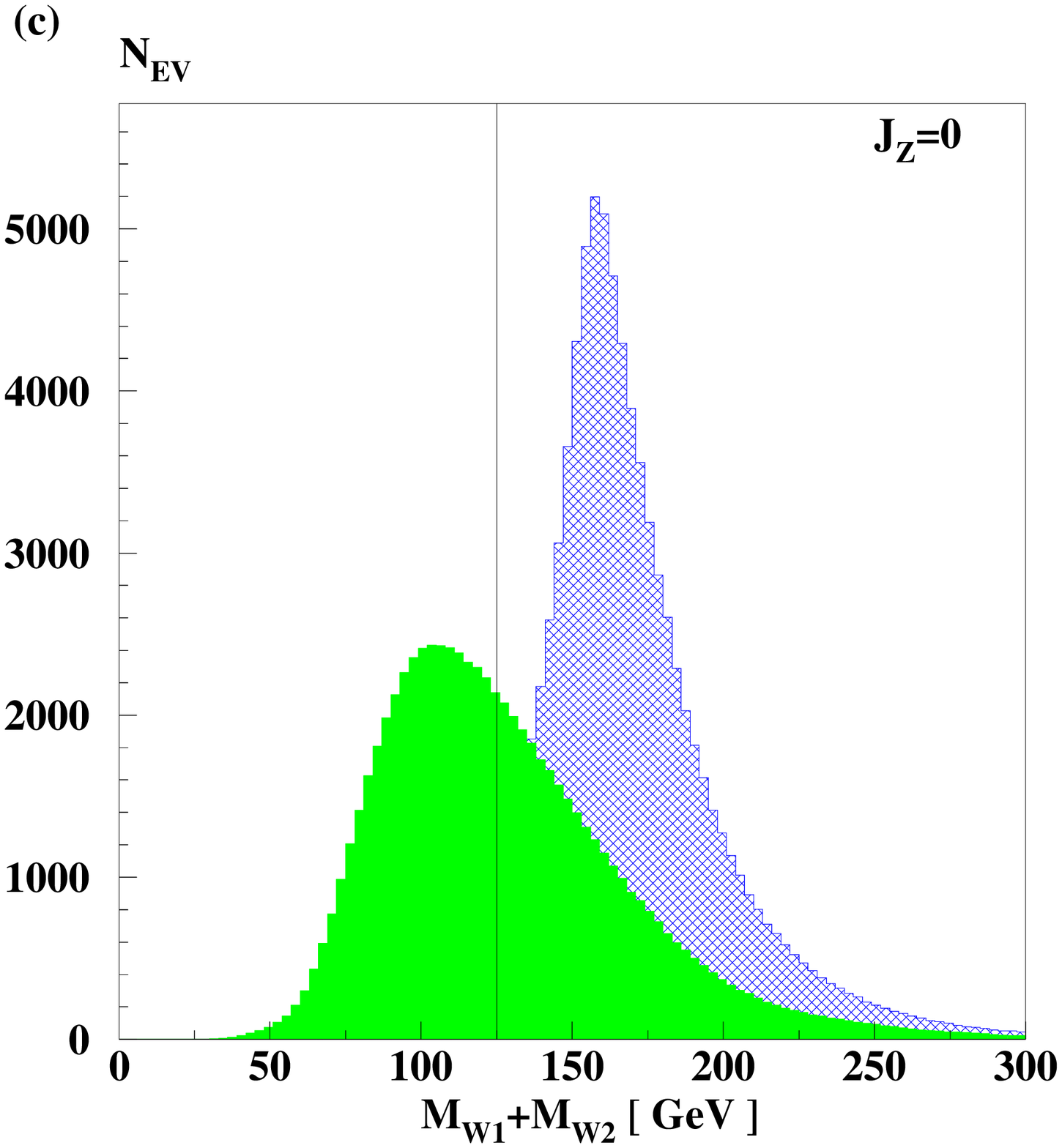}
\epsfxsize=3.0in
\epsfysize=3.0in
\epsfbox{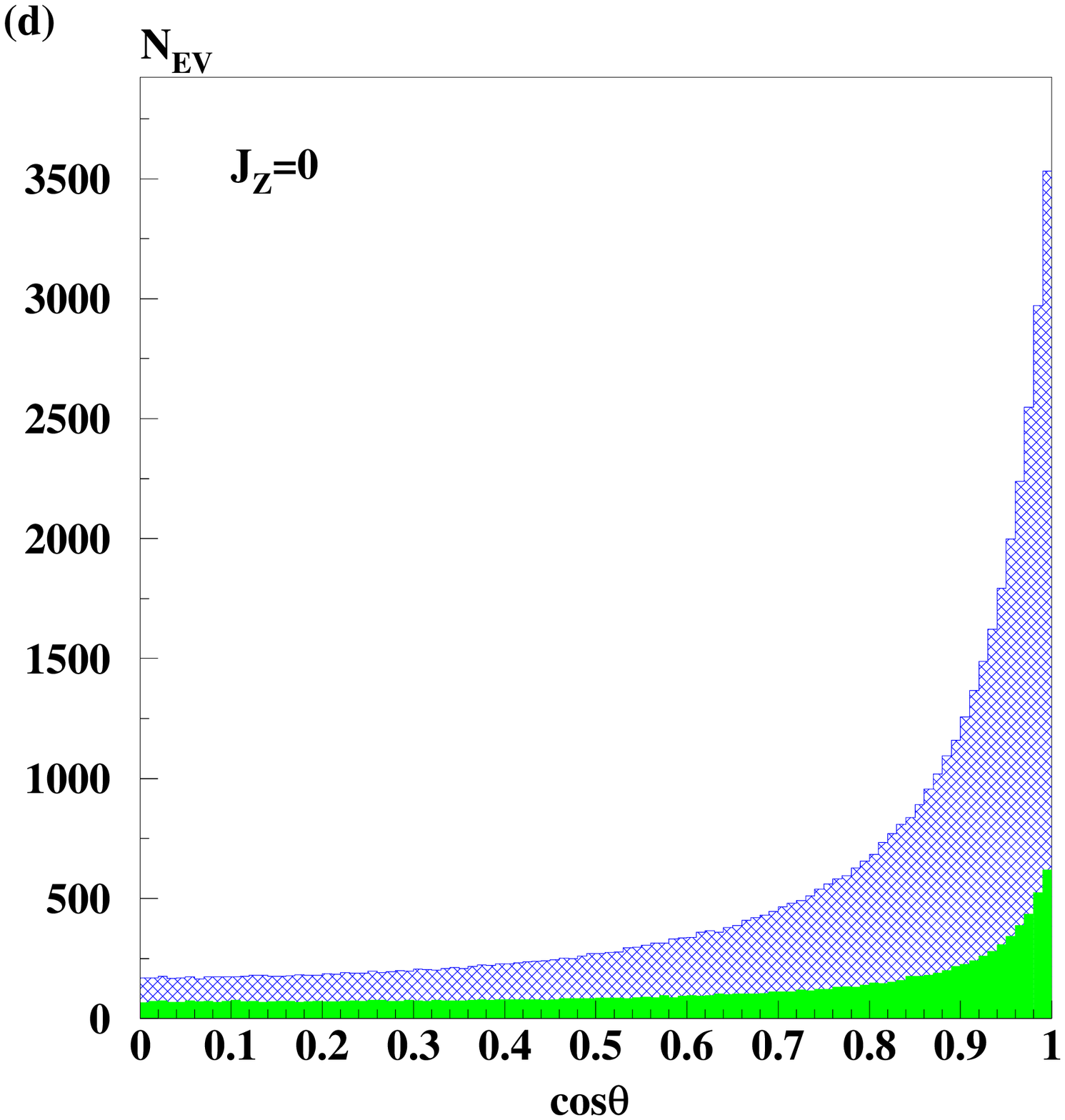}
\caption[bla]{(\textit{a-b}): QCD background of ${\cal O}(\alpha_{s}^{2})$ i.e. $\gamma\gamma \rightarrow q\bar{q}gg$ and $\gamma\gamma \rightarrow q\bar{q}g\rightarrow(q\bar{q}q\bar{q})$ in the pure $J_{Z}=0$ state. Different separation steps from (\textit{a-c}) result in the final angular distribution in (\textit{d}). In the plot (\textit{c}) the signal (blue) and ($\gamma\gamma \rightarrow q\bar{q}$ + QCD) background events (green) are compared. (\textit{d}): Final angular distribution of background events from $\gamma\gamma \rightarrow q\bar{q}$ + QCD (green) compared with signal events (blue). The events are normalized to the same luminosity.}
\label{fig:qcd_gg0}
\end{center}
\end{figure}
\par
The QCD contribution of ${\cal O}(\alpha_{s}^{2})$ in the pure $J_{Z}=0$ state from the two gluon emission $\gamma\gamma \rightarrow q\bar{q}gg$ and from the production of 'secondary' quarks $\gamma\gamma \rightarrow q\bar{q}(g\rightarrow)q\bar{q}$ from gluon splitting, simulated with MadGraph, gives four distinct jets unlike in the previous case where the radiated gluons (from Lund parton shower) are mainly soft and collinear with a parent particle. Since the QCD events simulated with MadGraph are fragmented using PYTHIA the gluon emission is double counted at some level. If the same cuts are applied for the QCD background rejection about 14.5$\%$ events survive all cuts in the pure $J_{Z}=0$ state. Fig.~\ref{fig:qcd_gg0}\,$d$ shows the comparison of the signal events with $\gamma\gamma \rightarrow q\bar{q}+QCD$ background events in the final angular distributions. The ratio of $N_{S}/N_{B}^{q\bar{q}+QCD}\approx$ 4.3 is obtained adding the two backgrounds together. 
\subsubsection{Top pair Production}
\begin{figure}[p]
\begin{center}
\epsfxsize=3.0in
\epsfysize=3.0in
\epsfbox{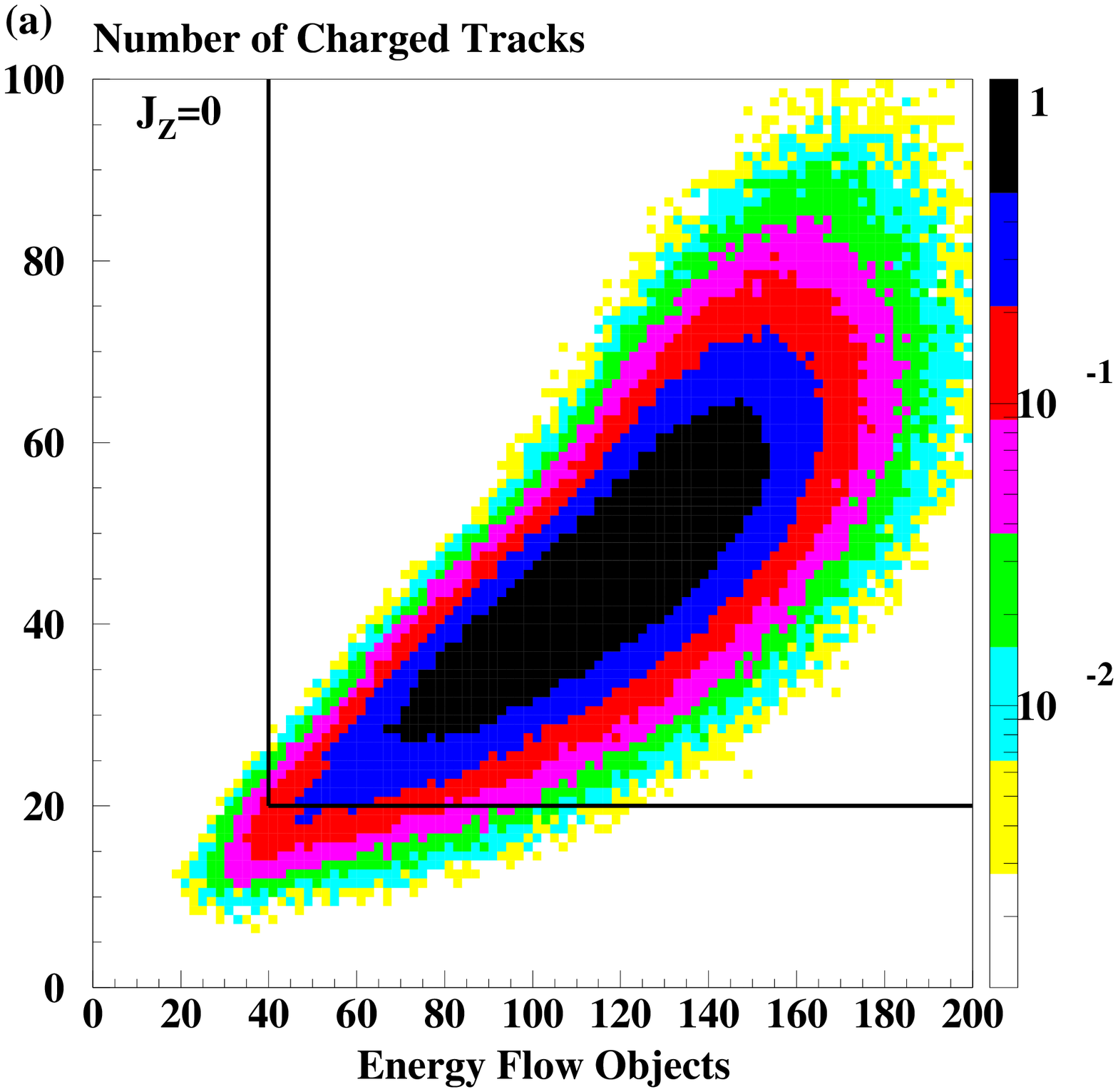}
\epsfxsize=3.0in
\epsfysize=3.0in
\epsfbox{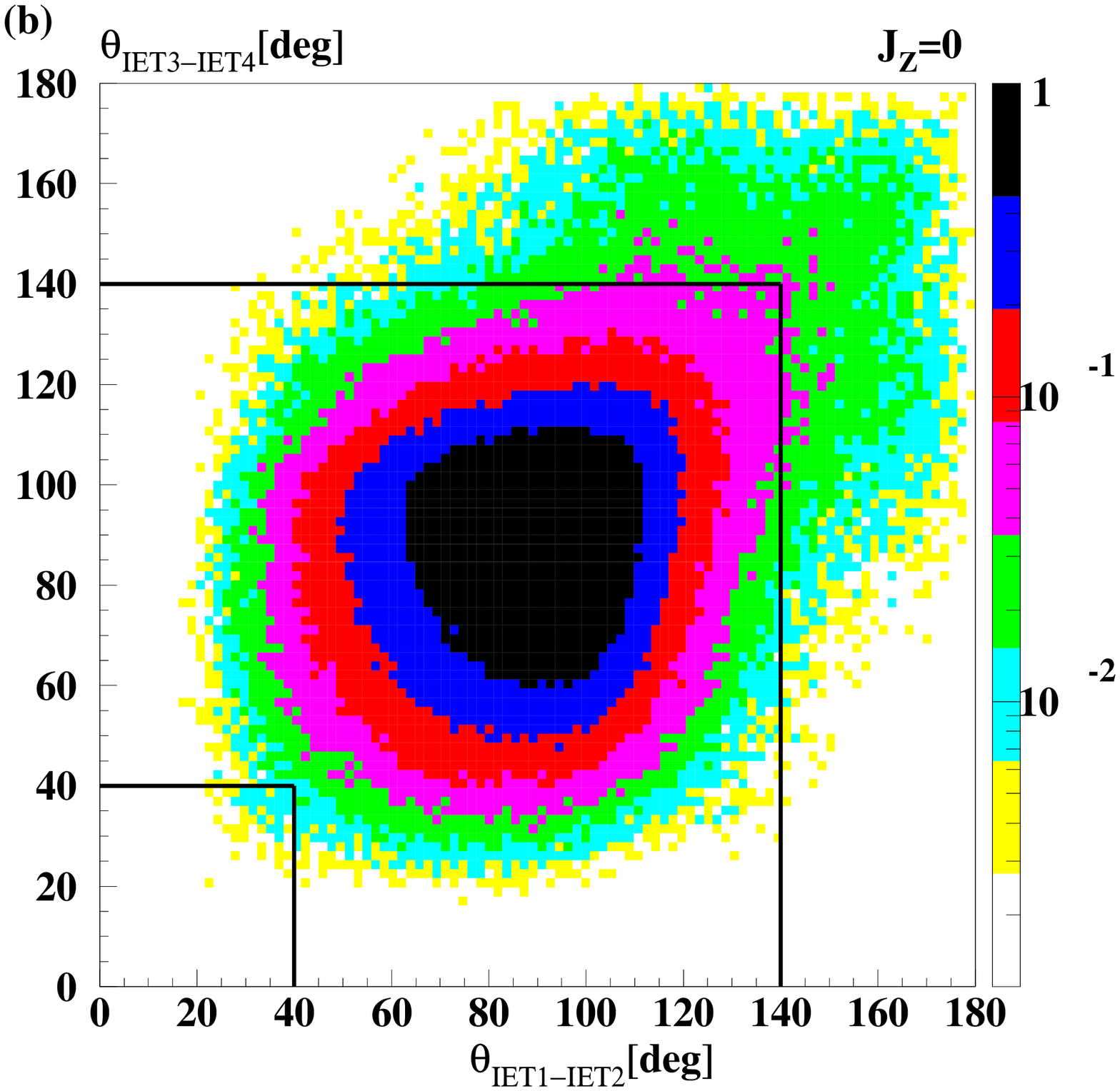}
\epsfxsize=3.0in
\epsfysize=3.0in
\epsfbox{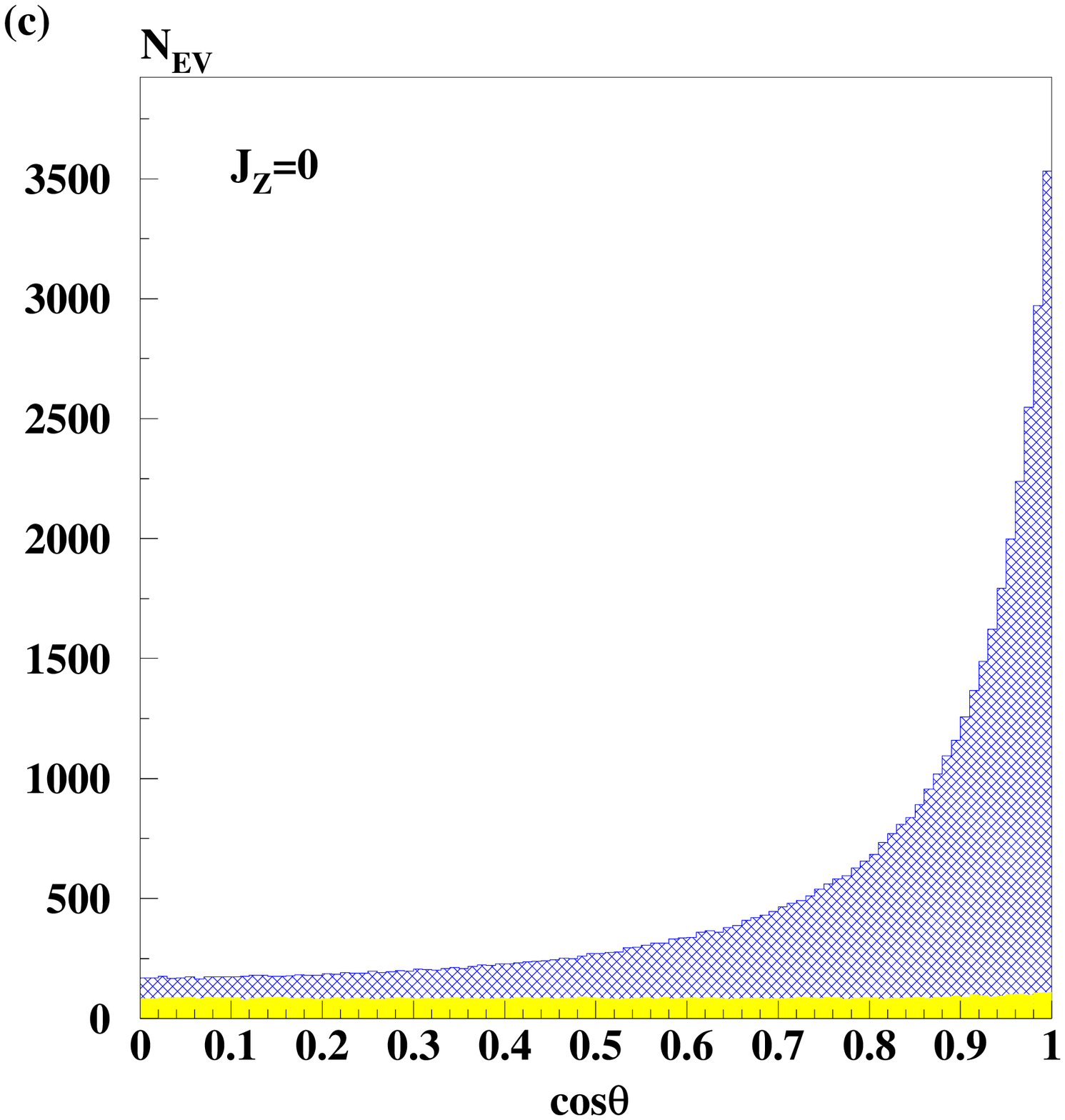}
\caption[bla]{(\textit{a-b}): Different steps in the separation procedure of $t\bar{t}$-pair events for $J_{Z}=0$. (\textit{c}): Final angular distributions of signal (blue) and $t\bar{t}$ (yellow) events, the $t\bar{t}$ contribution is multiplied by 100. Approximatly $11\%$ of $t\bar{t}$-pairs survive the applied cuts and the $N_{S}/N_{t\bar{t}}\approx 600$ after applied cuts.}
\label{fig:top}
\end{center}
\end{figure}

The $\gamma\gamma\rightarrow t\bar{t}$ background and the QCD contribution through $\gamma\gamma\rightarrow t\bar{t}(g\rightarrow)q\bar{q},t\bar{t}gg$ have been analyzed separately. $\gamma\gamma\rightarrow t\bar{t}$ events are generated with PYTHIA while the MadGraph cross-section is used for the normalization. In spite of top decays via $t^{\pm}\rightarrow W^{\pm}b^{\pm}$ which could considerably contribute as a background distributed over the same kinematical region as the signal (Fig.~\ref{fig:top}), it has been found that the $t\bar{t}$-pair contribution to the total $q\bar{q}$ background is negligible due to the small cross-sections \cite{top1} at the available center-of-mass energies provided by the simulated spectra. Concerning only the top pair production, the $J_{Z}=0$ configuration dominates over the $|J_{Z}|=2$ one below energies of 680 GeV while $|J_{Z}|=2$ starts to dominate at higher energies. Comparing with the cross-section for $q\bar{q}$ production ($u\bar{u},d\bar{d},c\bar{c},s\bar{s},b\bar{b}$) with $\sigma_{q\bar{q}}(M_{q\bar{q}}>50 \textrm{GeV})/\sigma_{t\bar{t}}=760$ represents only 0.13$\%$ of the total cross-section in the $J_{Z}=0$ state and even less in the $|J_{Z}|=2$ state (0.03$\%$). Comparing to the signal cross-sections $\sigma_{WW}/\sigma_{t\bar{t}}\approx$ 431 in the $|J_{Z}|=2$ state and $\approx$ 122 in the $J_{Z}=0$ state. Applying the same cuts as previously described, approximatively $11\%$ of $t\bar{t}$-pairs remain. That finally gives the ratio of signal to $t\bar{t}$ events of $N_{S}/N_{t\bar{t}}\approx 600$ in the $J_{Z}=0$ state at considered center-of-mass energies and thus, can be neglected.
\par
In the case where one photon originates from the bremsstrahlung as it was the case of the considered background in the real $\gamma e$ mode, the achievable $\sqrt{s}$ is even lower than in the collision of the two Compton photons, much more suppressing the top pair production.
%
\section{Monte Carlo Fit}
\begin{figure}[p]
\begin{center}
\epsfxsize=3.0in
\epsfysize=3.0in
\epsfbox{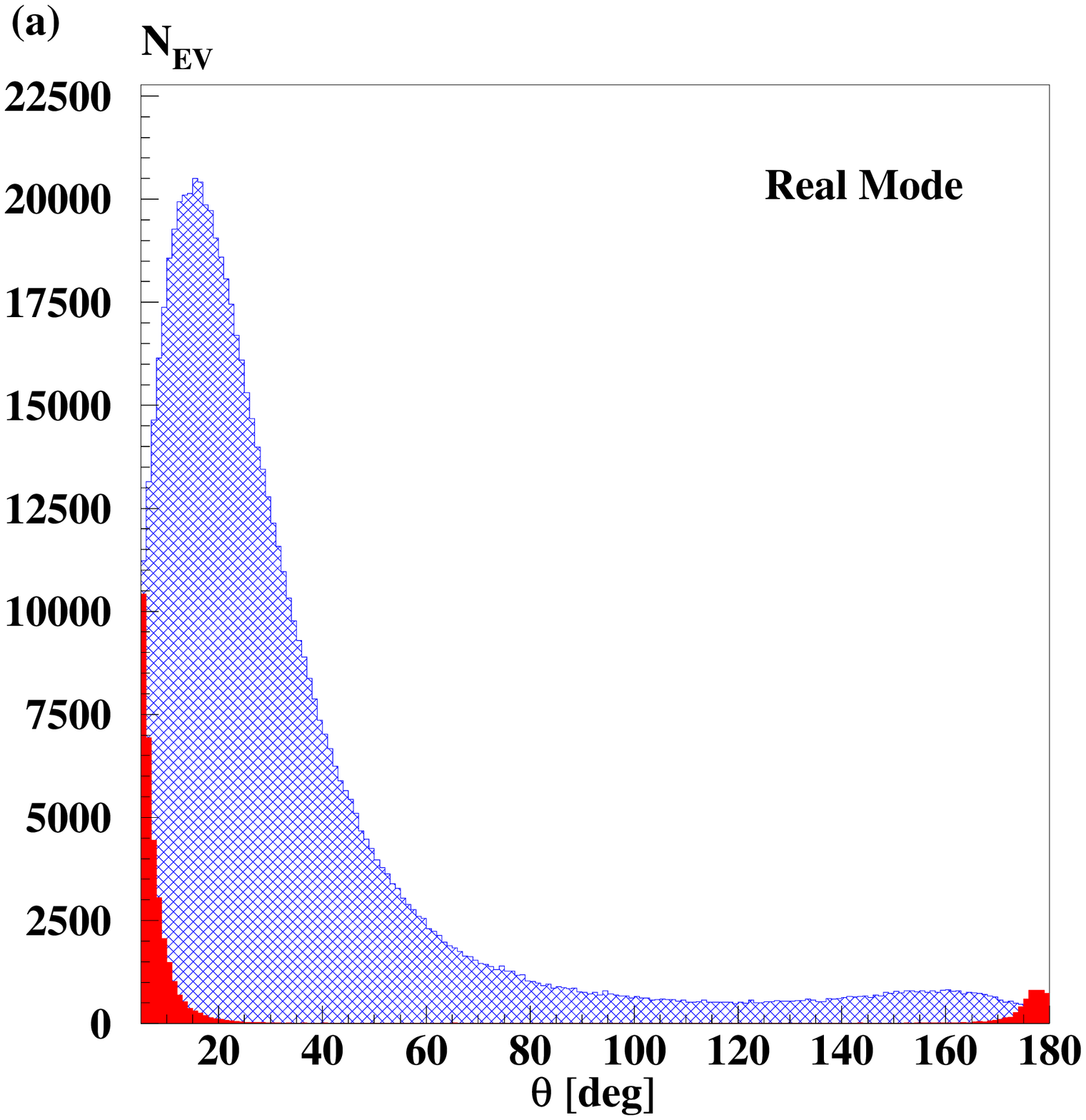}
\epsfxsize=3.0in
\epsfysize=3.0in
\epsfbox{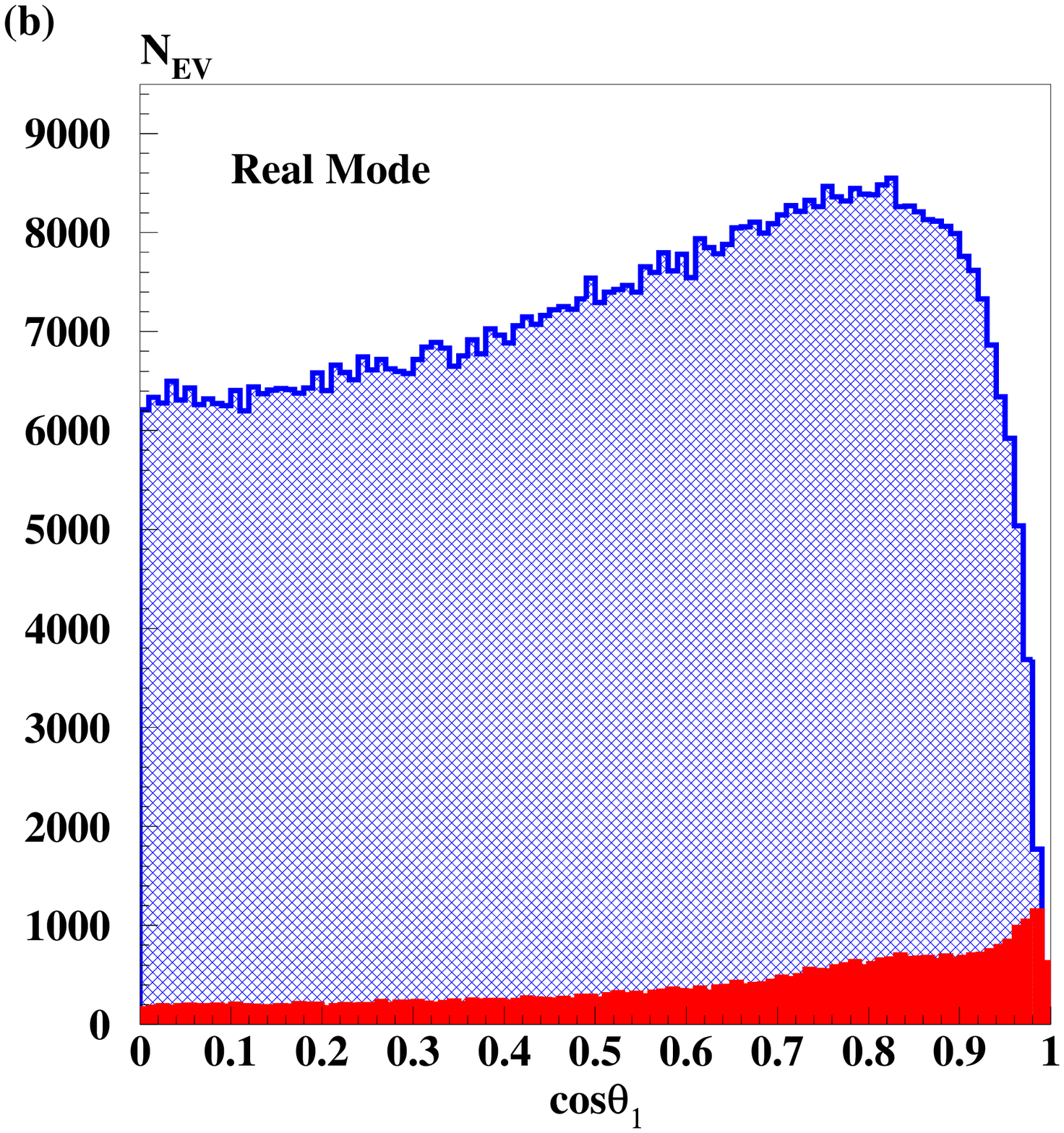}
\epsfxsize=3.0in
\epsfysize=3.0in
\epsfbox{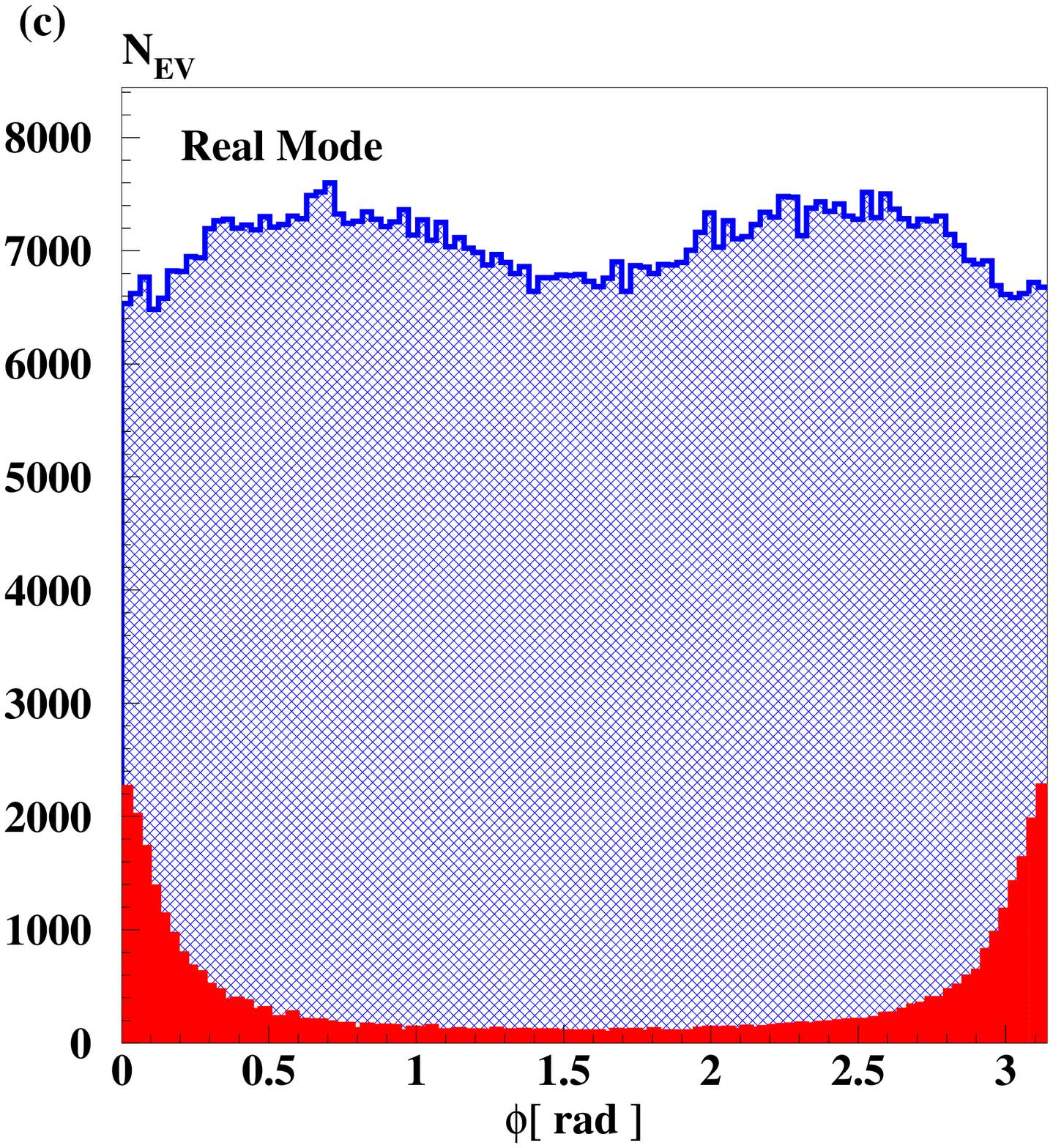}
\caption[bla]{Angular distributions of signal (blue) and background (red) events used in the fit in the real $\gamma e$ mode for (\textit{a}): the production angle $\theta$, (\textit{b}): the decay angle $\theta_{1}$ and (\textit{c}): the azimuthal angle $\phi$ after the detector simulation with included pileup events. A cut of 5$^{\circ}$ is applied.}
\label{fig:fitted_ge_real}
\end{center}
\end{figure}

\begin{figure}[p]
\begin{center}
\epsfxsize=3.0in
\epsfysize=3.0in
\epsfbox{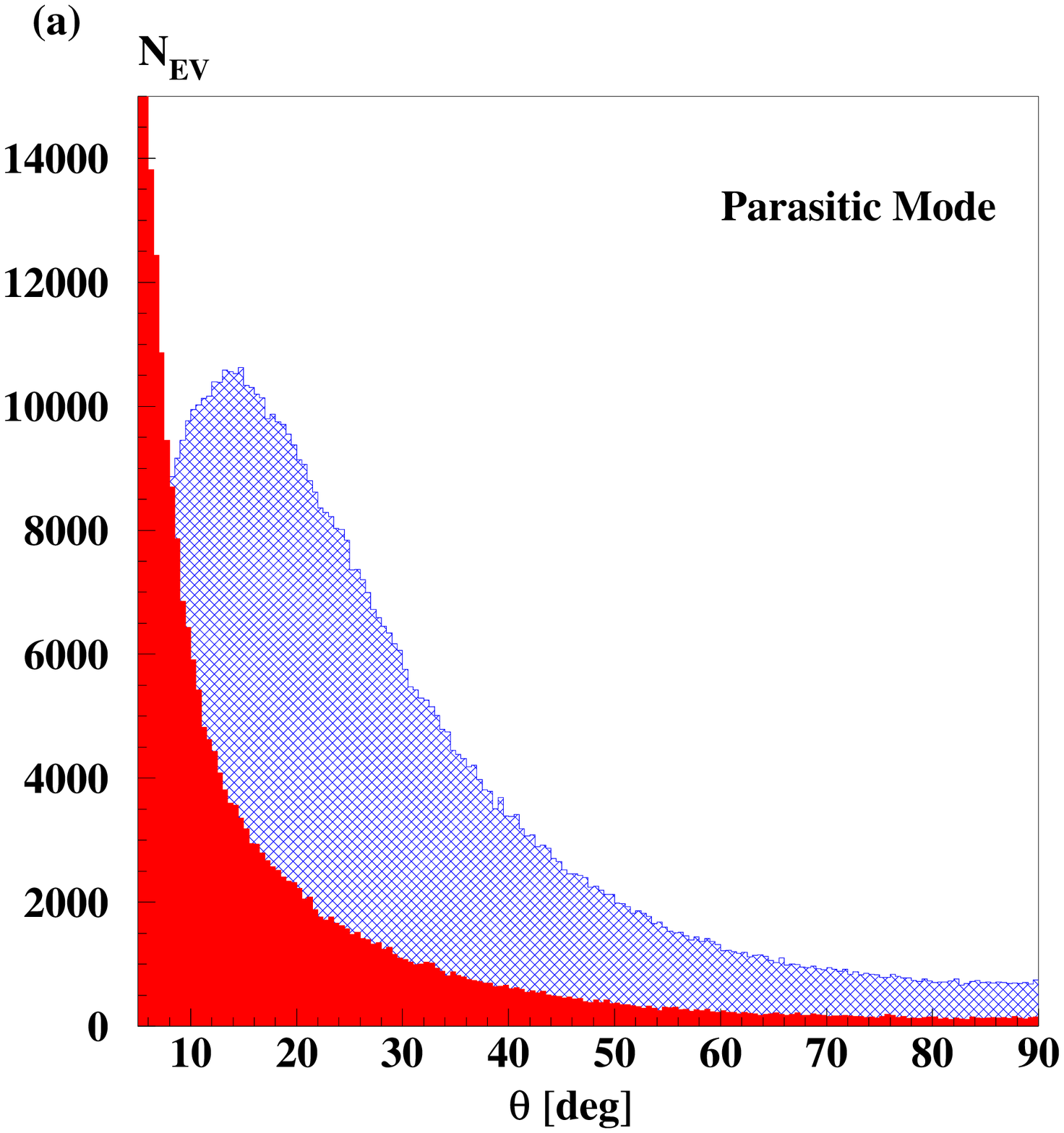}
\epsfxsize=3.0in
\epsfysize=3.0in
\epsfbox{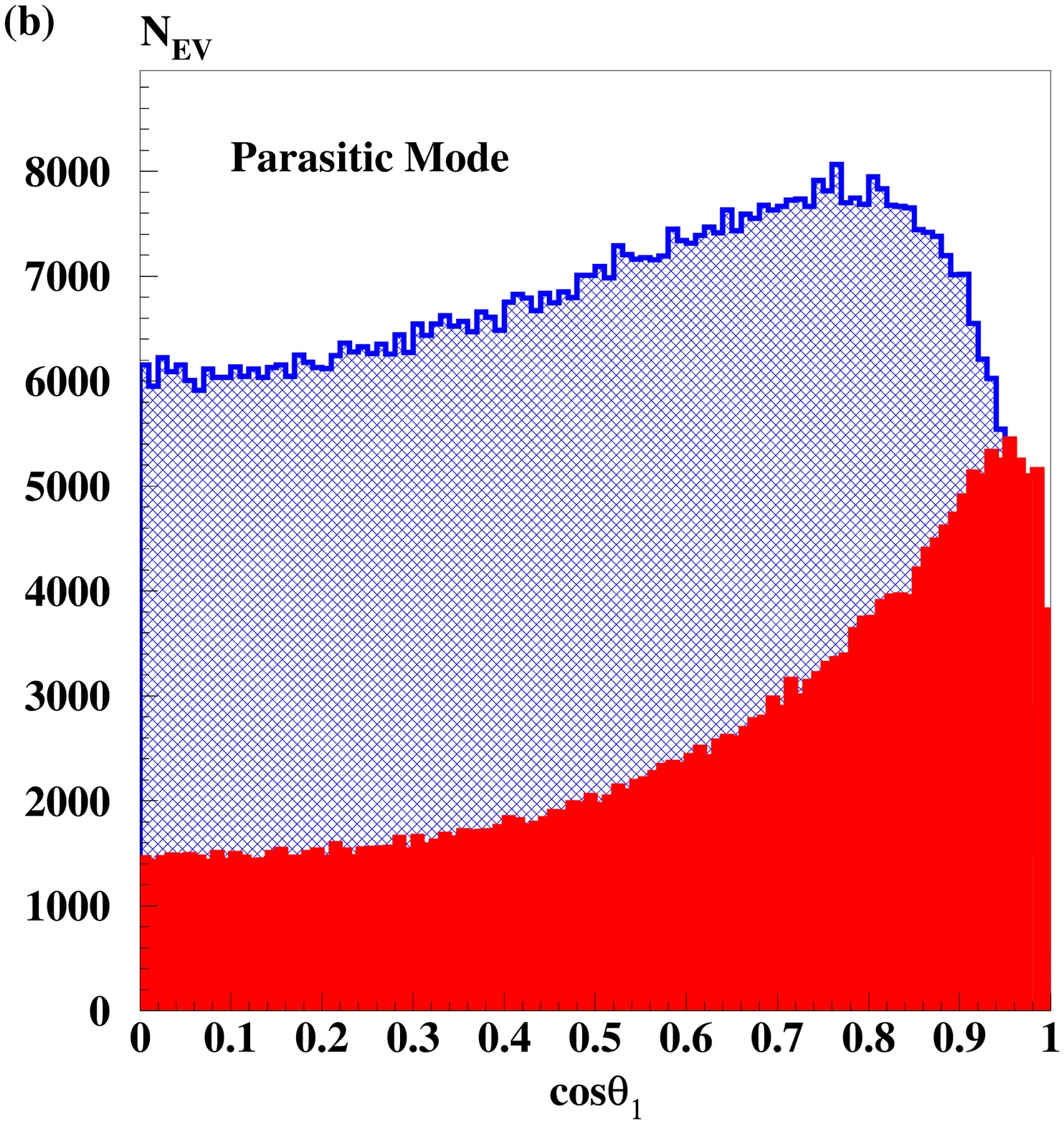}
\epsfxsize=3.0in
\epsfysize=3.0in
\epsfbox{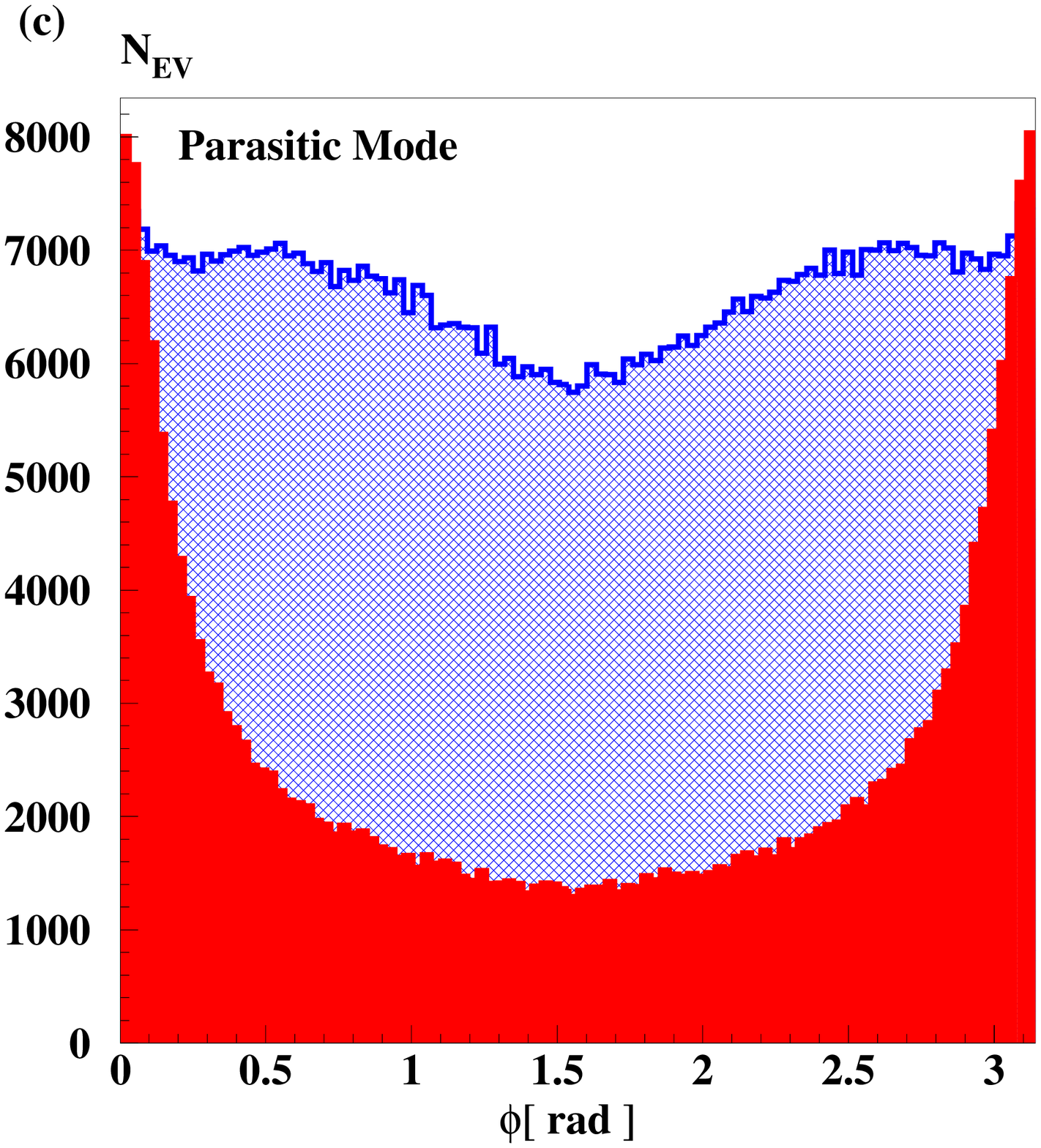}
\caption[bla]{Angular distributions of signal (blue) and background (red) events used in the fit in the parasitic $\gamma e$ mode for (\textit{a}): the production angle $\theta$, (\textit{b}): the decay angle $\theta_{1}$ and (\textit{c}): the azimuthal angle $\phi$ after the detector simulation with pileup events included. A cut of 5$^{\circ}$ is applied.}
\label{fig:fitted_ge_parasitic}
\end{center}
\end{figure}

For the extraction of the anomalous triple gauge boson couplings $\Delta\kappa_{\gamma}$ and $\Delta\lambda_{\gamma}$ from the reconstructed kinematical variables a ${\chi^{2}}$ and a binned maximum likelihood fit are used. Samples of ${10^{6}}$ SM signal events for single $W$ boson production and $2\cdot {10^{6}}$ signal events for $W$ boson pair production are generated and passed through the detector simulation (Fig.~\ref{fig:fitted_ge_real},~\ref{fig:fitted_ge_parasitic}). The number of signal events obtained after the detector simulation and after all cuts is normalized to the number of events expected after one year of running of an $\gamma e$ and $\gamma\gamma$ collider.
\par
Each single $W$ boson event is described reconstructing three kinematical variables: the cosine of the $W$ boson production angle ($\cos\theta$) with respect to the ${e^{-}}$ beam direction, the cosine of the $W$ boson polar decay angle $\cos{\theta_{1}}$ i.e. the angle of the fermion with respect to the $W$ boson flight direction measured in the $W$ boson rest frame where both fermions are back-to-back, and the azimuthal decay angle $\phi$ of the fermion with respect to a plane defined by the $W$ boson and the beam axis. In the single $W$ production the $\phi$ angle is measured modulo $\pi$.
\par
The two $W$ boson events are described reconstructing five kinematical variables: the cosine of the $W$ boson production angle ($\cos\theta$), the cosine of two polar decay angles $\cos\theta_{1,2}$ and two azimuthal angles $\phi_{1,2}$ for each $W$ boson, describing the direction of the fermion in the rest frame of the parent $W$ boson. The axes in the $W$ boson frame are defined such that the $z$-axis is along the parent $W$ boson flight direction and the $y$ axis is in the direction ($\vec{W}\times\vec{b}$) where $\vec{b}$ is the beam direction and $\vec{W}$ is the parent $W$ boson flight direction, and $\phi$ is the azimuthal angle of the fermion in the $x-y$ plane. The polar decay angle  is sensitive to the different $W$ boson helicity states while the azimuthal angle is sensitive to the interference between them. Since in hadronic $W$ boson decays the up- and down-type quarks cannot be separated, only $| \cos \theta_{1,2}|$ are measured as it is explained in Chapter 3. In the $W$ boson pair production the $\phi$ angle is measured in the $2\pi$ range providing an information about the interference between the two $W$ bosons.
\par
\begin{figure}[p]
\begin{center}
\epsfxsize=3.0in
\epsfysize=3.0in
\epsfbox{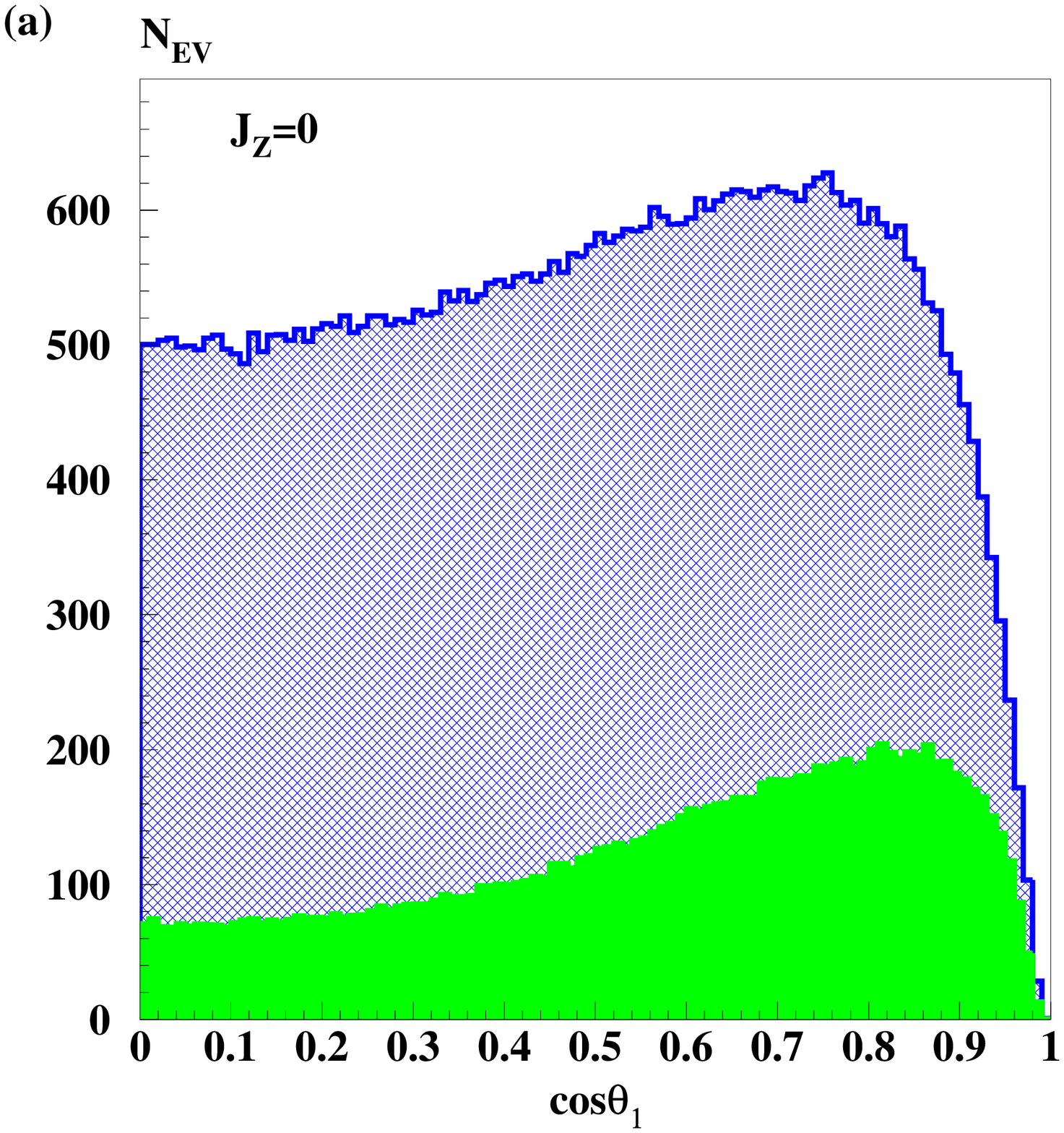}
\epsfxsize=3.0in
\epsfysize=3.0in
\epsfbox{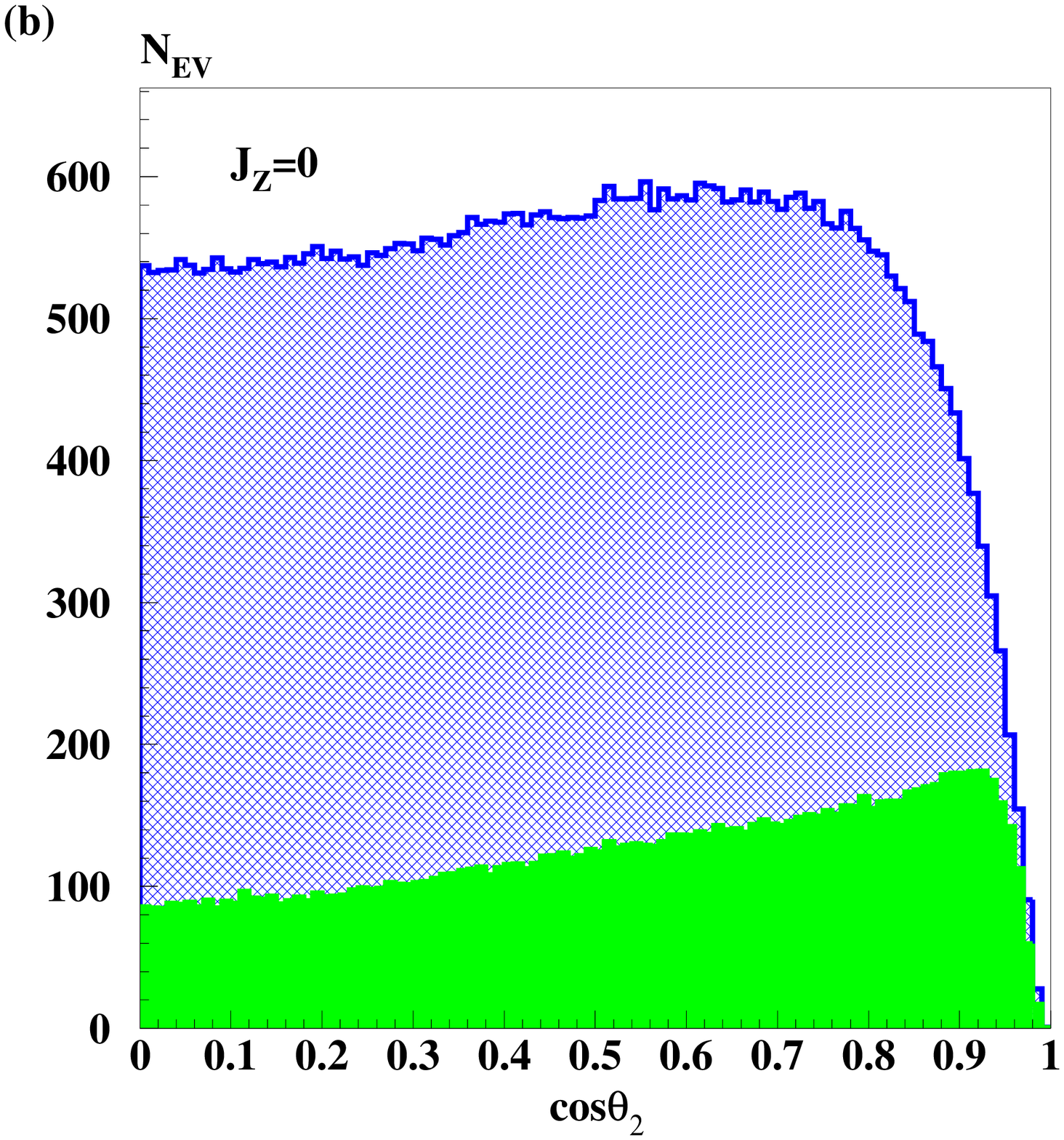}
\epsfxsize=3.0in
\epsfysize=3.0in
\epsfbox{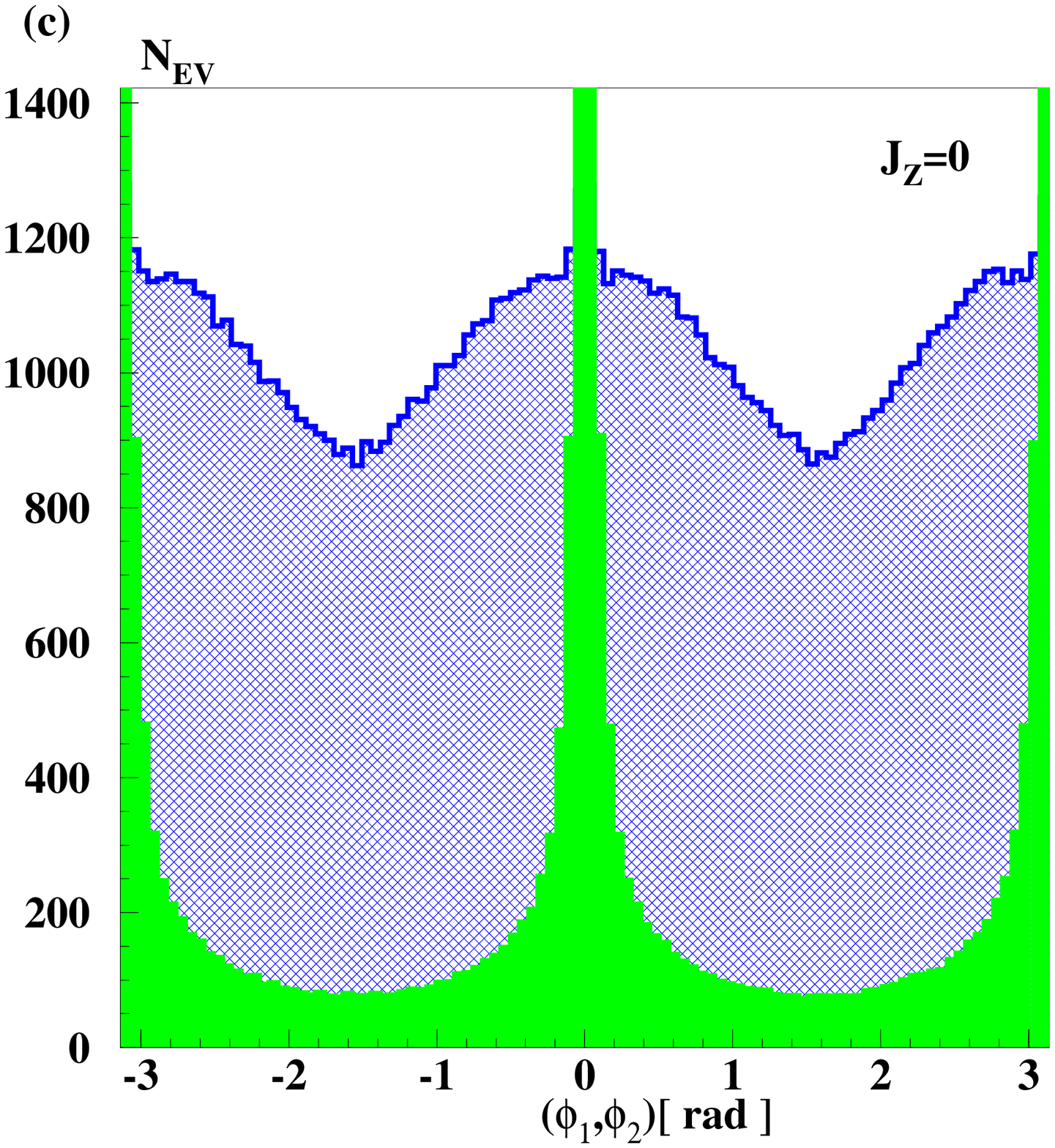}
\caption[bla]{Angular distributions of signal (blue) and background (green) events after the detector simulation used in the fit for the $J_{Z}=0$ state for (\textit{a}): the decay angle $\theta_{1}$ of $W_{F}$ (\textit{b}): the decay angle $\theta_{2}$ of $W_{B}$ (\textit{c}): the azimuthal angles ($\phi_{1},\phi_{2}$). Azimuthal distributions are affected by the presence of pileup tracks shown in Fig.~\ref{fig:pileup}.}
\label{fig:fitted_gg_j0}
\end{center}
\end{figure}
\begin{figure}[p]
\begin{center}
\epsfxsize=3.0in
\epsfysize=3.0in
\epsfbox{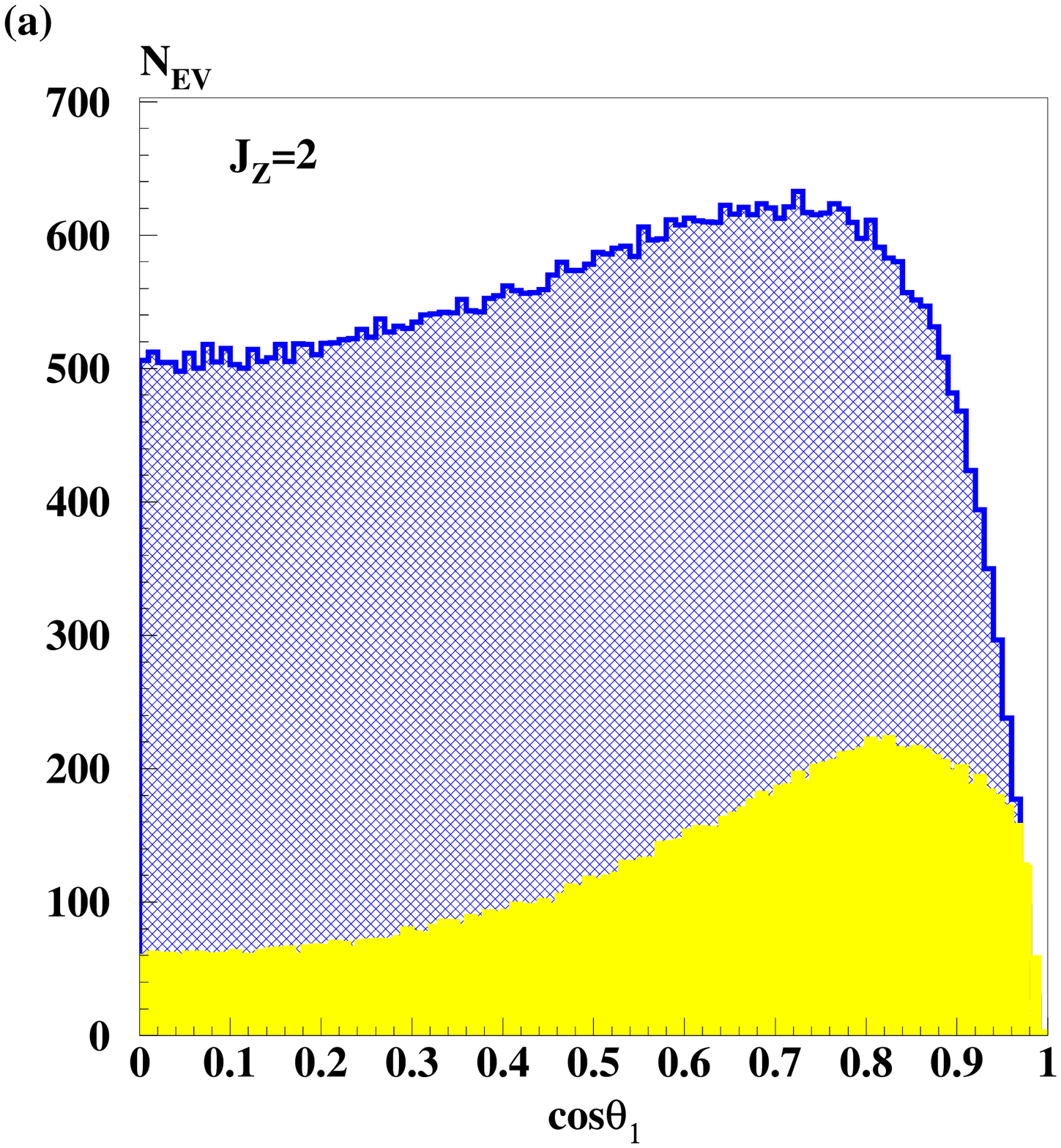}
\epsfxsize=3.0in
\epsfysize=3.0in
\epsfbox{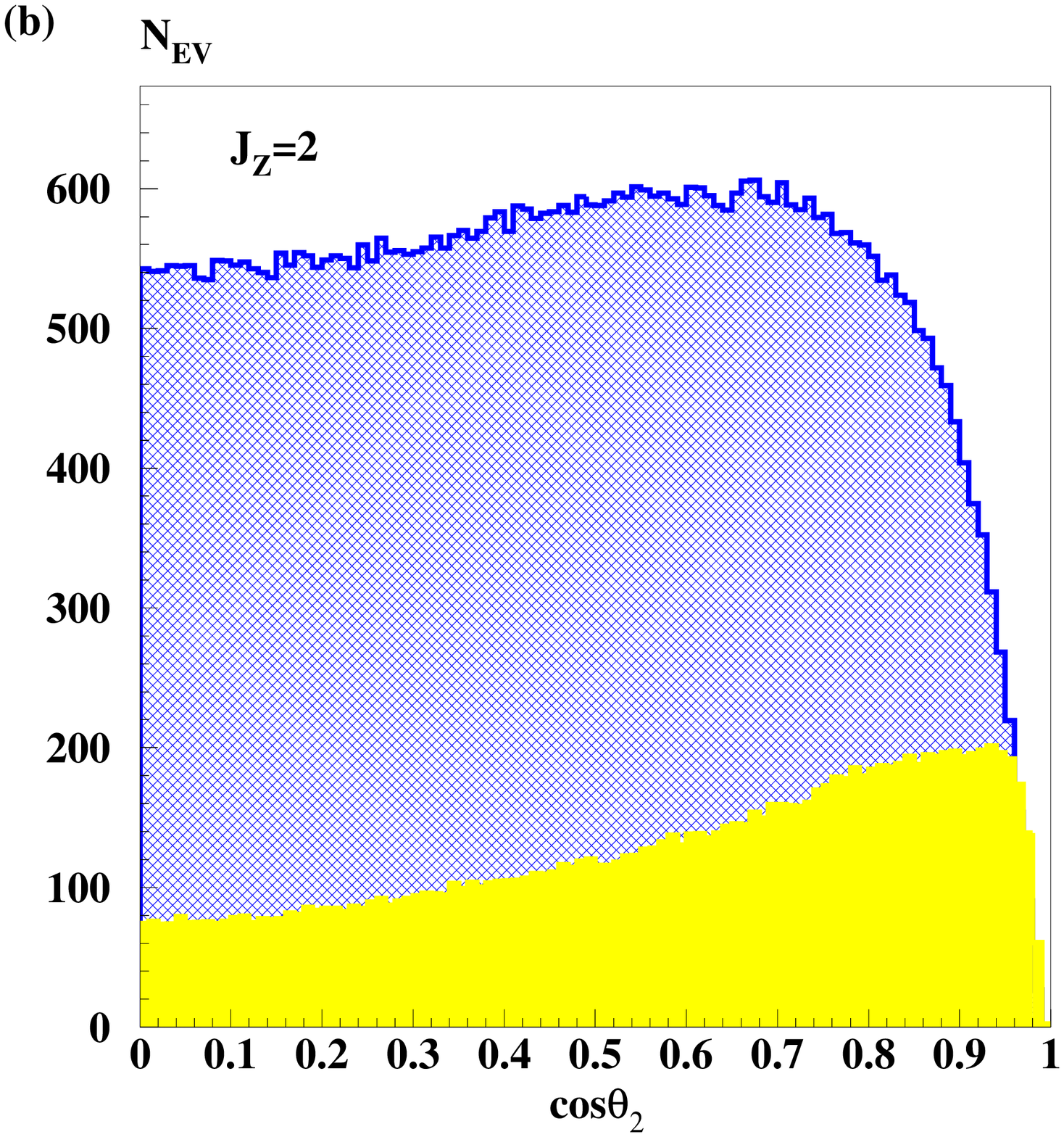}
\epsfxsize=3.0in
\epsfysize=3.0in
\epsfbox{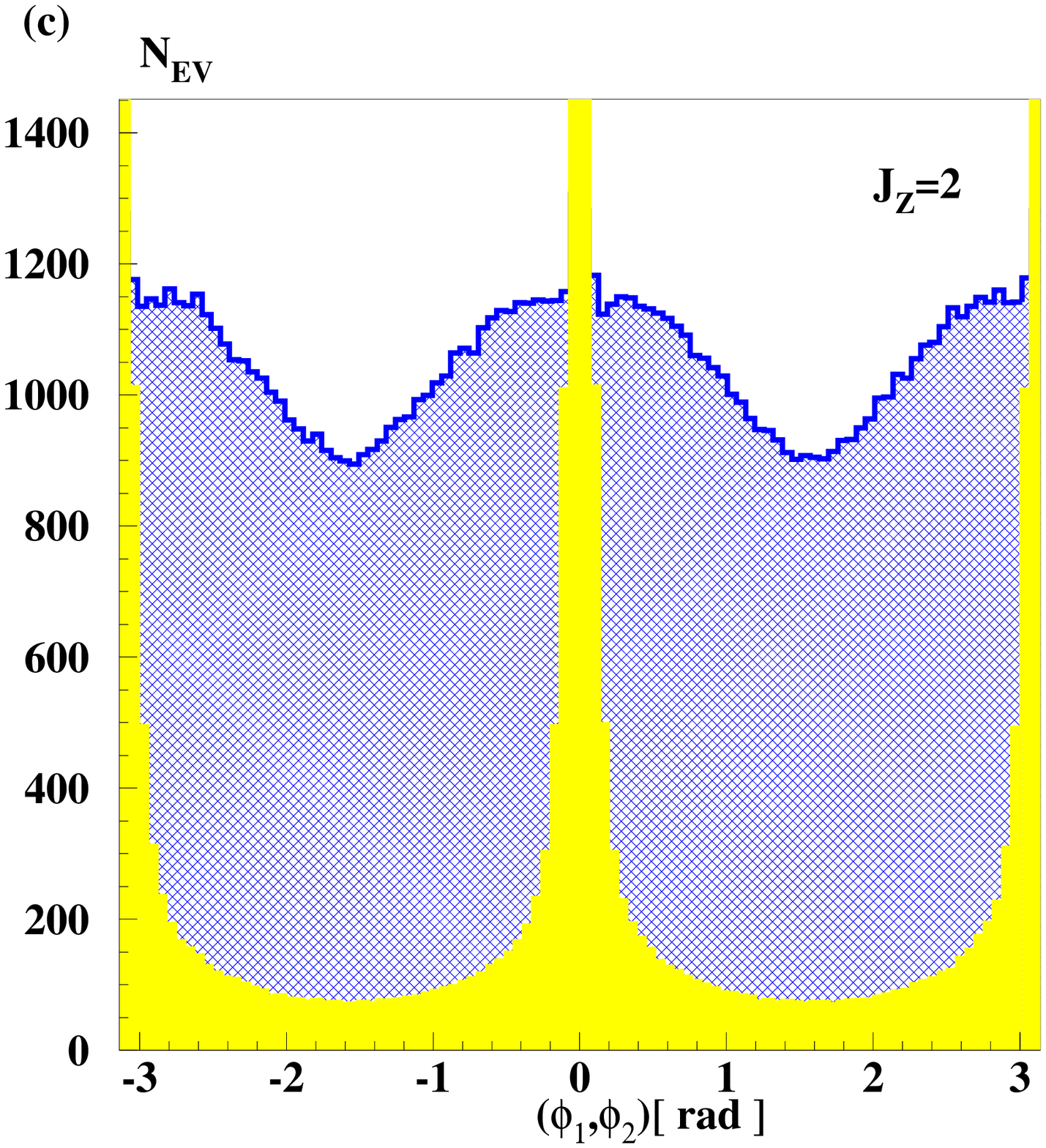}
\caption[bla]{Angular distributions of signal (blue) and background (yellow) events after the detector simulation used in the fit for the $|J_{Z}|=2$ state for (\textit{a}): the decay angle $\theta_{1}$ of $W_{F}$ (\textit{b}): the decay angle $\theta_{2}$ of $W_{B}$ (\textit{c}): the azimuthal angle ($\phi_{1},\phi_{2}$). Azimuthal distributions are affected by presence of pileup tracks shown in Fig.~\ref{fig:pileup}.}
\label{fig:fitted_gg_j2}
\end{center}
\end{figure}
\begin{figure}[p]
\begin{center}
\epsfxsize=3.0in
\epsfysize=3.0in
\epsfbox{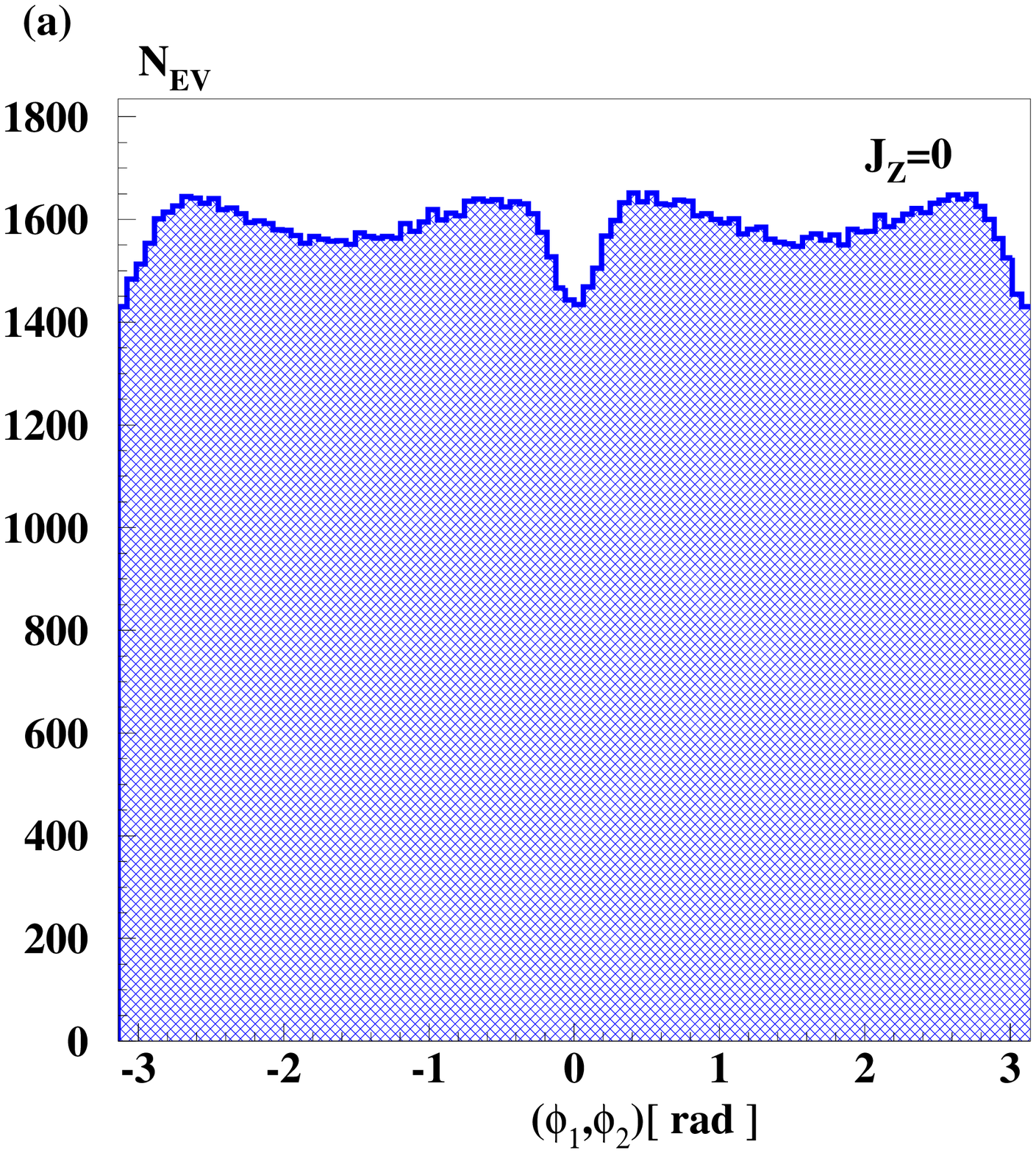}
\epsfxsize=3.0in
\epsfysize=3.0in
\epsfbox{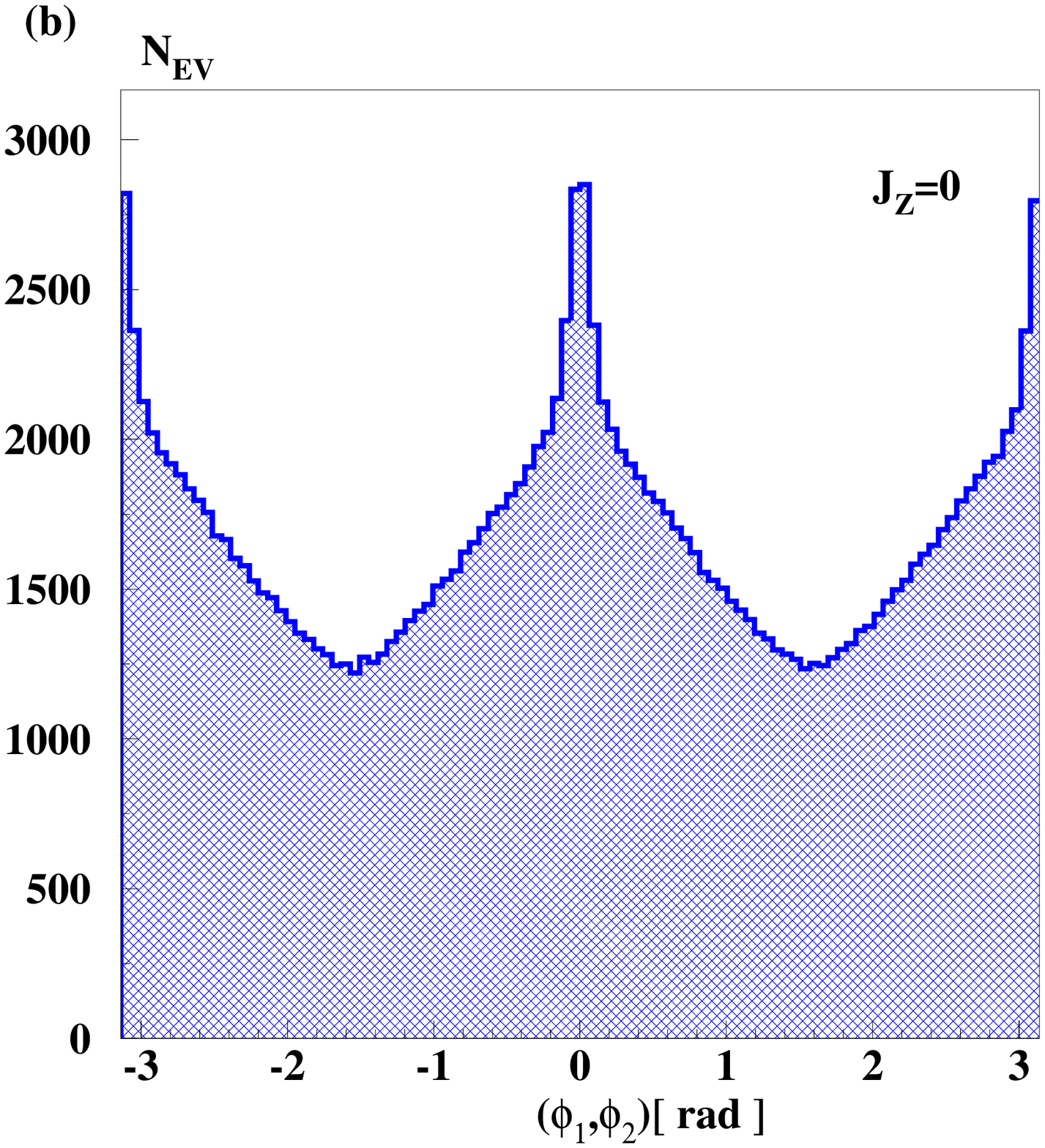}
\epsfxsize=3.0in
\epsfysize=3.0in
\epsfbox{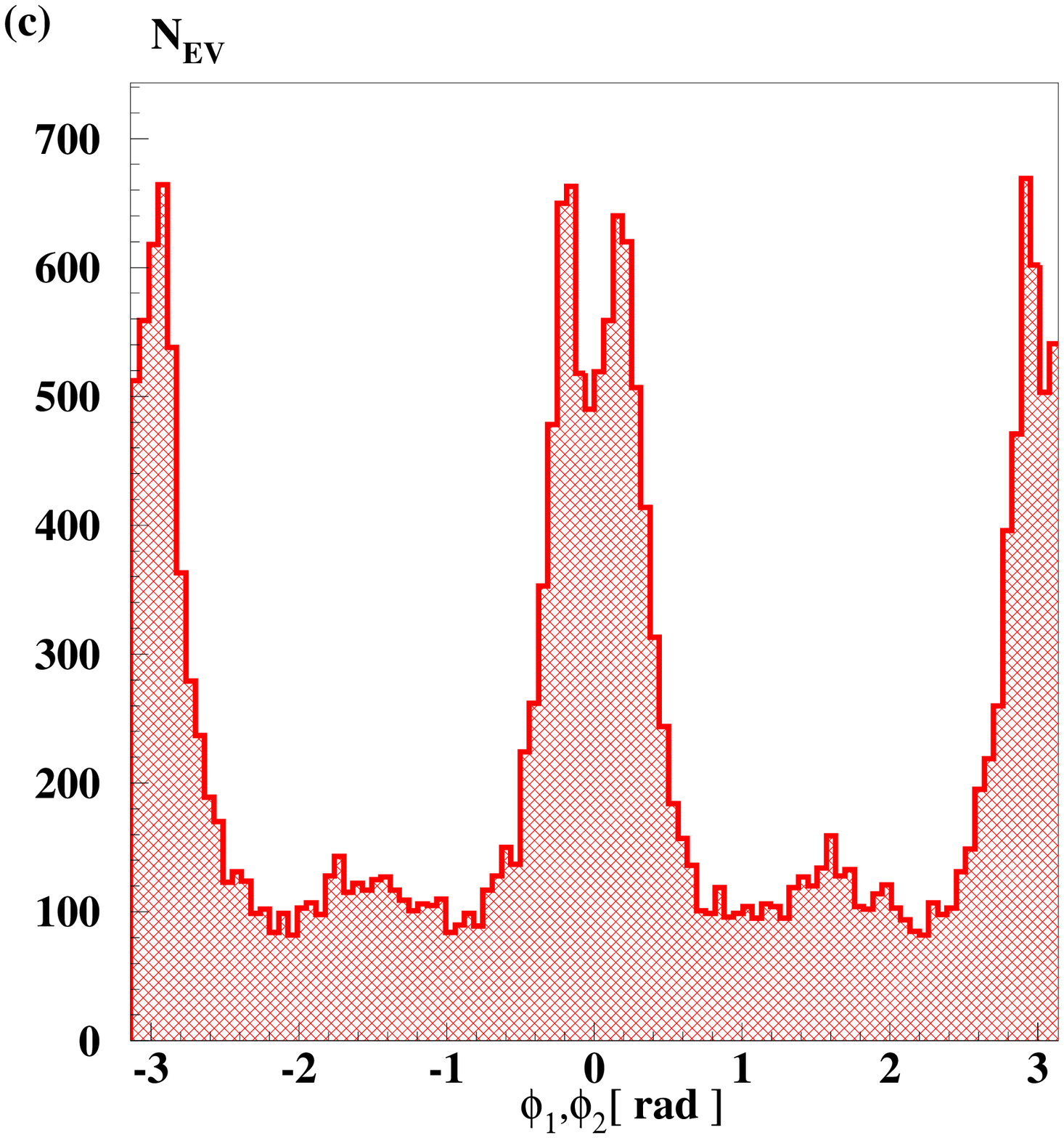}
\caption[bla]{(\textit{a}): The signal event distribution over the ($\phi_{1},\phi_{2}$) angles without any pileup contribution. (\textit{b}): The signal event distribution over the ($\phi_{1},\phi_{2}$) angles with pileup contribution. The pileup event distribution over ($\phi_{1},\phi_{2}$).}
\label{fig:pileup}
\end{center}
\end{figure}

The reconstructed variables after the detector simulation are plotted in Fig.~\ref{fig:fitted_gg_j0} and Fig.~\ref{fig:fitted_gg_j2}. The pileup tracks in the $W_{F,B}$ boson have tendencies to pull down the signal events close to the beam pipe (Fig.~\ref{fig:pileup}\,$b$) depending on their contribution to the $W$ boson. The pileup distributions for the $\phi$ angle (Fig.~\ref{fig:pileup}\,$c$) will be reflected in each $W$ boson $\phi$ distribution according to the pileup kinematics.
\par
The O'Mega matrix element calculations from WHIZARD are used to obtain weights to reweight the angular distributions as functions of the anomalous TGCs. Each Monte Carlo Standard Model event is weighted by a weight:
\[
{R({\Delta}{\kappa}_{\gamma},{\Delta\lambda}_{\gamma}) =
1+A{\Delta}{\kappa}_{\gamma}+B{{{\Delta}{\kappa}_{\gamma}}^{2}}+
C{\Delta\lambda}_{\gamma}+D{{{\Delta\lambda}_{\gamma}}^{2}}+
E{\Delta}{\kappa}_{\gamma}{\Delta\lambda}_{\gamma}},
\]
where ${\Delta}{\kappa}_{\gamma}$ and ${\Delta\lambda}_{\gamma}$ are the free parameters. The function ${R({\Delta}{\kappa}_{\gamma},{\Delta\lambda}_{\gamma})}$ describes the quadratic dependence of the differential cross-section on the coupling parameters and it is obtained in the following way: using the Standard Model events (${\Delta}{\kappa}_{\gamma}={\Delta\lambda}_{\gamma}=0$), the matrix elements of the events are recalculated for a set of five different combinations of ${\Delta}{\kappa}_{\gamma}$ and ${\Delta\lambda}_{\gamma}$ values (Table \ref{tab:set1}).
\begin{table}[h]
\begin{center}
\begin{tabular}{|c|c|c|c|c|c|} \hline
& $R_1$& $R_2$& $R_3$& $R_4$& $R_5$ \\ \hline
${\Delta}{\kappa}_{\gamma}$ & 0 & 0 & +0.001 & -0.001  & +0.001 \\ \hline
${\Delta\lambda}_{\gamma}$ & +0.001 & -0.001& 0 & 0 & +0.001 \\ \hline
\end{tabular}
\end{center}
\caption{${\Delta}{\kappa}_{\gamma}$, ${\Delta\lambda}_{\gamma}$ 
values used to calculate the reweighting coefficients.}
\label{tab:set1}
\end{table}
\par
The resulting recalculated events carry a weight which is given by the ratio of the new matrix element values compared to the Standard Model ones ($R_{i},i=1-5$). The particle momenta are left unchanged. According to the chosen ${\Delta}{\kappa}_{\gamma},{\Delta\lambda}_{\gamma}$ combinations from Table \ref{tab:set1} one gets:
\begin{eqnarray*}
 R_{1} & = & 
 1+C\mid{\Delta\lambda}_{\gamma}\mid+D\mid{\Delta\lambda}_{\gamma}^{2}\mid,\\
 R_{2} & = & 
 1-C\mid{\Delta\lambda}_{\gamma}\mid+D\mid{\Delta\lambda}_{\gamma}^{2}\mid,\\
 R_{3} & = & 
 1+A\mid{\Delta}{\kappa}_{\gamma}\mid+B\mid{\Delta}{\kappa}_{\gamma}^{2}\mid,\\
 R_{4} & = & 
 1-A\mid{\Delta}{\kappa}_{\gamma}\mid+B\mid{{{\Delta}
 {\kappa}_{\gamma}}^{2}\mid},\\
 R_{5} & = & 
 1+A\mid{\Delta}{\kappa}_{\gamma}\mid+B\mid{{{\Delta}
 {\kappa}_{\gamma}}^{2}\mid}+C\mid{\Delta\lambda}_{\gamma}\mid+
  D\mid{{{\Delta\lambda}_{\gamma}}^{2}\mid}+E\mid{\Delta}
  {\kappa}_{\gamma}\mid\mid{\Delta\lambda}_{\gamma}\mid, 
\end{eqnarray*}
where $|{\Delta}{\kappa}_{\gamma}|$=$|{\Delta\lambda}_{\gamma}|$=0.001. The coefficients \textit{A,B,C,D,E} are deduced for each event from the previous five equations.
\par
In the single $W$ boson production, four-dimensional ($\cos{\theta}$,$\cos{\theta_{1}},\phi$, energy) event distributions are fitted with MINUIT \cite{min}, minimizing ${\chi^{2}}$ as a function of $\kappa_\gamma$ and $\lambda_\gamma$ taking the Standard Model Monte Carlo sample as ``data'':
\[
\chi^{2} = \sum_{i,j,k,l}
\frac{ \left( z\cdot N^{SM}(i,j,k,l)- n \cdot z \cdot 
N^{{\Delta\kappa}_{\gamma},{\Delta\lambda}_{\gamma}}(i,j,k,l) \right)^{2}} 
{z \cdot \sigma^{2}(i,j,k,l)} +\frac{(n-1)^{2}}{(\Delta L^{2})},
\]
where \textit{i, j, k} and \textit{l} run over the reconstructed four-dimensional distribution of $\cos{\theta},\cos{\theta_{1},\phi}$ and the reconstructed $W$ boson energy, $N^{SM}(i,j,k,l)$ are the ``data'' which correspond to the Standard Model Monte Carlo sample, $N^{{\Delta\kappa}_{\gamma},{\Delta\lambda}_{\gamma}}(i,j,k,l)$ is the Standard Model event distribution weighted by the function $R({\Delta}{\kappa}_{\gamma},{\Delta\lambda}_{\gamma})$ and $\sigma(i,j,k,l)=\sqrt{N^{SM}(i,j,k,l)}$. For $W$ boson pair production, the six-dimensional ($\cos{\theta}$, $\cos{\theta_{1}}$,$\cos{\theta_{2}}$,$\phi_{1}$,$\phi_{2}$, center-of-mass energy) event distributions result in a Poisson distribution for the number of events per bin which cannot be approximated by a Gaussian distribution like in the $\gamma e$ case. Thus, they are fitted with MINUIT, minimizing the log-likelihood function:
\begin{eqnarray*}
{\log\cal L} & = & -\sum_{i,j,k,l,m,p}[
z\cdot N^{SM}(i,j,k,l,m,p)\cdot
\log \left(z\cdot n \cdot N^{{\Delta\kappa}_{\gamma},{\Delta\lambda}_{\gamma}}(i,j,k,l,m,p)\right) \\
 & & - z\cdot n \cdot N^{{\Delta\kappa}_{\gamma},{\Delta\lambda}_{\gamma}}(i,j,k,l,m,p)] + 
\frac{(n-1)^{2}}{2(\Delta L^{2})},
\end{eqnarray*}
where \textit{i,j,k,l,m} and \textit{p} run over the reconstructed six-dimensional distribution of $\cos{\theta}$, $\cos{\theta_{1}}$, $\cos{\theta_{2}}$, $\phi_{1}$, $\phi_{2}$ and the reconstructed center-of-mass energy, $N^{SM}(i,j,k,l,m,p)$ are the ``data'' which correspond to the Standard Model Monte Carlo sample and $N^{{\Delta\kappa}_{\gamma}{\Delta\lambda}_{\gamma}}$$(i$, $j$, $k$, $l$, $m$, $p)$ is the Standard Model event distribution weighted by the function $R({\Delta}{\kappa}_{\gamma}$, ${\Delta\lambda}_{\gamma})$. 
\begin{figure}[p]
\begin{center}
\epsfxsize=3.0in
\epsfysize=3.0in
\epsfbox{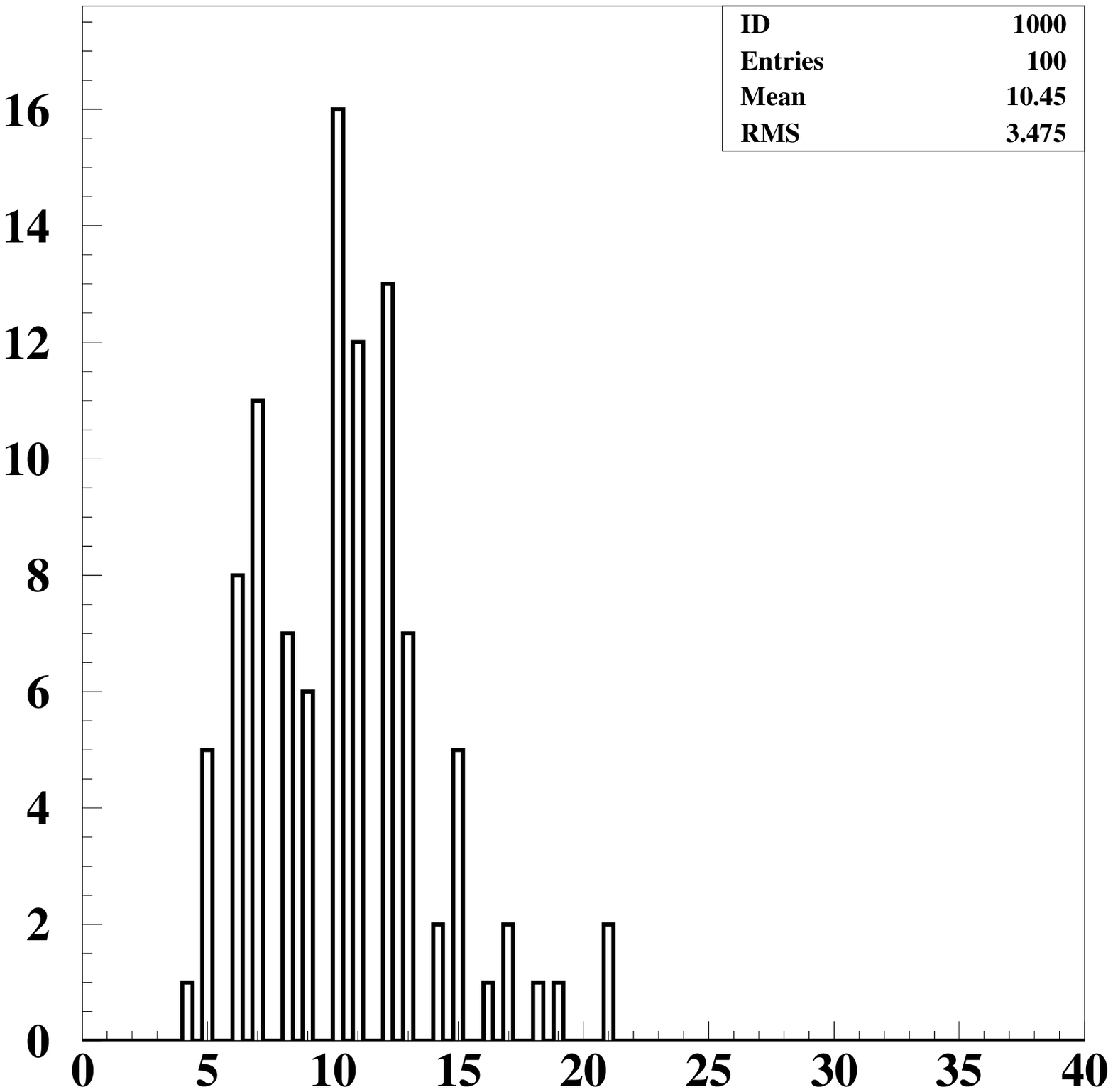}
\epsfxsize=3.0in
\epsfysize=3.0in
\epsfbox{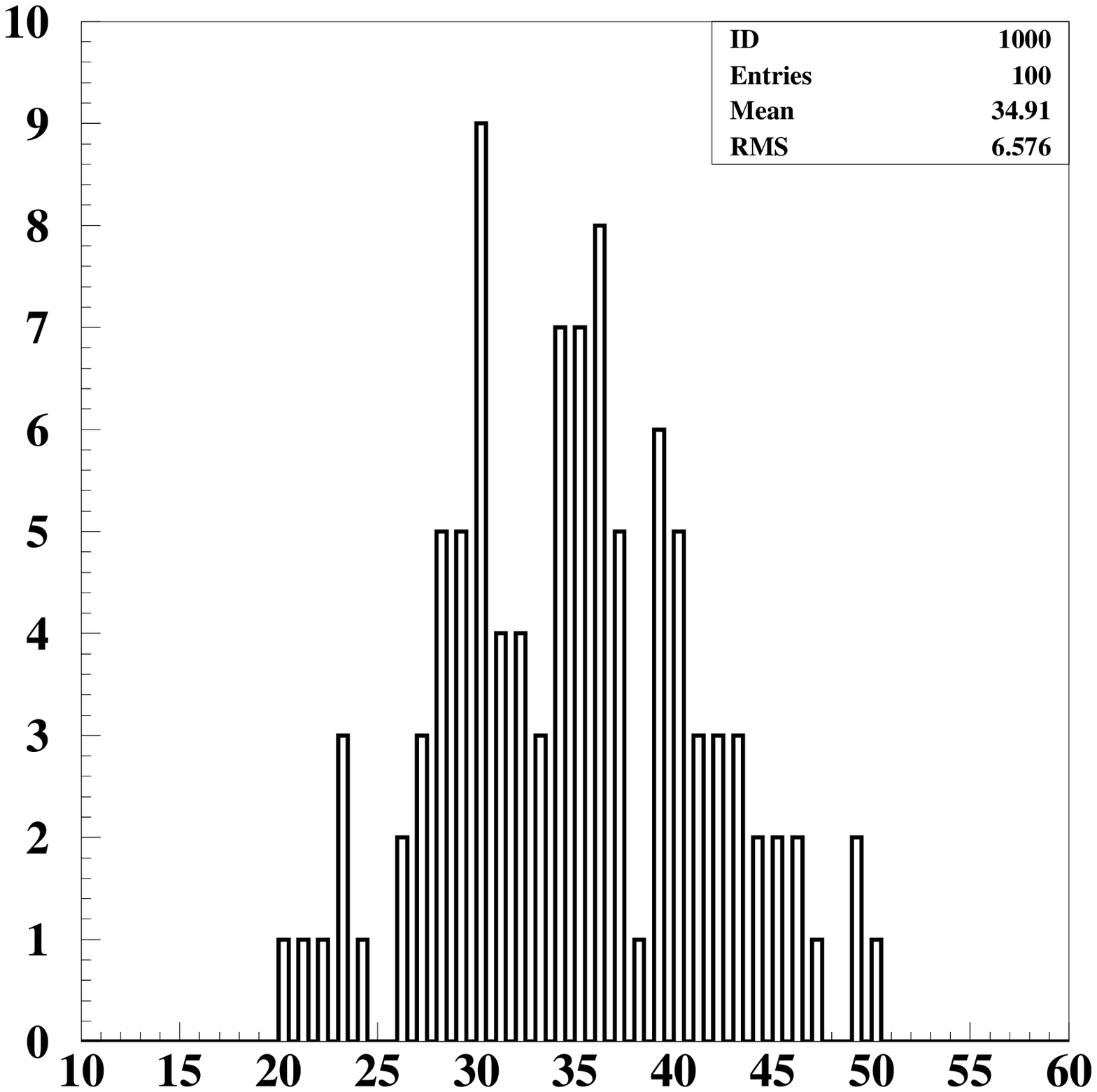}
\epsfxsize=3.0in
\epsfysize=3.0in
\epsfbox{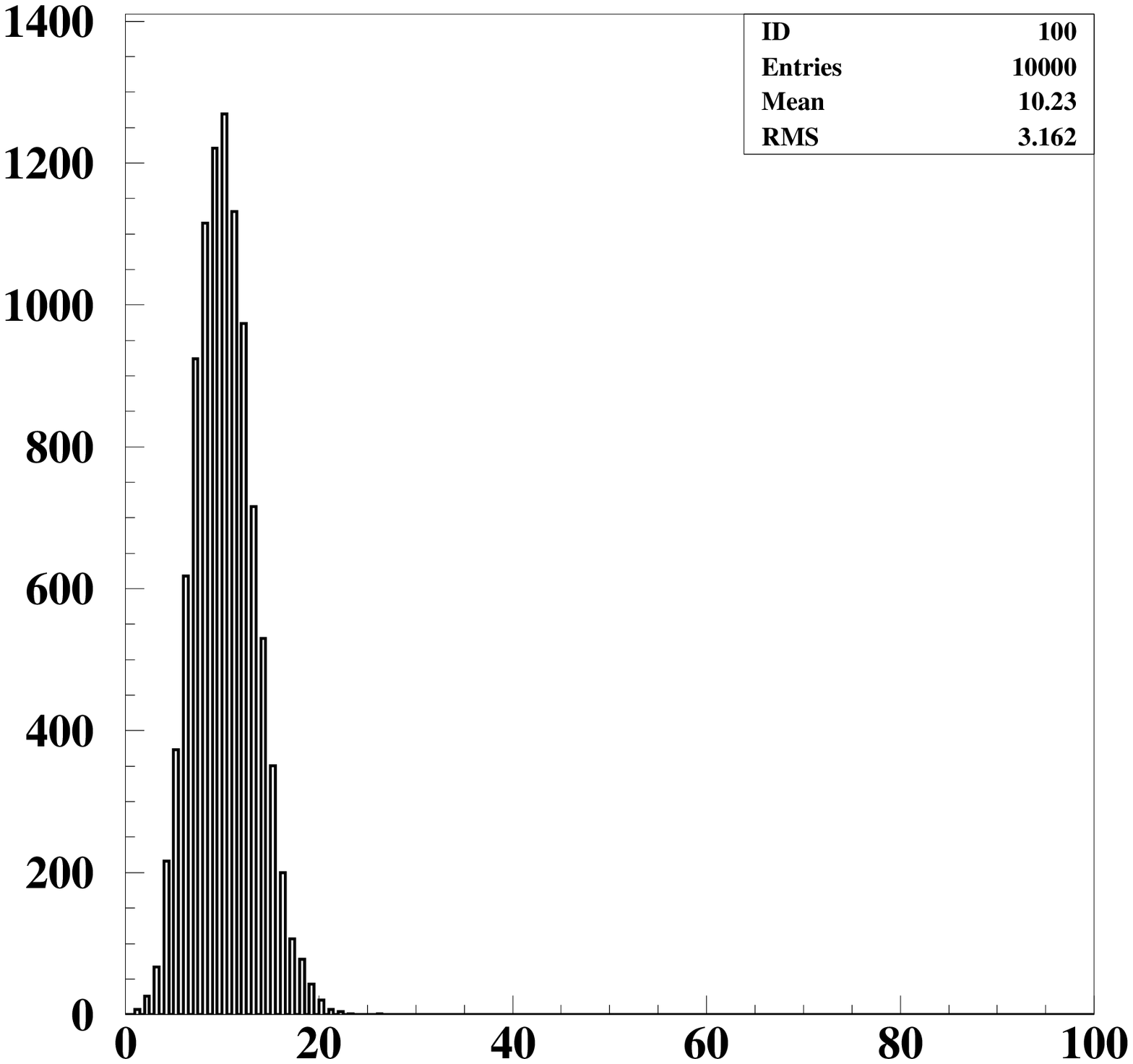}
\epsfxsize=3.0in
\epsfysize=3.0in
\epsfbox{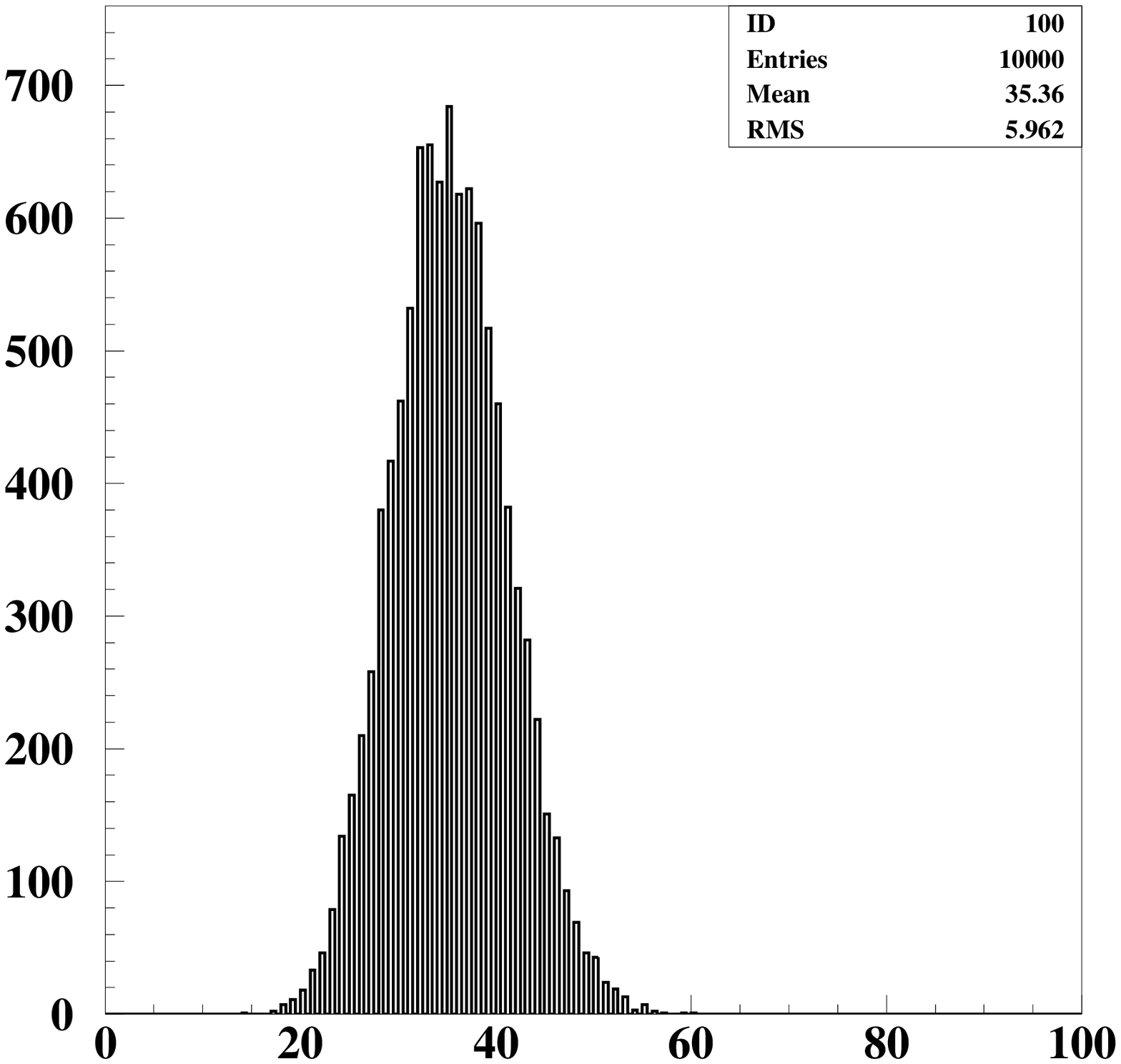}
\caption[bla]{$zN$ Poisson distributed events from one measurement (\textit{upper plots}) and 10000 different measurements (\textit{lower plots}) are shown for $\mu$=10 (\textit{left plot)} and for $\mu$=35 (\textit{right plot}).}
\label{fig:toy1}
\end{center}
\end{figure}
\par
\begin{figure}[p]
\begin{center}
\epsfxsize=3.0in
\epsfysize=3.0in
\epsfbox{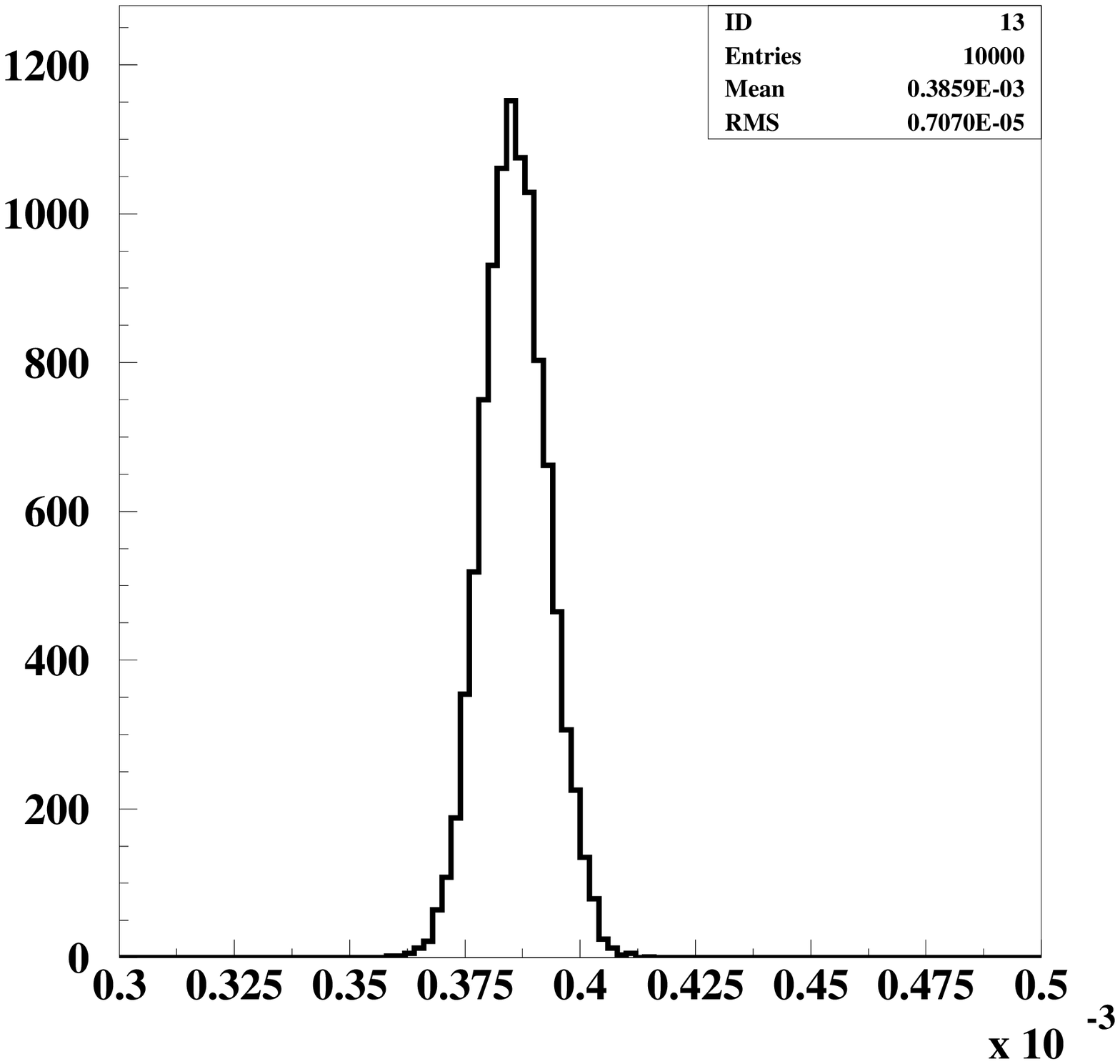}
\epsfxsize=3.0in
\epsfysize=3.0in
\epsfbox{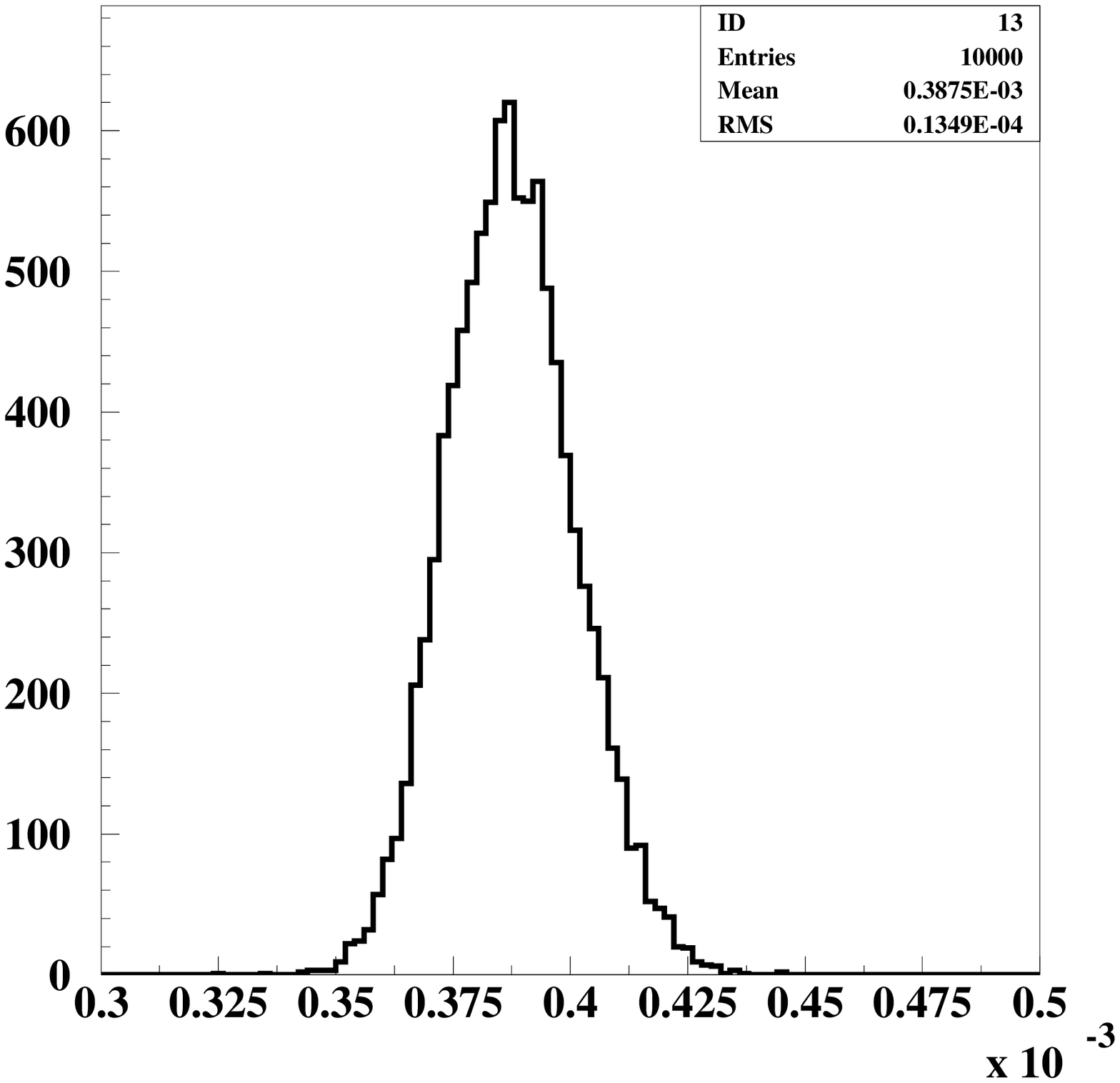}
\epsfxsize=3.0in
\epsfysize=3.0in
\epsfbox{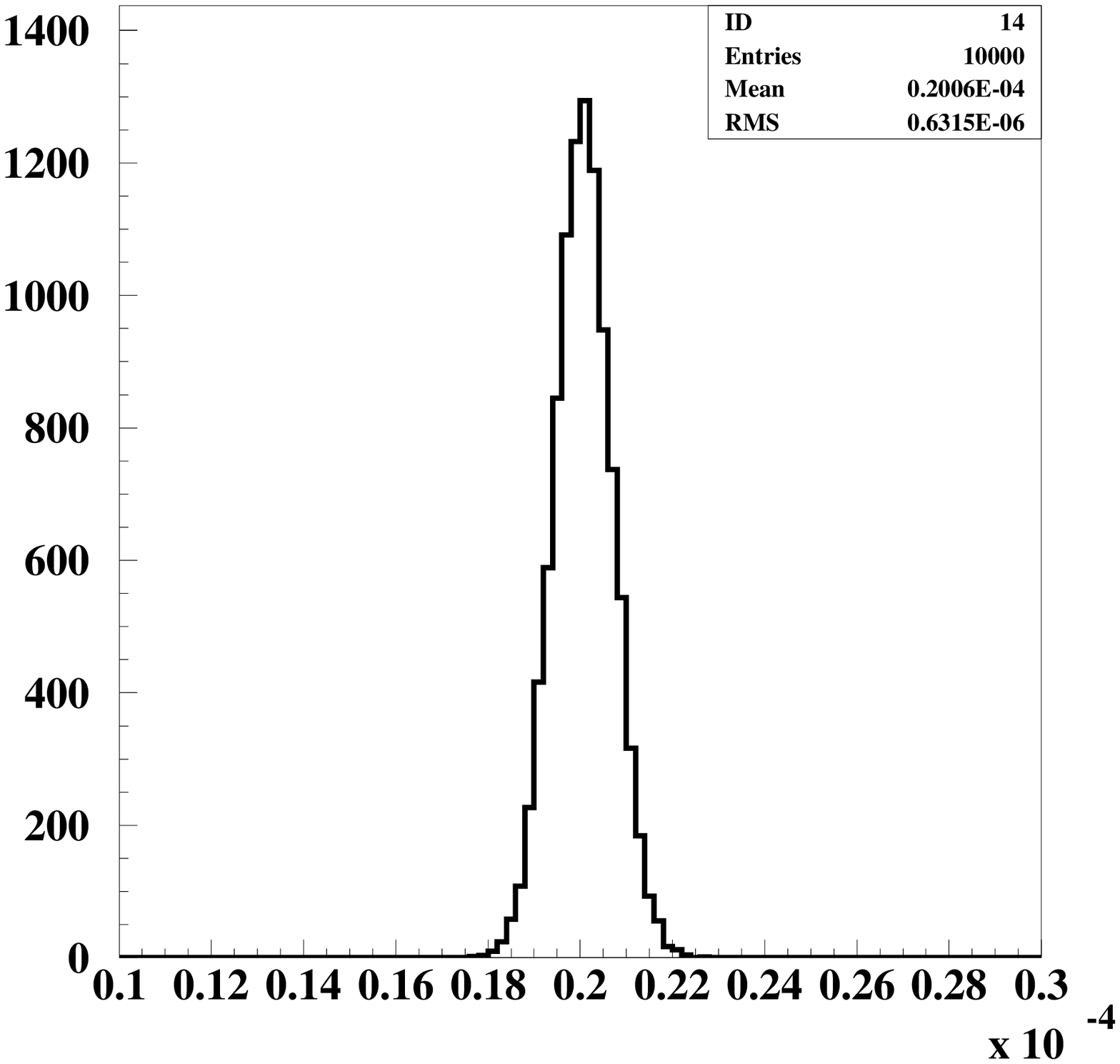}
\epsfxsize=3.0in
\epsfysize=3.0in
\epsfbox{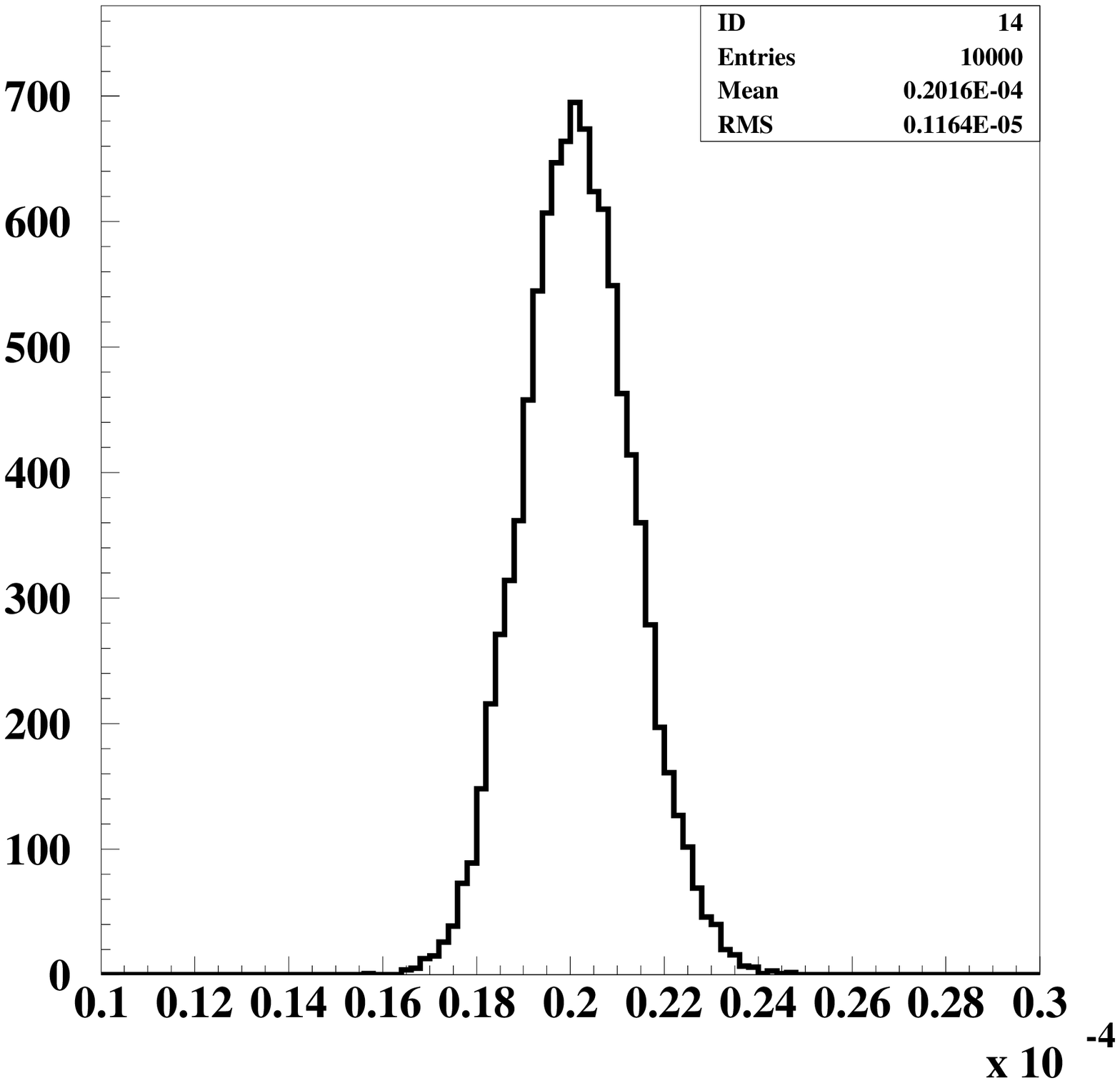}
\caption[bla]{\textit{Upper plots}: Estimated errors of the first parameter $P_{1}$ in a case (\textit{left plot}): without rescaling factor (0.3859$\cdot 10^{-3}$) and (\textit{right plot}): with a scaling factor $z$ (0.3875$\cdot 10^{-3}$) in the likelihood function. \textit{Lower plots}: Estimated errors of the second parameter $P_{2}$ in a case (\textit{left plot}): without rescaling factor (0.2006$\cdot 10^{-4}$) and (\textit{right plot}): with a scaling factor $z$ (0.2016$\cdot 10^{-4}$) in a likelihood function.}
\label{fig:toy2}
\end{center}
\end{figure}

Due to the large CPU power and the disk space needed to simulate the full statistics, the factor ${z}$ is used to set the number of signal events to the expected one after one year of running of a $\gamma e/\gamma\gamma$-collider. Thus, the number $zN$ represents the mean value of a Poisson distributed event sample expected after one year of running of a collider. The fact that events are rescaled in the fit does not influence the correct error estimation. This can be shown using the ``toy-model'' where both, ``data'' and Monte Carlo events, are simulated as two independent sets of $zN$ Poisson distributed events in 100 bins from 10000 measurements and fitted with log-likelihood function for two different cases. In the first case, the mean value is taken to be $\mu=$10 and the factor $z$ in the log-likelihood function (the function written above) is used to rescale the number of events. In the second case, the mean value is taken to be $\mu= z\cdot$10 ($\mu=35$) and the number of events in the log-likelihood function is not rescaled by the factor $z$ since this case represents the full statistics. Thus, the fitted functions in this ``toy-model'' represent 10000 one-dimensional histograms with $zN$ Poisson distributed events around the mean values $\mu =$10 and $\mu =$35 as it is shown in Fig.~\ref{fig:toy1}. The ``data'' are fitted to Monte Carlo with the same $\mu$, comparing two statistically different sets: if the ``data'' and Monte Carlo events are identical (Monte Carlo = ``data'') and if the ``data'' and Monte Carlo events are statistically independent (''data''$\neq$Monte Carlo). The Monte Carlo sample is weighted by a fit-function \footnote{The function is linear in the fit parameters and quadratic in the bin number.} depending on two fit-parameters $P_{1}$ and $P_{2}$ which represent two anomalous couplings $\Delta\kappa_{\gamma}$ and $\Delta\lambda_{\gamma}$.
The estimated errors of these parameters in Fig.~\ref{fig:toy2} are distributed around the same mean value and show a good agreement between the two different cases; $\mu =$10 (Fig.~\ref{fig:toy2}\, left plots) and  $\mu =$35 (Fig.~\ref{fig:toy2}\, right plots).
\par
\begin{figure}[p]
\begin{center}
\epsfxsize=3.0in
\epsfysize=3.0in
\epsfbox{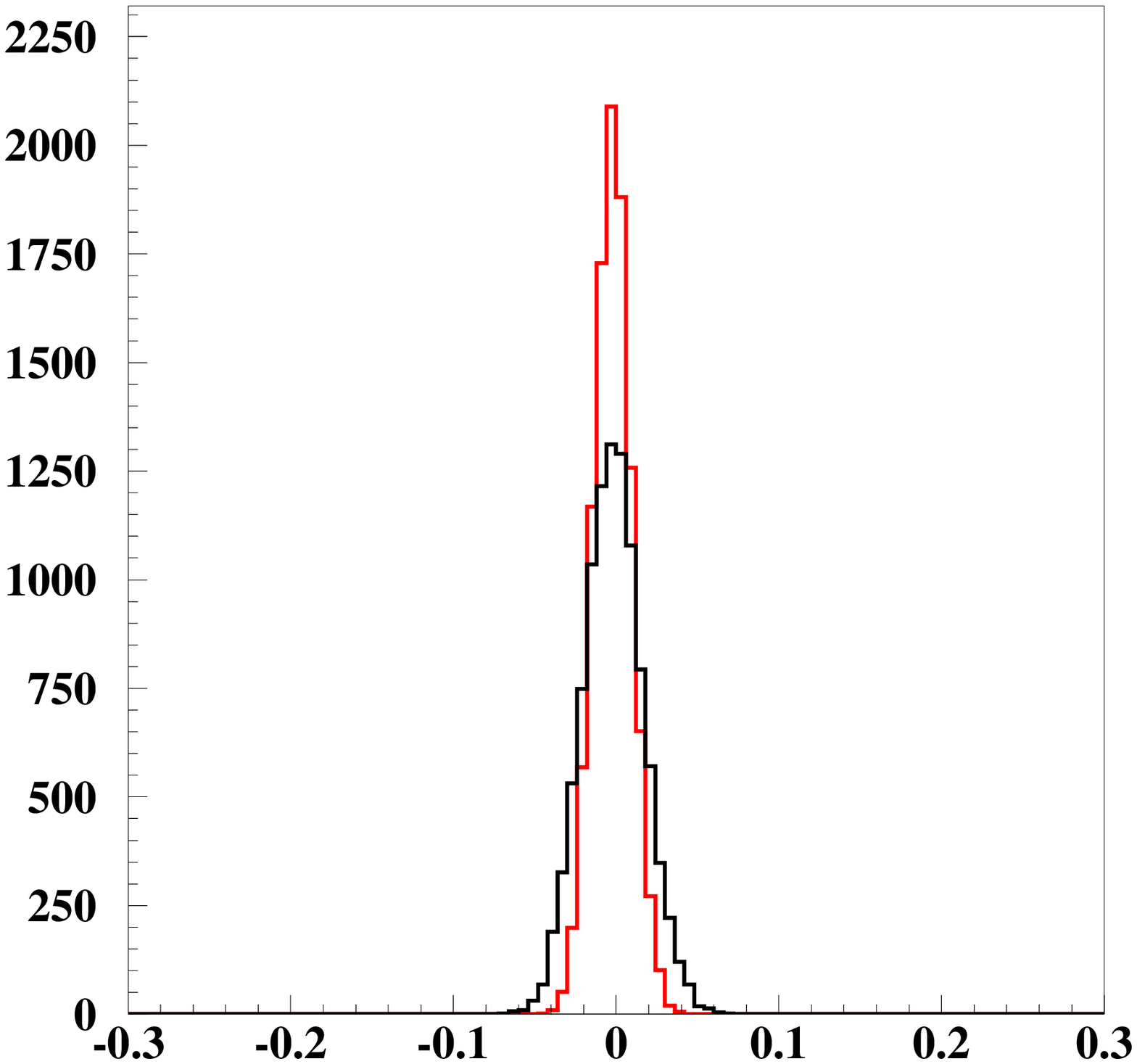}
\epsfxsize=3.0in
\epsfysize=3.0in
\epsfbox{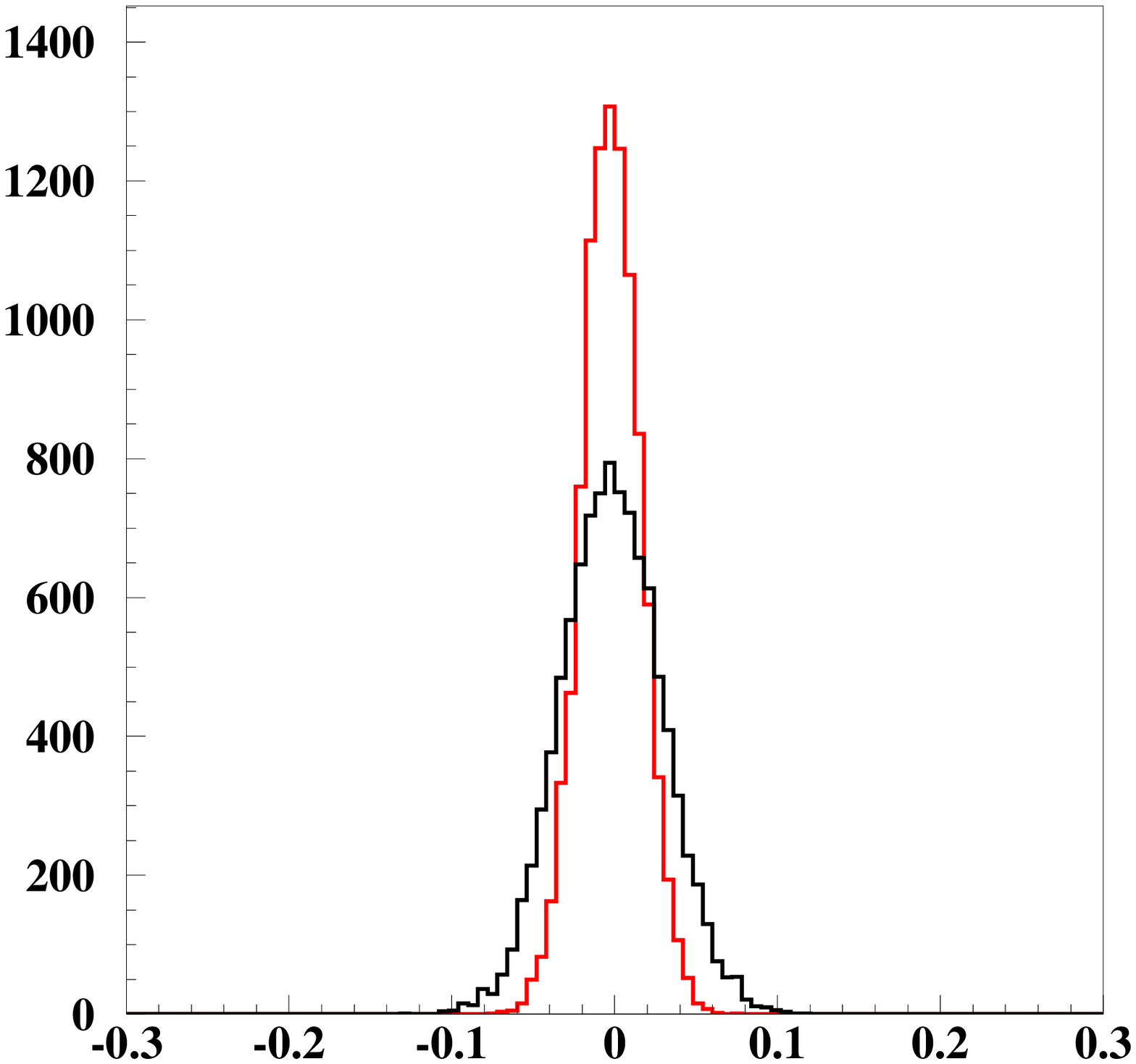}
\epsfxsize=3.0in
\epsfysize=3.0in
\epsfbox{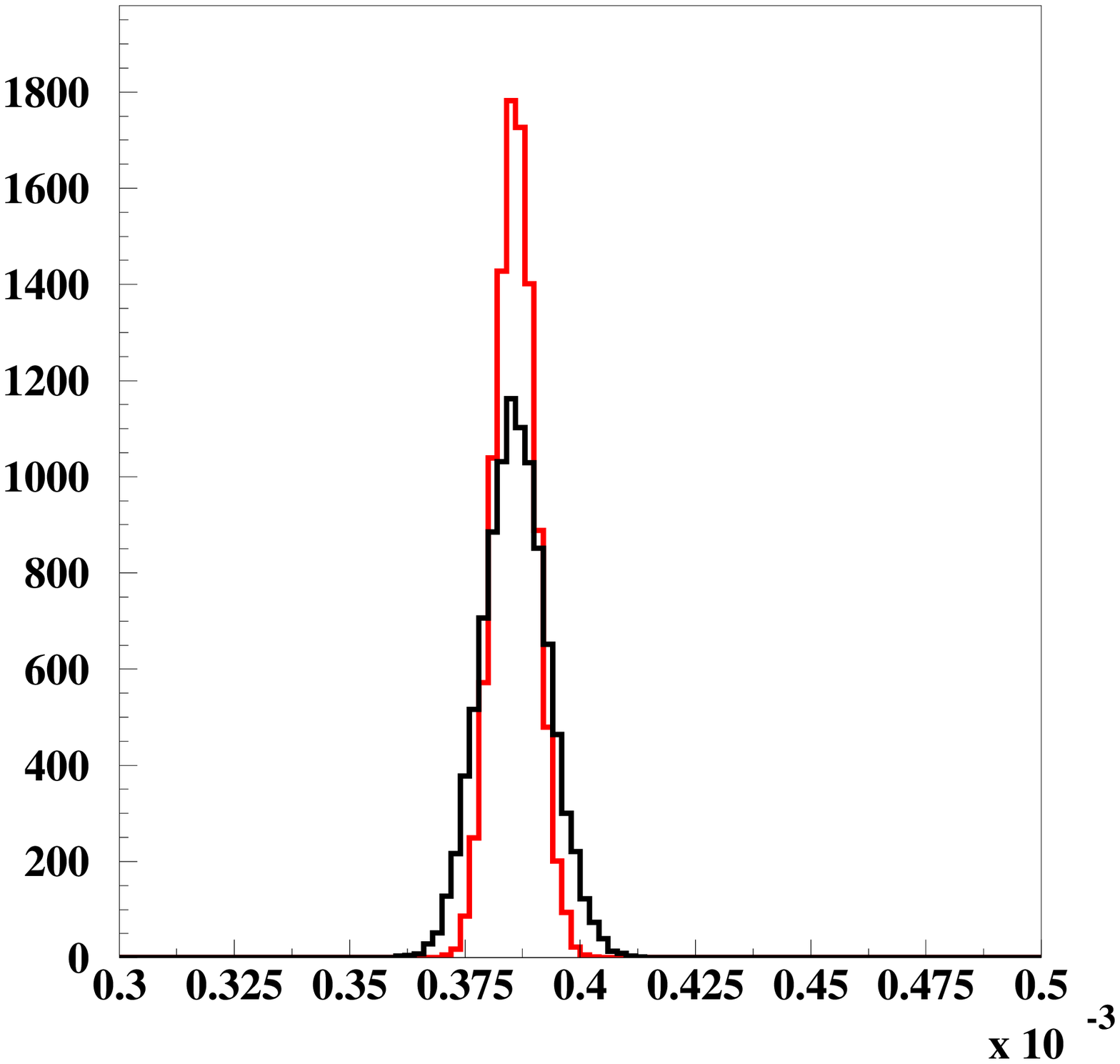}
\epsfxsize=3.0in
\epsfysize=3.0in
\epsfbox{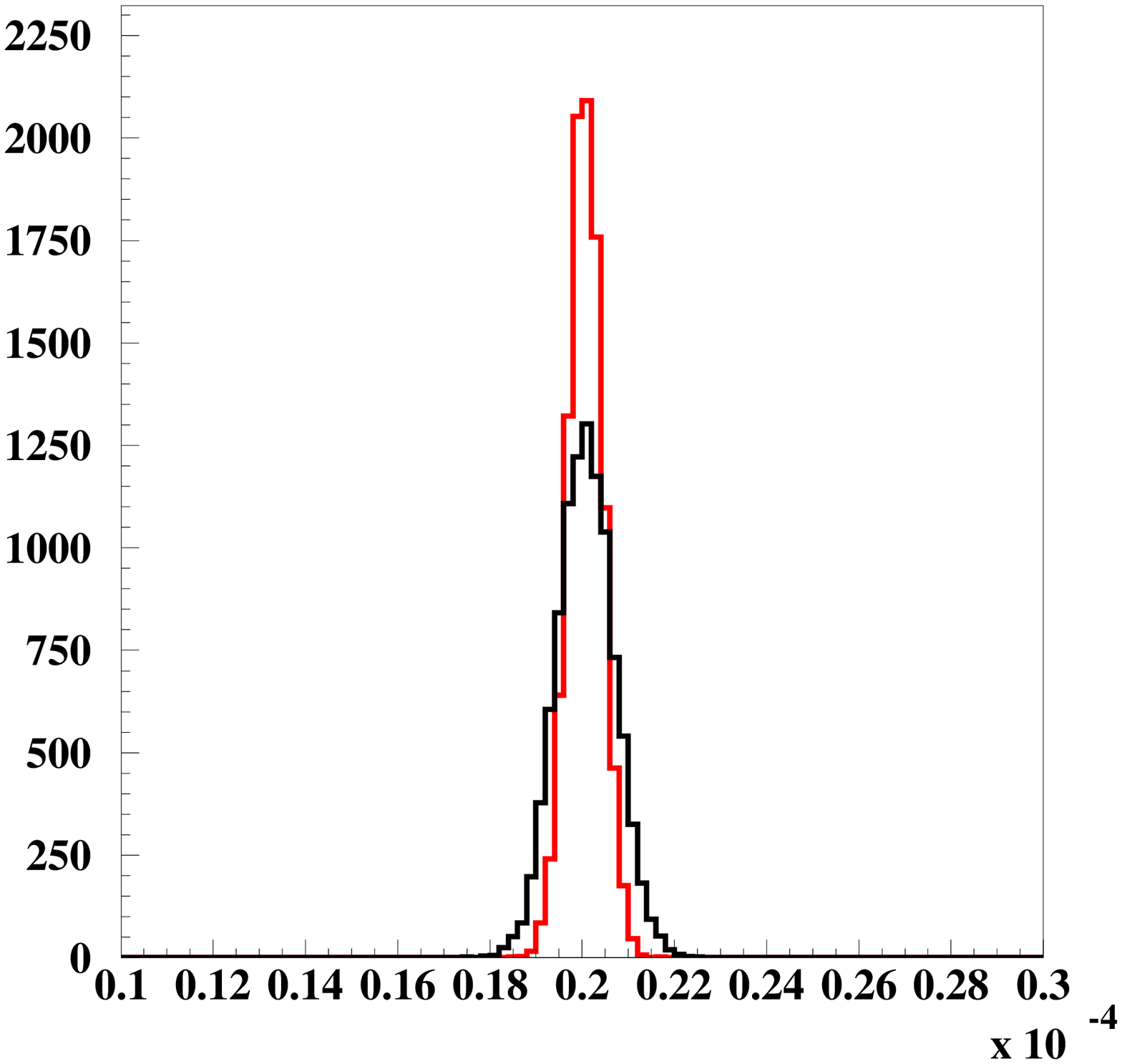}
\caption[bla]{\textit{Upper plots}: Relative deviation $\Delta$ in the fit-parameters $P_{1}$ (\textit{left plot}) and $P_{2}$ (\textit{right plot}) if the simulated events for ``data'' and Monte Carlo are statistically independent relative to the statistically identical events. $\Delta$=(N$_{MC=''data''}$-N$_{MC\neq''data''}$)/N$_{MC=''data''}$. \textit{Lower plots}: Estimated errors $P_{1}$ (\textit{left plot}) and $P_{2}$ (\textit{right plot}). The black solid line corresponds to the full statistics, i.e. $\mu$=35 and the red solid line corresponds to the rescaled case with Monte Carlo statistics increased by a factor 10 relative to the ``data'' statistics.}
\label{fig:toy3}
\end{center}
\end{figure}

In reality, the generated Monte Carlo sample is larger than the ``data'' sample and thus, normalized to the same number of events as the ``data'' sample. Having both samples, ``data'' and Monte Carlo, of same size the error is underestimated by a factor of $\sqrt{2}$, because Monte Carlo statistics is neglected. Increasing the Monte Carlo statistics this error tends to become negligible i.e. this improves in the error estimation as it is shown in Fig.~\ref{fig:toy3} (\textit{two upper plots}) where Monte Carlo statistics is ten times larger than the ``data''. The black line shows the relative deviation $\Delta$ in the $P_{1}$ (\textit{left plot}) and $P_{2}$ (\textit{right plot}) estimations if the sample size is different in ``data'' and Monte Carlo relative to the statistically identical events, i.e. [$\Delta$=(N$_{MC=''data''}$-N$_{MC\neq''data''}$)/N$_{MC=''data''}$], and if the number of events in both samples (``data'' and Monte Carlo) is the same. The black line in Fig.~\ref{fig:toy3} represents the deviation if the full statistics is simulated ($\Delta_{\mu=35}$) and narrower red line in Fig.~\ref{fig:toy3} corresponds to the case with 10 times increased Monte Carlo statistics, both located around zero. The two lower plots in Fig.~\ref{fig:toy3} represent the estimated errors on parameters $P_{1}$ (\textit{left plot}) and $P_{2}$ (\textit{right plot}).
\par
This model shows that the fit method applied in the present analysis is correct. If ``data'' and Monte Carlo samples are identical, the central value is correctly estimated since the dependence on the coupling is correctly taken into account. The error due to Monte Carlo statistics becomes negligible in the real experiment since a larger sample would be used.
\par
In the case where the background is included in the fit ${z}$ defines the sum of signal and background events and ${n}\cdot N^{{\Delta\kappa}_{\gamma},{\Delta\lambda}_{\gamma}} \rightarrow [{n} \cdot N_{signal}^{{\Delta\kappa}_{\gamma},{\Delta\lambda}_{\gamma}}+N_{bck}]$. The number of background events is normalized to the effective $W$ boson production cross-section in order to obtain the corresponding number of background events after one year of running of an $\gamma e/\gamma\gamma$-collider. It is assumed that the total normalization (efficiency, luminosity, electron polarization) is only known with a relative uncertainty $\Delta L$. Thus, $n$ is taken as a free parameter in the fit and constrained to unity with the assumed normalization uncertainty. Per construction the fit is bias-free and thus returns always exactly the Standard Model as central values. In all running modes $\Delta L=0.1\%$ is a realistic precision that can be achieved except for the $J_{Z}=0$ state where due to the small number of events the luminosity is expected to be measured with a larger error of $\Delta L=1\%$.
\newpage
\section{Error estimations}
\begin{figure}[htb]
\begin{center}
\epsfxsize=2.5in
\epsfysize=2.5in
\epsfbox{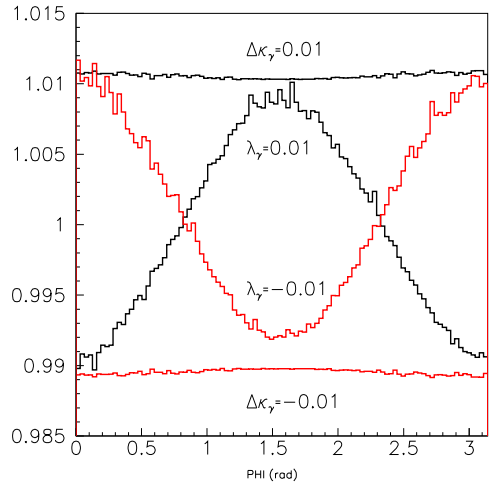}
\caption[bla]{Deviations from the Standard Model event distribution over the $\phi$ angle due to the anomalous couplings $\kappa_{\gamma}$ and $\lambda_{\gamma}$. The sensitivity to the anomalous ${\lambda}_{\gamma}$ is clearly visible in the shape of the distribution.}
\label{fig:phi}
\end{center}
\end{figure}

Table \ref{tab:bestge} shows the best estimates of statistical errors for the two couplings measured in $\gamma e$ interactions at fixed center-of-mass energy ${\sqrt{s_{\gamma e}}=450}\,{\rm GeV}$ at generator level. Accepting all events, two-dimensional ($\cos\theta,\cos|\theta_{1}|$) and three-dimensional ($\cos\theta,|\cos\theta_{1}|,\phi$) distributions are used in a two-parameter fit. A two-parameter fit means that both couplings are allowed to vary freely as well as the normalization \textit{n}. The number of events is normalized to the number that is expected to be collected with an integrated luminosity of 110 fb$^{-1}$ in the high energy peak. The 3D fit results in a higher sensitivity to ${\lambda}_{\gamma}$ due to the shape sensitivity of the $\phi$ event distribution to the anomalous ${\lambda}_{\gamma}$ values. The sensitivity of the $\phi$ distribution to the anomalous ${\kappa}_{\gamma}$ is negligible as it is clear from Fig.~\ref{fig:phi}. The estimated sensitivities to the $\kappa_{\gamma}$ and ${\lambda}_{\gamma}$ in the 2D two-parameter fit from the analytic formula are presented in Appendix B.
\begin{table}[htb]
\begin{center} 
\begin{tabular}{|l||c|c|c||c|c|c|} \hline
$|J_{Z}|=3/2$ & \multicolumn{3}{|c||}{2D fit} & \multicolumn{3}{c|}{3D fit}\\ \hline
${{\Delta}L}$ & 1$\%$ & 0.1$\%$ & 0 & 1$\%$ & 0.1$\%$ & 0 \\ \hline\hline
${\Delta}{\kappa}_{\gamma}{\cdot}10^{-3}$ & 4.3/5.1 & 1.0/1.1 & 0.4/0.5 & 
 3.3/3.4 & 1.0/1.0 & 0.4/0.4 \\ \hline
${\Delta}{\lambda}_{\gamma}{\cdot}10^{-4}$ & 18/30 & 15/23 & 15/22 & 2.9/3.1 & 2.7/2.9 & 2.7/2.9 \\ \hline
\end{tabular}
\end{center}
\caption{Estimated statistical errors for ${\kappa}_{\gamma}$ and ${\lambda}_{\gamma}$ from the two-dimensional (2D) and three-dimensional (3D) two-parameter fit at generator level for the real/parasitic $\gamma e$ mode at ${\sqrt{s_{\gamma e}}=450}$ GeV. The number of events in both modes is normalized to the expected one with integrated luminosity of 110 fb$^{-1}$ in the high energy peak.}
\label{tab:bestge}
\end{table}
\par
\begin{figure}[htb]
\begin{center}
\epsfxsize=3.5in
\epsfysize=3.5in
\epsfbox{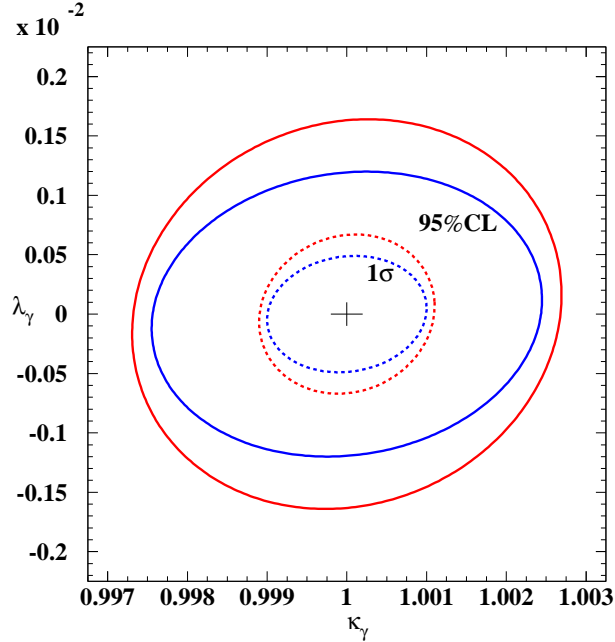}
\caption[bla]{1$\sigma$ (dashed lines) and 95$\%$ CL (solid lines) contours in the $\kappa_{\gamma}-\lambda_{\gamma}$ plane obtained from the 4D fit for ${{\Delta}L}=0.1\%$ for real (blue) and parasitic (red) $\gamma e$ modes. The cross denotes the Standard Model values of $\kappa_{\gamma}$ and $\lambda_{\gamma}$.}
\label{fig:contour1}
\end{center}
\end{figure}

More realistic errors are given in Table \ref{tab:set2} estimated using CIRCE2 spectra at $\sqrt{s_{e^{-}e^{-}}}$ = 500 GeV. A two-parameter four-dimensional (4D) fit at detector level, with and without pileup is performed. In this estimation the cut of $5^{\circ}$ on $W$ boson production angle is not applied. Including the background events and applying the cut of $5^{\circ}$ the obtained statistical errors are shown in Table \ref{tab:set3}. The $\kappa_{\gamma} - \lambda_{\gamma}$ contour plot for the two-parameter fit with pileup and background is shown in Fig.~\ref{fig:contour1} assuming a normalization error of 0.1${\%}$.
\begin{table}[htb]
\begin{center} 
\begin{tabular}{|l||c|c|c||c|c|c|} \hline
$|J_{Z}|=3/2$ & \multicolumn{3}{|c||}{without pileup} & \multicolumn{3}{c|}{with pileup}\\
\hline
${{\Delta}L}$ & 1$\%$ & 0.1$\%$ & 0 &
 1$\%$ & 0.1$\%$ & 0 \\ \hline\hline
${\Delta}{\kappa}_{\gamma}{\cdot}10^{-3}$ & 3.4/4.0 & 1.0/1.0 & 0.5/0.5 & 
 3.5/4.5 & 1.0/1.0 & 0.5/0.5 \\ 
\hline
${\Delta}{\lambda}_{\gamma}{\cdot}10^{-4}$ & 4.9/5.5 & 4.5/5.2 & 4.5/5.1 & 
5.2/6.7 & 4.9/6.4 & 4.9/6.4 \\ \hline
\end{tabular}
\end{center}
\caption{Estimated statistical errors for ${\kappa}_{\gamma}$ and ${\lambda}_{\gamma}$ from the two-parameter 4D fit at detector level for the real/parasitic $\gamma e$ mode at ${\sqrt{s_{e^{-}e^{-}}}=500}$ GeV, without and with pileup.}
\label{tab:set2}
\end{table}
\begin{table}[htb]
\begin{center}
\begin{tabular}{|l||c|c|c|} \hline
$|J_{Z}|=3/2$ & \multicolumn{3}{|c|}{pileup+background} \\ \hline
${{\Delta}L}$ & 1$\%$ & 0.1$\%$ & 0 \\ \hline\hline
${\Delta}{\kappa}_{\gamma}{\cdot}10^{-3}$ & 3.6/4.8 & 1.0/1.1 & 0.5/0.6 \\ 
\hline
${\Delta}{\lambda}_{\gamma}{\cdot}10^{-4}$ & 5.2/7.0 & 4.9/6.7 & 4.9/6.7 \\ 
\hline
\end{tabular}
\end{center}
\caption{Estimated statistical errors for ${\kappa}_{\gamma}$ and ${\lambda}_{\gamma}$ from the two-parameter 4D fit at detector level for the real/parasitic $\gamma e$ mode at ${\sqrt{s_{e^{-}e^{-}}}=500}$ GeV, with pileup and background events.}
\label{tab:set3}
\end{table}
\par
The main error on ${\kappa}_{\gamma}$ comes from the luminosity measurement while ${\lambda}_{\gamma}$ is not sensitive to that uncertainty. The two different $\gamma e$ modes give the same estimation for ${\Delta}{\kappa}_{\gamma}$ while ${\Delta}{\lambda}_{\gamma}$ is more sensitive to the different modes. The difference in the estimated ${\Delta}{\lambda}_{\gamma}$ for the two modes is a consequence of the ambiguity in the $W$ boson production angle which is present in the parasitic mode\footnote{In the parasitic mode only $|\cos\theta|$ has been reconstructed.} and due to the fact that the distance between the conversion region and the interaction point is larger in the real mode than in the parasitic mode. A smaller distance $b$ between the conversion and the interaction region, increases the luminosity at the price of a broader energy spectrum at lower energies\footnote{This is the pileup contribution.} in the parasitic $\gamma e$ mode. That decreases the sensitivity of the ${\lambda}_{\gamma}$ measurement.
\par
The pileup contribution is larger in the parasitic than in the real mode and therefore it influences the $W$ boson distributions (energy and angular) more than in the real mode. This leads to a decrease in sensitivity for ${\lambda}_{\gamma}$ of $\sim 10\,{\%}$ in the real and of $\sim 25\,{\%}$ in the parasitic mode while the influence on ${\Delta}{\kappa}_{\gamma}$ is negligible. The influence of the background is not so stressed as it is for the pileup. In the real mode it is almost negligible while it contributes to the parasitic mode decreasing the sensitivity of ${\lambda}_{\gamma}$ by less than $5\,{\%}$. {All comparisons in $\gamma e$ modes are done assuming ${{\Delta}L}=0.1\%$.
\par
The correlation between the fit parameters ${\Delta}{\kappa}_{\gamma}$ and ${\Delta}{\lambda}_{\gamma}$ is found to be negligible and it is shown in Table \ref{tab:set4} while ${\Delta}{\kappa}_{\gamma}$ strongly depends on \textit{n}.
\begin{table}[h]
\begin{center}
\begin{tabular}{|l||c|c|c|} \hline
& \multicolumn{3}{c|}{pileup+background}\\ \hline \hline
& ${\Delta}{\kappa}_{\gamma}$ & $\textit{n}$ & ${\Delta}{\lambda}_{\gamma}$ \\ \hline \hline
${\Delta}{\kappa}_{\gamma}$ & 1.000 & -0.857 & 0.122 \\ \hline
$\textit{n}$ & -0.857 & 1.000 & -0.094 \\ \hline
${\Delta}{\lambda}_{\gamma}$ & 0.122 & -0.094 & 1.000 \\ \hline
\end{tabular}
\end{center}
\caption{Correlation matrix for the two-parameter fit (${{\Delta}L}=0.1\%$) in the real $\gamma e$ mode.}
\label{tab:set4}
\end{table}
\par
The best estimates of statistical errors for the two couplings $\kappa_{\gamma}$ and $\lambda_{\gamma}$ measured in $\gamma\gamma$ interactions at fixed center-of-mass energy $\sqrt{s_{\gamma\gamma}}=400$ and 800 GeV in the pure\footnote{All events are produced either in the $J_{Z}=0$ state or in the $|J_{Z}|=2$ state without any contribution from other helicity combination.} $J_{Z}$ state at generator level are shown in Table \ref{tab:bestgg}. Accepting all events, five angular event distributions ($\cos\theta,|\cos\theta_{1}|,|\cos\theta_{2}|,\phi_{1},\phi_{2}$) are used in the two-parameter fit for both $J_{Z}$ states. The number of events is normalized to the expected number having an integrated luminosity of 110 fb$^{-1}$ in the high energy peak.
\begin{table}[htb]
\begin{center} 
\begin{tabular}{|l||c|c|c||c|c|c||c|c|c||c|c|c|} \hline
\multicolumn{1}{|c||}{110 fb$^{-1}$} & \multicolumn{6}{|c||}{$\sqrt{s_{\gamma\gamma}}=400$ GeV} & \multicolumn{6}{|c|}{$\sqrt{s_{\gamma\gamma}}=800$ GeV} \\ \hline
\multicolumn{1}{|c||}{5D fit} & \multicolumn{3}{|c||}{$J_{Z}=0$} & \multicolumn{3}{|c||}{$|J_{Z}|=2$} & \multicolumn{3}{|c||}{$J_{Z}=0$} & \multicolumn{3}{|c|}{$|J_{Z}|=2$} \\ \hline
${{\Delta}L}$ & 1$\%$ & 0.1$\%$ & 0 & 1$\%$ & 0.1$\%$ & 0 & 1$\%$ & 0.1$\%$ & 0 & 1$\%$ & 0.1$\%$ & 0 \\ \hline\hline
${\Delta}{\kappa}_{\gamma}{\cdot}10^{-4}$ & 14.4 & 5.4 & 2.6 & 20.1 & 6.2 & 3.8 & 7.2 & 4.5 & 2.4 & 8.1 & 4.6 & 2.6 \\ \hline
${\Delta}{\lambda}_{\gamma}{\cdot}10^{-4}$ & 3.0 & 3.0 & 3.0 & 1.6 & 1.6 & 1.6  & 1.3 & 1.3 & 1.3 & 0.63 & 0.58 & 0.56 \\ \hline
\end{tabular}
\end{center}
\caption{Estimated statistical errors for ${\kappa}_{\gamma}$ and ${\lambda}_{\gamma}$ from the five-dimensional (5D) two-parameter fit at generator level for the $J_{Z}=0$ and $|J_{Z}|=2$ at $\gamma\gamma$ collisions at $\sqrt{s_{\gamma\gamma}}=400$ and 800 GeV. The number of events for both $J_{Z}=0$ states is normalized to the expected one with integrated luminosity of 110 fb$^{-1}$ in the high energy peak.}
\label{tab:bestgg}
\end{table}
The estimated sensitivities are similar to the expected ones from the Chapter 6: the precision of the $\kappa_{\gamma}$ measurement is somewhat better in $J_{Z}=0$ state than in $|J_{Z}|=2$ while the opposite stands for the $\lambda_{\gamma}$ measurement. This is not valid any more for the $\kappa_{\gamma}$ measurement in the $|J_{Z}|=0$ state if $\Delta L=1\%$ which is the more realistic case. The increase of the center-of-mass energy from $\sqrt{s_{\gamma\gamma}}=400$ GeV to $\sqrt{s_{\gamma\gamma}}=800$ GeV leads to an improvement of the $\kappa_{\gamma}$ and $\lambda_{\gamma}$ sensitivities. While at $\sqrt{s_{\gamma\gamma}}=400$ GeV the information about $\kappa_{\gamma}$ is contained mainly in the cross-section at $\sqrt{s_{\gamma\gamma}}=800$ GeV the shape of the event distributions to provides the information about $\kappa_{\gamma}$.
\par
Event distributions simulated with CIRCE2 spectra at $\sqrt{s_{e^{-}e^{-}}}=500$ GeV and passed through the detector, with or without pileup, are fitted and the results of the two-parameter fit are shown in Table \ref{tab:set5}. To increase the sensitivity of the fit, the sixth distribution, over the center-of-mass energy, is included.
\par
As in the case of $\gamma e$, the main error on $\kappa_{\gamma}$ comes from the luminosity measurement while $\lambda_{\gamma}$ is not sensitive to that uncertainty. The larger sensitivity of $\lambda_{\gamma}$ than $\kappa_{\gamma}$ to the energy is reflected in the estimated values after the energy beam spectra are included into the simulation. On the other hand, there is always a mixture of both $J_{Z}$ states so that the contribution from the $J_{Z}=0$ state will increase the error on $\lambda_{\gamma}$ in $|J_{Z}|=2$ in addition. In the $|J_{Z}|=2$ for $\Delta L=0.1\%$, $\lambda_{\gamma}$ is also more influenced by pileup which decreases the sensitivity by 50$\%$ while $\Delta\kappa_{\gamma}$ is increased by 10$\%$. In the $J_{Z}=0$ for $\Delta L=1\%$, the pileup decreases the sensitivity in the $\lambda_{\gamma}$ measurement by 45$\%$ and in $\kappa_{\gamma}$ the error is increased by 35$\%$ due to a larger error on the luminosity measurement. Including in the fit the background events the estimated sensitivity in $\kappa_{\gamma}$ decreases $\approx 3\%$ and in $\lambda_{\gamma}$ decreases approximately 5$\%$ for both $J_{Z}$ states (Table \ref{tab:set6}) assuming $\Delta L=1\%$ for the $J_{Z}=0$ state and $\Delta L=0.1\%$ for the $|J_{Z}|=2$ state. The correlation between the fit parameters ${\Delta}{\kappa}_{\gamma}$ and ${\Delta}{\lambda}_{\gamma}$ is found to be stronger than in case of TGCs in $\gamma e$ as it is shown in Table \ref{tab:set7}; ${\Delta}{\kappa}_{\gamma}$ strongly depends on \textit{n}.
\begin{table}[htb]
\begin{center} 
\begin{tabular}{|l||c|c|c||c|c|c|} \hline
\multicolumn{1}{|c||}{6D fit} & \multicolumn{3}{|c||}{without pileup} & \multicolumn{3}{c|}{with pileup}\\ \hline
 & \multicolumn{3}{|c||}{$J_{Z}=0/|J_{Z}|=2$} & \multicolumn{3}{c|}{$J_{Z}=0/|J_{Z}|=2$} \\ \hline \hline
${{\Delta}L}$ & 1$\%$ & 0.1$\%$ & 0 & 1$\%$ & 0.1$\%$ & 0 \\ \hline\hline
${\Delta}{\kappa}_{\gamma}{\cdot}10^{-4}$ & 19.9/29.9 & 5.5/6.2 & 2.6/3.7 & 26.9/37.4 & 5.8/6.8 & 3.0/4.6 \\ \hline
${\Delta}{\lambda}_{\gamma}{\cdot}10^{-4}$ & 3.7/3.1 & 3.7/3.1 & 3.7/3.1 & 5.4/4.6 & 5.2/4.6 & 5.2/4.6 \\ \hline
\end{tabular}
\end{center}
\caption{Estimated statistical errors for ${\kappa}_{\gamma}$ and ${\lambda}_{\gamma}$ from the 6D fit at detector level for both $J_{Z}$ states in $\gamma\gamma$ collisions $\sqrt{s_{e^{-}e^{-}}}=500$ GeV, without and with pileup.}
\label{tab:set5}
\end{table}
\begin{table}[htb]
\begin{center}
\begin{tabular}{|l||c|c|c|} \hline
\multicolumn{1}{|c||}{6D fit} & \multicolumn{3}{|c|}{pileup+background} \\ \hline
 & \multicolumn{3}{|c|}{$J_{Z}=0/|J_{Z}|=2$} \\ \hline\hline
${{\Delta}L}$ & 1$\%$ & 0.1$\%$ & 0 \\ \hline\hline
${\Delta}{\kappa}_{\gamma}{\cdot}10^{-4}$ & 27.8/37.8 & 5.9/7.0 & 3.1/4.8 \\ \hline
${\Delta}{\lambda}_{\gamma}{\cdot}10^{-4}$ & 5.7/4.8 & 5.6/4.8 & 5.6/4.8 \\ \hline
\end{tabular}
\end{center}
\caption{Estimated statistical errors for ${\kappa}_{\gamma}$ and ${\lambda}_{\gamma}$ from the two-parameter 6D fit at detector level for both $J_{Z}$ states in $\gamma\gamma$ collisions $\sqrt{s_{e^{-}e^{-}}}=500$ GeV, with pileup and background events.}
\label{tab:set6}
\end{table}
\begin{figure}[htb]
\begin{center}
\epsfxsize=3.0in
\epsfysize=3.0in
\epsfbox{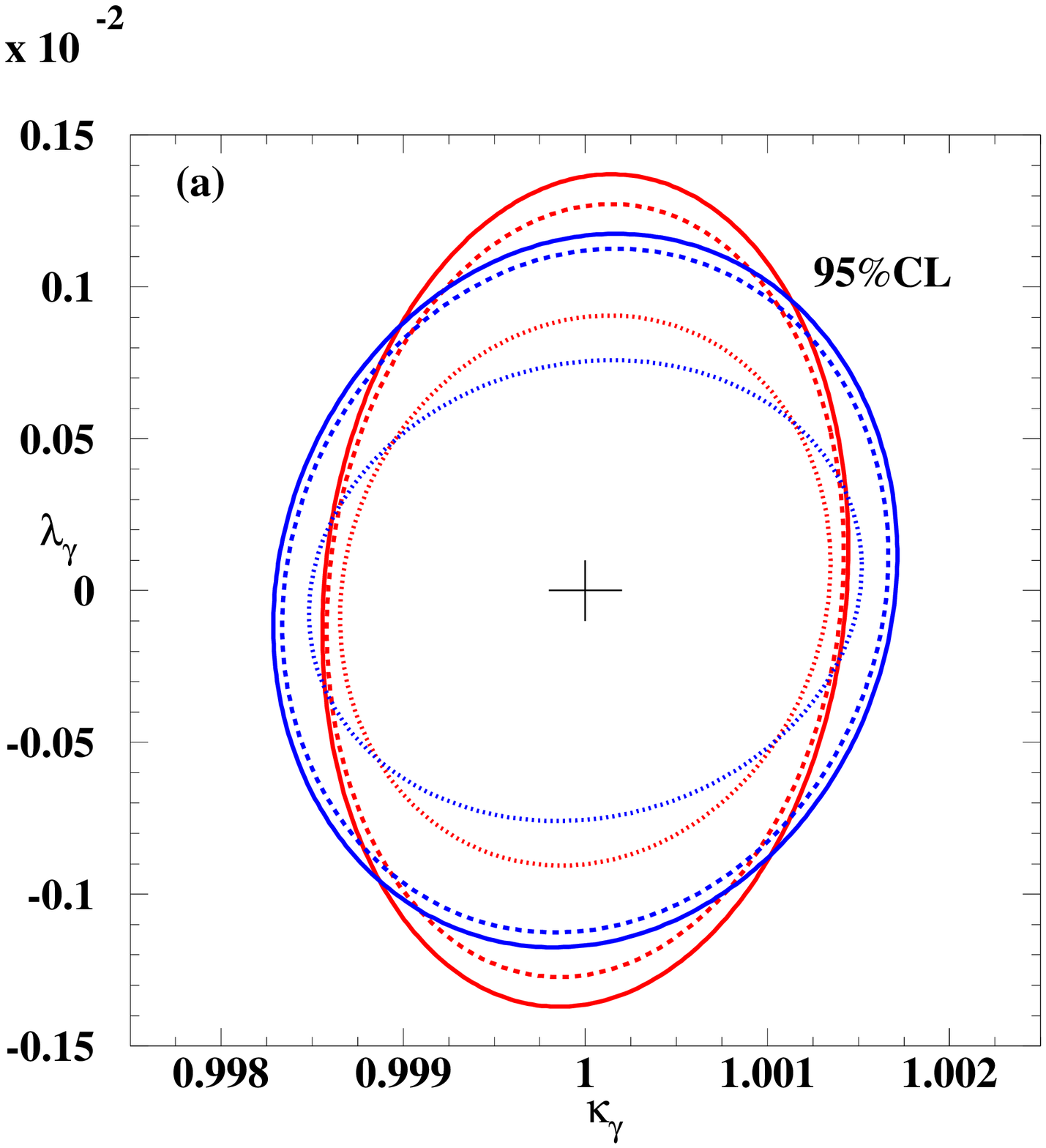}
\epsfxsize=3.0in
\epsfysize=3.0in
\epsfbox{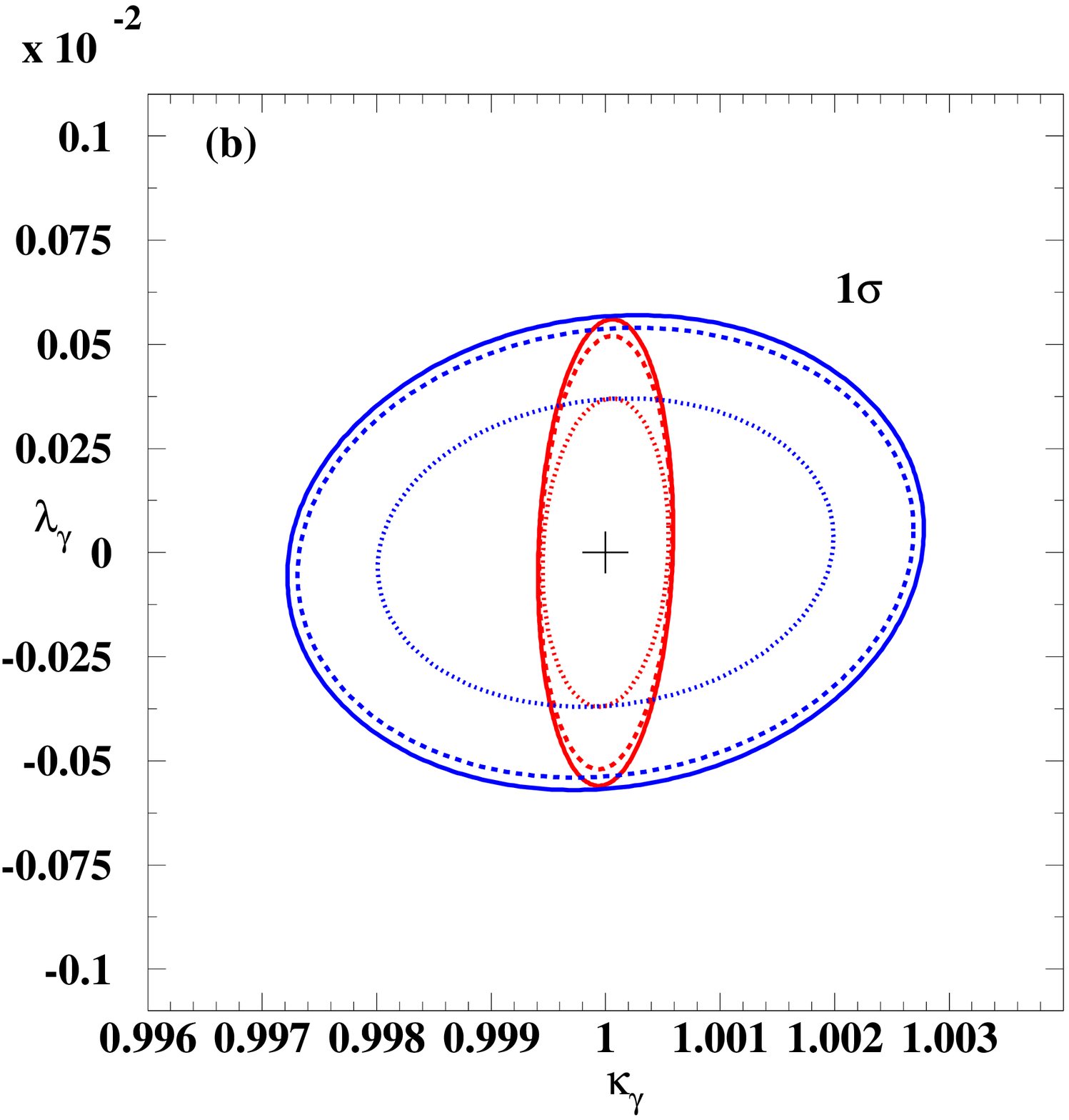}
\caption[bla]{(\textit{a}): 95$\%$ CL contours in the $\kappa_{\gamma}-\lambda_{\gamma}$ plane obtained from the 6D fit for ${{\Delta}L}=0.1\%$ for the $J_{Z}=0$ state (red) and $|J_{Z}|=2$ state (blue). Dotted lines correspond to the events without pileup, dashed lines correspond to the events with pileup and solid lines correspond to the events with pileup and background. The cross denotes the Standard Model values of $\kappa_{\gamma}$ and $\lambda_{\gamma}$. (\textit{b}): 1 $\sigma$ contours in the $\kappa_{\gamma}-\lambda_{\gamma}$ plane obtained from the 6D fit for ${{\Delta}L}=1\%$ (blue) and ${{\Delta}L}=0.1\%$ (red) for the $J_{Z}=0$ state.}
\label{fig:contour2}
\end{center}
\end{figure}
\begin{table}[h]
\begin{center}
\begin{tabular}{|l||c|c|c||c|c|c|} \hline
\multicolumn{1}{|c||}{6D fit} & \multicolumn{6}{|c|}{pileup+background} \\ \hline \hline
\multicolumn{1}{|c||}{} & \multicolumn{3}{|c||}{$J_{Z}=0$} & \multicolumn{3}{|c|}{$J_{Z}=2$} \\ \hline \hline
& ${\Delta}{\kappa}_{\gamma}$ & $\textit{n}$ & ${\Delta}{\lambda}_{\gamma}$ & ${\Delta}{\kappa}_{\gamma}$ & $\textit{n}$ & ${\Delta}{\lambda}_{\gamma}$ \\ \hline \hline
${\Delta}{\kappa}_{\gamma}$ & 1.000 & -0.994 & -0.173 & 1.000 & -0.719 & 0.554 \\ \hline
$\textit{n}$ & -0.994 & 1.000 & 0.225 & -0.719 & 1.000 & 0.010 \\ \hline
${\Delta}{\lambda}_{\gamma}$ & -0.173 & 0.225 & 1.000 & 0.554 & 0.010 & 1.000 \\ \hline
\end{tabular}
\end{center}
\caption{Correlation matrix for the two-parameter fit in the $J_{Z}=0$ state (${{\Delta}L}=1\%$) and in the $|J_{Z}|=2$ state (${{\Delta}L}=0.1\%$).}
\label{tab:set7}
\end{table}
%
\section{Systematic Errors}
Due to the large $W$ boson production cross-sections and achievable luminosities at the photon colliders the statistical errors are comparable with those estimated for the $e^{+}e^{-}$-collider and the main source of error comes from the systematics. Some sources of systematic errors have been investigated, assuming ${{\Delta}L}=0.1\%$ and in some cases ${{\Delta}L}=1\%$. It was found that the largest uncertainty in ${\kappa}_{\gamma}$ comes from uncertainties on the photon beam polarizations. Contrary to the $e^+ e^-$ case the luminosity and polarization measurements are not independent. In $\gamma e$ collisions, the dominant polarization state ($|J_Z| = 3/2$) can be measured accurately with $\gamma e \rightarrow e \gamma$ while the suppressed one ($|J_Z| = 1/2$) can only be measured with worse precision e.g. using $e Z \rightarrow e Z$ \cite{lumi}. To estimate the uncertainty on the TGCs therefore the dominant $|J_{Z}| = 3/2$ part is kept constant while the $|J_Z| = 1/2$ part is changed by 10\%, corresponding to a 1\% polarization uncertainty for ${\cal P}_\gamma=0.9$. This leads to a polarization uncertainty of $0.005$ for $\kappa_\gamma$, corresponding to five times the statistical error while the uncertainty on $\lambda_\gamma$ is negligible. The photon polarization thus needs to be known to 0.1$\%$ - 0.2$\%$ so that ${\kappa}_{\gamma}$ is not dominated by this systematic error. 
\par
In $\gamma\gamma$ collisions, having ${\cal P}_\gamma=0.9$, the dominant $J_{Z}=0$ or $|J_{Z}|=2$ part is kept constant while the $|J_{Z}|=2$ or $J_{Z}=0$ part is changed resulting in a 1\% polarization uncertainty, in the same way as for the $\gamma e$ mode. In the $J_{Z}=0$ state, the polarization uncertainty of $0.0021$ for $\kappa_\gamma$ corresponds to less than one statistical error assuming ${{\Delta}L}=1\%$ while in the $|J_{Z}|=2$ state, the polarization uncertainty of $0.0018$ for $\kappa_\gamma$ corresponds to less than three times the statistical error assuming ${{\Delta}L}=0.1\%$. The uncertainty on $\lambda_\gamma$ in both $J_{Z}$ states is negligible.
\par
In order to estimate the error coming from the $W$ boson mass measurement the data sample is recalculated with ${M_{W}}$ decreased/increased by $50\,{\rm MeV}$ (the expected ${\Delta}{M_{W}}$ at LHC is $\sim 15$ MeV) reweighting the Standard Model events. The nominal $W$ boson mass used for the Monte Carlo sample was $M_{W}=80.419\,{\rm GeV}$. As a result of the recalculation we get the ratios of matrix element values corresponding to the nominal $W$ boson mass and the mass $M_{W}^{'}= M_W \pm \Delta M_{W}$. The Monte Carlo sample (MC) is weighted by this ratio and fitted as fake data leaving the reference distributions unchanged. The resulting shift for TGCs is of the order of the statistical error for both coupling parameters for $\Delta M_W = 50\,{\rm MeV}$ and thus negligible with an improved $W$ boson mass measurement. 
\par
The nonlinear QED effects at the conversion region influence the Compton spectra of the scattered photons in such a way that increasing the nonlinearity $\xi^{2}$ the Compton spectrum becomes broader and shifted to lower energies. To estimate the error that comes from this effect the laser power is decreased changing $\xi^{2}$ from 0.3 to 0.15, increasing the peak energy by 2.5\%.  The ratio of the two Compton spectra is used as a weight function to obtain the ``data'' sample from the MC events.  The sample data obtained in that way are fitted leaving the reference distributions unchanged. It was found that the beam energy uncertainty influences the measurement of the coupling parameters only via the normalization $n$, and the errors ${\Delta}{\kappa}_{\gamma}$ and ${\Delta}{\lambda}_{\gamma}$ are considered as negligible since the value of $n$ is accessible from the luminosity measurement.
\par
The estimated systematic error for $\kappa_\gamma$ from background uncertainties is smaller than the statistical error if the background cross section is known to better than 3\% in the real $\gamma e$ mode and 1\% in the parasitic $\gamma e$ mode. For $\lambda_\gamma$ the background needs to be known only to 10\% in the parasitic $\gamma e$ mode while there are practically no restrictions in the real $\gamma e$ mode. In the $|J_{Z}|=2$ state the background cross section should be known better than 0.8\% i.e. better than 4\% in order to provide a systematic error smaller than the statistical error for $\kappa_\gamma$ i.e. for $\lambda_\gamma$, respectively. For $J_{Z}=0$ state, assuming ${{\Delta}L}=1\%$, the background cross section should be known better than 0.6\% in order to get the systematic error of $\lambda_\gamma$ measurement smaller than the statistical one. For $\kappa_\gamma$ estimations there are no restrictions on the background cross section due to the large error. Assuming ${{\Delta}L}=0.1\%$ in $J_{Z}=0$ state, the $\kappa_\gamma$ measurement gives a smaller systematic error than the statistical one if the background cross section is known better than 1.1\% while for $\lambda_\gamma$ it still should be known below 0.6\% (Table \ref{tab:set8}).
\begin{table}[h]
\begin{center}
\begin{tabular}{|l||c|c|c|c||c|c||c|} \hline
\multicolumn{1}{|c||}{} & \multicolumn{1}{|c|}{real $\gamma e$ mode} & \multicolumn{1}{|c|}{parasitic $\gamma e$ mode} &\multicolumn{2}{|c||}{$J_{Z}=0$} & \multicolumn{1}{|c|}{$|J_{Z}|=2$} \\ \hline \hline
$\Delta L$ & 0.1$\%$ & 0.1$\%$ & 1$\%$ & 0.1$\%$ & 0.1$\%$ \\ \hline \hline
$\kappa_{\gamma}$ & 3$\%$ & 1$\%$ & /$^{*}$ & 1.1$\%$ & 0.8$\%$ \\ \hline
$\lambda_{\gamma}$ & /$^{*}$ & 10$\%$ & 0.6$\%$ & 0.6$\%$ & 4$\%$ \\ \hline
\end{tabular}
\end{center}
\caption{Maximal allowed errors on background cross sections for estimation of the systematic errors on $\kappa_\gamma$ and $\lambda_\gamma$ measurements less than one statistical error in $\gamma e$ and $\gamma\gamma$ collisions. ($^{*}$) means that the estimated systematic errors are below one statistical error if the precision of the background cross section exceeds the level of 10\%.}
\label{tab:set8}
\end{table}

\chapter{Summary and Conclusions}
The sensitivity estimation of the measurement of the trilinear gauge couplings $\kappa_{\gamma}$ and $\lambda_{\gamma}$ at the International Linear Collider based on TESLA parameters has been studied. The considered interactions are the single $W$ boson production in $\gamma e\rightarrow W^{-}\nu_{e}$ ($W\rightarrow q\bar{q}^{'}$) and the $W$ boson pair production in $\gamma\gamma\rightarrow W^{+}W^{-}$ ($WW\rightarrow q\bar{q}^{'}q^{'}\bar{q}$) at $\sqrt{s_{e^{-}e^{-}}}=500$ GeV, where $\kappa_{\gamma}$ and $\lambda_{\gamma}$ characterize the $WW\gamma$ vertices with values $\kappa_{\gamma}=1$ and $\lambda_{\gamma}=0$ predicted by the Standard Model. Due to the possibility to produce high energetic and highly polarized photons, the two different initial states in $\gamma e$ and $\gamma\gamma$ collisions are available for a study of the TGCs in both channels: $|J_{Z}|=1/2$ or $3/2$ in $\gamma e$ collisions and $J_{Z}=0$ or $2$ in $\gamma\gamma$ collisions. In case of $\gamma e$ collisions the more sensitive initial state $|J_{Z}|=3/2$ is analyzed in two modes - the real and the parasitic. The hadronically decaying Standard Model signal and background events, overlayed with the pileup, are simulated and analyzed combining the different software packages developed for both, TESLA $e^{+}e^{-}$-collider and for the TESLA extension, the photon collider. To minimize the background contribution the set of consecutive cuts is applied in each production channel resulting in signal efficiencies of 70$\%$ in the real $\gamma e$ mode, 63$\%$ in the parasitic $\gamma e$ mode, 53$\%$ in the $|J_{Z}|=2$ state and 52$\%$ in the $J_{Z}=0$ state. The efficiencies for the $W$ boson pair production are somewhat less than for the single $W$ boson production due to the larger contribution of the pileup events remaining after rejection cuts. The final ratios of the signal to background events are $N_{S}/N_{B}=19$ in the real $\gamma e$ mode, $N_{S}/N_{B}=2.6$ in the parasitic $\gamma e$ mode and $N_{S}/N_{B}=4.3$ in $J_{Z}=0,2$ states. Selected Standard Model signal events are described with three (single $W$ boson events) or five ($W$ boson pair events) kinematical variables sensitive to $\kappa_{\gamma}$ and $\lambda_{\gamma}$, mainly with different angular $W$ boson distributions. A new method has been applied to simulate event angular distributions in the presence of anomalous couplings, $\Delta\kappa_{\gamma}\neq 0$ and $\Delta\lambda_{\gamma}\neq 0$, expanding the dependence of the differential cross-section up to quadratic term in the TGCs. This expression is used to define weighting function. The reweighted Standard Model multi-dimensional binned distributions dependent on $\Delta\kappa_{\gamma}$ and $\Delta\lambda_{\gamma}$ are fitted to the distributions predicted by the Standard Model including the error on the luminosity measurement $\Delta L/L$ via the normalization as a third fit parameter. 
\par
The sensitivities of the TGC measurement estimated by a $\chi^{2}$ fit in both $\gamma e$ modes at $\sqrt{s_{e^{-}e^{-}}}=500$ GeV with the integrated luminosities ${\cal L}_{\gamma e}^{real}\approx 160$ fb$^{-1}$ and ${\cal L}_{\gamma e}^{parasitic}\approx 230$ fb$^{-1}$, including the background events, are of order $\approx 1\cdot 10^{-3}$ for $\Delta\kappa_{\gamma}$ and $\approx 5-7\cdot 10^{-4}$ for $\Delta\lambda_{\gamma}$ assuming $\Delta L/L\approx 10^{-3}$. The sensitivity of the TGC measurement estimated by a Likelihood fit in both $\gamma\gamma$ initial states at $\sqrt{s_{e^{-}e^{-}}}=500$ GeV with the integrated luminosities ${\cal L}_{\gamma\gamma}\approx 1000$ fb$^{-1}$, including the background events, is of order $\approx 7\cdot 10^{-4}$ for $\Delta\kappa_{\gamma}$ and higher than $5\cdot 10^{-4}$ for $\Delta\lambda_{\gamma}$ in the $|J_{Z}|=2$ state assuming $\Delta L/L\approx 10^{-3}$. The state $J_{Z}=0$ takes into account a larger error on the luminosity measurement of $\Delta L/L\approx 10^{-2}$ resulting in a sensitivity to $\kappa_{\gamma}$ higher than $3\cdot 10^{-3}$ and to $\lambda_{\gamma}$ higher than $6\cdot 10^{-4}$. The luminosities are integrated over the whole energy spectrum and correspond to one year (10$^{7}$s) of running of an $\gamma e/\gamma\gamma$-collider.
\par
Some sources of systematic errors are considered, as the beam energy, the $W$ boson mass and the polarization measurement. While the TGC measurement in both $\gamma\gamma$ modes is not sensitive to previous uncertainties, the $\kappa_{\gamma}$ measurement in the $\gamma e$ mode is mainly influenced by the beam polarization. The influence of the background is different for each mode and coupling but in general, the knowledge about the background cross-section should be on the level of a few percents.
\par
The estimated precisions of the $\kappa_{\gamma}$ and $\lambda_{\gamma}$ measurements at a photon collider of $10^{-3}-10^{-4}$ are about one to two orders of magnitude higher than at LEP and Tevatron providing a measurement highly sensitive to the physics beyond the Standard Model. The comparison to $\kappa_{\gamma}$ and $\lambda_{\gamma}$ estimated at the TESLA $e^{+}e^{-}$-collider \cite{menges} at $\sqrt{s_{e^{-}e^{-}}}=500$ GeV is shown in Table \ref{tab:compare500} and at $\sqrt{s_{e^{+}e^{-},\gamma\gamma}}=800$ GeV is shown in Table \ref{tab:compare800}.
\begin{table}[htb]
\begin{center}
\begin{tabular}{|c||c|c||c|c||c|} \hline
 & \multicolumn{2}{|c||}{$\gamma e$} & \multicolumn{2}{|c||}{$\gamma\gamma$} & $e^{+}e^{-}$ \\ \hline\hline
 Mode & Real $|J_{Z}|=3/2$ & Parasitic $|J_{Z}|=3/2$ & $|J_{Z}|=2$ & $J_{Z}=0$ & $|J_{Z}|=1$\\ \hline
 $\int{\cal L}\Delta t$ & 160 fb$^{-1}$ & 230 fb$^{-1}$ & \multicolumn{2}{|c||}{1000 fb$^{-1}$} & 500 fb$^{-1}$\\ \hline
${{\Delta}L}$ & \multicolumn{2}{|c||}{0.1$\%$} & 0.1$\%$ & 1$\%$ & - \\ \hline\hline
${\Delta}{\kappa}_{\gamma}{\cdot}10^{-4}$ & 10.0 & 11.0 & 7.0 & 27.8 & 3.6$^*$ \\ \hline
${\Delta}{\lambda}_{\gamma}{\cdot}10^{-4}$ & 4.9 & 6.7 & 4.8 & 5.7 & 11.0$^*$ \\ \hline
\end{tabular}
\end{center}
\caption{Comparison of the $\kappa_{\gamma}$ and $\lambda_{\gamma}$ sensitivities at $\gamma e$-, $\gamma\gamma$- and $e^{+}e^{-}$-colliders estimated at $\sqrt{s_{e^{-}e^{-}}}=500$ GeV using polarized beams. In case of photon colliders, the background and the pileup are included. ($^*$) denotes estimations at the generator level.}
\label{tab:compare500}
\end{table}
\begin{table}[htb]
\begin{center}
\begin{tabular}{|c||c|c||c|} \hline
$\sqrt{s_{e^{+}e^{-},\gamma\gamma}}=800$ GeV & \multicolumn{2}{|c||}{$\gamma\gamma$} & $e^{+}e^{-}$ \\ \hline\hline
Mode & $|J_{Z}|=2$ & $J_{Z}=0$ & $|J_{Z}|=1$ \\ \hline
$\int{\cal L}\Delta t$ & \multicolumn{2}{|c||} {1000 fb$^{-1}$} & 1000 fb$^{-1}$\\ \hline
${{\Delta}L}$ & 0.1$\%$ & 1$\%$ & - \\ \hline\hline
${\Delta}{\kappa}_{\gamma}{\cdot}10^{-4}$ & 5.2 & 13.9 & 2.1$^*$ \\ \hline
${\Delta}{\lambda}_{\gamma}{\cdot}10^{-4}$ & 1.7 & 2.5 & 3.3$^*$ \\ \hline
\end{tabular}
\end{center}
\caption{Comparison of the $\kappa_{\gamma}$ and $\lambda_{\gamma}$ sensitivities at $\gamma\gamma$- and $e^{+}e^{-}$-colliders estimated at $\sqrt{s_{e^{+}e^{-},\gamma\gamma}}=800$ GeV using polarized beams. ($^*$) denotes estimations at the generator level. The sensitivities at a $\gamma\gamma$-collider are scaled for the background, pileup and the energy spectrum.}
\label{tab:compare800}
\end{table}
\par
The sensitivities to $\kappa_{\gamma}$ and $\lambda_{\gamma}$ in $\gamma\gamma\rightarrow W^{+}W^{-}$ at $\sqrt{s_{e^{-}e^{-}}}=1$ TeV including the variable energy spectrum, background and pileup events (Table \ref{tab:compare800}) are approximated scaling the estimated sensitivities at generator level (Table \ref{tab:bestgg}) by a factor obtained for $\sqrt{s_{e^{-}e^{-}}}=500$ GeV. Increasing the center-of-mass energy the sensitivity to $\kappa_{\gamma}$ increases by 25$\%$ and to $\lambda_{\gamma}$ by 65$\%$ in the $|J_{Z}|=2$ state, to $\kappa_{\gamma}$ by 50$\%$ and to $\lambda_{\gamma}$ by 56$\%$ in the $J_{Z}=0$ state. Since the influence of the considered systematics is negligible on ${\lambda}_{\gamma}$, a photon collider is a better place than an $e^{+}e^{-}$-collider for its accurate measurement. Thus, both collider types are complementary for the TGC measurements and in searches for the resonances within the strong EWSB scenario.  
\par
The optimization of the forward region of the $\gamma\gamma$-detector brings the amount of the low-energy background to the level of $e^{+}e^{-}$-collider providing a clean environment for the TGC measurements at a photon collider with previously estimated precisions.

\begin{appendix}
\chapter{$W$ boson Decay Functions}
\begin{equation}
\begin{array}{lcl}
D_{++}^{(s)}=\frac{1}{2}(1+\cos^{2}\theta), & & D_{++}^{(a)}=-\cos\theta; \\
D_{--}^{(s)}=\frac{1}{2}(1+\cos^{2}\theta), & & D_{--}^{(a)}=\cos\theta ; \\
D_{00}^{(s)}=\sin^{2}\theta, & & D_{00}^{(a)}=0; \\
D_{+-}^{(s)}=\frac{1}{2}(\sin^{2}\theta)e^{+2i\phi}, & & D_{+-}^{(a)}=0; \\
D_{+0}^{(s)}=\frac{1}{\sqrt{2}}(\cos\theta\sin\theta)e^{+i\phi}, & & D_{+0}^{(a)}=-\frac{1}{\sqrt{2}}\sin\theta e^{-i\phi}; \\
D_{-0}^{(s)}=-\frac{1}{\sqrt{2}}(\cos\theta\sin\theta)e^{-i\phi}, & & D_{-0}^{(a)}=-\frac{1}{\sqrt{2}}\sin\theta e^{-i\phi};
\label{eq:app1}
\end{array}{}
\end{equation}
\begin{figure}[htb]
\begin{center}
\epsfxsize=5.0in
\epsfysize=2.5in
\epsfbox{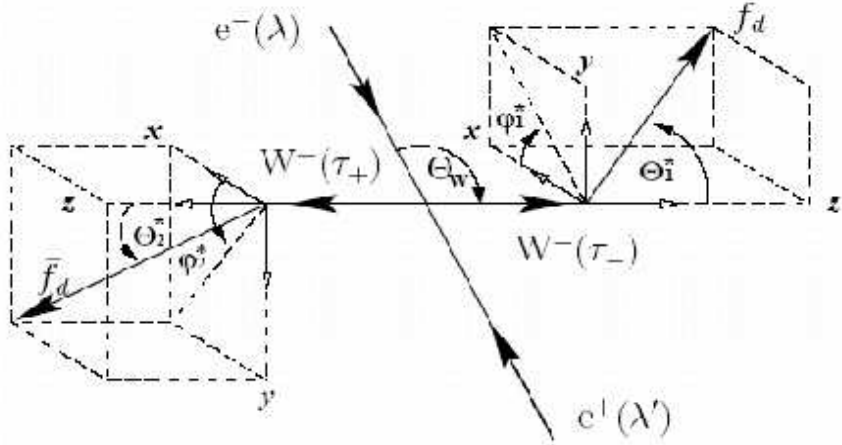}
\caption[bla]{Definition of angles and helicities in $e^{+}e^{-}\rightarrow W^{+}W^{-}$ event.}
\label{fig:topology}
\end{center}
\end{figure}
\chapter{TGC estimation in single $W$ boson production from an analytic formula}
As an alternative approach the two-dimensional event distribution over the production and decay angle of the $W$ boson is obtained using the analytical formula from \cite{denner} and \cite{yehu}. Production and decay angle event distributions obtained for a luminosity of $\sim110\,{{\rm fb}^{-1}}$ are divided into ${10}\times{10}$ bins respectively for the parasitic mode and ${20}\times{10}$ bins for the real $\gamma e$ mode and fitted with MINUIT, minimizing the ${\chi^{2}}$ function:
$$
\chi^{2}= \sum_{i=1}^{n} \sum_{j=1}^{n}\frac{\left( N^{SM}(i,j)- n \cdot
N^{{\Delta\kappa}_{\gamma},{\Delta\lambda}_{\gamma}}(i,j) \right)^2}
{\sigma^{2}(i,j)}+\frac{{(n-1)^{2}}}{({\Delta}L^{2})}
$$
where \textit{i} and \textit{j} run over the bins of the two-dimensional distribution. In Table \ref{tab:t5} we present the estimations of ${\Delta}{\kappa}_{\gamma}$ and ${\Delta}{\lambda}_{\gamma}$ obtained from the two-parameter fit for data generated from the analytical formula for the real $\gamma e$ mode. They agree reasonably well with the corresponding results from the reweighting fit (left block of table \ref{tab:bestge}). Increasing the integrated luminosity to 170 fb$^{-1}$, an improvement of $\approx$ 20$\%$ in $\Delta\lambda_{\gamma}$ is possible while $\Delta\kappa_{\gamma}$ is not sensitive, assuming $\Delta L=0.1\%$.
\begin{table}[h]
\begin{center}
\begin{tabular}{|l||c|c|c||c|c|c|} \hline
\multicolumn{1}{|c||}{} & \multicolumn{6}{|c|}{$\sqrt{s_{\gamma e}}=450$ GeV} \\ \hline
\multicolumn{1}{|c||}{2D fit} & \multicolumn{3}{|c||}{110 fb$^{-1}$} & \multicolumn{3}{|c|}{170 fb$^{-1}$} \\ \hline
${{\Delta}L}$ & 1$\%$ & 0.1$\%$ & 0 & 1$\%$ & 0.1$\%$ & 0  \\ \hline\hline
${\Delta}{\kappa}_{\gamma}{\cdot}10^{-3}$ & 4.9 & 1.1 & 0.4 & 4.2 & 1.0 & 0.4 \\ \hline
${\Delta}{\lambda}_{\gamma}{\cdot}10^{-4}$ & 22 & 16 & 15 & 18 & 13 & 12 \\ \hline
\end{tabular}
\end{center}
\caption{Estimated statistical errors for ${\kappa}_{\gamma}$ and ${\lambda}_{\gamma}$ for the real $\gamma e$ mode at $\sqrt{s_{\gamma e}}=450$ GeV, assuming 100$\%$ detector acceptance, from the two-parameter 2D fit of the data generated using the analytical formula. The number of events are normalized to the expected ones with integrated luminosities of 110 and 170 fb$^{-1}$ in the high energy peak.}
\label{tab:t5}
\end{table}
\end{appendix}

\boldmath

\end{document}